\documentclass[11pt,oneside]{book}
\usepackage{graphics}
\usepackage{makeidx}
\usepackage{nomencl}
\usepackage{xspace}
\usepackage{enumerate}
\usepackage{amssymb}
\usepackage{amsmath}
\usepackage{amscd}
\usepackage{psfrag}
\usepackage{latexsym}
\usepackage[dvips]{graphicx}
\usepackage{amsthm}
\usepackage[T2A,OT1]{fontenc}
\usepackage[ot2enc]{inputenc}
\usepackage[russian,english]{babel}
\usepackage{version}
\usepackage{indentfirst}
\usepackage{color}
\usepackage{changebar}
\definecolor{pink}{rgb}{1,0.5,0.5}
\usepackage[top=1cm, bottom=1cm, left=4cm, right=3cm]{geometry}

\newif\ifadjust  \adjustfalse

\newtheorem{thm}{Theorem}[section]

\theoremstyle{definition}	

\newtheorem{defn}[thm]{Definition}

\newcommand{\ez}{EZ-Freeze }

\newcommand{\dd}{\textrm{d}}
\newcommand{\dT}{\textrm{T}(g)}

\newcommand{\dTi}{\textrm{T}(g^{-1})}

\newcommand{\balp}{\boldsymbol{\alpha}}
\newcommand{\bbet}{\boldsymbol{\beta}}

\newcommand{\bmu}{\boldsymbol{\mu}}
\newcommand{\bphi}{\boldsymbol{\phi}}

\newcommand{\bpsi}{\boldsymbol{\psi}}
\newcommand{\bPsi}{\boldsymbol{\Psi}}

\newcommand{\bA}{\textbf{A}}
\newcommand{\ba}{\textbf{a}}
\newcommand{\bB}{\textbf{B}}
\newcommand{\bD}{\textbf{B}}
\newcommand{\bc}{\textbf{c}}
\newcommand{\bC}{\textbf{C}}
\newcommand{\bDD}{\textbf{D}}
\newcommand{\be}{\textbf{e}}
\newcommand{\bof}{\textbf{f}}
\newcommand{\boof}{\textbf{f}}
\newcommand{\booF}{\textbf{F}}
\newcommand{\boF}{\textbf{F}(\textbf{V})}
\newcommand{\boFU}{\textbf{F}(\textbf{U})}

\newcommand{\bFg}{\textbf{F}_{\mathcal{G}}(\bV)}
\newcommand{\bFm}{\textbf{F}_{\mathcal{M}}(\bV)}
\newcommand{\bg}{\textbf{g}}
\newcommand{\bG}{\textbf{G}}

\newcommand{\bh}{\textbf{h}}
\newcommand{\bH}{\textbf{H}}
\newcommand{\bHg}{\til{\bH}_{\mathcal{G}}(\bV,t)}
\newcommand{\bHm}{\til{\bH}_{\mathcal{M}}(\bV,t)}
\newcommand{\bI}{\textbf{I}}
\newcommand{\bk}{\textbf{k}}
\newcommand{\bM}{\textbf{M}}

\newcommand{\bn}{\textbf{n}}
\newcommand{\bP}{\textbf{P}}
\newcommand{\bp}{\textbf{p}}
\newcommand{\bq}{\textbf{q}}
\newcommand{\bQ}{\textbf{Q}}
\newcommand{\br}{\textbf{r}}
\newcommand{\bR}{\textbf{R}}
\newcommand{\bS}{\textbf{S}}
\newcommand{\bu}{\textbf{u}}
\newcommand{\bU}{\textbf{U}}
\newcommand{\bv}{\textbf{v}}
\newcommand{\bV}{\textbf{V}}
\newcommand{\bw}{\textbf{w}}
\newcommand{\bW}{\textbf{W}}
\newcommand{\bx}{\textbf{x}}

\newcommand{\til}{\tilde}

\newcommand{\Bh}{\hat{\partial}}
\newcommand{\hh}{\hat{\textbf{h}}}
\newcommand{\bHt}{\tilde{\textbf{H}}(\textbf{V},t)}

\newcommand{\bht}{\tilde{\textbf{h}}}
\newcommand{\btr}{\tilde{\textbf{r}}}

\newcommand{\tr}{\tilde{r}}
\newcommand{\tbr}{\tilde{\textbf{r}}}

\newcommand{\pderiv}[2]{\frac{\partial #1}{\partial #2}} 
\newcommand{\deriv}[2]{\frac{\dd #1}{\dd #2}}

\newcommand{\odRt}{\frac{\dd{R}}{\dd{t}}}
\newcommand{\odbRt}{\frac{\dd{\bR}}{\dd{t}}}
\newcommand{\odTt}{\frac{\dd{\dT}}{\dd{t}}}
\newcommand{\odTht}{\frac{\dd{\Theta}}{\dd{t}}}
\newcommand{\odThot}{\frac{\dd{\Theta_0}}{\dd{t}}}
\newcommand{\odThonet}{\frac{\dd{\Theta_1}}{\dd{t}}}

\newcommand{\odut}{\frac{\dd{\bU}}{\dd{t}}}

\newcommand{\odbVt}{\frac{\dd{\bV}}{\dd{t}}}
\newcommand{\odbVot}{\frac{\dd{\bV_0}}{\dd{t}}}
\newcommand{\odbVpt}{\frac{\dd{\bV_1}}{\dd{t}}}
\newcommand{\odvt}{\frac{\dd{\bV}}{\dd{t}}}

\newcommand{\odbVpit}{\frac{\dd{\bV}_{1pi}}{\dd{t}}}

\newcommand{\odbWt}{\frac{\dd{\bW}}{\dd{t}}}
\newcommand{\odXt}{\frac{\dd{X}}{\dd{t}}}
\newcommand{\odYt}{\frac{\dd{Y}}{\dd{t}}}

\newcommand{\pdbvx}{\frac{\partial{\textbf{v}}}{\partial{x}}}

\newcommand{\pdux}{\frac{\partial{u}}{\partial{x}}}
\newcommand{\pdvx}{\frac{\partial{v}}{\partial{x}}}

\newcommand{\pdvox}{\frac{\partial{\textbf{v}_0}}{\partial{x}}}
\newcommand{\pdbvox}{\frac{\partial{\textbf{v}_0}}{\partial{x}}}
\newcommand{\pdbtvox}{\frac{\partial{\til{\textbf{v}}_0}}{\partial{x}}}

\newcommand{\pdbvotx}{\frac{\partial{\textbf{v}_0}}{\partial{\til{x}}}}
\newcommand{\pdvoxx}{\frac{\partial^2{\textbf{v}_0}}{\partial{x}^2}}

\newcommand{\pdtxx}{\frac{\partial{\tilde{x}}}{\partial{x}}}
\newcommand{\pdtyx}{\frac{\partial{\tilde{y}}}{\partial{x}}}

\newcommand{\pdvoy}{\frac{\partial{\textbf{v}_0}}{\partial{y}}}
\newcommand{\pdbvoy}{\frac{\partial{\textbf{v}_0}}{\partial{y}}}
\newcommand{\pdbvoty}{\frac{\partial{\textbf{v}_0}}{\partial{\til{y}}}}

\newcommand{\pdvoyy}{\frac{\partial^2{\textbf{v}_0}}{\partial{y}^2}}

\newcommand{\pdvoth}{\frac{\partial{\textbf{v}_0}}{\partial{\theta}}}

\newcommand{\pdvothth}{\frac{\partial^2{\textbf{v}_0}}{\partial{\theta}^2}}

\newcommand{\pdbut}{\frac{\partial{\textbf{u}}}{\partial{t}}}
\newcommand{\pdvot}{\frac{\partial{\textbf{v}_0}}{\partial{t}}}

\newcommand{\pdut}{\frac{\partial{\textbf{u}}}{\partial{t}}}
\newcommand{\pdvt}{\frac{\partial{\textbf{v}}}{\partial{t}}}

\newcommand{\pdfa}{\frac{\partial{\textbf{f}}}{\partial{\alpha}}}

\newcommand{\pdfvov}{\frac{\partial{\textbf{f}(\textbf{v}_0)}}{\partial{\textbf{v}}}}
\newcommand{\pdgu}{\frac{\partial{\textbf{g}}}{\partial{\bV}}}

\newcommand{\pdvoxy}{\frac{\partial^2{\textbf{v}_0}}{\partial{x}\partial{y}}}

\newcommand{\pdvoyx}{\frac{\partial^2{\textbf{v}_0}}{\partial{y}\partial{x}}}

\newcommand{\pdvothx}{\frac{\partial^2{\textbf{v}_0}}{\partial{\theta}\partial{x}}}
\newcommand{\pdvothy}{\frac{\partial^2{\textbf{v}_0}}{\partial{\theta}\partial{y}}}

{
\newcounter{num}
\newcommand{\AF}[3]{%
  \refstepcounter{num}%
  \label{#1}
  \par\vspace*{2ex}\par
  \noindent\textbf{\color{red}AF.\thenum}\ p.#2 \textit{#3} \\ --- 
}

{
\newcounter{numex}
\newcommand{\EX}[2]{%
  \refstepcounter{numex}%
  \label{#1}
  \par\vspace*{2ex}\par
  \noindent\textbf{\color{red}EX.\thenumex} \textit{#2} \\ --- 
}

\ifadjust
\newcommand{\chg}[2][0]{{%
  \ifx0#1%
    \def\chgmk{{\color{red}\textrm{?}}}\color{red}%
  \else%
    \def\chgmk{{\color{blue}\textrm{\scriptsize{AF.\ref{#1}}}}}\color{blue}%
  \fi%
  \def\chgemptystring{}%
  \ifx#1\chgemptystring%
  \else%
    $^{\colorbox{yellow}{\chgmk}}$%
  \fi
  #2
}}
\else
\newcommand{\chg}[2][0]{#2}
\fi

\ifadjust
\newcommand{\chgex}[2][0]{{%
  \ifx0#1%
    \def\chgmk{{\color{red}\textrm{?}}}\color{red}%
  \else%
    \def\chgmk{{\color{blue}\textrm{\scriptsize{EX.\ref{#1}}}}}\color{blue}%
  \fi%
  \def\chgemptystring{}%
  \ifx#1\chgemptystring%
  \else%
    $^{\colorbox{green}{\chgmk}}$%
  \fi
  #2
}}
\else
\newcommand{\chgex}[2][0]{#2}
\fi

\newcommand{\panel}[1]{\shortstack[c]{%
      \includegraphics[width=0.2\linewidth]{#1}
    }}

\excludeversion{ajf}
\includeversion{ajfthesis}

\textheight=225mm
\textwidth=145mm
\topmargin=0mm
\headheight=0mm
\headsep=10mm

\oddsidemargin=15mm
\evensidemargin=15mm

\begin{document}

\baselineskip = 17pt

\title{Drift And Meander Of Spiral Waves}
\author{Thesis submitted in accordance with the requirements of \\
the University of Liverpool for the degree of Doctor in Philosophy \\
by \\
\\
Andrew J. Foulkes}
\date{March 2009}
\maketitle 
\frontmatter

\chapter{Abstract}

In this thesis, we are concerned with the dynamics of spiral wave solutions to Reaction-Diffsion systems of equations, and how they behave when subject to symmetry breaking perturbations. 

We present an asymptotic theory of the study of meandering (quasiperiodic spiral wave solutions) spiral waves which are drifting due to symmetry breaking perturbations. This theory is based on earlier theories: the 1995 Biktashev et al theory of drift of rigidly rotating spirals \cite{bik95}, and the 1996 Biktashev et al theory of meander of spirals in unperturbed systems \cite{bik96}. We combine the two theories by first rewriting the 1995 drift theory using the symmetry quotient system method of the 1996 meander theory, and then go on to extend the approach to meandering spirals by considering Floquet theory and using a singular perturbation method. We demonstrate the work of the newly developed theory on simple examples. 

We also develop a numerical implementation of the quotient system method, demonstrate its numerical convergence and its use in calculations which would be difficult to do by the standard methods, and also link this study to the problem of calculation of response functions of spiral waves.

\chapter{Declaration}

No part of the work referred to in this Thesis have been submitted in support of an
application for another degree or qualification of this or any other institution of learning.
However, some parts of the material contained herein have been previously published.

\tableofcontents
\addcontentsline{toc}{chapter}{Contents}

\listoffigures
\addcontentsline{toc}{chapter}{List of Figures}

\chapter{Acknowledgements}

There are many people who have supported me throughout my studies to which I am in debt.

Firstly, my supervisor Vadim Biktashev. Your support to me over the last three years has been fantastic, and I believe that without your assistance, intellect and understanding I would not be where I am today. I am particularly grateful for introducing me to the wonderful world of programming. I have discovered a talent, with your assistance, that I previously had no knowledge of. Finally, I would like to say {\cyr{boljwoe spasibo za vse Vy sdelali dlya menya za prowlye tri goda. Ya budu vsegda imet\cyrsftsn{} dolg k Vam.}}

Secondly, to the lecturers and postdocs from the University of Liverpool who have helped me throughout my studies. To Irina Biktasheva for your support and very useful constructive comments regarding my work and involving in your research project. To Bakhty Vasiev for many hours of your time discussing various numerical procedures. To Radostin Simitev for his help with general computer, programming and \LaTeX\, matters. Ozgur Selsil for being a good friend over the last three years and giving the support I need.

To Mike Nieves, thanks for putting up with me over the last three years, for the hours of discussions we've had and for being a good friend. To Stuart and Ibrahim, again many thanks for being such good friends and also for all the help you have given me over the last three years.

To my family, I thank you for believing in me and giving me all the support I have needed over the last three years. To my Mum especially, without you behind me I would not be here today writing this thesis. Words can't express just how much you mean to me. To my best mate Ste Winslow, I thank you for always being there and we will forever be best mates.

Finally, to my wife. You have had to put up with a lot of mixed emotions over the last three years. We have two lovely children and it has not been easy for us all, especially in the last year of my studies. I thank you from the bottom of my heart for being such a supportive wife and the mother of my lovely boys, Charlie and Sam.

\begin{center}
\vspace{14cm}
Dedicated to Charlie and Sammy-Jo, my whole world.
\end{center}

\section*{Publications}

Some of the work presented within this thesis has been shown in the following publications and presentations:

\subsection*{Publications}

\begin{enumerate}
 \item I.V. Biktasheva, D. Barkley, V.N. Biktashev, A.J. Foulkes, G.V. Bordyugov, \textit{Computation of the response functions of spiral waves in active media}, Phy. Rev. E, 2009 (in review)
 \item A.J. Foulkes, V.N. Biktashev, \textit{Spiral Wave Solutions in a Comoving Frame of Reference}, 2009 (in preparation)
 \end{enumerate}

\subsection*{Presentations}

\begin{enumerate}
 \item ``Dynamics of Spiral Waves Under Symmetry Breaking Perturbations'', \textit{2008 PhD Symposium}, Department of Mathematical Science, The University of Liverpool, \emph{21 May 2008} {\bf (talk)}.
 \item ``Group Theortical Approaches to Drift of Rigidly Rotating Spiral Waves'',\textit{Group Seminars}, Mathematical Cardiology Group, Department of Mathematical Science, The University of Liverpool, April 2008 - August 2008 {\bf (series of talks)}.
 \item ``Drift and Meander of Spiral Waves'', \textit{Seminar - Liverpool Mathematical Biology Group}, Department of Mathematical Science, The University of Liverpool, 10 \& 24 April 2008 {\bf (talk)}.
 \item ``Drift and Meander of Spiral Waves'', \textit{British Applied Mathematics Colloquium 2008}, The University of Manchester, 3 April 2008 {\bf(talk)}.
 \item ``Towards a Theory of the Drift of Meandering Spiral Waves'', \textit{2007 PhD Symposium}, Department of Mathematical Science, The University of Liverpool, 22 May 2007 {\bf (talk)}.
 \item ``Meandering and Drifting Spiral Waves due to Inhomogeneities'', \textit{British Applied Mathematics Colloquium 2007}, The University of Bristol, 18 April 2007 {\bf(talk)}.
 \item ``Drift of Meandering Spiral Waves due to Inhomogeneities'', \textit{Annual Research Day 2007}, The University of Liverpool,  14 March 2007 {\bf (poster - 1st prize, Dept. of Mathematical Sciences)}.
 \item ``Spiral Wave Meander'', \textit{Group Seminars}, Mathematical Cardiology Group, Department of Mathematical Science, The University of Liverpool, August 2006 - January 2007 {\bf (series of talks)}.
 \item ``Meander of Spiral Waves'', \textit{2006 PhD Symposium}, Department of Mathematical Science, The University of Liverpool, 23 May 2006 {\bf (talk)}.
 \end{enumerate}

\mainmatter

\chapter{Introduction}
\label{chap:1}

Spirals occur throughout Nature. From Spiral Galaxies that are many millions of light years in diameter, to snail shells that are maybe only a centimetre or two across. One important occurance of spirals in Nature is the presence of spiral wave in cardiac tissue, and in particular spiral waves have been largely linked to the onset of Cardiac Arrythmias. Hence, the study of the onset of these spiral waves, and how they behave, is an area of research that has attracted a lot of attention over the years. 

We are concerned with the dynamics of spiral wave solutions to Reaction-Diffusion systems of equations, and how they behave when subject to symmetry breaking perturbations. We present an asymptotic theory of the study of meandering (quasiperiodic spiral wave solutions) spiral waves which are drifting due to symmetry breaking perturbations, as well as a numerical method to solve the quotient system derived from the theory.

We also drew some motivation for the numerical studies from the initial work into frequency locked spiral waves. It was noted that the spiral wave solutions used in this initial analysis were conducted in a box which needed to be of a size such that the numerical simulations did not take a long time to generate. Therefore, the answer to this is to study the spiral wave solution in a frame of reference which is moving with the tip of the wave, thereby ensuring that the spiral wave never reaches the boundaries and enabling us to conduct the simulations for as long as would like, but using a relatively small box size.

The structure of the thesis is detailed below:

\begin{itemize}
\item {\bf Chapter \ref{chap:2}}: Here we shall give an introduction to spiral waves. We will introduce the main concepts and definitions, before giving a review of the main separate theories of meander and drift. We will also give a review of the software we shall use and the numerical methods implemented into the software.
\item {\bf Chapter \ref{chap:3}}: In this chapter, we shall show our asymptotic theory of the drift of meandering spiral waves. This is split into three distinct parts. The first part is concerned with the rewriting of the theory of drift of a rigidly rotating spiral wave using group theory as well as perturbation and asymptotic methods. The first part is concluded with three examples of drift: viz. resonant drift; electropheresis induced drift; and inhomogeneity induced drift. The extension of this theory to meandering spiral waves is then achieved using Floquet theory. We will review Floquet theory before applying it to our problem. We note that in applying Floquet theory to our problem, we discovered that we required a singular perturbation method in order to achieve boundedness in the perturbed part of the solution. This singular perturbation technique took the form of a correction to the time variable. We then end the chapter with some analysis of frequency locking and how we used the results from the singular perturbation method to obtain the Arnol'd Standard Mapping.
\item {\bf Chapter \ref{chap:2a}}: We will then show some of the initial numerical analysis we undertook in the early stages of this work. We investigated frequency locking using both inhomogeneity and electropheresis induced drift within Barkley's model. The results generated within this chapter motivated the work for Chap.\ref{chap:4}, since it was clear that we needed either a larger box size in which to study frequency locking or to study the spiral waves in a frame of reference comoving with the tip of the spiral wave.
\item {\bf Chapter \ref{chap:4}}: We shall then look at the numerical calculations of spiral waves in a frame of reference that is comoving with the tip of the spiral wave. We shall describe the methods that we use in this work and introduce the program that was generated from this work - EZ-Freeze. We shall then show the results of the testing of the program, before showing some examples and applications.
\item {\bf Chapter \ref{chap:5}}: This chapter is concerned with the numerical calculation of the response functions for spiral waves. The response functions are the eigenfunctions relating to the critical eigenvalues of the adjoint linearised Reaction-Diffusion system of equation. The analytical calculation of these is not possible and therefore they can only be studied numerically. A program called \verb|evcospi| is used in this work. We will also show how EZ-Freeze can be used in the generation of the initial conditions for \verb|evcospi|, and explain why using EZ-Freeze is more accurate than using other methods.
\item {\bf Chapter \ref{chap:6}}: Finally, will conclude this work and describe some of the future directions for this work.
\end{itemize}
\chapter{Literature Review}
\label{chap:2}

In this chapter we review the main papers upon which we build our research. We commence with an introduction to spiral waves, detailing the nature of the spiral wave, its features and its most important properties. The author will also include his own account of the various properties of spiral waves.

We will then move on to the dynamics of the spiral wave and the sort of motion that it can exhibit. We will start from the most basic type of spiral wave motion, \emph{Rigid Rotation}, before moving on to a more complicated type of motion known as \emph{Meander}. We will then round off the section with a review of a motion known as \emph{Drift}. All these reviews will be supported by the author's own numerical analysis wherever the author feels it necessary to help the reader.

The models used throughout our numerical work will then be discussed and several key properties of each model used will be scrutinised. We will also provide an analysis of the types of numerical methods used to solve such model (which are PDE's), and review the types of software that will be used to conduct the numerical methods.

Finally, we will conclude this chapter.

\section{Spiral Waves}

Spiral waves \chg[af]{can be mathematically studied as} spatio-temporal solutions to Reaction-Diffusion type systems of equations. That is not to say that all solutions, where they exist, to Reaction-Diffusion systems are spiral waves. \chg[af]{Also, they can appear in non-Reaction-Diffusion type systems such as the Hodgkin-Huxley system of equations.} They are, of course, dependent not only on initial conditions but also on the parameters of the Reaction-Diffusion model used (as we will show later on). So, the equations we will be considering are Reaction-Diffusion type equations:

\begin{equation}
\pderiv{\bu}{t} = \bDD\nabla^2\bu+\bof(\bu),\quad\mbox{where}\quad\bu,\bof\in\mathbb{R}^n;\bDD\in\mathbb{R}^{n\times n},\br=(x,y)\in\mathbb{R}^2
\end{equation}

The functions $\bof(\bu)$ may be nonlinear functions and, as we will see in later sections, we take this nonlinearity to be cubic, for reasons that will be explained.

\begin{figure}[tbp]
\begin{center}
\begin{minipage}{0.75\linewidth}
\centering
\includegraphics[width=\textwidth]{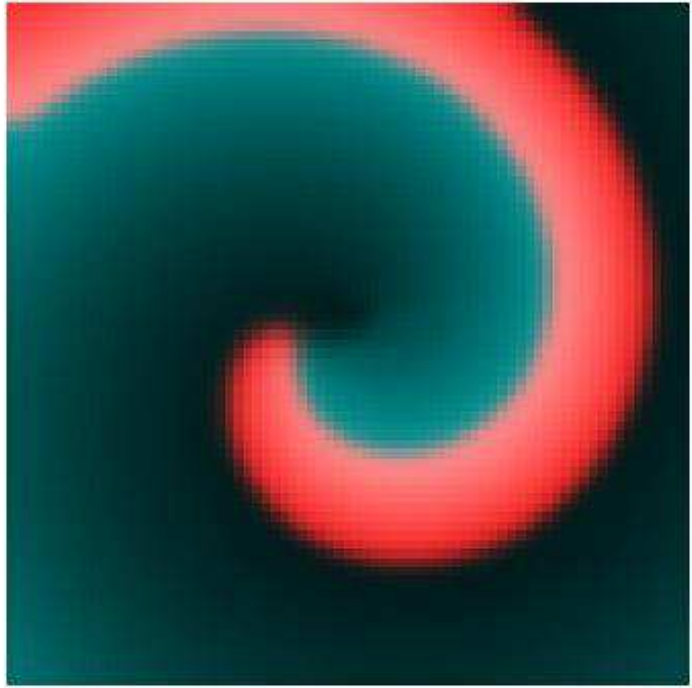}
\caption[A Spiral Wave]{A snapshot of a spiral wave solution to a Reaction-Diffusion system of equations in the $x$-$y$ plane}
\label{fig:spirals_typical}
\end{minipage}
\end{center}
\end{figure}

The spiral wave is a rotating wave whose shape takes a spiral form. A snapshot of a typical spiral wave is shown in Fig.(\ref{fig:spirals_typical}), where $\bu=(u,v)$.

We can see that the wave is defined by different colours. In this particular figure, we note that the red colour represents the excitation field (first component of $\bu$), and the inhibitor (second component of $\bu$) is shown in blue. Their interaction is what determines the spiral wave's behaviour. We also, note that for excitation, we always require that the inhibitor field always lags behind the excitation field.

\subsection{History}

Spiral waves as a mathematical object were first observed in 1946 by Wiener and Rosenblueth, who explained the idea of cardiac arrhythmias using the concept of excitable media \cite{wiener_ros}. In 1952, Turing introduced the Reaction-Diffusion system of equations \cite{turing}. It was noted that certain solutions to these equations were spiral wave solutions. In the 1960's, Belousov and Zhabotinski observed spiral patterns in light intensive chemical reactions, now known as the Belousov-Zhabotinski Reaction, or simply BZ-Reaction. It must be noted that they did not work together. Belousov initiated the work \cite{belousov_1959}, which was subsequently picked up by Zhabotinski several years later \cite{zhabot_1964a,zhabot_1964b}. However, their combined research is recognized in unison. This work was then made popular in the West by Arthur Winfree in the early 1970's, and it was to Winfree that the discovery of meandering spiral waves (to be discuss in the next section) is credited \cite{winf72}.

Since the 1970's there has been a massive surge in the amount of research that has taken place on spiral waves and their many physical occurrences - initiation of spiral waves; transition from rigid rotation to meander; drift of spiral waves; frequency locking within meandering waves and forced spiral waves, to name but a few. 

We are primarily interested in the dynamics of meandering and drifting spiral waves. As we progress through the following sections we will introduce the main concepts and theories that we will be using, as well as several paper reviews that we feel are necessary.

In general, there are several excellent books on spiral waves and patterns formed from Reaction-Diffusion systems. The first book is by Winfree and is a general introduction to this area \cite{winfree_time}. As Winfree states in his introduction, this book is primarily about patterns that involve time - memory and heartbeat as just two examples. He then splits the book into four parts. The first part is an introduction to arrhythmias, and considers temporal patterns without any spatial organization. He looks at circadian rhythms, as well as the heartbeat and other biological rhythms, and how these behave when interrupted. The second part considers spatially organised biological clocks, particular the heartbeat, and how the presence of rotating waves can prove lethal to the heartbeat. Then he extends the ideas introduced in the previous two parts to the three dimensional case - scroll waves. In part four, he summarizes what he has introduced and what questions remain outstanding. his book is extremely well written and supported by many pictures and diagrams, a lot of which are coloured.

The next book is by Murray \cite{murray}. In fact this is a collection of two books in Mathematical Biology gives an excellent introduction, at undergraduate level, to the mathematics behind spiral waves. The main text that we are interested is found in the second volume, Chapter 1, which describes various wave processes arising from Reaction-Diffusion type equations. The section of this chapter that becomes most interesting for our purposes is Section 1.6 et seq. Although the mathematical content relating to spiral waves in this text is aimed at undergraduate level, it provides a general background knowledge base for which to proceed into spiral wave research.

Another excellent publication is the book by Keener and Sneyd \cite{keener_sneyd}. This covers a vast array of areas relating to the physiology of the body. The sections of interest to us appear in Chaps.9\&10. Chap.9 introduces the mathematics behind a one dimensional wave train, and introduces two classic models which simulate wave propagation, viz. FitzHugh-Nagumo model and the Hodgkin-Huxley model. Also introduced is a singular perturbation approach to studying these phenomena. Finally, Chap.10 briefly describes the mathematics behind rotating waves (spiral waves) and how these waves can be studied analytically, even though the models used are PDE's which may not necessarily have general analytical solutions.

The next book is a more advanced text by Zykov \cite{zykov_88}. This, overall, is probably one of the best books that has been produced which concentrates on solely waves in excitable media. Zykov begins his book by describing the sort of systems that we may come across which support the propagation of excitation waves. He also places high importance on the use of computer modeling and simulation, and therefore taylors the text towards such work. To sumarise the book briefly, Chap.1 introduces the concept of excitable media and spiral waves. In particular, he introduces the sorts of equations that we are likely to come across in this area of work, including the famed Belousov-Zhabotinski Reaction. Chap.2 then gives a thorough review of the sorts of processes that we may come across in biological excitable tissues. Chaps.3\&4 \chg[p6spell1]{present} the methods to simulate wave propagation in one and two dimensions, with particular reference to the initiation of excitation waves. Chap.5 describes the simulation of waves in cardiac tissue, whilst in Chap.6 Zykov presents a simplified approach to studying spiral wave processes in two-dimensional excitable media in both bounded and unbounded dimensions. Chap.7 is an investigation into the general properties of excitation circulation in two-dimensions, and a simplified model is used in the simulations. Finally, Chap.8 deals with the issue of controlling excitation waves and in particular, how to control excitation waves in cardiac tissue. Overall, an excellent text for those wanting a thorough background reading into this area.

\subsection{Initiation}

\begin{figure}[tbp]
\begin{center}
\begin{minipage}[b]{0.22\linewidth}
\centering
\includegraphics[width=0.9\textwidth]{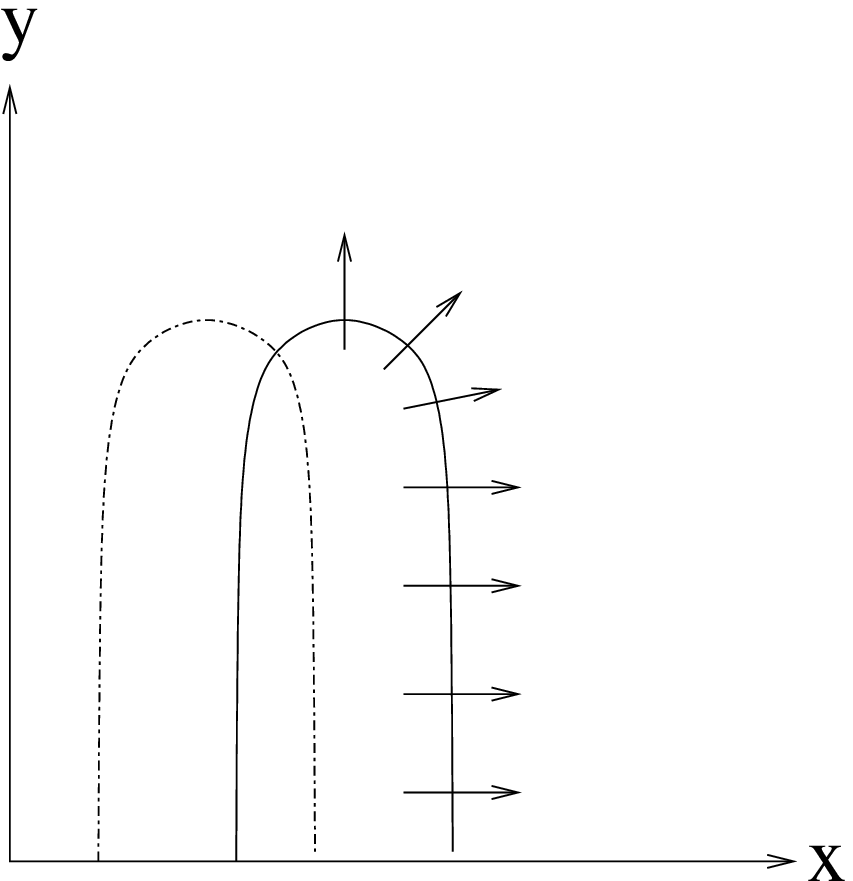}
\end{minipage}
\begin{minipage}[b]{0.22\linewidth}
\centering
\includegraphics[width=0.9\textwidth]{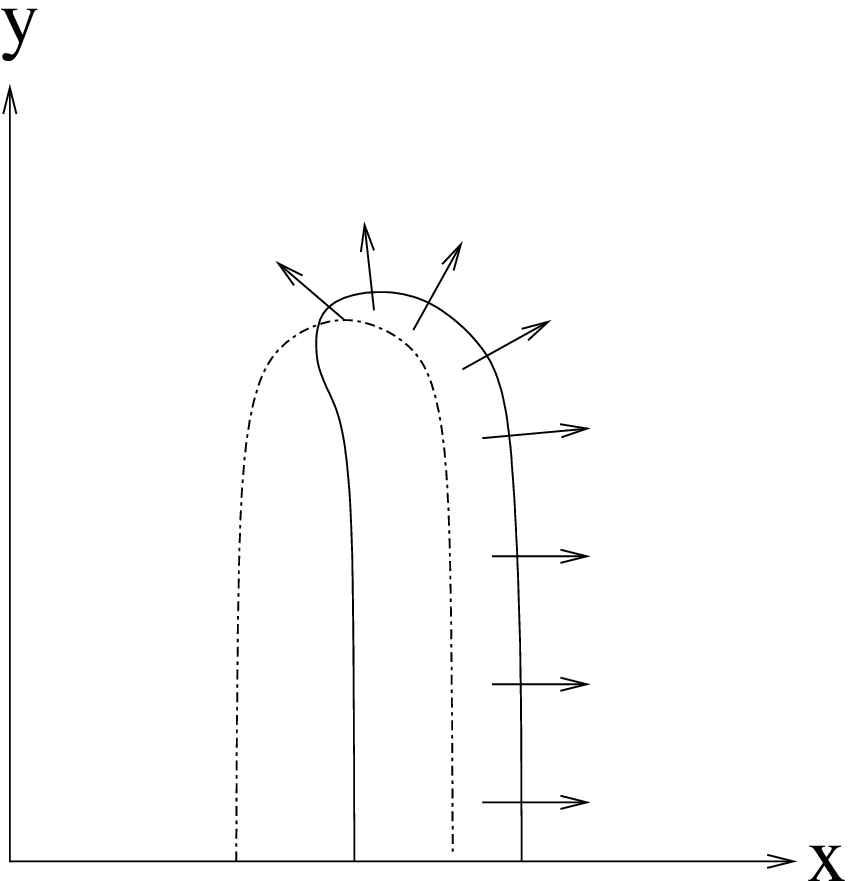}
\end{minipage}
\begin{minipage}[b]{0.22\linewidth}
\centering
\includegraphics[width=0.9\textwidth]{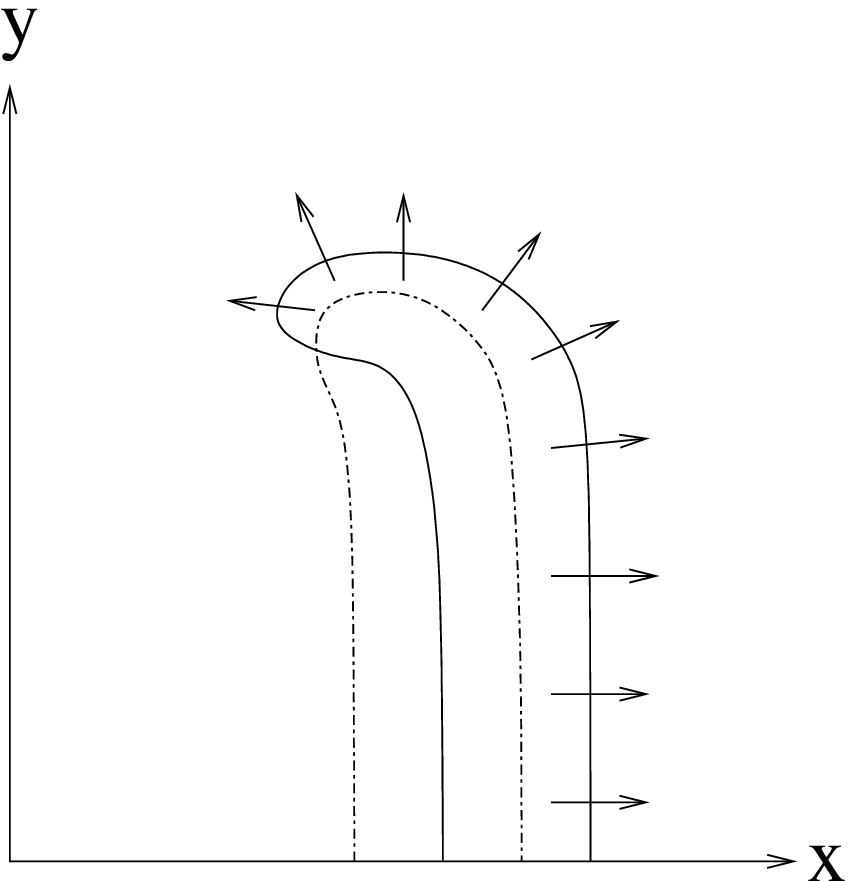}
\end{minipage}
\begin{minipage}[b]{0.22\linewidth}
\centering
\includegraphics[width=0.9\textwidth]{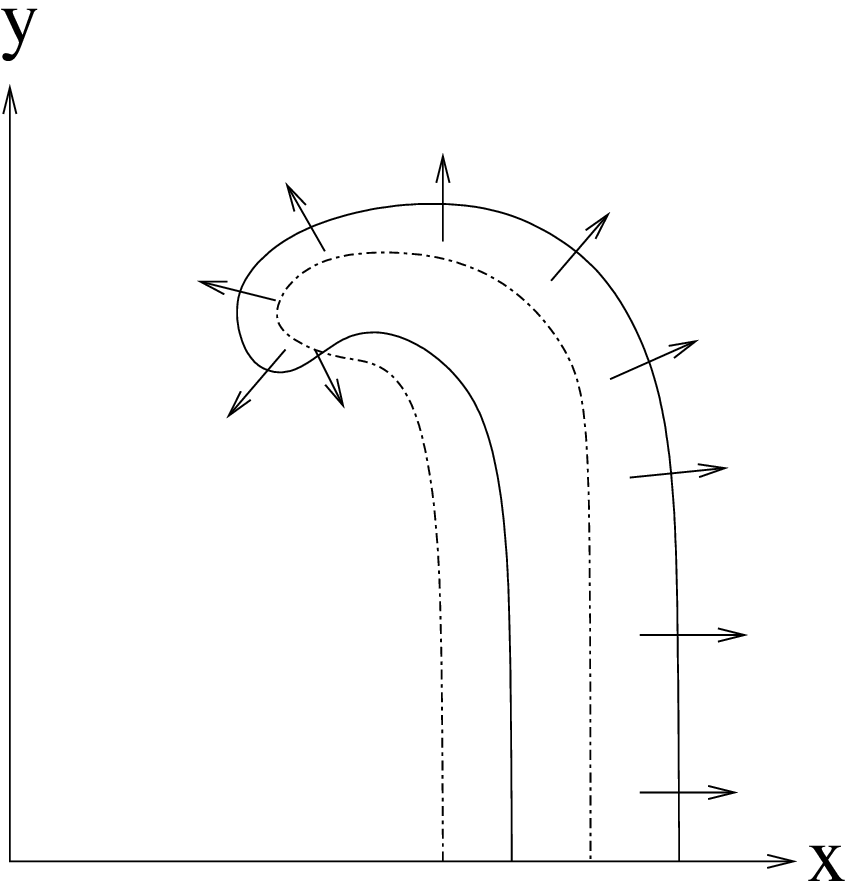}
\end{minipage}
\caption[Evolution of a Spiral Wave from a broken wave]{Evolution of a Spiral Wave from a broken wave \cite{zykov_88}.}
\label{fig:spiral_CFS}
\end{center}
\end{figure}

The subject of initiation of a spiral wave could be a Ph.D project in itself. Indeed, several authors have studied this in great details \cite{panf82,panf94}, and others have also looked at the 1-D case (travelling waves) \cite{ib08}, but not so much the 3-D case (scroll waves). The book by Zykov described in the previous section gives an excellent introduction into the initial process of excitation waves. For our purposes, we are not concerned with exactly how these \chg[p6spell2]{waves}are initiated, but how \chg[p6spell2]{they}evolve in time once they have been initiated.

We will, however, give a brief description of how these wave are initiated. The method which we will use, and which is implemented into the software we use, is known as \emph{Cross Field Stimulation}. This is when the spiral wave is initiated when two plane waves, travelling perpendicular to each other, collide, form a broken wave and \chg[p6spell3]{evolve}into a spiral. Of course, not every collision \chg[p6spell3]{result}in a spiral wave, but in our simulation, the initial conditions are chosen such that a spiral wave is formed in most instances.

So, once these waves have collided and a broken wave is formed, the spiral wave then takes shape, as shown in Fig.(\ref{fig:spiral_CFS}), where the arrows in each picture show the direction of motion of the excitation field. We also see that the inhibitor field is lagging behind the excitation field as described above. So, we can see that it is this interaction of the two fields that help form the wave.

\subsection{Tip of the Spiral Wave}

The location of the \emph{tip} of the spiral is extremely important. Throughout our work, we consider how the wave moves by studying the motion of the tip of the spiral wave. 

\begin{figure}[tbp]
\begin{center}
\begin{minipage}[b]{0.49\linewidth}
\centering
\psfrag{a}[l]{$y$}
\psfrag{b}[l]{$x$}
\psfrag{c}[l]{$u_*$}
\psfrag{d}[l]{$v_*$}
\includegraphics[width=0.9\textwidth]{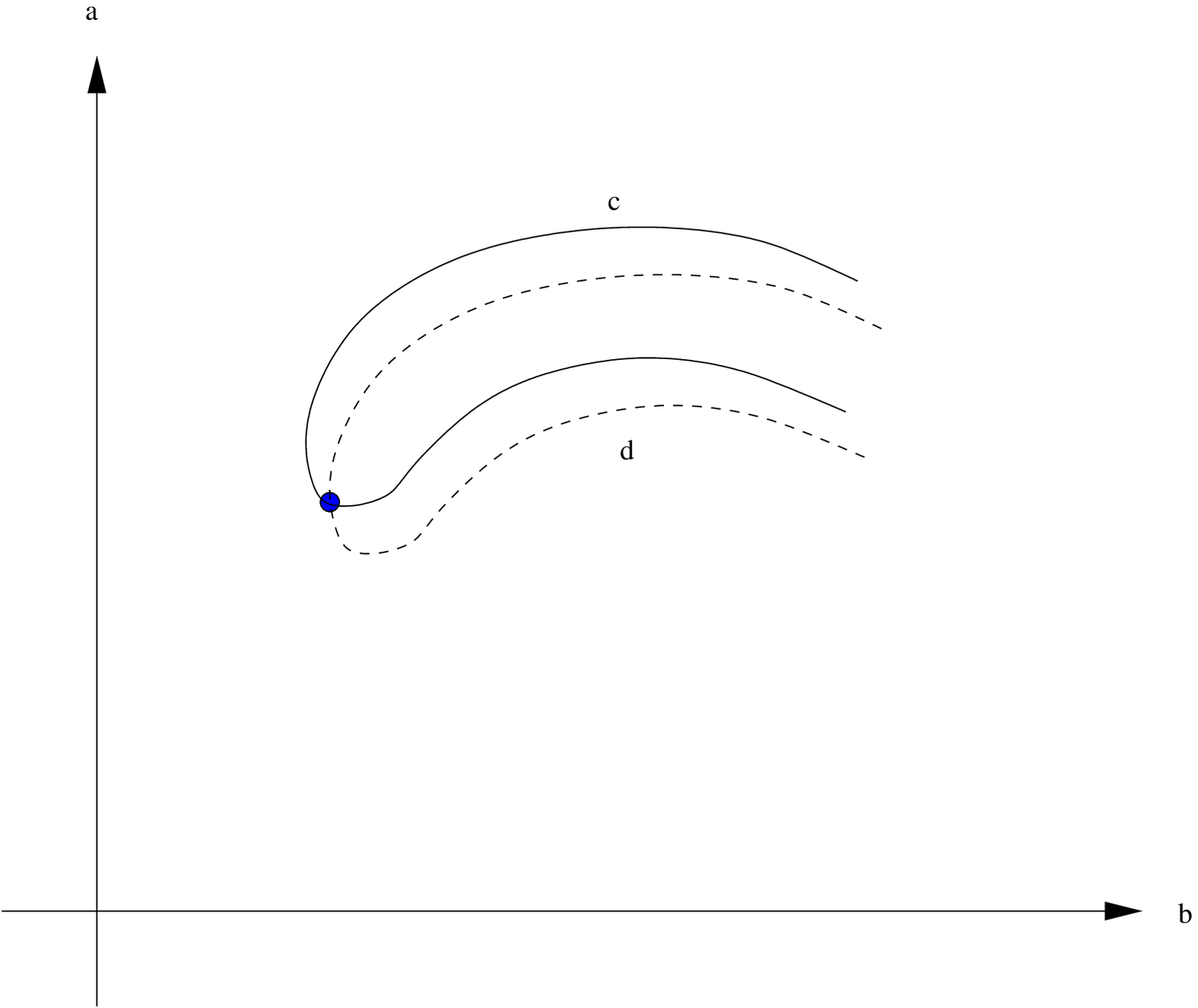}
\caption[Spiral wave tip: definition 1]{Spiral wave tip is the intersection (blue dot) of 2 isolines in the (x,y)-plane}
\label{fig:spiral_iso}
\end{minipage}
\begin{minipage}[b]{0.49\linewidth}
\centering
\psfrag{a}[l]{$\pderiv{u(\br,t)}{x} = 0$}
\psfrag{b}[l]{$x$}
\psfrag{c}[l]{$y$}
\psfrag{d}[l]{$u_*$}
\includegraphics[width=0.9\textwidth]{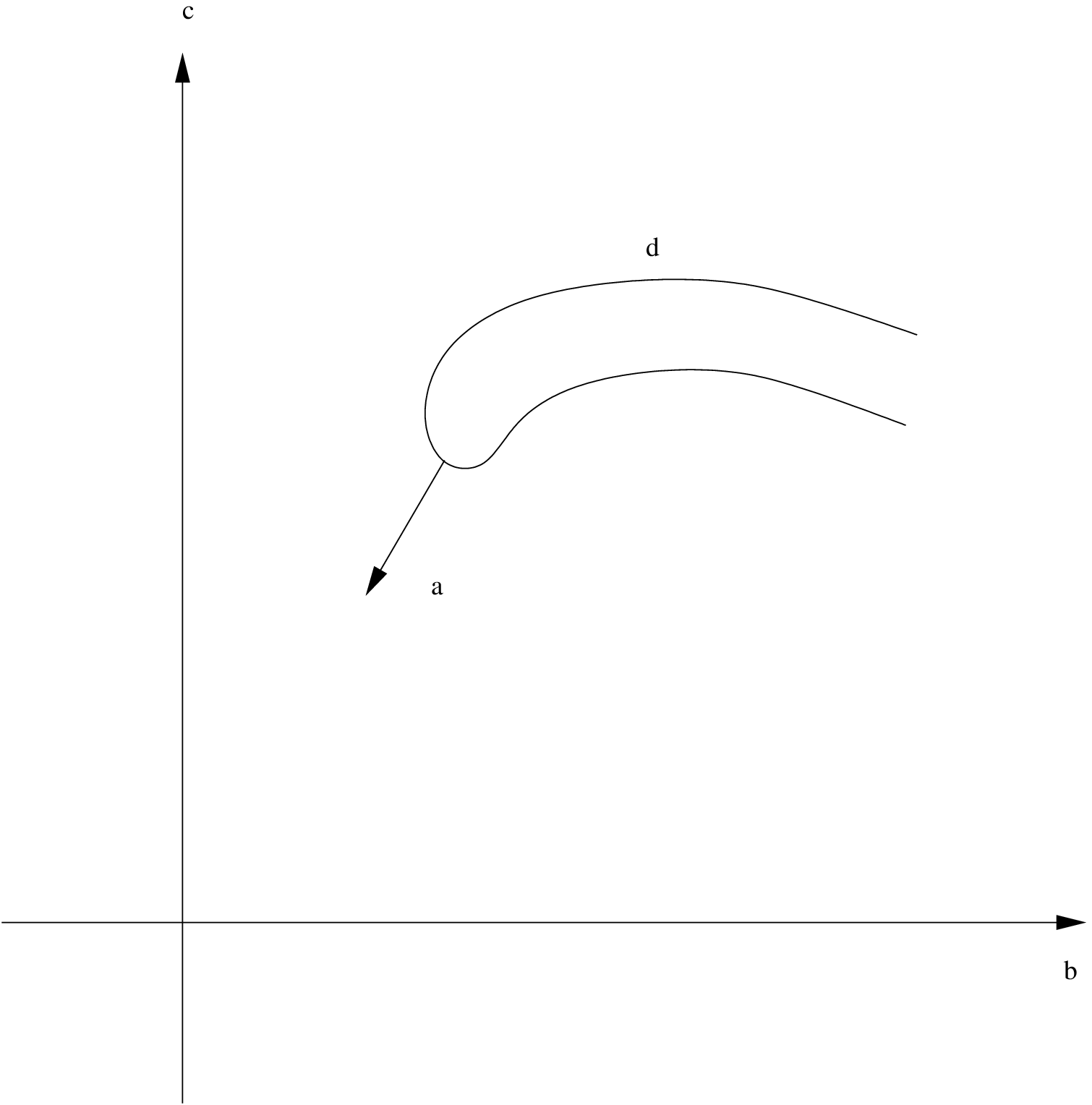}
\caption[Spiral wave tip: definition 2]{Spiral wave tip as defined by an isoline and a directional derivative}
\label{fig:spiral_iso_dir}
\end{minipage}
\end{center}
\end{figure}

The most common way to define the tip of a spiral wave is to define two isolines - one defined in the excitation field, and the other in the inhibitor field - and define the tip as the intersection of the two lines, \chg[p7]{Fig.}(\ref{fig:spiral_iso}):

\begin{eqnarray}
u(\br,t) &=& u_*\\
v(\br,t) &=& v_*
\end{eqnarray}

Another way is to define an isoline as described in the previous paragraph and fix the normal of that isoline at the tip to be in a particular direction.

\begin{eqnarray}
u(\br,t) &=& u_*\\
\label{eqn:spiral_tip_deriv}
\pderiv{u(\br,t)}{t} &=& 0
\end{eqnarray}

Of course the choice of $u_*$ and $v_*$ must be within certain limits (see Sec.(\ref{sec:models})). Also, Eqn.(\ref{eqn:spiral_tip_deriv}) does not necessarily have to be zero, it could be any value, but it seems more sensible to define this as zero.
\label{sec:rev_sw}

\section{Spiral Wave Dynamics}

We will now introduce the main types of motion of the spiral and review several key publications whose results we will refer to in this thesis.

There are three main types of motion of spiral waves. Each of these motions are most easily observed by tracking the trajectory of the tip of the spiral. 


\subsection{Rigid Rotation}

The most basic type of spiral wave motion is called \emph{Rigid Rotation}. When a spiral wave rigidly rotates, its shape remains constant, as its name suggests. This in turn leads to the fact that if the rigidly rotating spiral wave was observed in frame of reference which was moving with the tip of the spiral wave, then the wave would appear stationary. Some authors refer to rigid rotation as ``Relative Equilibria'', since when the solution is viewed in the space of group orbits, the solution can be represented as an equilibrium point \cite{chossat02}.

\begin{figure}[tbp]
\begin{center}
\begin{minipage}{0.32\linewidth}
\centering
\includegraphics[width=\textwidth]{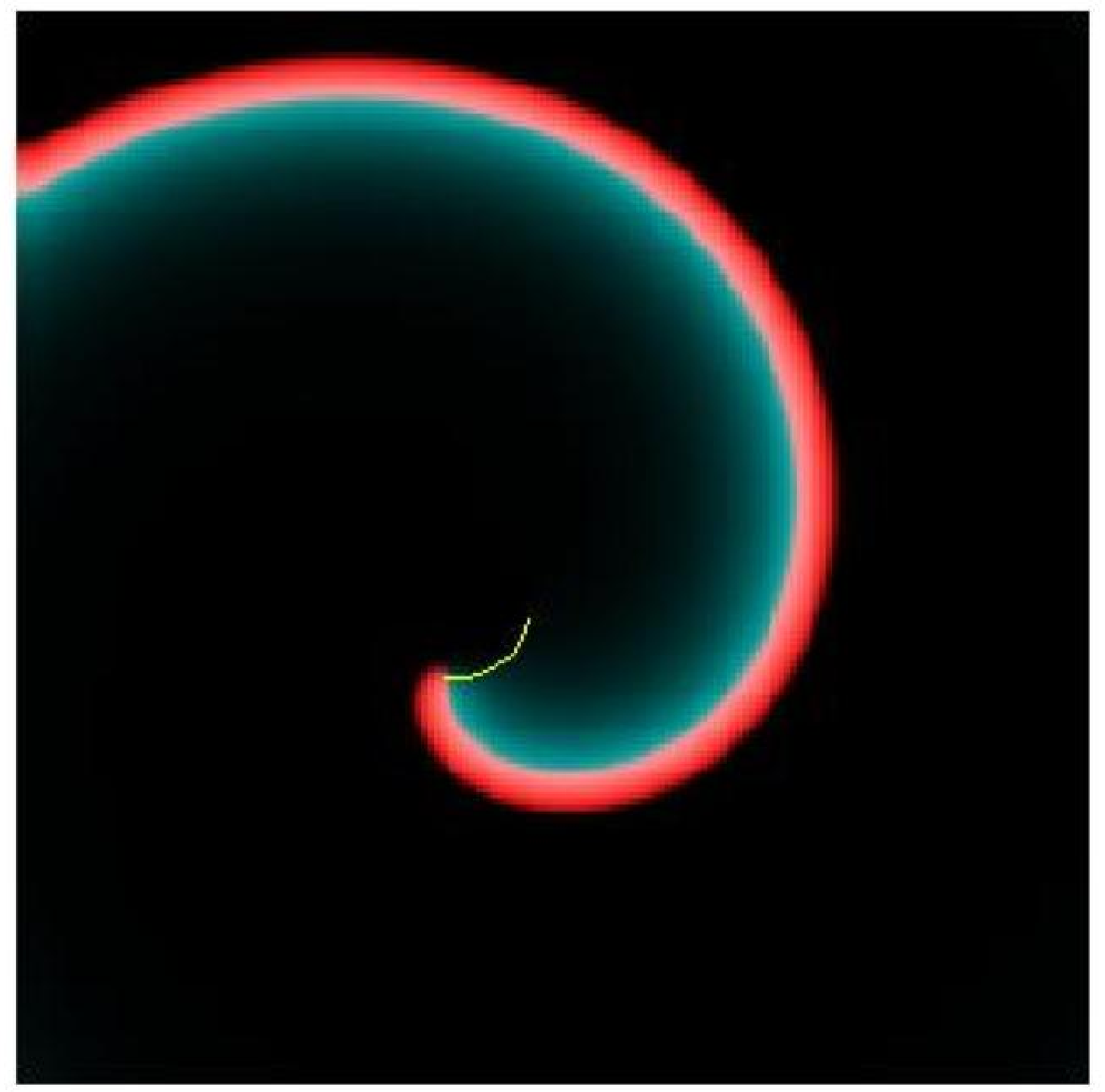}
\end{minipage}
\begin{minipage}{0.32\linewidth}
\centering
\includegraphics[width=\textwidth]{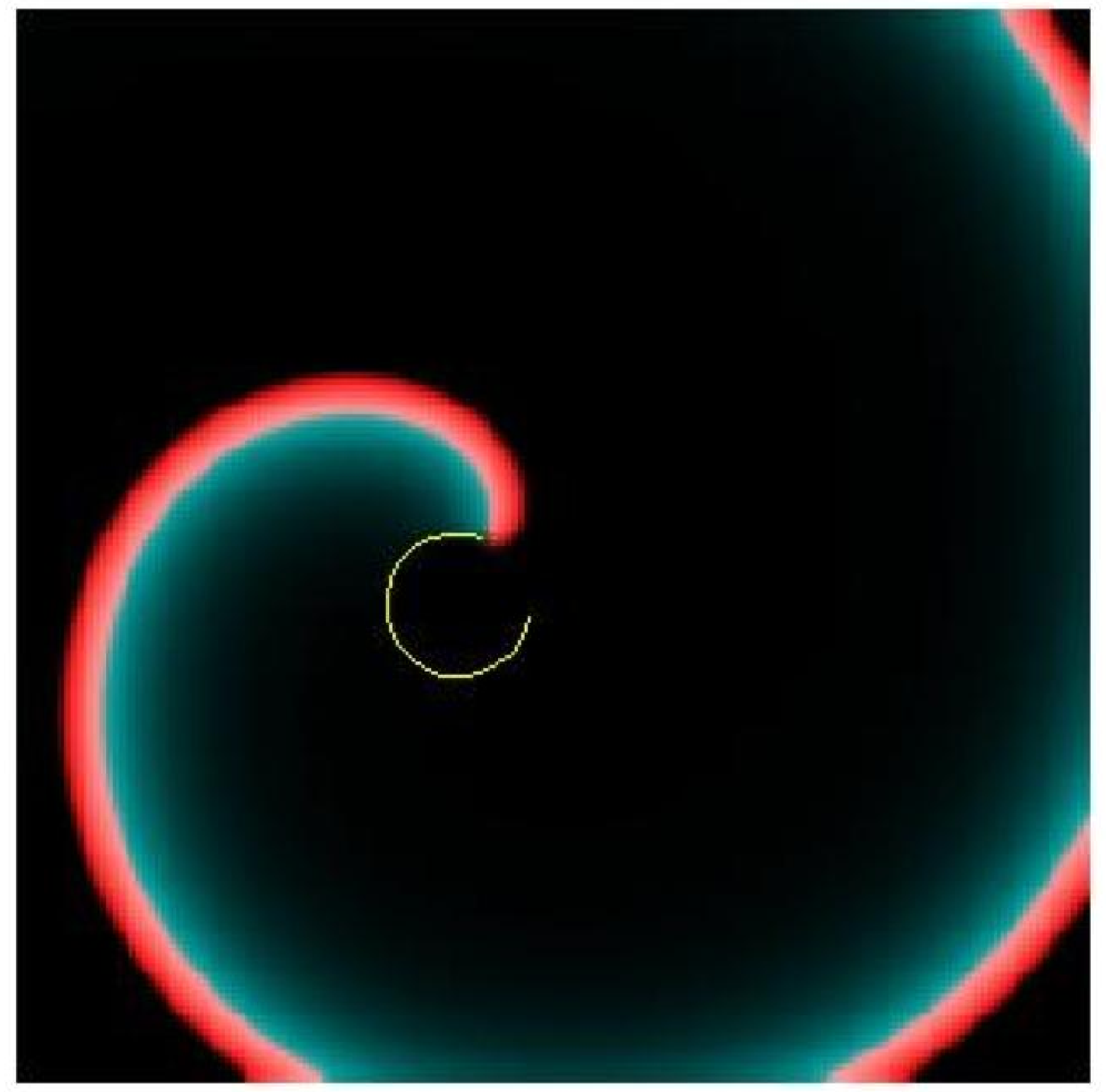}
\end{minipage}
\begin{minipage}{0.32\linewidth}
\centering
\includegraphics[width=\textwidth]{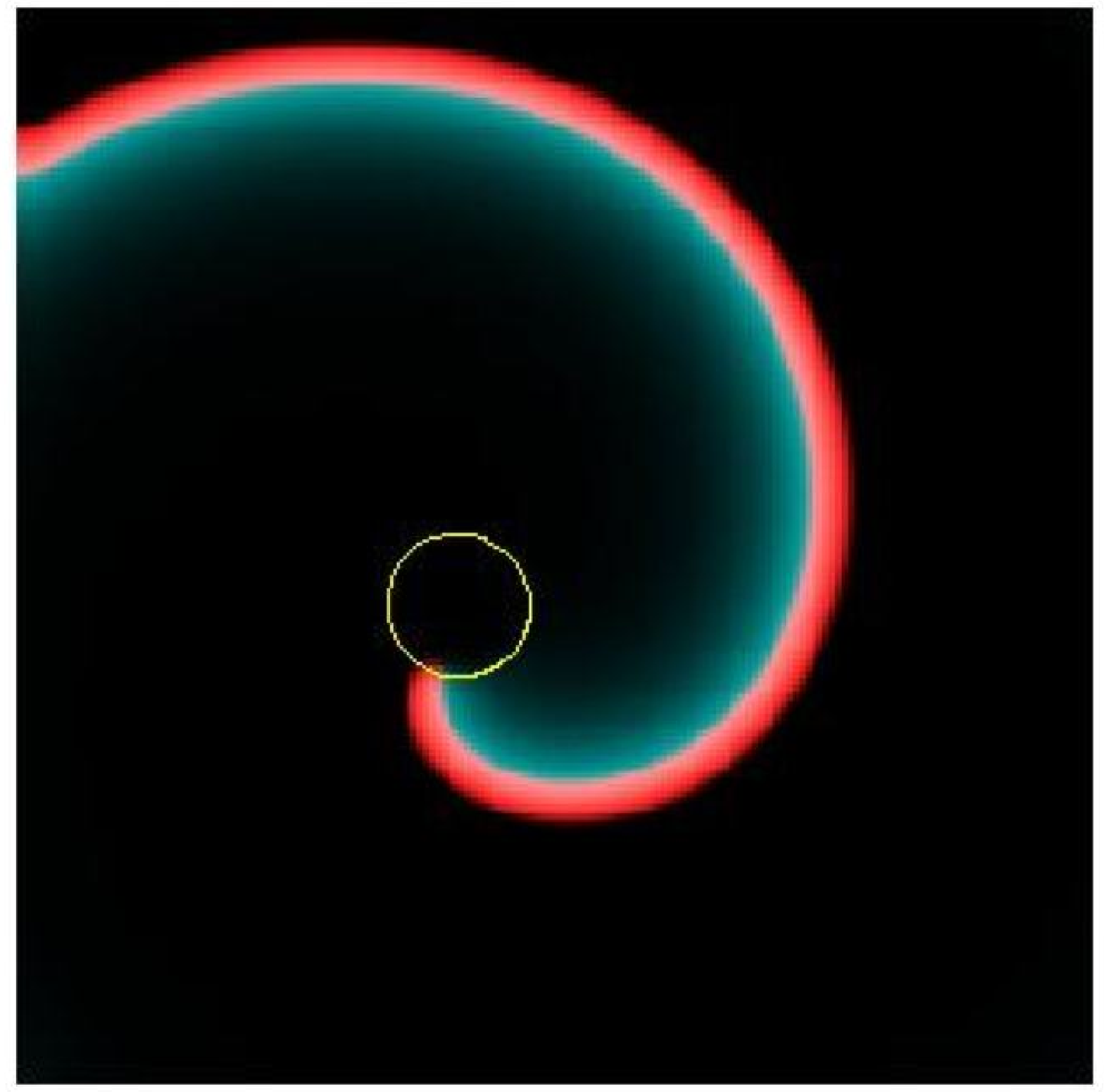}
\end{minipage}
\caption{A Rigidly Rotating Spiral Wave}
\label{fig:spiral_RW}
\end{center}
\end{figure}

Also, due to the rigidness of the solution, the tip of the spiral wave traces out a perfect circle as shown in Fig.(\ref{fig:spiral_RW}).

Another property of rigidly \chg[af]{rotating} spiral waves is that the spectrum of the linearised system has three critical eigenvalues, located on the imaginery axis. All other eigenvalues are assumed to have negative real part, i.e. the solution is stable. These three critical eigenvalues are 0, $\pm i\omega$, where $\omega$ is the angular velocity of the spiral wave. We will show in Chap.\ref{chap:3} that these relate directly to the symmetry of the system of Reaction-Diffusion equations.


\subsection{Meander}

Meander is a more complicated type of motion. It is, in essence, a quasiperiodic motion. The shape of the \chg[p9spell1]{arm of the}spiral wave, when it meanders, changes with time.

Also, we have that there are at least two angular velocities present within a meandering spiral wave solution. Throughout our work, we shall only consider simple meander, which means that there are only ever two angular velocities present (Euclidean frequency, $\omega_E$, and the Hopf Frequency, $\omega_H$). 

It has been known for some time that a Hopf Bifurcation is responsible for the transition from periodic to quasiperiodic motion. This was shown numerically by Lugosi \cite{Lugo89} and also by Skinner et al \cite{Skin89}. They both suggested that the bifurcation looked as if it was supercritical but it was Barkley et al in their 1990 paper who proved that this was actually the case \cite{Bark90}.

Now, in order to provide evidence of the presence of a supercritical Hopf bifurcation, Barkley et al computed the decay rates to simple rotations on one side of the bifurcation and on the other side, they computed the ratio of the amplitudes of the compound waves formed. Remember from the definitions section that meandering spiral waves (or compound waves, as they are sometimes referred to \cite{Bark90}) have two frequencies and are constructed by taking two circles - the primary circle with radius $r_1$ and secondary circle with $r_2$ - and, with the tip of the spiral located at a fixed point on the secondary circle, a flower pattern is traced out when the primary and secondary circles rotate with frequencies $\omega_1 \ \mbox{and} \ \omega_2$ respectively. The amplitudes of the compound waves are simply the radii of these circles. In reality, we find that the secondary circle is in fact an ellipse but is ``almost'' circular. Therefore, the radius $r_2$ is actually the maximum value of the radius\cite{Bark90}.

\begin{figure}[tbp]
\begin{center}
\begin{minipage}{0.7\linewidth}
\centering
\includegraphics[width=\textwidth]{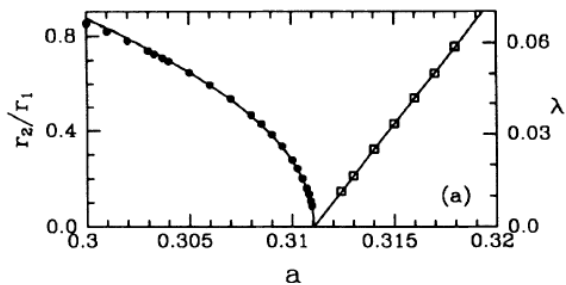}
\end{minipage}
\caption[Hopf bifurcation transition from rigid rotation to meander]{A plot of radius ratio $\frac{r_1}{r_2}$ against \chg[af]{parameter}$a$ and also, on the same plot decay rate $\lambda$ against parameter $a$\cite{Bark90}.}
\label{fig:bark_hopf}
\end{center}
\end{figure}

Barkley et al then \chg[af]{plotted} the ratio of the secondary radius to the primary radius ($r_1$/$r_2$) against the parameter $a$, and also on the same plot, the decay rate, $\lambda$, \chgex[ex]{against one of the parameters in Barkley's model, $a$,} therefore generating the plot shown in Fig.(\ref{fig:bark_hopf}) \cite{Bark90}.

As can be seen, the amplitude of the secondary mode grows from zero at the same point, $a_c$, where the decay rate goes through zero. It can also be seen that near the bifurcation point, the growth of the secondary radius is given by a power law $\approx \frac{1}{2}$.

Hence, these \chg[p9spell2]{observations}provide the first conclusive evidence that the transition from simple to compound rotation is via a supercritical Hopf Bifurcation.

\begin{figure}[tbp]
\begin{center}
\begin{minipage}{0.32\linewidth}
\centering
\includegraphics[width=\textwidth]{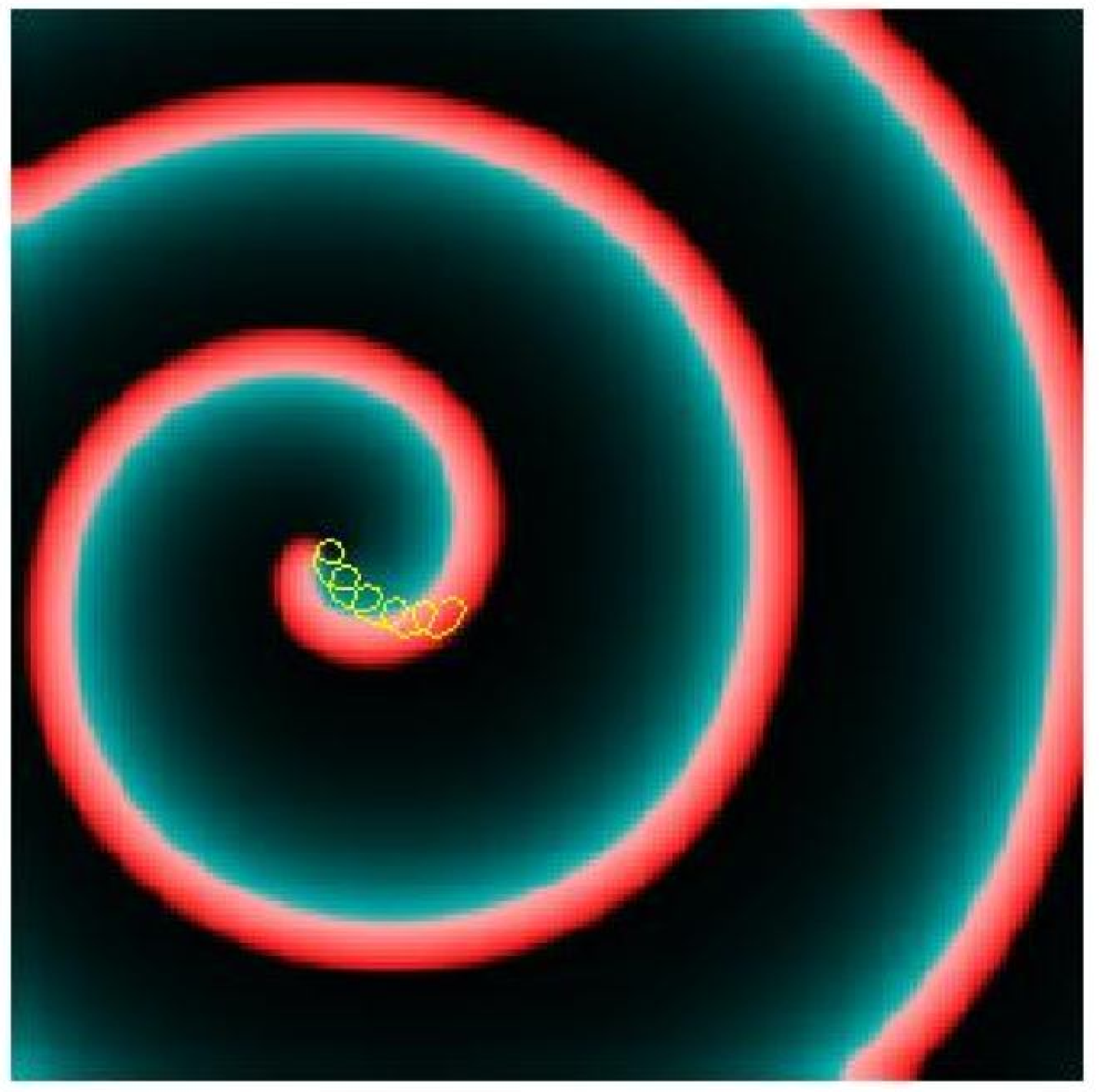}
\end{minipage}
\begin{minipage}{0.32\linewidth}
\centering
\includegraphics[width=\textwidth]{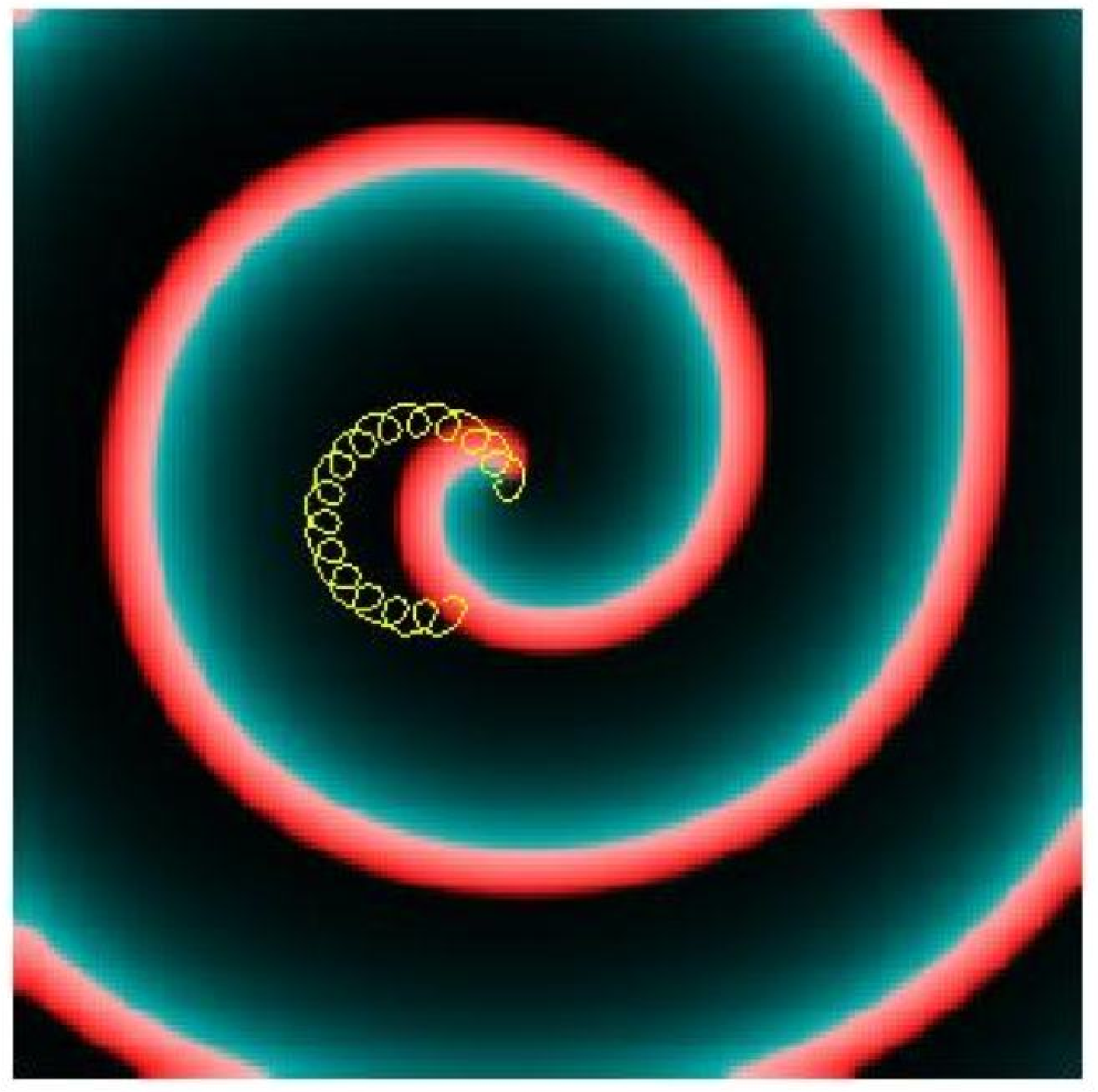}
\end{minipage}
\begin{minipage}{0.32\linewidth}
\centering
\includegraphics[width=\textwidth]{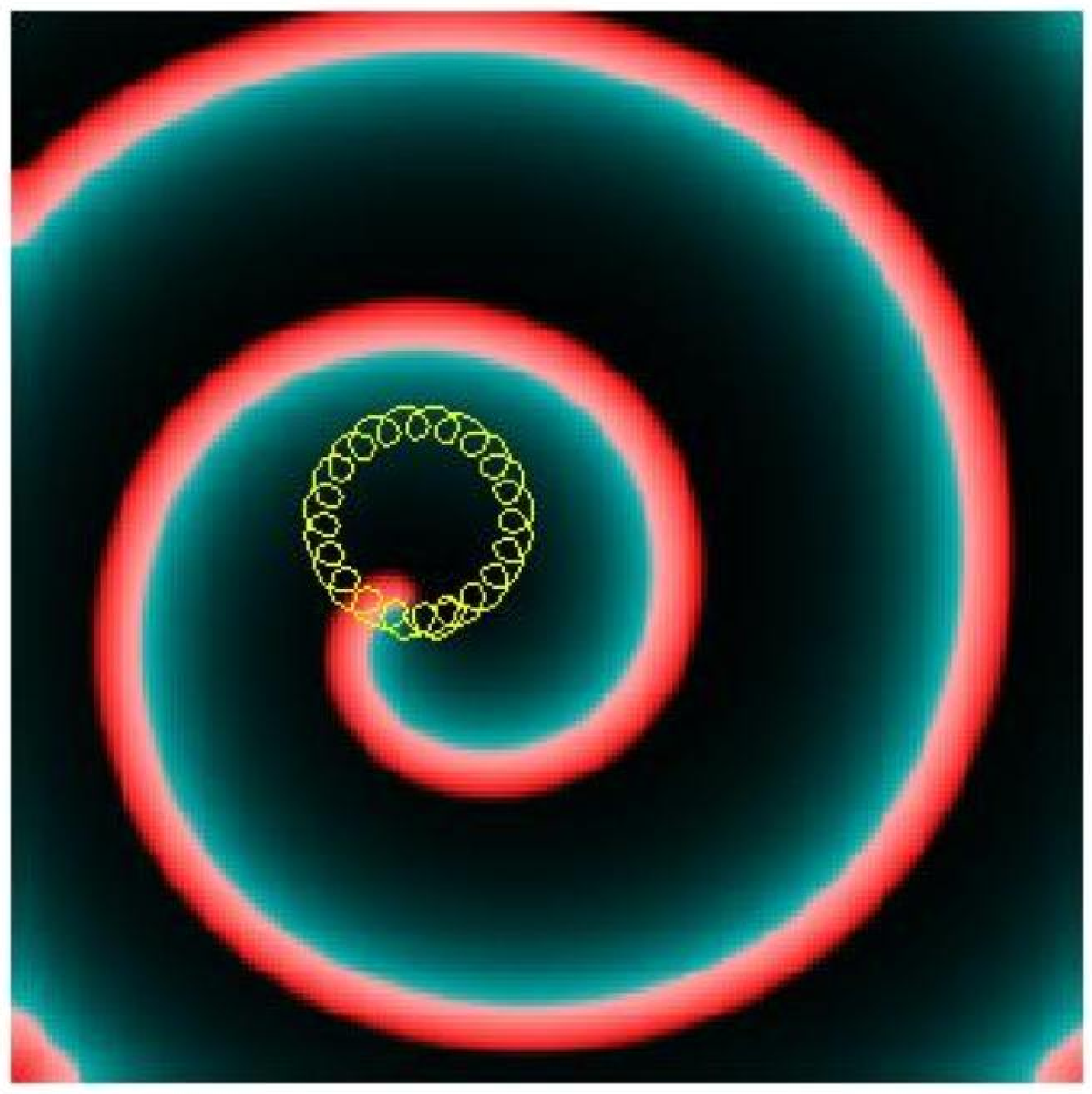}
\end{minipage}
\caption{A Meandering Spiral Wave with inward facing petals}
\label{fig:spiral_mrwin}
\end{center}
\end{figure}

\begin{figure}[tbp]
\begin{center}
\begin{minipage}{0.32\linewidth}
\centering
\includegraphics[width=\textwidth]{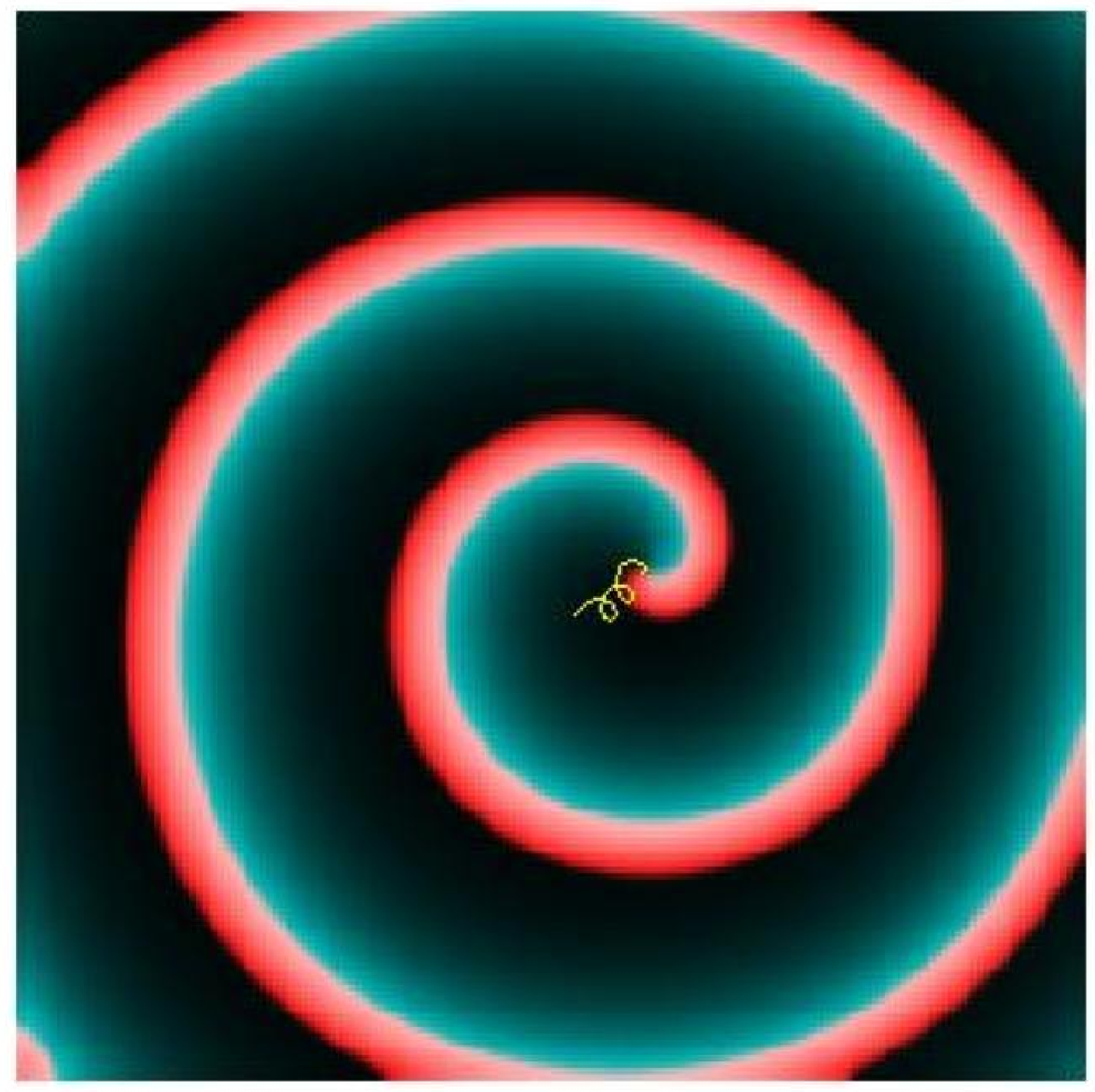}
\end{minipage}
\begin{minipage}{0.32\linewidth}
\centering
\includegraphics[width=\textwidth]{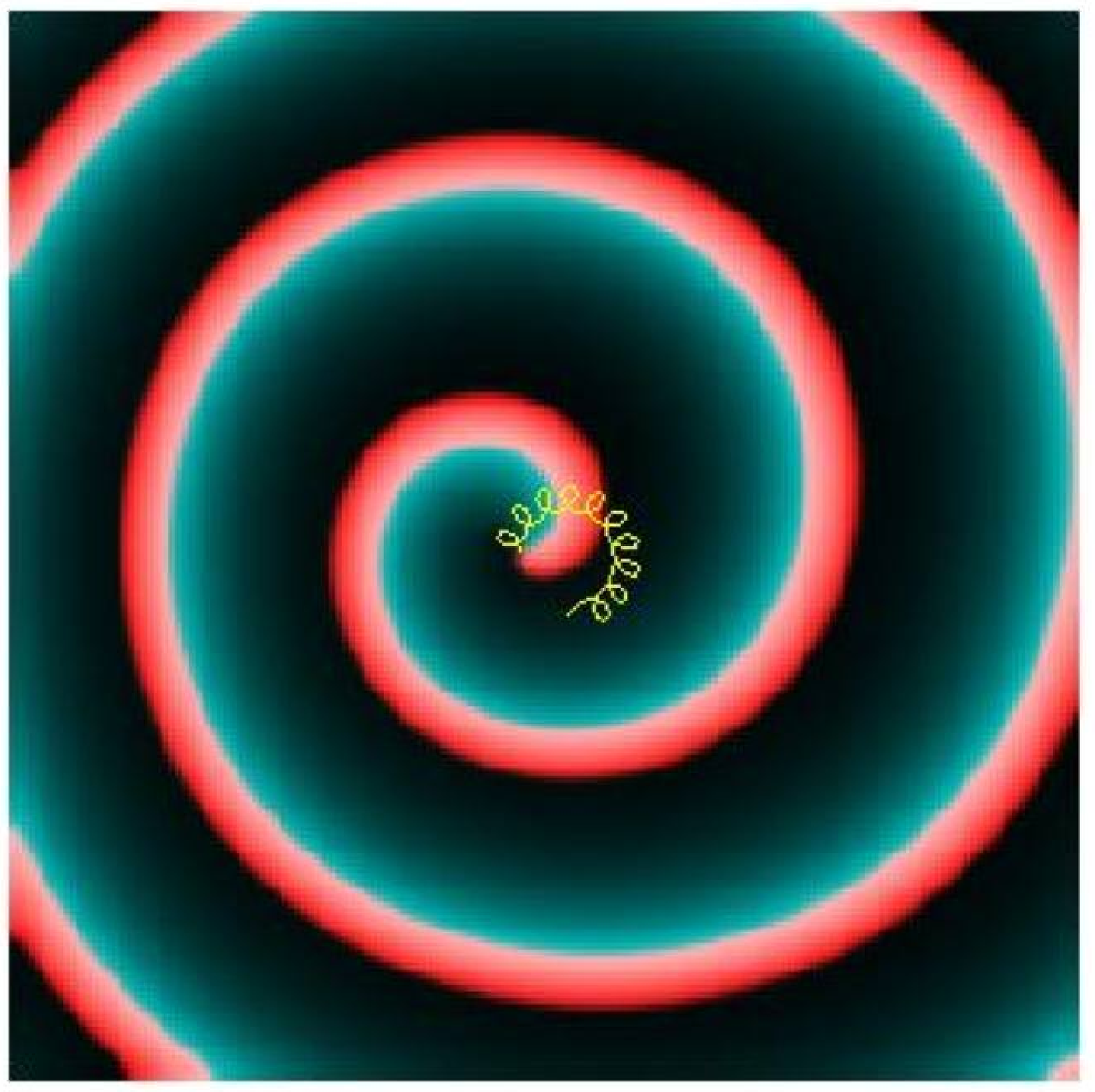}
\end{minipage}
\begin{minipage}{0.32\linewidth}
\centering
\includegraphics[width=\textwidth]{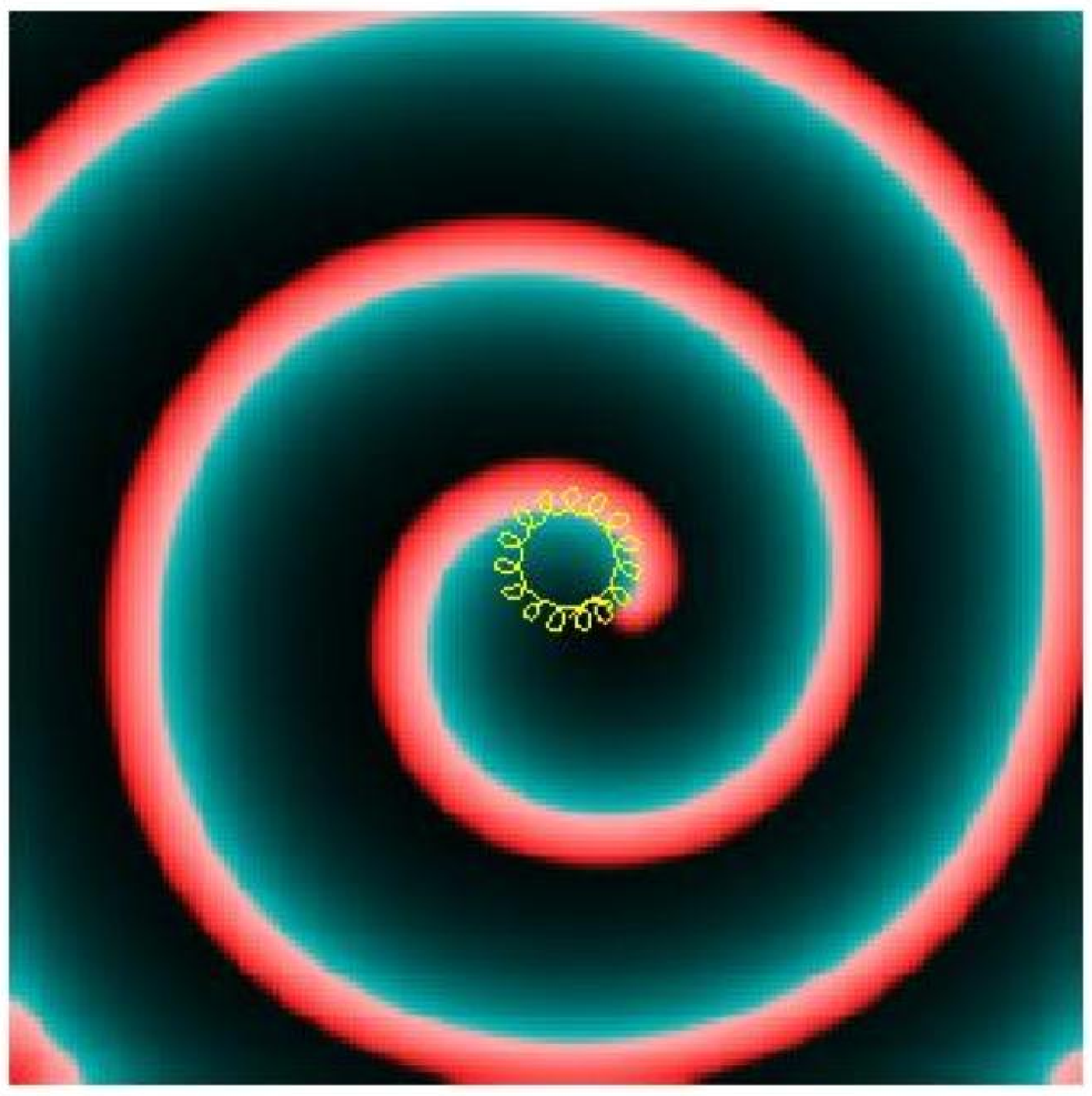}
\end{minipage}
\caption{A Meandering Spiral Wave with outward facing petals}
\label{fig:spiral_mrwout}
\end{center}
\end{figure}

Now, if we consider the trajectory of the tip of the spiral wave, we will see that the tip traces out a ``flower'' type pattern. These patterns have petals which face either inwards (Fig.(\ref{fig:spiral_mrwin})), or outwards (Fig.(\ref{fig:spiral_mrwout})).

The direction of the petality is determined by the values of the two frequencies. If $\omega_H<\omega_E$ then we have outward facing petals. \chg[af]{Similarly}, if $\omega_H>\omega_E$ then we have inward facing petals. So, if $\omega_E=\omega_H$ then we have spontaneous drift where the radius \chg[p10]{of the circle}about which the spiral wave \chg[af]{meanders}becomes infinite.

\begin{figure}[tbp]
\begin{center}
\begin{minipage}{0.49\linewidth}
\centering
\includegraphics[width=0.72\textwidth]{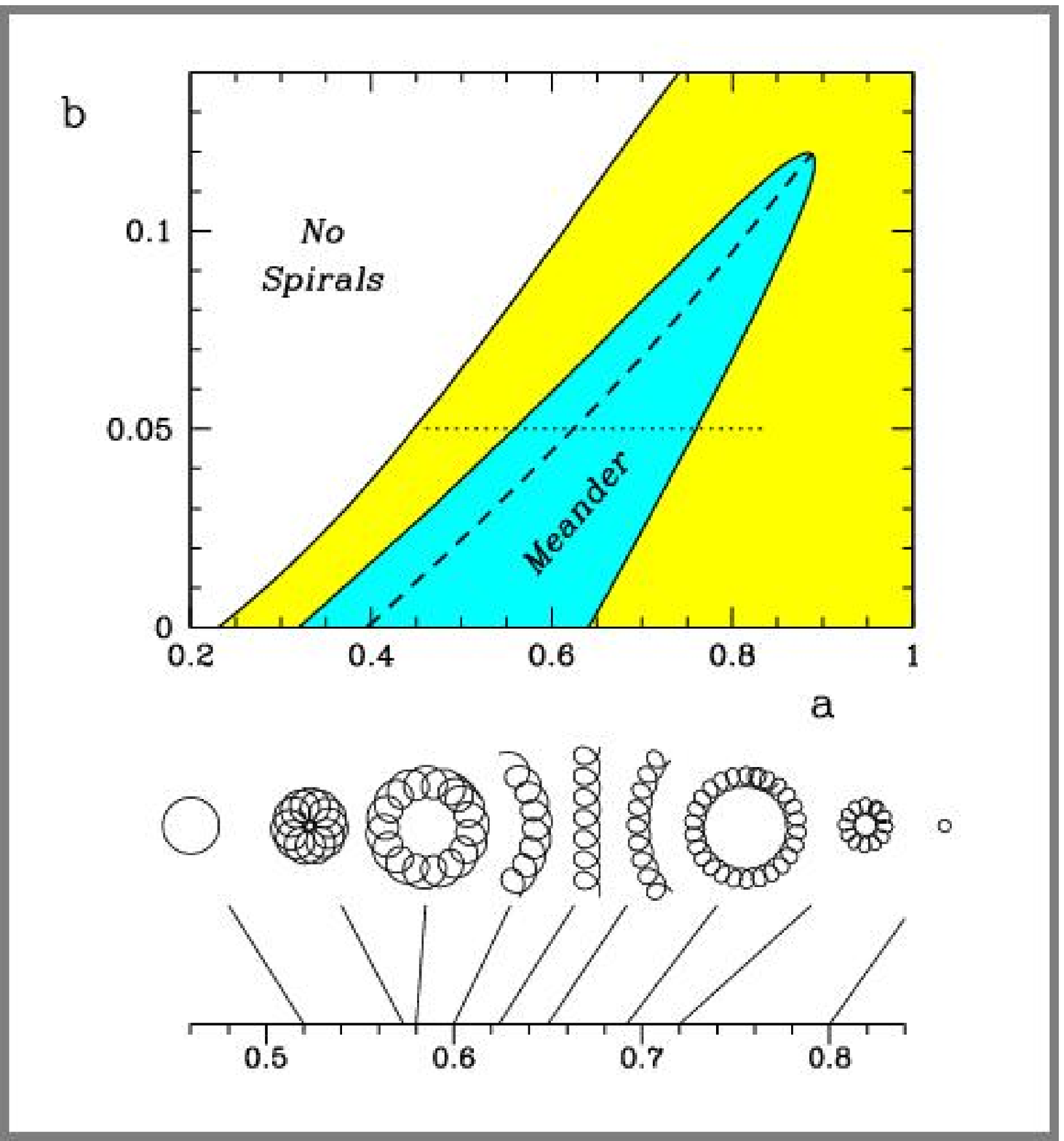}
\end{minipage}
\begin{minipage}{0.49\linewidth}
\centering
\includegraphics[width=1.2\textwidth]{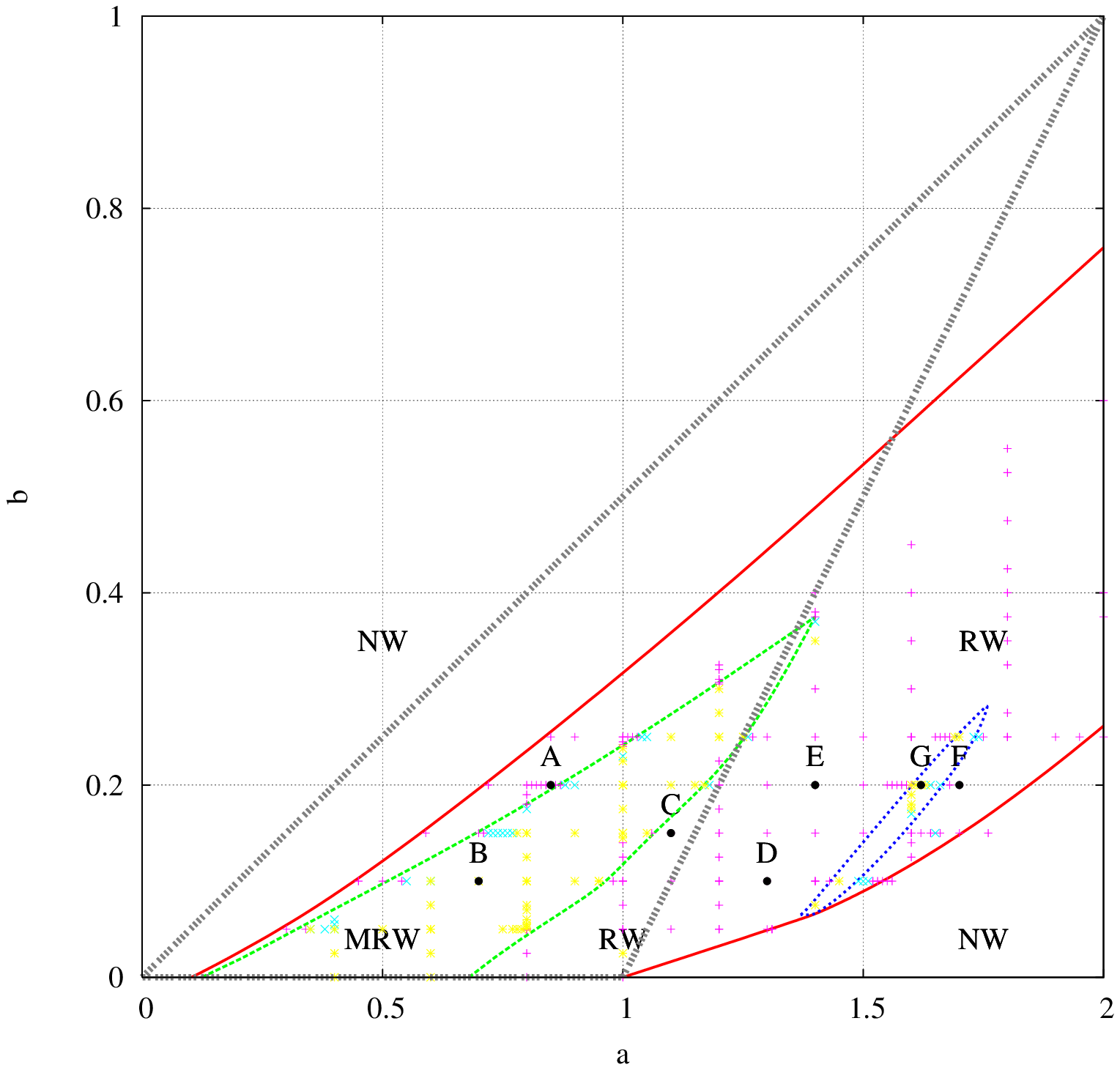}
\end{minipage}
\caption[Barkley's Flower garden]{Barkley's Flower garden for $\epsilon=0.02$ (left) \cite{bark_scholar}; flower garden for $\epsilon=0.01$ (right).}
\label{fig:spiral_bark_flower}
\end{center}
\end{figure}

In one of his 1994 papers, Barkley produced his ``Flower Garden'' \cite{Bark94a}. This is a \chg[af]{parametric}portrait and is based on Winfree's flower garden published in 1991\cite{Winf91}. We \chg[af]{reproduce}Barkley's flower garden in Fig.(\ref{fig:spiral_bark_flower}).

As an exercise in the author's first few months of this project, we reproduced this flower garden but this time with $\epsilon=0.01$. Our results are shown in Fig.(\ref{fig:spiral_bark_flower}).

Also, Barkley et al introduced a system of ODE's which, when numerically solved, produced patterns that were similar to the patterns produced by the PDE system, i.e Meandering patterns were observed \cite{Bark94a,Bark94b}. The system takes advantage of 2 particular properties of spiral wave solutions:

\begin{enumerate}
\item Equivariance under Euclidean Symmetry; and
\item Presence of a Supercritical Hopf Bifurcation.
\end{enumerate}

However, the system is \emph{a priori} in the sense that when Barkley et al introduced this system, the Bifurcation Theory at that time was insufficient to justify the production of a set of ODE's from the PDE system. They simply introduced this system, and indicated that it worked by showing an example. The system is as follows:

\begin{eqnarray*}
\dot{p} & = & v \\
\dot{v} & = & v\cdot \{f(|v|^2,w^2)+iw\cdot h(|v|^2,w^2)\}\\
\dot{w} & = & w\cdot g(|v|^2,w^2)
\end{eqnarray*}

As we can see, there are 3 dependent variables. $p$ is the position vector and is complex; $v$ is the velocity vector and is also complex; and $w$ is the frequency and is real. We now let $p=x+iy$ and $v=se^{i\phi}$, where $x,y$ are the position coordinates, $s$ is speed, and $\phi$ is the angle between the x-axis and the line from the origin to the point ($x,y$). We are therefore left with a 5-dimensional system of ODE's:

\begin{center}
$\dot{x}=s\cos(\phi)$, $\dot{y}=s\sin(\phi)$\\
$\dot{\phi}=w\cdot h(s^2,w^2)$, $\dot{s}=s\cdot f(s^2,w^2)$, $\dot{w}=w\cdot g(s^2,w^2)$
\end{center}

As you can see, we have 3 unknown functions, $f(s^2,w^2)$, $g(s^2,w^2)$ and $h(s^2,w^2)$, and it is up to the reader to define exactly what these functions are. Barkley et al took these functions and Taylor expanded them as follows:

\begin{eqnarray*}
f(s^2,w^2) & = & \alpha_0+\alpha_1s^2+\alpha_2w^2+\alpha_3s^4\\
g(s^2,w^2) & = & \beta_0+\beta_1s^2+\beta_2w^2\\
h(s^2,w^2) & = & \gamma_0
\end{eqnarray*}

After assigning specific values to $\alpha_0$, $\alpha_3$, $\beta_0$, $\beta_1$ and $\beta_2$, and letting $\xi=s^2$ and $\zeta=w^2$, we get the following reduced system:

\begin{eqnarray*}
\dot{\xi} & = & 2\xi f(\xi,\zeta)\\
\dot{\zeta} & = & 2\zeta g(\xi,\zeta)
\end{eqnarray*}

where:

\begin{eqnarray*}
f(\xi,\zeta) & = & -\frac{1}{4}+\alpha_1\xi+\alpha_2\zeta-\xi^2\\
g(\xi,\zeta) & = & \xi-\zeta-1\\
h(\xi,\zeta) & = & \gamma_0
\end{eqnarray*}

We can now take this reduced system and use our usual dynamical system and bifurcation analysis to shown that, for $\alpha_1=\frac{10}{3}$, there is a Hopf Bifurcation at $\alpha_2=-5$ and $\gamma_0=\sqrt{28}$.

\begin{figure}[htbp]
\begin{center}
\begin{minipage}[b]{0.3\linewidth}
\centering
\includegraphics[width=0.9\textwidth]{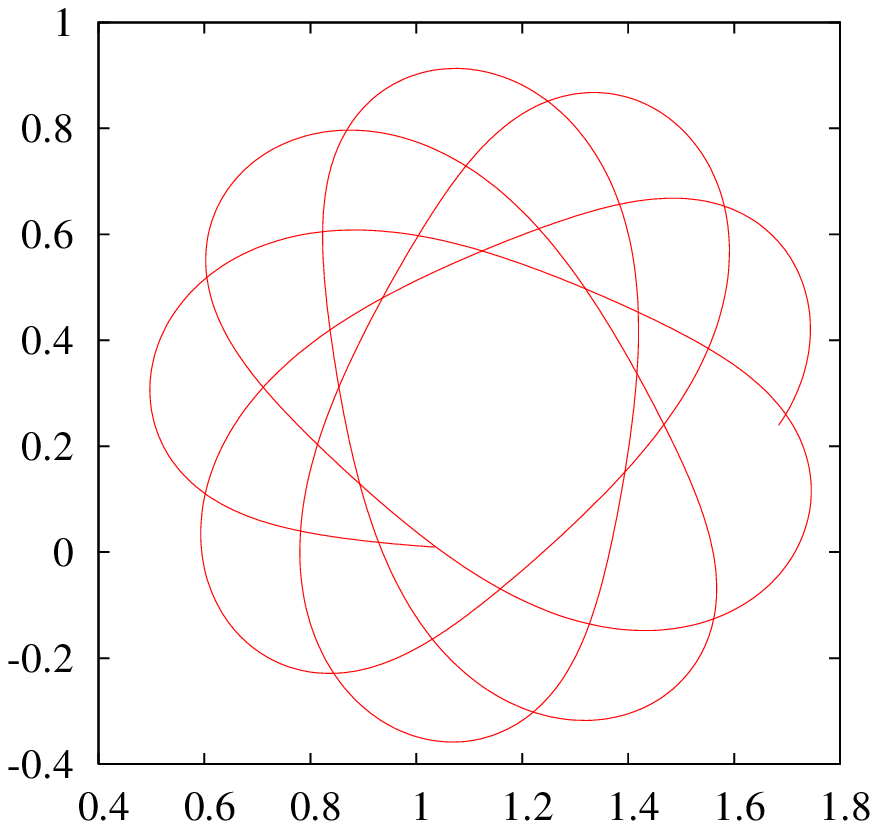}
\end{minipage}
\begin{minipage}[b]{0.3\linewidth}
\centering
\includegraphics[width=0.9\textwidth]{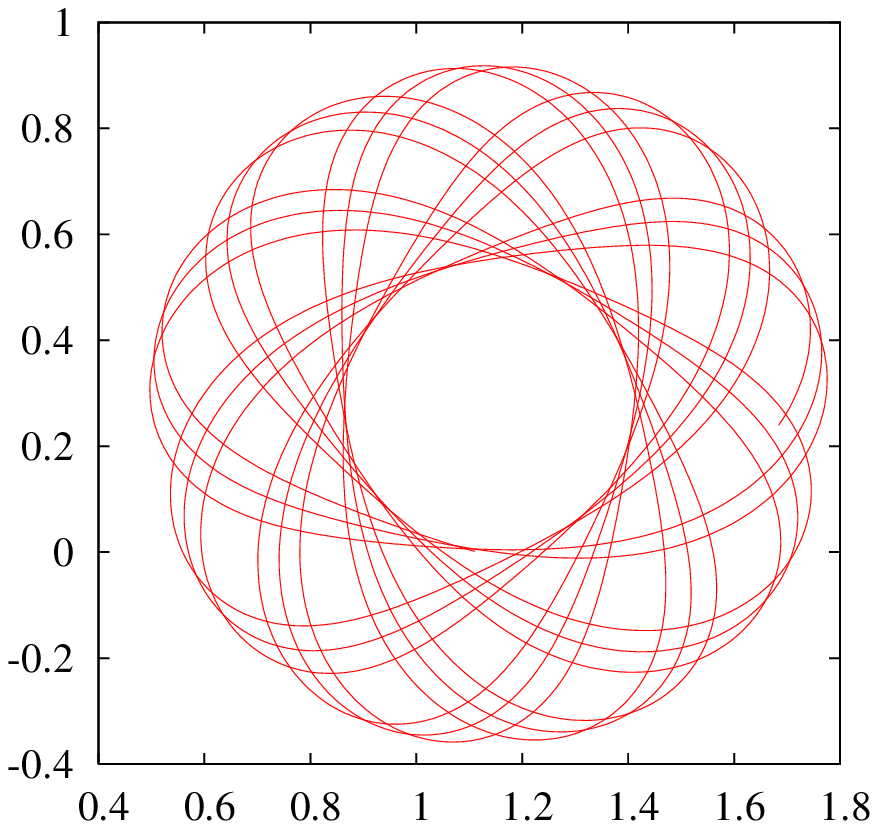}
\end{minipage}
\begin{minipage}[b]{0.3\linewidth}
\centering
\includegraphics[width=0.9\textwidth]{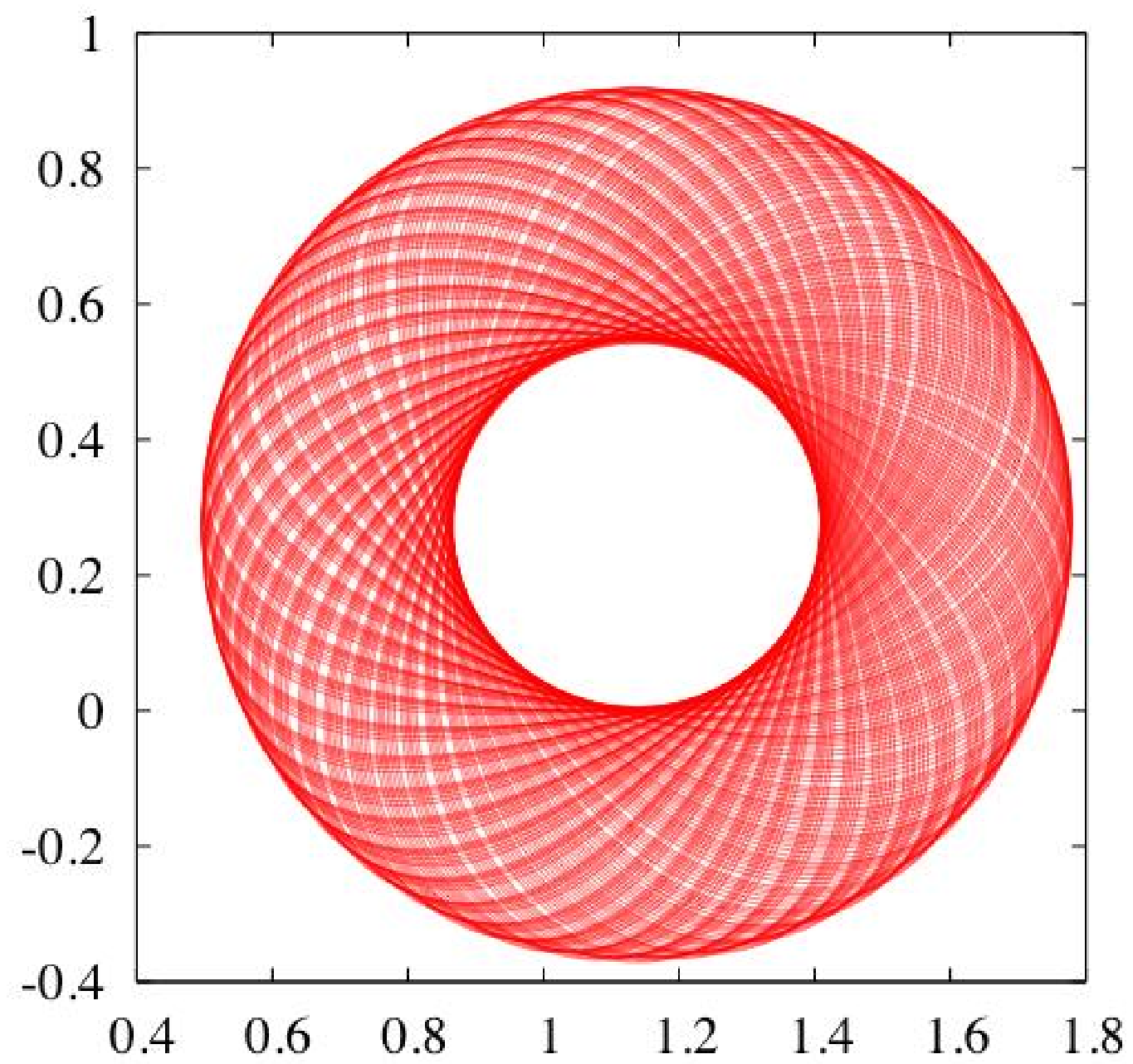}
\end{minipage}
\caption[Meandering spiral wave from ODE system]{Results from numerical analysis of the ODE system taken from the same tip file but at different time intervals}
\label{fig:ode}
\end{center}
\end{figure}

In order to verify that the patterns produced by the system of ODE's are the same as those produced by the PDE's, we created a simple C program which numerically solved the 5-dimensional system of ODE's with the parameters $\alpha_1$, $\alpha_2$ and $\gamma_0$ being the parameters that are varied, with the other parameters being kept at the values specified above. We observed that the patterns produced were the same as those produced by the PDE system. One of the patterns is shown in fig \ref{fig:ode}.

As we can see, a clear ``flower'' pattern develops. Furthermore, when we view the pattern at the latter stage, we can see that the trajectory is forming a dense orbit on the 2-Torus, which is then projected onto the 2-D plane.

This paper, though it contained some extremely groundbreaking facts, lacked a proof of exactly how the ODE system was determined. Of course, it was billed as a priori, but a more concrete derivation of the system was needed. 

This was then motivation for Biktashev et al \cite{bik96} to derive a system of equations the describe the dynamics of the tip of a meandering spiral wave. We will review this publication in more detail.


\subsubsection{\chg[p13title]{Theory of Meander} \cite{bik96}}

Biktashev et al considered the special Euclidean group $SE(2)$, together with the following reaction diffusion equation:

\begin{equation}
\label{eqn:meander_rde}
\partial_t{\bu} = \textbf{D}\nabla^2\bu+\bof(\bu)
\end{equation}
\\
for $\bu=\bu(\br,t)=(u^{(1)},u^{(2)},\hdots,u^{(l)})\in \mathbb{R}^l, l\geq2, \textbf{r}=(x,y)\in \mathbb{R}^2$. As mentioned, previously, this system is equivariant under transformations belonging to $SE(2)$. A proof of this is shown in the appendix (\chg[p13ref]{Sec.}(\ref{sec:euclid})). So, if $\bu(\textbf{r},t)$ is a solution then so too is $\tilde{\bu}(\textbf{r},t)$ such that:

\begin{equation}
\label{eqn:utilde}
\tilde{\bu}(\textbf{r},t) = T(g)\bu(\textbf{r},t), \quad \forall g\in SE(2)
\end{equation}
\\
where $T(g)$ is the group action of $g\in SE(2)$ on the function $\bu(\textbf{r},t)$. 

\begin{figure}[htbp]
\begin{center}
\begin{minipage}[b]{0.6\linewidth}
\centering
\includegraphics[width=0.9\textwidth]{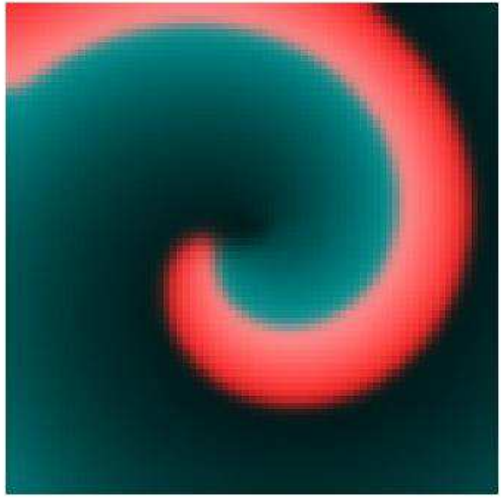}
\end{minipage}
\caption{A snap shot of a spiral wave.}
\label{fig:spiral}
\end{center}
\end{figure}

Let us for a moment think about the group $SE(2)$. This group is concerned with translations and rotations in the 2 dimensional plane, ($x,y$). Therefore, the elements of $SE(2)$ concerns only spatial transformations, not temporal transformations. Now, if we take a snap shot of a spiral wave as shown in Fig.(\ref{fig:spiral}) at a particular moment in time, then what $g\in SE(2)$ do we have which, when applied to this picture, give us the same picture?

Consider first of all rotations. If we locate the origin at the tip of the wave, then we must rotate by angle $\theta=2n\pi$, for $n=0,1,\ldots$. Any other angle would mean that we wouldn't get the exact same picture again. For translations, the only translation that would transform the wave back to itself in Fig.(\ref{fig:spiral}) is by $(X,Y)=(0,0)$. Therefore, the stabiliser for spiral waves can be said to be trivial. In \chg[af]{other words}, the \emph{Isotropy Subgroup} of spiral waves is trivial. Mathematically:

\begin{equation}
\label{eqn:isot}
T(g)\bu(\textbf{r},t) \neq\bu(\textbf{r},t),\quad \forall t,\forall g\neq I
\end{equation}
\\
where $I\in SE(2)$ is the trivial subgroup of $SE(2)$, i.e. it consists of \chgex[ex]{the identity element} in $SE(2)$.

Then, they took a solution to (\ref{eqn:meander_rde}) and represented it in a functional space, $\mathcal{B}$:

\chg[p15eqn1]{
\begin{equation}
\bu(\textbf{r},t) \mapsto \bU(t)
\end{equation}
}
Therefore, in $\mathcal{B}$, the PDE system (\ref{eqn:meander_rde}) can be represented as an ODE in the functional space:

\begin{equation}
\label{eqn:drift_rdeban}
\deriv{\bU}{t} = \booF(\bU)
\end{equation}

Equivariance is still present in this system and hence if \chg[p15eqn2]{$\bU(t)$} is a solution to (\ref{eqn:drift_rdeban}) then so too is $\tilde{\bU}(t)$ where:

\begin{equation}
\label{eqn:Utilde}
\tilde{\bU}(t) = T(g)\bU(t)
\end{equation}

\begin{ajf}
This now leads us to the solutions of (\ref{eqn:drift_rdeban}) being \emph{Equivariant} as follows:

\begin{eqnarray*}
\deriv{\tilde{\bU}}{t} &=& \booF(\tilde{\bU})\\
\Rightarrow \deriv{}{t}(T(g)\bU(t)) &=& \booF(T(g)\bU(t))\\
\Rightarrow T(g) \deriv{\bU}{t} &=& \booF(T(g)\bU(t))\\
\Rightarrow T(g) \booF(\bU) &=& \booF(T(g)\bU(t))
\end{eqnarray*}
\\
for fixed $T(g)$ and $\forall\bU\in \mathcal{B}$ and $\forall g\in SE(2)$. 
\end{ajf}
\begin{ajfthesis}
So, the condition of equivariance is:

\begin{eqnarray*}
T(g) \booF(\bU) &=& \booF(T(g)\bU(t))
\end{eqnarray*}
\\
for $\forall\bU\in \mathcal{B}$ and $\forall g\in SE(2)$. 
\end{ajfthesis}

In the appendix, we provide the definition of an \emph{Group Orbit} (\chg[p15ref1]{Sec}.(\ref{sec:orbit})). In our functional space, we take a solution, say $\bV(t)$, and apply \chg[af]{a particular transformation}from $SE(2)$ to this solution. This will give us our orbit. The orbit can therefore be viewed as being fixed in time. Also, all spiral wave solutions along this orbit have the same shape. If we took a solution along a particular orbit, then applying a transformation in $SE(2)$ to this solution, we get another solution along this orbit. However, the actual shape of the wave has not changed - only it's position and orientation. The only type of single armed spiral wave that corresponds to this situation is rigidly rotating spiral waves.\chg[p15sent]{}

So, what happens if we don't travel along the orbits, but transversally to the orbits? What we are observing in this case is that the wave is changing shape as we move across the orbits. Therefore, we get \emph{Meandering Spiral Waves} due to the shape of the wave changing \cite{sandstede01}.

\begin{figure}[tbp]
\begin{center}
\begin{minipage}[b]{0.6\linewidth}
\centering
\psfrag{a}[l]{$g$}
\psfrag{b}[l]{$\mathcal{G}$}
\psfrag{d}[l]{$\mathcal{B}$}
\psfrag{e}[l]{$\mathcal{M}$}
\psfrag{f}[l]{$U$}
\psfrag{g}[l]{$U'$}
\psfrag{h}[l]{$V$}
\psfrag{i}[l]{$V'$}
\psfrag{j}[l]{$g'$}
\includegraphics[width=0.9\textwidth]{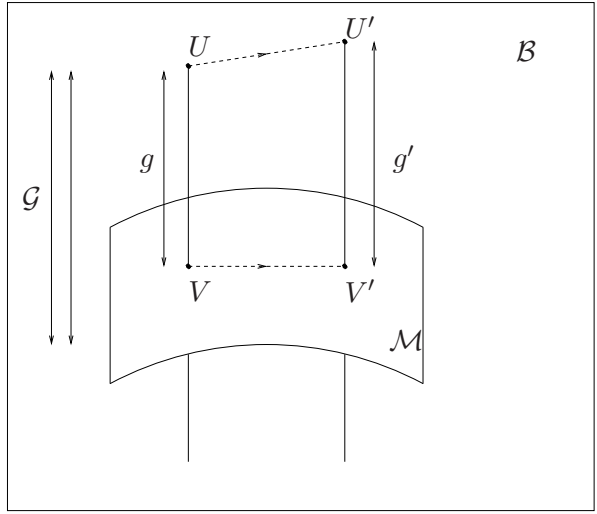}
\end{minipage}
\caption[Functional space representation of a spiral wave solution]{Decomposition of the movement in the functional space $\mathcal{B}$ onto movement along a manifold $\mathcal{M}$. We have that $V,V'\in\mathcal{M}$, $U,U''\in\mathcal{B}$ and $g,g'\in SE(2)$. $\mathcal{G}$ represents the group orbits in $\mathcal{B}$\cite{bik96}.}
\label{fig:ban1}
\end{center}
\end{figure}

With the above in mind, we therefore come to the picture as shown in Fig.(\ref{fig:ban1}). As we can see we have shown 2 particular orbits. Also, shown is a manifold, $\mathcal{M}$ which is a set of solutions such that it contains one and only one point from each orbit, and all \chg[af]{orbits}are transversal to the manifold. What is special about this manifold, is that if we know one solution, $V$, on the manifold, together with a corresponding transformation $g\in SE(2)$, then we can find any member of the orbit passing through $V$.

From Fig.(\ref{fig:ban1}), we also see that $V$ is related to $U$ as follows:

\begin{equation}
\label{eqn:UV}
U = T(g)V, \quad V\in\mathcal{M},g\in SE(2)
\end{equation}

Let us now consider how the solutions vary with time. Firstly, differentiate Eqn.(\ref{eqn:UV}) with respect to time:

\begin{ajf}
\begin{eqnarray*}
                     \deriv{\bU}{t} &=& \deriv{}{t}(T(g)\bV)\\
\Rightarrow              \booF(\bU) &=& \deriv{T(g)}{t}\bV+T(g)\deriv{\bV}{t}\\
\Rightarrow          \booF(T(g)\bV) &=& \deriv{T(g)}{t}\bV+T(g)\deriv{\bV}{t}\\
\Rightarrow          T(g)\booF(\bV) &=& \deriv{T(g)}{t}\bV+T(g)\deriv{\bV}{t}\\
\Rightarrow T(g^{-1})T(g)\booF(\bV) &=& T(g^{-1})\deriv{T(g)}{t}\bV+T(g^{-1})T(g)\deriv{\bV}{t}\\
\Rightarrow              \booF(\bV) &=& T(g^{-1})\deriv{T(g)}{t}\bV+\deriv{\bV}{t}
\end{eqnarray*}
\end{ajf}

\begin{ajfthesis}
\begin{eqnarray*}
\deriv{\bU}{t} &=& \deriv{}{t}(T(g)\bV)\\
\Rightarrow T(g)\booF(\bV) &=& \deriv{T(g)}{t}\bV+T(g)\deriv{\bV}{t}\\
\Rightarrow \booF(\bV) &=& T(g^{-1})\deriv{T(g)}{t}\bV+\deriv{\bV}{t}
\end{eqnarray*}
\end{ajfthesis}

Now, $\booF(\bV)$ determines the general motion of the solutions and therefore can be projected onto the group orbits and onto $\mathcal{M}$. Therefore, if $\booF(\bV)=\booF_{\mathcal{G}}(\bV)+\booF_{\mathcal{M}}(\bV)$, we get:

\begin{eqnarray}
\label{eqn:FG}
\booF_{\mathcal{G}}(\bV) &=& T(g^{-1})\frac{dT(g)}{dt}\bV\\
\label{eqn:FM}
\booF_{\mathcal{M}}(\bV) &=& \deriv{\bV}{t}
\end{eqnarray}

Therefore, we have reduced Eqn.(\ref{eqn:drift_rdeban}) to Eqn. (\ref{eqn:FM}), which lacks the symmetry of the first equation (i.e. it is generic). This is easy to see since the equation of motion along the manifold, $\mathcal{M}$, is independent of spatial coordinates, whereas the original equation is clearly dependent \chgex[ex]{on} spatial coordinates and therefore possesses a symmetry. Now, it is well documented \cite{DS1} that systems without symmetries are much easier to study than those which permit symmetric properties such as the reaction diffusion equations (\ref{eqn:meander_rde}) and (\ref{eqn:drift_rdeban}).

Before moving on, we must speak briefly about $\booF_{\mathcal{G}}$. If are considering meandering waves and therefore moving transversal to the group orbits, we note that we cross each group orbit at one point. We do not travel along an individual group orbit. If we did travel along a group, no matter how long, we would not get a meandering wave since this would mean that the shape of the wave is not changing. Therefore, we say that $\booF_{\mathcal{G}}$ is \chg[af]{in fact}the infinitesimal motion along the group.

We noted earlier that the Isotropy Subgroup for spiral waves is trivial. Now, we know that movement along the orbit corresponds to a spiral wave of a particular shape being acted upon by a transformation in $SE(2)$. Our manifold on the other hand is concerned with movement transversal to the orbits and therefore we must define the manifold in such a way that any point on the manifold ($\bV\in\mathcal{M}$) would be moved away from $\mathcal{M}$ by any non-identical transformations. Therefore, the following conditions are used as a definition of $\mathcal{M}$:

\begin{eqnarray}
\label{eqn:xtipiso}
v_1(0,0) &=& u_{10}\\
\label{eqn:ytipiso}
v_2(0,0) &=& u_{20}\\
\label{eqn:rotimp}
\partial_xv_1(0,0) &=& 0
\end{eqnarray}

For a moment, let us think about spiral waves and how we determine what motion they are displaying. To see how the wave is behaving we must locate its tip and observe how this tip moves. Locating the tip is one of the \chgex[ex]{most important tasks in} studying spiral waves as accurate numerical calculations can take a significant amount of time to perform.

\begin{figure}[tbp]
\begin{center}
\begin{minipage}[b]{0.6\linewidth}
\centering
\psfrag{a}[l]{$y$}
\psfrag{b}[l]{$x$}
\psfrag{c}[l]{$v_1(x,y)=u_{10}$}
\psfrag{d}[l]{$v_2(x,y)=u_{20}$}
\includegraphics[width=0.9\textwidth]{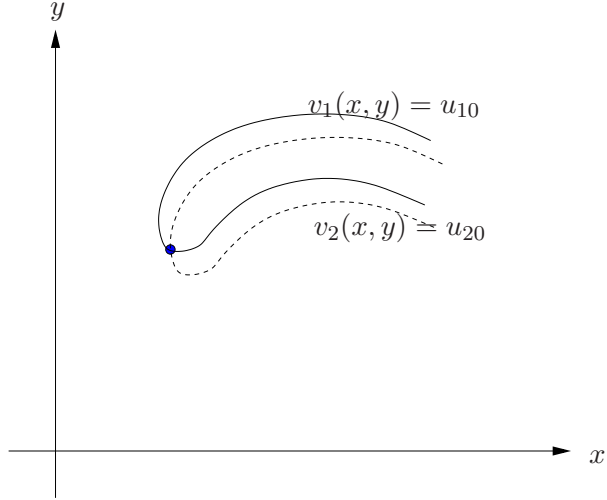}
\end{minipage}
\caption[Spiral wave tip location]{The intersection (blue dot) of 2 isolines in the plane}
\label{fig:iso}
\end{center}
\end{figure}

So how do we locate the tip of the spiral wave? One way is to define the tip as the intersection of 2 isolines (see Fig.(\ref{fig:iso})). We can therefore see how the wave evolves in time by locating the tip coordinates at each time step. Let us call the isolines $v_1$ and $v_2$:

\begin{eqnarray}
v_1(x,y) &=& u_{10}\\
v_2(x,y) &=& u_{20}
\end{eqnarray}
\\
where $u_{10}$ and $u_{20}$ are arbitrary constants. Now $u_{10}$ and $u_{20}$ need to be chosen carefully and in particular they must be located within the ``No Mans Land" as described in \cite{FHN1}, otherwise a spiral wave may appear to have no tip by this definition. 

Now, let us consider motion along the group orbit. We know, from the previous section, that the equation of motion is given by Eqn.(\ref{eqn:FG}). Biktashev et al then showed that by considering the motion along the group orbit we get:

\begin{equation}
\booF_{\mathcal{G}}(\bV) = -((\textbf{c},\nabla)+\omega\partial_\theta)\bV
\end{equation}
\\
for some $\bc\in\mathbb{R}^2$ and $\omega\in\mathbb{R}$.

Since $\booF(\bV)=\booF_{\mathcal{G}}(\bV)+\booF_{\mathcal{M}}(\bV)$, then the movement along the manifold is $\booF_{\mathcal{M}}(\bV)=\booF(\bV)-\booF_{\mathcal{G}}(\bV)$. 

Therefore, we get that in the PDE formulation, our reaction diffusion equation now becomes:

\begin{eqnarray}
\booF_{\mathcal{M}}(\bV) &=& \booF(\bV)-\booF_{\mathcal{G}}(\bV)\\
\Rightarrow \deriv{\bV}{t} &=& \booF(\bV)-\booF_{\mathcal{G}}(\bV)\\
\label{eqn:pderot}
\Rightarrow \pderiv{\bv}{t} &=& \textbf{D} \nabla^2\bv+\boof(\bv)+(\textbf{c},\nabla)\bv+\omega\partial_{\theta}\bv
\end{eqnarray}

So, what have we got in this case? Well, we now have a Reaction-Diffusion-Advection equation of the motion along the manifold, with each solution $\bv(\br,t)$ being located on the manifold. Therefore, the reaction diffusion equation is now in a comoving frame of reference (with the origin located at the tip of the spiral wave) and not a laboratory frame as in Eqn.(\ref{eqn:meander_rde}). In Eqn.(\ref{eqn:pderot}), we have that $\textbf{c}$ and $\omega$ are changing with time. Therefore, for each solution of Eqn.(\ref{eqn:pderot}), $\bv(\br,t)$, we have a corresponding unique value for $\textbf{c}$ and $\omega$. So by solving Eqn.(\ref{eqn:pderot}), we can find each $\bv,\textbf{c}$ and $\omega$, for each instant of time. Hence we can say that $\textbf{c}$ and $\omega$ can be found by whatever method we decide to use and so there exists a unique $\{\bv,\textbf{c},\omega\}$ for each point on the manifold.

Also, since each $\textbf{c}$ and $\omega$ change as each solution $\bv(x,y,t)$ changes, then Eqn.(\ref{eqn:pderot}) together with Eqns.(\ref{eqn:xtipiso})-(\ref{eqn:rotimp}) can be viewed as a dynamical system. Finally, Eqns. (\ref{eqn:xtipiso})-(\ref{eqn:rotimp}) can be viewed as the quotient system of \chgex[ex]{the} spiral wave solution to Eqn.(\ref{eqn:pderot}).

The next aim of the paper was to derive the equations of motion for the tip of the wave. Their analysis produced the following relation:

\begin{eqnarray}
\label{eqn:dynamics_T}
T(\{\textbf{c}dt,\omega dt\}) &=& {\rm{id}}-dt(\textbf{c},\nabla)-dt\omega\partial_\theta
\end{eqnarray}
\\
where $dt$ is the timestep. Also, by considering motion along the group orbit, i.e. Eqn.(\ref{eqn:FG}), they were able to establish another relation:

\begin{eqnarray}
\label{eqn:dynamics_T_split}
T(\{\textbf{R}+d\textbf{R},\Theta+d\Theta\}) &=& T(\{\textbf{R},\Theta\})T(\{dt\textbf{c},dt\omega\})
\end{eqnarray}
\\
where $d\bR$ and $d\Theta$ are infinitesimal shifts in $\bR$ and $\Theta$ respectively. Therefore, by taking Eqns.(\ref{eqn:dynamics_T}) and (\ref{eqn:dynamics_T_split}), and also considering,

\begin{equation}
T(g):\textbf{z}\mapsto \bR+\textbf{z}e^{\gamma\Theta}
\end{equation}
\\
where $\gamma$ is the rotational matrix given by:

\begin{equation*}
\gamma = \left(\begin{array}{cc}0 & -1\\ 1 & 0 \end{array}\right)
\end{equation*}
\\
then we have,

\begin{eqnarray}
T\{\bR+d\bR, \Theta+d\Theta\}\textbf{z} &=& T\{\bR,\Theta\}T\{dt\bc,dt\omega\}\textbf{z}\\
\Rightarrow \bR+d\bR+e^{\gamma(\Theta+d\Theta)}\textbf{z} &=& T\{\bR,\Theta\}(dt\bc+e^{\gamma dt\omega}\textbf{z})\\
\Rightarrow d\bR+e^{\gamma\Theta}e^{\gamma d\Theta}\textbf{z} &=& dte^{\gamma\Theta}\bc +e^{\gamma dt\omega}e^{\gamma\Theta}\textbf{z}
\end{eqnarray}
\\
which leads nicely to the equations of motion:

\begin{eqnarray}
\label{eqn:motionR1}
\deriv{\bR}{t} &=& e^{\gamma\Theta(t)}\bc(t)\\
\label{eqn:motionom1}
\deriv{\Theta}{t} &=& \omega(t)
\end{eqnarray}
\\
or, using complex notation:

\begin{eqnarray}
\label{eqn:motionR}
\deriv{R}{t} &=& c(t)e^{i\Theta(t)}\\
\label{eqn:motionom}
\deriv{\Theta}{t} &=& \omega(t)
\end{eqnarray}

It is easy that to see when we have rigid rotation, i.e. $c$ and $\omega$ are constant, then we can easily integrate Eqns.(\ref{eqn:motionR}) and (\ref{eqn:motionom}) to get:

\begin{eqnarray}
\label{eqn:dynamics_RW}
R &=& R_0-\frac{ic}{\omega}e^{i(\omega t+\Theta_0)}
\end{eqnarray}
\\
where $R=X+iY$ are the tip coordinates, $R_0=X_0+iY_0$ is the initial position of the tip of the wave, and $c=c_x+ic_y$ and $\omega$ are the translational and angular velocities. Of course, Eqn.(\ref{eqn:dynamics_RW}) is the equation of a circle, hence proving that the trajectory traced out by the tip of a rigidly rotating spiral wave is a perfect circle.

Let us now consider meandering spiral waves. It is well documented that there are 2 frequencies present in a simple meandering spiral wave (again, we name just a few of these for reference - \cite{Winf91}, \cite{Bark90}, \cite{Hakim97}, \cite{Lugo89}). It has also be shown that the transition from rigid rotation to meandering, is via a Hopf Bifurcation (\cite{Bark90}, \cite{karma90}). This Hopf Bifurcation affects the motion of the wave by introducing a new frequency into the system, the Hopf Frequency.

Biktashev et al proposed that the forms of $\bc$ and $\omega$ should be derived straight from Hopf Bifurcation Theory. They proposed to use the following form:


\begin{ajf}

Let us for a moment consider a Hopf Bifurcation and its definition. Consider the following dynamical system:

\begin{eqnarray}
\label{eqn:pderotds}
\frac{\partial{v}}{\partial{t}} &=& \textbf{D} \nabla^2v+f(v,\epsilon)+(\textbf{c},\nabla)v+\omega\partial_{\theta}v\\
\label{eqn:tip1}
v_1(0,0) &=& u_{10}\\
\label{eqn:tip2}
v_2(0,0) &=& u_{20}\\
\label{eqn:tip3}
\partial_xv_1(0,0) &=& 0
\end{eqnarray}
with $v\in\mathbb{R}^k$ and $v=\{v_1,v_2,\ldots,v_k\}$. Suppose that in this dynamical system there are fixed points, and associated with these fixed points are eigenvalues. If for one of these fixed points, $v_*$, 2 of these eigenvalues take the form:
\begin{equation}
\lambda_{1,2} = \epsilon\pm i\omega(\epsilon)
\end{equation}
then we see that for $\epsilon<0$ and with $Re\{\lambda_n\}<0$ for $n=3,...,k$, the fixed point will be stable, and at $\epsilon>0$, the fixed point will be unstable. So, we can see that as we vary the parameter $\epsilon$, the fixed point undergoes a change in stability.\\
\\
The Hopf Bifurcation Theorem (see \cite{kuznetsov} and also Section (\ref{sec:HB})) states that when a fixed point of a dynamical system undergoes a Hopf Bifurcation, 2 of its eigenvalues cross the imaginary axis and a limit cycle is generated in the system. Therefore, at the Hopf Bifurcation, there are 2 eigenvalues of that particular fixed point having the form $\lambda_{1,2} = \pm i\omega(0)=\pm i\omega_0$. It also states that a dynamical system possessing a Hopf Bifurcation can be reduced to the normal form of a Hopf Bifurcation:

\begin{equation}
\label{eqn:HBnormal}
\dot{z} = \alpha z-\beta z|z|^2
\end{equation}
for $\alpha,\beta\in\mathbb{C}$, and with $z$ being a limit cycle solution. Also, $\alpha$ is actually one of the  eigenvalues related to the the fixed point, $v_*$, whose real part changes sign after the Hopf Bifurcation (which one of these is irrelevant). $\beta$ on the other hand is known as the \emph{Lyapunov Quantity} and determines whether the system is \emph{supercritical} or \emph{subcritical}.\\
\\
Okay, bearing the above in mind let us consider the above dynamical system, Eqns.(\ref{eqn:pderotds})-(\ref{eqn:tip3}). Then, a solution of this system can be expressed as a linear combination of its eigenvalues and corresponding eigenvectors:
\begin{equation}
v = \sum_{n=1}^k{a_nw_ne^{\lambda_nt}}
\end{equation}
Also, we have that the spiral is changing shape since the speed of the wave and the frequency are no longer constant. Therefore, $c(t)$ and $\omega(t)$ and no longer constant, but are some function of $z$ (see Sec.(\ref{sec:HB})):

\end{ajf}


\begin{eqnarray}
\label{eqn:ct}
c(t) &=& c_0+c_1z+\bar{c}_1\bar{z}+O(|z|^2)\\
\label{eqn:omt}
\omega(t) &=& \omega_0+\omega_1z+\bar{\omega}_1\bar{z}+O(|z|^2)\\
\label{eqn:HBnormal}
\dot{z} &=& \alpha z-\beta z|z|^2
\end{eqnarray}

Therefore, using Eqns.(\ref{eqn:ct}) and (\ref{eqn:omt}), together with $z=re^{i(\omega_Ht+\phi)}$, we can integrate Eqns.(\ref{eqn:motionR}) and (\ref{eqn:motionom}) directly to get:

\begin{eqnarray}
R &=& R_0+A\left(\begin{array}{c}\sin(\alpha)\\-\cos(\alpha) \end{array}\right)\nonumber\\
\label{eqn:solxy}
&&+B
\left(\begin{array}{cc}\cos(\alpha) & -\sin(\alpha)\\ \sin(\alpha) & \cos(\alpha) \end{array}\right)\left(\begin{array}{c} m_1\sin(\beta)+n_2\cos(\beta) \\ m_2\sin(\beta)-n_1\cos(\beta) \end{array}\right)
\end{eqnarray}
\\
where:

\begin{eqnarray}
\alpha &=& \omega_0t+\theta_0\\
\beta &=& \omega_Ht+\phi\\
A &=& \frac{c_0}{\omega_0}\\
B &=& \frac{2r}{\omega_H(\omega_H^2-\omega_0^2)}\\
c_1 &=& c_{11}+ic_{12}\\
\omega_1 &=& \omega_{11}+i\omega_{12}\\
m_1 &=& \omega_H^2c_{11}-\omega_0c_0\omega_{11}\\
m_2 &=& \omega_H(c_0\omega_{12}-\omega_0c_{12})\\
n_1 &=& \omega_H(c_0\omega_{11}-\omega_0c_{11})\\
n_2 &=& \omega_H^2c_{12}-\omega_0c_0\omega_{12}
\end{eqnarray}

Clearly, if there is no limit cycle present, i.e. $r=0$, we get the equation of motion for the rigidly rotating waves, since we will have that $A=0$.

\begin{figure}[htbp]
\begin{center}
\begin{minipage}[htbp]{0.49\linewidth}
\centering
\includegraphics[width=0.8\textwidth]{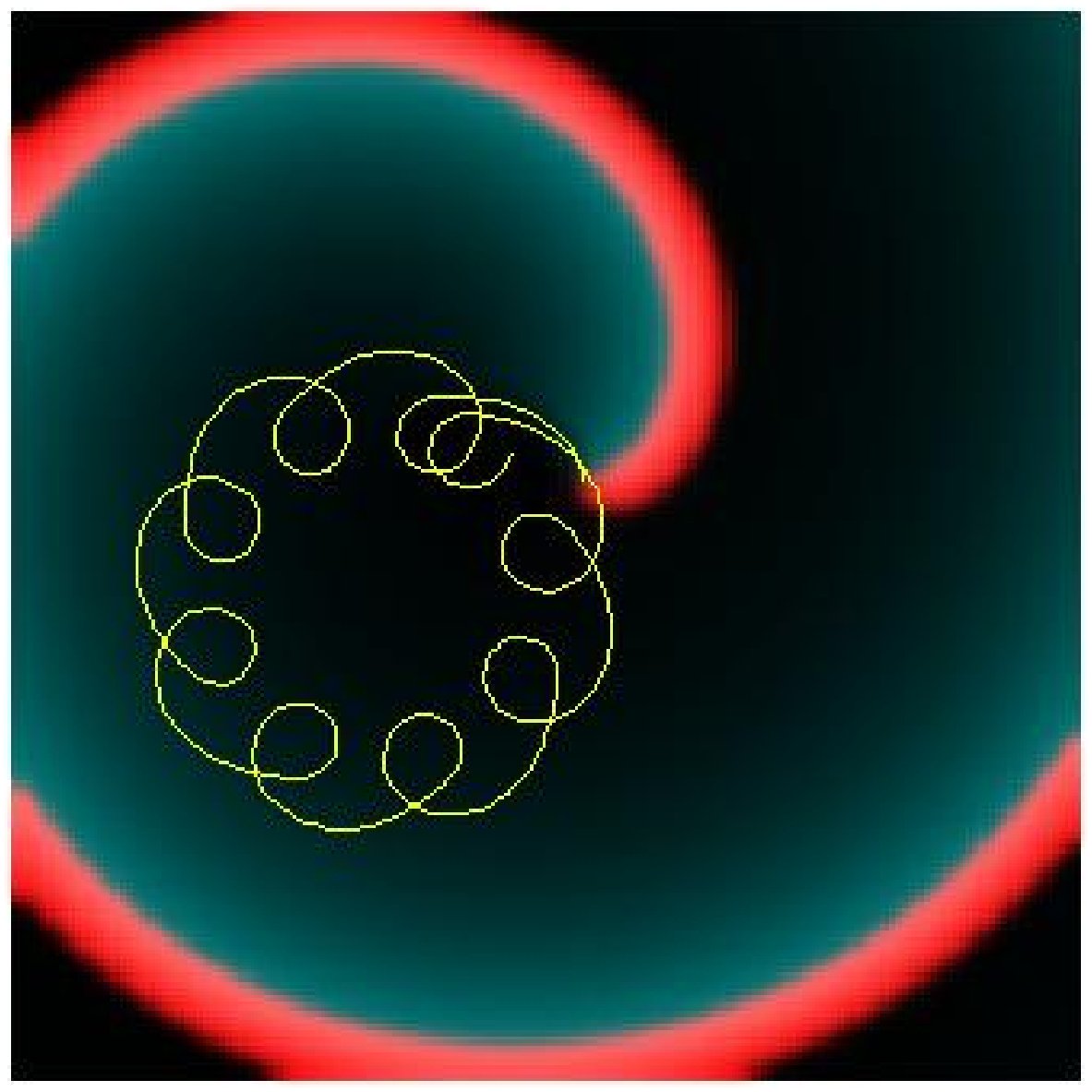}
\end{minipage}
\begin{minipage}[htbp]{0.49\linewidth}
\centering
\includegraphics[width=0.8\textwidth]{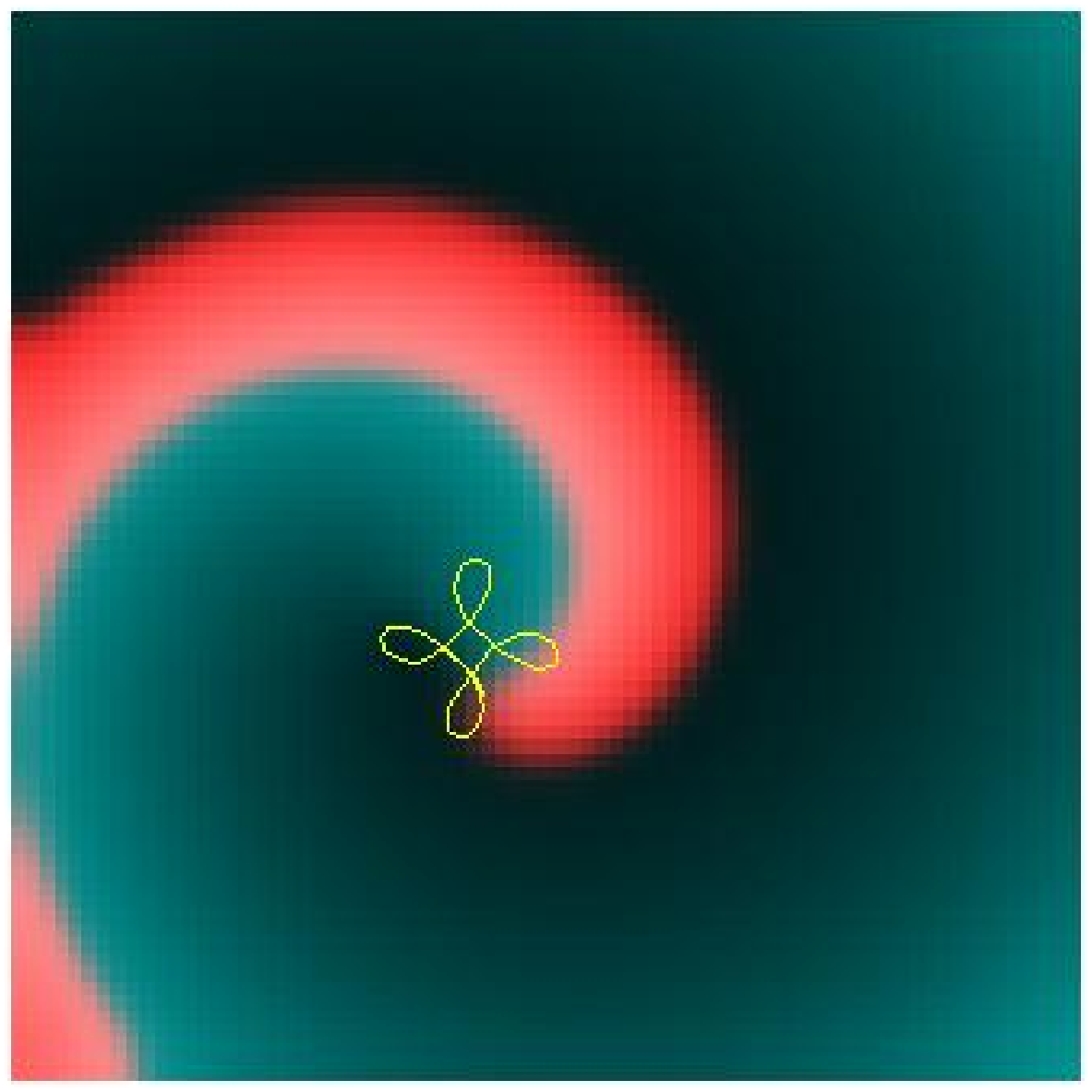}
\end{minipage}
\caption[Meandering spiral waves: numerical solution]{Meandering Spiral Waves. The picture on the left showing that the tip is tracing out inward flower patterns, whilst the picture on the right is showing outward petals.}
\label{fig:mrw}
\end{center}
\end{figure}

\begin{figure}[htbp]
\begin{center}
\begin{minipage}[t]{0.49\linewidth}
\centering
\includegraphics[width=1.0\textwidth]{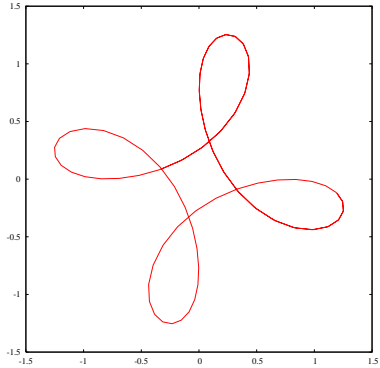}
\end{minipage}
\caption[Meandering spiral waves: analytical solution]{An outward meandering wave as produced by the analytical solution with $R=-0.530206$, $A=-0.0337636$, $\omega_0=-2.66298$, $\theta_0=10.4742$, $\omega_H=3.55141$, $\phi=-6.10354$, $m_1=5.54953$, $m_2=20.9893$, $n_1=7.1913$ and $n_2=20.9925$}
\label{fig:mrwode}
\end{center}
\end{figure}

Finally, we shown in Fig.(\ref{fig:mrw}) a typical 2 frequency meandering wave with the picture on the left showing inward facing petals and on the right outward facing. In Fig.(\ref{fig:mrwode}) we show a trajectory using Eqn.(\ref{eqn:solxy}), confirming that the trajectories produced in the pde system can be replicated in the ODE system.



\subsection{Drift}

Drift can be described as non-stationary rotation around a point of rotation which is moving around the medium. We shall be concentrating on drift caused by a symmetry breaking perturbation. The Reaction-Diffusion system that we now consider is:

\begin{equation}
\label{eqn:drift_rde_pert}
\partial_t{\bu} = \textbf{D}\nabla^2\bu+\bof(\bu)+\epsilon\bh(\bu,\nabla\bu,\br,t)
\end{equation}
\\
where $\bu=\bu(\br,t)=(u^{(1)},u^{(2)},\hdots,u^{(l)})\in \mathbb{R}^l, l\geq2, \textbf{r}=(x,y)\in \mathbb{R}^2,|\epsilon|\ll1$.

Without the perturbation, i.e. with $\epsilon=0$, we have a system of equations which are equivariant under the actions of elements belonging to the group $\mathbb{SE}(2)$. We choose the perturbation $\bh$ such that this symmetry is broken. From a physical point of view, the perturbation is a deviation from the unperturbed solution (i.e. when $\epsilon=0$). This perturbation could be dependent, on either spatial and/or temporal coordinates, and maybe even on the variable $u$. The important point to note is that the perturbation must be such to break the symmetry of the \chg[af]{original} Reaction-Diffusion system.

\begin{figure}[tbp]
\begin{center}
\begin{minipage}[t]{0.49\linewidth}
\centering
\includegraphics[width=1.0\textwidth]{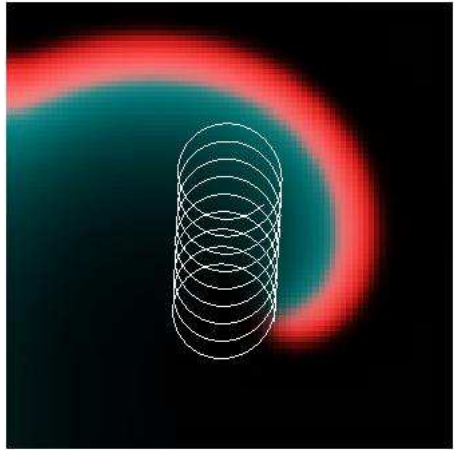}
\end{minipage}
\caption[Drifting rigidly rotating spiral wave]{A drifting rigidly rotating spiral wave in the plane using Barkley's model, where the drift is inhomogeneitity induced and the parameter $a=a_0+a_1x$.}
\label{fig:dynamics_drift}
\end{center}
\end{figure}

We are now looking at systems which contain small parameters. In this case, the small parameter is $\epsilon$. So, we will be using \emph{Perturbation Theory} to study this type of motion. Fig.(\ref{fig:dynamics_drift}) shows a rigidly rotating wave drifting due to inhomogeneitity induced drift (translational symmetry breaking perturbation).

\chgex[1]{}\chgex[]{
The concept of drift has been studied by many authors \cite{krinsky68,krinsky83,agladze87,dav88,pertsov88}. In 1988, Keener \& Tyson applied a singular perturbation technique to the study of the traveling waves in excitable media \cite{keener88a}, having noted that the concept of drift can be formulated as a perturbation problem. Taking advantage of this technique, Keener studied the dynamics of Scroll Waves (3D waves which rotate around a central filament, and whose cross section is a spiral), and in particular when the central filament is subject to drift \cite{keener88b}. In the equations of motion of the filament which Keener derived, there was a coefficient, $b_2=\langle\textbf{D}V_x,Y_x\rangle$, which Biktashev et al showed was in fact the filament tension \cite{bik94}. This coincided with an earlier observation by Panfilov et al that the tension of the central filament was related to the slow drift of the filament \cite{panf87}. However, in Keener's formulation little was known of the properties of the response functions, in particular that the response functions were localised at the core of the spiral wave.

In his formulation, Keener took a generic reaction-diffusion system of equations in the laboratory frame of reference and considered the two dimensional problem:

\begin{equation*}
\pderiv{u}{t} = D\nabla^2u+f(u)
\end{equation*}
\\
where $u=u(r,t)$, and $r=(x,y)$. They then considered a solution of the form:

\begin{equation*}
u(r,t) = U(r,t)+v(r,t)
\end{equation*}
\\ 
where $U$ is a known solution and $v$ is a small perturbation. This yields a linearised problem for $v$:

\begin{equation*}
L\alpha = \pderiv{\alpha}{t}- D\nabla^2\alpha-F(U)\alpha
\end{equation*}
\\
where $F(U)=\pderiv{f(u)}{u}\left.\right|_u=U$. The corresponding adjoint linearised equation was:

\begin{equation*}
L^*\beta = \pderiv{\beta}{t}+D\nabla^2\beta+F^T(U)\beta
\end{equation*}

Therefore, we note that in this frame of reference, the critical eigenvalues are all zero and that the eigenfunctions and response functions are all time dependent. The inner product used by Keener therefore took the form:

\begin{equation*}
\langle u,v\rangle = \lim_{r\rightarrow\infty}\frac{\int_0^P\int_0^{2\pi}\int_0^ru.vr\dd r\dd\theta\dd t}{\pi r^2P}
\end{equation*}

In 1995, Biktashev et al used used a similar perturbation technique to produce a theory of drift of a spiral wave \cite{bik95}. We shall review this in detail in the next section. However, we should note that Biktashev et al formulated their problem in a slightly different way, making use of the conjecture that the response functions were localised. This lead to a different formulation of the scalar product between the Goldstone modes (eigenfunction to the linearised reaction-diffusion equation) and the response functions, leading to a difference in the normalisation of the response functions. They also considered solutions in a rotating frame of reference, in which the rigidly rotating spiral wave solutions are stationary.} 


\subsubsection{Theory of Drift \cite{bik95}}

We shall review in some detail, the paper by Biktashev and Holden, published in 1995 on the theory of drift. Although the paper is entitled in such a way as to suggest that only resonant drift is considered, it actually considers just a symmetry breaking perturbation, with several excellent specific examples included for good measure.

They commence the theoretical work in this paper by considering the following generic Reaction-Diffusion Equation (Eqn.(\ref{eqn:drift_rde_pert})). They aimed to study spiral wave solutions of (\ref{eqn:drift_rde_pert}) and therefore sought solutions of the form:

\begin{equation}
\label{eqn:u}
u = u(\textbf{r},t) = u(\rho,\theta+\omega t)
\end{equation}
\\
where $\textbf{r}=(x,y)$ and $\rho=\rho(\textbf{r})$, $\theta=\theta(\textbf{r})$ are polar co-ordinates in the $(x,y)$ plane. In particular, we shall assume that the solution to (\ref{eqn:drift_rde_pert}) supports rigid rotation. This means that $\omega$ is constant. Furthermore, they \chg[af]{considered} the unperturbed situation and assumed that the solution to (\ref{eqn:drift_rde_pert}) is:

\begin{equation}
\label{eqn:drift_usol}
u = u_0(\textbf{r},t) = u_0(\rho,\theta+\omega t)
\end{equation}
\\
with $u_0$ being $2\pi$-periodic. This means that (\ref{eqn:drift_rde_pert}) becomes:

\begin{equation}
\label{eqn:Usys}
\partial_t{u_0} = \textbf{D}\nabla^2u_0+f(u_0)
\end{equation}

It is well known that spiral wave solutions to (\ref{eqn:drift_rde_pert}) with no perturbation are invariant under euclidean symmetry as well temporal transformations. This implies that (\ref{eqn:drift_usol}) is a solution to (\ref{eqn:drift_rde_pert}) with $\epsilon=0$ then so too is:

\begin{equation}
\label{eqn:drift_ushift}
u_{0;{\delta \textbf{r},\delta t}} = u_0(\textbf{r}-\delta \textbf{r},t-\delta t)
\end{equation}

It can \chg[af]{therefore} be said that equation (\ref{eqn:drift_ushift}) generates a 3 dimensional manifold of solutions to (\ref{eqn:drift_rde_pert}).

Also, since they know considered a rigidly rotating spiral waves, then the solution is stable. This means that any small deviation away from the solution eventual ends up in a small vicinity of the unperturbed solution. Hence we arrive at the following \textit{Stability Postulate}:

\textit{Stability Postulate}: Any solution to (\ref{eqn:drift_rde_pert}) at $\epsilon=0$ with initial conditions sufficiently close to (\ref{eqn:u}) tends to one of the solutions to (\ref{eqn:drift_ushift}) with some $\delta \textbf{r}$, $\delta t$ as $t\rightarrow\infty$, i.e. the family (\ref{eqn:drift_ushift}) is stable as a whole against small perturbations in initial conditions.

They firstly considered the system of solutions to (\ref{eqn:drift_rde_pert}) with $\epsilon=0$ and also with \chg[af]{a} perturbation present in the solution. A small deviation in the solution could arise from a small deviation in the initial conditions. It was assumed that the solution to (\ref{eqn:drift_rde_pert}) is of the form:

\begin{equation}
\label{eqn:upert}
u = u_0(\textbf{r},t)+v(\textbf{r},t) = u_0(\rho,\theta+\omega t)+v(\textbf{r},t)
\end{equation}
\\
where \chg[af]{$v(\textbf{r},t)=O(\epsilon)$} is the perturbation. Therefore, the initial conditions are:

\begin{equation}
\label{eqn:upertini}
u(\textbf{r},0) = u_0(\textbf{r},0)+v_0(\textbf{r})
\end{equation}

Substituting (\ref{eqn:upert}) into (\ref{eqn:drift_rde_pert}) and splitting out unperturbed from perturbed parts, we get:

\begin{eqnarray}
\label{eqn:vUsys}
\frac{\partial{u_0}}{\partial{t}} &=& \textbf{D}\nabla^2u_0+ f(u_0)\\
\label{eqn:vsys1}
\frac{\partial{v}}{\partial{t}} &=& Lv
\end{eqnarray}
\\
where linearised operator $L$ is given by:

\begin{equation}
\label{eqn:op}
L\alpha = \textbf{D}\nabla^2\alpha+\pderiv{f}{u}\alpha
\end{equation}

We see that Eqn.(\ref{eqn:vsys1})  is the equation of motion for the small perturbation from the original solution. $v(\textbf{r},t)$ can be thought of as the remainder terms if we are using Asymptotic Methods. Since they considered rigidly rotating spiral waves, it is more convenient to perform the analysis in a rotating frame of reference, i.e.:

\begin{eqnarray*}
t & \rightarrow & \tilde{t}=t\\
\textbf{r} & \rightarrow & \tilde{\textbf{r}}: \rho(\tilde{\textbf{r}})=\rho(\textbf{r}), \theta(\tilde{\textbf{r}})=\theta(\textbf{r})+\omega t
\end{eqnarray*}

Hence, (\ref{eqn:vsys1}) becomes:

\begin{equation}
\label{eqn:vsys}
\frac{\partial{v}}{\partial{\tilde{t}}} = \tilde{L}v
\end{equation}
\\
where:

\begin{equation}
\label{eqn:tilL}
\tilde{L}v = \textbf{D}\tilde{\nabla}^2v + v\textbf{F}(u_0(\tilde{\textbf{r}}))-\omega\frac{\partial{v}}{\partial{\tilde{\theta}}}
\end{equation}

Now, it was noted that the study of the linear stability of the operator, $\tilde{L}$, by studying the eigenvalues and eigenvectors of the linear system. It is well known that there are three critical eigenvalues who have with zero real parts for rigidly rotating spiral waves. These come from the symmetry of the unperturbed system. So, it can be said that the solution (\ref{eqn:drift_usol}) differs from (\ref{eqn:drift_ushift}) at small $\delta\textbf{r}$ and $\delta t$ by a function linearly composed of three linearly independent functions.


\begin{ajf}

Let us for a minute think about the perturbation $v(\textbf{r},t)=v(x,y,t)$. This perturbation is the difference between one solution, $u_0(x,y,t)$, and the next solution, $u_0(x',y',t')$. Let us consider the shift from one solution to the next along the $x$-axis. The first solution is $u_0(x,y,t)$ and the second will be $u_0(x-dx,y,t)$. The difference between these solutions is the $x$-component of $v(x,y,t)$, $v_x$ say:
\begin{equation}
\label{eqn:vx1}
v_x = u_0(x-dx,y,t)-u_0(x,y,t)
\end{equation}
Expanding the second solution about $dx=0$ to first order gives:
\begin{equation}
\label{eqn:vx2}
v_x = u_0(x,y,t)-dx\frac{du_0}{dx}-u_0(x,y,t) = -dx\frac{du_0}{dx}
\end{equation}
A Similar argument for shift in the $y$ direction and time yield the following set of equations:
\begin{equation}
\label{eqn:vx}
v_x = -dx\frac{du_0}{dx}
\end{equation}
\begin{equation}
\label{eqn:vy}
v_y = -dy\frac{du_0}{dy}
\end{equation}
\begin{equation}
\label{eqn:vt}
v_t = -dt\frac{du_0}{dt}
\end{equation}
The perturbation $v(x,y,t)$  can be thought of as a linear combination of the shifts in the $x,y$ directions as well as shifts in time for small enough shifts. Therefore:
\begin{eqnarray*}
v(x,y,t)&=&v_x+v_y+v_t\\
\Rightarrow v(x,y,t)&=&-dx\frac{du_0}{dx}-dy\frac{du_0}{dy}-dt\frac{du_0}{dt}
\end{eqnarray*}
\begin{equation}
\label{eqn:vfull}
\Rightarrow v(x,y,t)\ =\ V_xdx+V_ydy+V_tdt
\end{equation}
We mentioned earlier that it is more convenient to study spiral wave solutions from a rotating frame of reference. The reason for this will become apparent later on in our analysis. Let us consider how the change from a stationary frame to a rotating frame and look at the system in polar coordinates.\\
\\
The coordinates in the stationary frame of reference are:
\begin{equation}
\label{eqn:coord}
(x,y) = (\rho \cos(\theta),\rho \sin(\theta))
\end{equation}
Changing the frame of reference to the rotating frame, we come to the following change of coordinates:
\begin{equation}
\label{eqn:rotcoord}
(\tilde{x},\tilde{y}) = (\tilde{\rho} \cos(\tilde{\theta}-\omega\tilde{t}),\tilde{\rho} \sin(\tilde{\theta}-\omega\tilde{t}))
\end{equation}
Using the following set of relations:
\begin{eqnarray}
\label{eqn:tilderho}
\tilde{\rho} & = & \sqrt{\tilde{x}^2+\tilde{y}^2}\\
\label{eqn:tildeth}
\tan(\tilde{\theta}-\omega\tilde{t}) & = & \frac{\tilde{y}}{\tilde{x}}
\end{eqnarray}
we can differentiate (\ref{eqn:tilderho}) to obtain:
\begin{equation}
\label{eqn:rhox}
\frac{\partial{\tilde{\rho}}}{\partial{\tilde{x}}} = \frac{\tilde{x}}{\sqrt{\tilde{x}^2+\tilde{y}^2}} = \cos (\tilde{\theta}-\omega\tilde{t})
\end{equation}
\begin{equation}
\label{eqn:rhoy}
\frac{\partial{\tilde{\rho}}}{\partial{\tilde{y}}} = \frac{\tilde{y}}{\sqrt{\tilde{x}^2+\tilde{y}^2}} = \sin (\tilde{\theta}-\omega\tilde{t})
\end{equation}
Likewise, we can differentiate (\ref{eqn:tildeth}) to get the following:
\begin{eqnarray*}
\sec ^2(\tilde{\theta}-\omega\tilde{t}) \frac{\partial{\tilde{\theta}}}{\partial{\tilde{x}}} & = & -\frac{\tilde{y}}{\tilde{x}^2}\\
\Rightarrow \frac{\tilde{\rho}^2}{\tilde{x}^2} \frac{\partial{\tilde{\theta}}}{\partial{\tilde{x}}} & = & -\frac{\tilde{y}}{\tilde{x}^2}
\end{eqnarray*}
Hence:
\begin{equation}
\label{eqn:thetax}
\frac{\partial{\tilde{\theta}}}{\partial{\tilde{x}}} = -\frac{1}{\tilde{\rho}}\sin(\tilde{\theta}-\omega\tilde{t})
\end{equation}
By similar analysis:
\begin{equation}
\label{eqn:thetay}
\frac{\partial{\tilde{\theta}}}{\partial{\tilde{y}}} = \frac{1}{\tilde{\rho}}\cos(\tilde{\theta}-\omega\tilde{t})
\end{equation}
Going back to Eqns (\ref{eqn:vx}) - (\ref{eqn:vfull}) and using the facts that $\tilde{\rho}=\tilde{\rho}(\textbf{r})$ and $\tilde{\theta}=\tilde{\theta}(\textbf{r},t)$, we can arrive at expressions for $\tilde{V}_t$, $\tilde{V}_x$ and $\tilde{V}_y$ as shown below. Firstly, for $\tilde{V}_t$:
\begin{eqnarray*}
\tilde{V}_t & = & -\frac{\partial{u_0}}{\partial{t}}\\
& = & -\frac{\partial{u_0}}{\partial{\tilde{\theta}}}\frac{\partial{\tilde{\theta}}}{\partial{t}}\\
& = & -\omega\frac{\partial{u_0}}{\partial{\tilde{\theta}}}
\end{eqnarray*}
Next, $\tilde{V}_x$:
\begin{eqnarray*}
\tilde{V}_x & = & -\frac{\partial{u_0}}{\partial{x}}\\
& = & -\frac{\partial{u_0}}{\partial{\tilde{\rho}}}\frac{\partial{\tilde{\rho}}}{\partial{x}}-\frac{\partial{u_0}}{\partial{\tilde{\theta}}}\frac{\partial{\tilde{\theta}}}{\partial{x}}\\
& = & -\frac{\partial{u_0}}{\partial{\tilde{\rho}}}\cos(\tilde{\theta}-\omega\tilde{t})+\frac{1}{\tilde{\rho}}\frac{\partial{u_0}}{\partial{\tilde{\theta}}}\sin(\tilde{\theta}-\omega\tilde{t})\\
& = & -(\frac{\partial{u_0}}{\partial{\tilde{\rho}}}\cos(\tilde{\theta}-\omega\tilde{t})-\frac{1}{\tilde{\rho}}\frac{\partial{u_0}}{\partial{\tilde{\theta}}}\sin(\tilde{\theta}-\omega\tilde{t}))\\
\end{eqnarray*}
Similarly, $\tilde{V}_y$:
\begin{eqnarray*}
\tilde{V}_y & = & -\frac{\partial{u_0}}{\partial{y}}\\
& = & -\frac{\partial{u_0}}{\partial{\tilde{\rho}}}\frac{\partial{\tilde{\rho}}}{\partial{y}}-\frac{\partial{u_0}}{\partial{\tilde{\theta}}}\frac{\partial{\tilde{\theta}}}{\partial{y}}\\
& = & -\frac{\partial{u_0}}{\partial{\tilde{\rho}}}\sin(\tilde{\theta}-\omega\tilde{t})+\frac{1}{\tilde{\rho}}\frac{\partial{u_0}}{\partial{\tilde{\theta}}}\cos(\tilde{\theta}-\omega\tilde{t})\\
& = & -(\frac{\partial{u_0}}{\partial{\tilde{\rho}}}\sin(\tilde{\theta}-\omega\tilde{t})+\frac{1}{\tilde{\rho}}\frac{\partial{u_0}}{\partial{\tilde{\theta}}}\cos(\tilde{\theta}-\omega\tilde{t}))\\
\end{eqnarray*}

\begin{equation}
\label{eqn:Vt}
\tilde{V}_1=-\omega\frac{\partial{u_0}}{\partial{\tilde{\theta}}}
\end{equation}
\begin{equation}
\label{eqn:Vx}
\tilde{V}_x=-(\frac{\partial{u_0}}{\partial{\tilde{\rho}}}\cos(\tilde{\theta}-\omega\tilde{t})-\frac{1}{\tilde{\rho}}\frac{\partial{u_0}}{\partial{\tilde{\theta}}}\sin(\tilde{\theta}-\omega\tilde{t}))
\end{equation}
\begin{equation}
\label{eqn:Vy}
\tilde{V}_y=-(\frac{\partial{u_0}}{\partial{\tilde{\rho}}}\sin(\tilde{\theta}-\omega\tilde{t})+\frac{1}{\tilde{\rho}}\frac{\partial{u_0}}{\partial{\tilde{\theta}}}\cos(\tilde{\theta}-\omega\tilde{t}))
\end{equation}
We can note that Eqns. (\ref{eqn:Vx}) and (\ref{eqn:Vy}) can be rewritten as:
\begin{equation}
\left(\begin{array}{c} \tilde{V}_x \\ \tilde{V}_y \end{array}\right) = -\left(\begin{array}{cc} \cos(\tilde{\theta}-\omega\tilde{t}) & -\sin(\tilde{\theta}-\omega\tilde{t})\\
\sin(\tilde{\theta}-\omega\tilde{t}) & \cos(\tilde{\theta}-\omega\tilde{t}) \end{array}\right)\left(\begin{array}{c} \frac{\partial{u_0}}{\partial{\tilde{\rho}}} \\ \frac{1}{\tilde{\rho}}\frac{\partial{u_0}}{\partial{\tilde{\theta}}} \end{array}\right)
\end{equation}
Now, Eqn. (\ref{eqn:vfull}) is the solution to (\ref{eqn:vsys}) in the vicinity of a particular solution. Now the general solution to (\ref{eqn:vsys}) can be thought of as the linear combination of eigenvalues and their corresponding eigenvectors. We know that the transition from a traveling wave solution of (\ref{eqn:drift_rde_pert}) to a spiral wave solution is via a Hopf Bifurcation. This implies that there are three eigenvalues that lie on the imaginary axis. One lies at the origin whilst the other 2, due to symmetry in the solution, take the form $\pm i\omega$. Therefore, our solution now takes the form:
\begin{equation}
\label{eqn:vfulleigen}
v(x,y,t) = dt\tilde{V}_0e^{\lambda_tt}+dx\tilde{V}_1e^{\lambda_xt}+dy\tilde{V}_{-1}e^{\lambda_yt}+\sum_{i=2}^{l-1}a_iV_ie^{\lambda_it}
\end{equation}
where $\lambda_t=0$, $\lambda_x=i\omega$, and $\lambda_y=-i\omega$. We also assume that, due to the stability postulate, and also the Hopf Bifurcation Theorem, that $Re\{\lambda_{i\geq2}\}<0$. Therefore, Eqn. (\ref{eqn:vfulleigen}) becomes:
\begin{equation}
\label{eqn:vfulleigen1}
v(x,y,t) = dt\tilde{V}_0+dx\tilde{V}_1e^{i\omega t}+dy\tilde{V}_{-1}e^{-\omega t}+\sum_{i=2}^{l-1}a_iV_ie^{\lambda_it}
\end{equation} 
We now need to find expressions for $\tilde{V}_0$, $\tilde{V}_1$ and $\tilde{V}_{-1}$. Firstly, if we compare Eqn. (\ref{eqn:vfulleigen1}) to Eqn. (\ref{eqn:vfull}), we find that $\tilde{V}_0=-\omega\frac{\partial{u_0}}{\partial{\tilde{\theta}}}$. For $\tilde{V}_1$ and $\tilde{V}_{-1}$, this is a bit more tricky. Now, we note that the solution $v(x,y,t)$ can be written in the forms stated in Eqns. (\ref{eqn:vfull}) and (\ref{eqn:vfulleigen1}). The direct relation between $\tilde{V}_x$ and $\tilde{V}_1$ and $\tilde{V}_{-1}$, and also between $\tilde{V}_y$ and $\tilde{V}_1$ and $\tilde{V}_{-1}$, is not immediately obvious. However, we note that since we can write the solutions $v(x,y,t)$ in the forms stated in Eqns. (\ref{eqn:vfull}) and (\ref{eqn:vfulleigen1}), we can say that $\tilde{V}_x$ and $\tilde{V}_y$ are linear combinations of $\tilde{V}_1e^{-\omega t}$ and $\tilde{V}_{-1}e^{-i\omega t}$.\\
\\
Let us assume that $\tilde{V}_x$ and $\tilde{V}_y$ take the forms as shown below:
\begin{eqnarray}
\label{eqn:vx11}
\tilde{V}_x &=& Ae^{i\omega t}+Be^{-i\omega t}\\
\label{eqn:vy11}
\tilde{V}_y &=& Ce^{i\omega t}+De^{-i\omega t}
\end{eqnarray}
Consider firstly $\tilde{V}_x$. Let $A=\alpha e^{-i\theta}$ and $B=\beta e^{i\theta}$. This gives:
\begin{eqnarray}
\tilde{V}_x &=& \alpha e^{-i\theta}e^{i\omega t}+\beta e^{i\theta}e^{-i\omega t}\\
&=& \alpha e^{-i(\theta-\omega t)}+\beta e^{i(\theta-\omega t)}\\
&=& \alpha(\cos(\theta-\omega t)-i\sin(\theta-\omega t))+\beta(\cos(\theta-\omega t)+i\sin(\theta-\omega t))\\
&=& (\alpha+\beta)\cos(\theta-\omega t)-i(\alpha-\beta)\sin(\theta-\omega t)
\end{eqnarray}
Comparing this to Eqn. (\ref{eqn:Vx}) we get:
\begin{eqnarray}
\alpha+\beta &=& -\frac{\partial{u_0}}{\partial{\tilde{\rho}}}\\
-i(\alpha-\beta) &=& \frac{1}{\tilde{\rho}}\frac{\partial{u_0}}{\partial{\tilde{\theta}}}
\end{eqnarray}
Solving the above equations gives:
\begin{eqnarray}
\alpha &=& -\frac{1}{2}\left(\frac{\partial{u_0}}{\partial{\tilde{\rho}}}-i\frac{1}{\tilde{\rho}}\frac{\partial{u_0}}{\partial{\tilde{\theta}}}\right)\\
\beta &=& -\frac{1}{2}\left(\frac{\partial{u_0}}{\partial{\tilde{\rho}}}+i\frac{1}{\tilde{\rho}}\frac{\partial{u_0}}{\partial{\tilde{\theta}}}\right)
\end{eqnarray}
Therefore, $A$ and $B$ are now given by:
\begin{eqnarray}
A &=& -\frac{1}{2}\left(\frac{\partial{u_0}}{\partial{\tilde{\rho}}}-i\frac{1}{\tilde{\rho}}\frac{\partial{u_0}}{\partial{\tilde{\theta}}}\right)e^{-i\theta}\\
B &=& -\frac{1}{2}\left(\frac{\partial{u_0}}{\partial{\tilde{\rho}}}+i\frac{1}{\tilde{\rho}}\frac{\partial{u_0}}{\partial{\tilde{\theta}}}\right)e^{i\theta}
\end{eqnarray}
Now,let us consider $\tilde{V}_y$ and perform a similar analysis. Let $C=\gamma e^{-i\theta}$ and $D=\delta e^{i\theta}$. This gives:
\begin{eqnarray}
\tilde{V}_y &=& \gamma e^{-i\theta}e^{i\omega t}+\delta e^{i\theta}e^{-i\omega t}\\
&=& \gamma e^{-i(\theta-\omega t)}+\delta e^{i(\theta-\omega t)}\\
&=& \gamma(\cos(\theta-\omega t)-i\sin(\theta-\omega t))+\delta(\cos(\theta-\omega t)+i\sin(\theta-\omega t))\\
&=& (\gamma+\delta)\cos(\theta-\omega t)-i(\gamma-\delta)\sin(\theta-\omega t)
\end{eqnarray}
Comparing this to Eqn. (\ref{eqn:Vy}) we get:
\begin{eqnarray}
\gamma+\delta &=& -\frac{1}{\tilde{\rho}}\frac{\partial{u_0}}{\partial{\tilde{\theta}}}\\
i(\delta-\gamma) &=& -\frac{\partial{u_0}}{\partial{\tilde{\rho}}}
\end{eqnarray}
Solving the above equations gives:
\begin{eqnarray}
\gamma &=& -\frac{1}{2}\left(i\frac{\partial{u_0}}{\partial{\tilde{\rho}}}+\frac{1}{\tilde{\rho}}\frac{\partial{u_0}}{\partial{\tilde{\theta}}}\right)\\
\delta &=& \frac{1}{2}\left(i\frac{\partial{u_0}}{\partial{\tilde{\rho}}}-\frac{1}{\tilde{\rho}}\frac{\partial{u_0}}{\partial{\tilde{\theta}}}\right)
\end{eqnarray}
Therefore, $A$ and $B$ are now given by:
\begin{eqnarray}
C &=& -\frac{1}{2}\left(i\frac{\partial{u_0}}{\partial{\tilde{\rho}}}+\frac{1}{\tilde{\rho}}\frac{\partial{u_0}}{\partial{\tilde{\theta}}}\right)e^{-i\theta}\\
D &=& \frac{1}{2}\left(i\frac{\partial{u_0}}{\partial{\tilde{\rho}}}-\frac{1}{\tilde{\rho}}\frac{\partial{u_0}}{\partial{\tilde{\theta}}}\right)e^{i\theta}
\end{eqnarray}

Finally, since we know that the eigenvalues are $\pm i\omega$, we can say that the eigenvectors are:

\end{ajf}


They then considered the eigenfunctions to the linear operator $\til{L}$, and, after some analysis, it can be shown that these take the form:

\begin{eqnarray}
\tilde{V}_0 &=& -\omega\frac{\partial{u_0}}{\partial{\tilde{\theta}}},\quad\lambda_0=0\\
\tilde{V}_1 &=& -\frac{1}{2}\left(\frac{\partial{u_0}}{\partial{\tilde{\rho}}}-i\frac{1}{\tilde{\rho}}\frac{\partial{u_0}}{\partial{\tilde{\theta}}}\right)e^{-i\theta},\quad\lambda_1=i\omega\\
\tilde{V}_{-1} &=& -\frac{1}{2}\left(\frac{\partial{u_0}}{\partial{\tilde{\rho}}}+i\frac{1}{\tilde{\rho}}\frac{\partial{u_0}}{\partial{\tilde{\theta}}}\right)e^{i\theta},\quad\lambda_{-1}=-i\omega
\end{eqnarray}

Accordingly to the stability postulate, there should be no other eigenvalues on the imaginary axis. So, the solution to (\ref{eqn:vsys}) can be expanded in its eigenbasis as follows:

\begin{equation}
\label{eqn:vfull11}
v(\tilde{\textbf{r}},\tilde{t}) = c_0\tilde{V}_0(\tilde{\textbf{r}})+c_1e^{i\omega\til{t}}\tilde{V}_1(\tilde{\textbf{r}})+c_{-1}e^{-i\omega \til{t}}\tilde{V}_{-1}(\tilde{\textbf{r}})+o(1)
\end{equation}
\\
as $\tilde{t}\rightarrow\infty$. To determine the coefficients $c_i$, they took the scalar product of (\ref{eqn:vfull11}) with the eigenfunctions to the adjoint operator $\til{L}^+$ to get:

\begin{ajf}

The next question is how do we determine the coefficients $c_i$. Thinking about this in terms of Linear Algebra, we can see that the vector $v(x,y,t)$ consists of 3 linearly independent orthogonal vectors - $\tilde{V}_0$, $\tilde{V}_1e^{i\omega t}$ and $\tilde{V}_{-1}e^{-i\omega t}$ - and can be written as:
\begin{equation}
v(x,y,t) = c_0v_0+c_1v_1+c_{-1}v_{-1}
\end{equation}
Now, each constant $c_i$ can be obtained by taking the dot product of $v(x,y,t)$ with a vector, say $w_i$. This will yield:
\begin{equation}
\label{eqn:vwi}
(w_i,v) = c_0(w_i,v_0)+c_1(w_i,v_1)+c_{-1}(w_i,v_{-1})
\end{equation}
In order to find each individual $c_i$ , we must have that:
\[(w_i,v_j) = \left\{\begin{array}{ll} 1&\mbox{if}\quad i=j\\0&\mbox{if}\quad i\neq j\end{array}\right.\]
This is just the Kroneker Delta function and therefore can be written as:
\begin{equation}
\label{eqn:kron}
(w_i,v_j) = \delta_{ij}
\end{equation}
Now, going back to Eqn. (\ref{eqn:vfull11}), we can conclude that to find each $c_i$, we use:
\begin{equation}
(\tilde{W}_\lambda,\tilde{V}_\mu) = \delta_{\lambda\mu}
\end{equation}
and hence from Eqn. (\ref{eqn:vwi}):

\end{ajf}

\begin{equation}
c_{0,\pm 1} = (\tilde{W}_{0,\pm i\omega},v)
\end{equation}
\\
where $\tilde{W}_{0,\pm i\omega}$ satisfies:

\begin{equation}
\til{L}^+\tilde{W}_{0,\pm i\omega} = \til{\lambda}_{0,\pm i\omega}\tilde{W}_{0,\pm i\omega}
\end{equation}
\\
and the following orthogonality condition holds:

\begin{equation}
(\tilde{W}_i,\tilde{V}_j) = \delta_{ij}
\end{equation}

Now, the scalar products used above are the scalar products in a functional space and defined as:
\chg[af]{
\begin{eqnarray}
(W,V) &=& \int\int\left\langle W(\tilde{\textbf{r}}),V(\tilde{\textbf{r}})\right\rangle \dd^2\tilde{\textbf{r}}
\end{eqnarray}
\\
where the integration is over the whole plane.}
\begin{ajf}

We now show that $\tilde{W}_j$ is an eigenvector of of the adjoint linear differential operator to $L$ - call this $L^+$. First, note that since we are dealing with complex eigenvalues and eigenvectors, we must work in a Hermitian Space rather than an Euclidean Space. Therefore, the inner (dot/scalar) product must be semi-linear:
\[c(a,b) = (c^*a,b) = (a,cb)\]
where $*$ represents that complex conjugate. Now consider:
\begin{eqnarray*}
(\tilde{W}_\lambda,\tilde{L}v) &=& (\tilde{W}_\lambda,\partial_\tau v)\\
&=& \partial_\tau(\tilde{W}_\lambda,v)
\end{eqnarray*}
where $\partial_\tau$ represents partial derivative with respect to $\tau$. Also note that the second identity above has arisen due to $\tilde{W}$ not depending $\tau$. Now, we know that the eigenvalue problem is given by:
\begin{equation}
\tilde{L}v = \lambda v
\end{equation}
Therefore:
\begin{eqnarray*}
(\tilde{W}_\lambda,\tilde{L}v) &=& (\tilde{W}_\lambda,\lambda v)\\
&=& \lambda(\tilde{W}_\lambda,v)
\end{eqnarray*}
\[\Rightarrow (\tilde{W}_\lambda,\tilde{L}v) = (\tilde{W}_\lambda,\partial_\tau v) = \partial_\tau(\tilde{W}_\lambda,v) = \lambda(\tilde{W}_\lambda,v)\]
Now, we have:
\begin{eqnarray*}
(\lambda_j^*\tilde{W}_j,v_k) &=& \lambda_j\delta_{jk}\\
&=& \lambda_j(\tilde{W}_j,v_k)\\
&=& (\tilde{W}_j,\lambda_jv_k)\\
&=& (\tilde{W}_j,Lv_k)\\
&=& L(\tilde{W}_j,v_k)\\
&=& (L^+\tilde{W}_j,v_k)
\end{eqnarray*}
where $L^*$ is the adjoint operator to $L$. Therefore, since :
\begin{equation}
(\lambda_j^*\tilde{W}_j,v_k) = (L^+\tilde{W}_j,v_k)
\end{equation}
we have that $L^+\tilde{W}_j=\lambda\tilde{W}_j$ and therefore $\tilde{W}_j$ are eigenvectors to the adjoint operator, $L^+$, which is defined as:
\begin{equation}
L^+ = \textbf{D}\tilde{\nabla}^2+\omega\partial_\theta+\textbf{F}^+(u_0(\tilde{\textbf{r}}))
\end{equation}
with $\textbf{F}^+(u_0(\tilde{\textbf{r}}))$ denoting the transposed matrix to $\textbf{F}$. Furthermore, we can also say that $c_i$ are functionals and can be written in integral form:
\begin{eqnarray}
(\tilde{W}_\lambda,\cdot) &=& \int<\tilde{Y}_\lambda(\tilde{\textbf{r}}),\cdot>d^2\tilde{\textbf{r}}
\end{eqnarray}

\end{ajf}

They then considered the case for $\epsilon\neq0$, and firstly considered a regular perturbation technique.

\begin{equation}
u(\textbf{r},t) = u_0(\textbf{r},t)+\epsilon v(\textbf{r},t)+O(\epsilon^2)
\end{equation}

In the rotating frame of reference, we get:

\begin{equation}
\label{eqn:drift_rdepert}
\frac{\partial{v}}{\partial{\tilde{t}}} = \tilde{L}v+h(u_0(\tilde{\textbf{r}}),\tilde{\textbf{r}},\tilde{t})
\end{equation}
\\
which is similar to Eqn.(\ref{eqn:vsys}) but with the perturbation term added. It was then assumed that the solution to (\ref{eqn:drift_rdepert}) can be expressed as the linear combination of its eigenvectors and eigenvalues, and in particular that \chgex[ex]{its} eigenvectors form the span of solutions in the space of solutions:

\begin{equation}
\label{eqn:vspan}
v(\tilde{t}) = \sum_\lambda c_i(\tilde{t})V_i
\end{equation}
\\
where $V_i$ are the eigenvectors of the linearised system (\ref{eqn:vspan}). It was noted that $v$ is a function of $\tilde{t}$ and therefore, since $V_i$ is a vector constant, hence $c_i=c_i(\tilde{t})$. They then needed to find expressions for $c_i$. 

\begin{ajf}

Firstly, let us take the inner product of (\ref{eqn:vspan}) with the eigenvector of the adjoint operator, $L^+$:

\begin{eqnarray}
(\tilde{W}_j,v) &=& (\tilde{W}_j,\sum_\lambda c_iV_i)\\
&=& (\tilde{W}_j,c_jV_j)+(\tilde{W}_j,\sum_{\lambda, i\neq j} c_iV_i)\\
&=& c_j(\tilde{W}_j,V_j)+\sum_{\lambda, i\neq j}c_i(\tilde{W}_j,V_i)\\
&=& c_j
\end{eqnarray}
with the last relation coming from Eqn.(\ref{eqn:kron}). Next, let us rewrite (\ref{eqn:drift_rdepert}) as follows:
\begin{equation}
\frac{dc_j}{d\tilde{t}} = \lambda c_j+H(\tilde{t})
\end{equation}
where $H(t)=(\tilde{W},h)$. This can now be solved explicitly:
\begin{eqnarray}
\frac{dc_j}{d\tilde{t}}-\lambda c_j &=& H(\tilde{t})\\
e^{-\lambda{\tilde{t}}}\frac{dc_j}{d\tilde{t}}-e^{-\lambda{\tilde{t}}}\lambda c_j &=& e^{-\lambda{\tilde{t}}}H(\tilde{t})\\
\Rightarrow e^{-\lambda{\tilde{t}}}c_j &=& \int_{0}^{\tilde{t}} e^{-\lambda \tau}H(\tilde{\tau})d\tau\\
\Rightarrow c_j &=& e^{\lambda{\tilde{t}}}\int_{0}^{\tilde{t}} e^{-\lambda \tau}H(\tilde{\tau})d\tau
\end{eqnarray}
Hence, (\ref{eqn:vspan}) now becomes:

\end{ajf}

This can then be written as:

\begin{equation}
\label{eqn:vspan1}
v(\tilde{t}) = \sum_\lambda V_ie^{\lambda{\tilde{t}}}\int_{0}^{\tilde{t}} e^{-\lambda \tau}H(\tilde{\tau})d\tau
\end{equation}

In the knowledge that there are 3 eigenvalues with values $0,\pm i\omega$. They then considered the case when $\lambda=0$. They then showed that $(\tilde{W}_j,v)=c_j=e^{\lambda{\tilde{t}}}\int_{0}^{\tilde{t}} e^{-\lambda \tau}H(\tilde{\tau})d\tau$, noting that $j$ is the eigenvalue. Therefore, when $\lambda=0$:

\begin{eqnarray}
(\tilde{W}_0,v) &=& \int_{0}^{\tilde{t}}H(\tilde{\tau})d\tau\\
\label{eqn:Wvlam0}
\Rightarrow (\tilde{W}_0,v) &=& \int_{0}^{\tilde{t}}(\tilde{W}_0,h)d\tau
\end{eqnarray}

Now, as $\tilde{t}\rightarrow\infty$, they found that (\ref{eqn:Wvlam0}) grows to infinity, if, for example, they had that $(\tilde{W}_0,h)$ is a non-zero constant.

So, what does all this mean? Well, we have seen that if we have small perturbation, then as $\tilde{t}\rightarrow\infty$ the functions $c_i$ become very large. This in term means that $v(\tilde{t})$ also becomes large and therefore contradicts the stability postulate of a previous section. Hence, we can say that the Regular Perturbation technique implemented here does not work and so cannot describe drift. Therefore, Biktashev et al decided to look at Singular Perturbation methods.

They then decided to use a singular \chg[af]{perturbation} technique to study this particular system. This allowed to obtain solutions that are valid for large values of time at bounded values of $h$. They therefore considered solutions of the form:

\begin{eqnarray}
u(\textbf{r},t) &=& u_0\left(\textbf{r}-\textbf{R}(t),t-\frac{\Phi(t)}{\omega}\right)+\epsilon v(\textbf{r},t)\\
&=& u_0(\rho(\textbf{r}-\textbf{R}(t)),\theta(\textbf{r}-\textbf{R}(t))+\omega t-\Phi(t))+\epsilon v(\textbf{r},t)
\end{eqnarray}

By using similar techniques to the regular perturbation theory, they established the equation for $v$:
\chg[p27eqn1]{
\begin{footnotesize}
\begin{eqnarray}
\epsilon\partial_tv &=& \epsilon Lv+\epsilon h\left(u_0\left(\textbf{r}-\textbf{R},t-\frac{\Phi}{\omega}\right),\br,t\right)+\partial_\theta u_0\left(\textbf{r}-\textbf{R},t-\frac{\Phi}{\omega}\right)\nonumber\\
&& +\partial_xu_0\left(\textbf{r}-\textbf{R},t-\frac{\Phi}{\omega}\right)X'(t)+\partial_yu_0\left(\textbf{r}-\textbf{R},t-\frac{\Phi}{\omega}\right)Y'(t)+O(\epsilon)
\end{eqnarray}
\end{footnotesize}
}
In a frame of reference associated with the spiral wave, they got:
\chg[p27eqn1]{
\begin{eqnarray}
\epsilon\partial_{\til{t}}v &=& \epsilon\til{L}v+\epsilon h\left(u_0(\til{\br}),\br,t\right)+\frac{1}{\omega}\til{V}_0(\til{\br})\Phi'(t)\nonumber\\
\label{eqn:dynamics_sing}
&& +\left(\til{V}_1(\til{\br})e^{i(\omega t-\Phi)}+\til{V}_{-1}(\til{\br})e^{-i(\omega t-\Phi)}\right)X'(t)\nonumber\\
&& +\left(i\til{V}_1(\til{\br})e^{i(\omega t-\Phi)}-i\til{V}_{-1}(\til{\br})e^{-i(\omega t-\Phi)}\right)Y'(t)+O(\epsilon)
\end{eqnarray}
}
Therefore, by taking the scalar product of Eqn.(\ref{eqn:dynamics_sing}) with the eigenfunctions to the adjoint operator $\til{L}^+$, we have:
\chg[p27eqn1]{
\begin{eqnarray}
\label{eqn:drift_Phi}
\Phi'(t) &=& \epsilon\omega(\til{W}_0,h)+O(\epsilon)\\
\label{eqn:drift_X}
X'(t)    &=& \epsilon\rm{Re}\{e^{i(\omega t-\Phi)}(\til{W}_1,h)\}+O(\epsilon)\\
\label{eqn:drift_Y}
Y'(t)    &=& \epsilon\rm{Im}\{e^{i(\omega t-\Phi)}(\til{W}_1,h)\}+O(\epsilon)
\end{eqnarray}
}
Therefore, they have obtained the equations of motion along the manifold (\ref{eqn:drift_ushift}) under a generic perturbation $h(u,\br,t)$.

They then considered two examples; Resonant drift, where the perturbation is time dependent only; and Inhomogeneitity induced drift. We will review the simpler example in this thesis, which is Resonant drift.

So, consider a perturbation of the form:

\begin{equation}
h(u,\br,t) = H(\Omega t-\phi), \quad\mbox{such that}\quad H(\phi+2\pi)=H(\phi)
\end{equation}

Next, we shall consider the averaged motion of the spiral. Therefore, the equations of motion (\ref{eqn:drift_Phi})-(\ref{eqn:drift_Y}), become:
\chg[p27eqn1]{
\begin{eqnarray}
\label{eqn:drift_Phi_av}
\bar{\Phi}'(t) &=& \epsilon H_0+O(\epsilon)\\
\label{eqn:drift_X_av}
\bar{X}'(t)    &=& \epsilon|H_1|\cos((\omega t-\bar{\Phi})-(\Omega t-\phi)+\rm{arg}\{H_1\})+O(\epsilon)\\
\label{eqn:drift_Y_av}
\bar{Y}'(t)    &=& \epsilon|H_1|\sin((\omega t-\bar{\Phi})-(\Omega t-\phi)+\rm{arg}\{H_1\})+O(\epsilon)
\end{eqnarray}
\\
}where $H_n$ are given by:

\begin{equation}
H_n = \int\left\langle\int\til{W}_n\dd^2\br,H(\eta)\right\rangle e^{-in\eta}\dd{\eta}
\end{equation}

Introducing the notation:

\begin{eqnarray}
\varphi &=& (\omega t-\bar{\Phi})-(\Omega t-\phi)+\rm{arg}\{H_1\}
\end{eqnarray}
\\
we get:
\chg[p27eqn1]{
\begin{eqnarray}
\label{eqn:drift_Phi_AV}
\phi'(t) &=& \omega+\epsilon H_0-\Omega+O(\epsilon)\\
\label{eqn:drift_X_AV}
\bar{X}'(t)    &=& \epsilon|H_1|\cos(\phi)+O(\epsilon)\\
\label{eqn:drift_Y_AV}
\bar{Y}'(t)    &=& \epsilon|H_1|\sin(\phi)+O(\epsilon)
\end{eqnarray}
\\
}which, when integrated, give us the equation of a circle.
\label{sec:rev_swdyn}

\section{Models used in Numerical Analysis}
\label{sec:models}

In our numerical work, we will be using two models - the FitzHugh-Nagumo model \cite{FHN1,FHN2}, affectionately known as FHN, and Barkley's model \cite{Bark90,bark91}. Barkley's model is shown below
\chg[af]{
\begin{eqnarray}
\label{eqn:model_bark}
\frac{\partial{u^{(1)}}}{\partial{t}} & = & \nabla^2u^{(1)}+\frac{1}{\varepsilon}u^{(1)}(1-u^{(1)})\left[u^{(1)}-\frac{u^{(2)}+b}{a}\right]\\
\frac{\partial{u^{(2)}}}{\partial{t}} & = & D_v\nabla^2u^{(2)}+u^{(1)}-u^{(2)}
\end{eqnarray}}
\\
where $a, b$, and $\varepsilon$ are parameters, and also $u^{(i)}$ represents the $i$'th component of the two component system. In his original papers \cite{Bark90,bark91}, Barkley calls $u^{(1)}=u$ and $u^{(2)}=v$, but we will try to be consistent with notation throughout this thesis and therefore we adopt our notation as shown in (\ref{eqn:model_bark}).

FHN on the other hand, is similar to Barkley's model but takes a slightly different form:
\chg[af]{
\begin{eqnarray}
\label{eqn:model_fhn}
\frac{\partial{u^{(1)}}}{\partial{t}} & = & \nabla^2u^{(1)}+\frac{1}{\varepsilon}\left(u^{(1)}-\frac{(u^{(1)})^3}{3}-u^{(2)}\right)\\
\frac{\partial{u^{(2)}}}{\partial{t}} & = & D_v\nabla^2u^{(2)}+\varepsilon\left(u^{(1)}+\beta-\gamma u^{(2)}\right)
\end{eqnarray}}
\\
where the parameters in this model are $\beta$, $\gamma$ and $\varepsilon$.

It must be stressed at this stage that $\varepsilon$ in these models are not necessarily small quantities, unlike the $\epsilon$ that we will be using in Perturbation Theory. They are model parameters and in fact determine the slow and fast times within the models. In most cases they are taken to be small, but they must not get confused with the small parameters used in perturbation theory.

\subsection{Similarities in the FHN and Barkley's Models}

There are many \chg[af]{similarities} between the two models. For instance, in the evolution of the $u^{(1)}$ fields, both models have cubic local kinetics. Also in the $u^{(2)}$ fields, the local kinetics are linear. 

\begin{figure}[tbp]
\begin{center}
\begin{minipage}[b]{0.49\linewidth}
\centering
\psfrag{a}[l]{$u^{(2)}$}
\psfrag{b}[l]{$u^{(1)}$}
\psfrag{c}[l]{1}
\psfrag{d}[l]{0}
\psfrag{f}[l]{$l_1$}
\psfrag{g}[l]{$l_2$}
\includegraphics[width=\textwidth]{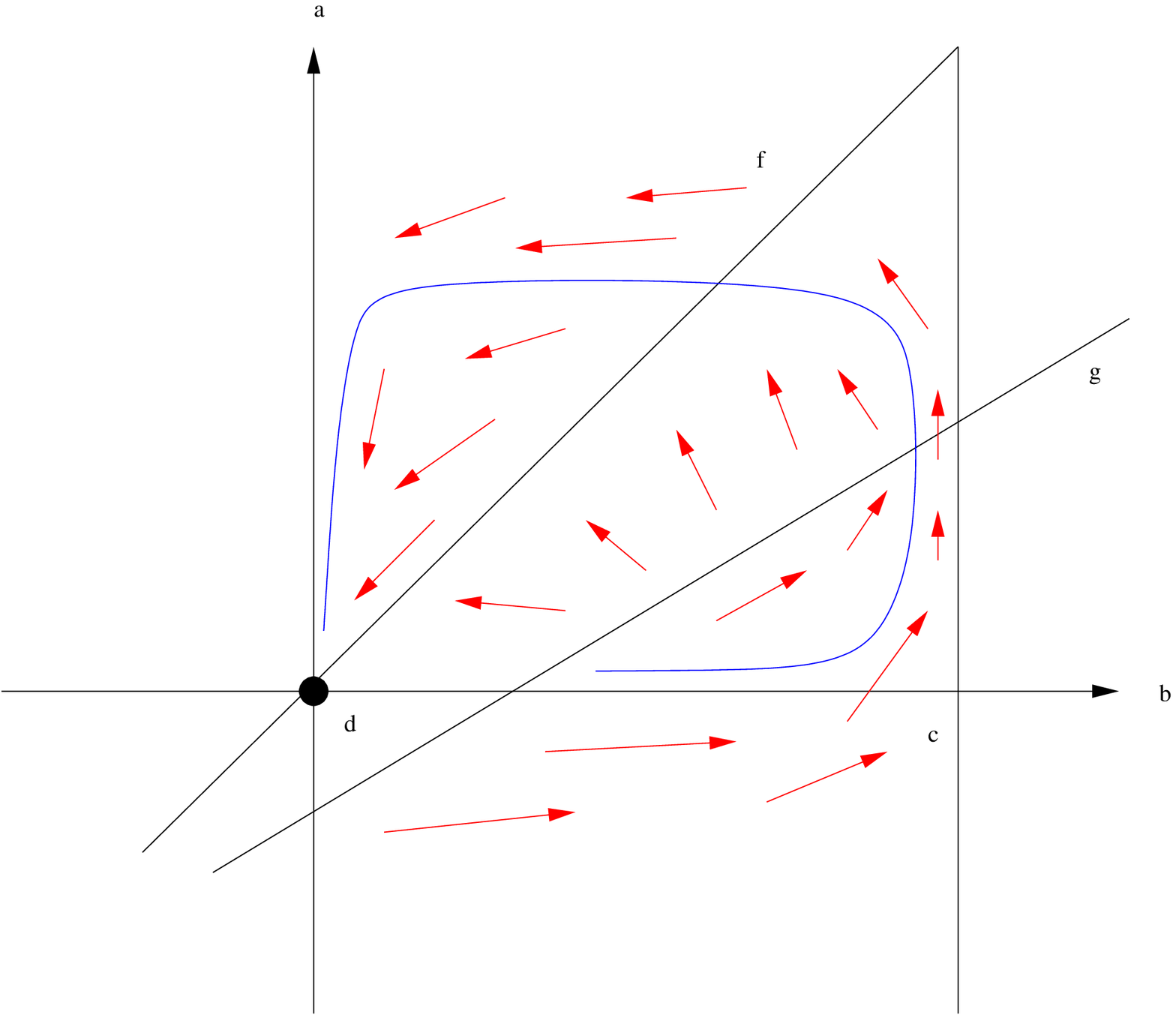}
\caption[Barkley's model: Phase Portrait]{Barkley's model:\newline Phase Portrait.}
\label{fig:models_bark_phase}
\end{minipage}
\begin{minipage}[b]{0.49\linewidth}
\centering
\psfrag{a}[l]{$u^{(2)}$}
\psfrag{b}[l]{$u^{(1)}$}
\psfrag{c}[l]{$l_1$}
\psfrag{d}[l]{$l_2$}
\includegraphics[width=0.8\textwidth]{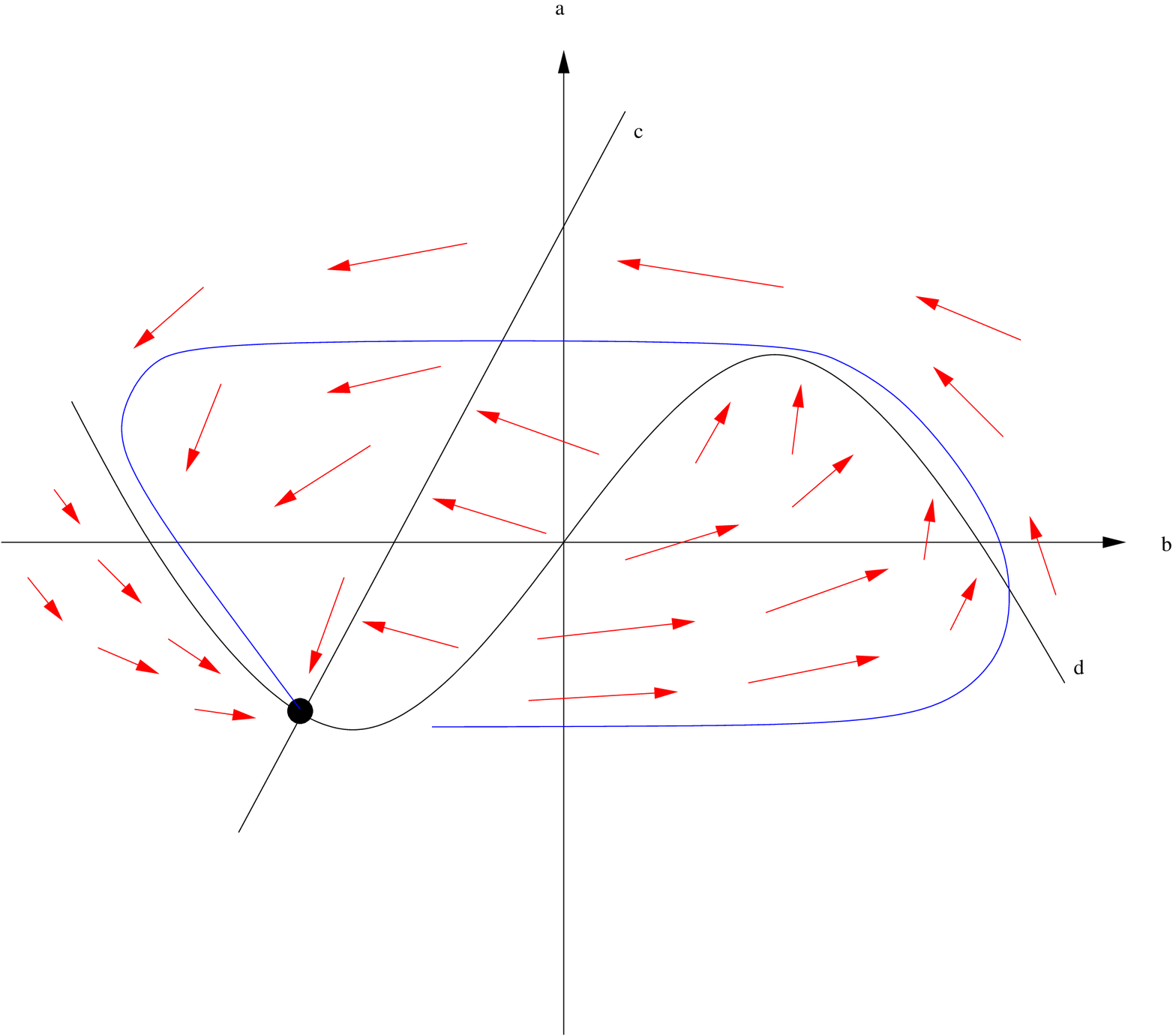}
\caption[FHN model: Phase Portrait]{FHN model:\newline Phase Portrait.}
\label{fig:models_fhn_phase}
\end{minipage}
\end{center}
\end{figure}

They also have very similar phase \chg[af]{portraits} as shown in Figs.(\ref{fig:models_bark_phase}) and (\ref{fig:models_fhn_phase}). 

For FHN, the nullclines are given by:

\begin{eqnarray}
l_1: u^{(2)} &=& \frac{1}{\gamma}u^{(1)}+\frac{\beta}{\gamma}\\
l_2: u^{(2)} &=& u^{(1)}-\frac{(u^{(1)})^3}{3}
\end{eqnarray}
\\
whereas in Barkley they are:
\chg[p30nul]{}
\begin{eqnarray}
\mbox{horizontal nullclines}:\quad u^{(2)} &=& u^{(1)}\\
\mbox{vertical nullclines}:\quad u^{(2)} &=& au^{(1)}-b,\quad \chg[]{u^{(1)}}=0,\quad u^{(1)}=1
\end{eqnarray}

Also we have that the intersections of the nullclines occur at $(0,0)$ \& $(\pm\sqrt{3},0)$ in FHN, and at $(0.0)$ \& \chg[p30fp]{$(1,1)$} in Barkley's model. 

In both phase portraits, we have detailed not only typical excitation trajectories (given in \textcolor{blue}{blue}) but have also shown the directions of the trajectories \chg[af]{throughout} the portrait \chg[p30spell]{given} by the \textcolor{red}{red} arrows. We have tried to use the length of these directional arrows to indicate the speed of the trajectory at those points; so a short arrow would indicate a slow speed, etc.

In each case, the stable fixed point is shown by a large black dot, and is located at $(0,0)$ in Barkley's model and $(x_S,y_S)$ in FHN where $x_S$ and $y_S$ are the coordinates of the intersection of the two isolines. Just thinking about the steady state in FHN, this state always depends on the values of the model parameters, $\beta$ and $\gamma$. However, in Barkley's model, the stable steady state is fixed, no matter what the values of the model parameters.

We can also go as far to say that Barkley's model is just a simplified version of the FHN model. Consider the \chg[af]{vertical} nullclines. In Barkley's model, two of these nullclines form the vertical lines of a backward ``N'' shape. These are the ``slow'' nullclines, meaning that the trajectory moves slowly along with respect to these nullclines, with the horizontal nullcline being the fast nullcline. The middle part of the ``N'' shape, is the horizontal nullcline and is the fast nullcline, so trajectories move quickly across the part of the plane spanned by this nullcline. In comparison to FHN, we have a cubic curve for the vertical nullclines. Again, this cubic curve is in fact the fast nullcline whilst the horizontal nullcline is the fast one.

The advantage that Barkley's model has over the FHN model, is that whilst still able to generate spiral wave solutions including meandering solutions, it can be much quicker to solve numerically. This is due to a ``trick'' that Barkley introduces in his 1991 paper \cite{bark91}.

\begin{figure}[tbp]
\begin{center}
\begin{minipage}[b]{0.7\linewidth}
\centering
\psfrag{a}[l]{$u^{(2)}$}
\psfrag{b}[l]{$u^{(1)}$}
\psfrag{c}[l]{1}
\psfrag{d}[l]{0}
\psfrag{e}[l]{$\delta$}
\psfrag{f}[l]{$l_1$}
\psfrag{g}[l]{$l_2$}
\includegraphics[width=0.8\textwidth]{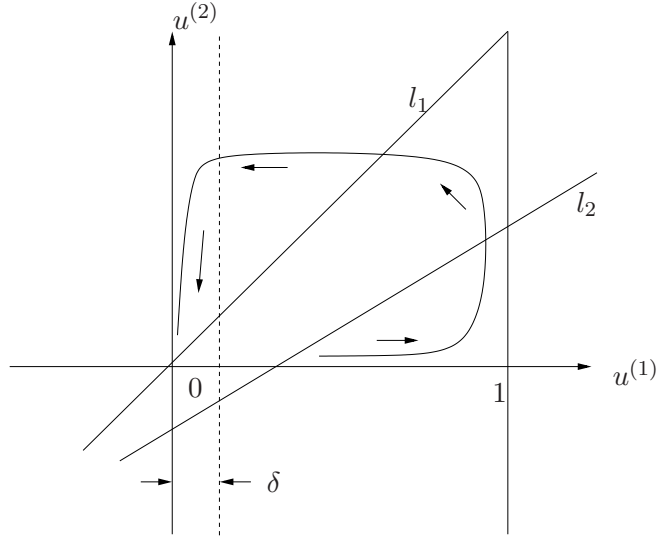}
\caption{Barkley's model: Phase Portrait showing the ``$\delta$'' line.}
\label{fig:models_bark_delta}
\end{minipage}
\end{center}
\end{figure}

Consider Fig.(\ref{fig:models_bark_delta}). Analysis of Barkley's model shows that the trajectory stays with the range $0\leq u^{(1)}\leq\delta$ for a substantial period of time during one cycle of its trajectory. Therefore, the trick that Barkley introduced, which we shall call the \emph{Delta Trick} from now on, is:

\begin{equation}
\mbox{if}\quad u^{(1)}<\delta,\quad \mbox{then}\quad u^{(1)}=0
\end{equation}

This makes the calculation very much faster. However, one downside to this is that the accuracy of the \chg[af]{original} calculations is diminished. But, if one wants to study spiral waves and one requires rapid calculations, then this Delta Trick will work.

Finally, after some analysis, it can be shown that the region of excitability within Barkley's model is given by the following inequalities:

\[\begin{array}{ccccc}
0 & < & a & < & 2\\
a-1 & < & b & < & \frac{a}{2}\\
\mbox{for} && b & > & 0
\end{array}\]
\label{sec:rev_models}

\section{Numerical Methods \& Software}

Throughout this thesis, we shall be conducting our numerical simulations through a program called \emph{EZ-Spiral}, which was developed by Barkley in 1991, to simulate spiral waves using his model. The program has somewhat evolved over the years and the current version (version 3.2, 2007) uses OpenGL graphics. It is \chg[p32spell]{available} as Freeware and can be downloaded from Barkley's website \cite{barkweb}.

We shall first of all describe the numerical procedures used in EZ-Spiral and then discuss how EZ-Spiral works and what it can do.

We must note also that Chap.4 of this thesis concerns the numerical solutions of spiral waves in a frame of reference comoving with the tip of the wave. A program was developed, which is based on EZ-Spiral and is called \emph{EZ-Freeze}. Some of the numerical procedures used in EZ-Spiral are implemented in EZ-Freeze, but we shall discuss this in Chap.4.


\subsection{Numerical Methods used in EZ-Spiral}

All the numerical methods used in EZ-Spiral are described in \cite{bark91}. We begin with Barkley's model, and note that all our simulations are conducted with diffusion present only in the $u^{(1)}$-field, whose diffusion coefficient is scaled to one. For the purpose of ease of notation and readability, and also to tie in to published literature, we shall denote the $u^{(1)}$-field as the $u$-field, and the $u^{(2)}$-field as the $v$-field:

\begin{eqnarray}
\label{eqn:numerical_bark_u}
\pderiv{u}{t} &=& \nabla^2u+f(u,v)\\
\label{eqn:numerical_bark_v}
\pderiv{v}{t} &=& g(u,v)
\end{eqnarray}
\\
where:

\begin{eqnarray}
f(u,v) &=& \frac{1}{\varepsilon}u(1-u)\left[u-\frac{v+b}{a}\right]\\
g(u,v) &=& u-v
\end{eqnarray}

Now, in his paper, Barkley introduces the notation,

\begin{equation}
u_{th} = \frac{v+b}{a}
\end{equation}
\\
to give:

\begin{eqnarray}
f(u,v) &=& \frac{1}{\varepsilon}u(1-u)\left[u-u_{th}\right]
\end{eqnarray}

So, we wish to solve the system of equations (\ref{eqn:numerical_bark_u})-(\ref{eqn:numerical_bark_v}) numerically. There are 4 main methods that are used, viz Explicit Forward Euler Method, Semi-Implicit Forward Euler Method, the 9-point Finite-Difference Laplacian Method and the 5-point Finite-Difference Laplacian Method.


\subsubsection{Explicit Forward Euler Method}

Firstly, let us consider Barkley's model without any diffusion. This gives us the following system:

\begin{eqnarray*}
\frac{\partial{u}}{\partial{t}} & = & f(u,v)\\
\frac{\partial{v}}{\partial{t}} & = & g(u,v)
\end{eqnarray*}

For the time being let us look just at the $u$-field. Let $u_t$ denote the value of $u$ at time $t$ and let $h_t$ denote the time step. Therefore, we can rewrite the equation for the $u$-field above as follows:

\begin{eqnarray*}
\frac{u_{t+h}-u_t}{h_t} & = & f(u_t,v_t)\\
\Rightarrow \frac{u_{t+h}-u_t}{h_t} & = & \frac{1}{\epsilon}u_t(1-u_t)(u_t-u_{th})\\
\Rightarrow u_{t+h} & = & u_t+h_t\frac{1}{\epsilon}u_t(1-u_t)(u_t-u_{th})
\end{eqnarray*}

Similarly, an equation for $v_{t+h}$ can also be found:

\begin{eqnarray*}
v_{t+h} & = & v_t+h_t(u_t-v_t)
\end{eqnarray*}

Given a suitable, small time-step, $h_t$, together with appropriate initial conditions, we can use this method to numerically calculate the values of $u_t$ \& $v_t$ for a set amount of steps.

This method, i.e. arranging the original PDE's into the above form, is known as the Explicit Forward Euler Method. Barkley, in applying this method to his model without diffusion, found it to be accurate in the fast regions as long as $h_t<\epsilon$.


\subsubsection{Implicit Forward Euler Method}

This particular method is very much similar to the explicit method. However, instead of using the form:

\begin{eqnarray*}
\frac{u_{t+h}-u_t}{h_t} & = & f(u_t,v_t)
\end{eqnarray*}

we now use:

\begin{eqnarray*}
\frac{u_{t+h}-u_t}{h_t} & = & f(u_{t+h},v_{t+h})
\end{eqnarray*}

This now gives us equations for $u_t$ \& $v_t$ in terms of $u_{t+h}$ \& $v_{t+h}$ and hence enables us to work backwards from a given set of values to the initial values.

However, we shall not being using or discussing the Implicit Method any further so we do not intend to spend any more time discussing this method.


\subsubsection{Semi-Implicit Forward Euler Method}

If we wish to take a large time-step then we need to use the Semi-Implicit form of the Euler method. In this instance, we use a mixture of both the implicit method and the explicit method as follows:

\begin{eqnarray*}
\frac{u_{t+h}-u_t}{h_t} & = & f(u_t,u_{t+h},u_{th})\\
u_{t+h} & = & \left\{\begin{array}{lrcl}
                           u_t + \frac{h_t}{\epsilon}u_{t+h}(1-u_t)(u_t-u_{th}) & \mbox{if} \ u_t & \leqslant & u_{th}\\
                           u_t + \frac{h_t}{\epsilon}u_t(1-u_{t+h})(u_t-u_{th}) & \mbox{if} \ u_t & > & u_{th}\end{array}\right.
\end{eqnarray*}

The conditions detailed underneath the above expressions for $u$ are clear to understand if we consider the rearrangement of the above as detailed below. We consider the above using $u_{t+h}=F(u_t,u_{th})$:

\begin{eqnarray*}
F(u_t,u_{th}) & = & \left\{\begin{array}{lrcl}
                                         \frac{u_t}{1-\frac{h_t}{\epsilon}(1-u_t)(u_t-u_{th})} & \mbox{if} \ u_t & \leqslant & u_{th}\\
                                         \frac{u_t+\frac{h_t}{\epsilon}u_t(u_t-u_{th})}{1+\frac{h_t}{\epsilon}u_t(u_t-u_{th})} & \mbox{if} \ u_t & > & u_{th}\end{array}\right.
\end{eqnarray*}

Now, \chg[p35inequal]{$u_{t+h}<1$} for all time in Barkley's model for waves to be generated. Therefore, it is clear from the first equation above that \chg[p35inequal2]{$\frac{h_t}{\epsilon}(1-u_{t+h})(u_t-u_{th})<0$} to guarantee that $u_{t+h}$ is to be positive. Now, as mentioned earlier we use the Semi-Implicit method if we want a larger time-step, in particular we use this method if $h_t > \epsilon \ \Rightarrow \ \frac{h_t}{\epsilon}>1 $. Hence, since $(1-u_t)>0$ then we must have that $(u_t-u_{th}) \leqslant 0 \ \Rightarrow \ u_t \leqslant u_{th}$.

A similar argument is used to show the second condition. Again, we require $u_{t+h}>0$. This therefore means that \chg[p35gram]{}the denominator is positive. Hence, to guarantee this condition, we need that $\frac{h_t}{\epsilon}u_t(u_t-u_{th})>0$. Now, $\frac{h_t}{\epsilon}>0$ and $u_t>0$ hence $u_t-u_{th}>0 \ \Rightarrow \ u_t>u_{th}$.

To \chg[af]{summarise} for the Euler Methods, we can see that for small $\frac{h_t}{\epsilon}$, we use the Explicit Method (note that the Explicit method can be derived from the expressions for the Semi-Implicit method by expanding the denominators in the expressions for $F(u_t,u_{th})$ above). For large $\frac{h_t}{\epsilon}$, we need to use a Semi-Implicit \chg[af]{Method} and find that we can simplify the expressions for $F(u_t,u_{th})$ as follows:

\[\begin{array}{rclc}
F(u_t,u_{th}) & = & 0 & \mbox{if} \ u_t < u_{th}\\
& = & u_{th} & \mbox{if} \ u_t  = u_{th}\\
& = & 1 & \mbox{if} \ u_t  > u_{th}
\end{array}\]


\subsubsection{Five-Point and Nine-Point Finite-Difference Laplacian Method}

The Forward Euler Methods as described above efficiently solve the system, numerically, in the \chg[af]{absence} of diffusion. Therefore, we now need to introduce a scheme which looks after the diffusion term.

We will use two schemes. The five-point scheme:

\begin{small}
\begin{eqnarray}
\nabla^2{u_{n,m}} & = & \frac{u_{n-1,m}+u_{n+1,m}+u_{n,m-1}+u_{n,m+1}-4u_{n,m}}{h_x^2}+O(h_x^3)
\end{eqnarray}
\end{small}
\\
and the nine-point scheme:

\begin{small}
\begin{eqnarray}
\nabla^2{u_{n,m}} &=& \frac{1}{h_x^2}\left(4\left(u_{n-1,m}+u_{n+1,m}+u_{n,m-1}+u_{n,m+1}\right)+u_{n-1,m-1}\right.\nonumber\\
&& \left.+u_{n+1,m-1}+u_{n-1,m+1}+u_{n-1,m+1}-20u_{n,m}\right)+O(h_x^3)
\end{eqnarray}
\end{small}

We note that both the five-point and nine-point schemes are second order accurate schemes and therefore whether we use a five-point scheme or a nine-point scheme makes no difference to the accuracy of the calculations.

Therefore, for the numerical work that we cover here, we shall be using the five-point scheme, unless stated otherwise. 


\subsubsection{Operator Splitting}

Another method used in the \chg[p36spell]{numerical} schemes is operator splitting. The user of EZ-Spiral has the choice as to whether they wish to use operator splitting or not. \chg[p36gram]{Let} us assume the evolution of the $u$-field takes the following form:

\begin{eqnarray}
\pderiv{u}{t} &=& \mathcal{R}(u,v)+\mathcal{D}(u)
\end{eqnarray}
\\
where $\mathcal{R}$ is the reaction terms ($f(u,v)$ in Barkley's model) and $\mathcal{D}$ is the diffusion term.

If operator splitting is used then for each step then the equation above can be solved as:

\begin{eqnarray}
u_{n+\frac{1}{2}} &=& u_n+h_t\mathcal{R}(u_n,v_n)\\
u_{n+1}           &=& u_{n+\frac{1}{2}}+h_t\mathcal{D}(u_{n+\frac{1}{2}})
\end{eqnarray}
\\
where $u_n$ is the current time step, $u_{n+1}$ is the next time step and $u_{n+\frac{1}{2}}$ is the half step between the current step and the next step.

We shall be using operator splitting in the schemes we use in Chap.4.


\subsection{EZ-Spiral}

We shall now provide some details on the program EZ-Spiral which solves Barkley's model using the above mentioned numerical schemes. However, it doesn't just solve it, it also provides a graphical interpretation of the data which it has generated.

The program is written in C. It has been made available as Freeware software on an open \chg[af]{license} from GNU. Users are encouraged to not only use the software for their research but to \chg[af]{amend} it as they feel fit.

The graphics, in the latest version, is provided through OpenGL and we shall be using the graphics part of the code within EZ-Spiral for the graphics in our own program EZ-Freeze.

The program is aimed at Linux/Unix based systems and therefore the graphical window is provided using X-Windows. The program can be run in non-graphical mode in MS Windows, but obviously, only data can be generated, not graphics.

Provided with the program files is a file called \verb|task.dat|. Within this file are a variety of parameters, including the model parameters, numerical parameters, and a variety of other parameters such as a parameter which determines whether a file is to be produced containing the tip coordinates, or a parameter which determines how much information is relayed to the user on the screen before, during and after the simulation. The purpose of providing this file is so that we can compile the program just once and if we need to amend any parameters, we can just do it easily through the \verb|task.dat| file and not through the code, thereby avoiding having to compile the program again.

So, once the program is made and is executed, an X-Window will appear. The simulation is then started by activating the X-Window (by either clicking in the window or hovering the mouse arrow over the window) and then pressing the \verb|space| bar. Using the parameters provided with the program when it is first downloaded, the user should see a wave rotating fairly fast (depending on the spec of the system being used).

A variety of key presses are provided with the program. We have come across \verb|space|. Another is \verb|p|, which pauses the simulation and \verb|q| quits the simulation. Once the simulation has been terminated using the \verb|q| key, then a final conditions file is generated containing a host of various bits of information, such as \chg[af]{parameters}used, and more importantly, the values of $u$ and $v$ for each grid point at the tip the program was terminated. The file produced is called \verb|fc.dat|. This file can be copied to an initial conditions file, \verb|ic.dat|, and the next simulations start from those initial conditions.

Other key presses that are important are \verb|t| and the arrow keys. The \verb|t| key switches tip plotting on and off. So, if we switch on tip plotting, then the tip trajectory will be plotted on the screen from the moment \verb|t| is pressed. The arrow keys moves the wave around the screen screen.

One other point to note is that the code is extremely well commented. So it is not hard for someone fluent in C to find their way around the code. 
\label{sec:rev_numerical}
\chapter{Asymptotic Theory of Drift and Meander}
\label{chap:3}

\section{Introduction}
\label{sec:theory_intro}
The aim of this chapter is to provide \chg[p38correc]{an asymptotic} theory of the drift of meandering spiral waves. Other authors have also worked on this area \cite{leblanc2000,leblanc2002,wulff96,Golub97}, but we provide an alternative approach to this subject and an approach which we think is more straightforward.

We shall base our theory on the papers that were described in detail in Chap.\ref{chap:2} - the theory of \chg[p38caps]{meander} \cite{bik96} and the theory of drift \cite{bik95}. Both theories were written using different techniques and we intend to rewrite one of the theories using similar techniques to those used in the other theory.

So, the first section of this chapter is devoted to the rewriting of the theory of drift using \chg[p38caps]{group theoretical approaches} as well as \chg[p38caps]{perturbation} techniques. We will concentrate, firstly, on a rigidly rotating spiral wave, and provide three examples of drift (Resonant Drift, Electrophoretic Induced Drift, and Inhomogeneity Induced Drift).

We will \chg[p38spell]{then} proceed to extend this theory to meandering spiral waves. We note that in a suitable functional space (in this case the functional space will be the space of all group orbits), the solutions are periodic. In fact they are limit cycle solutions. This has been proven numerically, with attempts to prove it analytically by proving the transition from rigid rotation to meander is via a Hopf bifurcation\cite{wulff96}. Therefore, we shall be using Floquet \chg[p38caps]{theory} to study the meandering part of the wave.

So, we will give a review of Floquet \chg[p38caps]{theory} as first presented in 1883 by \chg[af]{Gaston Floquet} \cite{floquet1883}, extending this to adjoint solutions, before applying Floquet theory to meandering spiral waves. One of the features of the theory we will be presenting, is that the singular perturbation method developed gives rise to an ODE which determines the shift in the limit cycle. The solution to this ODE can be put into the form of the Arnol'd Standard Mapping. Therefore, it is thought that frequency locking \chg[af]{within} meandering spiral waves which are drifting due to symmetry breaking perturbations can be observed.

We end the chapter with a conclusion and indication of further work to do in this area.

\section{Drift of a Rigidly Rotating Spiral Wave}
\label{sec:drift_rw}
\subsection{Formulation of the problem}

In this section, we present a study of the drift of a rigidly rotating spiral wave. We will take a general Reaction-Diffusion system of equations (RDS) in the plane which is subject to a symmetry breaking perturbation, and we apply perturbation and group theory techniques to develop a theory that will describe the dynamics of the spiral wave.

The system of equations we consider is shown below:

\begin{equation}
\label{eqn:rdeu}
\pdut = \bDD\nabla^2\bu+\bof(\bu)+\epsilon\bh(\bu,\nabla\bu,\br,t),\quad\bu,\bof,\bh\in\mathbb{R}^n,\quad\bDD\in\mathbb{R}^{n\times n}\quad\br\in\mathbb{R}^2
\end{equation}
\\
where \chg[p39eqns]{$\bu=\bu(\br,t)=\bu(x,y,t)=(u^{(1)},u^{(2)},\hdots,u^{(n)}$,} $\textbf{D}$ is the matrix of diffusion coefficients, $0<\epsilon\ll1$ and $n\geq 2$. If $\bh=\textbf{0}$, then we have that Eqn.(\ref{eqn:rdeu}) is equivariant with respect to Euclidean transformations of the plane.

The perturbation, $\bh$, could in practice be any perturbation, but for our purpose, we require it to be a symmetry breaking perturbation and also that it is bounded. Therefore, we have that Eqn.(\ref{eqn:rdeu}) is no longer equivariant under Euclidean symmetry. We will later on show three distinct examples using different perturbations of how our theory will work.

Now, we will be studying the spiral wave in a frame of reference that is moving with the tip of the wave. There are several advantages to doing this, including the derivation of a generic form of the equations of motion of the tip of the spiral wave from the analysis in transforming the laboratory frame of reference to the comoving frame. Therefore, we will need to define the tip of the wave in the comoving frame of reference as follows:
\chg[af]{}
\begin{eqnarray}
\label{eqn:tipcon1}
u^{(i)}(\bR,t) &=& u_*\\
\label{eqn:tipcon2}
u^{(j)}(\bR,t) &=& v_*\\
\label{eqn:tipcon3}
\pderiv{\chg[]{u^{(i)}}}{\til{x}}(\bR,t) &=& 0
\end{eqnarray}
\\
where $\bR=(X,Y)$ is the coordinate of the tip of the spiral wave in the $(x,y)$ plane, and $1\leq{i,j}\leq{n}$. Also, we note that (\ref{eqn:tipcon3}) is a derivative with respect to $\til{x}$ and not $x$. Now, $\til{x}$ is the transformed laboratory frame coordinate, $x$, to the comoving frame given as:

\begin{equation*}
\left(\begin{array}{c} x\\ y \end{array}\right)\mapsto \left(\begin{array}{c} \til{x}\\ \til{y} \end{array}\right)=\left(\begin{array}{cc} \cos(\Theta) & \sin(\Theta) \\ -\sin(\Theta) & \cos(\Theta) \end{array}\right)\left(\begin{array}{c}x-X\\ y-Y\end{array}\right)
\end{equation*}
\\
where $(X,Y)$ are the tip coordinates and $\Theta$ is the tip phase. Therefore, our tip conditions \chg[af]{in the laboratory frame of reference} now become:

\begin{eqnarray*}
\label{eqn:tipcon4}
u^{(i)}(\bR,t) &=& u_*\\
\label{eqn:tipcon5}
u^{(j)}(\bR,t) &=& v_*\\
\label{eqn:tipcon6}
\cos(\Theta)\pderiv{u^{(i)}}{x}(\bR,t)-\sin(\Theta)\pderiv{u^{(i)}}{y}(\bR,t) &=& 0
\end{eqnarray*}

Let us now define $\bv(\br,t)$ such that $\bv(\br,t)$ is a transformation of the solution $\bu(\br,t)$ so that the tip of the wave is located at the origin for all time, and also the phase of the tip of the wave is constant for all time. Therefore, we have that $\bv(\br,t)$ is the solution to a RDS which is in the frame of reference moving with the tip of the wave. In the next section, we will \chg[p40spell]{derive} the RDS in the moving frame of reference.

Now, we define $\bv$ such that it is related to $\bu$ as follows:

\begin{eqnarray*}
\label{eqn:vgroup1}
\bv(\br,t) &=& T(g^{-1})\bu(\br,t)\\
\label{eqn:vgroup2}
\Rightarrow \bu(\br,t) &=& T(g)\bv(\br,t)
\end{eqnarray*}
\\
where $g\in SE(2)$, $g=\{\bR,\Theta\}$, i.e. the group of all rotations and translations. $T(g)$ is the action of the group element $g\in SE(2)$ and is defined as:

\begin{equation*}
T(g)\bu(\br,t) = \bu(g^{-1}\br,t)
\end{equation*}

Now, we will be choosing $\bR$ and $\Theta$ such that they are the tip coordinates and phase. We note that if $(\til{x},\til{y})$ are the coordinates in the moving frame of reference, then:

\begin{equation*}
\left(\begin{array}{c}\til{x}\\ \til{y}\end{array}\right) = \left(\begin{array}{cc} \cos(\Theta) & \sin(\Theta) \\ -\sin(\Theta) & \cos(\Theta) \end{array}\right)\left(\begin{array}{c}x-X\\ y-Y\end{array}\right)
\end{equation*}

Clearly, we can see that in the moving frame of reference, Eqns.(\ref{eqn:tipcon1})-(\ref{eqn:tipcon3}) become:

\begin{eqnarray}
\label{eqn:tipconv1}
v^{(i)}(\textbf{0},t) &=& u_*\\
\label{eqn:tipconv2}
v^{(j)}(\textbf{0},t) &=& v_*\\
\label{eqn:tipconv3}
\pdvx^{(i)}(\textbf{0},t) &=& 0
\end{eqnarray}
\\
where $v^{(i)}$ and $v^{(j)}$ are the $i$'th and $j$'th components of $\bv$ respectively, and $1\leq{i,j}\leq{n}$, and we have taken the tip to be at the origin.

As we will see in the next section, by representing this system in a functional space we can rewrite it in the comoving frame of reference whose origin is located at the tip of the spiral wave.


\subsection{Representation in the comoving frame of reference}

We now represent Eqn.(\ref{eqn:rdeu}) in a functional Space, $\mathcal{B}$. The functional space in question is taken to be the space of all Euclidean group orbits. A formal construction of this space is found in \cite{wulff96}.

We therefore find that the RDS in this space is:

\begin{equation}
\label{eqn:rdeU}
\odut = \boFU+\epsilon\bH(\bU,t)
\end{equation}
\\
where $\bU,\textbf{F},\bH\in\mathcal{B}$, and $\bU(t)\leftrightarrow\bu(x,t)$, $\textbf{F}(\bU)\leftrightarrow\bof(\bu)+\nabla^2\bu$.

Next, we define a manifold, which we call a Representative Manifold, in $\mathcal{B}$ consisting of all functions which have the tip of the spiral wave at the origin of the frame of reference for all time. We restrict our attention to spiral wave solutions, which may be formally defined as those having just one tip. We also restrict the solutions to those spiral waves having just one arm. Let us denote such a set of solutions $\mathcal{S}\subset\mathcal{B}$. So, 

\begin{equation*}
\label{eqn:manifold}
\mathcal{M}:\forall\bU\in \mathcal{S},\quad\exists'g\in SE(2),\bV\in\mathcal{M}:\quad\bU=T(g)\bV
\end{equation*}

We will also assume that the set $\mathcal{S}$ with respect to (\ref{eqn:rdeu}) is open and invariant, i.e. spiral waves do not terminate spontaneously and do not break up into many spirals (if they do, then such behaviour is not either drift or meander, and is not considered by this theory).

\begin{figure}[tbp]
\begin{center}
\begin{minipage}[b]{0.6\linewidth}
\centering
\psfrag{a}[l]{$g$}
\psfrag{b}[l]{$\mathcal{G}$}
\psfrag{d}[l]{$\mathcal{B}$}
\psfrag{e}[l]{$\mathcal{M}$}
\psfrag{f}[l]{$U$}
\psfrag{g}[l]{$U'$}
\psfrag{h}[l]{$V$}
\psfrag{i}[l]{$V'$}
\psfrag{j}[l]{$g'$}
\includegraphics[width=0.9\textwidth]{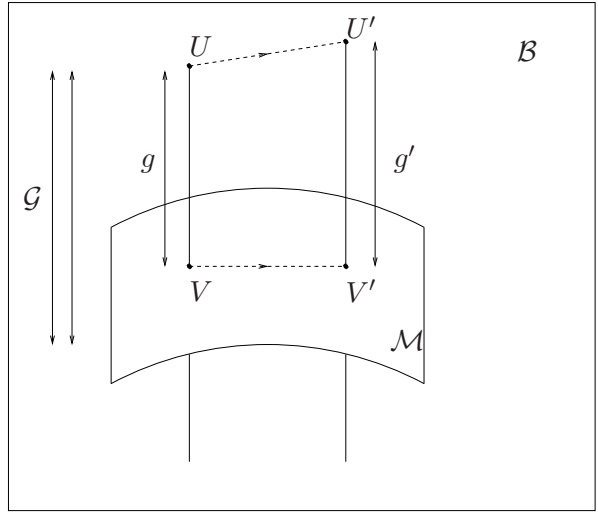}
\end{minipage}
\caption[Quotient system reduction in the functional space]{Decomposition of the movement in the functional space $\mathcal{B}$ onto movement along a manifold $\mathcal{M}$. We have that $V,V'\in\mathcal{M}$, $U,U''\in\mathcal{B}$ and $g,g'\in SE(2)$. $\mathcal{G}$ represents the group orbits in $\mathcal{B}$.}
\label{fig:theory_banach}
\end{center}
\end{figure}

We can decompose the motion in this space as motion along the Representative Manifold and motion along a group orbit which is transversal to this Representative Manifold. We show in Fig.(\ref{fig:theory_banach}) the representative manifold and also a spiral wave solution.

We have that:

\begin{equation}
\label{eqn:UTV}
\bU(t) = T(g)\bV(t)
\end{equation}

Substituting (\ref{eqn:UTV}) into (\ref{eqn:rdeU}), we get:

\begin{eqnarray}
\odTt\bV+\dT\odvt &=& \dT\boF+\epsilon\bH(\dT\bV,t)\nonumber\\
\label{eqn:rdeV}
\Rightarrow \dTi\odTt\bV+\odvt &=& \boF+\epsilon\dTi\bH(\dT\bV,t)
\end{eqnarray}

We note that the perturbation $\bH$ is not invariant under SE(2) symmetry and therefore equivariance does not apply here. For ease of notation, let $\til{\bH}(\bV,t) = \dTi\bH(\dT\bV,t)$.

Now split $\boF$ and $\bHt$ into their components along the Representative Manifold and the Group Orbit:

\begin{eqnarray}
\label{eqn:Fsplit}
\boF &=& \bFg+\bFm\\
\label{eqn:Hsplit}
\bHt &=& \bHg+\bHm
\end{eqnarray}

Using Eqns.(\ref{eqn:rdeV})-(\ref{eqn:Hsplit})we can then separate (\ref{eqn:rdeV}) into its equations along the Representative Manifold and also across the Group Orbit:

\begin{eqnarray}
\label{eqn:mani}
\odvt &=& \bFm+\epsilon\bHm\quad\quad\mbox{Manifold}\\
\label{eqn:group}
\dTi\odTt\bV &=& \bFg+\epsilon\bHg\quad\quad\mbox{Group Orbit}
\end{eqnarray}

Eqn.(\ref{eqn:mani}) can now be \chg[af]{rewritten} as follows using Eqns.(\ref{eqn:Fsplit}) and (\ref{eqn:Hsplit}):

\begin{equation*}
\label{eqn:man}
\odvt = \boF-\bFg+\epsilon\bHt-\epsilon\bHg
\end{equation*}

Using the equation along the Group Orbit (\ref{eqn:group}) together with Eqn.(\ref{eqn:UTV}), we can determine the generic forms of $\bHg$ and $\bFg$. 

To do this, we must first establish the generic forms \chg[p43spell1]{}of the equations of motion \chg[p43spell1]{along} the group orbits. Note that the representative manifold is made of spiral with their tip at the origin. Thus, the group transformations describe the position and orientation of the tip in the laboratory frame of reference.\chg[p43delsent]{}


\subsection{Generic Forms of the Equations of Motion}

Firstly, we consider the motion along the group orbit as given by Eqn.(\ref{eqn:group}) and, in particular, the form of the left hand side of this particular equation. We shall consider how the operator $\dTi\odTt$ acts on a function.

We recall from Group Theory that an action of $g\in SE(2)$ on a function $\bw(\br)$ is given by:

\begin{equation*}
\dT\bw(\br) = \bw(g^{-1}\br) = \bw(\tbr)
\end{equation*}
\\
where

\begin{equation*}
\tr = (r-R)e^{-i\Theta}
\end{equation*}
\\ 
for $r,R\in\mathbb{C}$ and $r=x+iy$, $R=X+iY$. In $\mathbb{R}^2$, we have that:

\begin{equation*}
\tbr = e^{-\gamma\Theta}(\br-\bR)
\end{equation*}
\\
where $\br,\bR\in\mathbb{R}^2$ and $\gamma$ is the rotational operator having direct correspondence to the Lie Group generators of $SE(2)$:

\begin{equation}
\label{eqn:theory_gamma_rot}
\gamma = \left(\begin{array}{cc} 0 & -1\\ 1 & 0 \end{array}\right)
\end{equation}

Therefore, we have that:

\begin{eqnarray}
\label{eqn:theory_Twr}
\dT\bw(\br) &=& \bw(e^{-\gamma\Theta}(\br-\bR))
\end{eqnarray}

Let us now differentiate\chg[p44eqn]{}Eqn.(\ref{eqn:theory_Twr}) with respect to time:

\begin{eqnarray*}
\odTt\bw(\br) &=& \frac{\partial}{\partial{t}}\bw(e^{-\gamma\Theta}(\br-\bR))\nonumber\\
              &=& \nabla\bw\cdot\frac{\partial}{\partial{t}}(e^{-\gamma\Theta}(\br-\bR))\nonumber\\
              &=& \nabla\bw\cdot(-e^{-\gamma\Theta}\dot{\bR}-\dot{\Theta}\gamma e^{-\gamma\Theta}(\br-\bR))\nonumber\\
              &=& \bW(\br)
\end{eqnarray*}

Now we apply the action of the inverse element of $g\in SE(2)$, bearing in mind that $g\br=\bR+e^{\gamma\Theta}\br$:

\begin{ajf}
\begin{eqnarray}
\dTi\odTt\bw(\br) &=& \dTi\bW(\br)\\
                  &=& \bW(g\br)\\
                  &=& \nabla\bw\cdot(-e^{-\gamma\Theta}\dot{\bR}-\gamma\dot{\Theta}e^{-\gamma\Theta}((\bR+e^{\gamma\Theta}\br)-\bR))\\
                  &=& \nabla\bw\cdot(-e^{-\gamma\Theta}\dot{\bR}-\gamma\dot{\Theta}e^{\gamma\Theta}e^{-\gamma\Theta}\br)\\
                  &=& \nabla\bw\cdot(-e^{-\gamma\Theta}\dot{\bR}-\gamma\dot{\Theta}\br)\\
                  &=& -\dot{\bR}e^{-\gamma\Theta}\cdot\nabla\bw-\gamma\dot{\Theta}\br\times\nabla\bw\\
                  &=& -(\dot{\bR}e^{-\gamma\Theta},\nabla)\bw-\dot{\Theta}\partial_\theta\bw\\
                  &=& -(\bc,\nabla)\bw-\omega\partial_\theta\bw
\end{eqnarray}
\\
where we have derived:

\begin{eqnarray}
\bc    &=& e^{-\gamma\Theta}\dot{\bR}\\
\omega &=& \dot{\Theta}
\end{eqnarray}
\end{ajf}

\begin{ajfthesis}
\begin{eqnarray*}
\dTi\odTt\bw(\br) &=& \dTi\bW(\br)\nonumber\\
                  &=& \bW(g\br)\nonumber\\
                  &=& \nabla\bw\cdot(-e^{-\gamma\Theta}\dot{\bR}-\dot{\Theta}\gamma e^{-\gamma\Theta}((\bR+e^{\gamma\Theta}\br)-\bR))\nonumber\\
                  &=& \nabla\bw\cdot(-e^{-\gamma\Theta}\dot{\bR}-\dot{\Theta}\gamma\br)
\end{eqnarray*}

Now, let us introduce the following notations:

\begin{eqnarray}
\label{eqn:theory_dR}
\bc    &=& e^{-\gamma\Theta}\dot{\bR}\\
\label{eqn:theory_dTh}
\omega &=& \dot{\Theta}
\end{eqnarray}

This therefore leads to:

\begin{eqnarray*}
\dTi\odTt\bw(\br) &=& \nabla\bw(-\bc-\omega\gamma\br)\nonumber\\
\label{eqn:theory_op_lhs}
                  &=& -\nabla\bw\cdot\bc-\omega\nabla\bw\gamma\br)
\end{eqnarray*}

Now, bearing in mind that $\gamma$ is the rotational matrix given by (\ref{eqn:theory_gamma_rot}), and letting:

\begin{eqnarray*}
\nabla\bw &=& \left(\pderiv{\bw}{x},\pderiv{\bw}{y}\right)\\
\bc &=& (c_x,c_y)^T\\
\br &=& (x,y)^T
\end{eqnarray*}
\\
we get that (\ref{eqn:theory_op_lhs}) becomes:

\begin{eqnarray}
\dTi\odTt\bw(\br) &=& -\left(\pderiv{\bw}{x},\pderiv{\bw}{y}\right)\left(\begin{array}{c}c_x\\c_y\end{array}\right)-\omega\left(\pderiv{\bw}{x},\pderiv{\bw}{y}\right)\left(\begin{array}{cc} 0 & -1\\ 1 & 0 \end{array}\right)\left(\begin{array}{c}x\\y\end{array}\right)\nonumber\\
&=& -\left(c_x\pderiv{}{x}+c_y\pderiv{}{y}\right)\bw-\omega\left(\pderiv{\bw}{x},\pderiv{\bw}{y}\right)\left(\begin{array}{c}-y\\x\end{array}\right)\nonumber\\
&=& -(\bc,\nabla)\bw-\omega\left(-y\pderiv{}{x}+x\pderiv{}{y}\right)\bw\nonumber\\
\label{eqn:theory_full_lhs_group}
&=& -(\bc,\nabla)\bw-\omega\pderiv{\bw}{\theta}
\end{eqnarray}
\\
where we now have that $(\bc,\nabla)$ is the scalar product between $\bc$ and $\nabla$. Now, we must consider that we are dealing with \chg[af]{solutions} in our functional space. Therefore, the spatial operators in (\ref{eqn:theory_full_lhs_group}) must be recognised as operators in the functional space, in which solutions depend on time. Therefore, we have that:

\begin{eqnarray*}
\dTi\odTt\bV &=& -(\bc,\hat{\nabla})\bV-\omega\hat{\partial}_\theta\bV\nonumber\\
\label{eqn:theory_full_lhs_banach}
\Rightarrow \bFg+\epsilon\bHg &=& -(\bc,\hat{\nabla})\bV-\omega\hat{\partial}_\theta\bV
\end{eqnarray*}
\\
where $\hat{\nabla}=(\hat{\partial}_x,\hat{\partial}_y)$, and the notation $\hat{\partial}$ denotes spatial \chg[p45gram]{partial differential} operations in the functional space.

\end{ajfthesis}

We can rearrange Eqns.(\ref{eqn:theory_dR}) and (\ref{eqn:theory_dTh}) and get the following generic forms for the equations of motion:

\begin{eqnarray*}
\dot{\bR}      &=& e^{\gamma\Theta}\bc\\
\dot{\Theta} &=& \omega
\end{eqnarray*}
\\
or, in complex terms:

\begin{eqnarray*}
\dot{R}      &=& ce^{i\Theta}\\
\dot{\Theta} &=& \omega
\end{eqnarray*}

We note that the value of $\omega$ and $\bc$ (or $c$) are defined from earlier considerations, namely that the dynamics of $\bV$ happens along the representative manifold.


\subsection{Motion along the Representative Manifold}

We will now consider Eqn.(\ref{eqn:mani}). We note that:
\chg[p46eqn1]{
\begin{eqnarray*}
\label{eqn:theory_1}
\bFm &=&\boF-\bFg\\
\label{eqn:theory_2}
\bHm &=& \bHt-\bHg
\end{eqnarray*}}

Therefore, using \chg[p46eqn2]{Eqns.(\ref{eqn:theory_1})-(\ref{eqn:theory_2}),} (Eqn.(\ref{eqn:mani}) becomes:

\begin{eqnarray}
\odvt &=& \bFm+\epsilon\bHm\nonumber\\
\Rightarrow \odvt &=& \boF+\epsilon\bHt-\bFg-\epsilon\bHg\nonumber\\
\label{eqn:mangen}
\Rightarrow \odvt &=& \boF+(\bc,\hat{\nabla})\bV+\omega\hat{\partial}_\theta\bV+\epsilon\bHt
\end{eqnarray}

In the original space, Eqn.(\ref{eqn:mangen}) can now be written as:

\begin{equation}
\label{eqn:maneuc}
\pdvt = \textbf{D}\nabla^2\bv+\bof(\bv)+(\bc,\nabla)\bv+\omega\partial_\theta\bv+\epsilon\bht(\bv,\br,t)
\end{equation}
\\
where $\nabla=(\partial_x,\partial_y)$, and the transformed perturbation $\bht$ is given by the explicit algorithm:

\begin{equation*}
\bht = \dTi\bh(\dT\bv,\br,t)
\end{equation*}

Let us consider for a moment what we have achieved, and what Eqn.(\ref{eqn:maneuc}) means. Eqn.(\ref{eqn:maneuc}) is not only a Reaction-Diffusion equation - it also contains Advection terms. The advection terms have evolved from the transformation from a laboratory (stationary) frame of reference to a frame of reference comoving with the tip of the spiral wave, i.e. in the functional space Eqn.(\ref{eqn:mangen}) is the motion along the Representative Manifold. So, we have taken a spiral wave solution, from the laboratory frame of reference and have simply transformed it using the Advection terms with carefully chosen advection coefficients, $\bc$ and $\omega$. 

Now, if we consider a meandering trajectory, for example, we would have that the trajectory along the representative manifold crosses the groups orbits passing transversally to the manifold. This means that, as expected, the shape of the wave is changing. Therefore, each group element $g$ whose group orbit the meandering trajectory passes, is changing with time. {\bf{Therefore, \chg[p47gram]{for meander,} the Advection coefficients, $\bc$ and $\omega$, are also changing with time and are also solutions to (\ref{eqn:maneuc})}}. We can therefore say that the Advection coefficients depend on the current value of the vector $\bV$ in the functional space or $\bv$ in terms of the PDE. This means that if $\bv\in\mathbb{R}^n$, then a solution will have $n+3$ components, $n$ of which are functions of $\br$ and the other 3 are just numbers. However, we only have $n$ equations. Now, Eqns.(\ref{eqn:tipconv1})-(\ref{eqn:tipconv3}) come from the definition of the manifold and we shall call these the \emph{tip pinning conditions}. Therefore, the full, closed system of equations is:

\begin{eqnarray*}
\label{eqn:theory_drift_101}
\pdvt &=& \textbf{D}\nabla^2\bv+\bof(\bv)+(\bc,\nabla)\bv+\omega\partial_\theta\bv+\epsilon\bht(\bv,\br,t)
\end{eqnarray*}
\begin{eqnarray*}
v^1(\textbf{0},t) &=& u_*\\
v^2(\textbf{0},t) &=& v_*\\
\label{eqn:theory_drift_102}
\pderiv{v^1(\textbf{0},t)}{x} &=& 0
\end{eqnarray*}

If we consider that the Advection coefficients are dependent on the solution $\bv$, the we can simply write the solution as:

\begin{eqnarray}
\label{eqn:theory_rw_fulleqn}
\pdvt &=& \textbf{D}\nabla^2\bv+\bof(\bv)+(\bc[\bv],\nabla)\bv+\omega[\bv]\partial_\theta\bv+\epsilon\bht(\bv,\br,t)
\end{eqnarray}

We shall see in Chap.\ref{chap:4} that it is possible to numerically solve this system. In fact, the whole of Chap.\ref{chap:4} is devoted to the numerical solution of such systems.

Finally, before moving onto the next part of the work, we must emphasise that a revised set of tip pinning conditions was found to be necessary and these are:

\begin{eqnarray*}
v^1(\textbf{0},t) &=& u_*\\
v^2(\textbf{0},t) &=& v_*\\
v^1(\bR_{inc},t) &=& u_*
\end{eqnarray*}
\\
where $\bR_{inc}=(x_{inc},y_{inc})$ is a point not at the origin, but some arbitrary point which the boundaries of the media in which we are studying this system. This set of pinning conditions will be discussed in full in Chap.\ref{chap:4}.

\subsection{Perturbation Theory}

We now have a system of equations which determines the evolution of the system along the manifold that we have defined. We know that since we are dealing with a perturbation problem, we must have a perturbed solution to this problem. Let us say that the solution is of the form:

\begin{equation}
\label{eqn:solper}
\bv = \bv_0+\epsilon\bv_1+O(\epsilon^2)
\end{equation}
\\
i.e. we are considering a regular perturbation technique. Also, we have from above that:

\begin{eqnarray}
\label{eqn:A1x}
c_{x} &=& c_{0x}+\epsilon c_{1x}\\
\label{eqn:A1y}
c_{y} &=& c_{0y}+\epsilon c_{1y}\\
\label{eqn:A2}
\omega &=& \omega_0+\epsilon\omega_1
\end{eqnarray}

Substituting Eqns.(\ref{eqn:solper}) and (\ref{eqn:A1x})-(\ref{eqn:A2}) into (\ref{eqn:maneuc}), we get:

\begin{eqnarray}
\label{eqn:rdeunpert}
\epsilon = 0 \Rightarrow \frac{\partial{\bv_0}}{\partial{t}} &=& \textbf{D}\nabla^2\bv_0+\bof(\bv_0)+(\bc_0,\nabla)\bv_0+\omega_0\partial_\theta\bv_0\\
\label{eqn:rdepert}
\epsilon \neq 0 \Rightarrow \frac{\partial{\bv_1}}{\partial{t}} &=& \textbf{D}\nabla^2\bv_1+\frac{\dd{\bof(\bv_0)}}{\dd{\bv}}\cdot\bv_1+(\bc_0,\nabla)\bv_1+\omega_0\partial_\theta\bv_1\nonumber\\
&& +(\bc_1,\nabla)\bv_0+\omega_1\bv_0+\bht(\bv_0,\br,t)+O(\epsilon)
\end{eqnarray}

If we let:

\begin{eqnarray}
\label{eqn:theory_L}
L\balp &=& \textbf{D}\nabla^2\balp+\frac{\dd{\bof(\bv_0)}}{\dd{\bv}}\cdot\balp+(\bc_0,\nabla)\balp+\omega_0\partial_\theta\balp\\
\label{eqn:h}
\mbox{and}\quad \hh &=& (\bc_1,\nabla)\bv_0+\omega_1\bv_0+\bht(\bv_0,\br,t)\nonumber
\end{eqnarray}
\\
we can rewrite (\ref{eqn:rdepert}) as:

\begin{equation*}
\label{eqn:rdepertL}
\frac{\partial{\bv_1}}{\partial{t}} = L(\bv_1)+\hh+O(\epsilon)
\end{equation*}

We also need to consider the tip pinning conditions to make the system closed. Therefore, we have that:

\[\begin{array}{rclcrcl}
  \epsilon=0: v_0^{(i)}(0,t) &=& u_* &,& \epsilon\neq0: v_1^{(i)}(0,t) &=& 0\\
   v_0^{(j)}(0,t) &=& v_* &,& v_1^{(j)}(0,t) &=& 0\\
   \pderiv{v_0^{(i)}(0,t)}{x} &=& 0 &,& \pderiv{v_1^{(i)}(0,t)}{x} &=& 0
  \end{array}\]

Furthermore, we note that the above conditions can be rewritten as:

\begin{eqnarray*}
(\bmu_k,\bv_0) &=& m_l\\
(\bmu_k,\bv_1) &=& 0
\end{eqnarray*}
\\
for $k=1,2,3$, and $l=i$ for $k=1,3$ and $l=j$ for $k=2$, where:

\begin{eqnarray*}
\bmu_k &=& \delta(\br)\be_l,\quad\mbox{for}\quad k=1,2\\
\bmu_3 &=& \partial_x\delta(\br)\be_i
\end{eqnarray*}
\\
and
\[\begin{array}{ccccccc}
\left(\begin{array}{c}m_1\\m_2\\m_3\end{array}\right) &=& \left(\begin{array}{c}u_*\\v_*\\0\end{array}\right) &,& \be_l &=& \left(\begin{array}{c}\delta_{1l} \\ \delta_{2l} \\ \cdot \\ \delta_{nl}\end{array}\right)\end{array}
\]

Also, we define the scalar product $(\balp,\bbet)$ as 

\begin{equation*}
(\balp,\bbet) = \int_{-\pi}^\pi\dd{\theta}\int_{-\infty}^\infty\int_{-\infty}^\infty\langle\bar{\balp},\bbet\rangle\dd{\br}
\end{equation*}

Therefore, two systems that we have are:

\begin{equation*}
\frac{\partial{\bv_0}}{\partial{t}} = \textbf{D}\nabla^2\bv_0+\bof(\bv_0)+(\bc_0,\nabla)\bv_0+\omega_0\partial_\theta\bv_0
\end{equation*}
\begin{eqnarray*}
(\bmu_j,\bv_0(0,t)) &=& m_j
\end{eqnarray*}
\\
for $j=1,2,3$, and
\begin{equation*}
\frac{\partial{\bv_1}}{\partial{t}} = L(\bv_1)+\hh+O(\epsilon)
\end{equation*}
\begin{eqnarray*}
(\bmu_j,\bv_1(0,t)) &=& 0
\end{eqnarray*}
\\
where $\bv_0,\bv_1,\boof\in\mathbb{R}^n$ and $\bDD\in\mathbb{R}^{n\times n}$


\subsection{Explicit forms of the eigenfunctions to $L$ \chg[p49title]{for rigidly rotating spiral waves}}

Before we proceed further, we will analytically derive explicit expressions for the eigenfunctions to the linear operator $L$.

Firstly, we note that the linear operator $L$ has an associated eigenvalue problem:

\begin{equation*}
\label{eqn:theory_L_eval_prob}
L\bphi_i = \lambda_i\bphi_i
\end{equation*}

Now, we also have an adjoint problem:

\begin{equation*}
\label{eqn:theory_L_eval_prob_adjoint}
L^+\bpsi_j = \bar{\lambda}_j\bpsi_j
\end{equation*}

Next, consider the equation for the unperturbed solution, i.e. Eqn.(\ref{eqn:rdeunpert}):

\begin{equation*}
\frac{\partial{\bv_0}}{\partial{t}} = \textbf{D}\nabla^2\bv_0+\bof(\bv_0)+(\bc_0,\nabla)\bv_0+\omega_0\partial_\theta\bv_0
\end{equation*}

We are looking primarily at rigidly rotating spiral wave solutions in the moving frame of reference. These solutions are actually stationary solutions in a frame of reference moving with the tip of the spiral wave. Therefore, we can take the partial derivative of $\bv_0$ with respect to time to be zero, giving us:

\begin{equation}
\label{eqn:rdscomov}
\textbf{D}\nabla^2\bv_0+\bof(\bv_0)+(\bc,\nabla)\bv_0+\omega_0\partial_\theta\bv_0 = 0
\end{equation}

\begin{ajf}
Let us now differentiate Eqn.(\ref{eqn:rdscomov}) with respect to $x$:

\begin{eqnarray}
\textbf{D}\nabla^2\pdvox+(\bc,\nabla)\pdvox+\omega_0\frac{\partial}{\partial{x}}(x\pdvoy-y\pdvox)+\pdfvov\pdvox &=& 0\nonumber\\
\textbf{D}\nabla^2\pdvox+(\bc,\nabla)\pdvox+\omega_0\pdvoy+\omega_0x\pdvoxy-\omega_0y\pdvoxx+\pdfvov\pdvox &=& 0\nonumber\\
\textbf{D}\nabla^2\pdvox+(\bc,\nabla)\pdvox+\omega_0\pdvoy+\omega_0x\frac{\partial}{\partial{y}}\pdvox-\omega_0y\frac{\partial}{\partial{x}}\pdvox+\pdfvov\pdvox &=& 0\nonumber\\
\textbf{D}\nabla^2\pdvox+(\bc,\nabla)\pdvox+\omega_0(x\frac{\partial}{\partial{y}}-y\frac{\partial}{\partial{x}})\pdvox+\pdfvov\pdvox &=& -\omega_0\pdvoy\nonumber\\
\textbf{D}\nabla^2\pdvox+(\bc,\nabla)\pdvox+\omega_0\partial_\theta\pdvox+\pdfvov\pdvox &=& -\omega_0\pdvoy\nonumber
\end{eqnarray}

Therefore, if we define $L$ to be:

\begin{equation}
L\alpha = \textbf{D}\nabla^2\alpha+(\bc,\nabla)\alpha+\omega_0\partial_\theta\alpha+\pdfvov\alpha
\end{equation}
\\
then we get:

\begin{equation}
\label{eqn:eigenx}
L\pdvox = -\omega_0\pdvoy
\end{equation}
\end{ajf}

\begin{ajfthesis}
Let us now differentiate Eqn.(\ref{eqn:rdscomov}) with respect to $x$. This gives:

\begin{eqnarray}
\textbf{D}\nabla^2\pdvox+(\bc,\nabla)\pdvox+\omega_0\partial_\theta\pdvox+\pdfvov\pdvox &=& -\omega_0\pdvoy\nonumber
\end{eqnarray}

Therefore, 

\begin{equation}
\label{eqn:eigenx}
L\pdvox = -\omega_0\pdvoy
\end{equation}
\\
where $L$ is defined in (\ref{eqn:theory_L}).
\end{ajfthesis}

Similarly, if we differentiate with respect to $y$ then we get:

\begin{equation}
\label{eqn:eigeny}
L\pdvoy = \omega_0\pdvox
\end{equation}

Now, if we multiply Eqn.(\ref{eqn:eigeny}) by $i$ and add to Eqn.(\ref{eqn:eigenx}) then we have:

\begin{eqnarray*}
L\bphi_1 &=& i\omega\bphi_1
\end{eqnarray*}
\\
where $\bphi_1$ is given by:

\begin{equation*}
\bphi_1 = \pdvox+i\pdvoy
\end{equation*}

Similarly:

\begin{eqnarray*}
L\bphi_{-1}       &=& -i\omega\bphi_{-1}
\end{eqnarray*}
\\
where:

\begin{equation*}
\bphi_{-1} = \pdvox-i\pdvoy
\end{equation*}

We can therefore see that there are 2 eigenvalues and their corresponding eigenfunctions for the operator $L$ given by:

\[\begin{array}{rclcrcl}
 \lambda_1    & = & i\omega_0  &,& \bphi_1    & = & \pdvox+i\pdvoy\\
 \lambda_{-1} & = & -i\omega_0 &,& \bphi_{-1} & = & \pdvox-i\pdvoy
\end{array}\]

There is one more eigenvalue that also appears on the imaginary axis. Consider differentiating Eqn.(\ref{eqn:rdscomov}) with respect to $\theta$:

\begin{eqnarray*}
\label{eqn:rdscomov2}
\textbf{D}\nabla^2\pdvoth+\frac{\partial}{\partial{\theta}}(\bc_0,\nabla)\bv_0+\omega_0\pdvothth+\pdfvov\pdvoth &=& 0
\end{eqnarray*}

\begin{ajf}
Let us consider the second and third terms of the above equation:

\begin{eqnarray}
\frac{\partial}{\partial{\theta}}(\bc_0,\nabla)\bv_0 &=& c_{0x}\pdvothx+c_{0y}\pdvothy\nonumber\\
&=& c_{0x}(x\pdvoyx-y\pdvoxx)+c_{0y}(x\pdvoyy-y\pdvoxy)
\end{eqnarray}

Now, consider the following equation:

\begin{eqnarray*}
(\bc_0,\nabla)\pdvoth &=& c_{0x}\frac{\partial}{\partial{x}}(\pdvoth)+c_{0y}\frac{\partial}{\partial{y}}(\pdvoth)\\
&=& c_{0x}\frac{\partial}{\partial{x}}(x\pdvoy-y\pdvox)+c_{0y}\frac{\partial}{\partial{y}}(x\pdvoy-y\pdvox)\\
&=& c_{0x}(\pdvoy+x\pdvoxy-y\pdvoxx)\\
&&  +c_{0y}(x\pdvoyy-y\pdvoyx-\pdvox)\\
&=& c_{0x}(x\pdvoxy-y\pdvoxx)+c_{0y}(x\pdvoyy-y\pdvoyx)+c_{0x}\pdvoy-c_{0y}\pdvox\\
&=& \frac{\partial}{\partial{\theta}}(\bc_0,\nabla)\bv_0+c_{0x}\pdvoy-c_{0y}\pdvox\\
\Rightarrow \frac{\partial}{\partial{\theta}}(\bc_0,\nabla)\bv_0 &=& (\bc_0,\nabla)\pdvoth-c_{0x}\pdvoy+c_{0y}\pdvox
\end{eqnarray*}

Therefore, Eqn.(\ref{eqn:rdscomov2}) can be written as:

\begin{eqnarray*}
\label{eqn:rdscomov3}
\textbf{D}\nabla^2\pdvoth+(\bc_0,\nabla)\pdvoth+c_{0x}\pdvoy-c_{0y}\pdvox+&&\\
\omega_0\pdvothth+\pdfvov\pdvoth &=& 0\nonumber\\
\textbf{D}\nabla^2\pdvoth+(\bc_0,\nabla)\pdvoth+\omega_0\pdvothth+\pdfvov\pdvoth &=& -c_{0x}\pdvoy+c_{0y}\pdvox
\end{eqnarray*}
\end{ajf}

\begin{ajfthesis}

\chg[p51gram]{We have:}

\begin{eqnarray*}
\frac{\partial}{\partial{\theta}}(\bc_0,\nabla)\bv_0 &=& (\bc_0,\nabla)\pdvoth-c_{0x}\pdvoy+c_{0y}\pdvox
\end{eqnarray*}

\end{ajfthesis}

Therefore, we get:

\begin{equation}
\label{eqn:eigenthno}
L\pdvoth = -c_{0x}\pdvoy+c_{0y}\pdvox
\end{equation}

Ideally, we wish to express Eqn.(\ref{eqn:eigenthno}) as an eigenvalue problem with eigenfunction $\bphi_0$.  If we look at the right hand side of Eqn.(\ref{eqn:eigenthno}), we can see that it can be expressed as follows:

\begin{equation*}
-c_{0x}\pdvoy+c_{0y}\pdvox = \mathrm{Im}\{\bar{c}_0\bphi_0\}
\end{equation*}

In fact, we can say that:

\begin{equation*}
-c_{0x}\pdvoy+c_{0y}\pdvox = \frac{1}{2i}(\bar{c}_0\bphi_1-c_0\bar{\bphi}_1)
\end{equation*}
\\
where $\bar{\bphi}_1=\bphi_{-1}$. So, we can see that the final eigenfunction we are looking for is a linear combination of $\pdvoth$, $\bphi_1$ and $\bar{\bphi}_1$:

\begin{ajf}

\begin{equation}
\bphi_0 = \pdvoth+\alpha\bphi_1+\beta_1a\bar{\bphi}_1
\end{equation}

We now apply the linear operator to the above to get:

\begin{eqnarray}
L\bphi_0 &=& L\pdvoth+\alpha L\bphi_1+\beta L\bar{\bphi}_1\\
L\bphi_0 &=& -c_{0x}\pdvoy+c_{0y}\pdvox+i\alpha\omega_0\bphi_1-i\beta\omega_0\bar{\bphi}_1\\
L\bphi_0 &=& \frac{1}{2i}(\bar{\bc}_0\bphi_1-\bc_0\bar{\bphi}_1)+i\alpha\omega_0\bphi_1-i\beta\omega_0\bar{\bphi}_1\\
L\bphi_0 &=& \bphi_1\left(\frac{\bar{\bc}_0}{2i}+i\alpha\omega_0\right)-\bar{\bphi}_1\left(\frac{\bc_0}{2i}+i\beta\omega_0\right)
\end{eqnarray}

Therefore we have that $\alpha$ and $\beta$ must have the following values:
\end{ajf}
\chg[p52eqn]{}
\begin{ajfthesis}
\begin{equation*}
\bphi_0 = \pdvoth+\alpha\bphi_1+\chg[]{\beta_1\bar{\bphi}_1}
\end{equation*}
\\
where
\end{ajfthesis}

\begin{eqnarray*}
\alpha &=& \frac{\bar{c}_0}{2\omega_0}\\
\beta  &=& \frac{c_0}{2\omega_0}
\end{eqnarray*}
\\
which then gives us:

\begin{equation*}
L\bphi_0 = 0
\end{equation*}
\\
implying that the eigenvalue in this instance is $\lambda_0=0$.

To summarise, we have shown that there are 3 eigenvalue that lie on the imaginary axis and these, together with their corresponding eigenfunctions, are:

\[\begin{array}{rclcrcl}
 \lambda_1    & = & i\omega_0  &,& \bphi_1    & = & \pdvox+i\pdvoy,\\
 \lambda_{-1} & = & -i\omega_0 &,& \bphi_{-1} & = & \pdvox-i\pdvoy,\\
 \lambda_0    & = & 0          &,& \bphi_0    & = & \pdvoth+\frac{\bar{\bc}_0}{2\omega_0}\bphi_1+\frac{\bc_0}{2\omega_0}\bar{\bphi}_1.
\end{array}\]

Finally, we can rearrange the above to get:

\begin{eqnarray}
\label{eqn:theory_pdvox}
\pdvox  &=& \frac{1}{2}(\bphi_1+\bphi_{-1}),\\
\label{eqn:theory_pdvoy}
\pdvoy  &=& \frac{1}{2i}(\bphi_1-\bphi_{-1}),\\
\label{eqn:theory_pdvoth}
\pdvoth &=& \bphi_0-\frac{\bar{\bc}_0}{2\omega_0}\bphi_1-\frac{\bc_0}{2\omega_0}\bar{\bphi}_1.
\end{eqnarray}


\subsection{Solvability Conditions}

Our next aim is to show that $\bv_1$ is bounded. To do this, we shall consider:

\begin{equation}
\label{eqn:theory_v1_evol}
\frac{\partial{\bv_1}}{\partial{t}} = L(\bv_1)+\hh+O(\epsilon)
\end{equation}
\\
and also consider $\bv_1$ expanded in the eigenbasis of $L$:

\begin{equation}
\label{eqn:theory_v1_basis}
\bv_1 = \sum_ia_i(t)\bphi_i(x)
\end{equation}
\\
where $\bphi_i$ satisfy:

\begin{equation}
\label{eqn:theory_L1}
L\bphi_i(x) = \lambda_i\bphi_i(x).
\end{equation}

Next, we assumed at the beginning of this theory that $\bh$ is bounded. Therefore, it follows that $\bht$ is also bounded. 

Let us now expand $\bht$ in its eigenbasis:

\begin{equation}
\label{eqn:theory_h_basis}
\bht = \sum_ih_i(t)\bphi_i(x)
\end{equation}

Associated to (\ref{eqn:theory_L1}) is an adjoint eigenvalue problem:

\begin{equation*}
\label{eqn:theory_Ladjoint1}
L^+\bpsi_j(x) = \bar{\lambda}_j\bpsi_j(x),
\end{equation*}
\\
where:

\begin{equation}
\label{eqn:theory_biorthog}
(\bpsi_j,\bphi_i) = \delta_{ij}.
\end{equation}

Let us now substitute Eqns.(\ref{eqn:theory_v1_basis}) and (\ref{eqn:theory_h_basis}) in (\ref{eqn:theory_v1_evol}):

\begin{equation*}
\sum_i\dot{a}_i\bphi_i = \sum_i(La_i\bphi_i+h_i\bphi_i)
\end{equation*}

Premultiplication by $\bpsi_j$ and using the biorthogonality condition (\ref{eqn:theory_biorthog}) gives us:

\begin{equation*}
\dot{a}_i = \lambda_ia_i+h_i.
\end{equation*}

Integration gives us:

\begin{equation*}
a_i(t) = a_i(0)e^{\lambda_it}+\int_0^th_i(\tau)e^{\lambda_i(t-\tau)}\dd{\tau}.
\end{equation*}
\\
or, if we define $\xi=t-\tau$:

\begin{equation}
\label{eqn:theory_ai}
a_i(t) = a_i(0)e^{\lambda_it}+\int_0^th_i(t-\xi)e^{\lambda_i\xi}\dd{\xi}
\end{equation}

We now wish to show that $\bv_1$ is bounded, which will be achieved if we can show that $a_i$ is bounded for all $i$.

We note from \cite{bik95} that the \emph{Stability Postulate} claims that for stable solutions we require that the eigenvalues $\lambda_i$ satisfy $\mathrm{Re}\{\lambda_i\}\leq0$. We know that for rigidly rotating spiral waves there are only three critical eigenvalues which lie on the imaginary axis \chg[af]{($\lambda_{0,\pm1}=0,\pm i\omega_0$).} Therefore, all other eigenvalues satisfy \chg[af]{${\rm Re}\{\lambda_{i\neq\{0,\pm1\}}\}<0$.}

Consider the absolute value of (\ref{eqn:theory_ai}):

\begin{eqnarray}
\label{eqn:theory_ai_abs}
|a_i(t)| &\leq& \left|a_i(0)e^{\lambda_it}\right|+\left|\int_0^th_i(t-\xi)e^{\lambda_i\xi}\dd{\xi}\right|
\end{eqnarray}

We now assume, and prove later on, that $h_i$ are bounded, $h_i\leq K$ where $K$ is a constant:
\chg[p54eqn]{}
\begin{eqnarray*}
|a_i(t)| &\leq& \left|a_i(0)e^{\lambda_it}\right|+K\left|\int_0^te^{\lambda_i\xi}\dd{\xi}\right|\nonumber\\
|a_i(t)| &\leq& a_i(0)\chg[]{e^{{\rm Re}\{\lambda_i\}t}}+\frac{K}{|\lambda_i|}\left(e^{\mathrm{Re}\{\lambda_i\}t}-1\right)
\end{eqnarray*}

Clearly, for $\mathrm{Re}\{\lambda_i\}<0$, i.e. for $i\neq0,\pm1$, we have that $a_i$ are bounded. For the case $i=0$ where $\lambda_0=0$, we are best referring back to Eqn.(\ref{eqn:theory_ai_abs}):

\begin{eqnarray*}
|a_0(t)| &\leq& \left|a_0(0)\right|+\left|\int_0^th_0(t-\xi)\dd{\xi}\right|
\end{eqnarray*}

This is guaranteed to be bounded if, for example, $h_0=0$. Similarly, it can be shown that $h_{\pm1}=0$ for bounded $a_{\pm1}$.

Now, we know that:

\begin{eqnarray*}
\hh &=& \sum_ih_i\bphi_i\nonumber\\
\Rightarrow h_i &=& (\bpsi_i,\hh)  
\end{eqnarray*}
\\
Therefore:

\begin{eqnarray*}
\label{eqn:theory_sol_cond}
\Rightarrow (\bpsi_{0,\pm1},\hh) &=& 0
\end{eqnarray*}
\\
is a solvability condition which we will use in the next section to help us determine the full equations of motion.

It remains for us to shown that the $h_i$ are indeed bounded. Consider $\hh$:

\begin{eqnarray*}
\hh = c_x\pderiv{\bv_0}{x}+c_y\pderiv{\bv_0}{y}+\omega\pderiv{\bv_0}{\theta}+\bht
\end{eqnarray*}

We note that the derivatives of $\bv_0$ with respect to $x$, $y$ and $\theta$ can be expressed as linear combinations of $\bphi_{0,\pm1}$. We also note that $\bht$ is bounded by the original assumption and can be expressed as:

\begin{equation*}
\bht = \sum_ik_i\bphi_i.
\end{equation*}

So, we can rewrite $\hh$ as:

\begin{eqnarray*}
\hh &=& \sum_i(A_ic_i+k_i)\bphi_i\\
\sum_ih_i\bphi_i &=& \sum_i(A_ic_i+k_i)\bphi_i
\end{eqnarray*}

Premultiplying by $\bpsi_j$, we get:

\begin{equation*}
h_i = A_ic_i+k_i
\end{equation*}

Consider the absolute value of $h_i$ above:

\begin{eqnarray*}
|h_i| &\leq& |A_ic_i|+|k_i|
\end{eqnarray*}

Now we know that $\bht$ is bounded, hence $k_i$ are also bounded. We also know that $c_i$ and $A_i$ are bounded constant terms. Therefore, we have that $h_i$ are bounded. QED.


\subsection{Equations of Motion}

Substitution of Eqn.(\ref{eqn:h}) into Eqn.(\ref{eqn:theory_sol_cond})yields:

\begin{eqnarray*}
\label{eqn:motion1}
c_{1x}(\bpsi_l,\frac{\partial{\bv_0}}{\partial{x}})+c_{1y}(\bpsi_l,\frac{\partial{\bv_0}}{\partial{y}})+\omega_1(\bpsi_l,\frac{\partial{\bv_0}}{\partial{\theta}})&=& -(\bpsi_l,\bht(\bv_0,\br,t))\quad\mbox{for}\quad l=0,\pm1
\end{eqnarray*}

Using Eqns.(\ref{eqn:theory_pdvox})-(\ref{eqn:theory_pdvoth}), we have that (\ref{eqn:motion1}) is transformed to:

\begin{eqnarray*}
c_{1x}(\bpsi_l,\frac{1}{2}(\bphi_1+\bphi_{-1}))+c_{1y}(\bpsi_l,\frac{1}{2i}(\bphi_1-\bphi_{-1}))&&\\
+\omega_1(\bpsi_l,\bphi_0-\frac{\bar{c}_0}{2\omega_0}\bphi_1-\frac{c_0}{2\omega_0}\bar{\bphi}_1) &=& -(\bpsi_l,\bht(\bv_0,\br,t))\\
(\frac{1}{2}c_{1x}-\frac{i}{2}c_{1y}-\frac{\bar{c}_0}{2\omega_0}\omega_1)(\bpsi_l,\bphi_1)+&&\\
(\frac{1}{2}c_{1x}+\frac{i}{2}c_{1y}-\frac{c_0}{2\omega_0}\omega_1)(\bpsi_l,\bar{\bphi}_1)+&&\\
\omega_1(\bpsi_l,\bphi_0) &=& -(\bpsi_l,\bht(\bv_0,\br,t))
\end{eqnarray*}

So, if $l=0$, we get:
\chg[af]{}
\begin{equation*}
\label{eqn:Omega}
\omega_1 = -(\bpsi_0,\bht(\bv_0,\br,t))
\end{equation*}

Now, we recall that we previously had that $\bphi_{-1}=\bar{\bphi}_1$. Hence, for $l=-1$:

\begin{eqnarray}
\frac{1}{2}(c_{1x}+ic_{1y}-\frac{c_0}{\omega_0}\omega_1) &=& -(\bar{\bpsi}_1,\bht(\bv_0,\br,t))\nonumber\\
c_{1x}+ic_{1y} &=& -2(\bar{\bpsi}_1,\bht(\bv_0,\br,t))+\frac{c_0}{\omega_0}\omega_1\nonumber\\
\label{eqn:theory_6}
c_1          &=& -2(\bar{\bpsi}_1,\bht(\bv_0,\br,t))+\frac{c_0}{\omega_0}(\bpsi_0,\bht(\bv_0,\br,t))
\end{eqnarray}

We previously noted that the equations of motion are:

\begin{eqnarray*}
\odRt &=& ce^{i\Theta}\\
\odTht &=& \omega
\end{eqnarray*}
\\
which implies that:

\begin{eqnarray*}
\odRt &=& (c_0+\epsilon c_1)e^{i\Theta}\\
\odTht &=& \omega_0+\epsilon\omega_1
\end{eqnarray*}

Therefore, our full equations of motion become:
\chg[p56eqn]{}
\begin{eqnarray}
\label{eqn:chap3_1}
\deriv{R}{t} &=& \left[c_0-\epsilon(2(\bar{\bpsi}_1,\bht(\bv_0,\br,t))+\frac{c_0}{\omega_0}(\bpsi_0,\bht(\bv_0,\br,t)))\right]e^{i\Theta}\\
\label{eqn:chap3_2}
\odTht &=& \omega_0-\epsilon (\bpsi_0,\bht(\bv_0,\br,t))
\end{eqnarray}


\subsection{\chgex[2]{Comparison to the Biktashev Approach}}
\chgex[]{
We shall now compare our results from this section to the previous results of Biktashev et al \cite{bik95}.

The main and obvious difference is that we consider the motion of the tip of the spiral, whereas Biktashev et al consider the motion of the center of rotation. Other differences include the form of the Goldstone Modes, in particular the translational Goldstone Mode corresponding to the eigenvalue $\lambda_0=0$. However, we note that the eigenvalues, in both the comoving frame of reference and rotating frame of reference are exactly the same.

We note that we can reformulate Eqns.(\ref{eqn:chap3_1})\&(\ref{eqn:chap3_2}), as follows. Firstly, consider the orientation, $\Theta$ of the tip of the spiral wave. We call the orientation of the center of rotation, $\Phi$, and note that for small perturbations, its derivative , $\Phi$ is slowly varying, i.e. its derivative is small, $\dot{\Phi}=O(\epsilon)$. Therefore, we can relate $\dot{\Phi}$ to $\dot{\Theta}$ as:

\begin{equation*}
\dot{\Theta} = \dot{\Phi}+\dot{\theta}
\end{equation*}

Next, in order to transform the coordinates of the tip to the coordinates of the center of rotation, we perform a sliding average of the tip equations, where the average is taken over the period of the unperturbed spiral. If we represent the coordinates of the center of rotation as $\bar{R}=\bar{X}+i\bar{Y}$, then we have:

\begin{eqnarray*}
\deriv{\bar{R}}{t} &=& \frac{1}{T}\int_{-\frac{T}{2}}^{\frac{T}{2}}\dd\tau\left[c_0-\epsilon(2(\bar{\bpsi}_1,\bht(\bv_0,\br,t))+\frac{c_0}{\omega_0}(\bpsi_0,\bht(\bv_0,\br,t)))\right]e^{i\Theta}\\
\deriv{\bar{\Phi}}{t} &=& -\frac{\epsilon}{T}\int_{-\frac{T}{2}}^{\frac{T}{2}}(\bpsi_0,\til{h})\dd\tau
\end{eqnarray*}
\\
where $T=\frac{2\pi}{\omega_0}$. We find that the resulting equations are:

\begin{eqnarray}
\label{eqn:theory_4}
\deriv{\bar{R}}{t} &=& -\frac{2\epsilon}{T}\int_{-\frac{T}{2}}^{\frac{T}{2}}\int_{\mathbb{R}^2}e^{i\theta}\langle\bar{\bpsi}_1,\til{h}\rangle\dd\br\dd\tau\\
\deriv{\bar{\Phi}}{t} &=& -\frac{\epsilon}{T}\int_{-\frac{T}{2}}^{\frac{T}{2}}\int_{\mathbb{R}^2}\langle\bpsi_0,\til{h}\rangle\dd\br\dd\tau\nonumber
\end{eqnarray}

If we take the equations of motion from Biktashev et al and represent them in their averaged coordinates then we find that we have:

\begin{eqnarray}
\label{eqn:theory_5}
\deriv{\bar{R}}{t} &=& \frac{\epsilon}{T}\int_{-\frac{T}{2}}^{\frac{T}{2}}\int_{\mathbb{R}^2}e^{i(\omega t-\Phi)}\langle\bpsi_1,h\rangle\dd\br\dd\tau\\
\deriv{\bar{\Phi}}{t} &=& \frac{\epsilon}{T}\int_{-\frac{T}{2}}^{\frac{T}{2}}\int_{\mathbb{R}^2}\langle\bpsi_0,h\rangle\dd\br\dd\tau\nonumber
\end{eqnarray}

We can see that the equations are very similar. One difference is the factor of two in the coordinates equations. This can be traced back to Eqns.(12) and (13) in \cite{bik95}, where the factor of $\frac{1}{2}$ is included in (12) and therefore does not appear explicitly in (13). Therefore, the factor of two is actually there but it is ``built in'' to the Goldstone Modes.

Another obvious difference is the use of the response function $\bar{\bpsi}_1$ in (\ref{eqn:theory_4}) as opposed to $\bpsi_1$ in (\ref{eqn:theory_5}). This is, of course, down to the choice of the Goldstone mode which we have decided to use in order to determine the forms of the first order perturbed parts of the translational and angular velocities ($c$ and $\omega$ respectively). We could have chosen to use the $\bphi_1$ Goldstone mode, rather than its comjugate, $\bar{\bphi}_1$.

We shall present three examples in the next section. Before we show the specifics of these examples, we would like to provide one more similarity between the theory presented here and Keener's theory. In \cite{keener88b}, Keener showed that the coefficient $b_2$, which Biktashev et al showed was equal to the filament tension of a scoll wave, is given by:

\begin{equation*}
b_2 = \langle\bDD V_x,Y_x\rangle
\end{equation*}

We also note from \cite{panf87} that the tension is equal to the velocity of electrophoretic induced drift. We show in the examples that our theory indicates that the velocity of the drift is given by:

\begin{equation*}
A_{-1,-1} = (\bar{\bpsi}_1,A\bar{\bphi}_1)
\end{equation*}
\\
which is of course comparable to what Keener derived.}

\section{Drift of a Rigidly Rotating Spiral Wave: Examples}
\label{sec:drift_rw_examples}
In this section, we will show three different examples of how our revised theory works. Each example will involve a different type of perturbation. 

We start with the simplest type of perturbation, which is a perturbation depending only on time. This gives rise to Resonant Drift. The next example is Electrophoretic induced drift, which is a rotational symmetry breaking perturbation. We conclude the section with an example of Inhomogeneity indiced drift, which gives rise to translational symmetry breaking.

In each case we shall calculate the transformed perturbations before deriving explicit expressions for the tip trajectory. We shall then show a comparison of the trajectories produced using our theory and those from actual numerical calculations.

\subsection{Resonant Drift}

We saw \chg[rw_examples_resonant]{in} Sec.(\ref{sec:drift_rw_examples}) that the equations of motion for the tip of a spiral wave that is drifting are:

\begin{eqnarray*}
\odRt  &=& c_0e^{i\Theta}-2\epsilon(\bar{\bpsi}_1,\bht(\bv_0,\br,t))e^{i\Theta}+\frac{\epsilon c_0}{\omega_0}(\bpsi_0,\bht(\bv_0,\br,t))e^{i\Theta}\\
\odTht &=& \omega_0+\epsilon (\bpsi_0,\bht(\bv_0,\br,t))
\end{eqnarray*}
\\
where $R=X+iY$ and $\Theta$ are the tip coordinate and phase of the spiral, $c_0\in\mathbb{C}$ is the unperturbed velocity of the spiral, $\omega\in\mathbb{R}$ is the unperturbed frequency, $\bht$ is the transformed symmetry breaking perturbation (this is what causes the spiral to drift), $\epsilon$ is a small parameter, and $\bpsi_i$ are the eigenfunctions to the adjoint operator of $L$ defined as:

\begin{equation*}
L\boldsymbol{\alpha} = \textbf{D}\nabla^2\boldsymbol{\alpha}+(c_0,\nabla)\boldsymbol{\alpha}+\omega_0\partial_\theta\boldsymbol{\alpha}+\pdfvov\boldsymbol{\alpha}
\end{equation*}

We are considering resonant drift, which means that the perturbation is \chg[af]{dependent} only on time. We take the perturbation to be:

\begin{equation}
\label{eqn:hreson}
\epsilon\bh = \epsilon\bh(t) = \bA\cos(\Omega t+\xi)
\end{equation}
\\
where $\Omega$ is the frequency of the perturbation, and $\bA=\epsilon\ba$ is a real valued $n$-component vector:

\begin{equation*}
\bA = (A_1,A_2,\cdots,A_n)^T
\end{equation*}
\\
whose elements are $O(\epsilon)$. Now, $\bht$ is defined to be:

\begin{equation*}
\bht = \dTi\textbf{h}(\dT\bv_0,\br,t)
\end{equation*}

Since $\bh$, as defined as in Eqn.(\ref{eqn:hreson}), is not \chg[af]{dependent} on spatial coordinates, then we have that:

\begin{equation*}
\label{eqn:hresontrans}
\til{\bh} = \bh(t) = \bA\cos(\Omega t+\xi)
\end{equation*}

Our equations of motion are now:

\begin{eqnarray*}
\odRt  &=& c_0e^{i\Theta}-\epsilon e^{i\Theta}(2(\bar{\bpsi}_1,\bh(t))+\frac{c_0}{\omega_0}(\bpsi_0,\bh(t)))\\
\odTht &=& \omega_0+\epsilon (\bpsi_0,\bh(t))
\end{eqnarray*}

We now consider the inner products $(\bpsi_i,\bh(t))$. By definition we have:

\begin{eqnarray*}
(\bpsi_i,\bh(t)) &=& (\bpsi_i,\bA\cos(\Omega t+\xi))\nonumber\\
								 &=& (\bpsi_i,\bA)\cos(\Omega t+\xi)\nonumber\\
								 &=& \beta_i\cos(\Omega t+\xi)
\end{eqnarray*}
\\
where $\beta_i=O(\epsilon)$ and is a constant since we are dealing with Rigid Rotation. We note that $\beta_i$ is either complex or real depending on $\bpsi_i$ - if $\bpsi_i\in\mathbb{C}$ then $\beta_i\in\mathbb{C}$ and vice versa. So, for $i=\pm1$, $\beta_{\pm1}$ is complex, whilst for $i=0$, $\beta_0\in\mathbb{R}$. Therefore, our equations of motion are:
\chg[rw_examples_resonant]{}
\begin{eqnarray}
\label{eqn:theory_rw_ex_reson_odeR}
\odRt  &=& c_0e^{i\Theta}-\cos(\Omega t+\xi)e^{i\Theta}(2\bar{\beta}_1+\frac{c_0}{\omega_0}\beta_0)\\
\label{eqn:theory_rw_ex_reson_odeTh}
\odTht &=& \chg[]{\omega_0-\beta_0\cos(\Omega t+\xi)}
\end{eqnarray}

Since we have only time dependency within Eqns.(\ref{eqn:theory_rw_ex_reson_odeR}) \& (\ref{eqn:theory_rw_ex_reson_odeTh}) then we can integrate them directly.

Consider firstly Eqn.(\ref{eqn:theory_rw_ex_reson_odeTh}):
\chg[rw_examples_resonant]{}
\begin{eqnarray}
\odTht &=& \chg[]{\omega_0-\beta_0\cos(\Omega t+\xi)}\nonumber\\
\label{eqn:theory_rw_ex_reson_Th}
\Rightarrow \Theta(t) &=& \chg[]{\Theta_0+\omega_0t-\frac{\beta_0}{\Omega}\sin(\Omega t+\xi)}
\end{eqnarray}

Let us now introduce the following notation:

\begin{equation*}
\gamma = 2\bar{\beta}_1+\frac{c_0}{\omega_0}\beta_0 = O(\epsilon)
\end{equation*}
\\
and substitute Eqn.(\ref{eqn:theory_rw_ex_reson_Th}) into Eqn.(\ref{eqn:theory_rw_ex_reson_odeR}):

\begin{ajf}
\begin{eqnarray}
\odRt  &=& c_0e^{i\Theta}-\gamma\cos(\Omega t+\xi)\chg[]{e^{i-\Theta}}\nonumber\\
\Rightarrow \odRt  &=& c_0e^{i(\Theta_0+\omega_0t+\frac{\beta_0}{\Omega}\sin(\Omega t+\xi))}-\gamma\cos(\Omega t+\xi)e^{i(\Theta_0+\omega_0t+\frac{\beta_0}{\Omega}\sin(\Omega t+\xi))}\nonumber\\
\Rightarrow \odRt  &=& c_0e^{i(\Theta_0+\omega_0t)}e^{i\frac{\beta_0}{\Omega}\sin(\Omega t+\xi)}-\gamma\cos(\Omega t+\xi)e^{i(\Theta_0+\omega_0t)}e^{(\frac{\beta_0}{\Omega}\sin(\Omega t+\xi))}\nonumber
\end{eqnarray}
\end{ajf}
\chg[rw_examples_resonant]{}
\begin{ajfthesis}
\begin{eqnarray*}
\odRt  &=& c_0e^{i(\Theta_0+\omega_0t)}\chg[]{e^{-i\frac{\beta_0}{\Omega}\sin(\Omega t+\xi)}}-\gamma\cos(\Omega t+\xi)e^{i(\Theta_0+\omega_0t)}\chg[]{e^{-i(\frac{\beta_0}{\Omega}\sin(\Omega t+\xi))}}\nonumber
\end{eqnarray*}
\end{ajfthesis}

We note that $\beta_0=O(\epsilon)$, and so we have:

\begin{ajf}
\begin{eqnarray}
\odRt  &=& c_0e^{i(\Theta_0+\omega_0t)}(1+i\frac{\beta_0}{\Omega}\sin(\Omega t+\xi))-\gamma\cos(\Omega t+\xi)e^{i(\Theta_0+\omega_0t)}+O(\epsilon^2)\nonumber\\
\odRt  &=& c_0e^{i(\Theta_0+\omega_0t)}+i\frac{\beta_0c_0}{\Omega}\sin(\Omega t+\xi)e^{i(\Theta_0+\omega_0t)}-\gamma\cos(\Omega t+\xi)e^{i(\Theta_0+\omega_0t)}+O(\epsilon^2)\nonumber
\end{eqnarray}
\end{ajf}
\chg[rw_examples_resonant]{}
\begin{ajfthesis}
\begin{eqnarray*}
\odRt  &=& c_0e^{i(\Theta_0+\omega_0t)}\chg[]{-i\frac{\beta_0c_0}{\Omega}}\sin(\Omega t+\xi)e^{i(\Theta_0+\omega_0t)}-\gamma\cos(\Omega t+\xi)e^{i(\Theta_0+\omega_0t)}+O(\epsilon^2)\nonumber
\end{eqnarray*}
\end{ajfthesis}
\\
or:

\begin{ajf}
\begin{eqnarray}
\odRt  &=& c_0e^{i(\Theta_0+\omega_0t)}\chg[]{-i\frac{\beta_0c_0}{2i\Omega}}(e^{i(\Omega t+\xi)}-e^{-i(\Omega t+\xi)})e^{i(\Theta_0+\omega_0t)}\nonumber\\
&& -\frac{\gamma}{2}(e^{i(\Omega t+\xi)}+e^{-i(\Omega t+\xi)})e^{i(\Theta_0+\omega_0t)}+O(\epsilon^2)\nonumber\\
\odRt  &=& c_0e^{i(\Theta_0+\omega_0t)}+i\frac{\beta_0c_0}{2i\Omega}(e^{i((\omega_0+\Omega)t+\xi+\Theta_0)}-e^{i((\omega_0-\Omega)t+\xi-\Theta_0)})\nonumber\\
&& -\frac{\gamma}{2}(e^{i((\omega_0+\Omega)t+\xi+\Theta_0)}+e^{i((\omega_0-\Omega)t+\xi-\Theta_0)})+O(\epsilon^2)\nonumber\\
\odRt  &=& c_0e^{i(\Theta_0+\omega_0t)}+\left(i\frac{\beta_0c_0}{2i\Omega}-\gamma\right)e^{i((\omega_0+\Omega)t+\xi+\Theta_0)}\nonumber\\
&& -\left(i\frac{\beta_0c_0}{2i\Omega}+\gamma\right)e^{i((\omega_0-\Omega)t+\xi-\Theta_0)}+O(\epsilon^2)\nonumber\\
\label{eqn:theory_rw_ex_reson_odRr}
\odRt  &=& c_0e^{i(\Theta_0+\omega_0t)}+\gamma_-e^{i((\omega_0+\Omega)t+\xi+\Theta_0)}-\gamma_+e^{i((\omega_0-\Omega)t+\xi-\Theta_0)}+O(\epsilon^2)
\end{eqnarray}
\end{ajf}
\chg[rw_examples_resonant]{}
\begin{ajfthesis}
\begin{eqnarray}
\label{eqn:theory_rw_ex_reson_odRr}
\odRt  &=& c_0e^{i(\Theta_0+\omega_0t)}\chg[]{-\gamma_-}e^{i((\omega_0+\Omega)t+\xi+\Theta_0)}-\gamma_+e^{i((\omega_0-\Omega)t+\xi-\Theta_0)}+O(\epsilon^2)
\end{eqnarray}
\end{ajfthesis}
\\
where \chg[rw_examples_resonant]{$\gamma_+=\frac{1}{2}\left(\frac{\beta_0c_0}{\Omega}+\gamma\right)$} and \chg[rw_examples_resonant]{$\gamma_-=\frac{1}{2}\left(\frac{\beta_0c_0}{2i\Omega}-\gamma\right)$}. Let us now integrate this equation:
\chg[rw_examples_resonant]{}
\begin{eqnarray}
R &=& R_0-\frac{ic_0}{\omega_0}e^{i(\Theta_0+\omega_0t)}\chg[]{+\frac{i\gamma_-}{\omega_0+\Omega}}e^{i((\omega_0+\Omega)t+\xi+\Theta_0)}\nonumber\\
\label{eqn:theory_rw_ex_reson_R}
&& \chg[]{-\frac{i\gamma_+}{\omega_0-\Omega}}e^{i((\omega_0-\Omega)t+\xi-\Theta_0)}+O(\epsilon^2)
\end{eqnarray}
\\
provided that $\omega_0^2\neq\Omega^2$. The first two terms in (\ref{eqn:theory_rw_ex_reson_R}) determine the initial position of the tip of the spiral wave ($R_0=X_0+iY_0$) and also the unperturbed trajectory ($-\frac{ic_0}{\omega_0}e^{i(\Theta_0+\omega_0t)}$). The other two terms determine the perturbation to this trajectory.

To demonstrate this perturbation, let us consider the following situation.

\subsection*{Perfect Resonance: $\omega_0=\Omega$}

Let us consider the case for $\omega_0=\Omega$. To see exactly what is happening here, it is best to go back to the ODE (\ref{eqn:theory_rw_ex_reson_odRr}):
\chg[rw_examples_resonant]{}
\begin{eqnarray*}
\odRt  &=& c_0e^{i(\Theta_0+\omega_0t)}\chg[]{-\gamma_-}e^{i(\chg[]{2\Omega t +\xi+\Theta_0})}\chg[]{+\gamma_+}e^{i(\xi-\Theta_0)}+O(\epsilon^2)
\end{eqnarray*}

We now see that the last term is actually just a complex constant, so integration gives us:

\chg[rw_examples_resonant]{
\begin{eqnarray}
R  &=& R_0-\frac{ic_0}{\Omega}e^{i(\Theta_0+\Omega t)}+\frac{i\gamma_-}{2\Omega}e^{i(2\Omega t+\xi+\Theta_0)}+\gamma_+te^{i(\xi-\Theta_0)}+O(\epsilon^2)\nonumber\\
\label{eqn:theory_rw_ex_perfectreson_R}
\Rightarrow R  &=& R_0-\frac{i|c_0|}{\Omega}e^{i(\Theta_0+\Omega t+{\rm arg}\{c_0\})}+\frac{i|\gamma_-|}{2\Omega}e^{i(2\Omega t+\xi+\Theta_0{\rm arg}\{\gamma_-\})}\nonumber\\
&& +\gamma_+te^{i(\xi-\Theta_0+{\rm arg}\{\gamma_+\})}+O(\epsilon^2)
\end{eqnarray}
}

So, what we now have is our original unperturbed trajectory, which is drifting in a straight line according to the drift velocity given by $\gamma_+e^{i(\xi-\Theta_0)}$. This straight line drift is in accordance with the original theory of drift given in \cite{bik95}.

\chg[rw_examples_resonant]{
\subsection*{Comparison of the Analytical Predictions to Simulations}

To show that the trajectory given by (\ref{eqn:theory_rw_ex_perfectreson_R}) is indeed correct, we fitted Eqns.(\ref{eqn:theory_rw_ex_reson_Th}) \& (\ref{eqn:theory_rw_ex_perfectreson_R}), to data obtained using an amended version of EZ-Spiral. The amendments to EZ-Spiral involved implementing the perturbation (\ref{eqn:hreson}) into the code. Once the data had been obtained, we used the fitting procedure in Gnuplot to fit the equations of motion to the data.

We used Barkley's model for the simulations with model parameters set at $a=0.52$, $b=0.05$ and $\varepsilon=0.02$. The physical and numerical parameters were chosen as $L=40$, $\Delta_x=0.25$, $\Delta_t=7.8125\times10^{-3}$ (the timestep corresponds to the parameters} \verb|ts| \chg[]{in EZ-Spiral being} \verb|ts|=0.5). \chg[]{We also note that we use the 5-point Laplacian.

\begin{figure}[btp]
\begin{center}
\begin{minipage}[htbp]{0.49\linewidth}
\centering
\includegraphics[width=0.7\textwidth, angle=-90]{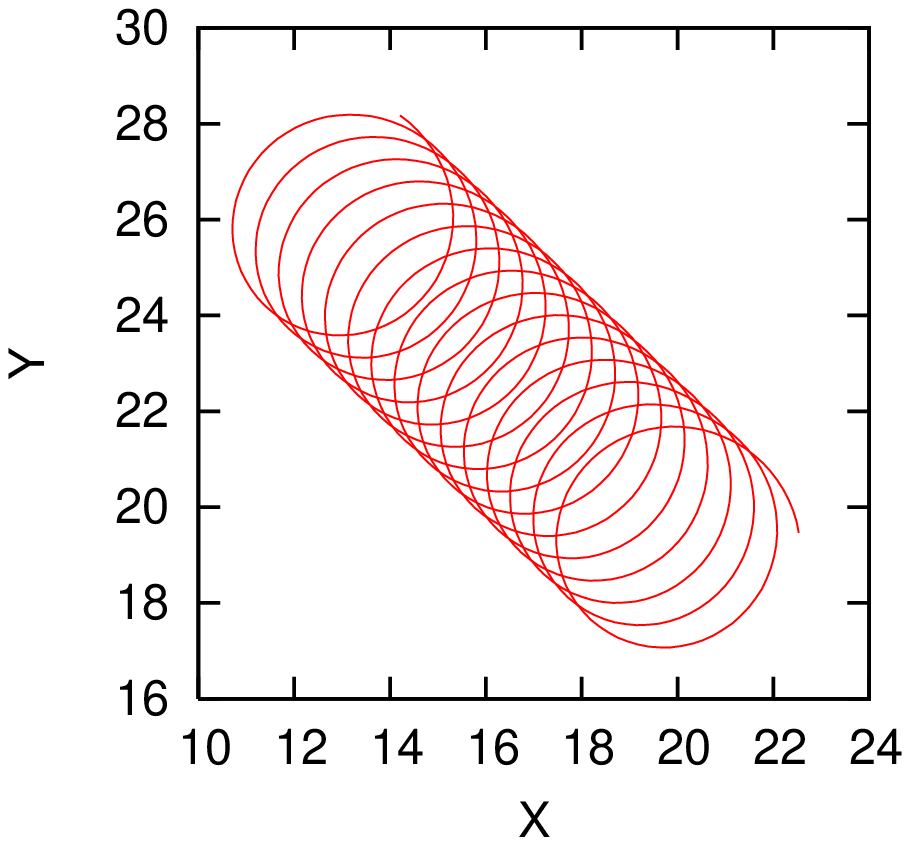}
\end{minipage}
\begin{minipage}[htbp]{0.49\linewidth}
\centering
\includegraphics[width=0.7\textwidth]{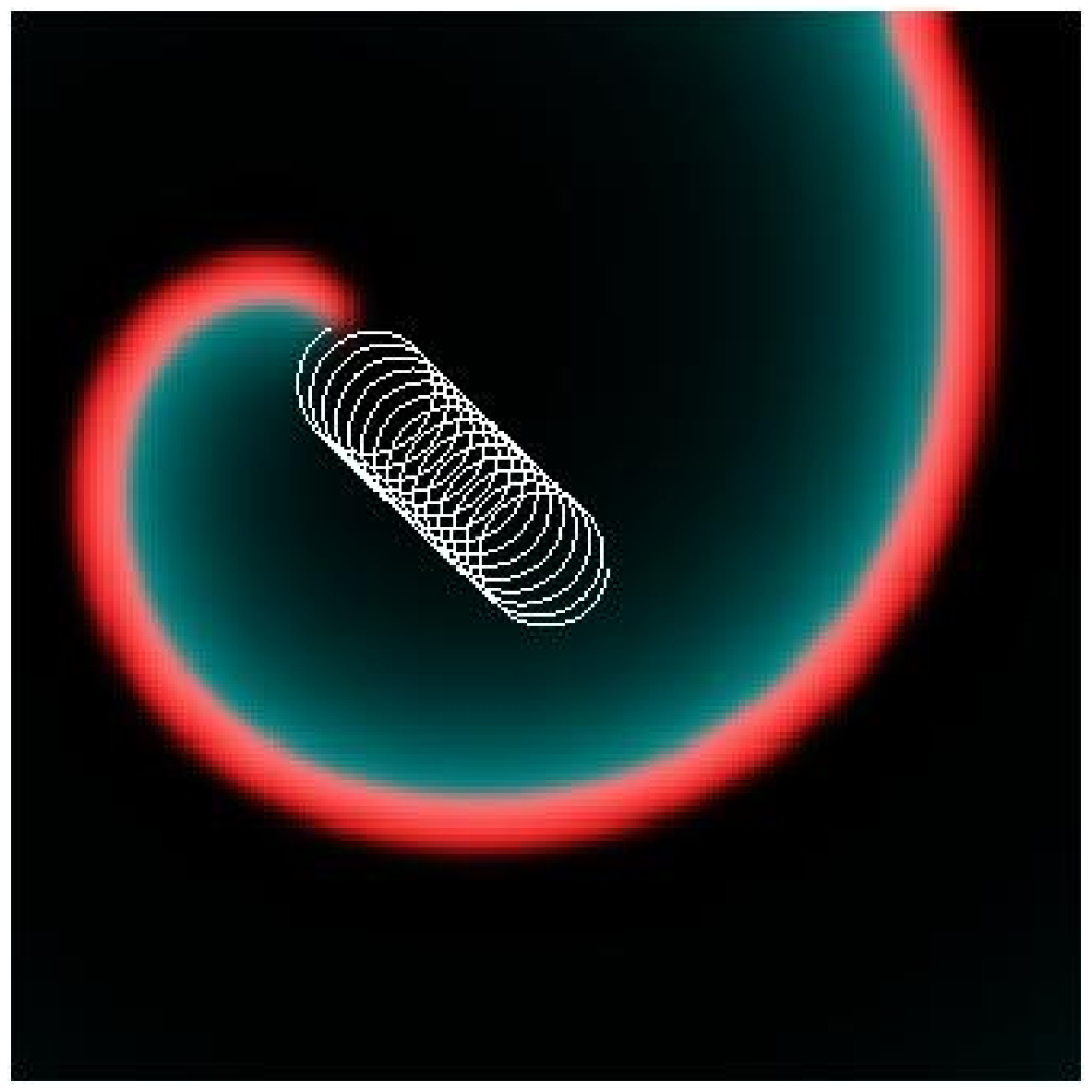}
\end{minipage}
\caption[Resonant drift: comparison of the analytical solution to numerical solution]{(Left) The trajectory of a drifting rigidly rotating spiral wave as determined by the asymptotic theory for resonant drift; (right) the spiral wave and its trajectory as generated using EZ-Spiral.}
\label{fig:theory_example_reson_a}
\end{center}
\end{figure}

Our results are shown in Fig.(\ref{fig:theory_example_reson_a}). We also note that values of the parameters found in Eqn.(\ref{eqn:theory_rw_ex_perfectreson_R}), were:

\begin{eqnarray*}
X_0 &=& 20.1112\\
Y_0 &=& 19.1824\\
\frac{|c_0|}{\Omega} &=& 2.4208\\
\Omega &=& 0.864343\\
\Theta_0+{\rm arg}\{c_0\} &=& -1.47035\\
\frac{|\gamma_-|}{2\Omega} &=& 0.0371196\\
\Theta_0+\xi+{\rm arg}\{\gamma_-\} &=& -0.307962\\
|\gamma_+|e^{i(\xi-\Theta_0+{\rm arg}\{\gamma_+\})} &=& -0.066517+i0.0639205
\end{eqnarray*}

This set of data can then be used to find the values of the scalar products $(\bpsi_j,\bht)$, which can then be used to check against the values derived using the program EVCOSPI (see Chap.(6)).
}


\subsection{Electrophoretic Induced Drift}

We now consider a perturbation which dependent on the gradient of $\bv$. We take the perturbation to be:

\begin{equation*}
\label{eqn:hreson1}
\bh = \bA\pdbvx,
\end{equation*}
\\
where $\bA$ is a $n\times n$ matrix:

\begin{equation*}
\bA = \left(
      \begin{array}{ccccc} A_1   & 0     & \cdot & \cdot & 0\\
                           0     & A_2   & 0     & \cdot & \cdot\\
                           \cdot & 0     & A_3   & 0     & \cdot\\
                           \cdot & \cdot & \cdot & \cdot & 0\\
                           0     & \cdot & \cdot & 0     & A_n \end{array} \right).   
\end{equation*}

We need to determine the transformed version of $\bh$. We know $\bht$ is defined to be:

\begin{equation*}
\bht = \dTi\textbf{h}(\dT\bv_0,r,t).
\end{equation*}

Firstly, we bear in mind how the transformation works. Given $g\in SE(2)$, where $g=(R,\Theta)$, then:

\begin{eqnarray}
\label{eqn:transg}
g: \left(\begin{array}{c} x\\y\end{array}\right) \mapsto   \left(\begin{array}{c} \til{x}\\ \til{y}\end{array}\right) &=& \left(\begin{array}{c} X\\Y\end{array}\right)+\left(\begin{array}{cc} \cos(\Theta) & -\sin(\Theta)\\ \sin(\Theta) & \cos(\Theta)\end{array}\right) \left(\begin{array}{c} x\\y\end{array}\right),\\
\label{eqn:transginv}
g^{-1}: \left(\begin{array}{c} x\\y\end{array}\right) \mapsto   \left(\begin{array}{c} \til{x}\\ \til{y}\end{array}\right) &=& \left(\begin{array}{cc} \cos(\Theta) & \sin(\Theta)\\ -\sin(\Theta) & \cos(\Theta)\end{array}\right) \left(\begin{array}{c} x-X\\y-Y\end{array}\right).
\end{eqnarray}

Next, we consider $\dT\bv_0$ (note that the transformations do not affect time dependency; therefore, the dependence on time is omitted from now on):

\begin{eqnarray*}
\dT\bv_0(r) &=& \bv_0(g^{-1}r)\\
            &=& \bv_0(\til{r})\\
            &=& \til{\bv}_0(r)
\end{eqnarray*}

Therefore, we have:

\begin{eqnarray*}
\epsilon h(T(g)\bv_0(r)) &=& \epsilon h(\til{\bv_0}(r))\\
                         &=& \bA\pdbtvox(r)\\
                         &=& \bA\pdbvox(\til{r})\\
                         &=& \bA\left(\pdbvotx(\til{r})\pdtxx+\pdbvoty(\til{r})\pdtyx\right)
\end{eqnarray*}

We now recall from Eqn.(\ref{eqn:transginv}) expressions for $\til{x}$ and $\til{y}$, hence:

\begin{eqnarray*}
\epsilon h(T(g)\bv_0(r)) &=& \bA\left(\pdbvotx(\til{r})\pdtxx+\pdbvoty(\til{r})\pdtyx\right)\\
                         &=& \bA\left(\cos(\Theta)\pdbvotx(\til{r})-\sin(\Theta)\pdbvoty(\til{r})\right)\\
                         &=& \hat{\bv}_0(\til{r})\\
                         &=& \hat{\bv}_0(\til{x},\til{y})
\end{eqnarray*}

We now consider the action of the element $g^{-1}$ on $\hat{\bv}_0(\til{x},\til{y})$:

\begin{eqnarray*}
\epsilon T(g^{-1})h(T(g)\bv_0(r)) &=& T(g^{-1})\hat{\bv}_0(\til{r})\\
                                  &=& \hat{\bv}_0(g(\til{r}))\\
                                  &=& \hat{\bv}_0(X+\til{x}\cos(\Theta)-\til{y}\sin(\Theta),\\
                                  & & \quad Y+\til{x}\sin(\Theta)+\til{y}\cos(\Theta))
\end{eqnarray*}

We now note that:

\begin{eqnarray*}
\til{x} &=& (x-X)\cos(\Theta)+(y-Y)\sin(\Theta)\\
\til{y} &=& -(x-X)\sin(\Theta)+(y-Y)\cos(\Theta)
\end{eqnarray*}

Therefore, we have:

\begin{eqnarray*}
X+\til{x}\cos(\Theta)-\til{y}\sin(\Theta) &=& X+((x-X)\cos(\Theta)+(y-Y)\sin(\Theta))\cos(\Theta)-\nonumber\\
                                          && (-(x-X)\sin(\Theta)+(y-Y)\cos(\Theta))\sin(\Theta)\\
                                          &=& X+(x-X)\cos^2(\Theta)+(y-Y)\sin(\Theta)\cos(\Theta)\nonumber\\
                                          && +(x-X)\sin^2(\Theta)-(y-Y)\sin(\Theta)\cos(\Theta)\\
                                          &=& X+(x-X)(\sin^2(\Theta)+\cos^2(\Theta))\\
                                          &=& x
\end{eqnarray*}

Similarly, it can be shown that:

\begin{equation*}
Y+\til{x}\sin(\Theta)+\til{y}\cos(\Theta) = y
\end{equation*}

Therefore, we have:

\begin{eqnarray*}
\epsilon\til{\bh} &=& \hat{\bv}_0(x,y)\\
                  &=& \hat{\bv}_0(r)\\
                  &=& \bA\left(\cos(\Theta)\pdbvox(r)-\sin(\Theta)\pdbvoy(r)\right)
\end{eqnarray*}

\chg[rw_examples_anis]{The} equations of motion are now:
\chg[rw_examples_anis]{}
\begin{eqnarray*}
\odRt  &=& c_0e^{i\Theta}-e^{i\Theta}\left[2\left(\bar{\bpsi}_1,\bA\left(\cos(\Theta)\pdbvox(r)-\sin(\Theta)\pdbvoy(r)\right)\right)\right.\nonumber\\
&& \left.+\frac{c_0}{\omega_0}\left(\bpsi_0,\bA\left(\cos(\Theta)\pdbvox(r)-\sin(\Theta)\pdbvoy(r)\right)\right)\right]\\
\odTht &=& \chg[]{\omega_0-}\left(\bpsi_0,\bA\left(\cos(\Theta)\pdbvox(r)-\sin(\Theta)\pdbvoy(r)\right)\right)
\end{eqnarray*}

We now consider the inner products $(\bpsi_i,\bA(\cos(\Theta)\pdbvox(r)-\sin(\Theta)\pdbvoy(r)))$:

\begin{eqnarray*}
(\bpsi_i,\bA(\cos(\Theta)\pdbvox(r)-\sin(\Theta)\pdbvoy(r))) &=& \cos(\Theta)(\bpsi_i,\bA\pdbvox(r))\nonumber\\
                                                             && -\sin(\Theta)(\bpsi_i,\bA\pdbvoy(r))\\
                                                             &=& (\bpsi_i,\epsilon\til{\bh})
\end{eqnarray*}

We know from previous analysis that $\partial_x\bv_0$ and $\partial_y\bv_0$ can be expressed as a linear combination of the eigenfunctions to the linear operator $L$. 

\begin{eqnarray*}
\partial_x\bv_0 &=& \frac{1}{2}(\bphi_1+\bar{\bphi}_1)\\
\partial_y\bv_0 &=& \frac{1}{2i}(\bphi_1-\bar{\bphi}_1)
\end{eqnarray*}

Therefore:
\chg[rw_examples_anis]{}
\begin{eqnarray*}
(\bpsi_i,\epsilon\til{\bh}) &=& \frac{\cos(\Theta)}{2}(\bpsi_i,\bA(\bphi_1+\bar{\bphi}_1))\chg[]{-\frac{\sin(\Theta)}{2i}}(\bpsi_i,\bA(\bphi_1-\bar{\bphi}_1))\\
&=& \frac{\cos(\Theta)}{2}(\bpsi_i,\bA\bphi_1)+\frac{\cos(\Theta)}{2}(\bpsi_i,\bA\bar{\bphi}_1)\nonumber\\
&& \chg[]{-\frac{\sin(\Theta)}{2i}(\bpsi_i,\bA\bphi_1)+\frac{\sin(\Theta)}{2i}(\bpsi_i,\bA\bar{\bphi}_1)}
\end{eqnarray*}

We note that the choice of $\bA$ is such that each element in the diagonal of $\bA$ could be non-zero and also they could all be different. The choice of these is arbitrary but such that they are \chg[rw_examples_anis]{``small".} This therefore means that $(\bpsi_i,A\bphi_j)\neq\delta_{ij}$. What we can do is denote each of these scalar products using:

\begin{equation*}
(\bpsi_i,A\bphi_j) = A_{i,j}
\end{equation*}

Therefore:

\begin{ajf}
\begin{eqnarray}
(\bpsi_i,\epsilon\til{\bh}) &=& \frac{\cos(\Theta)}{2}A_{i,1}+\frac{\cos(\Theta)}{2}A_{i,-1}+\frac{\sin(\Theta)}{2i}A_{i,1}-\frac{\sin(\Theta)}{2i}A_{i,-1}\\
&=& \frac{\cos(\Theta)}{2}A_{i,1}+\frac{\cos(\Theta)}{2}A_{i,-1}-\frac{i\sin(\Theta)}{2}A_{i,1}+\frac{i\sin(\Theta)}{2i}A_{i,-1}\\
&=& \frac{1}{2}A_{i,1}(\cos(\Theta)-i\sin(\Theta))+\frac{1}{2}A_{i,-1}(\cos(\Theta)+i\sin(\Theta))\\
&=& \frac{1}{2}A_{i,1}e^{-i\Theta}+\frac{1}{2}A_{i,-1}e^{i\Theta}
\end{eqnarray}
\end{ajf}
\chg[rw_examples_anis]{}
\begin{ajfthesis}
\begin{eqnarray*}
(\bpsi_i,\epsilon\til{\bh}) &=& \frac{1}{2}A_{i,1}\chg[]{e^{i\Theta}}+\frac{1}{2}A_{i,-1}\chg[]{e^{-i\Theta}}
\end{eqnarray*}
\end{ajfthesis}

\begin{ajf}
Our equations of motion are now:

\begin{eqnarray}
\odRt  &=& c_0e^{i\Theta}-e^{i\Theta}(A_{-1,1}e^{-i\Theta}+A_{-1,-1}e^{i\Theta}+\frac{c_0}{2\omega_0}(A_{0,1}e^{-i\Theta}+A_{0,-1}e^{i\Theta}))\\
\odTht &=& \omega_0-\frac{1}{2}(A_{0,1}e^{-i\Theta}+A_{0,-1}e^{i\Theta})
\end{eqnarray}
\end{ajf}

Therefore:
\chg[rw_examples_anis]{}
\begin{eqnarray*}
\label{eqn:theory_ex_1}
\odRt  &=& c_0e^{i\Theta}-\left(\frac{c_0}{2\omega_0}\chg[]{A_{0,1}}+\chg[]{A_{-1,1}}\right)e^{2i\Theta}-\left(\frac{c_0}{2\omega_0}\chg[]{A_{0,-1}}+\chg[]{A_{-1,-1}}\right)\\
\label{eqn:theory_ex_2}
\odTht &=& \omega_0-\frac{1}{2}(A_{0,1}\chg[]{e^{i\Theta}}+A_{0,-1}\chg[]{e^{-i\Theta})}
\end{eqnarray*}

We will now integrate Eqns.(\ref{eqn:theory_ex_1}) \& (\ref{eqn:theory_ex_2}) using asymptotic techniques, taking advantage of the fact that $A_{i,j}$ are small. 

Firstly, we consider Eqn.(\ref{eqn:theory_ex_2}) and note that $A_{0,1}$ and $A_{0,-1}$ are complex number, and in fact we note that $\bar{A}_{0,1}=A_{0,-1}$.

\begin{ajf}
\begin{eqnarray*}
A_{0,-1} &=& \int<\psi_0,A\bar\phi_1>\\
         &=& \int\sum_j\psi_{0j}(A\bar{\phi}_1)_j\dd{r}\\
         &=& \int\sum_j\psi_{0j}\bar{(A\phi)_1}_j\dd{r}\\
         &=& \int\sum_j\bar{\psi_{0j}(A\phi_1)_j}\dd{r}\\
         &=& \int\bar{<\psi_0,A\phi_1>}\\
         &=& \bar{A_{0,1}}
\end{eqnarray*}
\end{ajf}

We also assume that because $A_{0,1}$ and $A_{0,-1}$ are small, then:

\begin{eqnarray*}
A_{0,1}  &=& \epsilon a_{0,1}\\
A_{0,-1} &=& \epsilon a_{0,-1}\\
\Theta   &=& \Theta_0+\epsilon\Theta_1
\end{eqnarray*}

\begin{ajf}
Our problem now becomes:

\begin{equation}
\odThot+\epsilon\odThonet = \omega_0-\frac{\epsilon}{2}(a_{0,1}e^{-i(\Theta_0+\epsilon\Theta_1)}+a_{0,-1}e^{i(\Theta_0+\epsilon\Theta_1)})+O(\epsilon^2)
\end{equation}

Now, for a general function $f(\Theta_0+\epsilon\Theta_1)$, we have that:

\begin{equation}
f(\Theta) = f(\Theta_0+\epsilon\Theta_1) = f(\Theta_0)+\left.\epsilon\Theta_1\frac{\partial{f}}{\partial{\Theta}}\right|_{\Theta_0}+O(\epsilon^2)
\end{equation}

This therefore gives us:

\begin{equation}
e^{i(\Theta_0+\epsilon\Theta_1)} = e^{i\Theta_0}-i\epsilon\Theta_1e^{i\Theta_0}+O(\epsilon^2)
\end{equation}

Hence, we now have:

\begin{equation}
\odThot+\epsilon\odThonet = \omega_0-\frac{\epsilon}{2}(a_{0,1}e^{-i\Theta_0}+a_{0,-1}e^{i\Theta_0})+O(\epsilon^2)
\end{equation}
\end{ajf}
\begin{ajfthesis}
Our problem now becomes:
\chg[rw_examples_anis]{}
\begin{eqnarray*}
\odThot+\epsilon\odThonet &=& \omega_0-\frac{\epsilon}{2}(a_{0,1}\chg[]{e^{i(\Theta_0+\epsilon\Theta_1)}}+a_{0,-1}\chg[]{e^{-i(\Theta_0+\epsilon\Theta_1)}})+O(\epsilon^2)\nonumber\\
\Rightarrow\odThot+\epsilon\odThonet &=& \omega_0-\frac{\epsilon}{2}(a_{0,1}\chg[]{e^{i\Theta_0}}+a_{0,-1}\chg[]{e^{-i\Theta_0}})+O(\epsilon^2)
\end{eqnarray*}
\end{ajfthesis}

Splitting out the different orders in $\epsilon$, we have:
\chg[rw_examples_anis]{}
\begin{eqnarray*}
\label{eqn:Thun}
\epsilon^0:\quad\odThot &=& \omega_0\\
\label{eqn:Thper}
\epsilon^1:\quad\odThonet &=& -\frac{1}{2}(a_{0,1}\chg[]{e^{i\Theta_0}}+a_{0,-1}\chg[]{e^{-i\Theta_0}})
\end{eqnarray*}

We can see that (\ref{eqn:Thun}) gives us:

\begin{equation*}
\label{eqn:Th0}
\Theta_0 = \omega_0t+\Theta_0(0)
\end{equation*}

Substituting (\ref{eqn:Th0}) into (\ref{eqn:Thper}), we get:
\chg[rw_examples_anis]{}
\begin{eqnarray*}
\odThonet &=& -\frac{1}{2}(a_{0,1}\chg[]{e^{i(\omega_0t+\Theta_0(0))}}+a_{0,-1}\chg[]{e^{-i(\omega_0t+\Theta_0(0))}})\\
\Theta_1  &=& \chg[]{\Theta_1(0)+}\frac{ia_{0,1}}{2\omega_0}\chg[]{e^{i(\omega_0t+\Theta_0(0))}-}\frac{ia_{0,-1}}{2\omega_0}\chg[]{e^{-i(\omega_0t+\Theta_0(0))}}
\end{eqnarray*}

Hence, our final expression for $\Theta$ is:

\chg[rw_examples_anis]{
\begin{equation*}
\Theta = \Theta(0)+\omega_0t+\epsilon\left(\frac{ia_{0,1}}{2\omega_0}e^{i(\omega_0t+\Theta_0(0))}-\frac{-ia_{0,-1}}{2\omega_0}e^{i(\omega_0t+\Theta_0(0))}\right)
\end{equation*}
}

\begin{ajf}
We now consider Eqn.(\ref{eqn:theory_ex_1}):

\begin{eqnarray}
\odRt  &=& c_0e^{i\Theta}-\left(\frac{c_0}{2\omega_0}A_{0,-1}+A_{-1,-1}\right)e^{2i\Theta}-\left(\frac{c_0}{2\omega_0}A_{0,1}+A_{-1,1}\right)
\end{eqnarray}

First thing to note is that the coefficient to the double angle exponential and also the constant term are just complex number, and in fact these complex number have a small magnitude. Therefore, define the following complex numbers:
\end{ajf}
\begin{ajfthesis}
We now consider Eqn.(\ref{eqn:theory_ex_1}). The first thing to note is that the coefficient to the double angle exponential and also the constant term are just complex number, and in fact these complex number have a small magnitude. Therefore, define the following complex numbers:
\end{ajfthesis}

\begin{eqnarray*}
\epsilon\xi   &=& \frac{c_0}{2\omega_0}A_{0,-1}+A_{-1,-1}\\
\epsilon\zeta &=& \frac{c_0}{2\omega_0}A_{0,1}+A_{-1,1}
\end{eqnarray*}

Therefore, we get:
\chg[rw_examples_anis]{}
\begin{eqnarray}
\label{eqn:3}
\odRt &=& c_0e^{i\Theta}-\epsilon\chg[]{\zeta} e^{2i\Theta}-\epsilon\chg[]{\xi}+O(\epsilon^2)
\end{eqnarray}

We also note that the initial conditions for $\Theta$, say $\Theta(0)=\Theta_*$ give:

\chg[rw_examples_anis]{
\begin{equation*}
\Theta = (\omega_0t+\Theta_*)+\frac{i\epsilon}{2\omega_0}\left(a_{0,1}e^{i(\omega_0t+\Theta_*)}-a_{0,-1}e^{-i(\omega_0t+\Theta_*)}\right)
\end{equation*}
}
\begin{ajf}
Now, consider $e^{i\Theta}$:

\begin{eqnarray*}
e^{i\Theta} &=& e^{i\left((\omega_0t+\Theta_*)-\frac{i\epsilon}{2\omega_0}\left(a_{0,1}e^{-i(\omega_0t+\Theta_*)}+a_{0,-1}e^{i(\omega_0t+\Theta_*)}\right)\right)}\\
&=& e^{i(\omega_0t+\Theta_*)}e^{\frac{\epsilon}{2\omega_0}\left(a_{0,1}e^{-i(\omega_0t+\Theta_*)}+a_{0,-1}e^{i(\omega_0t+\Theta_*)}\right)}\\
&=& e^{i(\omega_0t+\Theta_*)}\left(1+\frac{\epsilon}{2\omega_0}\left(a_{0,1}e^{-i(\omega_0t+\Theta_*)}+a_{0,-1}e^{i(\omega_0t+\Theta_*)}\right)+O(\epsilon^2)\right)\\
&=& e^{i(\omega_0t+\Theta_*)}+\frac{\epsilon a_{0,1}}{2\omega_0}+\frac{\epsilon a_{0,-1}}{2\omega_0}e^{2i(\omega_0t+\Theta_*)}+O(\epsilon^2)
\end{eqnarray*}

Similarly, we can obtain an expression for $e^{2i\Theta}$:

\begin{eqnarray*}
e^{2i\Theta} &=& e^{2i\left((\omega_0t+\Theta_*)-\frac{i\epsilon}{2\omega_0}\left(a_{0,1}e^{-i(\omega_0t+\Theta_*)}+a_{0,-1}e^{i(\omega_0t+\Theta_*)}\right)\right)}\\
&=& e^{2i(\omega_0t+\Theta_*)}e^{\frac{\epsilon}{\omega_0}\left(a_{0,1}e^{-i(\omega_0t+\Theta_*)}+a_{0,-1}e^{i(\omega_0t+\Theta_*)}\right)}\\
&=& e^{2i(\omega_0t+\Theta_*)}+\frac{\epsilon a_{0,1}}{\omega_0}e^{i(\omega_0t+\Theta_*)}+\frac{\epsilon a_{0,-1}}{\omega_0}e^{3i(\omega_0t+\Theta_*)}+O(\epsilon^2)
\end{eqnarray*}
\end{ajf}
\begin{ajfthesis}
Now, consider $e^{i\Theta}$:
\chg[rw_examples_anis]{}
\begin{eqnarray*}
e^{i\Theta} &=& e^{i(\omega_0t+\Theta_*)}\chg[]{-\frac{\epsilon a_{0,-1}}{2\omega_0}-\frac{\epsilon a_{0,1}}{2\omega_0}}e^{2i(\omega_0t+\Theta_*)}+O(\epsilon^2)
\end{eqnarray*}

Similarly:

\chg[rw_examples_anis]{
\begin{eqnarray*}
e^{2i\Theta} &=& e^{2i(\omega_0t+\Theta_*)}+O(\epsilon)
\end{eqnarray*}}
\end{ajfthesis}

Using these expression for the exponentials, Eqn.(\ref{eqn:3}) now becomes:

\begin{ajf}
\begin{eqnarray}
\odRt &=& c_0\left(e^{i(\omega_0t+\Theta_*)}+\frac{\epsilon a_{0,1}}{2\omega_0}+\frac{\epsilon a_{0,-1}}{2\omega_0}e^{2i(\omega_0t+\Theta_*)}\right)\nonumber\\
&& -\epsilon\xi\left(e^{2i(\omega_0t+\Theta_*)}+\frac{\epsilon a_{0,1}}{\omega_0}e^{i(\omega_0t+\Theta_*)}+\frac{\epsilon a_{0,-1}}{\omega_0}e^{3i(\omega_0t+\Theta_*)}\right)\nonumber\\
&& -\epsilon\zeta+O(\epsilon^2)\\
&=& c_0e^{i(\omega_0t+\Theta_*)}+\epsilon\left(\frac{c_0a_{0,-1}}{2\omega_0}-\xi\right)e^{2i(\omega_0t+\Theta_*)}\\
&& +\epsilon\left(\frac{c_0a_{0,1}}{2\omega_0}-\zeta\right)\\
&=& c_0e^{i(\omega_0t+\Theta_*)}+\epsilon a_{-1,-1}e^{2i(\omega_0t+\Theta_*)}+\epsilon a_{-1,1}\\
\Rightarrow \odbRt &=& c_0e^{i(\omega_0t+\Theta_*)}-A_{-1,-1}e^{2i(\omega_0t+\Theta_*)}-A_{-1,1}
\end{eqnarray}
\end{ajf}
\begin{ajfthesis}
\chg[rw_examples_anis]{}
\begin{eqnarray*}
\odRt &=& c_0e^{i(\omega_0t+\Theta_*)}\chg[]{-\epsilon}\left(\frac{c_0\chg[]{a_{0,1}}}{2\omega_0}\chg[]{+\zeta}\right)e^{2i(\omega_0t+\Theta_*)}\nonumber\\
&& +\epsilon\left(\frac{c_0\chg[]{a_{0,-1}}}{2\omega_0}-\chg[]{\xi}\right)+O(\epsilon^2)\\
\Rightarrow \odRt &=& c_0e^{i(\omega_0t+\Theta_*)}\chg[]{-\left(\frac{c_0}{\omega_0}A_{0,1}+A_{-1,-1}\right)}e^{2i(\omega_0t+\Theta_*)}+\chg[]{A_{-1,1}}+O(\epsilon^2)
\end{eqnarray*}
\end{ajfthesis}

Hence, we can integrate to get:
\chg[rw_examples_anis]{
\begin{eqnarray*}
R &=& R(0)-\frac{ic_0}{\omega_0}e^{i(\omega_0t+\Theta_*)}+\frac{i}{2\omega_0}\left(\frac{c_0}{\omega_0}A_{0,1}+A_{-1,-1}\right)e^{2i(\omega_0t+\Theta_*)}\\
&& -\chg[]{A_{-1,-1}}t+O(\epsilon^2)\nonumber\\
\label{eqn:theory_rw_ex_anis}
\Rightarrow R &=& R(0)-\frac{i|c_0|}{\omega_0}e^{i(\omega_0t+\Theta_*+{\rm arg}\{c_0\})}+\frac{i}{2\omega_0}|\til{A}|e^{2i(\omega_0t+\Theta_*+{\rm arg}\{\til{A}\})}\\
&&-|A_{-1,-1}|e^{i{\rm arg}\{A_{-1,-1}\}}t+O(\epsilon^2)
\end{eqnarray*}}
\\
where $\til{A}=\left(\frac{c_0}{\omega_0}A_{0,1}+A_{-1,-1}\right)$
\chg[rw_examples_anis]{
\subsection*{Comparison with numerical simulations}

As we did for resonant drift, we shall show an example of just how accurate our theory is, by comparing the data from the numerical simulation (again using an amended version of EZ-Spiral) with the analytical predictions.

As with the resonant drift example, we used Barkley's model for the simulations with model parameters set at $a=0.52$, $b=0.05$ and $\varepsilon=0.02$. The physical and numerical parameters were chosen as $L=40$, $\Delta_x=0.25$, $\Delta_t=7.8125\times10^{-3}$ (the timestep corresponds to the parameters} \verb|ts| \chg[]{in EZ-Spiral being} \verb|ts|=0.5). \chg[]{We also note that we use the 5-point Laplacian.

We show our results in Fig.(\ref{fig:theory_example_anis})

\begin{figure}[btp]
\begin{center}
\begin{minipage}[htbp]{0.49\linewidth}
\centering
\includegraphics[width=0.7\textwidth, angle=-90]{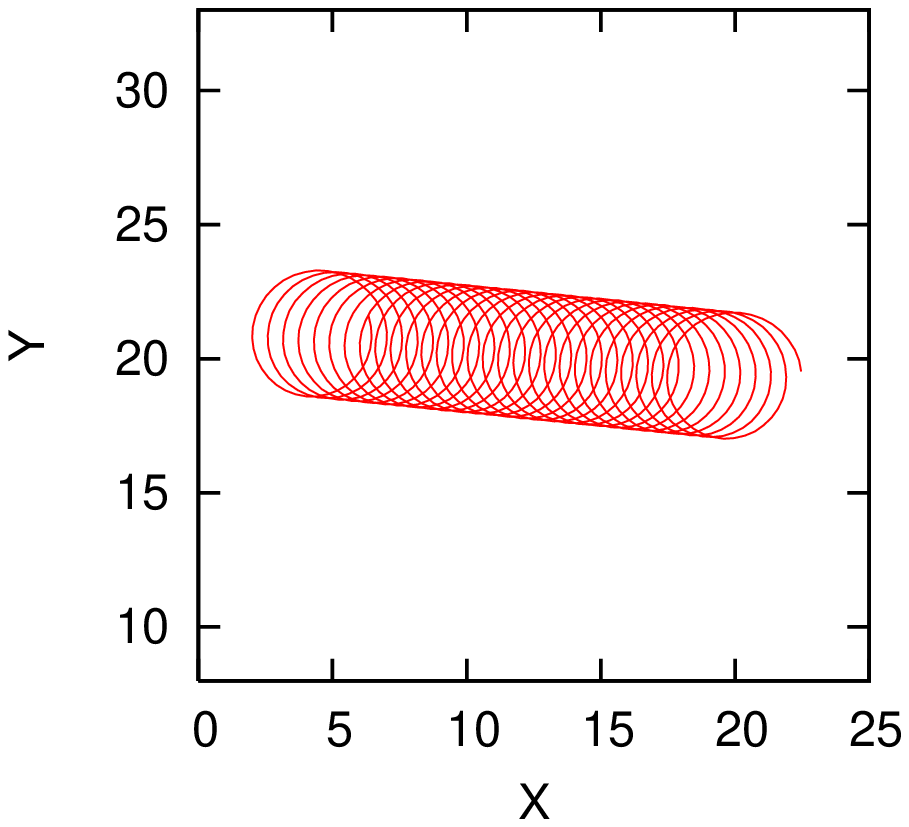}
\end{minipage}
\begin{minipage}[htbp]{0.49\linewidth}
\centering
\includegraphics[width=0.7\textwidth]{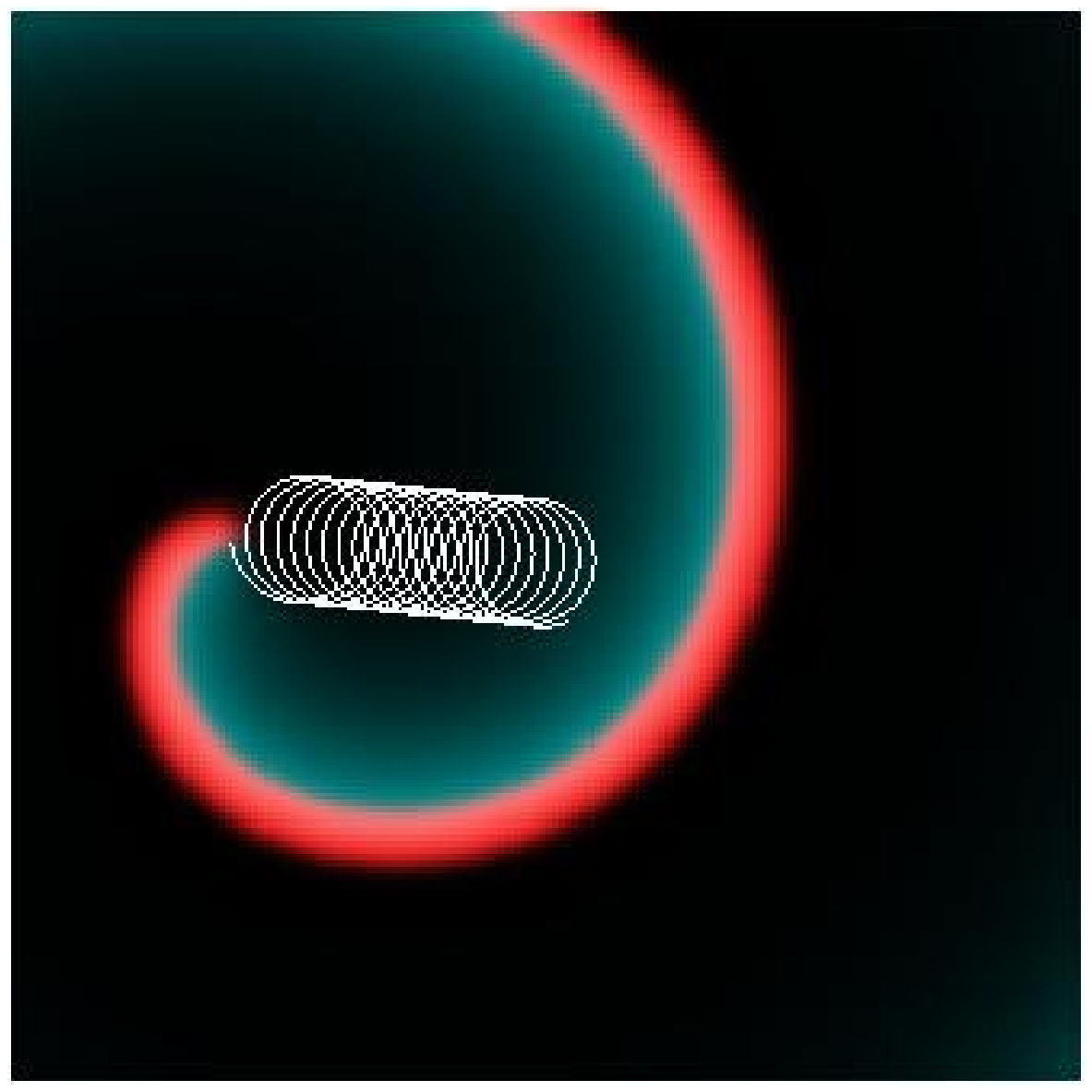}
\end{minipage}
\caption[Electrophoretic drift: comparison of the analytical and numerical solutions]{(Left) The trajectory of a drifting rigidly rotating spiral wave as determined by the asymptotic theory for electrophoretic induced drift; (right) the spiral wave and its trajectory as generated using EZ-Spiral.}
\label{fig:theory_example_anis}
\end{center}
\end{figure}

We also note that values of the parameters found in Eqn.(\ref{eqn:theory_rw_ex_anis}), were:

\begin{eqnarray*}
X_0 &=& 20.0081\\
Y_0 &=& 19.3437\\
\frac{|c_0|}{\omega_0} &=& 2.36589\\
\omega_0 &=& 0.880615\\
\Theta_*+{\rm arg}\{c_0\} &=& -1.49119\\
\frac{|\til{A}|}{2\omega_0} &=& 0.00136484\\
\Theta_*+{\rm arg}\{\til{A}\} &=& =-1.25518\\
A_{-1,-1} &=& -0.0802271+i0.008143       
\end{eqnarray*}}


\subsection{Inhomogeneity Induced Drift}

A spiral wave drifts due to inhomogeneities when the parameter(s) in the system depend on the spatial coordinates. We consider the case where one parameter depends on just one of the spatial coordinates and the relationship between them is linear:

\begin{equation*}
\alpha = \alpha_0+\alpha_1x
\end{equation*}
\\
where, for $x=0$ we have $\alpha(0)=\alpha_0$.

This leads to a symmetry breaking perturbation causing the wave to drift. We have the following RDS:

\begin{equation*}
\pdbut = \bDD\nabla^2\bu+\bof(\bu,\alpha(x))
\end{equation*} 

Expanding this using Taylor Series we get:

\begin{eqnarray*}
\pdbut &=& D\nabla^2\bu+\bof(\bu,\alpha_0)+\alpha_1x\pdfa(\bu,\alpha_0)+O(\alpha_1^2)
\end{eqnarray*} 

We see that the perturbation is now:
\chg[rw_examples_inhom]{}
\begin{equation*}
\epsilon\bh = \alpha_1x\pdfa(\bu,\alpha_0)\chg[]{+O(\epsilon^2)}
\end{equation*} 
\\
where $\epsilon=\alpha_1$ is gradient of the inhomogeneity and:

\begin{equation*}
\bh = x\pdfa(\bu,\alpha_0)
\end{equation*} 

We now have that $\bh = \bh(\bv_0,\br)$ and we next need to calculate $\bht$. Firstly, consider:

\begin{eqnarray*}
\bh(T(g)\bv_0,\br) &=& \bh(T(g)\bv_0(\br),\br)\nonumber\\
                   &=& \bh(\bv_0(g^{-1}\br),\br)\nonumber\\
                   &=& \bh(\bv_0(\btr),\br)
\end{eqnarray*}
\\
where $\btr$ is given by:

\begin{eqnarray*}
\left(\begin{array}{c} \til{x}\\ \til{y}\end{array}\right) &=& \left(\begin{array}{cc} \cos(\Theta) & \sin(\Theta)\\ -\sin(\Theta) & \cos(\Theta)\end{array}\right) \left(\begin{array}{c} x-X\\y-Y\end{array}\right)
\end{eqnarray*}

Therefore:

\begin{eqnarray*}
\bh(T(g)\bv_0,\br) &=& \bh(\bv_0(\btr),\br)\nonumber\\
                   &=& x\pdfa(\bv_0(\btr),\alpha_0\nonumber)\\
                   &=& \hat{\bv}(\br)
\end{eqnarray*}

Now, we apply the action of the inverse group element $g^{-1}$ on $\bh(T(g)\bv_0,\br)$:

\begin{eqnarray*}
\bht &=& T(g^{-1})\bh(T(g)\bv_0,\br)\\
     &=& T(g^{-1})\hat{\bv}(\br)\\
     &=& \hat{\bv}(g\br)\\
     &=& \hat{\bv}(X+x\cos(\Theta)-y\sin(\Theta),Y+x\sin(\Theta)+y\cos(\Theta))\\
     &=& (X+x\cos(\Theta)-y\sin(\Theta))\pdfa(\bv_0(\br),\alpha_0)
\end{eqnarray*}

\chg[rw_examples_inhom]{The} equations of motion are now:

\chg[rw_examples_inhom]{
\begin{eqnarray*}
\odRt  &=& c_0e^{i\Theta}-e^{i\Theta}\left[2\left(\bar{\bpsi}_1,(X+x\cos(\Theta)-y\sin(\Theta))\pdfa\right)\right.\nonumber\\
&& \left.-\frac{c_0}{\omega_0}\left(\bpsi_0,(X+x\cos(\Theta)-y\sin(\Theta))\pdfa\right)\right],\\
\odTht &=& \omega_0-\left(\bpsi_0,(X+x\cos(\Theta)-y\sin(\Theta))\pdfa\right).
\end{eqnarray*}
}
We now consider the inner products $(\bpsi_i,(X+x\cos(\Theta)-y\sin(\Theta))\pdfa)$:
\chg[rw_examples_inhom]{
\begin{eqnarray*}
\left(\bpsi_i,(X+x\cos(\Theta)-y\sin(\Theta))\pdfa\right) &=& X\left(\bpsi_i,\pdfa\right)\nonumber\\
                                                             && +\cos\left(\Theta)(\bpsi_i,x\pdfa\right)\nonumber\\
                                                             && -\sin\left(\Theta)(\bpsi_i,y\pdfa\right)
\end{eqnarray*}

Let us now define the following:

\begin{eqnarray*}
C_{i,0}        &=& \left(\bpsi_i,\pdfa\right),\\
C_{i,x}        &=& \left(\bpsi_i,x\pdfa\right),\\
C_{i,y}  			 &=& \left(\bpsi_i,y\pdfa\right)
\end{eqnarray*}

Hence, we have:

\begin{eqnarray*}
\left(\bpsi_i,\bht\right) &=& XC_{i,0}+\cos C_{i,x}-\sin C_{i,y}
\end{eqnarray*}
}
Therefore, we can write our equations of motion as follows. Firstly, $\odRt$:
\begin{ajf}
\begin{eqnarray}
\odRt &=& c_0e^{i\Theta}-\alpha_1e^{i\Theta}(2(X(\bar{\bpsi}_1,\pdfa)+\cos(\Theta)(\bar{\bpsi}_1,x\pdfa)-\sin(\Theta)(\bar{\bpsi}_1,y\pdfa))\nonumber\\
&& +\frac{c_0}{\omega_0}(X(\bpsi_0,\pdfa)+\cos(\theta)(\bpsi_0,x\pdfa)-\sin(\Theta)(\bpsi_0,y\pdfa)))\\
&=& c_0e^{i\Theta}-\alpha_1e^{i\Theta}(2(XC_{-1,0}+\cos(\Theta)C_{-1,x}-\sin(\Theta)C_{-1,y})\nonumber\\
&& +\frac{c_0}{\omega_0}(XC_{0,0}+\cos(\theta)C_{0,x}-\sin(\Theta)C_{0,y}))\\
&=& c_0e^{i\Theta}-\alpha_1e^{i\Theta}(X(2C_{-1,0}+\frac{c_0}{\omega_0}C_{0,0})+\cos(\Theta)(2C_{-1,x}+\frac{c_0}{\omega_0}C_{0,x})\nonumber\\
&& -\sin(\Theta)(2C_{-1,y}+\frac{c_0}{\omega_0}C_{0,y}))
\end{eqnarray}
\end{ajf}
\begin{ajfthesis}
\begin{eqnarray*}
\odRt &=& c_0e^{i\Theta}-\alpha_1e^{i\Theta}\left(X(2C_{-1,0}+\frac{c_0}{\omega_0}C_{0,0})+\cos(\Theta)(2C_{-1,x}+\frac{c_0}{\omega_0}C_{0,x})\right.\nonumber\\
&& \left.-\sin(\Theta)(2C_{-1,y}+\frac{c_0}{\omega_0}C_{0,y})\right)
\end{eqnarray*}
\end{ajfthesis}

Furthermore, if we define:

\begin{eqnarray*}
B^0   &=& 2C_{-1,0}+\frac{c_0}{\omega_0}C_{0,0},\\
2B^x  &=& 2C_{-1,x}+\frac{c_0}{\omega_0}C_{0,x},\\
2iB^y &=& 2C_{-1,y}+\frac{c_0}{\omega_0}C_{0,y},
\end{eqnarray*}
\\
then:
\begin{ajf}
\begin{eqnarray*}
\odRt &=& c_0e^{i\Theta}-\alpha_1e^{i\Theta}(XB^0+2\cos(\Theta)B^x-2i\sin(\Theta)B^y)\\
      &=& c_0e^{i\Theta}-\alpha_1e^{i\Theta}(XB^0+(e^{i\Theta}+e^{-i\Theta})B^x-(e^{i\Theta}-e^{-i\Theta})B^y)\\
      &=& (c_0-\alpha_1XB^0)e^{i\Theta}-\alpha_1(B^x-B^y)e^{2i\Theta}+\alpha_1(B^x+B^y)
\end{eqnarray*}
\end{ajf}
\begin{ajfthesis}
\chg[rw_examples_inhom]{}
\begin{eqnarray*}
\odRt &=& (c_0-\alpha_1XB^0)e^{i\Theta}-\alpha_1(B^x-B^y)e^{2i\Theta}\chg[]{-\alpha_1}(B^x+B^y).
\end{eqnarray*}
\end{ajfthesis}

Furthermore, if:

\begin{eqnarray*}
B^+ &=& B^x+B^y,\\
B^- &=& B^x-B^y.
\end{eqnarray*}
\\
then:
\chg[rw_examples_inhom]{}
\begin{eqnarray*}
\label{eqn:dRdtinhom}
\odRt &=& (c_0-\alpha_1XB^0)e^{i\Theta}-\alpha_1B^-e^{2i\Theta}\chg[]{-\alpha_1}B^+.
\end{eqnarray*}

Similarly we have that:
\chg[rw_examples_inhom]{}
\begin{eqnarray*}
\odTht &=& \omega_0\chg[]{-\alpha_1}(\bpsi_0,(X+x\cos(\Theta)-y\sin(\Theta))\pdfa)),\nonumber\\
\label{eqn:dThdtinhom}
       &=& \omega_0\chg[]{-\alpha_1}XC_{0,0}\chg[]{-\alpha_1}\cos(\Theta)C_{0,x}\chg[]{+\alpha_1}\sin(\Theta)C_{0,y}.
\end{eqnarray*} 

Finally, let us introduce further notation:

\begin{eqnarray*}
C_{i,1}   &=& (\bpsi_i,r\pdfa),\\
C_{i,-1}  &=& (\bpsi_i,\bar{r}\pdfa),
\end{eqnarray*}
\\
where $r=x+iy$. It can be seen that:

\begin{eqnarray*}
C_{i,x} &=& \frac{1}{2}(C_{i,1}+C_{i,-1})\\
C_{i,y} &=& \frac{1}{2i}(C_{i,1}-C_{i,-1})
\end{eqnarray*}

Therefore:
\begin{ajf}
\begin{eqnarray*}
\cos(\Theta)C_{0,x}-\sin(\Theta)C_{0,y} &=& \frac{1}{4}\left((C_{0,1}+C_{0,-1})(e^{i\Theta}+e^{-i\Theta})\right.\\
&& \left.+(C_{0,1}-C_{0,-1})(e^{i\Theta}-e^{-i\Theta})\right)\\
&=& \frac{1}{4}\left((C_{0,1}e^{i\Theta}+C_{0,-1}e^{i\Theta}+C_{0,1}e^{-i\Theta}+C_{0,-1}e^{-i\Theta})\right.\\
&& +\left.(C_{0,1}e^{i\Theta}-C_{0,-1}e^{i\Theta}-C_{0,1}e^{-i\Theta}+C_{0,-1}e^{-i\Theta})\right)\\
&=& \frac{1}{2}(C_{0,1}e^{i\Theta}+C_{0,-1}e^{-i\Theta})
\end{eqnarray*}
\end{ajf}
\begin{ajfthesis}
\begin{eqnarray*}
\cos(\Theta)C_{0,x}-\sin(\Theta)C_{0,y} &=& \frac{1}{2}(C_{0,1}e^{i\Theta}+C_{0,-1}e^{-i\Theta})
\end{eqnarray*}
\end{ajfthesis}

Hence:
\chg[rw_examples_inhom]{}
\begin{equation*}
\label{eqn:dThdtinhom1}
\odTht = \omega_0\chg[]{-\alpha_1}XC_{0,0}\chg[]{-\frac{\alpha_1C_{0,1}}{2}}e^{i\Theta}\chg[]{-\frac{\alpha_1C_{0,-1}}{2}}e^{-i\Theta}
\end{equation*}

Before we integrate the equations, we will firstly expand the solutions $R$ and $\Theta$ in orders of $\alpha_1$:

\begin{eqnarray*}
R &=& R_0+\alpha_1R_1+O(\alpha_1^2)\\
\Theta &=& \Theta_0+\alpha_1\Theta_1+O(\alpha_1^2)
\end{eqnarray*}

This leads to the equations:
\chg[rw_examples_inhom]{}
\begin{eqnarray}
\label{eqn:theory_ex_a}
\deriv{R_0}{t} &=& c_0e^{i\Theta_0}\\
\label{eqn:theory_ex_b}
\deriv{R_1}{t} &=& ic_0\Theta_1e^{i\Theta_0}-X_0B^0e^{i\Theta_0}-B^-e^{2i\Theta_0}\chg[]{-B^+}\\
\label{eqn:theory_ex_c}
\deriv{\Theta_0}{t} &=& \omega_0\\
\label{eqn:theory_ex_d}
\deriv{\Theta_1}{t} &=& \chg[]{-X_0}C_{0,0}\chg[]{-\frac{C_{0,1}}{2}}e^{i\Theta_0}\chg[]{-\frac{C_{0,-1}}{2}}e^{-i\Theta_0}
\end{eqnarray}

We will now integrate Eqns.(\ref{eqn:theory_ex_c}), (\ref{eqn:theory_ex_a}), (\ref{eqn:theory_ex_d}) and (\ref{eqn:theory_ex_b}), in that order. 

Firstly, Eqn.(\ref{eqn:theory_ex_c}) gives us, when integrated:

\begin{eqnarray*}
\Theta_0 &=& \Theta_*+\omega_0t
\end{eqnarray*}

Next, we integrate Eqn.(\ref{eqn:theory_ex_a}) to get:

\begin{eqnarray*}
R_0 &=& R_*-\frac{ic_0}{\omega_0}e^{i(\Theta_*+\omega_0t)}
\end{eqnarray*}
\\
which gives us:

\begin{eqnarray*}
X_0 &=& X_*-\frac{i}{2\omega_0}\left(c_0e^{i\Theta_0}-\bar{c}_0e^{-i\Theta_0}\right)
\end{eqnarray*}

Next, we consider (\ref{eqn:theory_ex_d}):
\chg[rw_examples_inhom]{
\begin{eqnarray*}
\deriv{\Theta_1}{t} &=& -\left(X_*-\frac{i}{2\omega_0}\left(c_0e^{i\Theta_0}-\bar{c}_0e^{-i\Theta_0}\right)\right)C_{0,0}-\frac{C_{0,1}}{2}e^{i\Theta_0}-\frac{C_{0,-1}{2}}e^{-i\Theta_0}\nonumber\\
\deriv{\Theta_1}{t} &=& -X_*C_{0,0}-\left(\frac{C_{0,1}}{2}-\frac{ic_0C_{0,0}}{2\omega_0}\right)e^{i\Theta_0}-\left(\frac{C_{0,-1}}{2}+\frac{i\bar{c}_0C_{0,0}}{2\omega_0}\right)e^{-i\Theta_0}
\end{eqnarray*}}

Integration gives:
\chg[rw_examples_inhom]{}
\begin{eqnarray*}
\Theta_1 &=& \Theta_+\chg[]{-X_*}C_{0,0}t+Pe^{i\Theta_0}+\bar{P}e^{-i\Theta_0}\
\end{eqnarray*}
\\
where:
\chg[rw_examples_inhom]{}
\begin{eqnarray*}
P &=& \chg[]{\frac{i}{2\omega_0}}\left(C_{0,1}-\frac{ic_0C_{0,0}}{\omega_0}\right)
\end{eqnarray*}

Before we move on to the final equation, we note that the perturbed part of $\Theta$, i.e. $\Theta_1$ must be bounded for our theory to work. This must mean that $C_{0,0}=0$, otherwise $\Theta_1$ is not bounded:

\begin{eqnarray*}
\Theta_1 &=& \Theta_+Pe^{i\Theta_0}+\bar{P}e^{-i\Theta_0}
\end{eqnarray*}
\\
where $P$ is no now given by:

\begin{eqnarray*}
P &=& -\frac{iC_{0,1}}{2\omega_0}
\end{eqnarray*}

Finally, we can consider Eqn.(\ref{eqn:theory_ex_b}):

\begin{eqnarray*}
\deriv{R_1}{t} &=& ic_0\Theta_1e^{i\Theta_0}-X_0B^0e^{i\Theta_0}-B^-e^{2i\Theta_0}+\alpha_1B^+\nonumber\\
\deriv{R_1}{t} &=& (ic_0-X_*B^0)e^{i\Theta_0}+\left(ic_0P+\frac{ic_0B^0}{2\omega_0}-B^-\right)e^{2i\Theta_0}\nonumber\\
&& +\left(ic_0\bar{P}-\frac{i\bar{c}_0B^0}{2\omega_0}+B^+\right)
\end{eqnarray*}

Integration gives:

\begin{eqnarray*}
R_1 &=& R_+-\frac{i}{\omega_0}(ic_0-X_*B^0)e^{i\Theta_0}-\frac{i}{2\omega_0}\left(ic_0P+\frac{ic_0B^0}{2\omega_0}-B^-\right)e^{2i\Theta_0}\nonumber\\
&& +\left(ic_0\bar{P}-\frac{i\bar{c}_0+B^+}{2\omega_0}-B^-\right)t\nonumber\\
R_1 &=& R_++A_1e^{i\Theta_0}+A_2e^{2i\Theta_0}+A_3t
\end{eqnarray*}
\\
where:

\begin{eqnarray*}
A_1 &=& -\frac{i}{\omega_0}(ic_0-X_*B^0)\\
A_2 &=& -\frac{i}{2\omega_0}\left(ic_0P+\frac{ic_0B^0}{2\omega_0}-B^-\right)\\
A_3 &=& ic_0\bar{P}-\frac{i\bar{c}_0+B^+}{2\omega_0}-B^-
\end{eqnarray*}

We also note that the velocity of the drift, $A_3$, can be expressed as:

\begin{eqnarray*}
A_3 &=& \frac{ic_0C_{-1,0}}{\omega_0}+C_{-1,-1}
\end{eqnarray*}

Therefore, the final full equations motion of the tip of a rigidly rotating spiral wave which is drifting due to an inhomogeneity induced drift, are given by:

\begin{eqnarray*}
R(t) &=& R(0)-\frac{ic_0}{\omega_0}e^{i(\Theta_*+\omega_0t)}+\alpha_1A_1e^{i(\Theta_*+\omega_0t)}\nonumber\\
&& +\alpha_1A_2e^{2i(\Theta_*+\omega_0t)}+\alpha_1A_3t+O(\alpha_1^2)\\
\Theta(t) &=& \Theta(0)+\omega_0t+\alpha_1Pe^{i(\Theta_*+\omega_0t)}+\alpha_1\bar{P}e^{-i(\Theta_*+\omega_0t)}+O(\alpha_1^2)
\end{eqnarray*}

\section{Floquet Theory - Periodic Solutions}
\label{sec:floquet}
\subsection{Introduction}

In the next stage of our work, we consider the limit cycle solutions in the \chg[af]{functional space} corresponding to \chg[af]{spiral wave}solutions. Some authors denote this phenomenon as \emph{Relative Periodic Solutions}. In general, limit cycle solutions correspond to \emph{Meandering Spiral Waves}.

The stability of these limit cycles is of utmost importance. \chg[p71para]{Unstable solution can lead to multiple solutions, for instance, and also solutions with \chg[af]{multiple} tips and arms, which are not covered by this theory. We consider only stable limit cycle solutions.}

We saw in Sec.(\ref{sec:drift_rw}) that the equations of motion for a drifting rigidly rotating wave were of the form:
\chg[p72eqns1]{}
\begin{eqnarray*}
\odRt  &=& \left[\chg[]{c_0}-\epsilon(2(\bar{\bpsi}_1,\bht(\bv_0,\br,t))+\frac{\chg[]{c_0}}{\omega_0}(\bpsi_0,\bht(\bv_0,\br,t)))\right]e^{i\Theta}\\
\odTht &=& \omega_0+\epsilon (\bpsi_0,\bht(\bv_0,\br,t))
\end{eqnarray*}
\\
where \chg[p72gram]{$c_0$} and $\omega_0$ are constant and form the Euclidean Projection part of the quotient \chg[p72gram]{solution}. Therefore, since they are constant, we have an equilibrium in the quotient space.

We now wish to consider non constant $\bc_0(t)$ and $\omega_0(t)$. We therefore wish to study the solutions in the quotient system to the Reaction-Diffusion-Advection system of equations (\ref{eqn:theory_rw_fulleqn}) in the functional space:

\chg[p72para]{
\begin{equation*}
\label{eqn:lcs_V_system}
\deriv{\bV}{t} = \booF(\bV)+(\bc[\bV],\hat{\partial}_\br)\bV+\omega[\bV]\hat{\partial}_\theta\bV+\epsilon\bHt(\bV,t)
\end{equation*}

We now consider a general system of equations as shown below which will be discussed in the next section:

\begin{equation*}
\label{eqn:odeut1}
\deriv{\bV}{t} = \bg(\bV)+\epsilon\bk(\bV,t)
\end{equation*}
\\
where $\bV$, $\bg$, $\bk\in\mathbb{R}^n$. The system (\ref{eqn:lcs_V_system}) is a functional space analogue of (\ref{eqn:odeut1})}


\subsection{General Floquet Theory}

Firstly, consider system (\ref{eqn:odeut1}) with $\bk=0$ with the following form:

\begin{eqnarray}
\label{eqn:odeut3}
\odbVt &=& \bg(\bV)
\end{eqnarray}

We assume that the solutions to (\ref{eqn:odeut3}) are periodic and that they are perturbed with the following form:

\begin{equation}
\label{eqn:pertsol}
\bV(t) = \bV_0(t)+\epsilon\bV_1(t)
\end{equation}

We also assume that the unperturbed solution $\bV_0(t)$ is a known limit cycle solution of period $T$:

\begin{equation*}
\label{Usol}
\bV_0(t+T) = \bV_0(t)
\end{equation*}

Substituting (\ref{eqn:pertsol}) into (\ref{eqn:odeut3}) and splitting out the orders of $\epsilon$, we get:

\begin{eqnarray}
\label{eqn:odUt}
\epsilon^0:\quad\odbVot &=& \bg(\bV_0)\nonumber\\
\label{eqn:odvt}
\epsilon^1:\quad\odbVpt &=& \bG(t)\bV_1
\end{eqnarray}
\\
where $\bG(t)=\left.\pdgu\right|_{\bV=\bV_0(t)}$ is a matrix of first order partial derivatives of $\bg(\bV)$ evaluated at $\bV_0(t)$.

Since we assume that we know the solution $\bV_0(t)$, we will consider for the remainder of this section the solution $\bV_1(t)$, and its evolution as determined by Eqn.(\ref{eqn:odvt})

Consider $\bG(t)$. This matrix of functions is periodic with period $T$ since it is evaluated at $\bV_0(t)$ which is periodic with period $T$:

\begin{equation*}
\bG(t+T) = \bG(t)
\end{equation*}

Now, any solution to Eqn.(\ref{eqn:odvt}) can be expressed as a linear combination of $n$-linearly independent solutions to (\ref{eqn:odvt}), where the $n$-linearly independent solutions form the columns of a matrix, $\bQ(t)$, known as the \textbf{Fundamental Matrix}. Since any solution to (\ref{eqn:odvt}) can be expressed as a linear combination of linearly independent solutions to (\ref{eqn:odvt}), then any solution can be expressed as:

\begin{equation*}
\bV_1(t) = \bQ(t)\bC
\end{equation*}
\\
where $\bC$ is a constant vector and is equal to the initial value of $\bV_1(t)$, i.e.:

\begin{equation*}
\bV_1(t) = \bQ(t)\bV_1(0)
\end{equation*}

Now, since the columns of $\bQ(t)$ are independent solutions to (\ref{eqn:odvt}), then $\bQ(t)$ itself must satisfy (\ref{eqn:odvt}):

\begin{equation*}
\dot{\bQ}(t) = \bG(t)\bQ(t)
\end{equation*}
\\
where the dot notation represents time derivatives. Now if $\bQ(t)$ is a solution to (\ref{eqn:odvt}) then so too is $\bQ(t+T)$:

\begin{eqnarray*}
\dot{\bQ}(t+T) &=& \bG(t+T)\bQ(t+T)\nonumber\\
\dot{\bQ}(t+T) &=& \bG(t)\bQ(t+T)
\end{eqnarray*}

So, if $\bQ(t)$ is a fundamental matrix, then so too is $\bQ(t+T)$.

\begin{thm}{\cite{nodejordan}}
Given a fundamental matrix $\bQ(t)$, then $\bQ(t+T)$ is linearly dependent on $\bQ(t)$. Furthermore, if:
\begin{equation}
\bQ(t) = [q_{ij}(t)]\nonumber
\end{equation}
then we have that:
\begin{equation}
\label{eqn:QtT}
\bQ(t+T) = [q_{ij}(t+T)] = \left[\sum_{k=1}^n{q_{ik}(t)b_{kj}}\right] = \bQ(t)\bD
\end{equation}
\\
where $b_{kj}$ are constant and $\bD$ is a constant matrix with $\bD = [b_{kj}]$.
\end{thm}

\begin{defn}
The \emph{Monodromy Matrix}, $\bM$ is defined as the matrix $\bQ$ evaluated at an initial time, $t_0$:
\begin{equation*}
\bM = \bQ^{-1}(t_0)\bQ(t_0+T)
\end{equation*}
\end{defn}

This implies that the constant matrix $\bD$ above is actually equal to the Monodromy Matrix. Also, we generally have that $t_0=0$ and also that $\bQ(0)=\bI$, where $\bI$ is an identity matrix. Hence, the definition of the Monodromy Matrix reduces to:

\begin{equation*}
\bM = \bQ(T)
\end{equation*}

An immediate consequence of this is that if we consider Eqn.(\ref{eqn:QtT}):

\begin{ajf}
\begin{eqnarray*}
\bQ(t+T)  &=& \bQ(t)\bM\\
\bQ(T)    &=& \bM\\
\bQ(T+T)  &=& \bQ(T)\bM\\
\Rightarrow \bQ(T+T) &=& \bM^2\\
\bQ(2T+T) &=& \bQ(2T)\bM\\
\Rightarrow \bQ(2T+T)&=& \bM^3\\
\end{eqnarray*}

So generally:
\end{ajf} 

\begin{equation*}
\bQ(nT) = \bM^n
\end{equation*}

Which implies that any solution evaluated at $t=T$ is:

\begin{equation*}
\label{eqn:solnT}
\bV_1(nT) = \bM^n\bV_1(0)
\end{equation*}

It follows that the stability of the solutions $\bV_1$ to (\ref{eqn:odvt}) is done by studying the eigenvalues of $\bM$. These eigenvalues are called the \emph{Floquet Multipliers} of the system (\ref{eqn:odvt}), and are given by:

\begin{eqnarray}
\label{eqn:floq_mult}
\textrm{det}(\bM-\mu\bI) = 0,\quad \mbox{for}\quad\mu=(\mu_1,\cdots,\mu_n)
\end{eqnarray}

We note that if $\left|\mu_i\right|<1\quad\forall i=1,\hdots,n$ then the system is stable. Else, if $\exists\mu_i>1$ for some $i$, then the system is unstable. There always exists one multiplier $\mu_*=1$. The reason for this is that we require periodic solutions, and if there does not exist a multiplier equal to one, then we will not get periodic solutions. We shall show a proof of this in Sec.(\ref{sec:theory_sing})\chg[p75punct]{.} This will become more evident when we state Thm.(\ref{thm:multi}). We assume that we are dealing with stable solutions and therefore we must have that $\left|\mu_i\right|<1\quad\forall i=1,\hdots,n$ and $i\neq *$.

Furthermore, we come to the following definition:

\begin{defn}
Let $\mu$ be the Floquet Multipliers as defined by (\ref{eqn:floq_mult}) corresponding to the period $T$ of $\bG(t)$. Then, the corresponding \emph{Floquet Exponent}, $\rho$, is defined as:

\begin{equation*}
e^{\rho T} = \mu.
\end{equation*}
\end{defn}

We now come to the following fundamental theorem:

\begin{thm}\cite{nodejordan}
\label{thm:multi}
Suppose that $\bM$ has $n$ distinct multipliers, $\mu_i$ for $i=1,..,n$. Then Eqn.(\ref{eqn:odvt}) has $n$ linearly independent solutions of the form:

\begin{equation*}
\bq_i(t) = \bp_i(t)e^{\rho_it}
\end{equation*}
\\
where the $\bp_i(t)$ are functions with period $T$. The $\bq_i(t)$ are the columns of the Fundamental Matrix which therefore has the following form:

\begin{equation*}
\bQ(t) = \bP(t)e^{\bR t} 
\end{equation*}
\\
where $\bR$ is a constant matrix, known as the \textbf{Indicator Matrix}.

\end{thm}

From this Theorem, it immediately follows that at $t=0$ we have:

\begin{ajf}
\begin{eqnarray*}
\bQ(0) &=& \bP(0)e^{0}\\
\bQ(0) &=& \bP(0)\\
\bI    &=& \bP(0)
\end{eqnarray*}
\\
and also at time $t=T$, we get:
\begin{eqnarray*}
\bQ(T) &=& \bP(T)e^{\bR T}\\
\bQ(T) &=& \bP(0)e^{\bR T}\\
\bQ(T) &=& e^{\bR T}
\end{eqnarray*}
\\
due to the $T$-periodicity of $\bP(t)$. We know that the Monodromy Matrix, $\bM$, is the Fundamental Matrix evaluated at time $t=T$, and so we come to another important derivation:
\end{ajf}
\begin{ajfthesis}
\begin{eqnarray*}
\bI    &=& \bP(0)
\end{eqnarray*}
\\
and also at time $t=T$, we get:
\begin{eqnarray*}
\bQ(T) &=& e^{\bR T}
\end{eqnarray*}
\\
due to the $T$-periodicity of $\bP(t)$. We know that the Monodromy Matrix, $\bM$, is the Fundamental Matrix evaluated at time $t=T$, and so we come to another important derivation:
\end{ajfthesis}

\begin{equation*}
\bM = e^{\bR T} 
\end{equation*}

Now, any solution to Eqn.(\ref{eqn:odvt}) can be therefore be expressed as:
\chg[p77eqn]{
\begin{equation*}
\bV_{1i}(t) = \bQ(t)\bC = \bP(t)e^{\bR t}\bV_1(0) = \sum_i\bp_i(t)e^{\rho_it}\chg[]{V_{1i}(0)}
\end{equation*}}
\\
\chg[af]{where $\bV_1(t)=(V_{11}(t),V_{12}(t),\hdots,V_{1n}(t))$.}

So, at time $T$, we have:
\begin{ajf}
\begin{eqnarray*}
\bV_1(T)             &=& \bP(T)e^{\bR T}\bV_1(0)\\
\Rightarrow \bV_1(T) &=& \bP(0)e^{\bR T}\bV_1(0)\\
\Rightarrow \bV_1(T) &=& e^{\bR T}\bV_1(0)\\
\Rightarrow \bV_1(T) &=& \bM\bV_1(0)
\end{eqnarray*}
\end{ajf}
\begin{ajfthesis}
\begin{equation*}
\bV_1(T) = \bP(T)e^{\bR T}\bV_1(0) = \bP(0)e^{\bR T}\bV_1(0) = \bM\bV_1(0)
\end{equation*}
\end{ajfthesis}
\\
with the second equation coming from the periodicity of $P$. So, generally, for time $nT$, we have:

\begin{ajf}
\begin{eqnarray*}
\bV_1(nT)             &=& \bP(nT)e^{n\bR T}\bV_1(0)\\
\Rightarrow \bV_1(nT) &=& \bP(0)(e^{\bR T})^n\bV_1(0)\\ 
\Rightarrow \bV_1(nT) &=& (e^{\bR T})^n\bV_1(0)\\
\Rightarrow \bV_1(nT) &=& \bM^n\bV_1(0)
\end{eqnarray*}
\end{ajf}
\begin{ajfthesis}
\begin{eqnarray}
\bV_1(nT) &=& \bM^n\bV_1(0)
\end{eqnarray}
\end{ajfthesis}

To complete the picture, let us consider the solution at time $t=t+T$:

\begin{equation*}
\bV_1(t+T) = \bQ(t+T)\bV_1(0) = \bQ(t)\bQ(T)\bV_1(0) = \bQ(t)\bM\bV_1(0)
\end{equation*}

Now, let us introduce the multiplier problem for $\bM$:

\begin{equation*}
(\bM-\mu\bI)\bbet = 0
\end{equation*}
\\
where $\bbet$ is the eigenvector corresponding to the the Floquet multiplier of $\bM$.

Generally, there is a solution $\bV_1$, called the \textbf{eigensolution}, which is a linear combination of the columns of the Fundamental Matrix, such that $\bV_1(0)=\bbet$:

\begin{equation*}
\bV_1(t) = \bQ(t)\bbet
\end{equation*}

Therefore:

\begin{equation*}
\bV_1(t+T) = \bQ(t)\bM\bbet = \bQ(t)\mu\bbet = \mu\bQ(t)\bbet = \mu\bV_1(t)
\end{equation*}

So, in general it can be shown that:

\begin{eqnarray*}
\bV_1(t+nT) &=& \mu^n\bV_1(t)
\end{eqnarray*}

Let us now claim the following. We know that the eigenvalues to $\bR$ are $\rho$ and satisfy:

\begin{equation*}
\textrm{det}(\bR-\rho_i\bI)=0
\end{equation*}

The eigenvalue problem is:

\begin{equation*}
\bR\balp_i = \rho_i\balp_i
\end{equation*}
\\
where $\balp_i$ is an eigenvector of $\bR$. This therefore means that:

\begin{equation}
\label{eqn:expobalp}
e^{\bR t}\balp_i = e^{\rho_it}\balp_i
\end{equation}

Then there will exist a solution to (\ref{eqn:odvt}) of the form:
\chg[p77eqn]{}
\begin{equation*}
\chg[]{\bV_{1i}(t)} = \bQ(t)\balp_i = \bP(t)e^{\bR t}\balp_i\ = \bP(t)e^{\rho_it}\balp_i = e^{\rho_it}\bP(t)\balp_i = e^{\rho_it}\bphi_i(t)
\end{equation*}
\\
where $\bphi_i(t)$ is periodic and defined as the Eigenfunction corresponding to the Floquet Exponent, $\rho_i$. We shall call this eigenfunction the {\bf Floquet Eigenfunction}:

\begin{equation*}
\bphi_i(t) = \bP(t)\balp_i
\end{equation*}

Let us now derive the eigenvalue problem relating to these eigenfunctions. From Eqn.(\ref{eqn:expobalp}) we have:

\begin{eqnarray*}
e^{\bR t}\balp_i                &=& e^{\rho_it}\balp_i\\
\bP e^{\bR t}\balp_i            &=& e^{\rho_it}\bP\balp_i\\
\bP e^{\bR t}\bP^{-1}\bP\balp_i &=& e^{\rho_it}\bP\balp_i\\
\bP e^{\bR t}\bP^{-1}\bphi_i    &=& e^{\rho_it}\bphi_i\\
\bQ\bP^{-1}\bphi_i              &=& e^{\rho_it}\bphi_i
\end{eqnarray*}
\\
Therefore, we have an operator defined as $\bQ\bP^{-1}$ for which the above equation holds true.

Let us now consider the solution \chg[p78eqn1]{$\bV_{1i}(t)=e^{\rho_it}\bphi_i(t)$,}and substitute this into Eqn.(\ref{eqn:odvt}):

\begin{eqnarray*}
\rho_ie^{\rho_it}\bphi_i(t)+e^{\rho_it}\dot{\bphi_i}(t) &=& \bG(t) e^{\rho_it}\bphi_i(t)\nonumber\\
\Rightarrow \rho_i\bphi_i(t)+\dot{\bphi_i}(t) &=& \bG(t)\bphi_i(t)\nonumber\\
\Rightarrow \dot{\bphi_i}(t) &=& (\bG(t)-\rho_i\bI)\bphi_i(t)
\end{eqnarray*}


\subsection{Floquet Eigenfunctions corresponding to Meander}

We will now consider the unperturbed Reaction-Diffusion-Advection system of equations, whose solution is now a meandering spiral wave. This time, let us consider the system in our original space:
\chg[p78eqn2]{}
\begin{equation}
\label{eqn:app_rda_pde}
\pdvot = \textbf{D}\nabla^2\bv_0+\bof(\bv_0)+(\bc(\bv_0),\nabla)\bv_0+\omega(\bv_0)\partial_\theta\bv_0
\end{equation}

Let us now differentiate Eqn.(\ref{eqn:app_rda_pde}) with respect to $x$:

\begin{eqnarray*}
\pderiv{}{x}\left(\pdvot\right) &=& \textbf{D}\nabla^2\left(\pdvox\right)+\pderiv{\bof(\bv_0)}{\bv_0}\left(\pdvox\right)+c_x(\bv_0)\pderiv{^2\bv_0}{x^2}+c_y(\bv_0)\pderiv{^2\bv_0}{x\partial{y}}\nonumber\\
&& +\omega(\bv_0)\pderiv{}{x}\left(x\pderiv{\bv_0}{y}-y\pderiv{\bv_0}{x}\right)+O(\epsilon^2)\nonumber\\
\pderiv{}{t}\left(\pdvox\right) &=& \textbf{D}\nabla^2\left(\pdvox\right)+\bof'(\bv_0)\left(\pdvox\right)+c_x(\bv_0)\pderiv{}{x}\left(\pderiv{\bv_0}{x}\right)+c_y(\bv_0)\pderiv{}{y}\left(\pderiv{\bv_0}{x}\right)\nonumber\\
&& +\omega(\bv_0)\pderiv{\bv_0}{y}+\omega(\bv_0)\left(x\pderiv{}{y}\pderiv{\bv_0}{x}-y\pderiv{}{x}\pderiv{\bv_0}{x}\right)+O(\epsilon^2)\nonumber\\
\pderiv{}{t}\left(\pdvox\right) &=& \textbf{D}\nabla^2\left(\pdvox\right)+\bof'(\bv_0)\left(\pdvox\right)+(\bc(\bv_0),\nabla)\left(\pderiv{\bv_0}{x}\right)+\omega(\bv_0)\pderiv{}{\theta}\left(\pdvox\right)\nonumber\\
&& +\omega(\bv_0)\left(\pderiv{\bv_0}{y}\right)+O(\epsilon^2)\nonumber\\
\pderiv{}{t}\left(\pdvox\right) &=& \bG(t)\left(\pdvox\right)+\omega(\bv_0)\left(\pderiv{\bv_0}{y}\right)+O(\epsilon^2)
\end{eqnarray*}
\\
where $G(t)$ is given by:

\begin{equation}
\label{eqn:app_linear_x}
\bG(t)\alpha = \textbf{D}\nabla^2\alpha+\bof'(\bv_0)\alpha+(\bc(\bv_0),\nabla)\alpha+\omega(\bv_0)\pderiv{\alpha}{\theta}
\end{equation}

Similarly, by differentiating Eqn.(\ref{eqn:app_rda_pde}) with respect to $y$, we get:

\begin{eqnarray}
\label{eqn:app_linear_y}
\pderiv{}{t}\left(\pdvoy\right) &=& \bG(t)\left(\pdvoy\right)-\omega(\bv_0)\left(\pdvox\right)+O(\epsilon^2)
\end{eqnarray}

We see that if we consider (\ref{eqn:app_linear_x})+i(\ref{eqn:app_linear_y}), then:

\begin{eqnarray*}
\pderiv{}{t}\left(\pdvox+i\pdvoy\right) &=& \bG(t)\left(\pdvox+i\pdvoy\right)-i\omega(\bv_0)\left(\pdvox+i\pdvoy\right)+O(\epsilon^2)\nonumber\\
\left(\bG(t)-\pderiv{}{t}\right)\til{\bphi}_1 &=& i\omega(\bv_0)\til{\bphi}_1
\end{eqnarray*}
\\
where $\til{\bphi}_1$ is the eigenfunction in a comoving frame of reference given by:

\begin{equation*}
\til{\bphi}_1 = \pdvox+i\pdvoy
\end{equation*}
\\
and therefore the eigenfunction in the laboratory frame of reference needs to be transformed:

\begin{equation}
\label{eqn:app_lab_eigen_a}
\bphi_1 = e^{i\Theta_0}\til{\bphi}_1
\end{equation}
\\
where $\Theta_0$ satisfies:

\begin{equation*}
\dot{\Theta}_0 = \omega(\bv_0)
\end{equation*}

Differentiating Eqn.(\ref{eqn:app_lab_eigen_a}) with respect to $t$, we get:

\begin{eqnarray*}
\pderiv{\bphi_1}{t} &=& i\dot{\Theta}_0e^{i\Theta_0}\til{\bphi}_1+e^{i\Theta_0}\pderiv{\til{\bphi}_1}{t}\\
\pderiv{\bphi_1}{t} &=& i\omega(\bv_0)e^{i\Theta_0}\til{\bphi}_1+e^{i\Theta_0}\bG(t)\til{\bphi}_1-i\omega(\bv_0)e^{i\Theta_0}\til{\bphi}_1\\
\pderiv{\bphi_1}{t} &=& e^{i\Theta_0}\bG(t)\til{\bphi}_1\\
\pderiv{\bphi_1}{t} &=& \bG(t)\bphi_1\\
\left(\bG(t)-\pderiv{}{t}\right)\bphi_1 &=& 0
\end{eqnarray*}

This is equivalent to the Floquet exponent problem given by:

\begin{equation*}
\left(\bG(t)-\pderiv{}{t}\right)\bphi_i = \rho_i\bphi_i
\end{equation*}

Hence, the eigenfunction $\bphi_1$ corresponds to the zero Floquet exponent. Similarly, it can be shown that the complex conjugate of $\bphi_1=\bphi_{-1}$ is given by,
\chg[af]{}
\begin{equation*}
\bphi_{-1} = \chg[]{e^{-i\Theta_0}}\til{\bphi}_{-1}
\end{equation*}
\\
where,

\begin{equation*}
\til{\bphi}_{-1} = \pdvox-i\pdvoy
\end{equation*}
\\
and satisfies:

\begin{equation*}
\left(\bG(t)-\pderiv{}{t}\right)\bphi_{-1} = 0
\end{equation*}

Again, we see that $\bphi_{-1}$ corresponds to the zero Floquet exponent.

Finally, let us consider differentiating Eqn.(\ref{eqn:app_rda_pde}) with respect to $\theta$. After some analysis, we get:

\begin{equation*}
\pderiv{}{t}\left(\pdvoth\right) = \bG(t)\left(\pdvoth\right)-c_x(\bv_0)\left(\pdvoy\right)+c_y(\bv_0)\left(\pdvox\right)+O(\epsilon^2)
\end{equation*}

Let us define the eigenfunction $\til{\bphi}_0$ as:

\begin{equation*}
\til{\bphi}_0 = \pdvoth+\alpha\til{\bphi}_1+\beta\til{\bphi}_{-1}
\end{equation*}

Now, consider the operation of $\bG(t)$ on $\til{\bphi}_0$:

\begin{eqnarray*}
\bG(t)\til{\bphi}_0 &=& \bG(t)\pdvoth+\alpha\bG(t)\til{\bphi}_1+\beta\bG(t)\til{\bphi}_{-1}\nonumber\\
\bG(t)\til{\bphi}_0 &=& \pderiv{}{t}\left(\pdvoth\right)+c_x(\bv_0)\left(\pdvoy\right)-c_y(\bv_0)\left(\pdvox\right)\nonumber\\
&& +\alpha\left(i\omega(\bv_0)\til{\bphi}_1+\pderiv{}{t}\til{\bphi}_1\right)+\beta\left(-i\omega(\bv_0)\til{\bphi}_{-1}+\pderiv{}{t}\til{\bphi}_{-1}\right)\nonumber\\
\bG(t)\til{\bphi}_0 &=& \pderiv{}{t}\left(\pdvoth\right)+\frac{c_x(\bv_0)}{2i}\left(\til{\bphi}_1-\til{\bphi}_{-1}\right)-\frac{c_y(\bv_0)}{2}\left(\til{\bphi}_1+\til{\bphi}_{-1}\right)\nonumber\\
&& +\alpha\left(i\omega(\bv_0)\til{\bphi}_1+\pderiv{}{t}\til{\bphi}_1\right)+\beta\left(-i\omega(\bv_0)\til{\bphi}_{-1}+\pderiv{}{t}\til{\bphi}_{-1}\right)\nonumber\\
\bG(t)\til{\bphi}_0 &=& \pderiv{}{t}\left(\pdvoth+\alpha\til{\bphi}_1+\beta\til{\bphi}_{-1}\right)+\til{\bphi}_1\left(\frac{c_x(\bv_0)}{2i}-\frac{c_y(\bv_0)}{2}+\alpha i\omega(\bv_0)\right)\nonumber\\
&& -\til{\bphi}_{-1}\left(\frac{c_x(\bv_0)}{2i}+\frac{c_y(\bv_0)}{2}+\beta i\omega(\bv_0)\right)\nonumber\\
\bG(t)\til{\bphi}_0 &=& \pderiv{}{t}\til{\bphi}_0+\til{\bphi}_1\left(-\frac{i\bar{\bc}(\bv_0)}{2}+\alpha i\omega(\bv_0)\right)-\til{\bphi}_{-1}\left(-\frac{i\bc(\bv_0)}{2}+\beta i\omega(\bv_0)\right)
\end{eqnarray*}

So, if:

\begin{eqnarray*}
\alpha &=& \frac{\bar{\bc}(\bv_0)}{2\omega(\bv_0)}\\
\beta  &=& \frac{\bc(\bv_0)}{2\omega(\bv_0)}
\end{eqnarray*}
\\
then we have:
\chg[af]{}
\begin{eqnarray*}
\til{\bphi}_0 &=& \pdvoth+\frac{\bar{\bc}(\bv_0)}{2\omega(\bv_0)}\til{\bphi}_1+\frac{\bc(\bv_0)}{2\omega(\bv_0)}\til{\bphi}_{-1}\\
\Rightarrow \til{\bphi}_0 &=& \pdvoth+\frac{\bar{\bc}(\bv_0)}{2\omega(\bv_0)}\bphi_1\chg[]{e^{-i\Theta_0}}+\frac{\bc(\bv_0)}{2\omega(\bv_0)}\bphi_{-1}\chg[]{e^{i\Theta_0}}
\end{eqnarray*}
\\
which in turn implies that:

\begin{eqnarray*}
\bG(t)\til{\bphi}_0 &=& \pderiv{}{t}\til{\bphi}_0\\
\left(\bG(t)-\pderiv{}{t}\right)\til{\bphi}_0 &=& 0
\end{eqnarray*}

Clearly, we have that:

\begin{equation*}
\bphi_0 = \til{\bphi}_0
\end{equation*}
\\
and $\bphi_0$ satisfies:

\begin{equation*}
\left(\bG(t)-\pderiv{}{t}\right)\bphi_0 = 0
\end{equation*}

implying that $\bphi_0$ is another eigenfunction of the Floquet exponent problem for the zero exponent.

To \chg[af]{summarise, we have in the laboratory frame of reference:}
\chg[af]{}
\begin{equation}
\label{eqn:gm_lab}
{\begin{array}{rclcrclcrcl}
   \bphi_* &=& \pderiv{\bv_0}{\tau} &,& \rho_* &=& 0\\
   \bphi_1 &=& e^{i\Theta_0}\left(\pderiv{\bv_0}{x}+i\pderiv{\bv_0}{y}\right) &,& \rho_1 &=& 0\\
   \bphi_{-1} &=& e^{-i\Theta_0}\left(\pderiv{\bv_0}{x}-i\pderiv{\bv_0}{y}\right) &,& \rho_{-1} &=& 0\\
   \bphi_0 &=& \pdvoth+\frac{\bar{\bc}(\bv_0)}{2\omega(\bv_0)}\bphi_1e^{-i\Theta_0}+\frac{\bc(\bv_0)}{2\omega(\bv_0)}\bphi_{-1}e^{i\Theta_0} &,& \lambda_0 &=& 0
  \end{array}}
\end{equation}
  
In the comoving frame of reference, we have:

\chg[af]{}
\begin{equation}
\label{eqn:gm_com}
{\begin{array}{rclcrclcrcl}
   \bphi_* &=& \pderiv{\bv_0}{\tau} &,& \rho_* &=& 0\\
   \bphi_1    &=& \pderiv{\bv_0}{x}+i\pderiv{\bv_0}{y} &,& \rho_1 &=& i\omega(\bv_0)\\
   \bphi_{-1} &=& \pderiv{\bv_0}{x}-i\pderiv{\bv_0}{y} &,& \rho_{-1} &=& -i\omega(\bv_0)\\
   \bphi_0 &=& \pdvoth+\frac{\bar{\bc}(\bv_0)}{2\omega(\bv_0)}\bphi_1+\frac{\bc(\bv_0)}{2\omega(\bv_0)}\bphi_{-1} &,& \rho_0 &=& 0\\
  \end{array}}
\end{equation}


\subsection{Floquet Theory: The Adjoint Problem}

We summarize several results from the previous section. Firstly, we note that for the system $\dot{\bV}_1=\bG(t)\bV_1$, we have the Fundamental Matrix defined as:

\begin{equation*}
\bQ(t) = \bP(t)e^{\bR t}
\end{equation*}
\\
where $\bR$ is a constant matrix known as the Indicator Matrix and $\bP(t+T)=\bP(t)$. The eigenvalues, $\rho_i$, to $\bR$ are known as the Floquet Exponents and satisfy:

\begin{equation*}
\bR\balp_i = \rho_i\balp_i
\end{equation*}
\\
where $\balp_i$ are the corresponding eigenvectors. Also, the Fundamental Matrix evaluated at $T$, where $T$ is the period of the solution to the system, is known as the Monodromy Matrix, \bM, whose eigenvalues are \chg[p82gram]{the} Floquet Multipliers.

Now, \bQ(t) consists of $n$ independent fundamental solutions, $\bq_i$, which are defined as:

\begin{equation}
\label{eqn:theory_floquet_qi}
\bq_i(t) = \bp_i(t)e^{\rho_it} = \bQ(t)\balp_i
\end{equation}

We then defined the Floquet Eigenfunctions as:

\begin{equation*}
\bphi_i(t) = \bP(t)\balp_i
\end{equation*}
\\
which satisfy:

\begin{equation*}
\left(\bG(t)-\frac{\dd}{\dd{t}}\right)\bphi_i = \rho_i\bphi_i\\
\Rightarrow L\bphi_i = \rho_i\bphi_i
\end{equation*}
\\
where:
\chg[p83eqn1]{
\begin{eqnarray*}
L\alpha &=& \bG(t)\alpha-\frac{\dd\alpha}{\dd{t}}
\end{eqnarray*}
}
We can now define the adjoint problem. The adjoint system of equation to $\dot{\bV}_1=\bG(t)\bV_1$ is given by:

\begin{eqnarray}
\label{eqn:adjw}
\Rightarrow \odbWt &=& -\bG^+(t)\bW
\end{eqnarray}

The eigenfunction generated by this system are $\bpsi_j$, which are periodic and satisfy:

\begin{equation*}
\label{eqn:lcs_adjoint_eprob}
\left(\bG^+(t)+\frac{\dd}{\dd{t}}\right)\bpsi_j = \sigma_j\bpsi_j
\end{equation*}
\\
Also they take the following form:

\begin{equation*}
\bpsi_j(t) = \til{\bP}(t)\bbet_j
\end{equation*}

Using:
\chg[p83eqn2]{
\begin{eqnarray*}
L^+\beta &=& \bG^+(t)\beta+\frac{\dd\beta}{\dd{t}}
\end{eqnarray*}}
\\
our adjoint eigenproblem is :

\begin{equation*}
L^+\bpsi_j = \sigma_j\bpsi_j
\end{equation*}

The eigenfunctions, $\bpsi_j$, are called the \emph{Floquet response functions}.


\subsection{Biorthogonality Conditions}

There exists a biorthogonality condition between the Floquet response functions, $\bpsi_j$, and the eigenfunctions, $\phi_i$, to the linear operator $L$. Let us define the scalar product between these in two different ways:

\begin{eqnarray}
\label{eqn:lcs_scalar_1}
(1)\quad (\bpsi_j,\bphi_i) &=& \langle\bpsi_j,\bphi_i\rangle\\
\label{eqn:lcs_scalar_2}
(2)\quad (\bpsi_j,\bphi_i) &=& \int_{-\frac{T}{2}}^\frac{T}{2}\langle\bpsi_j(\tau),\bphi_i(\tau)\rangle\dd{\tau}
\end{eqnarray}
\\
where definition (1) is local in time, with definition (2) being averaged over the period of the spiral, i.e. global in time, and  $\langle\balp_i,\bbet_j\rangle$ is the usual Hermitian scalar product.

The first definition means that the biorthogonality condition takes the form of the scalar product of the two functions in question for \emph{any} given time.

It would be beneficial for us if we could use definition (1) due to the fact that we would not have to integrate nor would we have to consider a range of times, but just one particular \chg[af]{moment.}

Let us consider definition Eqn.(\ref{eqn:lcs_scalar_1}) and define the following:

\begin{equation}
\label{eqn:LCS_scalar_def}
<\bpsi_j,\bphi_i> = g_{ji}(t)
\end{equation}

Let us now differentiate (\ref{eqn:LCS_scalar_def}):

\begin{eqnarray*}
\frac{\dd{g_{ji}}}{\dd{t}} &=& <\dot{\bpsi}_j,\bphi_i>+<\bpsi_j,\dot{\bphi}_i>
\end{eqnarray*}

Using the following:

\begin{eqnarray*}
\dot{\bphi}_i &=& (\bG(t)-\rho_i)\bphi_i\\
\dot{\bpsi}_j &=& -(\bG(t)^+-\sigma_j)\bpsi_j
\end{eqnarray*}
\\
we get:
\chg[p85eqn1]{
\begin{eqnarray*}
\frac{\dd{g_{ji}}}{\dd{t}} &=& \langle-\bG(t)^+\bpsi_j,\bphi_i\rangle+\langle\sigma_j\bpsi_j,\bphi_i\rangle\\
                            && +\langle\bpsi_j,\bG(t)\bphi_i\rangle-\langle\bpsi_j,\rho_i\bphi_i\rangle\\
\frac{\dd{g_{ji}}}{\dd{t}} &=& -\langle\bpsi_j,\bG(t)\bphi_i\rangle+\bar{\sigma}_j\langle\bpsi_j,\bphi_i\rangle\\
                            && +<\bpsi_j,\bG(t)\bphi_i\rangle-\rho_i\langle\bpsi_j,\bphi_i\rangle\\
\frac{\dd{g_{ji}}}{\dd{t}} &=& \bar{\sigma}_j\langle\bpsi_j,\bphi_i\rangle-\rho_i\langle\bpsi_j,\bphi_i\rangle\\
\frac{\dd{g_{ji}}}{\dd{t}} &=& (\bar{\sigma}_j-\rho_i)\langle\bpsi_j,\bphi_i\rangle\\
\frac{\dd{g_{ji}}}{\dd{t}} &=& (\bar{\sigma}_j-\rho_i)g_{ji}
\end{eqnarray*}}

\begin{ajf}
Now let us consider the definition of the adjoint operator:

\begin{eqnarray}
                                  (\bpsi_j,L\bphi_i) &=& (L^+\bpsi_j,\bphi_i)\nonumber\\
                 \Rightarrow (\bpsi_j,\rho_i\bphi_i) &=& (\sigma_j\bpsi_j,\bphi_i)\nonumber\\
                 \Rightarrow \rho_i(\bpsi_j,\bphi_i) &=& \bar{\sigma}_j(\bpsi_j,\bphi_i)\nonumber\\
\Rightarrow (\rho_i-\bar{\sigma}_j)(\bpsi_j,\bphi_i) &=& 0
\end{eqnarray}

For biorthogonality, we require that $(\bpsi_j,\bphi_i)=0$ for $i\neq j$ and that $(\bpsi_j,\bphi_i)\neq0$ for $i=j$. This therefore means that:
\end{ajf}

Since  $L^+$ is an adjoint operator to $L$ with respect to the global scalar product, we have:
\chg[p85eqn2]{}
\[\begin{array}{ccrcl}
\mbox{for} & i=j     &            \rho_j &=& \bar{\sigma}_j\\
\mbox{for} & i\neq j & \chg[]{\langle\bpsi_j,\bphi_i\rangle} &=& 0
\end{array}\]

Going back to our differential equation in $g_{ji}$, we have:

\begin{eqnarray*}
\frac{\dd{g_{ji}}}{\dd{t}} &=& (\bar{\sigma}_j-\rho_i)g_{ji}\nonumber\\
\Rightarrow         g_{ji}(t) &=& g_{ji}(0)e^{(\bar{\sigma}_j-\rho_i)t}
\end{eqnarray*}

So, if $i=j$:

\begin{eqnarray*}
g_{ii}(t) &=& g_{ii}(0)
\end{eqnarray*}

We know biorthogonality means that:

\begin{equation*}
(\bpsi_j,\bphi_i) = \int_{-\frac{T}{2}}^\frac{T}{2}g_{ij}(\tau)\dd\tau = \delta_{ji}
\end{equation*}

This therefore implies that:

\begin{ajf}

WRONG!!!!
\begin{eqnarray}
\int_{-\frac{T}{2}}^\frac{T}{2}g_{ij}(0)\dd\tau &=& \delta_{ji}\nonumber\\
Tg_{ij}(0)                                      &=& \delta_{ji}\nonumber\\
g_{ij}(0)                                       &=& \frac{\delta_{ji}}{T}
\end{eqnarray}

We can see straightaway that for $i=j$:

\begin{equation}
g_{ii}(t) = g_{ii}(0) = \frac{1}{T}
\end{equation}
\\
and for $i\neq j$ we have:

\begin{equation}
g_{ji}(t) = g_{ji}(0)e^{(\bar{\sigma}_j-\rho_i)t}
\end{equation}
\\
which is zero if and only if $g_{ji}(0)=0$. To summerise:

\begin{eqnarray}
g_{ji}(t)                     &=& \frac{\delta_{ji}}{T}\nonumber\\
\Rightarrow <\bpsi_j,\bphi_i> &=& \frac{\delta_{ji}}{T}\nonumber\\
\Rightarrow (\bpsi_j,\bphi_i) &=& \frac{\delta_{ji}}{T}
\end{eqnarray}
\end{ajf}

\begin{eqnarray*}
\langle\bpsi_j,\bphi_i\rangle &=& \frac{\delta_{ji}}{T}
\end{eqnarray*}


\subsection{Solvability Conditions}
\label{sec:floquet_Solv_Conds}

Now, we have that $\bphi_i$ and $\bpsi_j$ satisfy:

\begin{eqnarray*}
L\bphi_i   &=& \rho_i\bphi_i\\
L^+\bpsi_j &=& \bar{\rho}_j\bpsi_j
\end{eqnarray*}

Let us consider the fundamental matrix of (\ref{eqn:adjw}). We assume that it takes the following form:

\begin{equation*}
\til{\bQ}(t) = \til{\bP}e^{\til{\bR}t}
\end{equation*}

So, $\til{\bQ}(t)$ also satisfies (\ref{eqn:adjw}) and gives:
\chg[af]{}
\begin{equation*}
\frac{\dd{\til{\bQ}}}{\dd{t}} = \chg[]{\bG^+(t)\til{\bQ}}
\end{equation*}

Now, we know that the eigenvalues to $\til{\bR}$ are $\sigma_j=\bar{\rho}_j$. So let us assume that the eigenvectors are $\bbet$ and satisfy;

\begin{equation*}
\til{\bR}\bbet_j = \bar{\rho}_j\bbet_j
\end{equation*}

Now if the eigenfunctions are orthogonal, then the eigenvectors must also be orthogonal:

\begin{equation}
\label{eqn:theory_floquet_orth}
(\bbet_j,\balp_i) = \delta_{ji}
\end{equation}

\begin{ajf}
CORRECT IN IDEA, WRONG IN CONTENT:

So, if we substitute the definitions of $\bphi_i$ and $\bpsi_j$ into $(\bpsi_j,\bphi_i) = \delta_{ji}$, then we get:

\begin{eqnarray}
(\bpsi_j,\bphi_i)                     &=& \delta_{ji}\nonumber\\
(\til{\bP}(t)\bbet_j,\bP(t)\balp_i)   &=& \delta_{ji}\nonumber\\
\til{\bP}^+(t)\bP(t)(\bbet_j,\balp_i) &=& \delta_{ji}
\end{eqnarray}

So if $i\neq j$ then $(\bbet_j,\balp_i)=0$, and if $i=j$, then:

\begin{eqnarray}
\til{\bP}^+(t)\bP(t)       &=& \bI\nonumber\\
\Rightarrow \til{\bP}^+(t) &=& \bP^{-1}(t)\nonumber\\
\Rightarrow \til{\bP}(t) &=& (\bP^{-1})^+(t)
\end{eqnarray}

This means that the eigenfunctions to the adjoint operator are given by:

\begin{equation}
\bpsi_j(t) = (\bP^{-1})^+(t)\bbet_j
\end{equation}
\end{ajf}
 
\begin{ajfthesis}
So, if we substitute the definitions of $\bphi_i$ and $\bpsi_j$ into $(\bpsi_j,\bphi_i) = \delta_{ji}$ and use (\ref{eqn:theory_floquet_orth}), then we get:

\begin{equation*}
\bpsi_j(t) = (\bP^{-1})^+(t)\bbet_j
\end{equation*}
\end{ajfthesis}

Therefore, we now have a complete algorithm to find the Floquet response functions.

We now need to define, a suitable solvability condition. In order to have a unique solution for $\bV_1$, let us impose the following orthogonality condition:

\begin{equation*}
(\bpsi_*,\bV_1) = 0
\end{equation*}
\\
where $\bpsi_*$ is the the adjoint floquet eigenfunction to the adjoint linear operator $L^+$ corresponding to the unit multiplier, i.e. $\mu_*=1\quad\Rightarrow\quad\rho_*=0$.

Let us differentiate $(\bpsi_*,\bV_1)$ with respect to time:

\begin{eqnarray*}
\frac{\dd}{\dd{t}}(\bpsi_*,\bV_1)       &=& 0\\
(\bpsi_*,\dot{\bV}_1)+(\dot{\bpsi}_0,\bV_1)  &=& 0\\
\Rightarrow (\bpsi_*,\dot{\bV}_1)      &=& -(\dot{\bpsi}_0,\bV_1)
\end{eqnarray*}

Using Eqn.(\ref{eqn:lcs_adjoint_eprob}), we have:

\begin{eqnarray*}
(\bpsi_*,\dot{\bV}_1)             &=& -(-\bG(t)^+\bpsi_*+\sigma_*\bpsi_*,\bV_1)\nonumber\\
\Rightarrow (\bpsi_*,\dot{\bV}_1) &=& (\bG(t)^+\bpsi_*,\bV_1)\nonumber\\
\Rightarrow (\bpsi_*,\dot{\bV}_1) &=& (\bpsi_*,\bG(t)\bV_1)
\end{eqnarray*}

Hence, we now have two solvability conditions:

\begin{eqnarray*}
(\bpsi_*,\bV_1 ) &=& 0\\
(\bpsi_*,\dot{\bV}_1) &=& (\bpsi_*,\bG(t)\bV_1)
\end{eqnarray*}


\subsection{Stability in \chgex[ex]{a} Perturbed System; Regular Perturbation Techniques.}
\label{sec:theory_reg}

We now consider the following perturbed system:

\begin{eqnarray}
\label{eqn:odeut4}
\odbVt &=& \bg(\bV)+\epsilon\bk(\bV,t)
\end{eqnarray}

Like the unperturbed case (\ref{eqn:odeut3}), we assume that $\bV$ has the regular form (\ref{eqn:pertsol}):
\chg[p87eqn1]{}
\begin{equation}
\label{eqn:pertsol2}
\bV(t) = \bV_0(t)+\epsilon\bV_1(t)+\chg[]{O(\epsilon^2)}
\end{equation}

Substituting Eqn.(\ref{eqn:pertsol2}) into (\ref{eqn:odeut4}), using Taylor expansion and splitting out into orders of $\epsilon$, we get:

\begin{eqnarray}
\label{eqn:odeUt2}
\epsilon^0:\quad\odbVot &=& \bg(\bV_0)\\
\label{eqn:odevt2}
\epsilon^1:\quad\odbVpt &=& \bG(t)\bV_1+\bk(\bV_0,t)
\end{eqnarray}

As before, we assume that $\bV_0$ is a limit cycle solution. Therefore, we concentrate our efforts on (\ref{eqn:odevt2}). Furthermore, we see that, as before, $\bG(t)$ is a periodic function of period $T$ and also we now have that $\bk(\bV_0,t)$ must be a periodic function also of period $T$.

Since we know the solution to (\ref{eqn:odeUt2}), we shall concentrate on the solutions to (\ref{eqn:odevt2}). Firstly, we recall that $\bG(t)$ is a matrix of partial derivatives of $\bg(\bV)$:

\begin{equation*}
\label{eqn:F}
\bG(t) = \left[\frac{\partial\bg_i}{\partial\bV_j}\right]
\end{equation*}

We also see that $\bG(t)$ is periodic with period $T$. We therefore want to solve the following ODE:

\begin{equation*}
\label{eqn:vode1}
\odbVpt-\bG(t)\bV_1 = \bk(t)
\end{equation*}

We do this by firstly considering the homogeneous equation to find a Complementary Function, and then use a suitable ansatz for the Particular Integral, which will enable us to find the General Solution:
\chg[p88eqn1]{}
\begin{equation*}
\bV_1 = \chg[]{\bV_{1,CF}}+\bV_{1,PI}
\end{equation*}

So, for the complimentary function, consider the homogeneous equation:

\begin{equation}
\label{eqn:vode}
\deriv{V_{1,CF}}{t} = \bG(t)\bV_{1,CF}
\end{equation}

We know from Floquet Theory that this has the solution:

\begin{eqnarray*}
\label{eqn:inhom}
\bV_{1,CF} &=& \bQ(t)\bV_1(0)
\end{eqnarray*}
\\
where we have an explicit expression for the Fundamental Matrix $\bQ(t)$:

\begin{equation*}
\bQ(t) = e^{\int{\bG(t)\dd{t}}}
\end{equation*}

For a particular integral, we use the Method of Variation of Parameters \cite{nayfeh}:

\begin{equation}
\label{eqn:pi1}
\bV_{1,PI} = \bQ(t)\bA(t)
\end{equation}

This time, we assume that the vector premultiplied by $\bQ(t)$ is time dependent. Substituting (\ref{eqn:pi1}) into (\ref{eqn:vode}) we get:
\begin{ajf}
\begin{eqnarray}
\odbVpit-\bG(t)\bV_{1,PI}                                &=& \bk(t)\nonumber\\
\frac{\dd}{\dd{t}}(\bQ(t)\bbet(t))-\bG(t)\bQ(t)\bbet(t) &=& \bk(t)\nonumber\\
\dot{\bQ}\bbet+\bQ\dot{\bbet}-\bG(t)\bQ(t)\bbet(t)      &=& \bk(t)\nonumber\\
(\dot{\bQ}-\bG(t)\bQ(t))\bbet+\bQ\dot{\bbet}            &=& \bk(t)\nonumber\\
\bQ\dot{\bbet}                                          &=& \bk(t)\nonumber\\
\dot{\bbet}                                             &=& \bQ^{-1}(t)\bk(t)\nonumber\\
\bbet                                                   &=& \int^t_0{\bQ^{-1}(\tau)\bk(\tau)}\dd\tau
\end{eqnarray}

Therefore, the Particular Integral is :

\begin{equation}
\label{eqn:pi}
\bV_{1pi} = \bQ(t)\int^t_0{\bQ^{-1}(\tau)\bk(\tau)}\dd\tau
\end{equation}

Hence, the general solution is:

\begin{eqnarray}
\bV_1 &=& \bV_{1cf}+\bV_{1pi}\\
\bV_1 &=& \bQ(t)\bC+\bQ(t)\int^t_0{\bQ^{-1}(\tau)\bk(\tau)}\dd\tau
\end{eqnarray}
\end{ajf}

\begin{ajfthesis}
\begin{eqnarray*}
\bA(t) &=& \int^t_0{\bQ^{-1}(\eta)\bk(\eta)}\dd\eta
\end{eqnarray*}

Therefore, the Particular Integral is :

\begin{equation*}
\label{eqn:pi}
\bV_{1,PI} = \bQ(t)\int^t_0{\bQ^{-1}(\eta)\bk(\eta)}\dd\eta
\end{equation*}
\\ \end{ajfthesis}
giving the full solution to be:
\chg[p89eqn1]{}
\begin{equation}
\label{eqn:full_reg}
\bV = \bV_0(t)+\epsilon\bQ(t)\left(\bV_1(0)+\int^t_0{\bQ^{-1}(\eta)\bk(\eta)}\dd{\eta}\right)\chg[]{+O(\epsilon^2)}
\end{equation}

A property that we require from the solutions $\bV$ is that they are bounded. Numerical results show that for large time periods \chg[p89gram1]{(Chap.\ref{chap:4}),} the spiral wave solution in the quotient system remains bounded. The purpose of this \chg[p89gram2]{part of the work} is to analytically study the evolution of the limit cycles that are evident in the quotient system. Therefore, we require that the limit cycle solutions remain bounded.

Let us show that, in general, the solution shown in (\ref{eqn:full_reg}) is NOT in fact bounded. Consider the first order of $\epsilon$ part of (\ref{eqn:full_reg}) and let us assume that the vector function \chg[af]{$\bk(t)$} can be expressed as:

\begin{equation*}
\bk(t) = \sum_ik_i(t)\bphi_i(t)
\end{equation*}

Therefore, \chg[p89gram3]{the first order part of} (\ref{eqn:full_reg}) now becomes:

\begin{eqnarray*}
\bV_1(t) &=& \bQ(t)\bV_1(0)+\sum_i\bQ(t)\int_0^t{\bQ^{-1}(\eta)k_i(\eta)}\bphi_i(\eta)\dd{\eta}
\end{eqnarray*}

Now, we know that $\bphi_i(t)=\bP(t)\balp_i$, so therefore:
\chg[p89eqn2]{}
\begin{eqnarray*}
\bV_1(t) &=& \bQ(t)\bV_1(0)+\sum_i\bQ(t)\int_0^t{\bQ^{-1}(\eta)k_i(\eta)}\bP(\eta)\balp_i\dd{\eta}\\
&=& \bQ(t)\bV_1(0)+\sum_i\bQ(t)\int_0^t{k_i(\eta)}e^{-\bR\eta}\bP^{-1}(\eta)\bP(\eta)\balp_i\dd{\eta}\\
&=& \bQ(t)\bV_1(0)+\sum_i\bQ(t)\int_0^t{k_i(\eta)}e^{-\bR\eta}\balp_i\dd{\eta}\\
&=& \bQ(t)\bV_1(0)+\sum_i\bQ(t)\int_0^t{k_i(\eta)}e^{-\rho_i\eta}\balp_i\dd{\eta}\\
&=& \bQ(t)\bV_1(0)+\sum_i\bQ(t)\balp_i\int_0^t{k_i(\eta)}e^{-\rho_i\eta}\dd{\eta}
\end{eqnarray*}

Now we know that $\balp_i$ are eigenvectors and that $\bQ(t)\balp_i=\bq_i(t)=\bp_i(t)e^{\rho_it}$ from (\ref{eqn:theory_floquet_qi}):

\begin{eqnarray*}
\bV_1(t)             &=& \bQ(t)\bV_1(0)+\sum_i\bp_i(t)e^{\rho_it}\int_0^t{k_i(\eta)e^{-\rho_i\eta}}\dd{\eta}\\
 &=& \bQ(t)\bV_1(0)+\sum_i\bp_i(t)\int_0^t{k_i(\eta)e^{-\rho_i(\eta-t)}}\dd{\eta}
\end{eqnarray*}

Now, we require that the elements $k_i$ of the perturbation are bounded. Also, we note that $t\geq\eta$ and since we require stable limit cycles (meaning that $\rho_i\leq0$) then \chg[p90gram]{$e^{-\rho_i(\eta-t)}\ll1$}, and using the fact that $\rho_0=0$ and $\rho_{i\neq0}<0$:
\chg[p90eqn]{}
\begin{eqnarray*}
\bV_1(t)             &=& \bQ(t)\bV_1(0)+\sum_{i\neq0}\bp_i(t)\int_0^t{k_i(\eta)e^{-\rho_i(\eta-t)}}\dd{\eta}+\bp_0(t)\int^t_0{k_0(\eta)}\dd{\eta}\\
\Rightarrow \bV_1(t) &\chg[]{\approx}& \bQ(t)\bV_1(0)+\bp_0(t)\int_0^t{k_0(\eta)}\dd{\eta}
\end{eqnarray*}
\\
for large time. Therefore we see that there will be a linear growth in $\bV_1(t)$ as time grows and hence $\bV_1(t)$ is not bounded, unless of course if $\int_0^t{k_0(\eta)}\dd{\eta}$ converges. Therefore, if $\int_0^t{k_0(\eta)}\dd{\eta}$ does not converge, a singular perturbation technique is required.


\subsection{Stability in an Perturbed System; Singular Perturbation Techniques.}
\label{sec:theory_sing}

Given the following system:

\begin{eqnarray}
\label{eqn:odeut5}
\odbVt &=& \bg(\bV)+\epsilon\bk(\bV)
\end{eqnarray}
\\
we assume a solution to (\ref{eqn:odeut5}) takes the following form:

\begin{equation*}
\bV(t) = \bV_0(t+\theta(t))+\epsilon\bV_1(t+\theta(t))
\end{equation*}
\\
where $\theta(t)$ is a shift to time and can be thought of as a perturbation of the limit cycle's phase. Using the method of Strained Coordinates \cite{nayfeh}, let us introduce the following notation:

\begin{equation*}
 \tau = t+\theta(t)
\end{equation*}

This means that our solution can be expressed as:

\begin{equation}
\label{eqn:sing}
\bV(t) = \bV_0(\tau)+\epsilon\bV_1(\tau)
\end{equation}

Substituting (\ref{eqn:sing}) into (\ref{eqn:odeut5}) we get:
\chg[p91eqn1]{}
\begin{eqnarray*}
(1+\dot{\theta})\deriv{\bV_0}{\tau}+\epsilon(1+\dot{\theta})\deriv{\bV_1}{\tau} &=& \bg(\bV_0)+\chg[]{\epsilon\bG(\tau)\bV_1}+\epsilon\bk(\bV_0)
\end{eqnarray*}
\\
where $\bG(\tau)=\left.\pderiv{\bg}{\bV}\right|_{\bV=\bV_0}$.

Splitting out the unperturbed and perturbed parts, we obtain:

\begin{eqnarray*}
\label{eqn:unpertsing}
\mbox{Unperturbed}\quad\deriv{\bV_0}{\tau} &=& \bg(\bV_0)\\
\label{eqn:expandsing}
\mbox{Perturbed}\quad\dot{\theta}\deriv{\bV_0}{\tau}+\epsilon(1+\dot{\theta})\deriv{\bV_1}{\tau} &=& \epsilon\bG(\tau)\bV_1+\epsilon \bk(\bV_0)
\end{eqnarray*}

This therefore means that $\dot{\theta}$ must be of the order of epsilon, $\dot{\theta}=O(\epsilon)$. \chg[p91gram]{Let $\dot{\theta}$ take the following form:}

\begin{equation}
\label{eqn:dotth}
\dot{\theta} = \epsilon A
\end{equation}

Before moving on to find an expression for $A$, we note that:
\chg[p91eqn2]{}
\begin{eqnarray*}
 \deriv{\theta}{t} &=& \deriv{\theta}{\tau}\deriv{\tau}{t}\\
 \deriv{\theta}{t} &=& \deriv{\theta}{\tau}(1+\dot{\theta})\\
 \deriv{\theta}{t} &=& \deriv{\theta}{\tau}\frac{1}{\chg[]{1-\deriv{\theta}{\tau}}}\\
 \deriv{\theta}{t} &=& \deriv{\theta}{\tau}+O(\epsilon^2)
\end{eqnarray*}

Now, on substituting (\ref{eqn:dotth}) into (\ref{eqn:expandsing}) and bearing in mind that $\epsilon\dot{\theta}\bV_1'=O(\epsilon^2)$, we get:

\begin{eqnarray*}
A\deriv{\bV_0}{\tau}+\deriv{\bV_1}{\tau} &=& \bG(\tau)\bV_1+\bk(\bV_0)+O(\epsilon)
\end{eqnarray*}

If $\bS=\bk(\bV_0)-A\bV_0'$, then we get:

\begin{equation*}
\label{eqn:v1}
\deriv{\bV_1}{\tau} = \bG(\tau)\bV_1+\bS+O(\epsilon)
\end{equation*}

\begin{ajf}
WRONG IN PRACTICE

From Sec.(\ref{sec:floquet_Solv_Conds}) we see that we have the following two solvability conditions:

\begin{eqnarray}
\label{eqn:solvcond1}
(\bpsi_*,\bV_1)  &=& 0\\
\label{eqn:solvcond2}
(\bpsi_*,\deriv{\bV_1}{\tau}) &=& (\bpsi_*,\bG(\tau)\bV_1)
\end{eqnarray}

So, if we take the scalar product of (\ref{eqn:v1}) with $\bpsi_*$, then we have:

\begin{eqnarray}
(\bpsi_*,\deriv{\bV_1}{\tau}) &=& (\bpsi_*,\bG(\tau)\bV_1)+(\bpsi_*,\bS)+O(\epsilon)\\
0              		      &=& (\bpsi_*,\bS)+O(\epsilon)
\end{eqnarray}
\end{ajf}

\begin{ajfthesis}
Consider this equation and premultiply it with the Floquet \chg[af]{response} functions corresponding to the unit \chg[af]{multiplier:}

\begin{eqnarray}
 \left(\bpsi_j,\deriv{\bV_1}{\tau}\right) &=& \left(\bpsi_j,\bG(\tau)\bV_1\right)+\left(\bpsi_j,\bS\right)\nonumber\\
-\left(\dot{\bpsi}_j,\bV_1\right)         &=& \left(\bpsi_j,\bG(\tau)\bV_1\right)+\left(\bpsi_j,\bS\right)\nonumber\\
 \left(\bG^+\bpsi_j,\bV_1\right) 					&=& \left(\bpsi_j,\bG(\tau)\bV_1\right)+\left(\bpsi_j,\bS\right)\nonumber\\
 \left(\bpsi_j,\bG(\tau)\bV_1\right) 			&=& \left(\bpsi_j,\bG(\tau)\bV_1\right)+\left(\bpsi_j,\bS\right)\nonumber\\
 \label{eqn:S_cond}
 \left(\bpsi_j,\bS\right) 								&=& 0
\end{eqnarray}
\\
for $j=*,0,\pm1$.
\end{ajfthesis}

We now have the following relation:

\begin{eqnarray*}
(\bpsi_*,\bS)                                      &=& 0\nonumber\\
\Rightarrow (\bpsi_*,\bk(\bV_0)-A\bV_0')           &=& 0\nonumber\\
\Rightarrow (\bpsi_*,\bk(\bV_0))-(\bpsi_*,A\bV_0') &=& 0\nonumber\\
\Rightarrow (\bpsi_*,\bk(\bV_0))                   &=& (\bpsi_*,A\bV_0')\nonumber\\
\label{eqn:A}
\Rightarrow A(\bpsi_*,\bV_0')                      &=& (\bpsi_*,\bk(\bV_0))\\
\Rightarrow A(\bpsi_*,\bphi_*)                     &=& (\bpsi_*,\bk(\bV_0))\\
\Rightarrow A                                      &=& (\bpsi_*,\bk(\bV_0))
\end{eqnarray*}

So the full closed system of equations which governs the dynamics of the perturbed limit cycle is:

\begin{eqnarray}
\deriv{\bV_0}{\tau}  &=& \bg(\bV_0)\nonumber\\
\label{eqn:V1_sing}
\deriv{\bV_1}{\tau}  &=& \bG(\tau)\bV_1+\bS+O(\epsilon)\\
\label{eqn:lcs_thdot}
\deriv{\theta}{\tau} &=& \epsilon(\bpsi_*,\bk(\bV_0))\nonumber
\end{eqnarray}
\\
where $\bS=\bk(\bV_0)-(\bpsi_*,\bk)\bV_0'$. Now, we note that Eqn.(\ref{eqn:V1_sing}) is equivalent to Eqn.(\ref{eqn:odevt2}), and therefore we will perform a similar exercise to see if $\bV_1$ is now bounded.

Firstly, we note that the solution to (\ref{eqn:V1_sing}) can be split out as the sum of a complementary function and a particular integral. 
\chg[p92eqn]{
\begin{equation*}
\bV_1 = \bV_{1,CF}+\bV_{1,PI}
\end{equation*}}

The complementary function is:

\begin{eqnarray*}
\bV_{1,CF}  &=& \bQ(\tau)\bV_1(0)
\end{eqnarray*}
\\
and the particular integral turns out to be:
\chg[p93eqn1]{}
\begin{equation*}
\label{eqn:PI}
\chg[]{\bV_{1,PI}} = \bQ(\tau)\int_0^\tau{\bQ^{-1}(t)\bS(t)}\dd t
\end{equation*}

Therefore, the full solution is:

\begin{equation}
\label{eqn:full_reg1}
\bV(\tau) = \bV_0(\tau)+\epsilon\bQ(\tau)\bV_1(0)+\epsilon\bQ(\tau)\int_0^\tau{\bQ^{-1}(t)\bS(t)}\dd t
\end{equation}

We now show that, in general, the solution shown in (\ref{eqn:full_reg1}) is bounded. Consider the first order of $\epsilon$ part of (\ref{eqn:full_reg1}) and let us assume that the vector function $\bS(t)$ can be expanded in its eigenbasis as:

\begin{equation*}
\bS(t) = \sum_is_i(t)\bphi_i(t)
\end{equation*}

It follows that:
\chg[p93eqn2]{}
\begin{eqnarray*}
\bS(\tau) &=& \sum_is_i(\tau)\bphi_i(\tau)\nonumber\\
\Rightarrow (\bpsi_i,\bS(\tau)) &=& s_i(\tau)\nonumber\\
\Rightarrow s_i(\tau) &=& (\bpsi_i,\bk-(\bpsi_*,\bk)\bV_0')\nonumber\\
\Rightarrow s_i(\tau) &=& (\bpsi_i,\bk)-(\bpsi_*,\bk)(\bpsi_i,\bV_0')\nonumber\\
\Rightarrow s_i(\tau) &=& (\bpsi_i,\bk)-(\bpsi_*,\bk)\chg[]{(\bpsi_i,\bphi_*)}\nonumber\\
\Rightarrow s_i(\tau) &=& (\bpsi_i,\bk)-(\bpsi_*,\bk)\chg[]{\delta_{i,*}}
\end{eqnarray*}

Using a technique similar to that used in Sec.(\ref{sec:theory_reg}), we find that:

\begin{eqnarray}
\label{eqn:bV1}
\bV_1(\tau) &=& \bQ(\tau)\bV_1(0)+\sum_i\bp_i(\tau)\int_0^\tau{s_i(t)e^{-\rho_i(t-\tau)}}\dd t
\end{eqnarray}

Consider the right hand side of Eqn.(\ref{eqn:bV1}). We know that $\bQ(\tau)$ is periodic and hence bounded. Therefore, since $\bV_1(0)$ is a constant vector, then $\bQ(\tau)\bV_1(0)$ is a vector whose components are all bounded. The second term on the right hand side is also bounded which is seem by taking the absolute value of the integral and using the \emph{Absolute Value Integral Inequality}:

\begin{eqnarray*}
\left|\sum_i\bp_i(\tau)\int_0^\tau{s_i(t)e^{-\rho_i(t-\tau)}}\dd t\right| &\leq& \sum_i\left|\bp_i(\tau)\right|\left|\int_0^\tau{s_i(t)e^{-\rho_i(t-\tau)}}\dd t\right|\nonumber\\
 &\leq& \sum_i\left|\bp_i(\tau)\right|\int_0^\tau\left|{s_i(t)e^{-\rho_i(t-\tau)}}\right|\dd t\nonumber\\
 &\leq& \sum_i\left|\bp_i(\tau)\right|\int_0^\tau{\left|s_i(t)\right|e^{-\rho_i(t-\tau)}}\dd t
\end{eqnarray*}

Now, we know that $\bp_i(\tau)$ are periodic and hence bounded by, say, $p_i$, and also if we assume that the perturbation $\bk$ is bounded then we also have that $s_i(\tau)$ are also bounded:

\begin{equation*}
 ||\bk(\tau)|| \leq K \quad\Rightarrow\quad |s_i(\tau)|<c_i
\end{equation*}
\\
\chg[p94eqn]{for some $K$ and $c_i$.} Therefore, we have:

\begin{eqnarray*}
\left|\sum_i\bp_i(\tau)\int_0^\tau{s_i(t)e^{-\rho_i(t-\tau)}}\dd t\right| &<& \sum_ip_i\int_0^\tau{c_ie^{-\rho_i(t-\tau)}}\dd t\nonumber\\
&<& \sum_ip_ic_i\int_0^\tau{e^{-\rho_i(t-\tau)}}\dd t\nonumber\\
&<& \sum_ip_ic_i\left[-\frac{1}{\rho_i}{e^{-\rho_i(t-\tau)}}\right]_0^\tau\nonumber\\
&<& \sum_i\frac{p_ic_i}{\rho_i}\left(1-e^{\rho_i\tau}\right)
\end{eqnarray*}

We note that $\rho_i\leq0$. Therefore,\chg[af]{}we have:

\begin{eqnarray*}
\left|\sum_i\bp_i(\tau)\int_0^\tau{s_i(t)e^{-\rho_i(t-\tau)}}\dd t\right| 
&<& \sum_i\frac{p_ic_i}{\rho_i}
\end{eqnarray*}
\\
which shows that provided the sum on the right hand side above converges, we have that $\bV_1$ is bounded.

\section{Application to the Quotient System}
\label{sec:drift_mrw}
We will now apply the theory reviewed and developed in Sec.(\ref{sec:floquet}) to our own particular problem, which is the study of meandering spiral waves that are the subject of symmetry breaking perturbations. We know from \cite{Bark94a,bik96,wulff96,Golub97} that in a suitable functional \chg[p94spell]{space,} we have limit cycle solutions. Therefore, we will now apply Floquet Theory techniques to this specific problem.


\subsection{Application to the Quotient System}

In a suitable functional space, which we can think of as the space of group orbits, we saw that we have the following equations, the solutions to which are spiral waves in a frame of reference comoving with the tip of the spiral wave.

\begin{eqnarray*}
\label{eqn:app_original1}
\odbVt &=& \boF+(\bc,\Bh_\br)\bV+\omega\Bh_\theta\bV+\epsilon\bHt
\end{eqnarray*}
\begin{eqnarray}
\label{eqn:app_original_cond1}
V^{(1)}(t) &=& u_*\\
V^{(2)}(t) &=& v_*\\
\label{eqn:app_original_cond3}
\hat{\partial}_xV^{(1)} &=& 0
\end{eqnarray}

We note that we can reformulate this system of equations by resolving (\ref{eqn:app_original_cond1})-(\ref{eqn:app_original_cond3}) with respect to \chg[p95spell]{$\bc$} and $\omega$, so that $\bc$ and $\omega$ depend on $\bV$:

\begin{eqnarray}
\label{eqn:app_original2}
\odbVt &=& \boF+(\bc(\bV),\Bh_\br)\bV+\omega(\bV)\Bh_\theta\bV+\epsilon\bHt
\end{eqnarray}
\\
We will use (\ref{eqn:app_original2}) for the remainder of this theory.

In Sec.(\ref{sec:floquet}) we took a general equation of the form:

\begin{equation*}
\label{eqn:app_banach}
\odbVt = \bg(\bV)+\epsilon\bk(\bV,t)
\end{equation*}
\\
and we saw that we needed to consider shifts in time and a new time variable was introduced:

\begin{equation*}
\tau = t+\theta(t)
\end{equation*}
\\
which can be written as:

\begin{equation*}
t = \tau-\theta(\tau)
\end{equation*}

We also note that we can expand our solutions in powers of $\epsilon$ as follows:

\begin{eqnarray*}
\bV(\tau)    &=& \bV_0(\tau)+\epsilon\bV_1(\tau)+O(\epsilon)\\
\bc(\tau)    &=& \bc_0(\tau)+\epsilon\bc_1(\tau)+O(\epsilon)\\
\omega(\tau) &=& \omega_0(\tau)+\epsilon\omega_1(\tau)+O(\epsilon)
\end{eqnarray*}
\\
where $\bc_0$, $\bc_1$, $\omega_0$ and $\omega_1$ are given by:

\begin{eqnarray*}
\bc_0(\tau)    &=& \bc(\bV_0)\\
\bc_1(\tau)    &=& \deriv{\bc(\bV_0)}{\bV}\bV_1\\
\omega_0(\tau) &=& \omega(\bV_0)\\
\omega_1(\tau) &=& \deriv{\omega(\bV_0)}{\bV}\bV_1
\end{eqnarray*}

We therefore find that we get the following system of equations:

\begin{eqnarray*}
\label{eqn:app_V0_deriv}
\deriv{\bV_0}{\tau}  &=& \bg(\bV_0)\\
\label{eqn:app_V1_deriv}
\deriv{\bV_1}{\tau}  &=& L\bV_1+\bS+O(\epsilon)\\
\label{eqn:app_theta_deriv}
\deriv{\theta}{\tau} &=& \epsilon(\bpsi_*,\bk)
\end{eqnarray*}
\\
where \chg[p96eqn]{}
\begin{eqnarray}
\bg(\bV_0) &=& \booF(\bV_0)+(\bc_0,\Bh_\br)\bV_0+\omega_0\Bh_\theta\bV_0\nonumber\\
L\bV_1     &=& \deriv{\booF(\bV_0)}{\bV}+(\bc_0,\Bh_\br)\bV_1+\omega_0\Bh_\theta\bV_1\nonumber\\
\label{eqn:theory_S}
\bS        &=& (\bc_1,\Bh_\br)\bV_0+\omega_1\Bh_\theta\bV_0+\bHt-\chg[]{(\bpsi_0,\til{\bH})}\deriv{\bV_0}{\tau}
\end{eqnarray}

Finally, in order to make $\bV_1$ a unique solution, we require an additional condition which guarantees uniqueness. We choose this to be:

\begin{equation*}
(\bpsi_*,\bV_1) = 0
\end{equation*}

An immediate consequence of this condition is the solvability condition (\ref{eqn:S_cond}).

  
\subsection{Full Equations of Motion}

Let us take the scalar product of (\ref{eqn:theory_S}) with $\bpsi_i$:

\begin{eqnarray}
(\bpsi_i,\bS) &=& \bc_{1x}(\bpsi_i,\hat{\partial}_x\bV_0)+\bc_{1y}(\bpsi_i,\hat{\partial}_y\bV_0)\nonumber\\
\label{eqn:app_scalar}
 && +\omega_1(\bpsi_i,\Bh_\theta\bV_0)+(\bpsi_i,\bHt)-(\bpsi_*,\til{\bH})(\bpsi_i,\deriv{\bV_0}{\tau})
\end{eqnarray}

Using the expressions for the eigenfunctions $\bphi_i$ in the comoving frame of reference, (\ref{eqn:gm_com}), the solvability condition (\ref{eqn:S_cond})and also also Eqn.(\ref{eqn:app_scalar}), we have:
\chg[p97eqn]{}
\begin{eqnarray}
\label{eqn:theory_example_2}
i=0:\quad\omega_1 &=& -(\bpsi_0,\bHt)\\
i=1:\quad\frac{\bar{\bc}_1}{2}-\frac{\omega_1\bar{\bc}_0}{2\omega_0} &=& -(\bpsi_1,\bHt)\nonumber\\
i=-1:\quad\frac{\bc_1}{2}-\frac{\omega_1\bc_0}{2\omega_0} &=& -(\bpsi_{-1},\bHt)\nonumber\\
\label{eqn:theory_example_3}
\Rightarrow \bc_1 &=& \frac{\omega_1\bc_0}{\omega_0}-2(\bpsi_{-1},\bHt)
\end{eqnarray}

Therefore, our full equations of motion are:
\chgex[3]{
\begin{eqnarray*}
\deriv{R}{t} &=& \left[c_0(t)-\epsilon\frac{c_0(t)}{\omega_0(t)}(\bpsi_0(t+\theta(t)),\bHt)-2\epsilon(\bpsi_{-1}(t+\theta(t)),\bHt)\right]e^{i\Theta}+O(\epsilon^2)\\
\deriv{\Theta}{t} &=& \omega_0(t)-\epsilon(\bpsi_0(t+\theta(t)),\bHt)+O(\epsilon^2)\\
\deriv{\theta}{t} &=& \epsilon(\bpsi_*(t+\theta(t)),\bHt)+O(\epsilon^2)
\end{eqnarray*}
\\
where $\bV=\bV(\tau)=\bV(t+\theta(t))$, and $\bpsi_j=\bpsi_j(\tau)=\bpsi_j(t+\theta(t))$.

We note that if the amplitude of meander vanishes, then these equations are similar to the equations for the rigidly rotating spiral wave without meander. This is as expected. However, the difference in this instance is that the velocities $c_0$ and $\omega_0$, and also the response functions, are no longer constant but are dependent on time. The theory so far developed in this thesis can equally apply to rigidly rotating spiral waves and also meandering spiral, both of which are subject to drift. 

However, we cannot say the same about the theories developed by Keener and also Biktashev. They were developed specifically for rigidly rotating spiral waves. No periodic solutions were assumed in both those theories and therefore we cannot compare the theory developed in this section to those theories. If we wanted to do a full comparison, then further work will need to be done, which is beyond the scope of this thesis.

Finally, we show in the next section an example of a meandering spiral wave that is subject to resonant drift.
}

\section{\chgex[3]{Drift \& Meander Example: Resonant Drift}}
\label{sec:theory_example}
Let us now consider the equations of motion for a meandering spiral wave which is drifting due to a time dependent perturbation - i.e. resonant drift. As stated in Sec.(\ref{sec:theory_freq_lock}), our perturbation will take the form:

\begin{equation*}
\label{eqn:theory_example_1}
\bH = \bA\cos(\Omega t+\phi_r)
\end{equation*}
\\
where $\bA$ is a where $\bA$ is an $n$-dimensional column vector. Also, since this is a time dependent perturbation, then we also have that $\til{\bH}=\bH$.

We recall that the equations of motion for the meandering and drifting spiral wave are:

\begin{eqnarray*}
\deriv{R}{t} &=& \left[c_0(t)-\epsilon\frac{c_0(t)}{\omega_0(t)}(\bpsi_0(t+\theta(t)),\bHt)-2\epsilon(\bpsi_{-1}(t+\theta(t)),\bHt)\right]e^{i\Theta}+O(\epsilon^2)\\
\deriv{\Theta}{t} &=& \omega_0(t)-\epsilon(\bpsi_0(t+\theta(t)),\bHt)+O(\epsilon^2)\\
\deriv{\theta}{t} &=& \epsilon(\bpsi_*(t+\theta(t)),\bHt)+O(\epsilon^2)
\end{eqnarray*}
\\
with $R,c_0\in\mathbb{C}$. Let us for a moment consider $c_0$ and $\omega_0$, which are now time dependent. We will take advantage of the fact that a Hopf bifurcation has occurred in the transition from rigid rotation to meander. Therefore, from Hopf bifurcation theory, we can express $c$ and $\omega$ as:

\begin{eqnarray}
c_0     &=& \bc_*+\bc_{\til{1}}z+\bar{\bc}_{\til{1}}\bar{z}+O(|z|^2)\nonumber\\
\omega_0  &=& \omega_{*}+\omega_{\til{1}}z+\bar{\omega}_{\til{1}}\bar{z}+O(|z|^2)\nonumber\\
\label{eqn:app_hopf_normal}
\dot{z} &=& \alpha z+\beta z|z|^2+O(|z|^2)
\end{eqnarray}
\\
where $z$ is a limit cycle solution satisfying (\ref{eqn:app_hopf_normal}), which is the Hopf Normal Form, and also we assume that we are close to the Hopf Bifurcation. If we take the limit cycle $z$ to have the form:

\begin{equation*}
z = \epsilon e^{i(\nu t+\eta)}
\end{equation*}
\\
then the forms for $c_0$ and $\omega_0$ become:

\begin{eqnarray}
\label{eqn:app_hopf_c}
c_0     &=& c_*+2\epsilon|c_{\til{1}}|\cos(\nu t+\xi)+O(\epsilon^2)\\
\label{eqn:app_hopf_omega}
\omega_0  &=& \omega_{*}+2\epsilon|\omega_{\til{1}}\cos(\nu t+\zeta)+O(\epsilon^2)+O(|z|^2)
\end{eqnarray}
\\
where $\xi=\eta+{\rm arg}\{c_{\til{1}}\}$ and $\zeta=\eta+{\rm arg}\{\omega_{\til{1}}\}$.

Next, let us consider the scalar products $(\bpsi_i,\bHt)$. We see that:

\begin{eqnarray*}
(\bpsi_i,\bHt) &=& (\bpsi_i,\bA\cos(\Omega t+\phi_r))\nonumber\\
\Rightarrow(\bpsi_i,\bHt) &=& \cos(\Omega t+\phi_r)(\bpsi_i,\bA)\nonumber\\
\Rightarrow(\bpsi_i,\bHt) &=& \alpha_i(\tau)\cos(\Omega t+\phi_r)
\end{eqnarray*}
\\
where $\alpha_i(\tau)=(\bpsi_i,\bA)$. We note also that $\alpha_i$ is dependent on $\tau$, not $t$. This is due to $\bpsi_i$ being functions of $\tau$ and not $t$, which is stated in Sec.(\ref{sec:floquet}). We see that since we consider Hopf bifurcations, then we can assume that $\alpha_i(\tau)$ takes the form:

\begin{equation*}
\alpha_i(\tau) = \alpha_i\cos(\nu\tau+\phi_r)\\
\Rightarrow \alpha_i(t) = \alpha_i\cos(\nu(t+\theta(t))+\phi_r)
\end{equation*}
\\
where $\nu$ is the hopf frequency and the $\alpha_i$ are constants. Therefore:

\begin{eqnarray*}
(\bpsi_i,\bHt) &=& \alpha_i\cos(\nu(t+\theta(t))+\phi_r)\cos(\Omega t+\phi_r)
\end{eqnarray*}
\\
Furthermore, if we let $\theta(t)=\theta_0+\epsilon\beta(t)+O(\epsilon^2)$, then:

\begin{eqnarray*}
(\bpsi_i,\bHt) &=& \alpha_i\cos(\nu t+\phi_*)\cos(\Omega t+\phi_r)
\end{eqnarray*}
\\
where $\phi_*=\nu\theta_0+\phi_r$. So, the equations of motion are now given by:

\begin{eqnarray}
\label{eqn:theory_examples_4}
\deriv{R}{t} &=& c_0e^{i\Theta}-\epsilon \left(\frac{c_0\alpha_0}{\omega_0}+2\alpha_{-1}\right)\cos(\nu t+\phi_*)\cos(\Omega t+\phi_r)e^{i\Theta}\\
\label{eqn:theory_examples_5}
\deriv{\Theta}{t} &=& \omega_0-\epsilon\alpha_0\cos(\nu t+\phi_*)\cos(\Omega t+\phi_r)
\end{eqnarray}

Using Eqns.(\ref{eqn:app_hopf_c})\&(\ref{eqn:app_hopf_omega}), we find that (\ref{eqn:theory_examples_4})\&(\ref{eqn:theory_examples_5}) become:

\begin{eqnarray}
\label{eqn:theory_examples_6}
\deriv{R}{t} &=& (c_*+2\epsilon|c_{\til{1}}|\cos(\nu t+\xi))e^{i\Theta}-\epsilon\hat{\alpha}\cos(\nu t+\phi_*)\cos(\Omega t+\phi_r)e^{i\Theta}\\
\label{eqn:theory_examples_7}
\deriv{\Theta}{t} &=& \omega_*-\epsilon\alpha_0\cos(\nu t+\phi_*)\cos(\Omega t+\phi_r)
\end{eqnarray}
\\
where $\hat{\alpha}=\frac{c_*\alpha_0}{\omega_*}+2\alpha_{-1}$.

Let us consider (\ref{eqn:theory_examples_7}). This can be rewritten as:

\begin{eqnarray*}
\deriv{\Theta}{t} &=& \omega_*-\frac{\epsilon\alpha_0}{2}\cos((\nu+\Omega)t+\phi_*+\phi_r)-\frac{\epsilon\alpha_0}{2}\cos((\nu-\Omega)t+\phi_*-\phi_r)
\end{eqnarray*}

Integration gives:

\begin{eqnarray*}
\Theta(t) &=& \Theta_0+\omega_*t-\frac{\epsilon\alpha_0}{2(\nu+\Omega)}\sin((\nu+\Omega)t+\phi_*+\phi_r)\\
&&-\frac{\epsilon\alpha_0}{2(\nu-\Omega)}\sin((\nu-\Omega)t+\phi_*-\phi_r)+O(\epsilon^2)
\end{eqnarray*}

Now consider (\ref{eqn:theory_examples_6}). This can be rewritten as:

\begin{footnotesize}
\begin{eqnarray*}
\deriv{R}{t} &=& c_*e^{i(\Theta_0+\omega_*t)}\\
&&+\frac{\epsilon c_*\alpha_0}{2}\left(\frac{1}{\nu+\Omega}\sin((\nu+\Omega)t+\phi_*+\phi_r)-\frac{1}{\nu-\Omega}\sin((\nu-\Omega)t+\phi_*-\phi_r)\right)e^{i(\Theta_0+\omega_*t)}\\
&& +2\epsilon|c_{\til{1}}|\cos(\nu t+\xi))e^{i(\Theta_0+\omega_*t)}-\epsilon\hat{\alpha}\cos(\nu t+\phi_*)\cos(\Omega t+\phi_r)e^{i(\Theta_0+\omega_*t)}\\
\deriv{R}{t} &=& c_*e^{i(\Theta_0+\omega_*t)}+2\epsilon|c_{\til{1}}|\cos(\nu t+\xi))e^{i(\Theta_0+\omega_*t)}\\
&&+\frac{\epsilon c_*\alpha_0}{2}\left(\frac{1}{\nu+\Omega}\sin((\nu+\Omega)t+\phi_*+\phi_r)-\frac{1}{\nu-\Omega}\sin((\nu-\Omega)t+\phi_*-\phi_r)\right)e^{i(\Theta_0+\omega_*t)}\\
&& -\frac{\epsilon\hat{\alpha}}{2}\left(\cos((\nu+\Omega)t+\phi_*+\phi_r)+\cos((\nu-\Omega)t+\phi_*-\phi_r)\right)e^{i(\Theta_0+\omega_*t)}
\end{eqnarray*}
\end{footnotesize}

After some algebra, we get:

\begin{eqnarray*}
\deriv{R}{t} &=& c_*e^{i(\Theta_0+\omega_*t)}+\epsilon|c_{\til{1}}|e^{i((\omega_*+\nu)t+\Theta_0+\xi)}+\epsilon|c_{\til{1}}|e^{i((\omega_*-\nu)t+\Theta_0-\xi)}\\
&& -\til{a}_1e^{i((\omega_*+\nu+\Omega)t+\Theta_0+\phi_*+\phi_r)}-\til{a}_2e^{i((\omega_*-\nu-\Omega)t+\Theta_0-\phi_*-\phi_r)}\\
&& -\til{b}_1e^{i((\omega_*-\nu+\Omega)t+\Theta_0-\phi_*+\phi_r)}-\til{b}_2e^{i((\omega_*+\nu-\Omega)t+\Theta_0+\phi_*-\phi_r)}
\end{eqnarray*}
\\
where we have defined:

\begin{eqnarray*}
\til{a}_1 &=& \frac{\epsilon}{4}\left(\hat{\alpha}+\frac{ic_*\alpha_0}{\nu+\Omega}\right)\\
\til{a}_2 &=& \frac{\epsilon}{4}\left(\hat{\alpha}-\frac{ic_*\alpha_0}{\nu+\Omega}\right)\\
\til{b}_1 &=& \frac{\epsilon}{4}\left(\hat{\alpha}+\frac{ic_*\alpha_0}{\nu-\Omega}\right)\\
\til{b}_2 &=& \frac{\epsilon}{4}\left(\hat{\alpha}-\frac{ic_*\alpha_0}{\nu-\Omega}\right)
\end{eqnarray*}

Integrating with the initial condition $R(0)=R_0$, we get:

\begin{footnotesize}
\begin{eqnarray}
R &=& R_0-\frac{ic_*}{\omega_*}e^{i(\Theta_0+\omega_*t)}-\epsilon\frac{i|c_{\til{1}}|}{\omega_*+\nu}e^{i((\omega_*+\nu)t+\Theta_0+\xi)}-\epsilon\frac{i|c_{\til{1}}|}{\omega_*-\nu}e^{i((\omega_*-\nu)t+\Theta_0-\xi)}\nonumber\\
&& +\frac{i\til{a}_1}{\omega_*+\nu+\Omega}e^{i((\omega_*+\nu+\Omega)t+\Theta_0+\phi_*+\phi_r)}+\frac{i\til{a}_2}{\omega_*-\nu-\Omega}e^{i((\omega_*-\nu-\Omega)t+\Theta_0-\phi_*-\phi_r)}\nonumber\\
\label{eqn:theory_examples_10}
&& +\frac{i\til{b}_1}{\omega_*-\nu+\Omega}e^{i((\omega_*-\nu+\Omega)t+\Theta_0-\phi_*+\phi_r)}+\frac{i\til{b}_2}{\omega_*+\nu-\Omega}e^{i((\omega_*+\nu-\Omega)t+\Theta_0+\phi_*-\phi_r)}
\end{eqnarray}
\end{footnotesize}

Let us consider each term on the right hand side. Firstly, we have the initial position vector, $R_0$. The next term determines the ``core'' trajectory. By that, we mean the trajectory of the underlying, unperturbed spiral. We can see that if we set $\epsilon=0$, then we would get the trajectory of the rigidly rotating spiral, which is of course a perfect circle with radius $\frac{c_*}{\omega_*}$. The next two terms actually determine the ``petals'' of the trajectory. These can be rewritten as follows:

\begin{eqnarray*}
T_{petals} &=& -\epsilon\frac{i|c_{\til{1}}|}{\omega_*+\nu}e^{i((\omega_*t+\nu)t+\Theta_0+\xi)}-\epsilon\frac{i|c_{\til{1}}|}{\omega_*-\nu}e^{i((\omega_*t-\nu)t+\Theta_0-\xi)}\\
&=& -ae^{i(\omega_*t+\Theta_0)}e^{i(\nu t+\xi)}-be^{i(\omega_*t+\Theta_0)}e^{-i(\nu t+\xi)}
\end{eqnarray*}
\\
where

\begin{eqnarray*}
a &=& \epsilon\frac{i|c_{\til{1}}|}{\omega_*+\nu}\\
b &=& \epsilon\frac{i|c_{\til{1}}|}{\omega_*-\nu}
\end{eqnarray*}

If we split out the exponentials into their complex equivalents, and gather the same trigonometric functions, we find that we can write the equations of the trajectory of the petals, in matrix terms, as:

\begin{eqnarray*}
T_{petals} &=& -\left(\begin{array}{cc}\cos(\omega_*t+\Theta_0) & -\sin(\omega_*t+\Theta_0)\\\sin(\omega_*t+\Theta_0) & \cos(\omega_*t+\Theta_0)\end{array}\right)\left(\begin{array}{c} -m\sin(\nu t+\xi)\\n\cos(\nu t+\xi)\end{array}\right)
\end{eqnarray*}
\\
where

\begin{eqnarray*}
n &=&  \epsilon|c_{\til{1}}|\left(\frac{1}{\omega_*+\nu}+\frac{1}{\omega_*-\nu}\right)\\
m &=&  \epsilon|c_{\til{1}}|\left(\frac{1}{\omega_*+\nu}-\frac{1}{\omega_*-\nu}\right)
\end{eqnarray*}

The other four terms all describe the trajectory with respect to the perturbation, i.e. the drift. Using the technique we used for the petal trajectories we get that the trajectory is:

\begin{footnotesize}
\begin{eqnarray*}
T_{drift} &=& \left(\begin{array}{cc}\cos((\omega_*+\nu)t+\Theta_0+\phi_*) & -\sin((\omega_*+\nu)t+\Theta_0+\phi_*)\\ \sin((\omega_*+\nu)t+\Theta_0+\phi_*) & \cos((\omega_*+\nu)t+\Theta_0+\phi_*)\end{array}\right)\left(\begin{array}{c} -p_1\sin(\Omega t+\phi_*)\\p_2\cos(\Omega t+\phi_*)\end{array}\right)\\
&& + \left(\begin{array}{cc}\cos((\omega_*-\nu)t+\Theta_0-\phi_*) & -\sin((\omega_*-\nu)t+\Theta_0-\phi_*)\\ \sin((\omega_*-\nu)t+\Theta_0-\phi_*) & \cos((\omega_*-\nu)t+\Theta_0-\phi_*)\end{array}\right)\left(\begin{array}{c} -p_3\sin(\Omega t+\phi_*)\\p_4\cos(\Omega t+\phi_*)\end{array}\right)
\end{eqnarray*}
\end{footnotesize}
\\
where

\begin{eqnarray*}
p_1 &=&  \frac{i\til{a}_1}{\omega_*+\nu+\Omega}-\frac{i\til{b}_2}{\omega_*+\nu-\Omega}\\
p_2 &=&  \frac{i\til{a}_1}{\omega_*+\nu+\Omega}+\frac{i\til{b}_2}{\omega_*+\nu-\Omega}\\
p_3 &=&  \frac{i\til{b}_1}{\omega_*-\nu+\Omega}-\frac{i\til{a}_2}{\omega_*-\nu-\Omega}\\
p_4 &=&  \frac{i\til{b}_1}{\omega_*-\nu+\Omega}+\frac{i\til{a}_2}{\omega_*-\nu-\Omega}
\end{eqnarray*}

The final overall trajectory, is given by:
\begin{footnotesize}
\begin{eqnarray*}
R &=& R_0-\frac{ic_*}{\omega_*}e^{i(\Theta_0+\omega_*t)}\\
&& -\left(\begin{array}{cc}\cos(\omega_*t+\Theta_0) & -\sin(\omega_*t+\Theta_0)\\\sin(\omega_*t+\Theta_0) & \cos(\omega_*t+\Theta_0)\end{array}\right)\left(\begin{array}{c} -m\sin(\nu t+\xi)\\n\cos(\nu t+\xi)\end{array}\right)\\
&& +\left(\begin{array}{cc}\cos((\omega_*+\nu)t+\Theta_0+\phi_*) & -\sin((\omega_*+\nu)t+\Theta_0+\phi_*)\\ \sin((\omega_*+\nu)t+\Theta_0+\phi_*) & \cos((\omega_*+\nu)t+\Theta_0+\phi_*)\end{array}\right)\left(\begin{array}{c} -p_1\sin(\Omega t+\phi_*)\\p_2\cos(\Omega t+\phi_*)\end{array}\right)\\
&& + \left(\begin{array}{cc}\cos((\omega_*-\nu)t+\Theta_0-\phi_*) & -\sin((\omega_*-\nu)t+\Theta_0-\phi_*)\\ \sin((\omega_*-\nu)t+\Theta_0-\phi_*) & \cos((\omega_*-\nu)t+\Theta_0-\phi_*)\end{array}\right)\left(\begin{array}{c} -p_3\sin(\Omega t+\phi_*)\\p_4\cos(\Omega t+\phi_*)\end{array}\right)
\end{eqnarray*}
\end{footnotesize}
It all looks a little complicated, and can in fact be viewed as a type of chaotic motion. However, we can see what happens better if we consider when the frequencies are in resonance. The problem here is that we have three different frequencies and therefore the question is which of them shall we have in resonance. Let us consider the case when $\Omega=\omega_*+\nu$. We can see, from Eqn.(\ref{eqn:theory_examples_10}), that the speed of the drift will be given by:

\begin{eqnarray*}
S &=& \left|\til{b}_2e^{i(\Theta_0+\phi_*-\phi_r)}\right|\\
\Rightarrow S &=& \left|\frac{\epsilon}{4}\left(\hat{\alpha}-\frac{ic_*\alpha_0}{\nu-\Omega}\right)\right|
\end{eqnarray*}

However, the drift is obviously not as straightforward for meandering spirals, compared to the rigidly rotating spirals. The motion is quite chaotic and will be a very interesting area of research in the future. We can see that this cannot be directly compared to the drift equations obtained for rigidly rotating.

\section{Frequency Locking}
\label{sec:theory_freq_lock}
In this section, we consider resonant drift, and shall consider whether we can detect in particular frequency locking using the theory developed so far together with the techniques from Arnol'd \cite{arnold65}.

\subsection{The Arnol'd Standard Mapping for Resonant Drift}
\label{sec:theory_lock_arnold}

We now introduce a particular form of the symmetry breaking perturbation, which we choose to be Resonant Drift:

\begin{equation}
\label{eqn:lcs_pert}
\bH = \bA\cos(\Omega t+\phi_r)
\end{equation}
\\
where $\bA$ is an $n$-dimensional column vector, $\Omega$ is the frequency of the perturbation, and $\phi_r$ is the phase of the perturbation. Also, since this is a time dependent perturbation, then we also have that $\til{\bH}=\bH$. 

We will first of all look at a method to detect whether a meandering and drifting spiral wave is exhibiting frequency locking. Consider Eqn.(\ref{eqn:lcs_pert}):

\begin{ajf}
\begin{eqnarray}
\deriv{\theta}{\tau} 	         &=& \epsilon(\bpsi_*(\tau),\bk(t))+O(\epsilon^2)\nonumber\\
\Rightarrow \deriv{\theta}{\tau} &=& \epsilon(\bpsi_*(\tau),\bA\cos(\Omega t+\phi_r))+O(\epsilon^2)\nonumber\\
\Rightarrow \deriv{\theta}{\tau} &=& \epsilon\cos(\Omega t+\phi_r)(\bpsi_*(\tau),\bA)+O(\epsilon^2)\nonumber\\
\label{eqn:app_lock_dthdt}
\Rightarrow \deriv{\theta}{\tau} &=& \epsilon\alpha(\tau)\cos(\Omega t+\phi_r)+O(\epsilon^2)
\end{eqnarray}
\end{ajf}
\begin{ajfthesis}
\begin{eqnarray}
\deriv{\theta}{\tau} 	         &=& \epsilon(\bpsi_*(\tau),\bk(t))+O(\epsilon^2)\nonumber\\
\label{eqn:app_lock_dthdt}
\Rightarrow \deriv{\theta}{\tau} &=& \epsilon\alpha(\tau)\cos(\Omega t+\phi_r)+O(\epsilon^2)
\end{eqnarray}
\end{ajfthesis}
\\
where $\alpha(\tau)=(\bpsi_*(\tau),\bA)$.

We now introduce the following variables:

\begin{eqnarray*}
\sigma &=& \frac{\Omega t}{2\pi}\\
\rho   &=& \frac{\nu\tau}{2\pi}
\end{eqnarray*}

We also see that $\sigma$ is related to $\rho$ by introducing a correction term $\xi$:

\begin{equation*}
\sigma = \rho+\xi(\tau)
\end{equation*}

This therefore leads to:
\begin{ajf}
\begin{eqnarray}
\sigma &=& \rho+\xi(\tau)\\
\frac{\Omega t}{2\pi} &=& \frac{\nu\tau}{2\pi}+\xi(\tau)\nonumber\\
\frac{\Omega}{2\pi}(\tau-\theta(\tau)) &=& \frac{\nu\tau}{2\pi}+\xi(\tau)\nonumber\\
\frac{\Omega}{2\pi}\theta(\tau) &=& \frac{\Omega\tau}{2\pi}-\frac{\nu\tau}{2\pi}-\xi(\tau)\nonumber\\
\theta(\tau) &=& \tau-\frac{\nu}{\Omega}\tau-\frac{2\pi\xi(\tau)}{\Omega}\nonumber\\
\theta(\tau) &=& \frac{\Omega-\nu}{\Omega}\tau-\frac{2\pi\xi(\tau)}{\Omega}
\end{eqnarray}

Therefore, Eqn.(\ref{eqn:app_lock_dthdt}), now becomes:

\begin{eqnarray}
\frac{\Omega-\nu}{\Omega}-\frac{2\pi}{\Omega}\deriv{\xi}{\tau} &=& \epsilon\alpha(\tau)\beta(t)+O(\epsilon^2)\nonumber\\
\Rightarrow \deriv{\xi}{\tau} &=& \frac{\Omega-\nu}{2\pi}-\frac{\epsilon\Omega}{2\pi}\alpha(\tau)\beta(t)+O(\epsilon^2)
\end{eqnarray}
\\
where $\beta(t)=\cos(\Omega t+\phi_r)$. We can further arrange this \chg[af]{equation,} by changing variable:

\begin{eqnarray}
\deriv{\xi}{\rho}\deriv{\rho}{\tau} &=& \frac{\Omega-\nu}{2\pi}-\frac{\epsilon\Omega}{2\pi}\alpha(\rho)\beta(\sigma)+O(\epsilon^2)\nonumber\\
\deriv{\xi}{\rho}\frac{\nu}{2\pi}   &=& \frac{\Omega-\nu}{2\pi}-\frac{\epsilon\Omega}{2\pi}\alpha(\rho)\beta(\rho+\xi)+O(\epsilon^2)\nonumber\\
\deriv{\xi}{\rho}		 &=& \left(\frac{\Omega-\nu}{\nu}\right)-\frac{\epsilon\Omega}{\nu}\alpha(\rho)\beta(\rho+\xi)+O(\epsilon^2)\nonumber\\
\label{eqn:app_lock_dxidrho}
\deriv{\xi}{\rho}		 &=& A-K'\alpha(\rho)\beta(\rho+\xi)+O(\epsilon^2)
\end{eqnarray}
\\
where:

\begin{eqnarray}
A  &=& \left(\frac{\Omega-\nu}{\nu}\right)\\
K' &=& \frac{\epsilon\Omega}{\nu}
\end{eqnarray}
\end{ajf}

\begin{ajfthesis}
\begin{eqnarray*}
\theta(\tau) &=& \frac{\Omega-\nu}{\Omega}\tau-\frac{2\pi\xi(\tau)}{\Omega}
\end{eqnarray*}

So, Eqn.(\ref{eqn:app_lock_dthdt}), now becomes:

\begin{eqnarray*}
\deriv{\xi}{\tau} &=& \frac{\Omega-\nu}{2\pi}-\frac{\epsilon\Omega}{2\pi}\alpha(\tau)\beta(t)+O(\epsilon^2)
\end{eqnarray*}
\\
where $\beta(t)=\cos(\Omega t+\phi_r)$. We can further arrange this \chg[af]{equation,} by changing variable:

\begin{eqnarray}
\label{eqn:app_lock_dxidrho}
\deriv{\xi}{\rho}		 &=& A-K'\alpha(\rho)\beta(\rho+\xi)+O(\epsilon^2)
\end{eqnarray}
\\
where:

\begin{eqnarray*}
A  &=& \left(\frac{\Omega-\nu}{\nu}\right)\\
K' &=& \frac{\epsilon\Omega}{\nu}
\end{eqnarray*}
\end{ajfthesis}

We now consider the solutions to Eqn.(\ref{eqn:app_lock_dxidrho}) as an iterated scheme:

\begin{eqnarray*}
\xi^{(0)}(\rho) &=& \xi_0+A\rho\\
\xi^{(n+1)}(\rho) &=& \xi_0+A\rho-K'\int^\rho_0\alpha(\eta)\beta(\eta+\xi^{(n)}(\eta))\dd{\eta}+O(\epsilon^2)
\end{eqnarray*}

We wish to study this system for one full period of the limit cycle, so therefore we have:

\begin{eqnarray*}
\xi^{(0)}(1) &=& \xi_0+A\\
\xi^{(1)}(1) &=& \xi_0+A-K'\int^1_0\alpha(\eta)\beta(\eta+\xi^{(0)}(\eta))\dd{\eta}+O(\epsilon^2)
\end{eqnarray*}

Let, $\xi_1=\xi^{(1)}(1)$:

\begin{eqnarray}
\xi_1 &=& \xi_0+A-K'\int^1_0\alpha(\eta)\beta(\eta+\xi_0+A\eta)\dd{\eta}+O(\epsilon^2)\nonumber\\
\label{eqn:app_lock_sol_period}
\Rightarrow \xi_1 &=& \xi_0+A-K'\int^1_0\alpha(\eta)\beta(\xi_0+(1+A)\eta)\dd{\eta}+O(\epsilon^2)
\end{eqnarray}

Thus, Eqn.(\ref{eqn:app_lock_sol_period}) above is a solution to (\ref{eqn:app_lock_dxidrho}), for the first iteration over the period of the limit cycle.

Let us consider a specific example. We note that several authors have shown numerically that the transition from a rigidly rotating spiral to a meandering spiral wave is via a Hopf Bifurcation. Therefore, if we are near the point at which the Hopf Bifurcation has \chg[af]{occurred,} then the projection, $\alpha(\tau)$, onto the limit cycle can be thought of as harmonic. Hence, we use the following:

\begin{eqnarray*}
\alpha(\tau) &=& \alpha_*\cos(\nu\tau+\phi_c)\\
\beta(t)     &=& \cos(\Omega t+\phi_r)
\end{eqnarray*}
\\
where $\nu$ is the Hopf frequency. This now needs to be rewritten in terms of $\rho$ and $\sigma$:

\begin{eqnarray*}
\alpha(\rho)  &=& \cos(2\pi\rho+\phi_c)\\
\beta(\sigma) &=& \alpha_*\cos(2\pi\sigma+\phi_r)
\end{eqnarray*}
\\
which leads to:

\begin{ajf}
\begin{eqnarray}
\xi_1 &=& \xi_0+A-K'\alpha_*\int^1_0\cos(2\pi\eta+\phi_c)\cos(2\pi(\xi_0+(1+A)\eta)+\phi_r)\dd{\eta}+O(\epsilon^2)\nonumber\\
\Rightarrow \xi_1 &=& \xi_0+A-K'\alpha_*\int^1_0\cos(2\pi\eta+\phi_c)\cos\left(2\pi\xi_0+\frac{2\pi\Omega}{\nu}\eta+\phi_r\right)\dd{\eta}+O(\epsilon^2)\nonumber\\
\Rightarrow \xi_1 &=& \xi_0+A-\frac{K'\alpha_*}{2}\int^1_0\left[\cos\left(2\pi\eta\left(\frac{\Omega+\nu}{\nu}\right)+2\pi\xi_0+\phi_r+\phi_c\right)\right.\nonumber\\
&& \left.+\cos\left(2\pi\eta\left(\frac{\Omega-\nu}{\nu}\right)+2\pi\xi_0+\phi_r-\phi_c\right)\right]\dd{\eta}+O(\epsilon^2)\nonumber\\
\Rightarrow \xi_1 &=& \xi_0+A-\frac{K'\alpha_*}{2}\int^1_0\sum_{\pm}\cos\left(2\pi\eta\left(\frac{\Omega\pm\nu}{\nu}\right)+2\pi\xi_0+\phi_r\pm\phi_c\right)\dd{\eta}+O(\epsilon^2)\nonumber\\
\Rightarrow \xi_1 &=& \xi_0+A-\frac{K'\alpha_*\nu}{2(\Omega^2-\nu^2}\left[\sum_{\pm}(\Omega\mp\nu)\sin\left(2\pi\eta\left(\frac{\Omega\pm\nu}{\nu}\right)+2\pi\xi_0+\phi_r\pm\phi_c\right)\right]^1_0+O(\epsilon^2)\nonumber\\
\Rightarrow \xi_1 &=& \xi_0+A-\frac{K'\alpha_*\nu}{2(\Omega^2-\nu^2}\sum_{\pm}\left[(\Omega\mp\nu)\sin\left(2\pi\left(\frac{\Omega\pm\nu}{\nu}\right)+2\pi\xi_0+\phi_r\pm\phi_c\right)\right.\nonumber\\
&& \left.-(\Omega\mp\nu)\sin\left(2\pi\xi_0+\phi_r\pm\phi_c\right)\right]+O(\epsilon^2)\nonumber\\
\Rightarrow \xi_1 &=&\xi_0+A-\frac{K'\alpha_*\nu}{2(\Omega^2-\nu^2}\sum_{\pm}\left[(\Omega\mp\nu)\sin\left(\frac{2\pi\Omega}{\nu}+2\pi\xi_0+\phi_r\pm\phi_c\right)\right.\nonumber\\
&& \left.-(\Omega\mp\nu)\sin\left(2\pi\xi_0+\phi_r\pm\phi_c\right)\right]+O(\epsilon^2)\nonumber\\
\Rightarrow \xi_1 &=&\xi_0+A-\frac{K'\alpha_*\nu}{2(\Omega^2-\nu^2}\sum_{\pm}(\Omega\mp\nu)\cos\left(\frac{\pi\Omega}{\nu}+2\pi\xi_0+\phi_r\pm\phi_c\right)\sin\left(\frac{\pi\Omega}{\nu}\right)+O(\epsilon^2)\nonumber\\
\Rightarrow \xi_1 &=&\xi_0+A-\frac{K'\alpha_*\nu}{2(\Omega^2-\nu^2}\sin\left(\frac{\pi\Omega}{\nu}\right)\sum_{\pm}(\Omega\mp\nu)\cos\left(\frac{\pi\Omega}{\nu}+2\pi\xi_0+\phi_r\pm\phi_c\right)+O(\epsilon^2)\nonumber
\end{eqnarray}
\end{ajf}
\begin{ajfthesis}
\begin{eqnarray}
\xi_1 &=&\xi_0+A-\frac{K'\alpha_*\nu}{2(\Omega^2-\nu^2}\sin\left(\frac{\pi\Omega}{\nu}\right)\sum_{\pm}(\Omega\mp\nu)\cos\left(\frac{\pi\Omega}{\nu}+2\pi\xi_0+\phi_r\pm\phi_c\right)+O(\epsilon^2)\nonumber
\end{eqnarray}
\end{ajfthesis}

Consider now the sum on the right hand side and denote this by $\mathcal{S}$:

\begin{ajf}
\begin{eqnarray}
\mathcal{S} &=& (\Omega-\nu)\cos\left(\frac{\pi\Omega}{\nu}+2\pi\xi_0+\phi_r+\phi_c\right)+(\Omega+\nu)\cos\left(\frac{\pi\Omega}{\nu}+2\pi\xi_0+\phi_r-\phi_c\right)\nonumber\\
\mathcal{S} &=& (\Omega-\nu)\left(\cos\left(\frac{\pi\Omega}{\nu}+2\pi\xi_0+\phi_r\right)\cos(\phi_c)-\sin\left(\frac{\pi\Omega}{\nu}+2\pi\xi_0+\phi_r\right)\sin(\phi_c)\right)\nonumber\\
&& +(\Omega+\nu)\left(\cos\left(\frac{\pi\Omega}{\nu}+2\pi\xi_0+\phi_r\right)\cos(\phi_c)+\sin\left(\frac{\pi\Omega}{\nu}+2\pi\xi_0+\phi_r\right)\sin(\phi_c)\right)\nonumber\\
\mathcal{S} &=& 2\Omega\cos\left(\frac{\pi\Omega}{\nu}+2\pi\xi_0+\phi_r\right)\cos(\phi_c)+2\nu\sin\left(\frac{\pi\Omega}{\nu}+2\pi\xi_0+\phi_r\right)\sin(\phi_c)
\end{eqnarray}
\end{ajf}

\begin{ajfthesis}
\begin{eqnarray*}
\mathcal{S} &=& 2\Omega\cos\left(\frac{\pi\Omega}{\nu}+2\pi\xi_0+\phi_r\right)\cos(\phi_c)+2\nu\sin\left(\frac{\pi\Omega}{\nu}+2\pi\xi_0+\phi_r\right)\sin(\phi_c)
\end{eqnarray*}
\end{ajfthesis}

Now, let:

\begin{eqnarray*}
\mathcal{C} &=& 2\Omega\cos(\phi_c)\\
\mathcal{B} &=& 2\nu\sin(\phi_c)
\end{eqnarray*}

Hence, we have:

\begin{eqnarray*}
\mathcal{S} &=& \mathcal{C}\cos\left(\frac{\pi\Omega}{\nu}+2\pi\xi_0+\phi_r\right)+\mathcal{B}\sin\left(\frac{\pi\Omega}{\nu}+2\pi\xi_0+\phi_r\right)
\end{eqnarray*}

Next, we let $\mathcal{C}$ and $\mathcal{B}$ be:

\begin{eqnarray*}
\mathcal{C} &=& \sqrt{\mathcal{C}^2+\mathcal{B}^2}\sin(\gamma)\\
\mathcal{B} &=& \sqrt{\mathcal{C}^2+\mathcal{B}^2}\cos(\gamma)
\end{eqnarray*}
\\
which leads to:

\begin{ajf}
\begin{eqnarray}
\mathcal{S} &=& \sqrt{\mathcal{C}^2+\mathcal{B}^2}\left(\sin(\gamma)\cos\left(\frac{\pi\Omega}{\nu}+2\pi\xi_0+\phi_r\right)+\cos(\gamma)\sin\left(\frac{\pi\Omega}{\nu}+2\pi\xi_0+\phi_r\right)\right)\nonumber\\
\mathcal{S} &=& \sqrt{\mathcal{C}^2+\mathcal{B}^2}\sin\left(\frac{\pi\Omega}{\nu}+2\pi\xi_0+\phi_r+\gamma\right)
\end{eqnarray}
\end{ajf}
\begin{ajfthesis}
\begin{eqnarray*}
\mathcal{S} &=& \sqrt{\mathcal{C}^2+\mathcal{B}^2}\sin\left(\frac{\pi\Omega}{\nu}+2\pi\xi_0+\phi_r+\gamma\right)
\end{eqnarray*}
\end{ajfthesis}

Hence, our full solution is now:

\begin{ajf}
\begin{eqnarray}
\xi_1 &=&\xi_0+A-\frac{K'\alpha_*\nu\sqrt{\mathcal{C}^2+\mathcal{B}^2}}{2(\Omega^2-\nu^2}\sin\left(\frac{\pi\Omega}{\nu}\right)\sin\left(\frac{\pi\Omega}{\nu}+2\pi\xi_0+\phi_r+\gamma\right)+O(\epsilon^2)\nonumber\\
\xi_1 &=&\xi_0+A-\frac{K}{2\pi}\sin\left(\frac{\pi\Omega}{\nu}+2\pi\xi_0+\phi_r+\gamma\right)+O(\epsilon^2)
\end{eqnarray}
\\
where:

\begin{eqnarray}
A &=& \frac{\Omega-\nu}{\nu}\\
K &=& \frac{K'\alpha_*\pi\nu\sqrt{\mathcal{C}^2+\mathcal{B}^2}}{\Omega^2-\nu^2}\sin\left(\frac{\pi\Omega}{\nu}\right)\nonumber\\
\Rightarrow K &=& \frac{\epsilon\alpha_*\pi\Omega\sqrt{\mathcal{C}^2+\mathcal{B}^2}}{\Omega^2-\nu^2}\sin\left(\frac{\pi\Omega}{\nu}\right)\\
\sqrt{\mathcal{C}^2+\mathcal{B}^2} &=& 2\sqrt{\Omega^2\cos^2(\phi_c)+\nu^2\sin^2(\phi_c)}
\end{eqnarray}
\end{ajf}
\begin{ajfthesis}
\begin{eqnarray*}
\xi_1 &=&\xi_0+A-\frac{K}{2\pi}\sin\left(\frac{\pi\Omega}{\nu}+2\pi\xi_0+\phi_r+\gamma\right)+O(\epsilon^2)
\end{eqnarray*}
\\
where:

\begin{eqnarray*}
A &=& \frac{\Omega-\nu}{\nu}\\
K &=& \frac{K'\alpha_*\pi\nu\sqrt{\mathcal{C}^2+\mathcal{B}^2}}{\Omega^2-\nu^2}\sin\left(\frac{\pi\Omega}{\nu}\right)\nonumber\\
\Rightarrow K &=& \frac{\epsilon\alpha_*\pi\Omega\sqrt{\mathcal{C}^2+\mathcal{B}^2}}{\Omega^2-\nu^2}\sin\left(\frac{\pi\Omega}{\nu}\right)\\
\sqrt{\mathcal{C}^2+\mathcal{B}^2} &=& 2\sqrt{\Omega^2\cos^2(\phi_c)+\nu^2\sin^2(\phi_c)}
\end{eqnarray*}
\end{ajfthesis}

Now, let:

\begin{equation*}
2\pi\eta_i = 2\pi\xi_i+\frac{\pi\Omega}{\nu}+\phi_r+\gamma
\end{equation*}

This therefore gives us:

\begin{equation*}
\boxed{\eta_1 = \eta_0+A-\frac{K}{2\pi}\sin\left(2\pi\eta_0\right)+O(\epsilon^2)}
\end{equation*}

This is exactly the form for Arnold's Standard Mapping \cite{arnold65}, up to the $O(\epsilon^2)$ terms.


\subsection{Arnol'd's Tongues \& Locking}

Frequency locking occurs when two of the frequencies within the system being studied are rationally related. In our case, we consider the relationship between the Hopf frequency, $\nu$, and the forcing frequency, $\Omega$.

We have seen in the above subsection, that we can derive the Arnol'd Standard Mapping from the correction term to the time variable in our singular perturbation theory. In his breakthrough paper \cite{arnold65}, Arnol'd described how, for mappings in the form of his Standard Mapping, frequency locking in a periodic system subject to external periodic forcing can, for a range of parameters, exhibit frequency locking. Within the parameter space, we can draw different types of tongues which relate to different types of locking, whether they are 1:1 locking or 2:1 locking etc. 

In his book, Wiggins then describes how we can determine the analytical descriptions of the boundaries of the Arnol'd Tongues from the parameters of the Standard Mapping \cite{wiggins}. Therefore, we will use the results of Wiggins to determine the boundaries of these Arnol'd tongues and see if our results from the above subsection can be applied here. In Wiggins' book Sec.21.6, he details the nature of the standard mapping and how to determine the boundaries of the Arnol'd tongues. Several other authors have also determined methods for calculating the boundaries of tongues \cite{hall84,bush73}, but we will utilise the results from Wiggins.

Let us take the 2:1 resonance tongue. From analytical considerations, the boundaries of this tongue are given by:

\begin{equation*}
 A = \frac{1}{2}\pm\frac{K^2}{8\pi}+O(K^3)
\end{equation*}

\begin{figure}[tbp]
\begin{center}
\begin{minipage}{0.7\linewidth}
\centering
\includegraphics[width=0.75\textwidth,angle=-90]{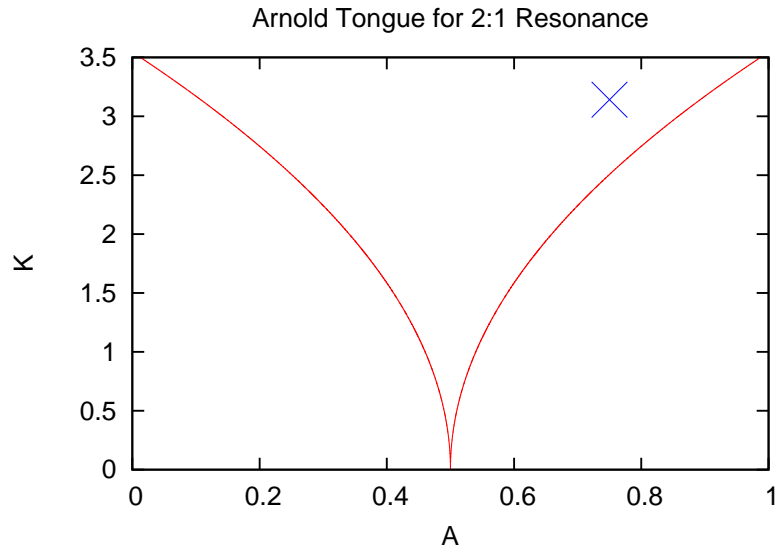}
\end{minipage}
\caption[2:1 Arnol'd Tongue]{2:1 Arnol'd Tongue in the AK-plane. The point indicated by the {\color{blue}{blue cross}} is the point that we consider for frequency locking, having coordinates (0.75,$\pi$).}
\label{fig:theory_freq_2to1}
\end{center}
\end{figure}

We show how this tongue looks like in the $AK$-plane in Fig.(\ref{fig:theory_freq_2to1}).

From the Sec.(\ref{sec:theory_lock_arnold}), we have that $A$ and $K$ take the form:

\begin{eqnarray}
\label{eqn:theory_lock_1}
A &=& \frac{\Omega-\nu}{\nu}\\
\label{eqn:theory_lock_2}
K &=& \frac{2\epsilon\alpha_*\pi\Omega\sqrt{\Omega^2\cos^2(\phi_c)+\nu^2\sin^2(\phi_c)}}{\Omega^2-\nu^2}\sin\left(\frac{\pi\Omega}{\nu}\right)
\end{eqnarray}

Therefore, Eqn.(\ref{eqn:theory_lock_1}) implies that:

\begin{equation}
\label{eqn:theory_lock_3}
 \Omega = (A+1)\nu
\end{equation}

Hence we can obtain an expression for $\epsilon\alpha_*$ using Eqns.(\ref{eqn:theory_lock_2}) and (\ref{eqn:theory_lock_3}) to give:

\begin{equation*}
\label{eqn:theory_lock_4}
\epsilon\alpha_* = \frac{K((A+1)^2-1)}{2\pi(A+1)\sin(\pi(A+1))\sqrt{(A+1)^2\cos^2(\phi_c)+\sin^2(\phi_c)}}
\end{equation*}

Let us consider a non-trivial point within the tongue, and by non-trivial we mean that $A\neq\frac{1}{2}$. We choose the point $(A,K)=(0.75,\pi)$ as \chg[af]{indicated} in Fig.(\ref{fig:theory_freq_2to1}). This implies that, from Eqn.(\ref{eqn:theory_lock_3}) we have:

\begin{equation*}
\label{eqn:theory_lock_5}
\Omega = \frac{7}{4}\nu
\end{equation*}

Also, if we take that $\phi_c=\pi$, then Eqn.(\ref{eqn:theory_lock_4}) gives us:

\begin{eqnarray*}
\label{eqn:theory_lock_6}
\epsilon\alpha_* &=& \frac{K((A+1)^2-1)}{2\pi(A+1)^2\sin(\pi(A+1))}\nonumber\\
\Rightarrow \epsilon\alpha_* &=& \frac{33}{49\sqrt{2}}\nonumber\\
\Rightarrow \epsilon\alpha_* &\approx& -0.476215
\end{eqnarray*}

So, if our singular perturbation is indeed correct, we should find that the Poincare mapping of equation governing the evolution of the correction to the time variable gives us one point. This equation is:
\chg[p103eqn1]{}
\begin{eqnarray}
\deriv{\theta}{\tau} &=& \epsilon\alpha_0\cos(\Omega t+\phi_r)+O(\epsilon^2)\nonumber\\
\label{eqn:theory_lock_7}
\Rightarrow \deriv{\theta}{t} &=& \epsilon\alpha_*\cos(\chg[]{\nu(t+\theta(t))}+\phi_c)\cos(\Omega t+\phi_r)+O(\epsilon^2)
\end{eqnarray}

Numerically, Eqn.(\ref{eqn:theory_lock_7}) is solved using:
\chg[p103eqn2]{}
\begin{eqnarray}
\theta_{n+1} &=& \theta_n+\Delta t\epsilon\alpha_*\cos(\chg[]{\nu(t+\theta_n)}+\phi_c)\cos(\Omega t+\phi_r)
\end{eqnarray}
\\
where $\Delta t$ is the timestep.

\begin{figure}[tbp]
\begin{center}
\begin{minipage}{0.7\linewidth}
\centering
\psfrag{a}[l]{$\theta_{n+1}$}
\psfrag{b}[l]{$\theta_{n}$}
\includegraphics[width=0.75\textwidth,angle=-90]{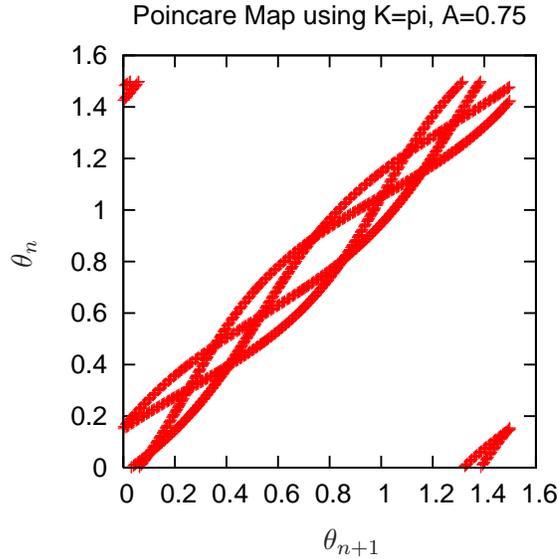}
\end{minipage}
\caption[Poincare mapping]{Poincare map for $\nu=\frac{4\pi}{3}$, $\Omega=\frac{7\pi}{3}$, $\theta_0=0.2$, $\phi_r=0.2$.}
\label{fig:theory_freq_pcm}
\end{center}
\end{figure}

We show in Fig.(\ref{fig:theory_freq_pcm}), the Poincare mapping for this iterative scheme with the parameters as detailed. We can clearly see that we do not get a frequency locked solution. We must therefore conclude that we can not detect frequency locking using the order of accuracy implemented into the above scheme.

\section{Conclusion \& Further Work}
\label{sec:theory_conc}
To conclude this chapter, we have seen that the theory of drift can be rewritten using group theory and perturbation techniques. We have shown that by studying spiral wave solutions in a frame of reference comoving with the tip of the wave, it is possible to extract the exact equations of motion of the tip of the wave, for both rigidly rotating and meandering spiral waves which are drifting due to the presence of symmetry breaking perturbations within the dynamical systems. 

We have also seen that by applying Floquet Theory to the meandering solutions, we are able to not only determine what the equations of motion should be, but also what perturbation techniques we should apply. We have seen that in the case of meandering spiral waves that are drifting, a singular perturbation technique is required to provide us with bounded solutions. 

By considering the correction term to the time variable in the singular perturbation method we employed, we saw this correction term can be transformed into the Arnol'd Standard Mapping. We also showed that a first order approximation is not sufficient to detect frequency locking. We can therefore conclude that at least a second order approximation is required to detect frequency locking here.

There are many open ended questions that have arisen from this work, which require attention in the furture. We feel that the method of frequency locking described in this chapter will be a clean and accurate method which will be applicable to any perturbation. However, more work is required in this direction including expanding the system to include second order terms.
\chapter{Initial Numerical Analysis}
\label{chap:2a}

\section{Introduction}
In this chapter, we shall discuss the initial numerical analysis of the drift and meander of spiral waves. The role of the analysis is to motivate subsequent chapters.

This chapter is concerned mainly with the drift of spiral waves. We shall split it into two main parts: Inhomogeneity Induced Drift; and Electrophoresis Induced Drift. We also note that we will use only Barkley's model in this analysis.

For Inhomogeneity induced drift, we shall test the analytical theory that the velocity of the drift of a rigidly rotating spiral wave should be linearly proportional to the drift parameter \cite{bik95}. We will also investigate whether the generic forms of the equations of motion can be determined numerically for both rigidly rotating spiral waves and meandering spiral waves, and touch on whether we can detect frequency locking.

For Electrophoresis induced drift, we shall conduct some analysis into frequency locking within Barkley's model.

All simulations were conducted using EZ-Spiral, amended accordingly for the purposes of the study.

\label{sec:ini_num_intro}

\section{Inhomogeneity Induced Drift of Spiral Wave}
When a wave drifts due to Inhomogeneities the model parameter(s) is(are) dependent on the spatial coordinates. \chg[p112spell1]{Its} point of rotation is no longer stationary but drifts along a straight line \chg[af]{\cite{bik95}}. Therefore, as an example, we have that \chg[p112spell2]{$a=a(x)$} in Barkley's model. In our analysis, we \chg[af]{took} $a$ to be linearly dependent on $x$ to get:

\begin{equation}
 a = a_0+a_1x
\end{equation}
\\
where $a_1$ is known as the gradient of the drift. A formal analysis of this is done in Chap.(\ref{chap:3}).

We will be using Barkley's model throughout the simulations:

\begin{eqnarray}
\pderiv{u}{t} &=& \nabla^2u+\frac{1}{\varepsilon}u(1-u)\left[u-\frac{v+b}{a_0+a_1x}\right]\\
\pderiv{v}{t} &=& u-v
\end{eqnarray}

We will use EZ-Spiral throughout this analysis \cite{barkweb} and will firstly look at rigidly rotating spiral waves.


\subsection{Initial Analysis}

We decided choose the following values of model parameters:

\begin{eqnarray*}
a_0 & = & 0.7\\
a_1 & = & 0\\
b & = & 0.15\\
\varepsilon & = & 0.01
\end{eqnarray*}
\\
noting that we initially do not have any drift. Other numerical and physical parameters were chosen as follows:

\begin{eqnarray*}
N_x & = & 181\\
L_x & = & 60\\
\mbox{Steps per plot} & = & 16\\
ts & = & 0.8
\end{eqnarray*}
\\
\chgex[ex]{where $N_x$ is the number of grid points used for the numerical grid, $L-x$ is the length of the numerical box used, and $ts$ is timestep as a fraction of the diffusion stability limit. These are all specified in the} \verb|task.dat| \chgex[]{file used with EZ-Spiral.} This means that our space step and time step were $h_x=\frac{1}{3}$ and $h_t=\frac{1}{30}$ respectively. These steps were kept constant throughout our analysis. We also decided to use the Nine Point Laplacian formula for solving the diffusion terms.

First of all we need to show that our amended EZ-Spiral ran correctly, having implemented the changes for Inhomogeneity induced drift. We therefore ran the program for various values of $a_0$ with $a_1=0$. These runs were then compared to the original data using the original program. The results showed that the amended program gave exactly the same results as the original program. Early indications showed that the program was working correctly.

There is an established theory which states that for small gradients the speed of the wave is proportional to the gradient \cite{bik95}. Therefore, in order to test whether EZ-Spiral is certainly working correctly, we are to prove, numerically, this theory.

We ran the program for a range of gradients, $a_1$. As usual, a data file,  \verb|tip.dat|,  was produced which contained the tip coordinates and phase at particular timesteps. We then calculated the speed of the wave using the tip file by plotting $t$ vs $x$ and $t$ vs $y$ using Gnuplot and making Gnuplot include a line of best fit. The gradient of this line would then give us $\frac{dx}{dt}$ and $\frac{dy}{dt}$. Using $s=\sqrt{\dot{x}^2+\dot{y}^2}$, where $s$ is the speed of the wave and the ``dots'' represent the usual differentiation with respect to t, we then plotted $s$ vs gradient ($a_1$) and then observed what relationship(s), if any, are present.

\begin{figure}[tbp]
\begin{center}
\begin{minipage}[b]{0.49\linewidth}
\centering
\includegraphics[width=1.0\textwidth]{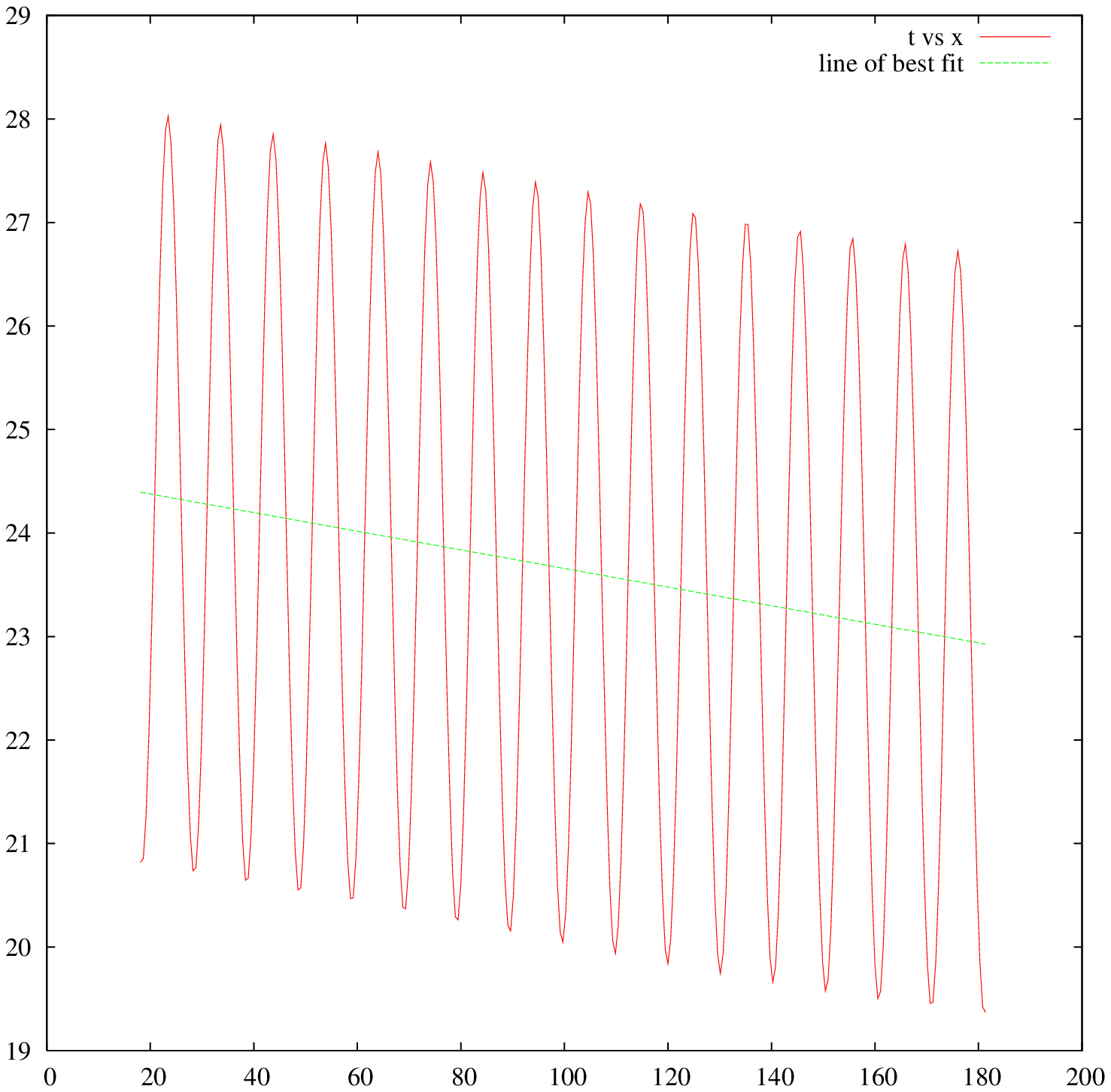}
\end{minipage}
\begin{minipage}[b]{0.49\linewidth}
\centering
\includegraphics[width=1.0\textwidth]{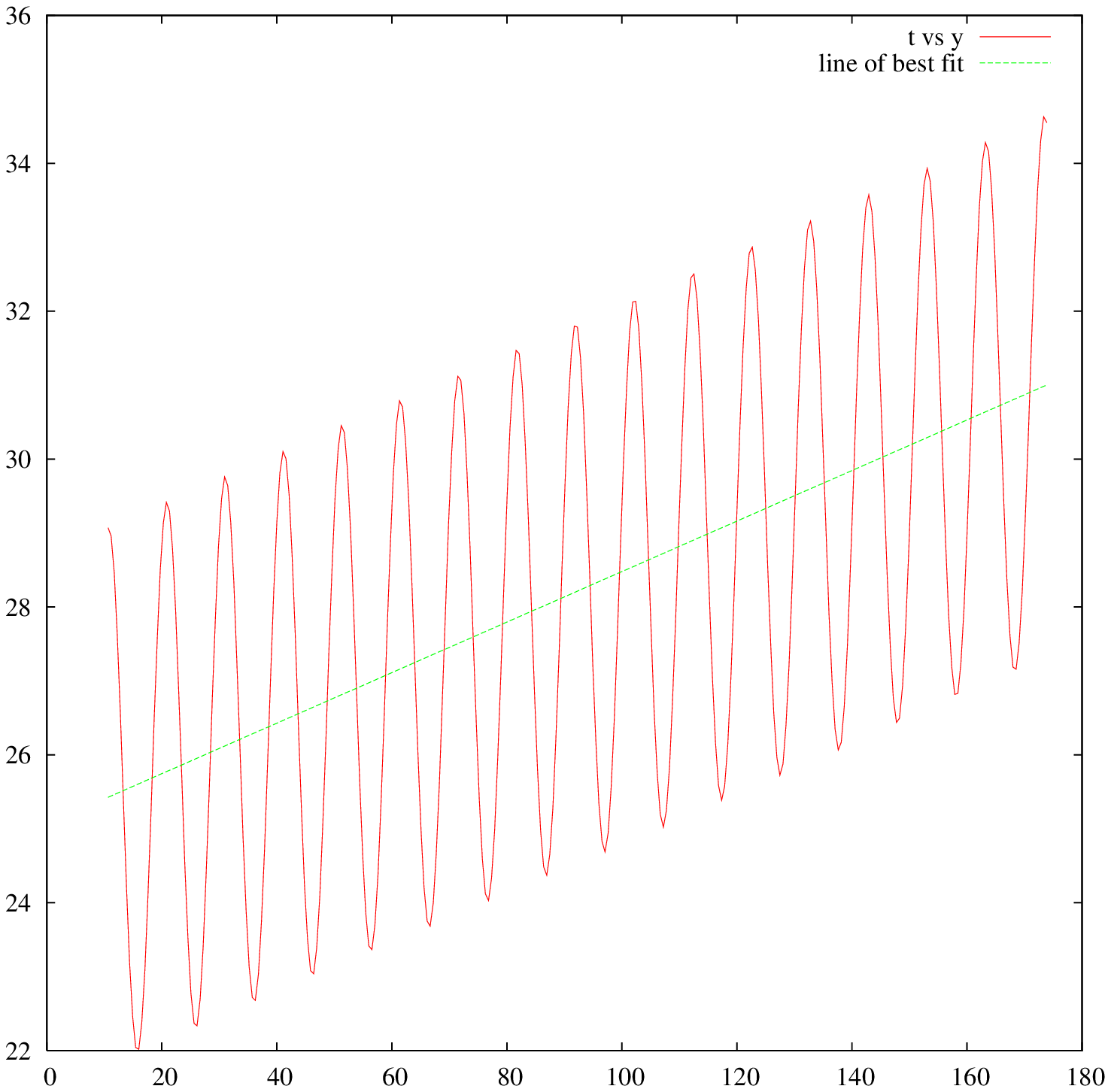}
\end{minipage}
\caption[Inhomogeneitiy induced drift: gradient=0.0008]{Graph of $t$ vs $x$ (left) $t$ vs $y$ (right) for $a_1$=0.0008.}
\label{fig:retest9}
\end{center}
\end{figure}

We show in Fig(\ref{fig:retest9}) an example of the fitting results produced by Gnuplot.

We also determined that in order to get the program working correctly, we needed to make sure that the following techniques were implemented:

\begin{itemize}
\item generation of initial initial conditions must be inverted for negative gradients, 
\item initial transient of the spiral wave is eliminated before the test is carried out,
\item the tip of the spiral wave must be located at the center of the box at the start of the tests.
\end{itemize}

\begin{figure}[tbp]
\begin{center}
\begin{minipage}[b]{0.6\linewidth}
\centering
\includegraphics[width=1.0\textwidth]{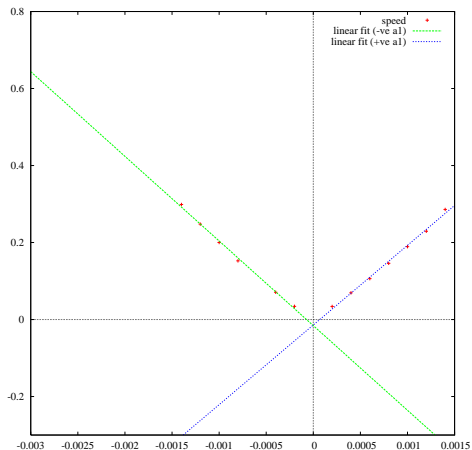}
\end{minipage}
\caption[Inhomogeneitiy induced drift: gradient vs. speed]{Graphs of $a_1$ vs $speed$ with a linear fit.}
\label{fig:fit_speed_late_grad}
\end{center}
\end{figure}

With these points in mind, we get results as shown in Table.(\ref{table:speed4}) and Fig.(\ref{fig:fit_speed_late_grad}).

\begin{table}[tbh]
\begin{center}
\begin{tabular}{|c|c|c|c|}
\hline
$a_1$ & $\dot{x}$ & $\dot{y}$ & speed\\
\hline
-0.0014 & 0.0722856 & -0.289977 & 0.298851\\
-0.0012 & 0.0622736 & -0.239769 & 0.247724\\
-0.001 & 0.0469367 & -0.19443 & 0.200015\\
-0.0008 & 0.0353927 & -0.148323 & 0.152487\\
-0.0004 & 0.0162417 & -0.0690147 & 0.0709001\\
-0.0002 & 0.00673007 & -0.033589 & 0.0342566\\
0.0002	& -0.0076846 & 0.0330261 & 0.0339084\\
0.0004	& -0.0154812 & 0.0675423 & 0.0692938\\
0.0006 & -0.0243027 & 0.10343 & 0.106247\\
0.0008 & -0.0348249 & 0.141664 & 0.145882\\
0.001 & -0.0453209 & 0.183785 & 0.189291\\
0.0012	& -0.0544602 & 0.222769 & 0.229329\\
0.0014 & -0.0690389 & 0.277411 & 0.285873\\
0.0016 & -0.0720951 & 0.314733 & 0.322885\\
0.0018 & -0.0983964 & 0.388473 & 0.400741\\
\hline
\end{tabular}
\caption{Results using refined initial conditions.}
\label{table:speed4}
\end{center}
\end{table}

We can see that these results fit the theory that the speed of the spiral wave which is subject to inhomogeneity induced drift is proportional to the gradient. Therefore, the inclusion of the drift numerics within EZ-Spiral appear to be working correctly.


\subsection{Frequency Locking}

We will now investigate some of the properties of meandering spiral waves that are drifting due to inhomogeneities. The idea behind this study is to see whether we can detect any frequency locking (or phase locking as it is sometimes known) in meandering waves subject to inhomogeneity induced drift. We therefore need to devise a method of measuring the frequencies (Euclidean Frequency and the Hopf Frequency) present within the solution.

We note from \cite{bik96} that for meandering spiral waves, the components of the quotient solution can take the following form:

\begin{eqnarray}
c &=& \bc_0+zc_1+\bar{z}\bar{c}_1+O(|z|^2)\\
\label{eqn:ini_num_om}
\omega &=& \omega_*+z\omega_++\bar{z}\bar{\omega}_++O(|z|^2)\\
\label{eqn:ini_num_hopf_normal}
\dot{z} &=& \alpha z-\beta z|z|^2+O(|z|^2)
\end{eqnarray}
\\
where Eqn.(\ref{eqn:ini_num_hopf_normal}) is the Hopf Normal Form, and $z$ is a limit cycle solution which can be put into the form:

\begin{equation}
z = re^{i(\omega_Ht+\phi_H)}
\end{equation}
\\
with $r$ being the amplitude of the limit cycle, and $\omega_H$ is the Hopf Frequency. We note that Eqn.(\ref{eqn:ini_num_om}) can be expressed as:

\begin{eqnarray}
\label{eqn:initial_hopf}
\omega &=& \omega_*-2r|\omega_+|\cos(\omega_Ht+\theta_0)
\end{eqnarray}
\\
where $\theta_0=\phi_H+\mbox{arg}\{\omega_+\}$. We know that this is true for a meandering wave which is not subject to any drift. 

From Chap.(\ref{chap:3}), we know that $\omega$ for a drifting and meandering spiral wave is given by:

\begin{eqnarray}
\omega &=& \omega_0+\epsilon\omega_1+O(\epsilon^2),
\end{eqnarray}
\\
where $\omega_0$ is given by Eqn.(\ref{eqn:initial_hopf}) and $\omega_1$ is dependent \chg[p116spell1]{on} the perturbation within the system. For our purposes, we shall only be interested in $\omega_0$, since we can extract the values of the Euclidean frequency ($\omega_*$) and the Hopf frequency ($\omega_H$) from the numerical data for omega. This is due to $\omega_*$ being the average value of $\omega$ and $\omega_H$ being determined by $\omega_H=\frac{2\pi}{T}$, where T is the period of the limit cycle (time difference between successive peaks in the \chg[p116spell2]{plots of time against $\omega$}).

We calculate the quotient system by considering the equations of motion for the tip of a spiral wave:
\chg[p116eqn]{}
\begin{eqnarray}
\deriv{R}{t} &=& \chg[]{ce^{i\Theta}}\\
\deriv{\Theta}{t} &=& \omega
\end{eqnarray}
\\
where, $R=X+iY$, and $c=c_x+ic_y$. Rearranging, we get that:

\begin{eqnarray}
c_x(t) &=& \deriv{X}{t}\cos(\Theta)+\deriv{Y}{t}\sin(\Theta)\\
c_y(t) &=& -\deriv{X}{t}\sin(\Theta)+\deriv{Y}{t}\cos(\Theta)\\
\omega(t) &=& \deriv{\Theta}{t}
\end{eqnarray}

So, after a simulation, we are given the tip coordinates, \chg[p117gram]{$(X,Y)$}, and the phase of the tip, $\Theta$, for a range of times. We can therefore, numerically differentiate the given tip data to find $\deriv{X}{t}$, $\deriv{Y}{t}$, and $\deriv{\Theta}{t}$, and hence find the corresponding values of $c_x$, $c_y$ and $\omega$.

However, we note that we have used a numerical scheme to generate this data, and with numerical schemes comes numerical errors (noise). Upon differentiating this data, which contains the numerical noise, we are effectively amplifying the noise.

We bypass this hurdle by using a Tikhonov Regularisation method to ``smoothen out'' the numerical data, before differentiating it. The type of regularisation method we use is The Double Sweep Method (or Progonka in Russian) \cite{progonka} to solve a finite differences boundary value problem. This is a very easy and fast method to implement into a C code.

Also, as with any regularisation method, there is a regularisation parameter. We fixed this parameter at $\lambda_{reg}=1.0$ throughout these calculations. This value gave us the desired results for $\omega$. One of the side affects of using a regularisation method, is the suppressing of the amplitude of the limit cycle. However, we are not, for this study, interested in the amplitude of the limit cycles, but just the values of $\omega_*$ and $\omega_H$, which are not affected by the regularisation technique.

So having regularised the data, we are able to differentiate it and use the results to calculate the quotient solution.

\begin{figure}[tb]
\begin{center}
\begin{minipage}{0.7\linewidth}
\centering
\includegraphics[width=0.7\textwidth, angle=-90]{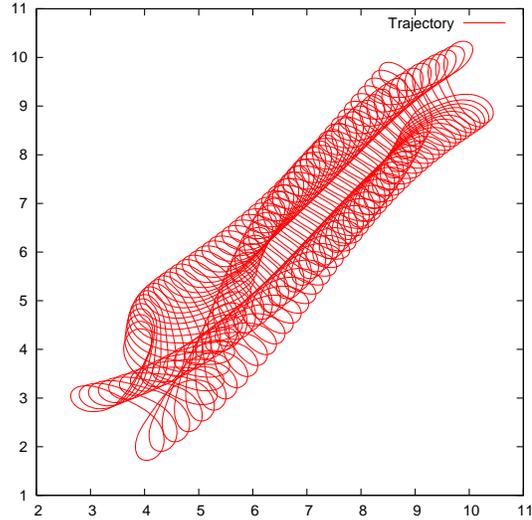}
\end{minipage}
\caption[Inhomogeneitiy induced drift of a meandering spiral wave]{Meandering Spiral Wave drifting due to inhomogeneities, passing through a point of resonance.}
\label{fig:ini_num_traj}
\end{center}
\end{figure}

\chgex[ex]{We} decided to investigate whether there is any frequency locking around a 4:3 resonance point. This is where the Hopf Frequency is four thirds larger than the Euclidean frequency. A typical trajectory around this point is shown in Fig.(\ref{fig:ini_num_traj}).

What is interesting about this is the apparent sudden change of direction around the area of the point $(4,5)$. We are unable to provide a concrete answer to this problem, \chgex[ex]{but we suggest that this could be connected to the boundary conditions.} If this is the case, then we provide a solution to this in Chap.(\ref{chap:4}) where we solve the system in a frame of reference which is comoving with the tip of the spiral wave. The methods described in that chapter enable us to study the spiral wave solution in the knowledge \chgex[ex]{that the boundaries} do not affect the dynamics of the spiral wave. Also, it provides us with a \chgex[ex]{tool which we can let} run for as long as we desire (\chg[af]{given} computer hardware constraints), since the tip of the spiral wave does not actually ever reach the boundary of the box.

\begin{figure}[tbp]
\begin{center}
\begin{minipage}{0.49\linewidth}
\centering
\includegraphics[width=0.7\textwidth, angle=-90]{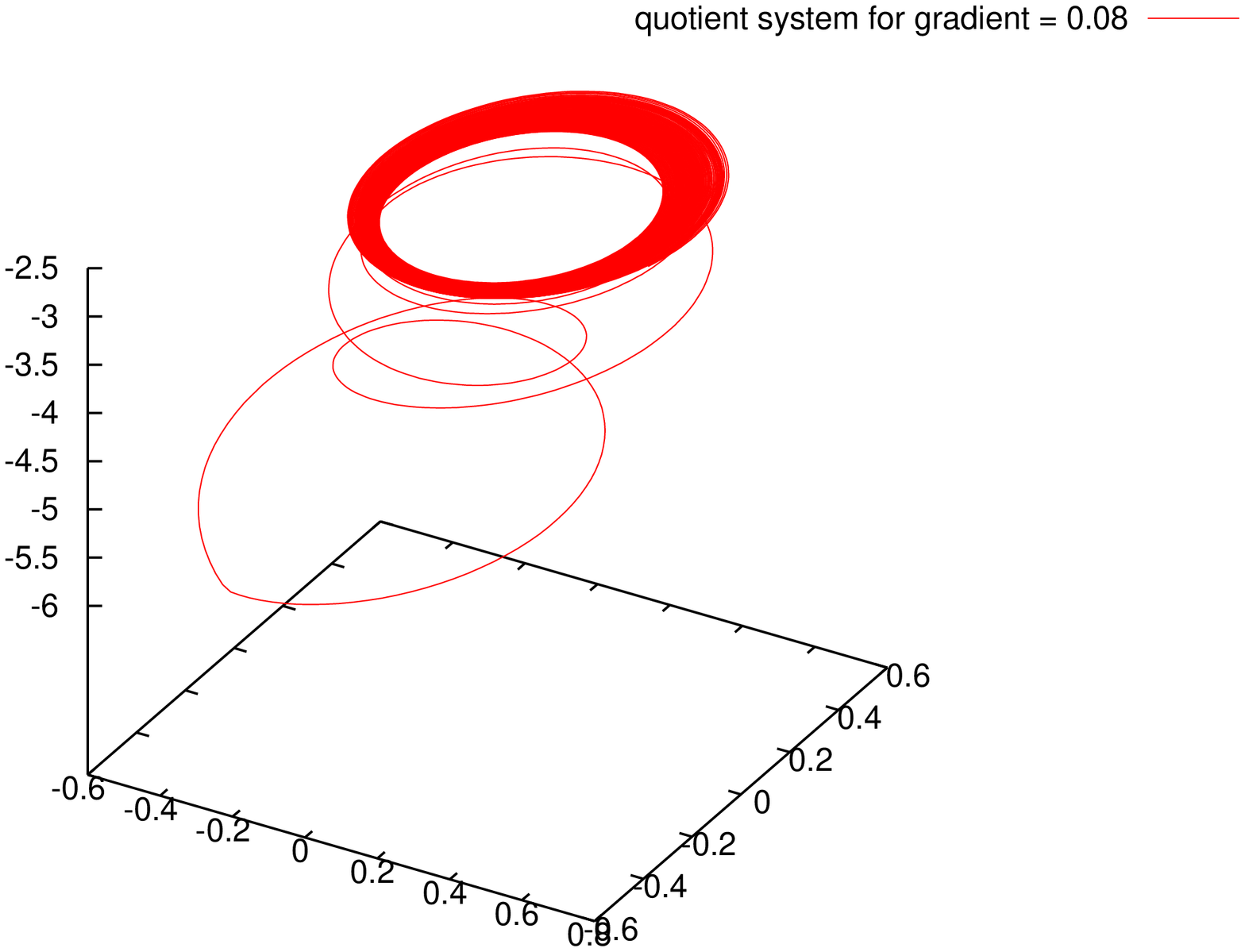}
\end{minipage}
\begin{minipage}{0.49\linewidth}
\centering
\includegraphics[width=0.7\textwidth, angle=-90]{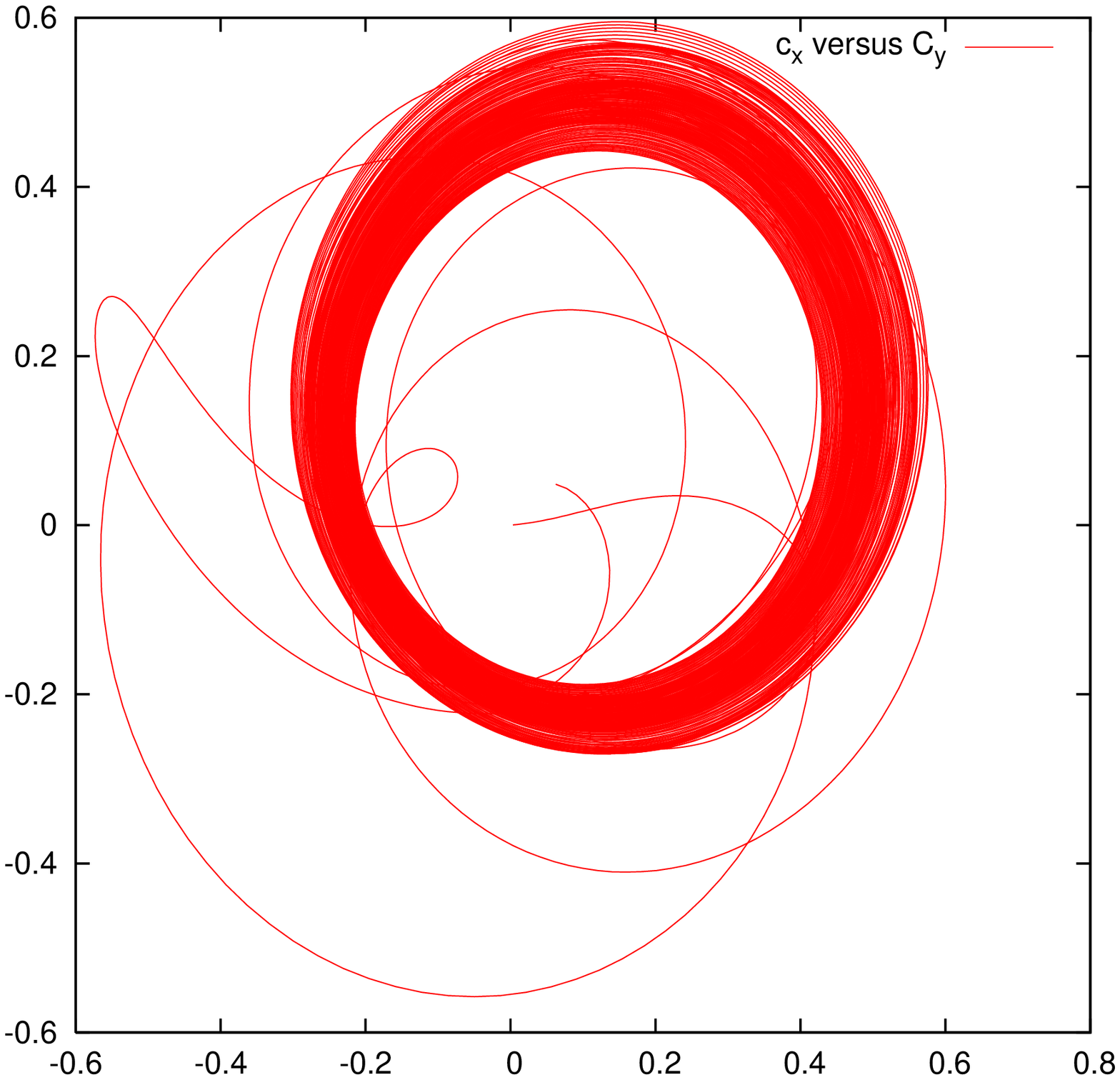}
\end{minipage}
\begin{minipage}{0.49\linewidth}
\centering
\includegraphics[width=0.7\textwidth, angle=-90]{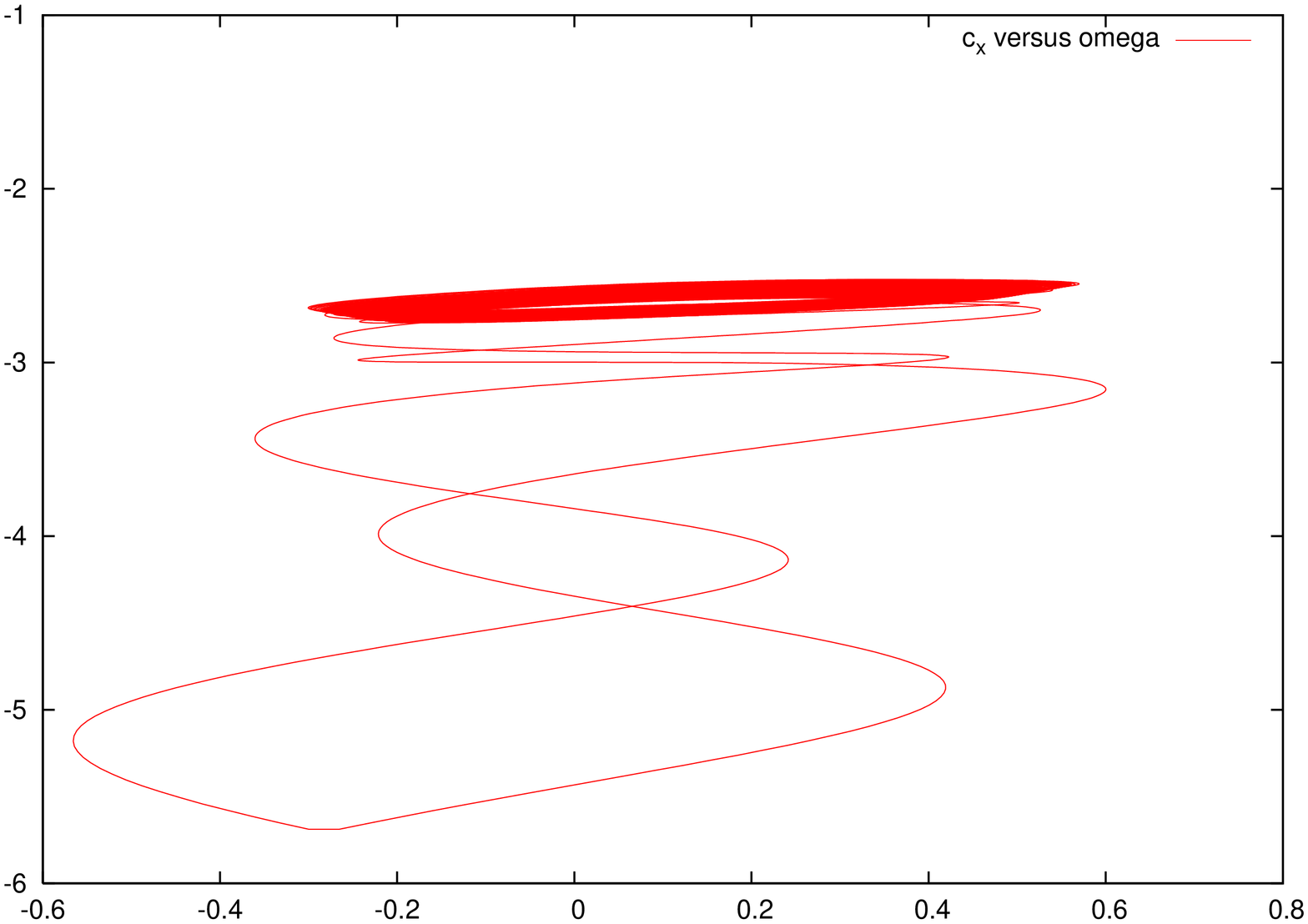}
\end{minipage}
\begin{minipage}{0.49\linewidth}
\centering
\includegraphics[width=0.7\textwidth, angle=-90]{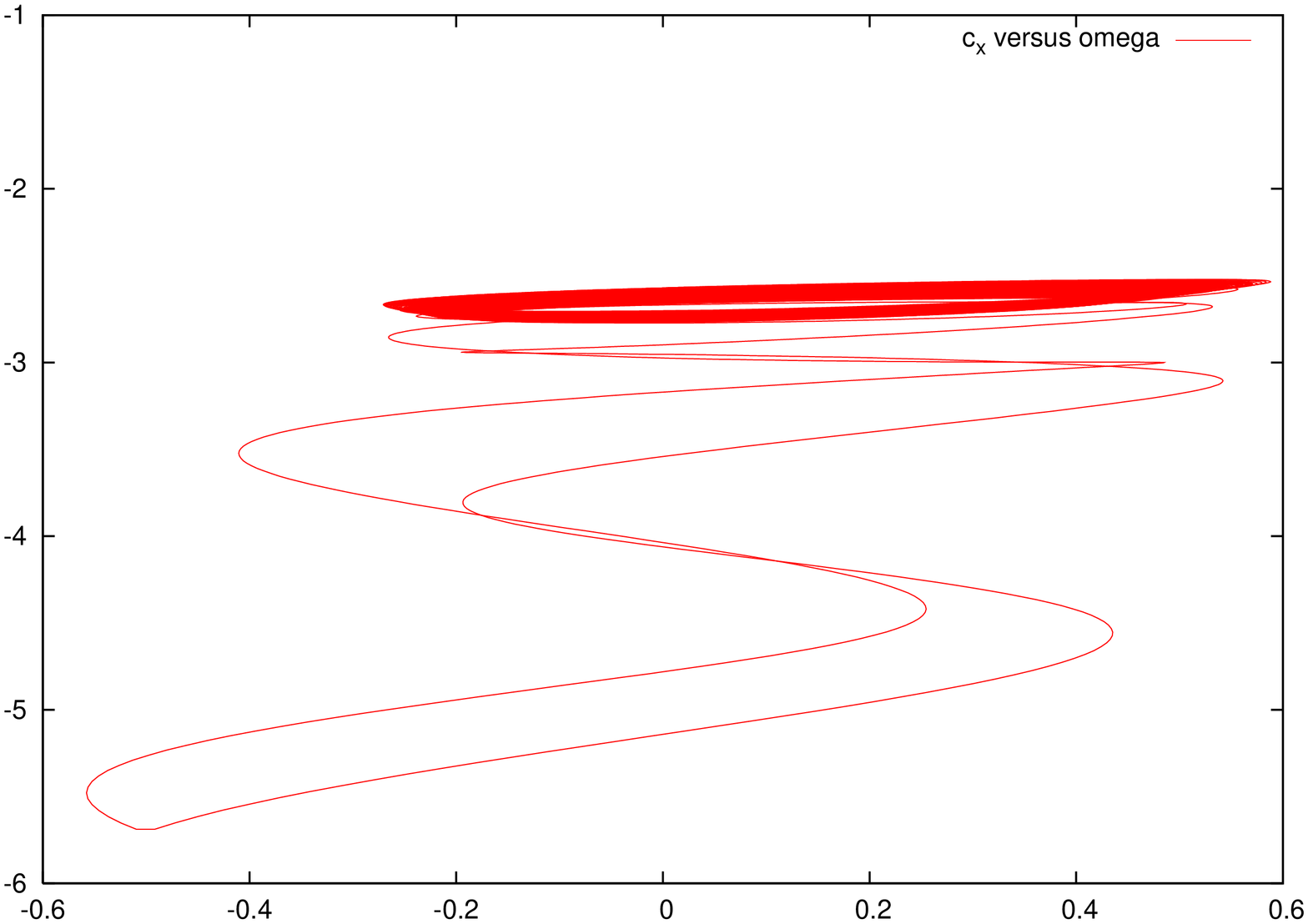}
\end{minipage}
\caption[Inhomogeneitiy induced drift: Quotient solution]{(Top left) The plot of the full quotient system; (top right) $c_x$ against $c_y$ on the right; (bottom left) $c_x$ against $\omega$; (bottom right) $c_y$ against $\omega$.}
\label{fig:ini_inhom_quot}
\end{center}
\end{figure}

\chg[p118para]{Regularising this data, we find that the quotient solution is shown in Fig.(4.4). One of the points to note here is the presence of the strange deviation from the main limit cycle. What we can rule out is that it is not associated with the initial transient of the spiral wave, but provides an interesting phenominem for further study.}

It is not very obvious whether there is any frequency locking present here. Therefore, we shall look at the direction of the translational velocities in the laboratory frame of reference (i.e. $\deriv{X}{t}$ and $\deriv{Y}{t}$). The idea behind this is that if there was any locking at all, then we would observe that the translational acceleration would be zero for a particular length of time.

\begin{figure}[tbp]
\begin{center}
\begin{minipage}{0.7\linewidth}
\centering
\includegraphics[width=0.7\textwidth, angle=-90]{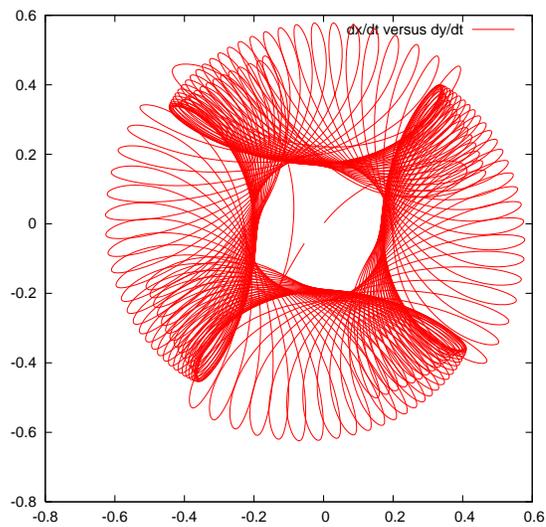}
\end{minipage}
\caption[Inhomogeneitiy induced drift: translational velocities]{$\frac{dx}{dt}$ versus $\frac{dy}{dt}$.}
\label{fig:ini_num_speed}
\end{center}
\end{figure}

We show the \chg[af]{velocities in} Fig.(\ref{fig:ini_num_speed}).

\begin{figure}[tbp]
\begin{center}
\begin{minipage}{0.7\linewidth}
\centering
\includegraphics[width=0.7\textwidth, angle=-90]{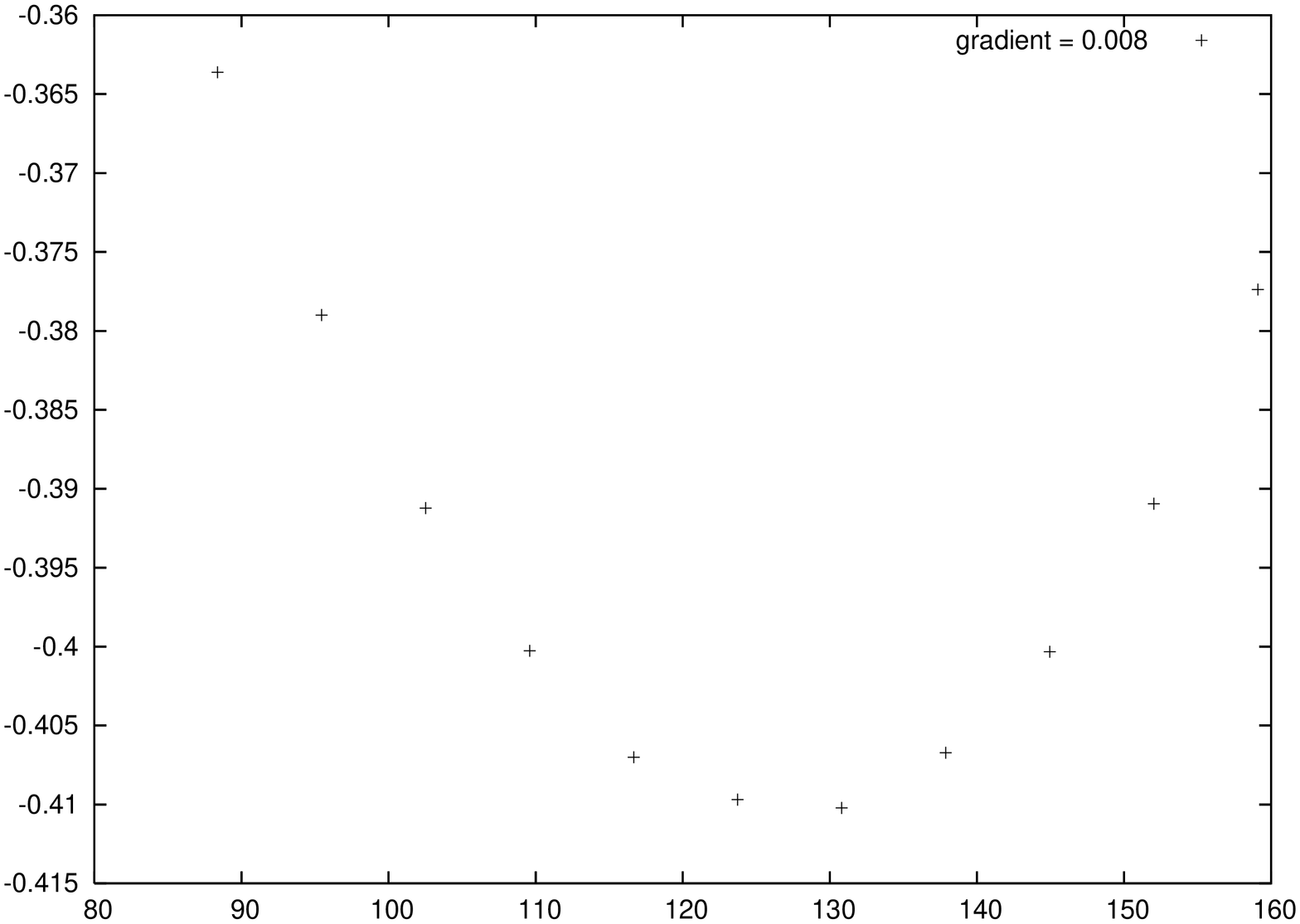}
\end{minipage}
\caption[Inhomogeneitiy induced drift: cross section of the translational velocities]{Time vs $\deriv{y}{t}$, with $\deriv{x}{t}=-0.3$.}
\label{fig:ini_num_pcs}
\end{center}
\end{figure}

In Fig.(\ref{fig:ini_num_pcs}), we show a cross section of the plot of $\deriv{x}{t}$ against $\deriv{y}{t}$ using $\deriv{x}{t}=-0.3$. The figure clearly shows how the values of $\deriv{y}{t}$ change with time as the trajectory crosses the line $\deriv{x}{t}=-0.3$. If there is frequency locking, we believe that we should observe a curve with a flat top, i.e. the value of $\deriv{y}{t}$ should not change for a certain amount of time.

As we can see, there is an instantaneous change of direction of the spiral wave. So, we can conclude that the technique used here does not detect frequency locking within Barkley's model in this particular instance. That is not to say that either our technique is incorrect, or there is frequency locking within the range of parameters we tried out in Barkley's model. It could be that the drift parameter used here is too large, or that the range of parameters in which we should observe frequency locking is very narrow - too narrow in our case.

Another explanation is that the trajectory we are observing is too short to draw any sort of meaningful conclusions. This is to say that we need to let the simulation from which the data is drawn, run for much longer. This would mean a much larger box size than what we are using at the present. The answer to this is to study the solutions in a comoving frame of reference, which is described in Chap.\ref{chap:4}.
\label{sec:ini_num_inhom}

\section{\chg[p120title]{Electrophoretic} Induced Drift of Spiral Wave}
Previously, we have considered whether there has been any frequency locking using Barkleys model and whether the behaviour around particular resonance points displays any unusual behaviour. We used Barkleys model in which the symmetry breaking perturbation contain inhomogenetic properties (the model parameters depend on the spatial coordinates). It was not obvious from the previous experiment whether frequency locking had \chg[ex]{occurred.}

We then decided to try a different symmetry breaking perturbation to see whether frequency locking is clearly observed. The perturbation introduced, breaks the rotational symmetry of the system therefore causing the spiral wave to drift. This is known as \emph{Electrophersis Induced Drift}. The system of equations we consider are:

\begin{eqnarray}
\pderiv{u}{t} & = & f(u,v)+\nabla^2u+A\pdux\\
\pderiv{v}{t} & = & g(u,v)
\end{eqnarray}
\\
where $A$ is a parameter which determines the strength of the drift, as well as the direction. We also note that $f(u,v)$ and $g(u,v)$ are given by:

\begin{eqnarray}
f(u,v) & = & \frac{1}{\varepsilon}u(1-u)(u-\frac{v+b}{a})\\
g(u,v) & = & u-v
\end{eqnarray}

Our aim is to determine whether there is a range of values of $A$ for which frequency locking is observed, and of course to see whether frequency locking actually does occur.


\subsection{Method}

We used an \chg[af]{amended} version of EZ-Spiral, which had been modified to include the symmetry breaking perturbation. From previous experiments using inhomogeneity induced drift, we found that there is a 4:3 resonance point in the region of the parameters $a=0.5465$, $b=0.01$ and $\varepsilon=0.01$. We decided to vary parameter $a$ in the range $0.5200\leq a \leq 0.5650$ and calculate the Euclidean frequency, $\omega_*$, and the Hopf \chg[p121spell]{frequency}, $\omega_H$.

To calculate these frequencies, we shall use the results stated in Sec.(\ref{sec:ini_num_inhom}):

\begin{eqnarray}
\label{eqn:ini_num_om_anis}
\omega &=& \omega_0-2r|\omega_1|\cos(\omega_Ht+\theta_0)
\end{eqnarray}

We also note from earlier observations that the data produced by EZ-Spiral contained noise which, when numerically differentiated, grew significantly and distorted the data. Therefore, we regularised the data before differentiating it, eventually finding the quotient solution.

This method is done for numerous experiments over the range of parameter $a$ as detailed above. We were then able to calculate the frequency ratio $\omega_0:\omega_H$ for each run and plot the graph of $a$ vs. $\omega_0:\omega_H$. If frequency locking is present, then we should observe that for a particular range of values of $a$, each $a$ should have the same frequency ratio.


\subsection{Observations}

We show below the results of the curve fitting exercise to show that the curve does fit for the values that we require. As we noted, we need the value of $\omega_0$, which is the average value of the graph, and $\omega_H$ which is \chg[af]{calculated} as:

\begin{equation}
\omega_H = \frac{2\pi}{T}
\end{equation}
\\
where $T$ is the time between consecutive``peaks" on the graph.

\begin{figure}[btp]
\begin{center}
\begin{minipage}{0.45\linewidth}
\centering
\includegraphics[width=1.0\textwidth]{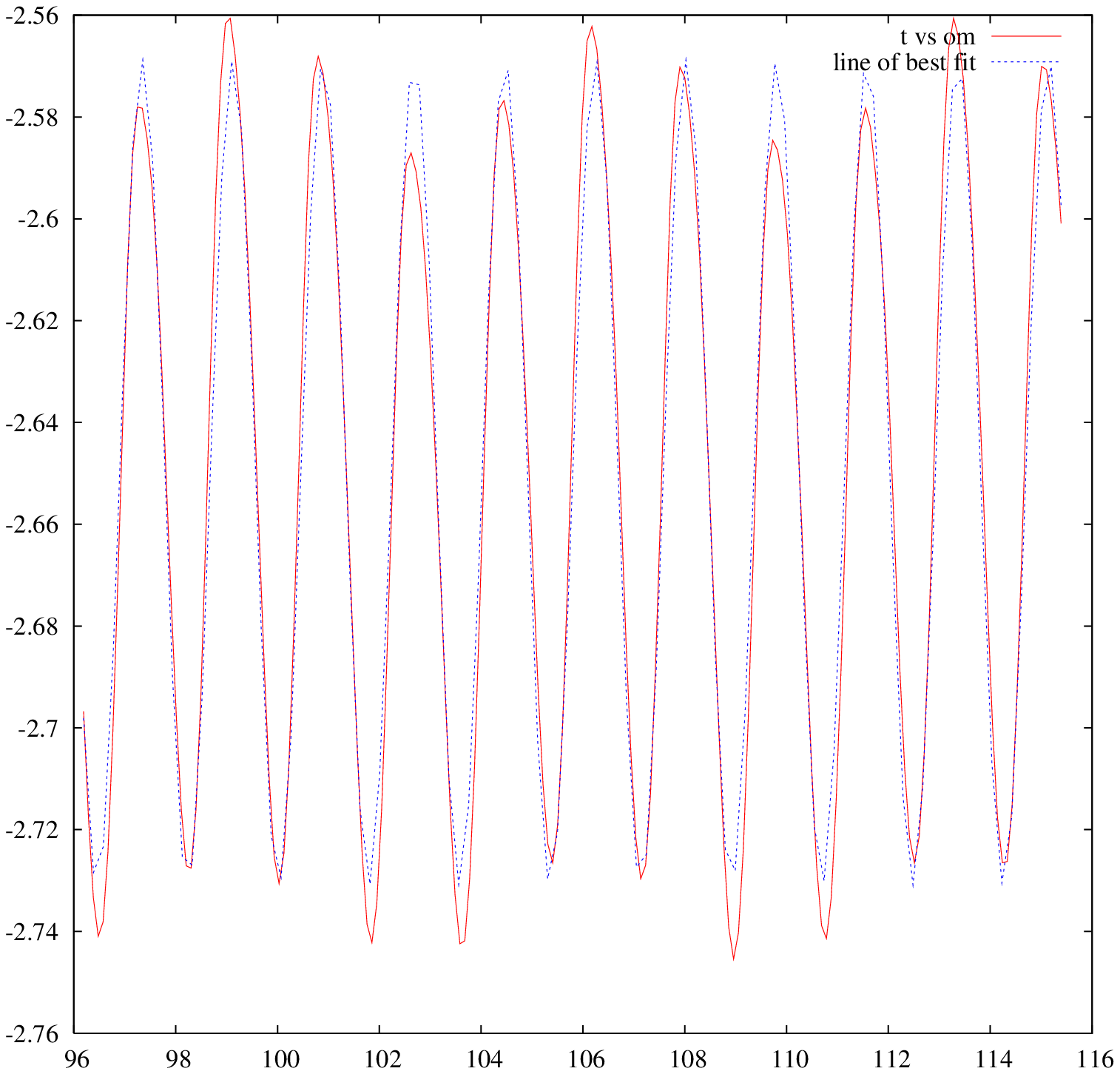}
\end{minipage}
\begin{minipage}{0.45\linewidth}
\centering
\includegraphics[width=1.0\textwidth]{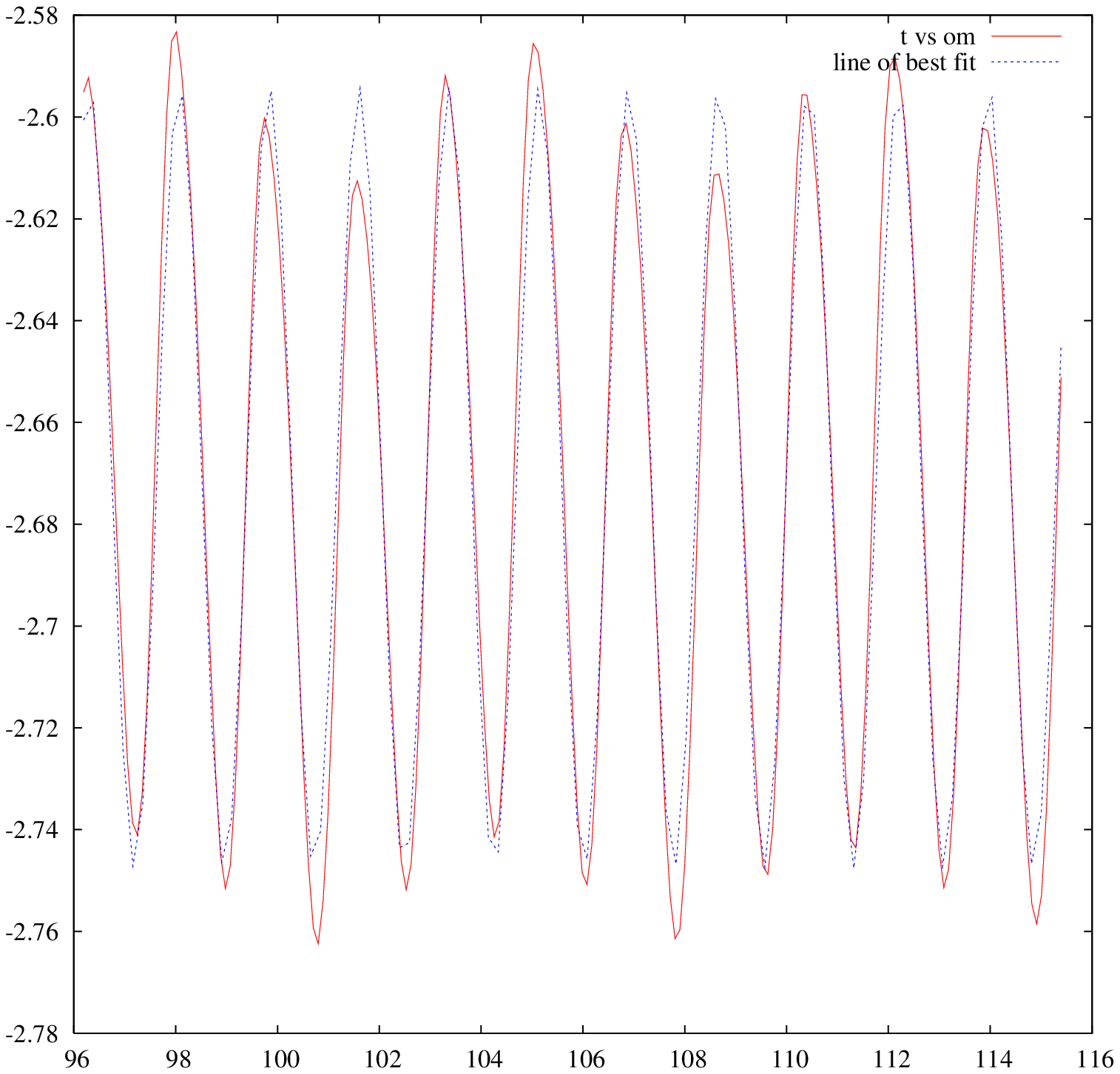}
\end{minipage}
\caption[Electrophoretis induced drift: time vs. $\omega$]{Plots of time vs. $\omega$, with $a_0$ = 0.5400, $\omega_0$ = -2.64982, $\omega_H$ = 3.52981 (left), and $a_0$ = 0.5530, $\omega_0$ = -2.67089, $\omega_H$ = 3.55742 (right).}
\label{fig:b}
\end{center}
\end{figure}

We show in table (\ref{table:05}) the values of the frequencies calculated and also their ratios for $A=0.05$

\begin{table}
\begin{center}
\begin{tabular}{|c|c|c|c|}
\hline
$a$ & $\omega_0$ & $\omega_H$ & $\omega_0:\omega_H$\\
\hline
0.5200 & -2.61630 & 3.47044 & 1.32647\\
0.5250 & -2.62663 & 3.47938 & 1.32466\\
0.5300 & -2.63481 & 3.49618 & 1.32692\\
0.5350 & -2.64208 & 3.51388 & 1.32997\\
0.5400 & -2.64982 & 3.52981 & 1.33209\\
0.5450 & -2.65869 & 3.53618 & 1.33005\\
0.5455 & -2.65965 & 3.53823 & 1.33034\\
0.5460 & -2.66023 & 3.54018 & 1.33078\\
0.5470 & -2.66078 & 3.54539 & 1.33246\\
0.5480 & -2.66259 & 3.54743 & 1.33232\\
0.5500 & -2.66593 & 3.55241 & 1.33252\\
0.5510 & -2.66744 & 3.55349 & 1.33217\\
0.5520 & -2.66920 & 3.55500 & 1.33186\\
0.5530 & -2.67089 & 3.55742 & 1.33192\\
0.5540 & -2.67242 & 3.56027 & 1.33223\\
0.5550 & -2.67417 & 3.56443 & 1.33291\\
0.5600 & -2.68225 & 3.58135 & 1.33520\\
0.5650 & -2.68961 & 3.58996 & 1.33475\\
\hline
\end{tabular}
\caption{Frequencies and their ratios}
\label{table:05}
\end{center}
\end{table}

\begin{figure}[btp]
\begin{center}
\begin{minipage}[htbp]{0.7\linewidth}
\centering
\includegraphics[width=0.7\textwidth, angle=-90]{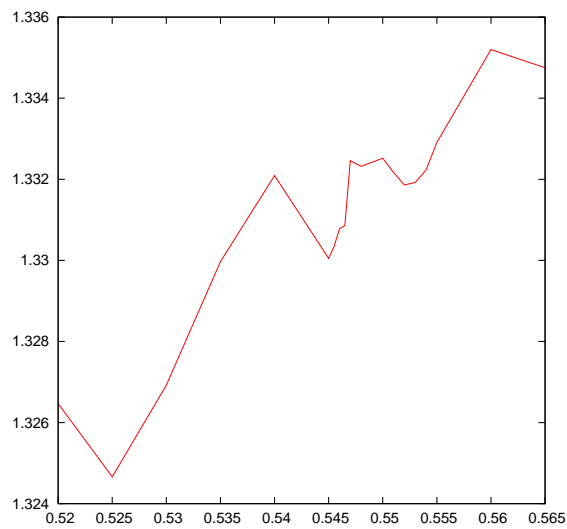}
\caption[Electrophoretis induced drift: $a$ vs. $\omega_0$:$\omega_H$]{The graph of $a$ vs. $\omega_0$:$\omega_H$.}
\label{fig:c}
\end{minipage}
\end{center}
\end{figure}

We \chg[af]{show} in Fig.(\ref{fig:c}) the plot of $a$ against the frequency ratio.

\begin{figure}[btp]
\begin{center}
\begin{minipage}[htbp]{0.45\linewidth}
\centering
\includegraphics[width=0.6\textwidth, angle=-90]{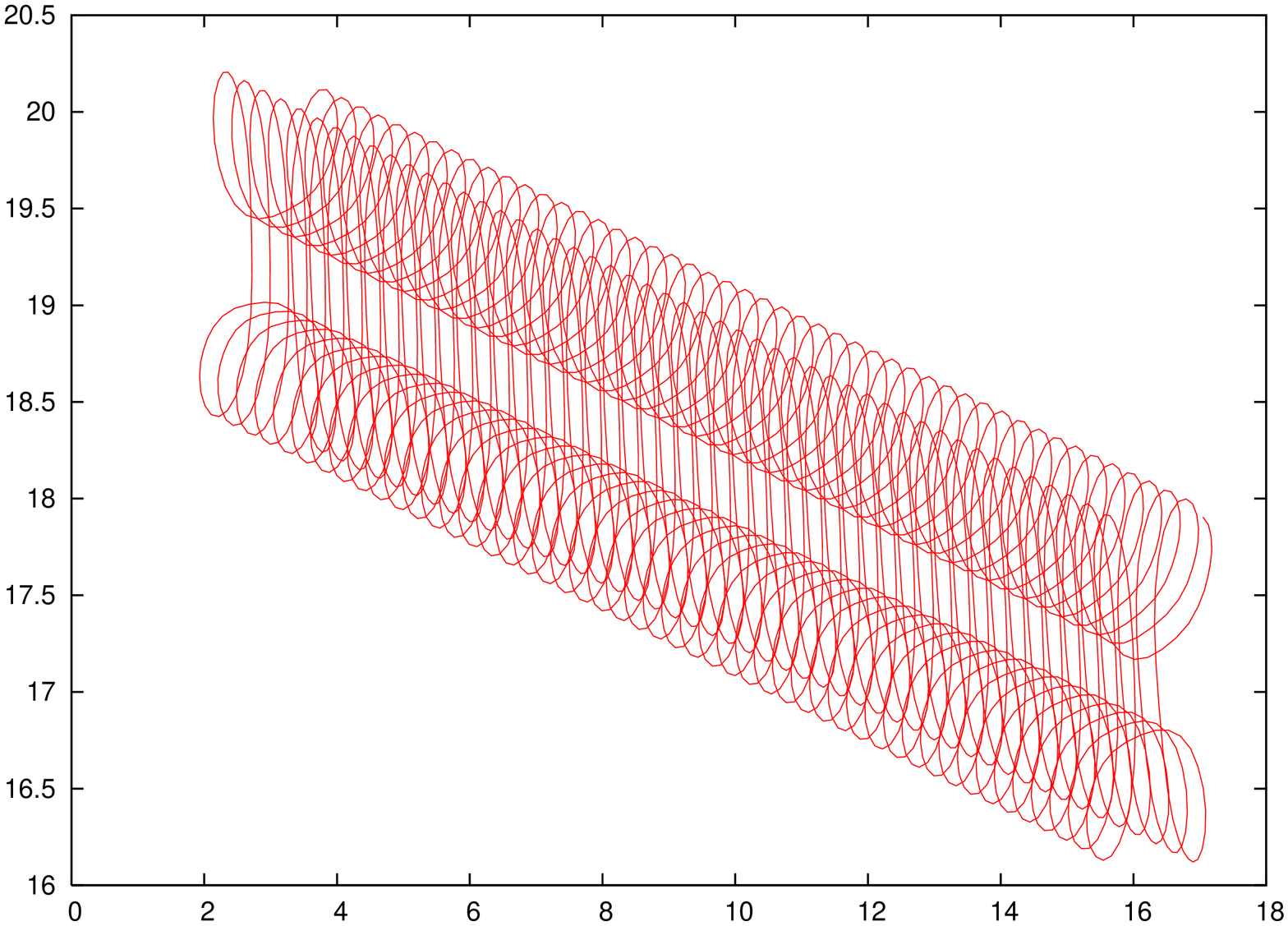}
\end{minipage}
\begin{minipage}[htbp]{0.45\linewidth}
\centering
\includegraphics[width=0.6\textwidth, angle=-90]{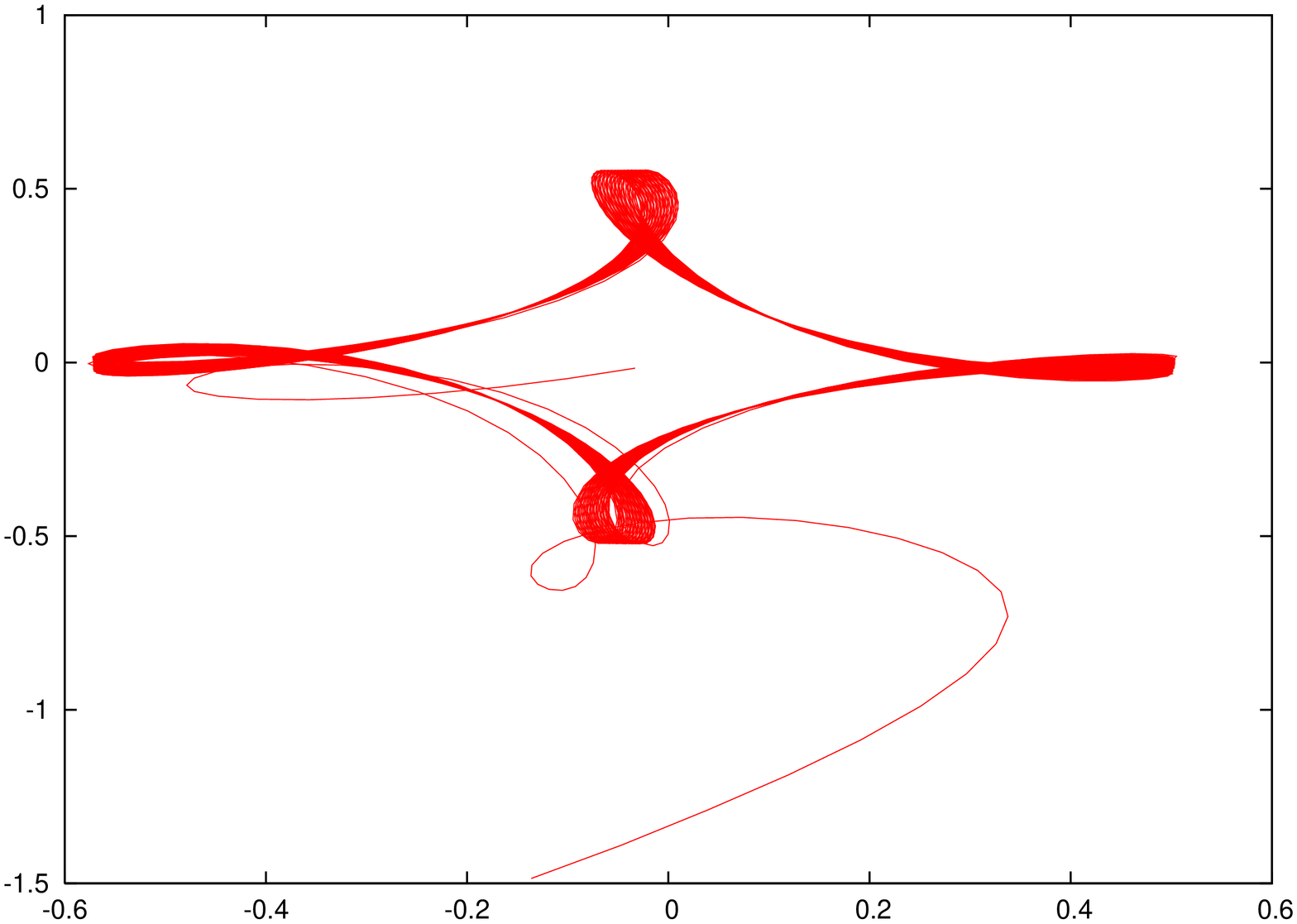}
\end{minipage}
\caption[Electrophoretis induced drift: Trajectory for $a$=0.5530]{Trajectory ($X$ vs $Y$) of the spiral wave for $a$=0.5530 (left), $\odXt$ vs. $\odYt$ (right).}
\label{fig:e}
\end{center}
\end{figure}

\chg[af]{Furthermore,} we can see in Fig(\ref{fig:e}) \chgex[ex]{that for $a=0.5530$ it} looks as though we have locking. In fact Fig.(\ref{fig:e}) shows that the wave is still slightly rotating.

\begin{figure}[btp]
\begin{center}
\begin{minipage}[htbp]{0.45\linewidth}
\centering
\includegraphics[width=0.6\textwidth, angle=-90]{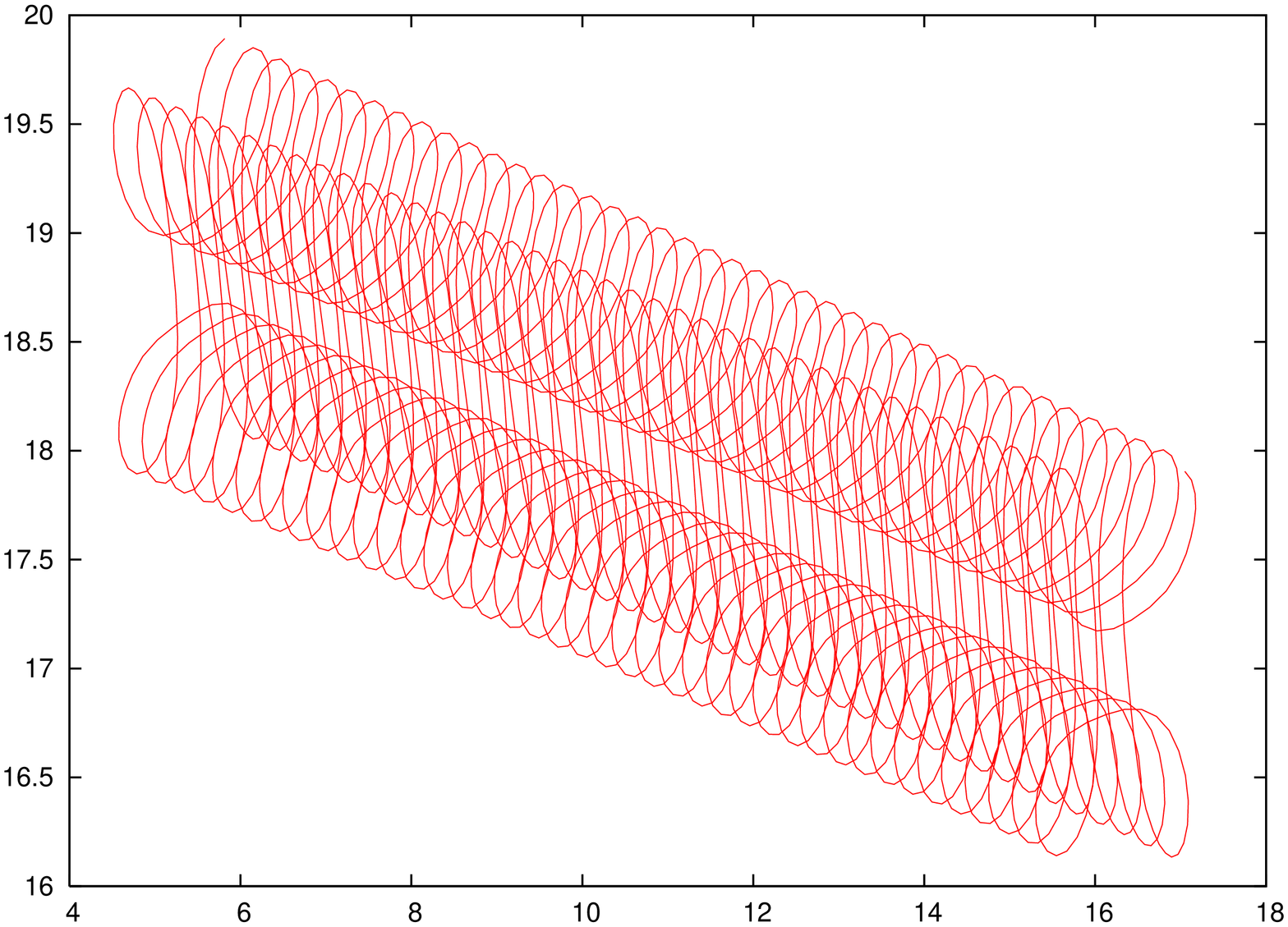}
\end{minipage}
\begin{minipage}[htbp]{0.45\linewidth}
\centering
\includegraphics[width=0.6\textwidth, angle=-90]{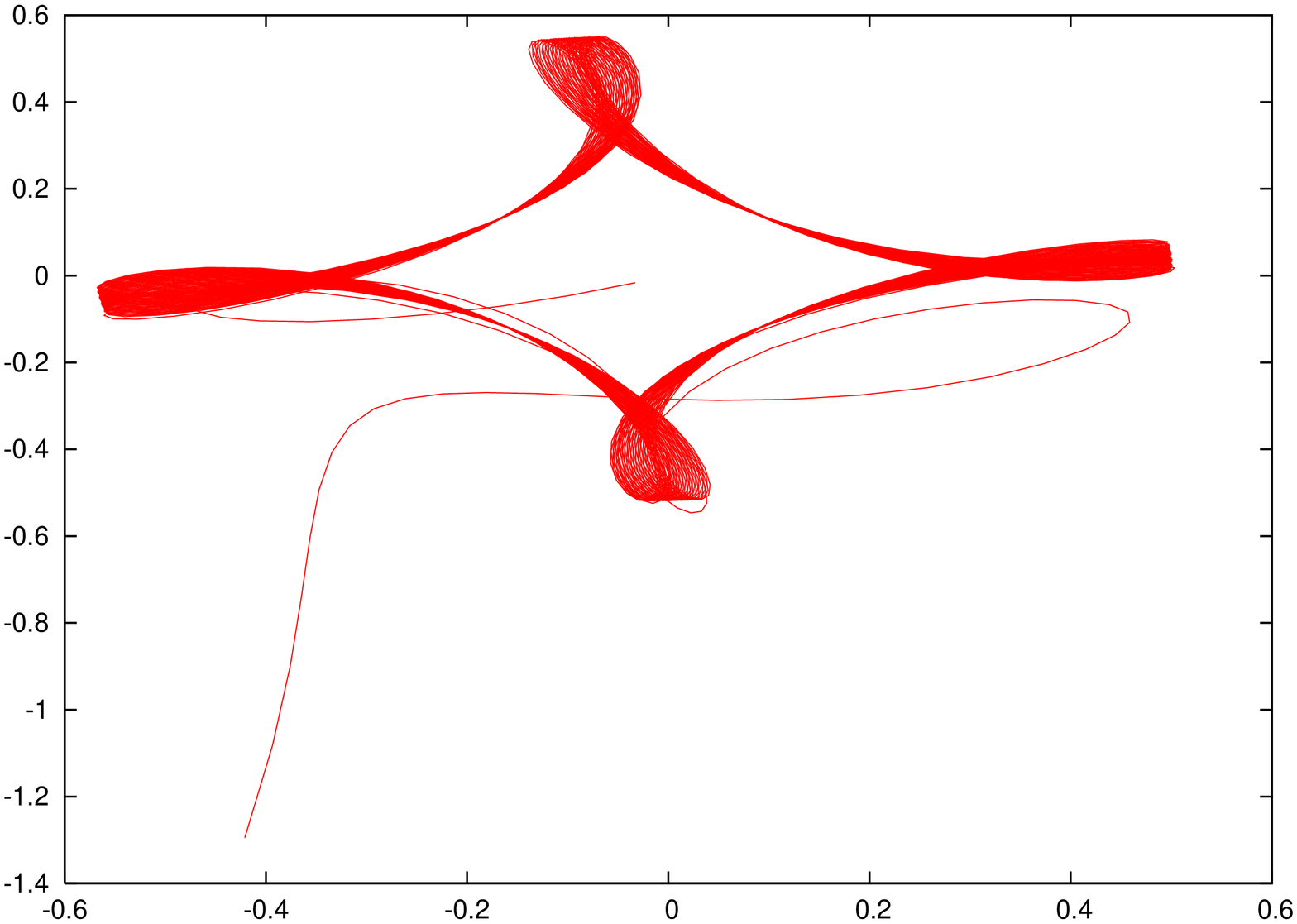}
\end{minipage}
\caption[Electrophoretis induced drift: Trajectory for $a$=0.5540]{Trajectory ($X$ vs $Y$) of the spiral wave for $a$=0.5540 (left), $\odXt$ vs. $\odYt$ (right).}
\label{fig:g}
\end{center}
\end{figure}

\begin{figure}[btp]
\begin{center}
\begin{minipage}[htbp]{0.7\linewidth}
\centering
\includegraphics[width=1.0\textwidth]{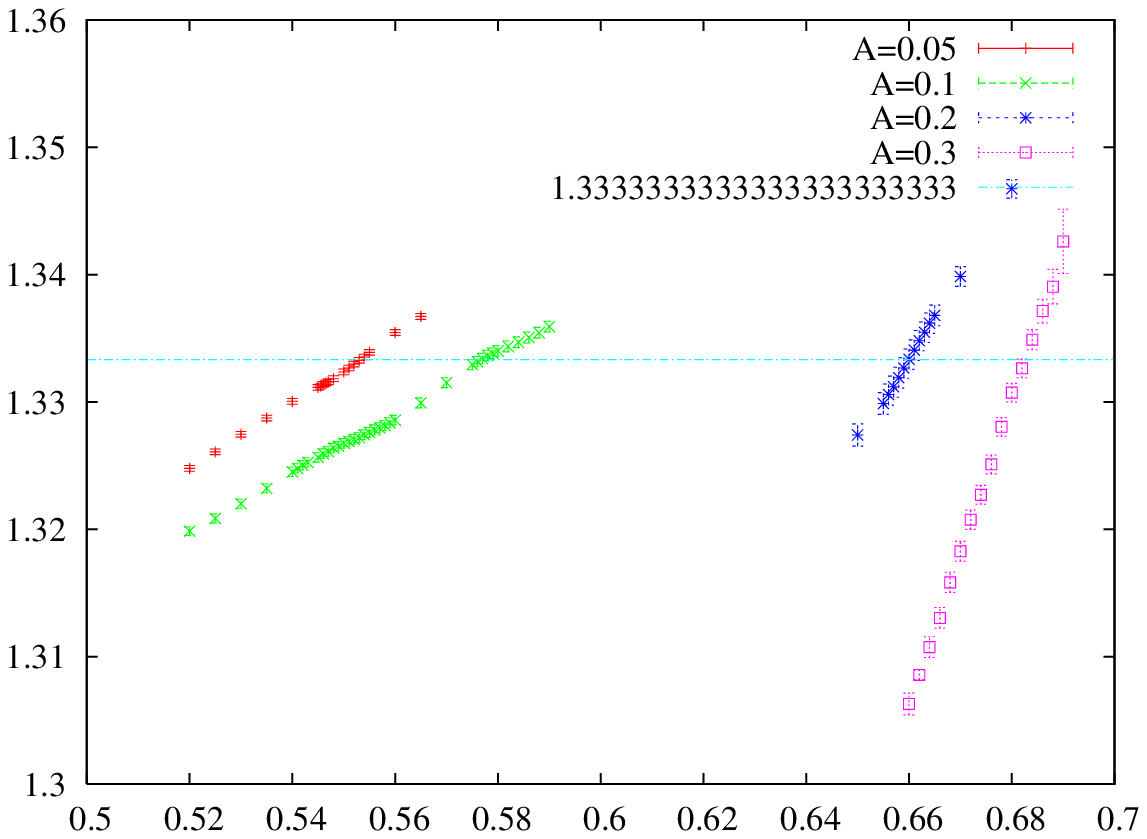}
\end{minipage}
\caption[Electrophoretis induced drift: $a$ vs. $\omega_0$:$\omega_H$]{$a$ vs. $\omega_0$:$\omega_H$}
\label{fig:ini_num_anis_all}
\end{center}
\end{figure}

To finish off this part of our work, we show in Fig.(\ref{fig:ini_num_anis_all}) the results from a range of values for $A$.

It is clear that for all these parameter values, we do not get frequency locking in our simulations. As with inhomogeneity induced drift, \chgex[af]{perhaps} evidence of locking could be observed for longer trajectories but that would mean an increase in box size and related increase in computational resources. Hence, a way around this would be to study the solutions in a frame of reference comoving with the tip of the spiral wave \chg[p122gram]{(Chap.(\ref{chap:4})).}
\label{sec:ini_num_anis}

\section{Generic Forms for the Equation of motion}
We will now show some work that was conducted prior to the development of theory as detailed in Chap.\ref{chap:3}. We shall show an estimate of the equations of motion for a meandering spiral wave trajectory, before showing an estimate of the equations for the trajectory of meandering spiral wave that is subject to inhomogeneity induced drift. We shall generate a spiral wave solution using EZ-Spiral and fit the estimated equations of motion to the data.

\subsection{Non-Drifting Meandering Waves}

We know that meandering spiral waves are quasiperiodic. Let us assume that the equations of motion are:

\begin{eqnarray}
\label{eqn:initial_mrw_111}
x(t) & = & x_0+R_x\cos(\omega_{1x}t+\phi_{1x})+r_x\cos(\omega_{2x}t+\phi_{2x})\\
\label{eqn:initial_mrw_222}
y(t) & = & y_0+R_y\sin(\omega_{1y}t+\phi_{1y})-r_y\sin(\omega_{2y}t+\phi_{2y})
\end{eqnarray}

We should observe that $R_x$ and $R_y$ are in fact the same (say $R$) as are $\omega_{1x}$ and $\omega_{1y}$ ($\omega_1$). Let us assume that $r_x$ and $r_y$ ($r$ say) are the same and also $\omega_{2x}$ and $\omega_{2y}$ are also the same ($\omega_2$). Note that $R$ and $\omega_1$ are the primary radius and Euclidean frequency respectively, and $r$ and $\omega_2$ are the secondary radius and Hopf frequency.

\begin{figure}[tbp]
\begin{center}
\begin{minipage}[b]{0.49\linewidth}
\centering
\includegraphics[width=\textwidth]{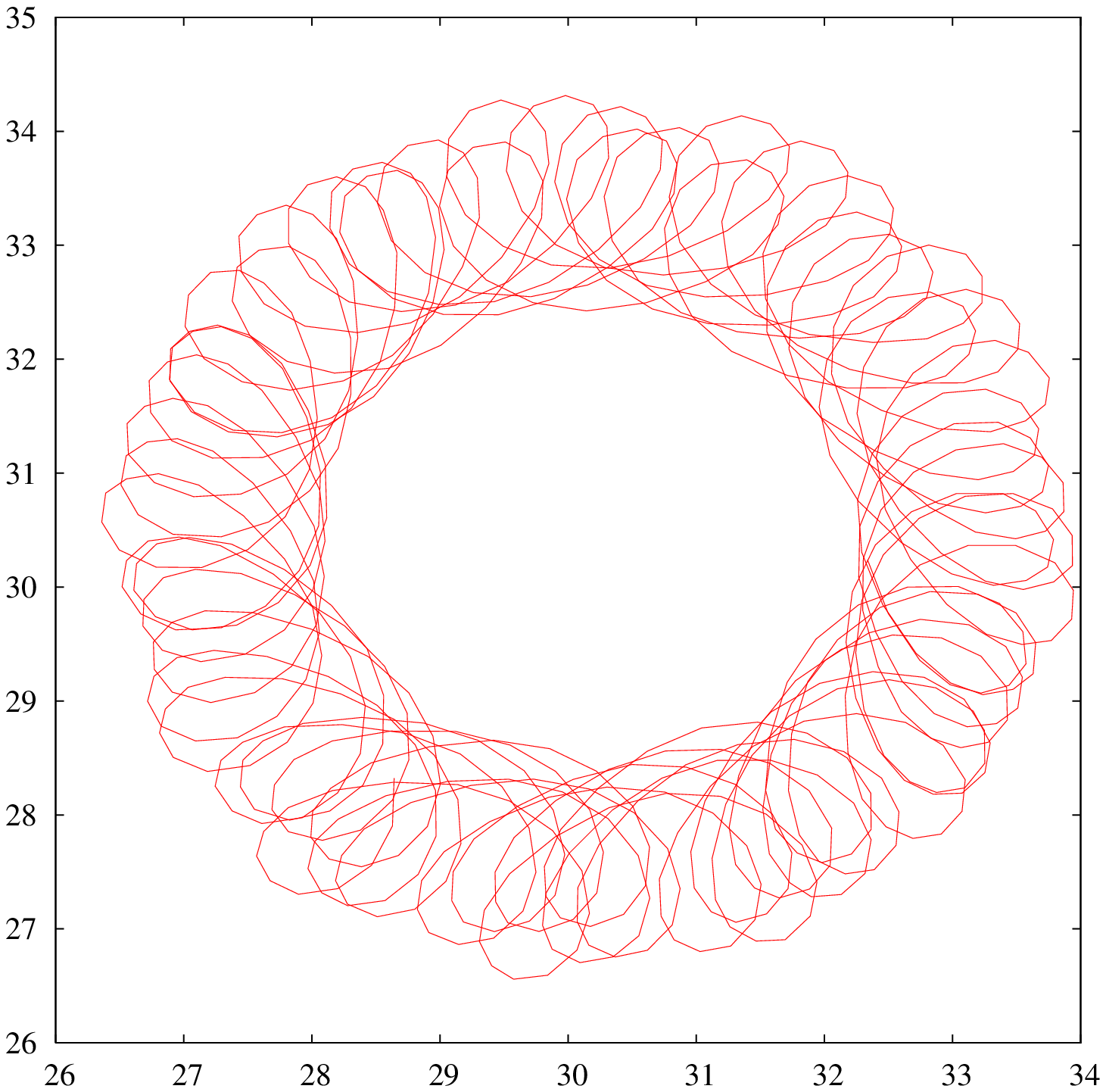}
\end{minipage}
\begin{minipage}[b]{0.49\linewidth}
\centering
\includegraphics[width=\textwidth]{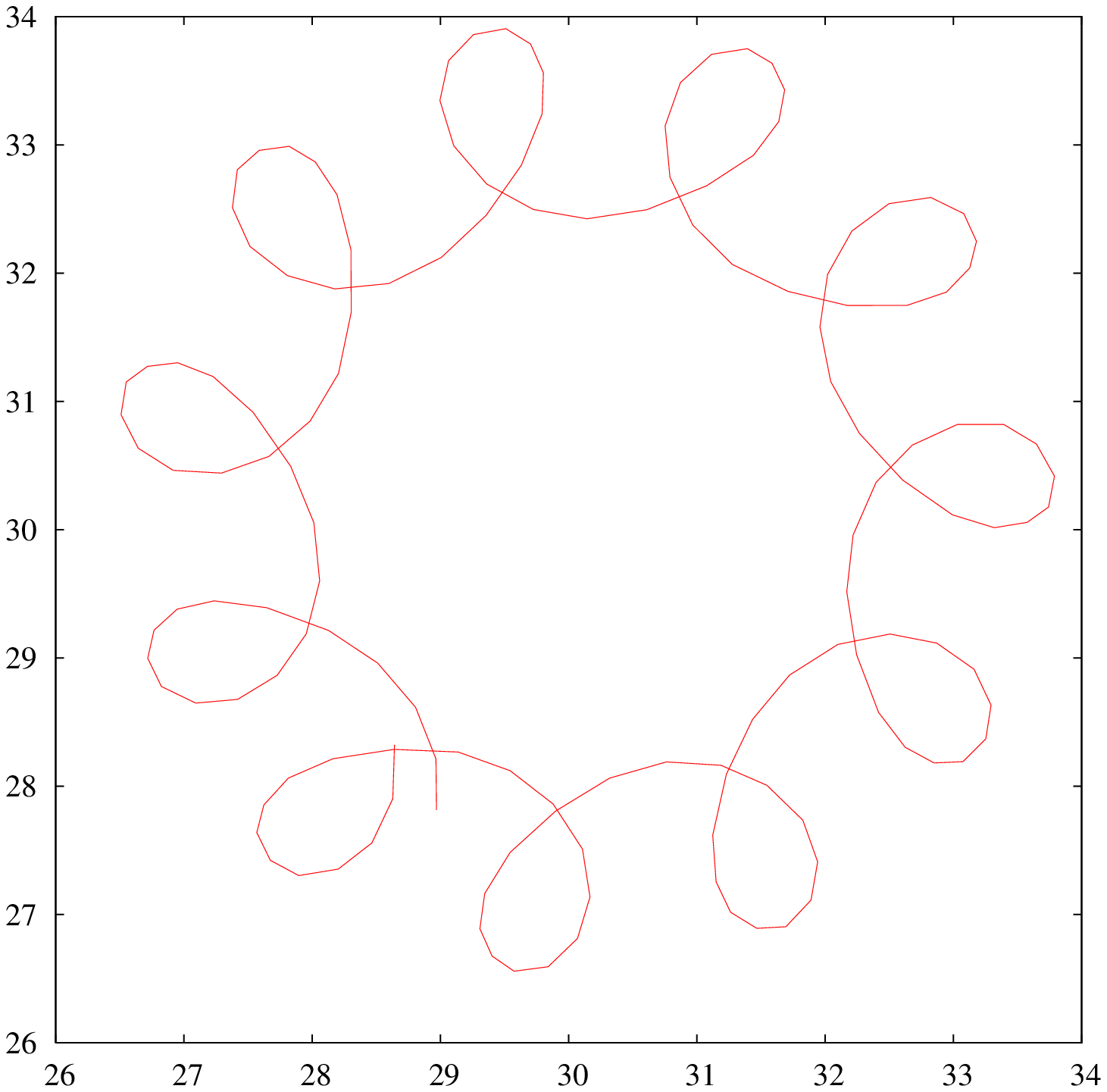}
\end{minipage}
\caption[Meandering spiral wave: trajectories]{Graphs of the trajectories for a non-drifting meandering wave with the full trajectory on the left and a part trajectory (1 full rotation) on the right.}
\label{fig:mrw41}
\end{center}
\end{figure}

Looking \chgex[ex]{at the data file which contains the numerical values of the tip coordinates at various timesteps,} and also at the graph of the trajectory of the wave (see figure \ref{fig:mrw41}) we get the following initial estimates of the various parameters:

\begin{eqnarray*}
x_0 & = & 28.73429 \\
y_0 & = & 28.50880 \\
R & = & 3.25 \\
\omega_1 & = & 0.162496 \\
\phi_1 & = & 1.00\\
r & = & 0.9\\
\omega_2 & = & 1.65\\
\phi_2 & = & 0.8  
\end{eqnarray*}

\begin{figure}[tbp]
\begin{center}
\begin{minipage}[b]{0.49\linewidth}
\centering
\includegraphics[width=\textwidth]{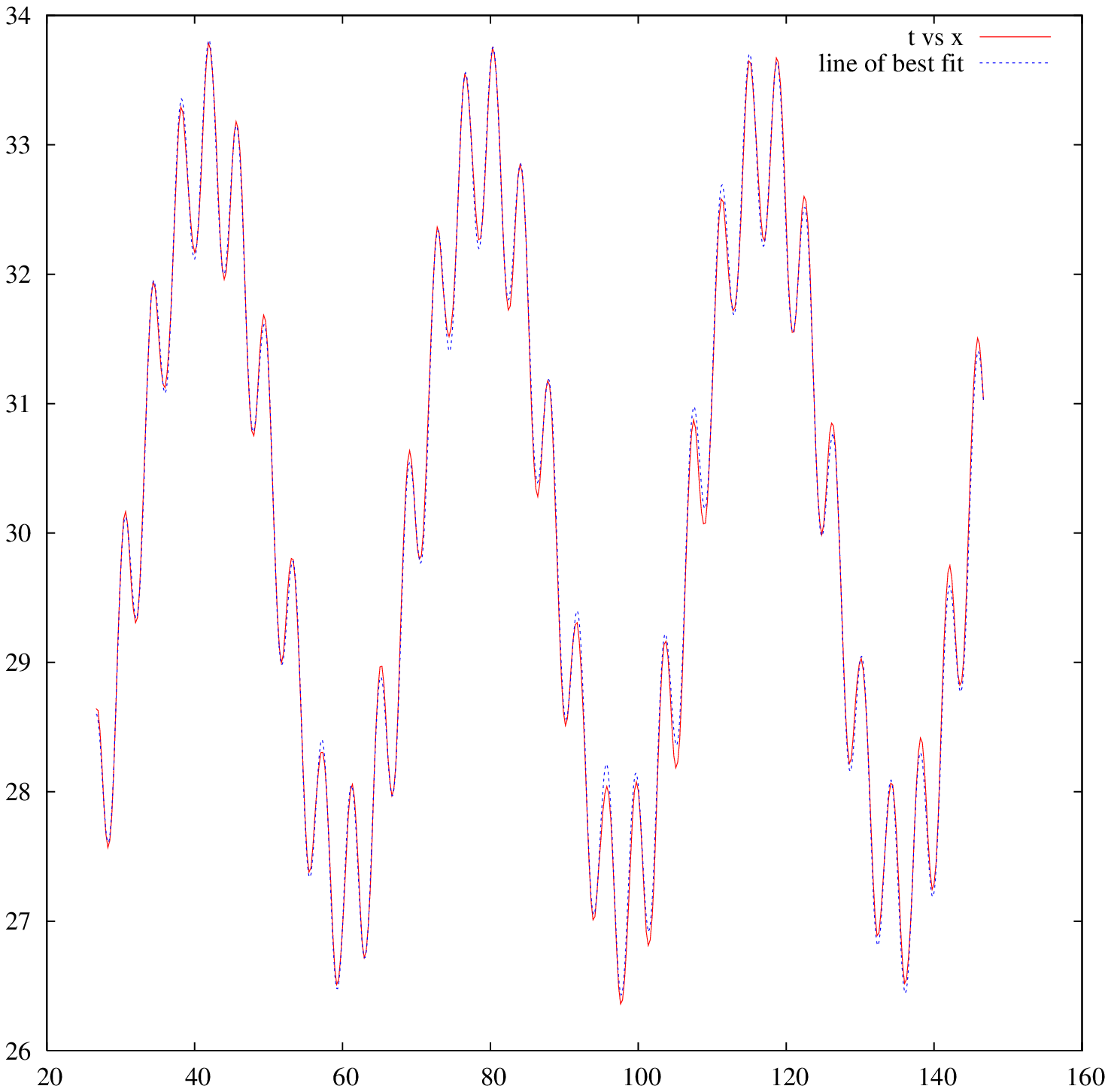}
\end{minipage}
\begin{minipage}[b]{0.49\linewidth}
\centering
\includegraphics[width=\textwidth]{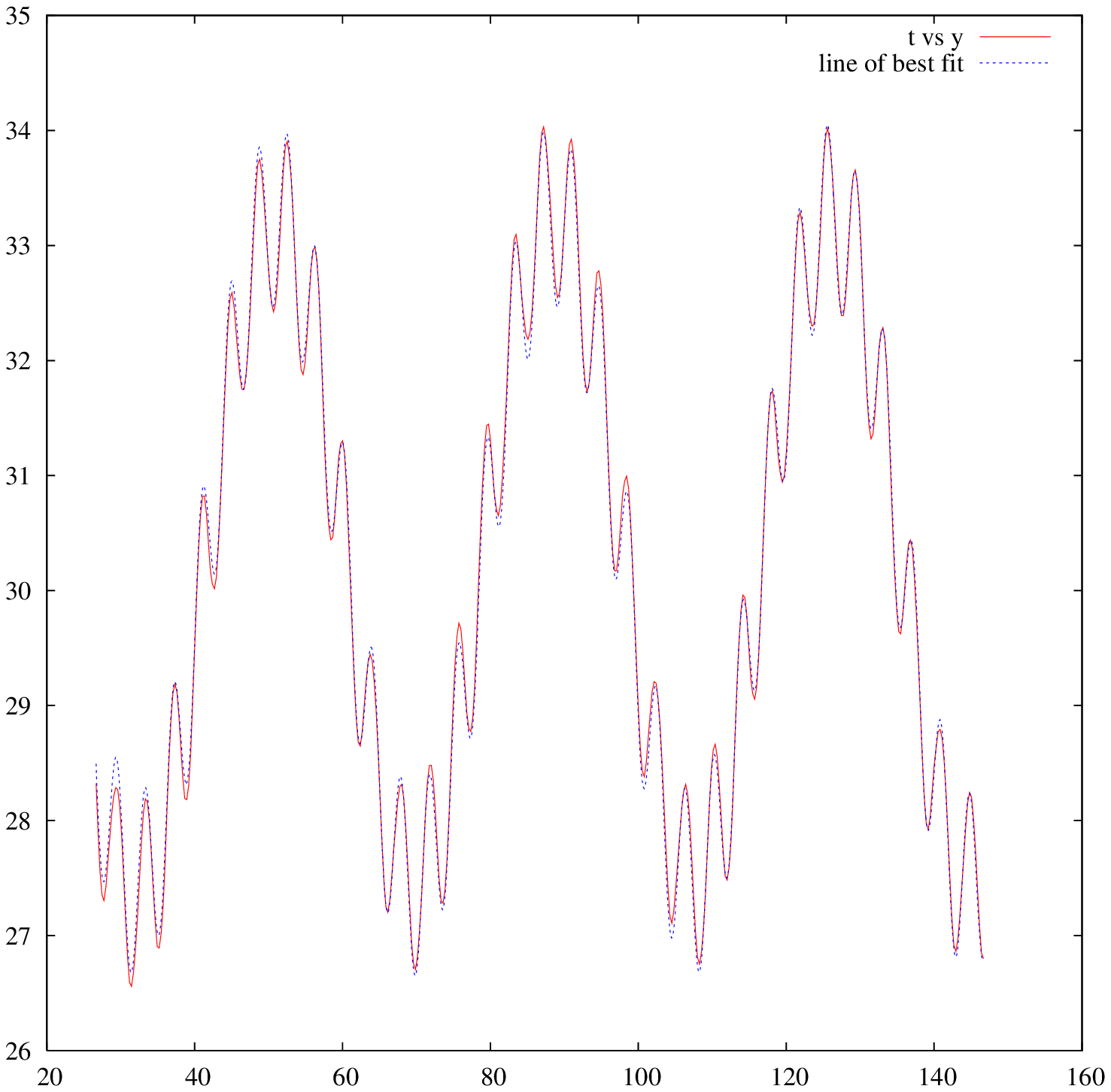}
\end{minipage}
\caption[Meandering spiral wave: $x$ and $y$ components]{Time vs $x(t)$ (left) and Time vs $y(t)$ (right) with the results of the fitting obtained using Gnuplot.}
\label{fig:mrw41fitmore}
\end{center}
\end{figure}

Preliminary results using these initial estimates by Gnuplot are shown in figure \ref{fig:mrw41fitmore}.

We can see that \chg[af]{Eqns.(\ref{eqn:initial_mrw_111})\&(\ref{eqn:initial_mrw_222})}	 fits the data and the equations are given by:

\begin{eqnarray*}
x(t) & = & 30.1212+2.89537\cos(0.167333t-0.674339)\\
&& +0.799068\cos(1.63446t+0.544719)\\
y(t) & = & 30.356+2.9127\sin(0.167285t-0.675932)\\
&& -0.794017\sin(1.63477t+0.531315)
\end{eqnarray*}

We can see that the parameters $R_x$, $r_x$, $\omega_{1x}$, $\omega_{2x}$, $\phi_{1x}$ and $\phi_{2x}$ to 3 signigificant figures can be said to be the same as their $y$ counterparts with the difference being less than 1\%. We conclude that for a meandering spiral wave with a stationary point of rotation we can generate the equation of the tip of the wave with the above equations.

\subsection{Drifting Meandering Wave}

We saw from the previous section that the equations of motion for the tip of a non-drifting meandering wave can be approximated using the following equations:

\begin{eqnarray*}
x(t) & = & x_0+R_x\cos(\omega_{1x}t+\phi_{1x})+r_x\cos(\omega_{2x}t+\phi_{2x})\\
y(t) & = & y_0+R_y\sin(\omega_{1y}t+\phi_{1y})-r_y\sin(\omega_{2y}t+\phi_{2y})
\end{eqnarray*}

We propose that the equations of drifting meandering waves are given by:

\begin{eqnarray*}
x(t) & = & x_0+s_xt+R_x\cos(\omega_{1x}t+\phi_{1x})+r_x\cos(\omega_{2x}t+\phi_{2x})\\
y(t) & = & y_0+s_yt+R_y\sin(\omega_{1y}t+\phi_{1y})-r_y\sin(\omega_{2y}t+\phi_{2y})
\end{eqnarray*}

The only difference being the velocity components for $x$ and $y$. To put this proposition to the test, let us take the data for when the gradient is $a_1$=0.0004. We will use the following initial estimates of the parameters.

\begin{eqnarray*}
x_0 & = & 28.73429 \\
y_0 & = & 28.50880 \\
s_x & = & -0.00763589\\
s_y & = & -0.0104151\\
R & = & 3.25 \\
\omega_1 & = & 0.162496 \\
\phi_1 & = & 1.00\\
r & = & 0.9\\
\omega_2 & = & 1.65\\
\phi_2 & = & 0.8  
\end{eqnarray*}

\begin{figure}[tbp]
\begin{center}
\begin{minipage}[b]{0.49\linewidth}
\centering
\includegraphics[width=\textwidth]{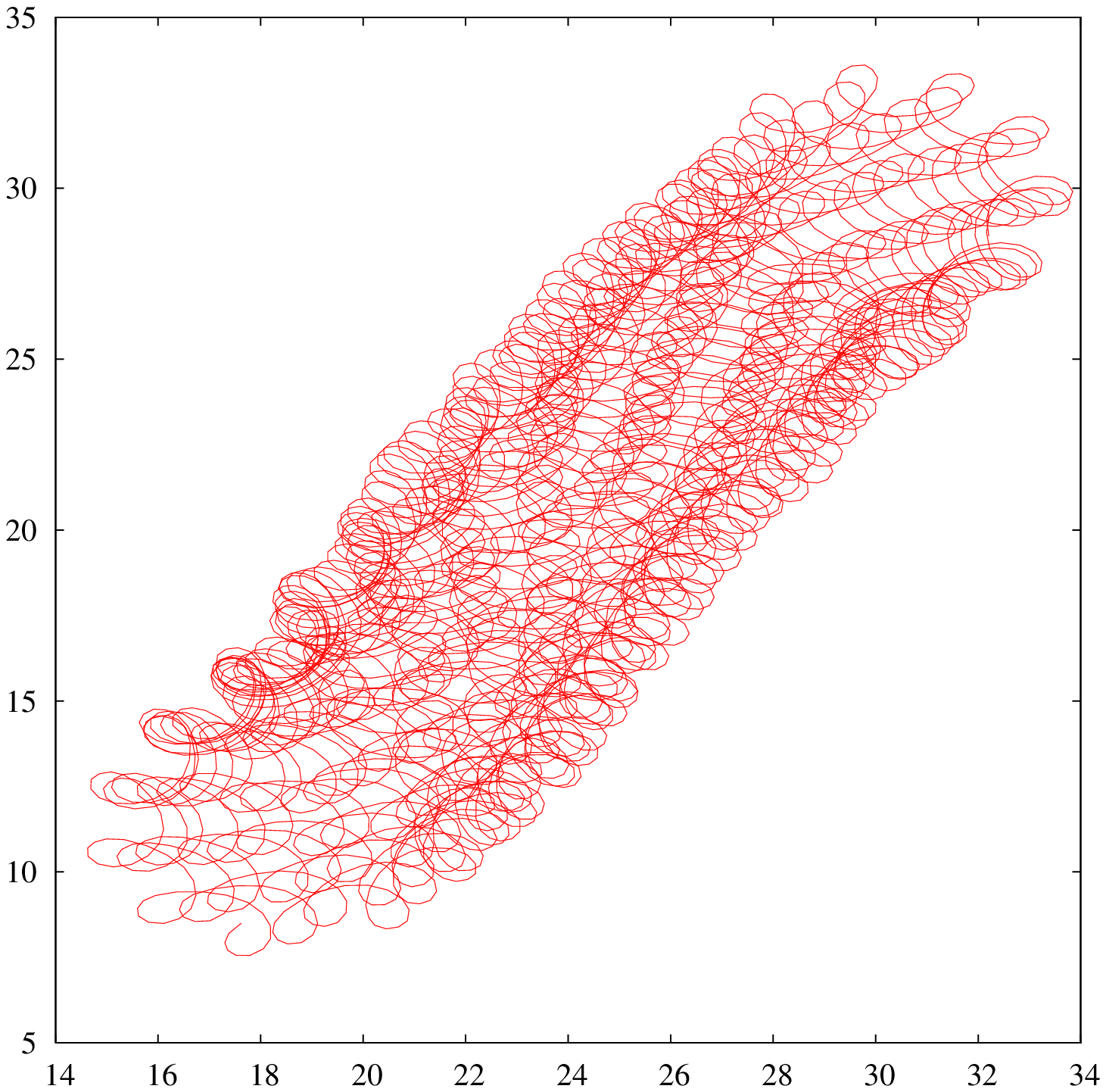}
\end{minipage}
\begin{minipage}[b]{0.49\linewidth}
\centering
\includegraphics[width=\textwidth]{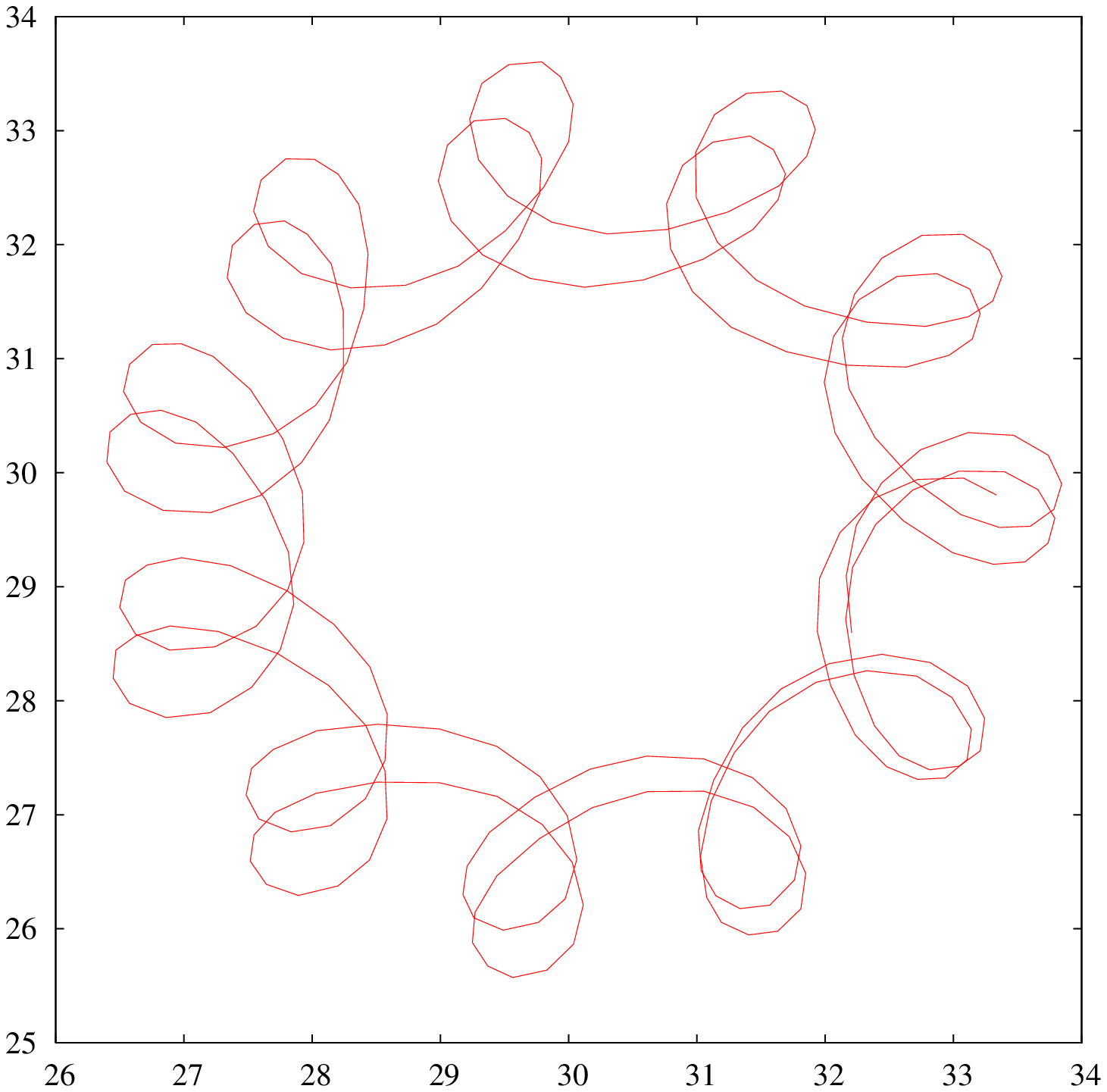}
\end{minipage}
\caption[Drifting and meandering spiral wave: trajectory]{Graphs of the trajectories for a drifting meandering wave with the full trajectory on the left and a part trajectory (2 full rotation) on the right.}
\label{fig:mrw002}
\end{center}
\end{figure}

We can see from \chg[af]{Fig.(\ref{fig:mrw002})} that the wave is moving to the bottom left and actually moves rather slowly. Let us see how Gnuplot fits these functions.

\begin{figure}[tbp]
\begin{center}
\begin{minipage}[b]{0.49\linewidth}
\centering
\includegraphics[width=\textwidth]{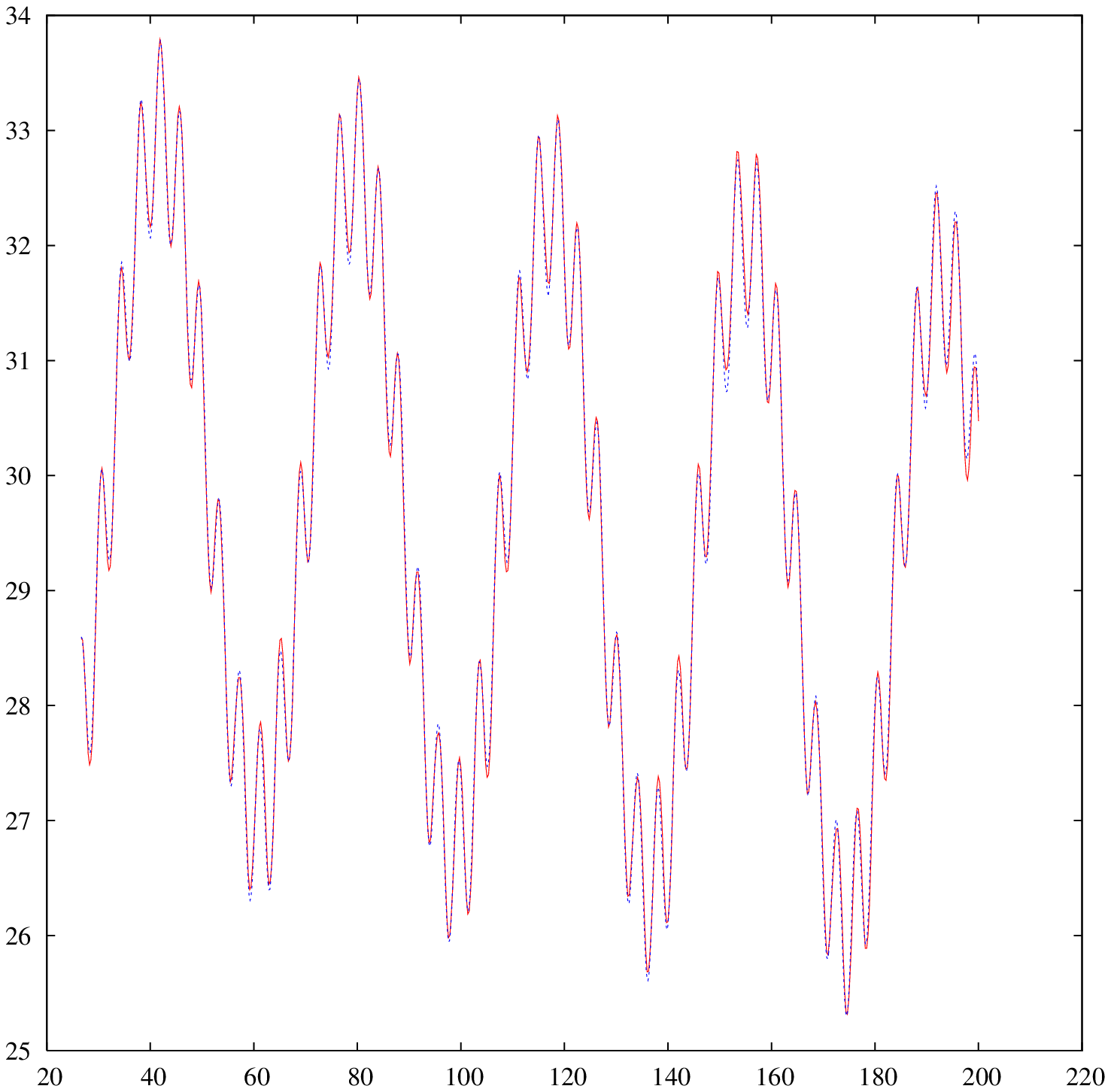}
\end{minipage}
\begin{minipage}[b]{0.49\linewidth}
\centering
\includegraphics[width=\textwidth]{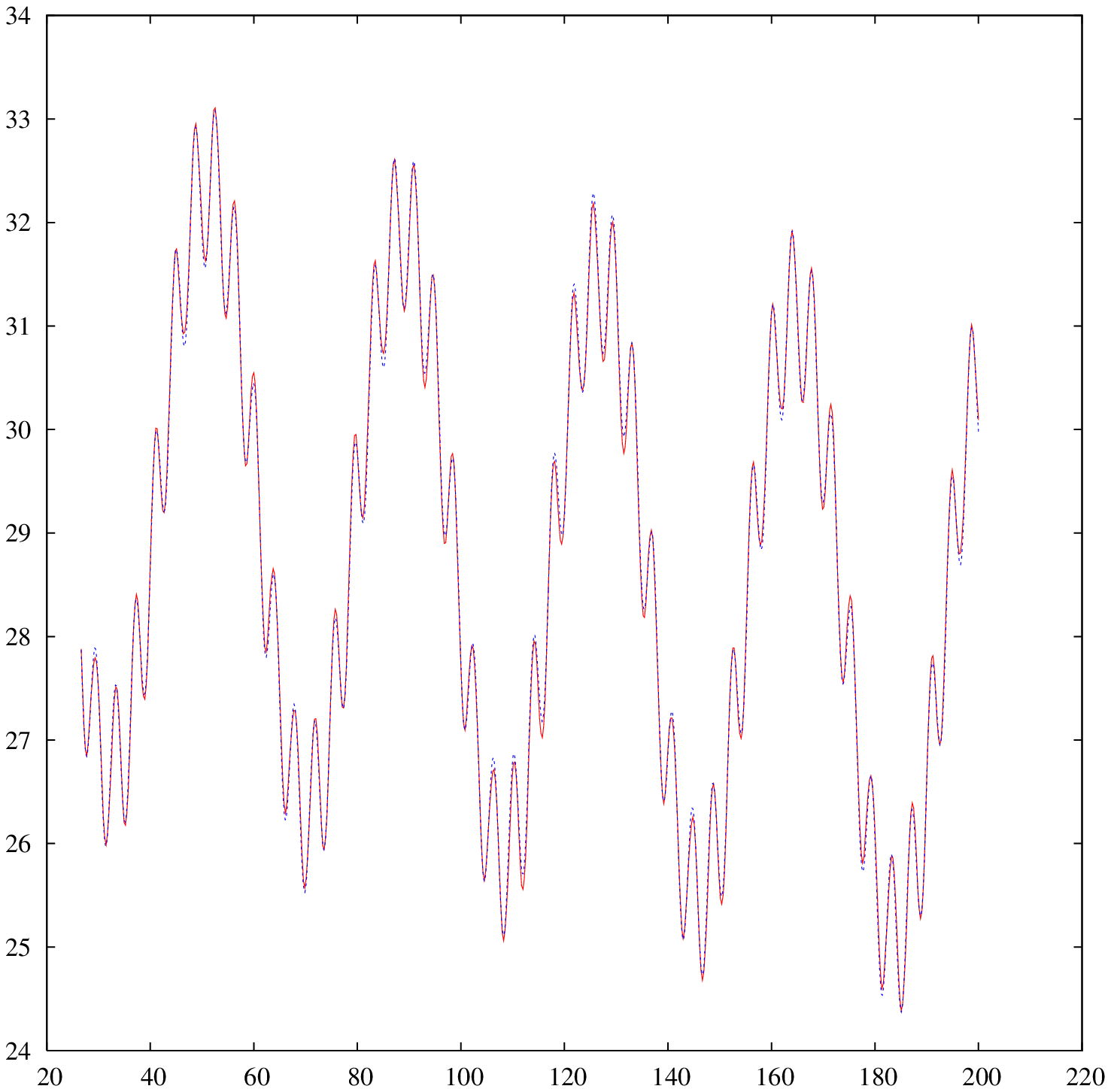}
\end{minipage}
\caption[Drifting and meandering spiral wave: $x$ and $y$ components]{The above graphs show how the functions were fitted to the data with $x(t)$ on the left and $y(t)$ on the right.}
\label{fig:mrw002fit}
\end{center}
\end{figure}

From the graphs in \chg[af]{Fig.(\ref{fig:mrw002fit})}, we can see that Gnuplot has fitted the functions quite well. The functions take the form:

\begin{eqnarray*}
x(t) & = & 30.4127-0.00805152t+2.91423\cos(0.166231t-0.676989)\\
&&+0.795599\cos(1.63444t+0.571694)\\
y(t) & = & 29.9896-0.0107487t+2.92391\sin(0.166176t-0.666409)\\
&& +0.793362\sin(1.63461t-2.59023)
\end{eqnarray*}

\subsection{Conclusion}

We can conclude from the results present in this section, that the general form of the equations of motion for the tip of a meandering wave that drifts is:

\begin{eqnarray*}
x(t) & = & x_0+s_xt+R_x\cos(\omega_{1x}t+\phi_{1x})+r_x\cos(\omega_{2x}t+\phi_{2x})\\
y(t) & = & y_0+s_yt+R_y\sin(\omega_{1y}t+\phi_{1y})-r_y\sin(\omega_{2y}t+\phi_{2y})
\end{eqnarray*}
\\
This was shown to be true numerically and therefore we need to show it is true analytically.
\label{sec:ini_num_eqns_motion}

\section{Conclusion}
We have seen that the systems studied in section \ref{sec:ini_num_inhom} are sensitive to initial conditions. The initial transient before the wave settles down plays a big part in determining the speed of the wave. Therefore, in order to generate accurate results we must get rid of the initial transient by initiating a non-drifting wave that rotates around a point as close to the center of the box as possible and then introduce a gradient to make the wave drift. This will eliminate any errors produced by the initial transient and therefore give us more accurate numerical results.

For frequency locking in the Electrophersis induced drift simulations, we can see that Fig.(\ref{fig:c}) does not give us conclusive evidence of frequency locking. We need to consider the range of values of $a$ between $0.5400\leq a \leq 0.5530$ to see exactly what is happening here.

It would also be useful to see what the graph would look like for other values of $A$ and hopefully produce the ``Arnold Tongue" for this resonance in Barkleys model, if it exists.
\label{sec:ini_num_conc}
\chapter{Numerical Solutions of Spiral Waves in a Moving Frame of Reference}
\label{chap:4}

\section{Introduction}
\label{sec:ezf_numerics_intro}
This chapter is concerned with the numerical solution to the Reaction-Diffusion-Advection system of equations for spiral wave solutions. As mentioned previously, such solutions are spiral wave solutions in a frame of reference that is comoving with the tip of the wave.

The motivation behind this work stemmed from our initial analysis into the drift of meandering spiral waves, subject to symmetry breaking perturbations. In our initial analysis in the laboratory frame of reference, we noted that in certain instances when the drift was ``fast'', the spiral wave reached the boundary very quickly. So, a much larger box size was needed in order to try to extract any meaningful results.

Therefore, our solution to this was to use a frame of reference that was comoving with the tip of the wave (referred to for the rest of this chapter as the ``comoving frame''). This would mean that the tip of the wave would \emph{never} reach the boundary of the box and therefore we could run the program for as long as we wished, safe in the knowledge that the wave would never reach the boundary. Also, we could afford a smaller box size in order to gain faster numerical calculations, provided of course that the box size is not too small such that the boundaries have an \chg[ex]{effect.}

This area has been regarded previously by Beyn \& Thummler, who used a Center Bundle Reduction approach to generated the equations of motion of the top of the spiral wave \cite{freeze}. However, although they managed to pin a rigidly rotating spiral wave, they found it very difficult to pin a periodic (meandering) solution. We will show that it is possible to pin a meandering solution and show this in the examples section.

In the first section of this chapter, we will discuss the numerical implementation of this problem, and discuss the hurdles we had to overcome and how we overcame them. This implementation resulted in a new program called ``EZ-Freeze'', which is an amended version of ``EZ-Spiral'' \cite{barkweb}. We will then show two examples using EZ-Freeze; one for a rigidly rotating spiral wave in Barkley's model, and one for a meandering spiral wave in FHN.

We then discuss the accuracy of EZ-Freeze by showing results from our convergence analysis, before moving onto to two further applications of EZ-Freeze; viz. studies into 1:1 resonance for meandering spiral waves, and large core rigidly rotating spiral waves.

We \chg[p131gram]{} conclude the chapter with a summary of the results, methods developed and questions for the future.

This chapter is based on results which are currently in preparation to be published \cite{Foul08}.

\section{Numerical Implementation}
\label{sec:ezf_numerics_imp}
We will be looking at the numerical solution to the Reaction-Diffusion-Advection system of equations as derived in Chap.\ref{chap:3}. The system of equations derived there (Eqns.(\ref{eqn:theory_drift_101})-(\ref{eqn:theory_drift_102})) is \chg[af]{reminded} below:
\chgex[ex]{}
\begin{eqnarray}
\label{eqn:ezfnum_rda}
\pderiv{\bv}{t} &=& \textbf{D}\nabla^2\bv+\boof(\bv)+(\bc,\nabla)\bv+\omega\pderiv{\bv}{\theta}+\epsilon\bht(\bv,\br,t)
\end{eqnarray}
\vspace{-1.05cm}
\begin{eqnarray}
\label{eqn:ezfnum_tc1b}
v_1(\bR,t)      &=& u_*\\
\label{eqn:ezfnum_tc2b}
v_2(\bR,t)      &=& v_*\\
\label{eqn:ezfnum_tc3b}
\pderiv{\chg[]{v_1(\bR,t)}}{x} &=& 0
\end{eqnarray}
\\
with $\bv,\bR,\boof,\bht\in\mathbb{R}^2$, and $\boof(\bv)$ are the local kinetics. We shall \chgex[ex]{}consider two models, the FHN model \cite{FHN1,FHN2}:
\chgex[ex]{}
\begin{eqnarray*}
\pderiv{u}{t} &=& \nabla^2u+\frac{1}{\varepsilon}\left(u-\frac{u^3}{3}-v\right)+(\bc,\nabla)u+\omega\pderiv{u}{\theta}+\epsilon h_u(u,v,x,y,t)\\
\pderiv{v}{t} &=& D_v\nabla^2v+\varepsilon(u+\beta-\gamma v)+(\bc,\nabla)v+\omega\pderiv{v}{\theta}+\epsilon h_v(u,v,x,y,t)
\end{eqnarray*}
\vspace{-1.05cm}
\begin{eqnarray*}
\label{eqn:ezfnum_tc1_fhnb}
u(\bR,t)      &=& u_*\\
\label{eqn:ezfnum_tc2_fhnb}
v(\bR,t)      &=& v_*\\
\label{eqn:ezfnum_tc3_fhnb}
\pderiv{\chg[]{u(\bR,t)}}{x} &=& 0
\end{eqnarray*}
\\
and Barkley's model \cite{Bark90}:
\chgex[ex]{}
\begin{eqnarray*}
\pderiv{u}{t} &=& \nabla^2u+\frac{1}{\varepsilon}u(1-u)\left[u-\frac{v+b}{a}\right]+(\bc,\nabla)u+\omega\pderiv{u}{\theta}+\epsilon h_u(u,v,x,y,t)\\
\pderiv{v}{t} &=& D_v\nabla^2v+(u-v)+(\bc,\nabla)v+\omega\pderiv{v}{\theta}+\epsilon h_v(u,v,x,y,t)
\end{eqnarray*}
\vspace{-1.05cm}
\begin{eqnarray*}
\label{eqn:ezfnum_tc1_barkb}
u(\bR,t)      &=& u_*\\
\label{eqn:ezfnum_tc2_barkb}
v(\bR,t)      &=& v_*\\
\label{eqn:ezfnum_tc3_barkb}
\pderiv{\chg[]{u(\bR,t)}}{x} &=& 0
\end{eqnarray*}
\\
where $h_u$ and $h_v$ are the $u$ and $v$ components of the perturbation $\bht$, i.e. $\bht=(h_u,h_v)$, and $D_v$ is the ratio of diffusion coefficients, which in all our numerical simulations, is taken to be $D_v=0$.

This system was derived in \chg[p132gram1]{Chap.(\ref{chap:3})} during our analytical work. In the first part of our work we shall consider the case when $\epsilon=0$, but we will also consider $\epsilon\neq0$ later on in this section.

We will call Eqns.(\ref{eqn:ezfnum_tc1b})-(\ref{eqn:ezfnum_tc3b}) the tip pinning conditions, and it is these three conditions which define the Representative Manifold that we introduced in Chap.\ref{chap:3}. The system of Eqns.(\ref{eqn:ezfnum_rda})-(\ref{eqn:ezfnum_tc3b}) will be called the Quotient System, with the phase space $\mathcal{M}=\{\bv,\bc,\omega\}$. For brevity, we will call the $\bc$ and $\omega$ components of the Quotient System Solution, the \emph{Quotient Solution}. If we had just Eqn.(\ref{eqn:ezfnum_rda}), then we would have just 2 equations for 5 unknowns. Therefore, the tip pinning conditions, together with Eqn.(\ref{eqn:ezfnum_rda}), give us a closed system of equations.

We shall see in Sec.(\ref{sec:ezfnum_tpc}) that the above tip pinning conditions do not give accurate calculations of $\omega$, and therefore we will be using refined conditions which will be introduced and described in detail in that section.

Let us consider the physical interpretation of the advection terms. The reaction and diffusion parts of the system, can give us a spiral wave solution (for particularly chosen parameters and initial conditions), and in the absence of advection, the solution will be in the laboratory frame of reference. The advection terms, for carefully chosen advection coefficients ($\bc, \omega$), actually move the frame of reference such that it is moving with the tip of the spiral wave. So, the crucial point to note here is that the spiral wave in the comoving frame is simply the spiral wave in the laboratory frame transformed so that the tip is in \chg[p132gram2]{}a particular position and with a particular phase (these are discussed below).

So, the idea we have in this part of our research, is to use operator splitting. We take a spiral wave solution as generated using the numerical methods adopted by Dwight Barkley in EZ-Spiral \cite{bark91,barkweb}, and then use the following methods detailed below to solve the advection terms. Therefore, the numerical equations to solve are:

\begin{eqnarray*}
\label{eqn:ezfnum_step1}
\bv^{n+\frac{1}{2}}_{i,j} &=& \bv^{n}_{i,j}+\Delta_t\mathcal{R}(\bv^{n}_{i,j})+O(\Delta_t^2)\\
\label{eqn:ezfnum_step2}
\bv^{n+1}_{i,j} &=& \bv^{n+\frac{1}{2}}_{i,j}+\Delta_t\mathcal{A}(\bv^{n+\frac{1}{2}}_{i,j})+O(\Delta_t^2)\
\end{eqnarray*}
\\
where $\bv^{n}_{i,j}$ is $\bv$ at the $n$-th time step and at the grid coordinate $(i,j)$, and $\mathcal{R}$ and $\mathcal{A}$ are the Reaction-Diffusion and Advection terms respectively, and $\Delta_t$ is the time step. We note here that we have used an explicit, forward Euler method to first order accuracy to calculate the time derivatives.

Let us consider the numerical methods implemented in the Reaction-Diffusion part of the system. We shall be using a first order accurate forward Euler method to calculate the temporal derivatives, and the Five Point Laplacian method for the Laplacian. These are all detailed extensively in \cite{bark91}.

Let $\bR=(X,Y)$ be the tip coordinates. Now, we choose $\bc$ and $\omega$ such that the tip of the wave remains in a fixed position and at a fixed orientation for all time. This means that $\bR$ is fixed and we are free to choose the fixed value of $\bR$ as we feel fit. We choose to fix $\bR$ at the center of the box. This would mean that the tip \chgex[ex]{is the maximum possible distance away} from any boundary. We also take the center of the box to be the origin, and therefore we have that $\bR=(0,0)$.

So, how exactly do we numerically solve the second step (\ref{eqn:ezfnum_step2}) above? The differential equation is:

\begin{equation}
\label{eqn:ezfnum_second_half_step}
\pderiv{\bv^{n+1}}{t} = c_x\pderiv{\bv^{n+\frac{1}{2}}}{x}+c_y\pderiv{\bv^{n+\frac{1}{2}}}{y}+\omega\pderiv{\bv^{n+\frac{1}{2}}}{\theta}
\end{equation}

There are two main calculations that we must do which are coupled. Firstly we must calculate the advection coefficients, $c_x$, $c_y$, and $\omega$. Secondly, we need to calculate the spatial derivatives which appear in (\ref{eqn:ezfnum_second_half_step}). Although the two are connected, we will describe the methods to solve these separately.

\subsection{Advection \chg[af]{Coefficients:} $c_x$, $c_y$, and $\omega$}
\label{sec:ezfnum_quot}

We devised two methods to calculate $c_x$, $c_y$ and $\omega$, which we will describe in detail in the following pages. The first method (Method 1) involves measuring how far the tip of the wave has traveled after the first half step and determining what $c_x$, $c_y$ and $\omega$ are needed in order to bring back the tip to the desired position. The second method (Method 2) is a more concrete and accurate method which involves the solution to a system of three linear equations \footnote{This method was suggested by Dr B. Vasiev, Mathematical Sciences, University of Liverpool}.

\subsubsection{Method 1: Shifts in the Euclidean Space.}

\begin{figure}[btp]
\begin{center}
\begin{minipage}[htbp]{0.32\linewidth}
\centering
\psfrag{a}[l]{$(X,Y)$}
\psfrag{b}[l]{$(X_1,Y_2)$}
\psfrag{c}[l]{$(x_c,y_c)$}
\includegraphics[width=0.8\textwidth]{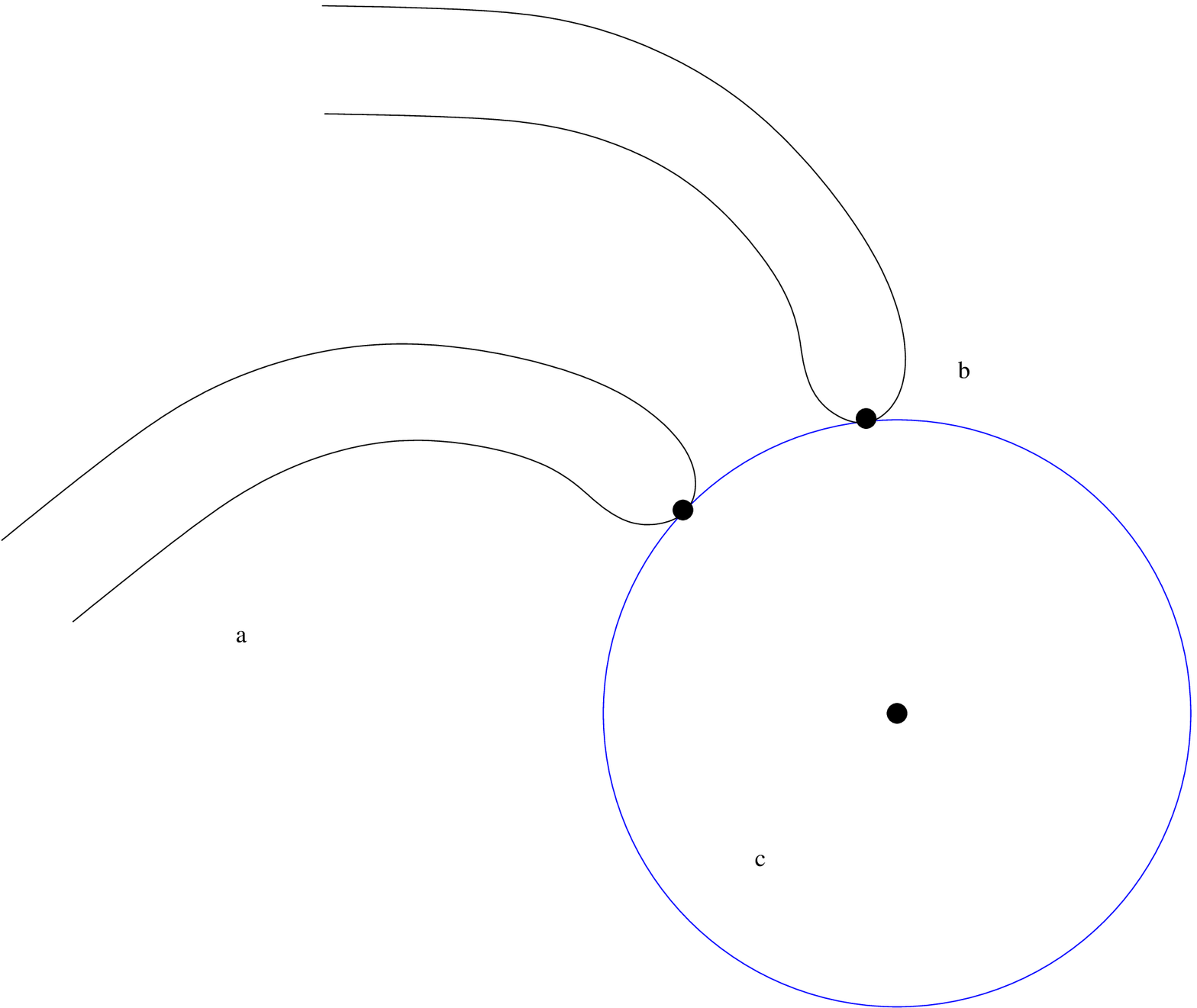}
\end{minipage}
\begin{minipage}[htbp]{0.32\linewidth}
\centering
\psfrag{b}[l]{$\Delta X$}
\psfrag{c}[l]{$\Delta Y$}
\includegraphics[width=0.8\textwidth]{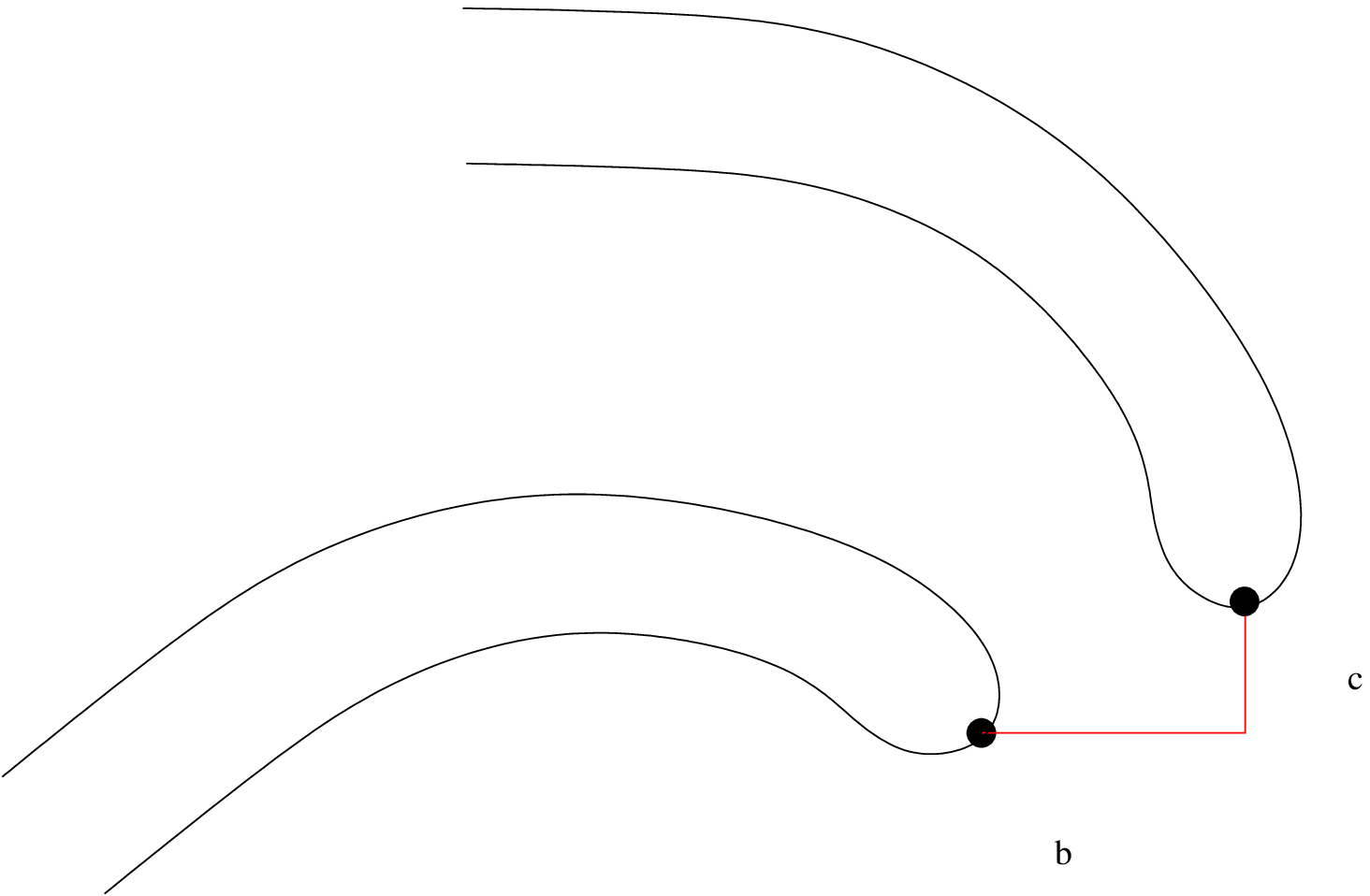}
\end{minipage}
\begin{minipage}[htbp]{0.32\linewidth}
\centering
\psfrag{c}[l]{$\Delta\theta$}
\includegraphics[width=0.8\textwidth]{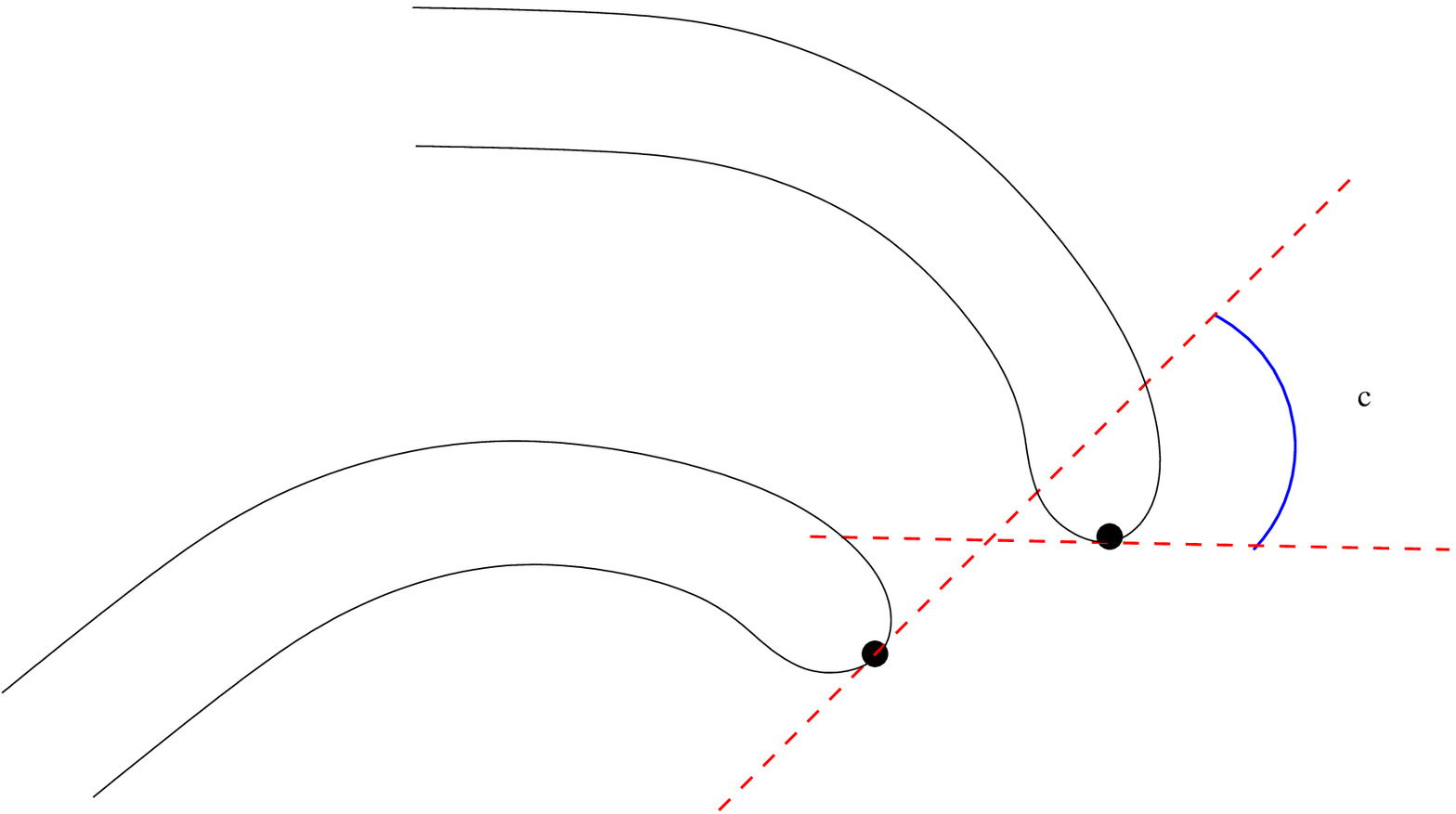}
\end{minipage}
\caption{Calculation of the translational and rotational shifts.}
\label{fig:ezf_num_meth1}
\end{center}
\end{figure}

Consider now the spiral wave solution in the first half step, and assume that the solution, $\bv^n$, i.e. the solution before the first half step is applied, had its tip at the desired position ($X$,$Y$), and with the desired orientation, $\Theta$, Fig.(\ref{fig:ezf_num_meth1}). \chg[p134gram]{}

Once the Reaction and Diffusion parts have been applied to our solution, the tip has moved to a new position, ($X_1$,$Y_1$), and has a new phase, $\Theta_1$. Therefore, the translational and rotational shifts are $\Delta X=X_1-X$, $\Delta Y=Y_1-Y$, $\Delta\Theta=\Theta_1-\Theta$.

Now, consider the equations of motion of the tip of the spiral wave:

\begin{eqnarray*}
\deriv{R}{t} &=& ce^{i\Theta}\\
\deriv{\Theta}{t} &=& \omega
\end{eqnarray*}
\\
where $R=X+iY$ and $c=c_x+ic_y$. These can be written in numerical terms, using an Explicit Forward Euler Method as:

\begin{eqnarray*}
\label{eqn:ezfnum_quot_1_c}
\frac{\Delta R}{\Delta_t} &=& ce^{i\Theta}\\
\label{eqn:ezfnum_quot_1_om}
\frac{\Delta\Theta}{\Delta_t} &=& \omega
\end{eqnarray*}

Therefore, an approximation for $\omega$ is given by:

\begin{eqnarray*}
\omega &=& \frac{\Theta_1-\Theta}{\Delta_t} 
\end{eqnarray*}
\\
where $\Delta_t$ is the time step.

\chg[p134spell]{Let} us consider $c$. We see that:

\begin{eqnarray*}
c &=& \frac{\Delta R}{\Delta_t}e^{-i\Theta}\\
\Rightarrow c &=& \frac{\Delta R}{\Delta_t}e^{-i(\Theta_1-\omega\Delta t)}
\end{eqnarray*}

Since we can choose $\Theta_1$ to be arbitrary, we choose $\Theta_1=0$. This give us:

\begin{eqnarray*}
\frac{\Delta\bR}{\Delta_t} &=& \bc e^{-i\omega\Delta t}\\
\frac{\Delta\bR}{\Delta_t} &=& \bc(1+O(\Delta_t))
\end{eqnarray*}

Hence we have:

\begin{equation*}
\bc = \frac{\bR_1-\bR}{\Delta_t}
\end{equation*}

Summarizing, we have:

\begin{eqnarray*}
\omega &=& \frac{\Theta_1-\Theta}{\Delta_t} \\
c_x    &=& \frac{X_1-X}{\Delta_t}\\
c_y    &=& \frac{Y_1-Y}{\Delta_t}
\end{eqnarray*}

Hence, we have a first order accurate scheme to calculate the quotient system, which is straight forward and quick. However, as we shall see in the examples following, it is unstable.

\subsubsection{Method 2: Solution to a System of Linear Equations in $c_x$, $c_y$ and $\omega$.}

We now consider creating a system of linear equations which can be solved to give us the quotient system. We have three unknowns; $c_x$, $c_y$ and $\omega$. Therefore, we require three linear equations in $c_x$, $c_y$ and $\omega$. We derive these equations from Eqn.(\ref{eqn:ezfnum_second_half_step}), and also consider the tip pinning conditions (\ref{eqn:ezfnum_tc1b})-(\ref{eqn:ezfnum_tc3b}). We note that in these tip pinning conditions, we consider when the tip is at the origin, $\bR=(0,0)$.

Consider for a moment the tip pinning conditions:

\begin{eqnarray*}
v_1(0,0,t)      &=& u_*\\
v_2(0,0,t)      &=& v_*\\
\pderiv{v_1(0,0,t)}{x} &=& 0
\end{eqnarray*}

The first two are fairly straight forward to understand. They say that the values of $v_1$ and $v_2$ at the desired position are fixed for all time. The third condition, says:

\begin{eqnarray*}
\pderiv{v_1(0,0,t)}{x} &=& 0\\
\frac{v_1(\Delta_x,0,t)-v_1(0,0,t)}{\Delta_x} &=& 0\\
\Rightarrow v_1(\Delta_x,0,t) &=& v_1(0,0,t)
\end{eqnarray*}

Therefore, the third pinning condition interprets as the value of $v_1$ at the grid point in the $x$-direction next to the desired point is the same as that at the desired point. So, our linear system of three equations considers the values of $v_1$ at the desired point and the point next to it, and also the value of $v_2$ at the desired point, giving:

\begin{eqnarray*}
\pderiv{v_1(0,0,t)}{t} &=& c_x\pderiv{v_1(0,0,t)}{x}+c_y\pderiv{v_1(0,0,t)}{y}+\omega\pderiv{v_1(0,0,t)}{\theta}\\
\pderiv{v_2(0,0,t)}{t} &=& c_x\pderiv{v_2(0,0,t)}{x}+c_y\pderiv{v_2(0,0,t)}{y}+\omega\pderiv{v_2(0,0,t)}{\theta}\\
\pderiv{v_1(\Delta_x,0,t)}{t} &=& c_x\pderiv{v_1(\Delta_x,0,t)}{x}+c_y\pderiv{v_1(\Delta_x,0,t)}{y}\\
&& +\omega\pderiv{v_1(\Delta_x,0,t)}{\theta}
\end{eqnarray*}

We note that the derivatives with respect to $\theta$ can be rewritten as:

\begin{equation*}
\partial_\theta = x\partial_y-y\partial_x
\end{equation*}
\\
and we also denote $v_1(\Delta_x,0,t)=\til{v}_1(0,0,t)$.

Next, we use the tip pinning conditions and find that the linear system now becomes:

\begin{eqnarray*}
\pderiv{v_1(0,0,t)}{t} &=& c_x\pderiv{v_1(0,0,t)}{x}+c_y\pderiv{v_1(0,0,t)}{y}\\
\pderiv{v_2(0,0,t)}{t} &=& c_x\pderiv{v_2(0,0,t)}{x}+c_y\pderiv{v_2(0,0,t)}{y}\\
\pderiv{\til{v}_1(0,0,t)}{t} &=& c_x\pderiv{\til{v}_1(0,0,t)}{x}+c_y\pderiv{\til{v}_1(0,0,t)}{y}+\omega\Delta{x}\pderiv{\til{v}_1(0,0,t)}{y}
\end{eqnarray*}

These are then easily solved using linear algebra to give:

\begin{eqnarray*}
c_x &=& \frac{\partial_x{v_1(0,0,t)}\partial_t{v_2(0,0,t)}-\partial_x{v_2(0,0,t)}\partial_t{v_1(0,0,t)}}{\partial_x{v_1(0,0,t)}\partial_y{v_2(0,0,t)}-\partial_y{v_1(0,0,t)}\partial_x{v_2(0,0,t)}}\\
c_y &=& \frac{\partial_y{v_1(0,0,t)}\partial_t{v_2(0,0,t)}-\partial_y{v_2(0,0,t)}\partial_t{v_1(0,0,t)}}{\partial_x{v_1(0,0,t)}\partial_y{v_2(0,0,t)}-\partial_y{v_1(0,0,t)}\partial_x{v_2(0,0,t)}}\\
\omega &=& \frac{\partial_t{\til{v}_1(0,0,t)}-c_x\partial_x{\til{v}_1(0,0,t)}-c_y\partial_y{\til{v}_1(0,0,t)}}{\Delta{x}\partial_y{\til{v}_1(0,0,t)}}
\end{eqnarray*}
\\
where $\partial_x$ represents partial differentiation with respect to $x$; similarly for $y$ and $t$.

As we have already mentioned, it is possible to solve this system since we already know the solution $\bv^{n+\frac{1}{2}}$.

In order to implement this system into our code, we must now discuss how we can approximate the spatial derivatives.

\subsection{Numerical Approximation of the Spatial Derivatives}

First, we implemented an explicit, upwind Euler method:

\begin{eqnarray*}
\alpha\pderiv{\bv(x,y,t)}{x} &=& \alpha\frac{\bv(x+\Delta_x,y,t)-\bv(x,y,t)}{\Delta_x}+O(\Delta_x^2)\quad\mbox{if}\quad\alpha>0\\
\alpha\pderiv{\bv(x,y,t)}{x} &=& \alpha\frac{\bv(x,y,t)-\bv(x-\Delta_x,y,t)}{\Delta_x}+O(\Delta_x^2)\quad\mbox{if}\quad\alpha<0
\end{eqnarray*}
\\
where $\Delta_x$ is the shift in the $x$-direction. Note that we do not need to consider when the advection coefficient is $\alpha=0$ in the above equations, since $\alpha=0$ means that the particular advection will not be present!

As we will show in the next section section, this approximation is not accurate enough and therefore we need a second order scheme. We therefore implement a second order accurate, explicit upwind scheme which is given below \cite{bik98}:

\begin{small}
\begin{eqnarray*}
               \left.\alpha\pderiv{u}{x}\right|_{(x,y)} &\approx& \frac{\alpha}{2\Delta_x}(-3u(x,y)+4u(x+\Delta_x,y)-u(x+2\Delta_x,y))\quad\alpha>0\\
\mbox{or}\quad \left.\alpha\pderiv{u}{x}\right|_{(x,y)} &\approx& \frac{\alpha}{2\Delta x}(3u(x,y)-4u(x+\Delta_x,y)+u(x-2\Delta_x,y))\quad\alpha<0
\end{eqnarray*}
\end{small}

\subsection{Tip Pinning Conditions}
\label{sec:ezfnum_tpc}

As mentioned earlier, we found that the calculation of the quotient system using the tip pinning conditions (\ref{eqn:ezfnum_tc1b})-(\ref{eqn:ezfnum_tc3b}) appeared to be less accurate than required. We will show in the next section, just how inaccurate this was. Therefore, a refined set of conditions was required.

Consider the conditions that we already have:

\begin{eqnarray*}
v_1(0,0,t)      &=& u_*\\
v_2(0,0,t)      &=& v_*\\
\pderiv{v_1(0,0,t)}{x} &=& 0
\end{eqnarray*}

As we have seen in Sec.(\ref{sec:ezfnum_quot}), the third condition can be interpreted numerically as:

\begin{eqnarray*}
v_1(\Delta_x,0,t) &=& u_*
\end{eqnarray*}
\\
where $\Delta_x$ is the spacestep in the $x$-direction. So, this means that the third pinning condition implies that the second pinning point is the point next to the origin in the $x$-direction. Hence, if the third condition is the condition which helps determine $\omega$, then the calculation of $\omega$ may not be as accurate as, for instance, if we had a pinning point which was further away.

\begin{figure}[btp]
\begin{center}
\begin{minipage}[htbp]{0.6\linewidth}
\centering
\psfrag{a}[l]{A}
\psfrag{b}[l]{B}
\psfrag{c}[l]{C}
\psfrag{d}[l]{$\theta$}
\includegraphics[width=0.7\textwidth]{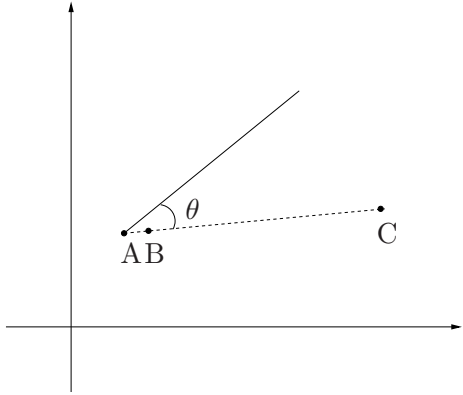}
\end{minipage}
\caption{Analogy behind the third pinning condition.}
\label{fig:ezfnum_analogy}
\end{center}
\end{figure}

To see why this is true, consider the following analogy. Say for instance that we wanted to calculate the angle between two lines. One line is referenced to the object that we are considering. The other is drawn between two points - one which is fixed and is the interception of the two lines, and the other is chosen arbitrarily. See Fig.(\ref{fig:ezfnum_analogy}).

So, if we were to draw the second line between points A and B, we would find it quite difficult to get an accurate measurement of the angle since the reference line would be very short, especially if we have that A and B are next to each other, \chgex[ex]{similar to} what we have in our spiral wave case. However, if we were to use the points A and C to draw the reference line then we would find that the angle could be drawn more accurately since the reference line is bigger and more prominent.

So, with this in mind, we propose to replace the third pinning with the following pinning condition:

\begin{equation}
\label{eqn:exfnum_tc3_new}
v_1(X_{inc},Y_{inc},t) = u_*
\end{equation}

This is a variant of the original condition, but this time we have that the second pinning point is now at $(X_{inc},Y_{inc})$ where $X_{inc}$ and $Y_{inc}$ are chosen arbitrarily, but cannot be zero simultaneously and must also remain within the box. 

However, it was noted in our simulations conducted for the results of the convergence testing in Sec.(\ref{sec:ezf_convergence}) that the optimum distance between the pinning points must not exceed one full wavelength of the spiral wave.

Let us revisit the calculation of the Quotient system. Our linear system is now:

\begin{footnotesize}
\begin{eqnarray*}
\pderiv{v_1(0,0,t)}{t} &=& c_x\pderiv{v_1(0,0,t)}{x}+c_y\pderiv{v_1(0,0,t)}{y}+\omega\pderiv{v_1(0,0,t)}{\theta}\\
\pderiv{v_2(0,0,t)}{t} &=& c_x\pderiv{v_2(0,0,t)}{x}+c_y\pderiv{v_2(0,0,t)}{y}+\omega\pderiv{v_2(0,0,t)}{\theta}\\
\pderiv{v_1(X_{inc},Y_{inc},t)}{t} &=& c_x\pderiv{v_1(X_{inc},Y_{inc},t)}{x}+c_y\pderiv{v_1(X_{inc},Y_{inc},t)}{y}\\
&& +\omega\pderiv{v_1(X_{inc},Y_{inc},t)}{\theta}
\end{eqnarray*}
\end{footnotesize}

Letting \chg[p139eqn]{}$\til{v}_1(0,0,t)=v_1(X_{inc},\chg[]{Y_{inc}},t)$, and using the tip pinning conditions we find that the system now becomes:

\begin{eqnarray*}
\pderiv{v_1(0,0,t)}{t} &=& c_x\pderiv{v_1(0,0,t)}{x}+c_y\pderiv{v_1(0,0,t)}{y}\\
\pderiv{v_2(0,0,t)}{t} &=& c_x\pderiv{v_2(0,0,t)}{x}+c_y\pderiv{v_2(0,0,t)}{y}\\
\pderiv{\til{v}_1(0,0,t)}{t} &=& c_x\pderiv{\til{v}_1(0,0,t)}{x}+c_y\pderiv{\til{v}_1(0,0,t)}{y}\\
                             && +\omega\left(X_{inc}\pderiv{\til{v}_1(0,0,t)}{y}-Y_{inc}\pderiv{\til{v}_1(0,0,t)}{x}\right)
\end{eqnarray*}

The solution to this system is therefore:

\begin{eqnarray*}
c_x &=& \frac{\partial_x{v_1(0,0,t)}\partial_t{v_2(0,0,t)}-\partial_x{v_2(0,0,t)}\partial_t{v_1(0,0,t)}}{\partial_{v_x1(0,0,t)}\partial_y{v_2(0,0,t)}-\partial_y{v_1(0,0,t)}{y}\partial_x{v_2(0,0,t)}}\\
c_y &=& \frac{\partial_y{v_1(0,0,t)}\partial_t{v_2(0,0,t)}-\partial_y{v_2(0,0,t)}\partial_t{v_1(0,0,t)}}{\partial_x{v_1(0,0,t)}\partial_y{v_2(0,0,t)}-\partial_y{v_1(0,0,t)}\partial_x{v_2(0,0,t)}}\\
\omega &=& \frac{\partial_t{\til{v}_1(0,0,t)}-c_x\partial_x{\til{v}_1(0,0,t)}-c_y\partial_y{\til{v}_1(0,0,t)}}{X_{inc}\partial_y{\til{v}_1(0,0,t)}-Y_{inc}\partial_x{\til{v}_1(0,0,t)}}
\end{eqnarray*}

Also, two constraints that we must impose is that if $X_{inc}=0$ then $Y_{inc}\neq0$, and vice versa. Also $|X_{inc}|<\frac{L_x}{2}$ and $|Y_{inc}|<\frac{L_y}{2}$, where $L_x$ and $L_y$ are the lengths of the box in the $x$ and $y$ direction respectively. 

\begin{figure}[btp]
\begin{center}
\begin{minipage}[htbp]{0.6\linewidth}
\centering
\psfrag{a}[l]{A}
\psfrag{b}[l]{B}
\includegraphics[width=0.7\textwidth]{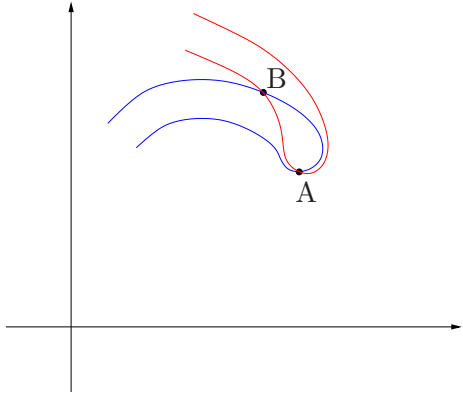}
\end{minipage}
\caption{Non-uniqueness of the revised tip pinning condition.}
\label{fig:ezfnum_unique1}
\end{center}
\end{figure}

Let us, for a moment, consider the pinning points. The solution to Eqns.(\ref{eqn:ezfnum_rda})-(\ref{eqn:ezfnum_tc2b}) \& (\ref{eqn:exfnum_tc3_new}) does not have a unique solution. Consider Fig.(\ref{fig:ezfnum_unique1}).

In Fig.(\ref{fig:ezfnum_unique1}), we see that the two positions shown satisfy the three tip pinning conditions. As we can see, we have two pinning points A and B. Although point A fixes the tip of the wave at the origin, the second point can fix a point on either the front of the wave (\textcolor{blue}{blue}) or on the tail (\textcolor{red}{red}). 

To overcome this we note that the values of the variable, whether it be the \chg[p140spell]{$v_1$ or $v_2$} variable, within the isoline is higher than the value of the variable outside the isoline. So, so if we want to attach the second point to the front or the tail of the wave then we impose the conditions:

\begin{eqnarray*}
\mbox{front}:\quad v_1(X_{inc}-\delta x,Y_{inc},t)>v_1(X_{inc}+\delta x,Y_{inc},t)\\
\mbox{tail}: \quad v_1(X_{inc}-\delta x,Y_{inc},t)<v_1(X_{inc}+\delta x,Y_{inc},t)
\end{eqnarray*}
\\
where $\delta x$ is a small perturbation from $X+X_{inc}$ along the $x$-axis.


\subsection{Physical Implementation \& Stability}

We must mention at this point just how the advection terms are implemented into the program EZ-Freeze and just how we overcome the question of instabilities.

Firstly, we note that if the tip position at the moment before the advection terms are activated is at a position which is not at the first pinning point (i.e. the desired position of the tip, which is taken as the point (0,0) in our implementation), then the initial values of $c_x$, $c_y$ and $\omega$ could possibly be very large. This in turn could lead to numerical instabilities and the program crashing.

Now the stability conditions for $c_x$, $c_y$ and $\omega$, were taken to be:

\begin{eqnarray*}
|c_x| &\leq& \frac{\Delta_x^2}{2\Delta_t}\\
|c_y| &\leq& \frac{\Delta_x^2}{2\Delta_t}\\
|\omega| &\leq& \frac{1}{N_X\Delta_t}
\end{eqnarray*}
\\
where $\Delta_x$ is the spacestep.

If the absolute values of $c_x$ and $c_y$ were greater than these limits then $c_x$ and $c_y$ were restricted to the maximum (or minimum if $c_x$ or $c_y$ were negative) values stated above. Also, we eliminated the need to restrict the values of $c_x$ and $c_y$ to their stability limits by getting the program to physically move the spiral wave solution so that the tip of the spiral wave is in the center of the box. 

For $\omega$, we implemented the restriction that if $|\omega|$ exceeded its maximum stability value, then $\omega=0$. This lead to less instabilities occurring in the solutions.

\subsection{Boundary Conditions}

In EZ-Spiral, the user can specify either Neumann boundary conditions or periodic boundary conditions. In EZ-Freeze, we use both Neumann boundary condition or Dirichlet boundary condition. No periodic boundary conditions are used.

Numerically, Dirichlet boundary conditions can be translated as:
\chg[af]{}
\begin{eqnarray*}
\bv\left(x            ,-\frac{L_y}{2},t\right)  &=& 0\\
\bv\left(x            ,\chg[]{\frac{L_y}{2}},t\right)  &=& 0\\\
\bv\left(-\frac{L_x}{2},y            ,t\right)  &=& 0\\
\bv\left(\frac{L_x}{2} ,y            ,t\right)  &=& 0
\end{eqnarray*}
\\
where $L_x$ and $L_y$ are the physical sizes of the boxes (remember that we take the center of the box to be at the \chg[af]{origin} $(0,0)$), and $-\frac{L_x}{2}\leq x\leq\frac{L_x}{2}$ and $-\frac{L_y}{2}\leq y\leq\frac{L_y}{2}$. Numerically we have:

\begin{eqnarray*}
\hat{\bv}(i ,0 ,t)  &=& 0\\
\hat{\bv}(i ,N_Y,t)  &=& 0\\\
\hat{\bv}(0 ,j ,t)  &=& 0\\
\hat{\bv}(N_X,j ,t)  &=& 0
\end{eqnarray*}
\\
where $\hat{\bv}(i,j,t)$ is the numerically approximation to $\bv(x,y,t)$\chg[p142gram]{,} $N_X$, $N_Y$ are the number of grid points in the numerical discretization, and $i,j\in\mathbb{Z}$ with $0\leq i\leq N_X$ and $0\leq j\leq N_Y$. We also assume that any point outside of the box (which we need for spatial derivatives for both diffusion and advection) are automatically zero.

On the other hand, Neumann boundary condition are given as:

\begin{eqnarray*}
\pderiv{\bv}{\br} &=& 0
\end{eqnarray*}
\\
which can be interpreted numerically as:

\begin{eqnarray*}
\hat{\bv}(i    ,-1   ,t) &=& \hat{\bv}(i    ,1    ,t)\\
\hat{\bv}(i    ,N_Y-1,t) &=& \hat{\bv}(i    ,N_Y+1,t)\\\
\hat{\bv}(-1   ,j    ,t) &=& \hat{\bv}(1    ,j    ,t)\\
\hat{\bv}(N_X-1,j    ,t) &=& \hat{\bv}(N_X+1,j    ,t)
\end{eqnarray*}
\\
where $i,j\in\mathbb{Z}$ with $0\leq i\leq N_X$ and $0\leq j\leq N_Y$.

When we come to use the second order scheme for spatial derivatives in the advection terms, we also use the following additional scheme.

\begin{eqnarray*}
\hat{\bv}(i    ,-2   ,t)  &=& \hat{\bv}(i    ,2    ,t)\\
\hat{\bv}(i    ,N_Y-2,t)  &=& \hat{\bv}(i    ,N_Y+2,t)\\\
\hat{\bv}(-2   ,j    ,t)  &=& \hat{\bv}(2    ,j    ,t)\\
\hat{\bv}(N_X-2,j    ,t)  &=& \hat{\bv}(N_X+2,j    ,t)
\end{eqnarray*}


\subsection{The case for $\epsilon\neq0$.}

We now describe when we have symmetry breaking perturbations within the system. Throughout this project we are concerned with only three types of perturbations: resonant drift (time dependent perturbations); Electrophoresis induced drift (rotational breaking perturbations); and Inhomogeneity induced drift (parameters of the system are dependent on the spatial coordinates).

In order to implement these three different types of drift into EZ-Freeze, we need to consider the transformed perturbations, the derivation of which are given in \chg[p143gram]{Chap.\ref{chap:3}}.

All perturbations are implemented into EZ-Freeze using the same technique. Again, we use operator splitting and instead of having just two steps as we did for the case when we had no perturbation ($\epsilon=0$), we now have three steps:

\begin{eqnarray*}
\label{eqn:ezfnum_step1_dr}
\bv^{n+\frac{1}{3}}_{i,j} &=& \bv^{n}_{i,j}+\Delta_t\mathcal{R}(\bv^{n}_{i,j})+O(\Delta_t^2)\\
\label{eqn:ezfnum_step2_dr}
\bv^{n+\frac{2}{3}}_{i,j} &=& \bv^{n+\frac{1}{3}}_{i,j}+\Delta_t\mathcal{D}(\bv^{n+\frac{1}{3}}_{i,j})+O(\Delta_t^2)\\
\label{eqn:ezfnum_step3_dr}
\bv^{n+1}_{i,j} &=& \bv^{n+\frac{2}{3}}_{i,j}+\Delta_t\mathcal{A}(\bv^{n+\frac{2}{3}}_{i,j})+O(\Delta_t^2)\
\end{eqnarray*}
\\
where, $\mathcal{R}$ are the Reaction-Diffusion terms, $\mathcal{D}$ are the perturbation terms, and $\mathcal{A}$ are the advection terms.

We shall briefly consider each of the three examples individually.

\subsubsection{Resonant Drift}

When we have resonant drift, the perturbation is dependent only on time, not space. Therefore, the transformed perturbation is exactly the same as the original perturbation, since the transformation only concerns spatial transformations. Therefore, we consider the perturbation as:

\begin{eqnarray*}
\epsilon\bht &=& \bA\cos(\Omega t+\xi)
\end{eqnarray*}
\\
where $\bA$ is a vector whose elements are small (i.e. $O(\epsilon)$), and is given by:

\begin{equation}
\bA = (A_1,A_2)^T
\end{equation}

These parameters, like all the drift parameters used in these numerical simulations, are specified within the file \verb|task.dat|.

\subsubsection{Electrophoresis Induced Drift}

The transformed perturbation for this particular example is given by:

\begin{eqnarray*}
\epsilon\til{\bh} &=& \bB\left(\cos(\Theta)\pdbvox(r)-\sin(\Theta)\pdbvoy(r)\right)
\end{eqnarray*}
\\
where $\bB$ is a $2\times2$ diagonalized matrix:

\begin{equation*}
\bB = \left(
      \begin{array}{ccccc} B_1   & 0     \\
                           0     & B_2\end{array} \right)    
\end{equation*}

The element $B_1$ and $B_2$ are both small quantities and are specified in the \verb|task.dat| file.

\subsubsection{Inhomogeneity Induced Drift}

Finally, the transformed perturbation for Inhomogeneity induced drift is give by:

\begin{eqnarray*}
\epsilon\bht &=& \alpha_1(X+x\cos(\Theta)-y\sin(\Theta))\pdfa(\bv_0(\br),\alpha_0)
\end{eqnarray*}
\\
where $\alpha_1$ is the drift gradient and \chgex[ex]{controls} the velocity of the drift.


\subsection{Tip Reconstruction}

We note that the coordinates for tip of the spiral wave are given:

\begin{eqnarray*}
\deriv{R}{t} &=& ce^{i\Theta}\\
\deriv{\Theta}{t} &=& \omega
\end{eqnarray*}

We can solve these numerically using the following numerical scheme:

\begin{eqnarray*}
\Theta^{n+1} &=& \Theta^n+\Delta_t\omega^n\\
X^{n+1} &=& X^n+\Delta_t(c_x\cos(\Theta^n)-c_y\sin(\Theta^n))\\
Y^{n+1} &=& Y^n+\Delta_t(c_x\sin(\Theta^n)+c_y\cos(\Theta^n))
\end{eqnarray*}

So, once we have found the $c_x$, $c_y$ and $\omega$, we can use these values to numerically reconstruct the trajectory of the tip of the spiral wave as viewed in the laboratory frame of reference.

\section{Examples: Rigidly Rotation and Meander}
\label{sec:ezf_examples}
In this section we shall show several examples. They will be split into two parts; those for rigidly rotating spiral waves and those for meandering waves. We shall use Barkley's model for the rigid rotation and FHN for meander.

We shall show, for each example, results using a first order scheme, second order scheme, Method 1 for calculating the quotient system, and also Method 2. We will also show the values of the quotient system and how they converge for rigid rotation and oscillate for meander. We shall also show the reconstructed tip trajectories.

For all simulations, we shall keep the following parameters fixed:

\begin{itemize}
 \item $L_x$ = $L_y$ = 20
 \item $N_x$ = $N_y$ = 101
 \item \verb|ts| = 0.1
\end{itemize}

\noindent where \verb|ts| is the ratio of the timestep to the diffusion stability limit. Therefore, the numerical parameters are:

\begin{itemize}
 \item $\Delta_x$ = 0.2
 \item $\Delta_t$ = 0.001
\end{itemize}


\subsection{Rigid Rotation}
\label{sec:ezf_rw}

\begin{figure}[bth]
\begin{center}
\begin{minipage}[htbp]{0.49\linewidth}
\centering
\includegraphics[width=0.7\textwidth, angle=-90]{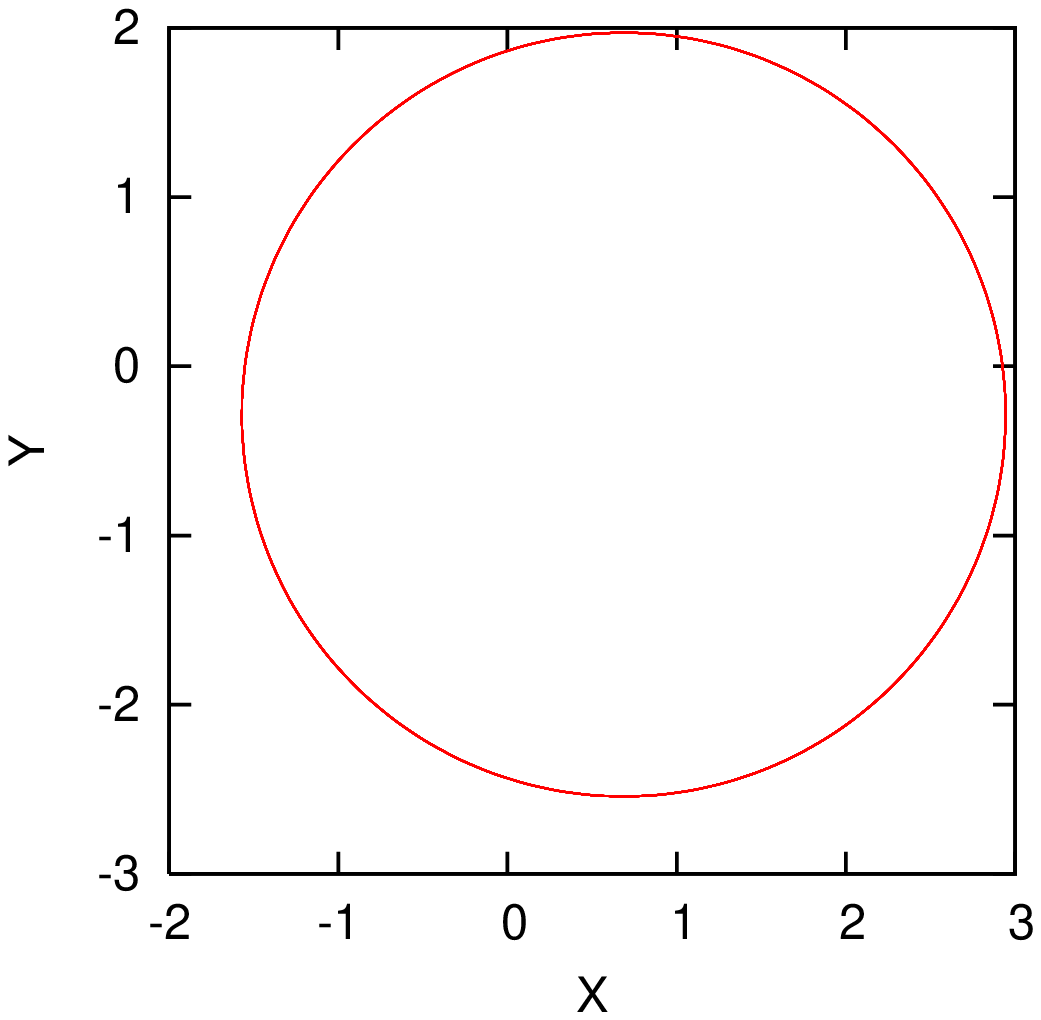}
\end{minipage}
\begin{minipage}[htbp]{0.49\linewidth}
\centering
\includegraphics[width=0.7\textwidth, angle=-90]{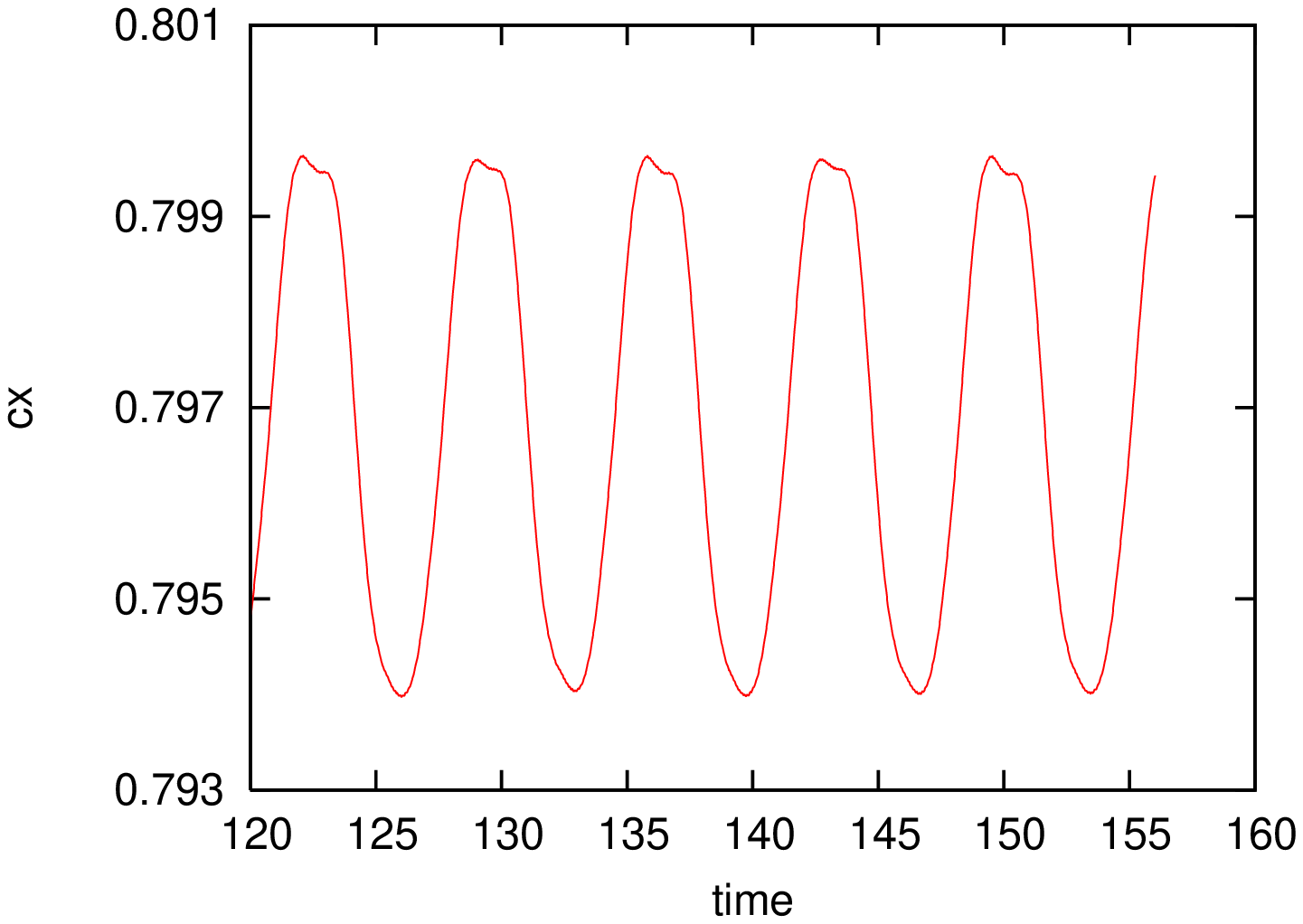}
\end{minipage}
\begin{minipage}[htbp]{0.49\linewidth}
\centering
\includegraphics[width=0.7\textwidth, angle=-90]{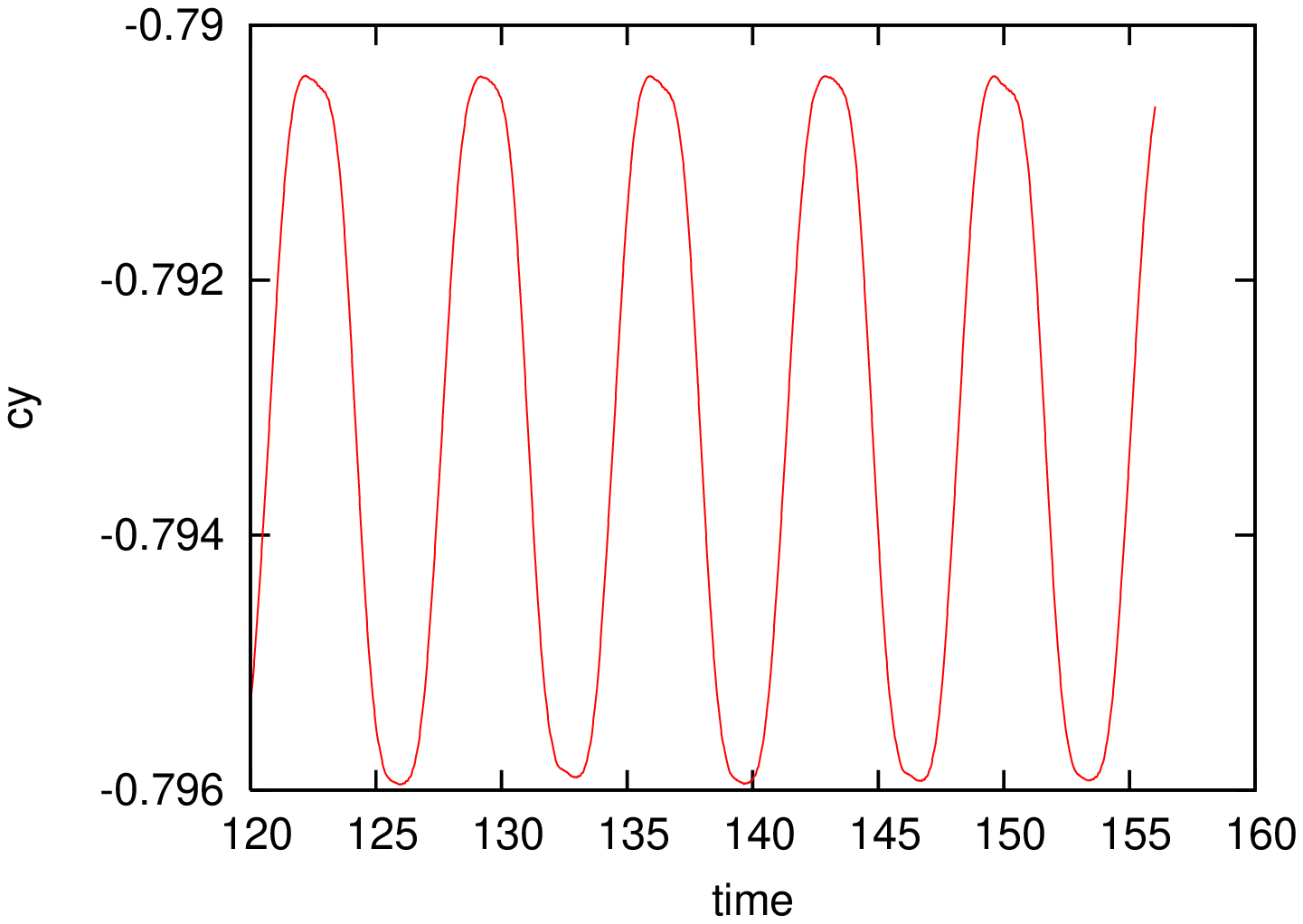}
\end{minipage}
\begin{minipage}[htbp]{0.49\linewidth}
\centering
\includegraphics[width=0.7\textwidth, angle=-90]{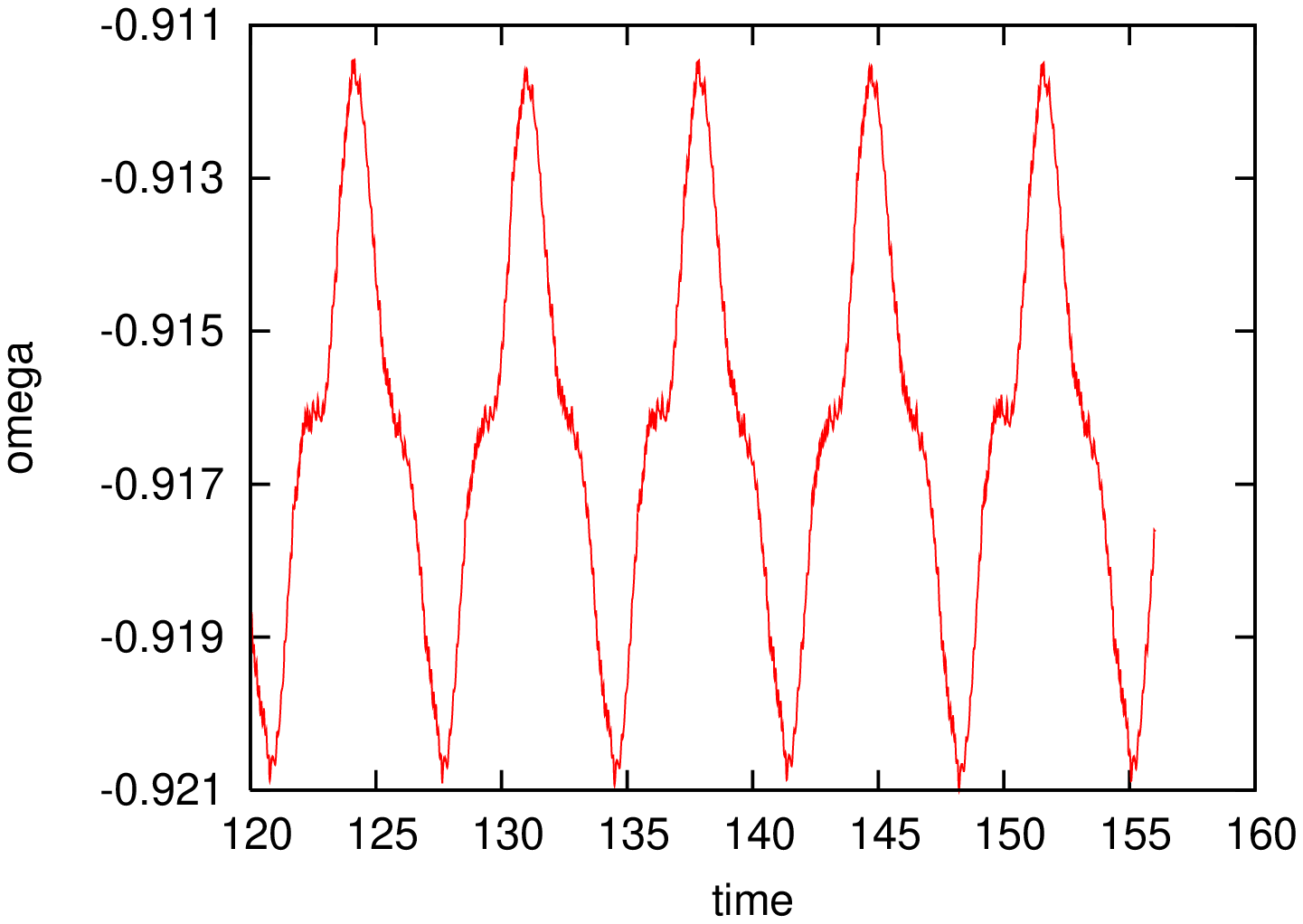}
\end{minipage}
\caption{Rigid Rotation in Barkley's model and in the laboratory frame of reference.}
\label{fig:ezf_ex_rw_reg}
\end{center}
\end{figure}

We use Barkley's model in this particular example with the following model parameters being fixed throughout:

\begin{itemize}
 \item $a$ = 0.52
 \item $b$ = 0.05
 \item $\varepsilon$ = 0.02
\end{itemize}
\chg[p145para]{}
Before we show these examples, we note that the trajectory in the laboratory frame of reference is shown in Fig.(\ref{fig:ezf_ex_rw_reg})

We can also determine averaged values of the components of the quotient system, by firstly using a Tikhonov Regularisation method and then solving the boundary value problem with the Double Sweep Method. These are described in Sec.(\ref{sec:rev_numerical}). The results are shown in Fig.(\ref{fig:ezf_ex_rw_reg}), \chg[p146para]{with the original data shown in the top left hand corner of the figure.}

We can see the values of $c_x$ , $c_y$ and $\omega$ do not converge to one particular value but oscillate with a period that is equal to the period of the spiral wave. This suggests that these small oscillations are due to either the influence of the boundaries or the discretization of the numerical methods. However, we note that the averaged values are $c_x=0.796796$, $c_y=-0.7931725$ and $\omega=-0.916312$. However, we must note that the values of $c_x$ and $c_y$ are distorted by the regularisation technique for high values of $\lambda$ (regularization parameter), but $\omega$ is not distorted .

\chgex[ex]{This} information is useful just to get an idea of what the quotient solution should look like. It is by no means a highly accurate calculation of the quotient solution. It is a means to depress the noise in the numerical data so that it can be differentiated.

\begin{figure}[p]
\begin{center}
\begin{minipage}{0.4\linewidth}
\centering
\includegraphics[width=0.7\textwidth, angle=-90]{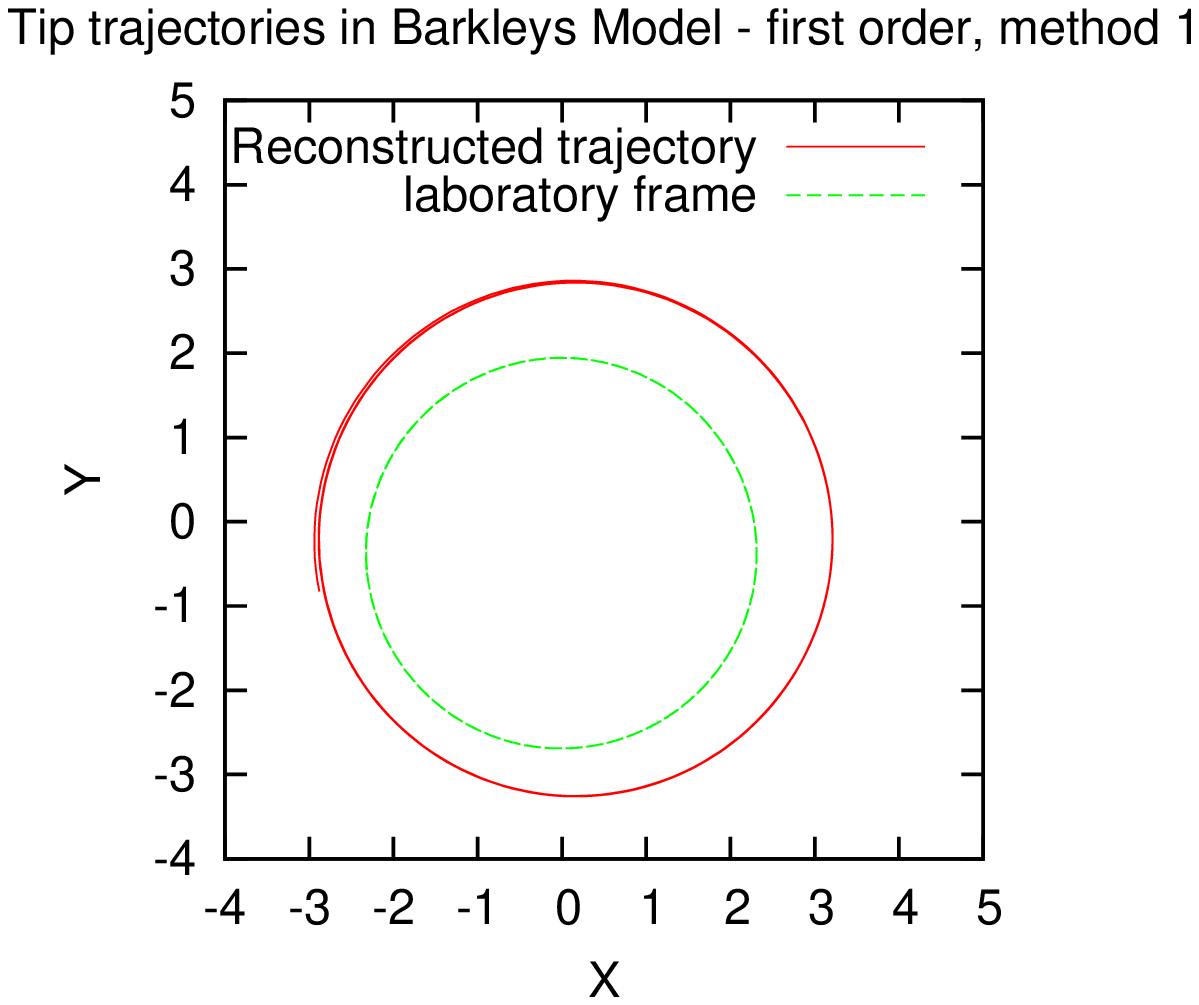}
\end{minipage}
\begin{minipage}{0.4\linewidth}
\centering
\includegraphics[width=0.7\textwidth, angle=-90]{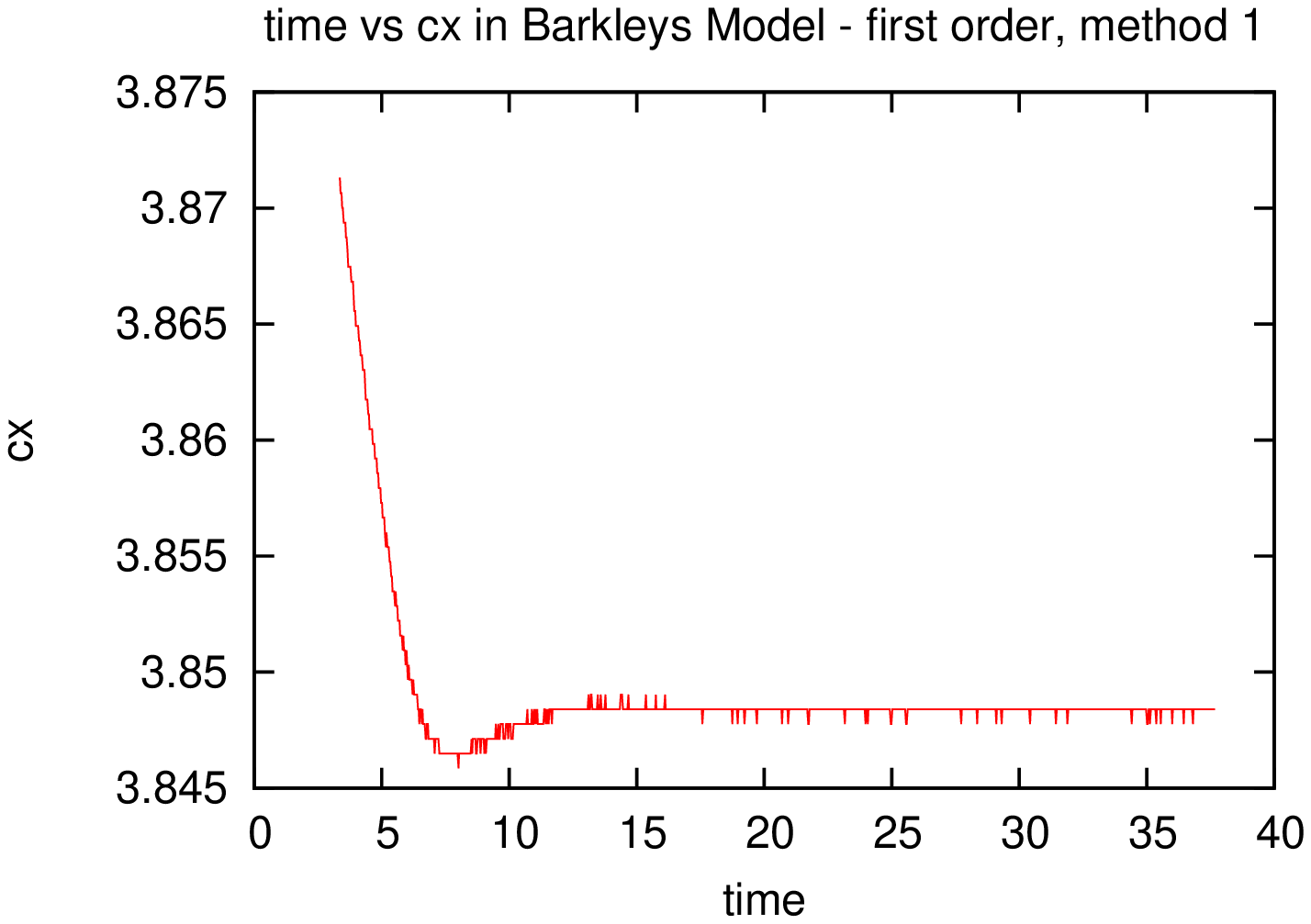}
\end{minipage}
\begin{minipage}{0.4\linewidth}
\centering
\includegraphics[width=0.7\textwidth, angle=-90]{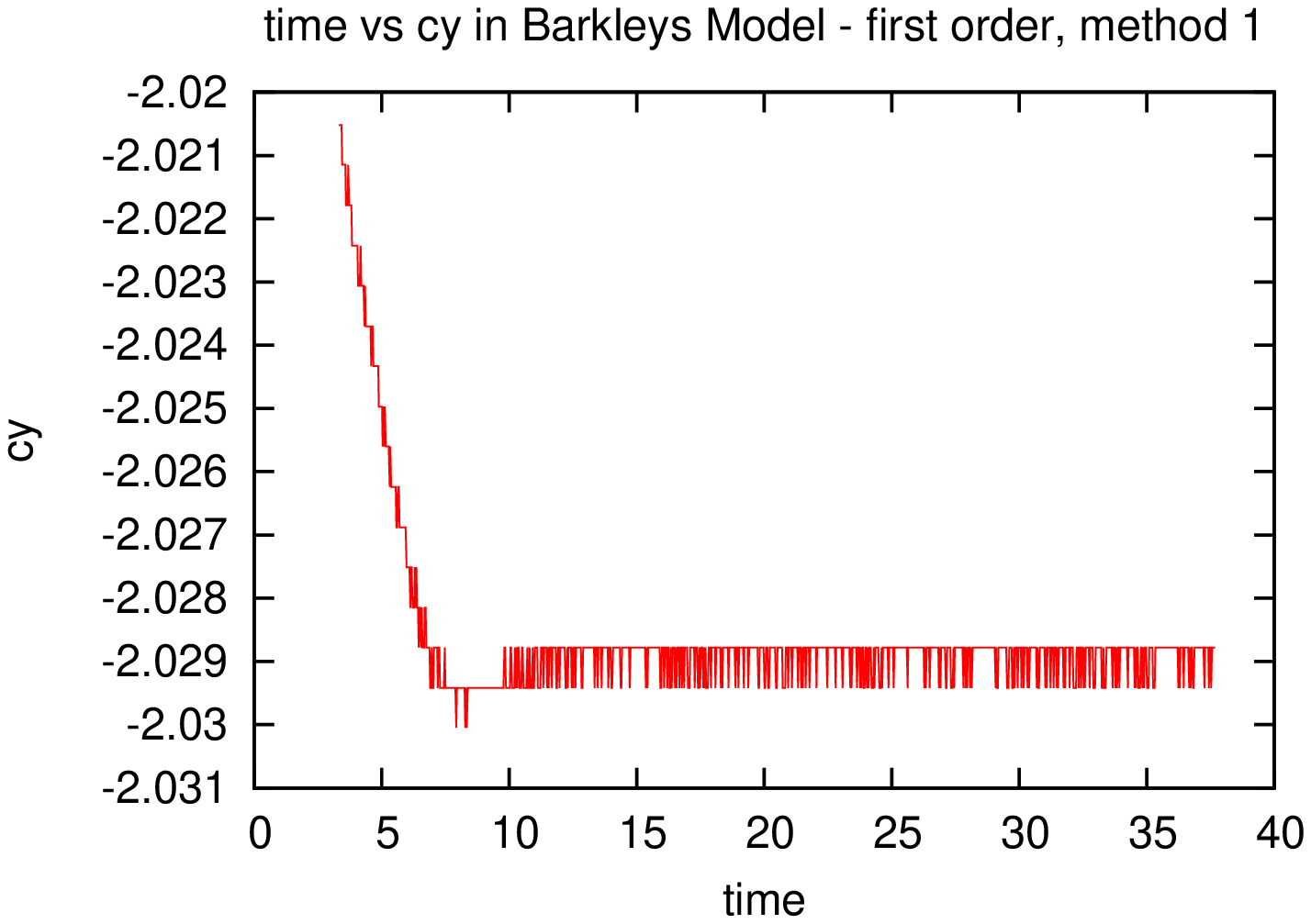}
\end{minipage}
\begin{minipage}{0.4\linewidth}
\centering
\includegraphics[width=0.7\textwidth, angle=-90]{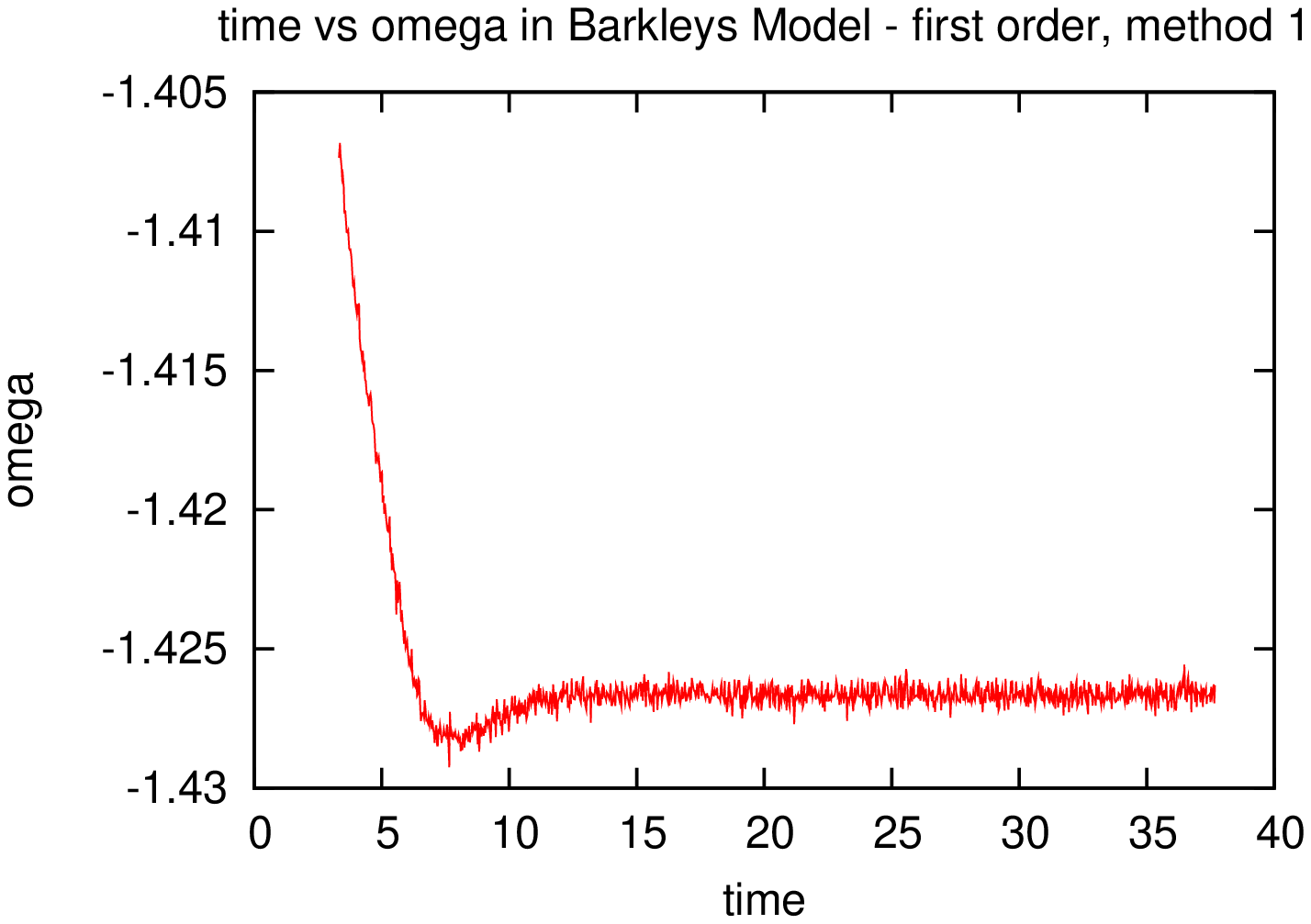}
\end{minipage}
\caption{Rigid rotation: Barkley's model, First order, Method 1.}
\label{fig:ezf_ex_rw_fir_m1}
\end{center}
\end{figure}

\begin{figure}[p]
\begin{center}
\begin{minipage}{0.4\linewidth}
\centering
\includegraphics[width=0.7\textwidth, angle=-90]{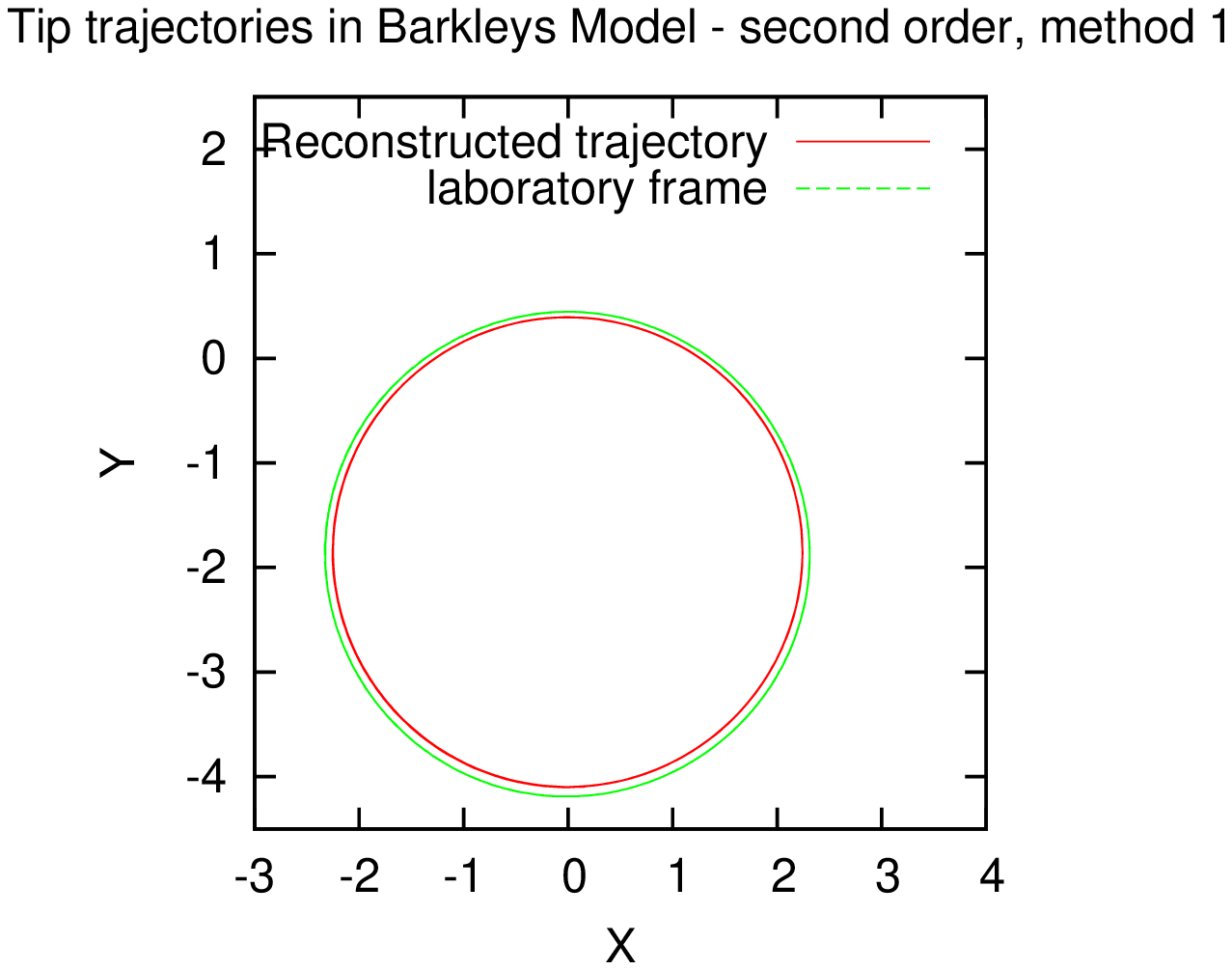}
\end{minipage}
\begin{minipage}{0.4\linewidth}
\centering
\includegraphics[width=0.7\textwidth, angle=-90]{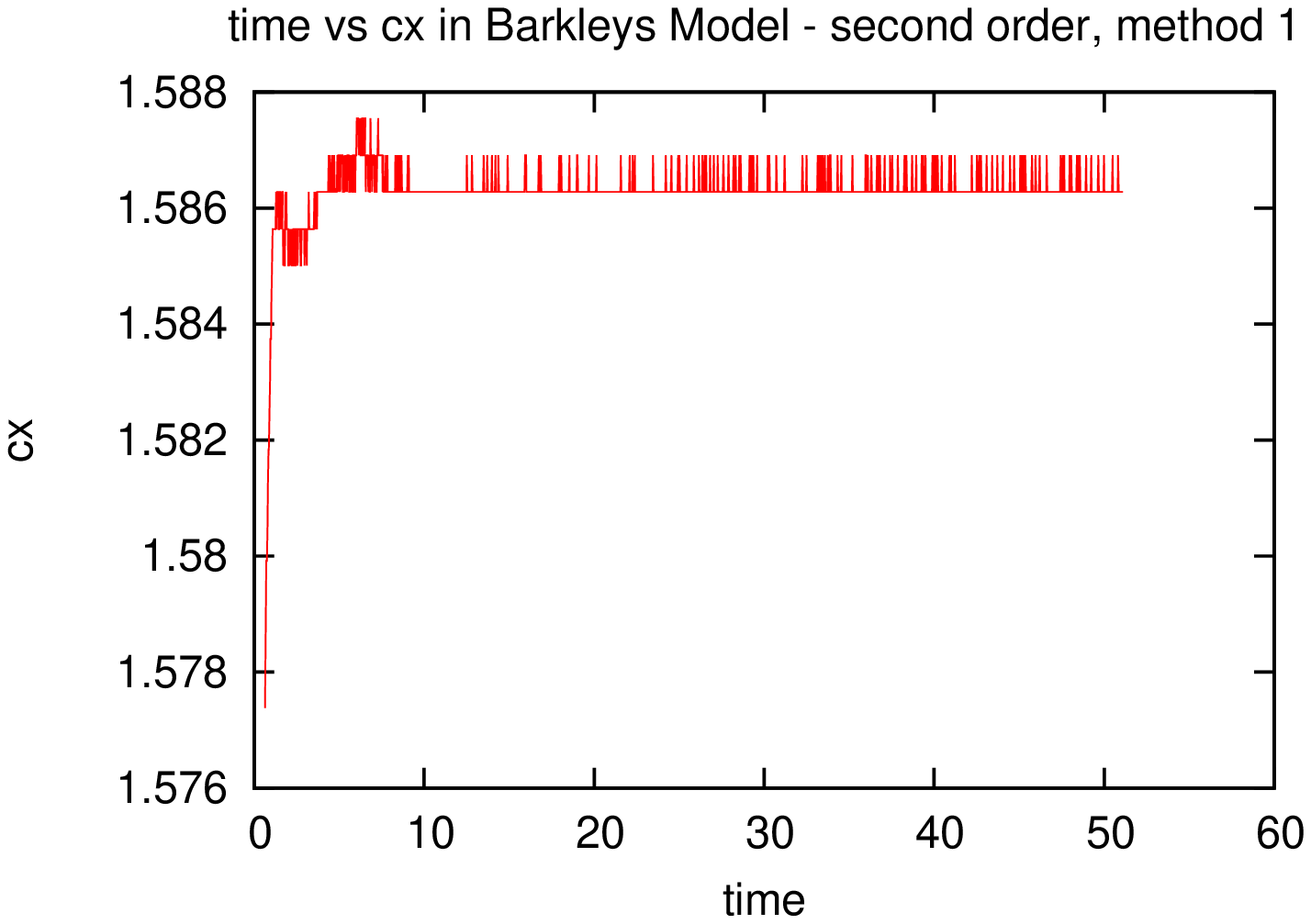}
\end{minipage}
\begin{minipage}{0.4\linewidth}
\centering
\includegraphics[width=0.7\textwidth, angle=-90]{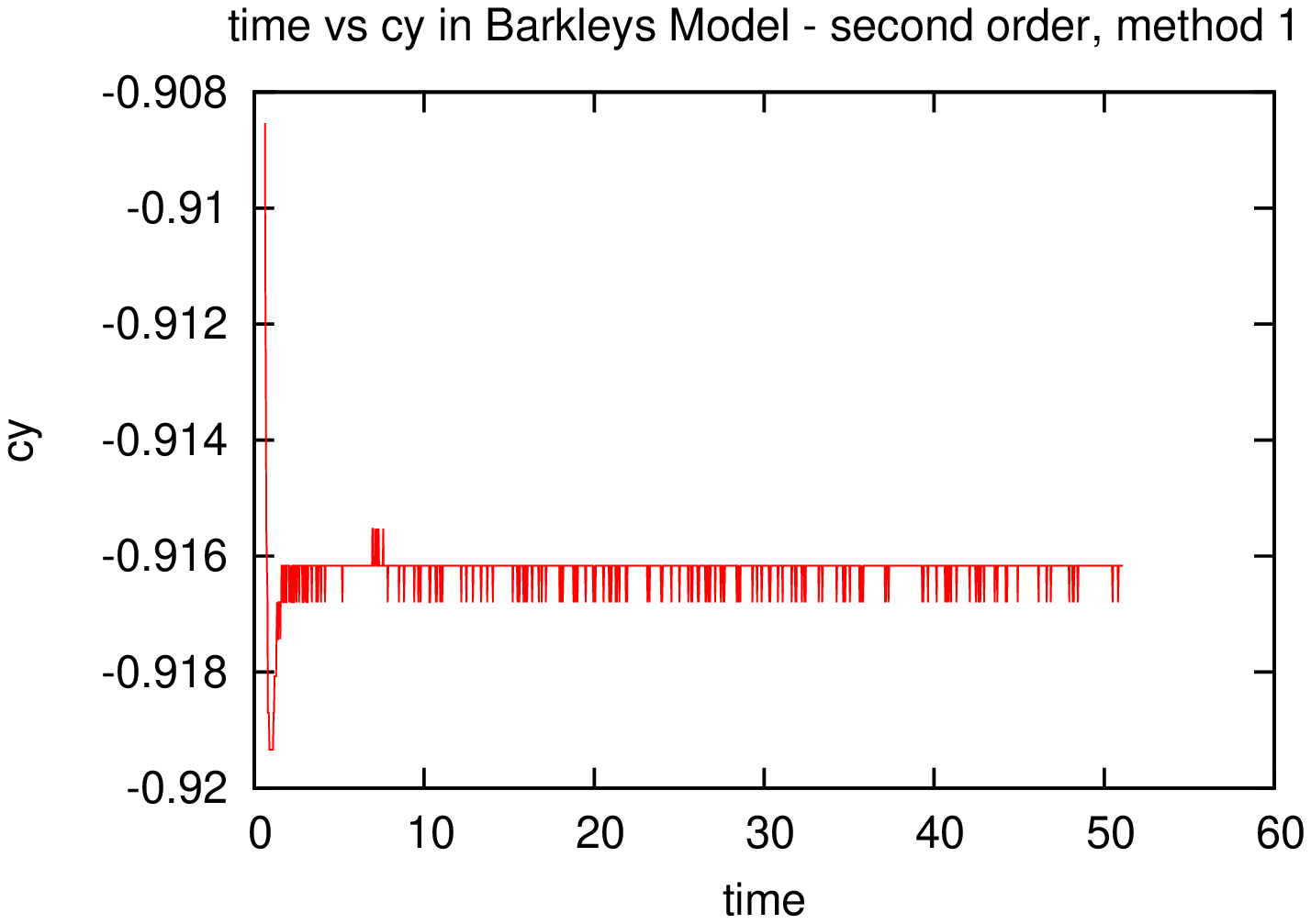}
\end{minipage}
\begin{minipage}{0.4\linewidth}
\centering
\includegraphics[width=0.7\textwidth, angle=-90]{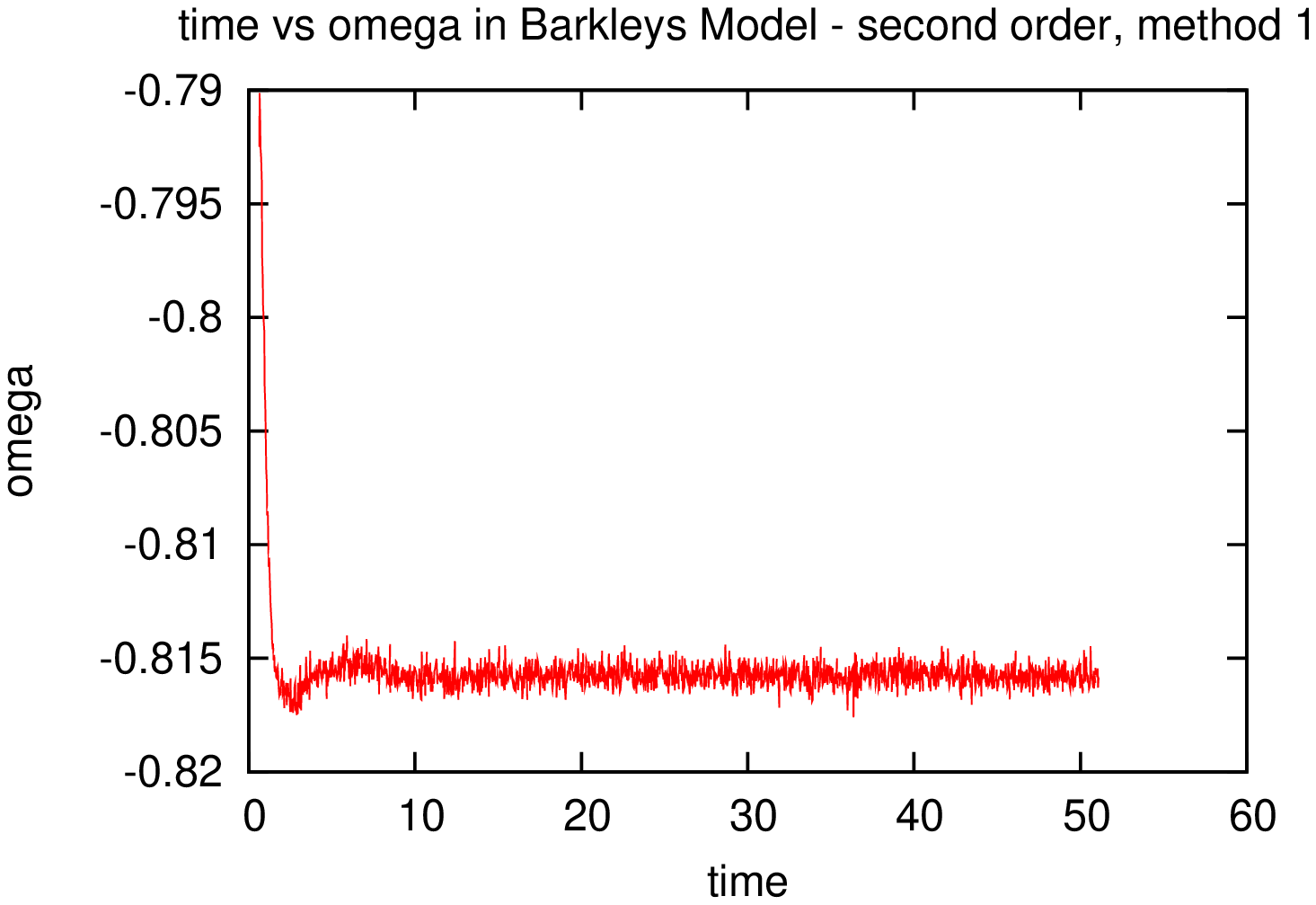}
\end{minipage}
\caption{Rigid rotation: Barkley's model, Second order, Method 1.}
\label{fig:ezf_ex_rw_sec_m1}
\end{center}
\end{figure}

\begin{figure}[p]
\begin{center}
\begin{minipage}{0.4\linewidth}
\centering
\includegraphics[width=0.7\textwidth, angle=-90]{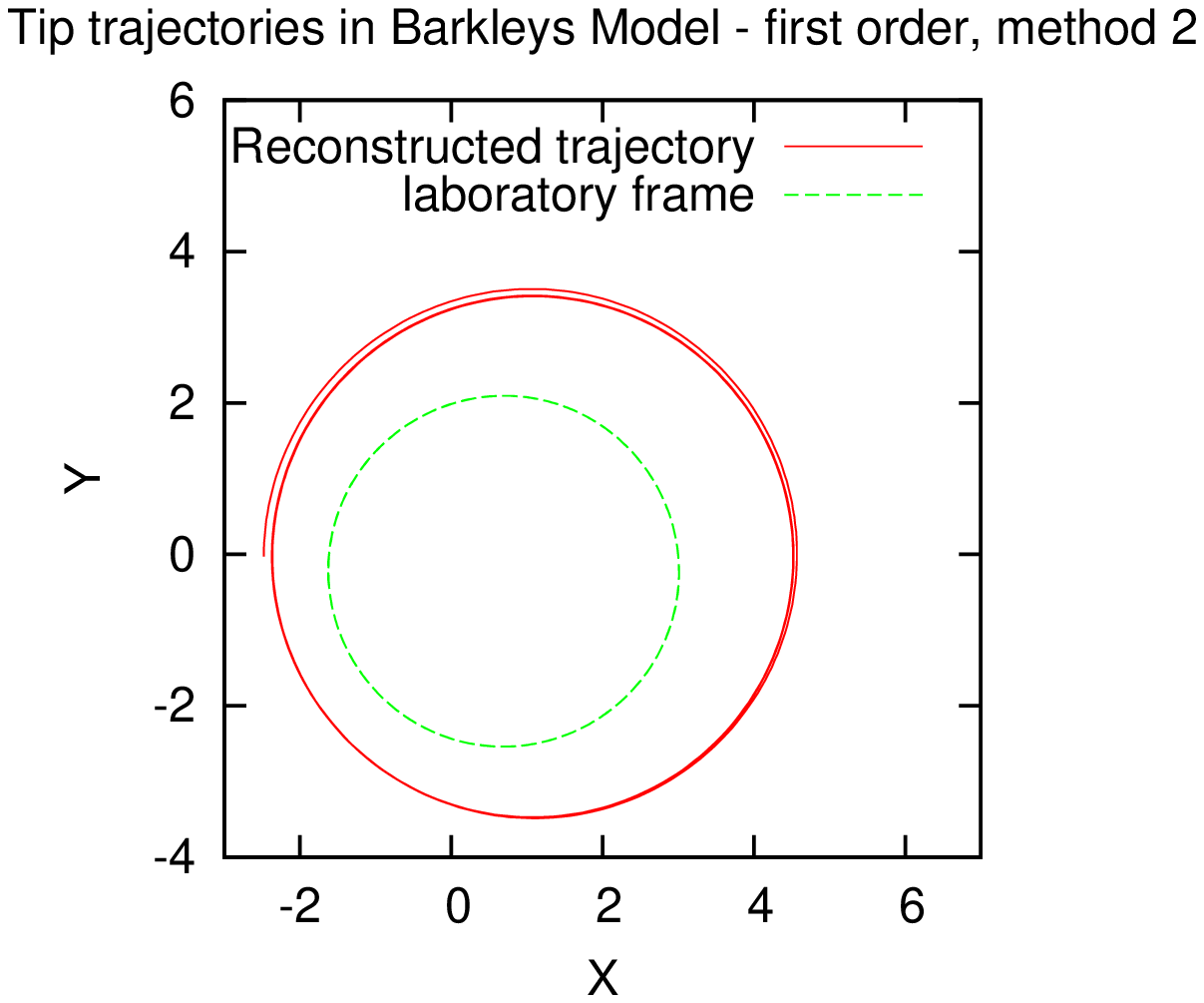}
\end{minipage}
\begin{minipage}{0.4\linewidth}
\centering
\includegraphics[width=0.7\textwidth, angle=-90]{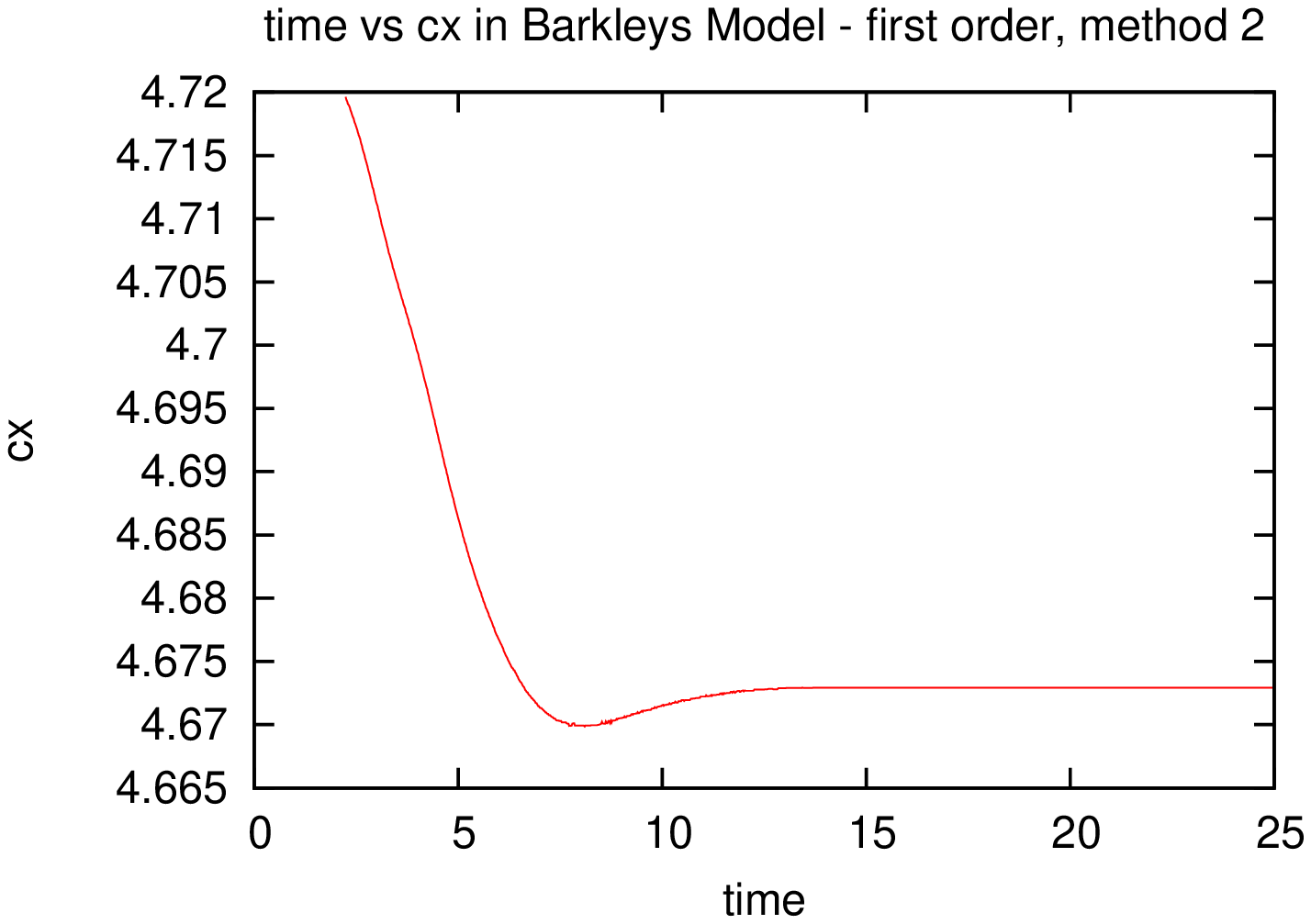}
\end{minipage}
\begin{minipage}{0.4\linewidth}
\centering
\includegraphics[width=0.7\textwidth, angle=-90]{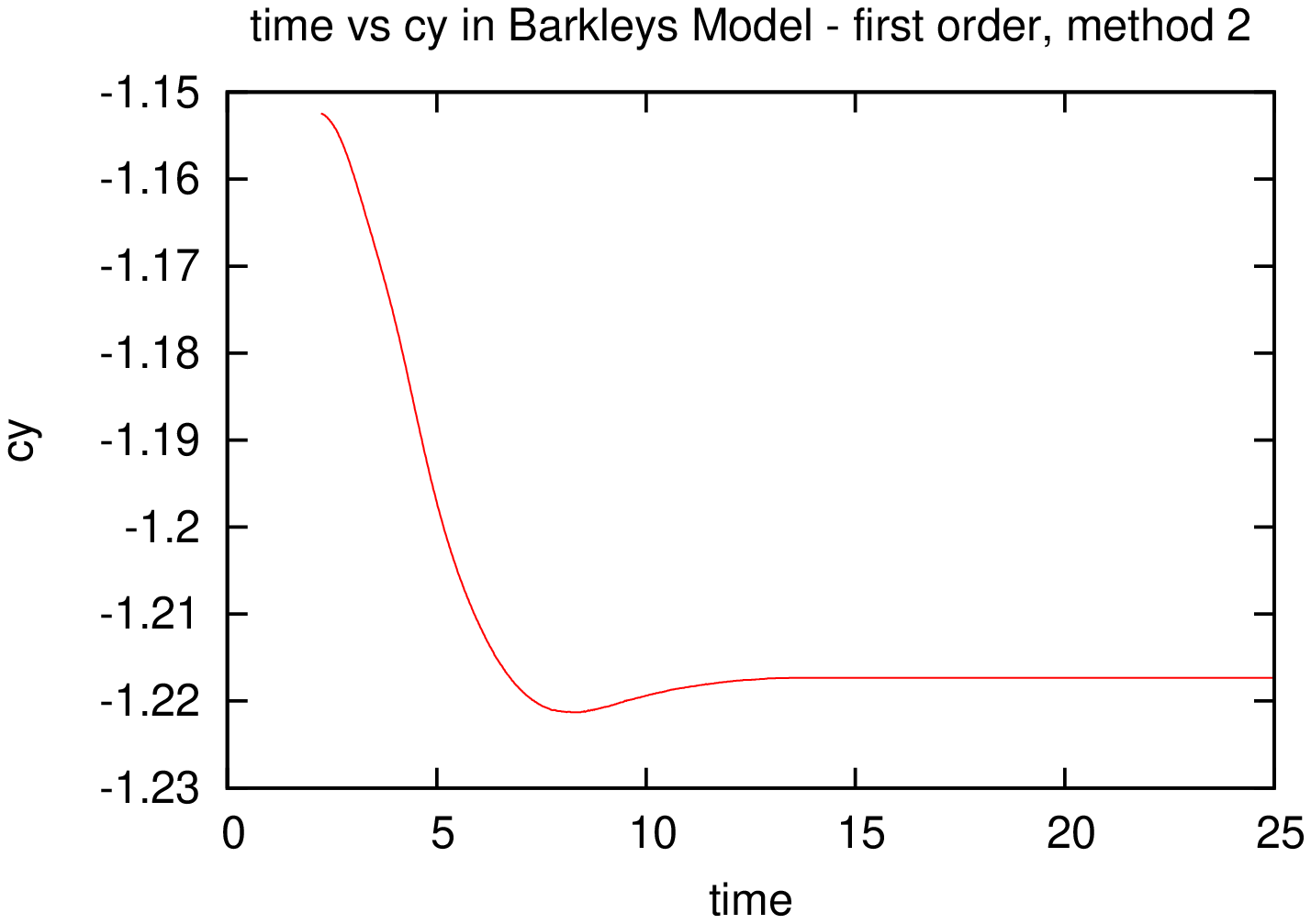}
\end{minipage}
\begin{minipage}{0.4\linewidth}
\centering
\includegraphics[width=0.7\textwidth, angle=-90]{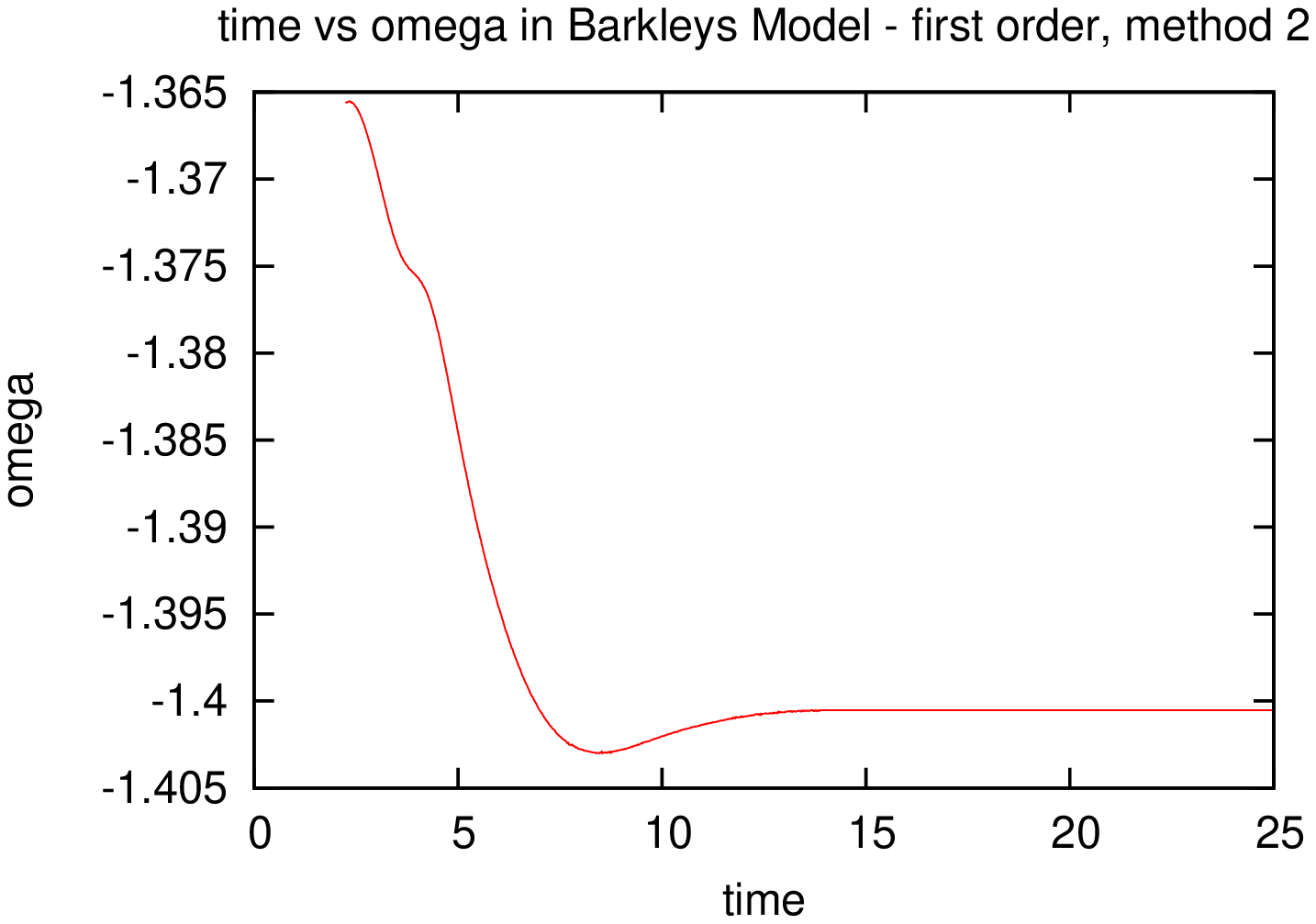}
\end{minipage}
\caption{Rigid rotation: Barkley's model, First order, Method 2.}
\label{fig:ezf_ex_rw_fir_m2}
\end{center}
\end{figure}

\begin{figure}[p]
\begin{center}
\begin{minipage}{0.4\linewidth}
\centering
\includegraphics[width=0.7\textwidth, angle=-90]{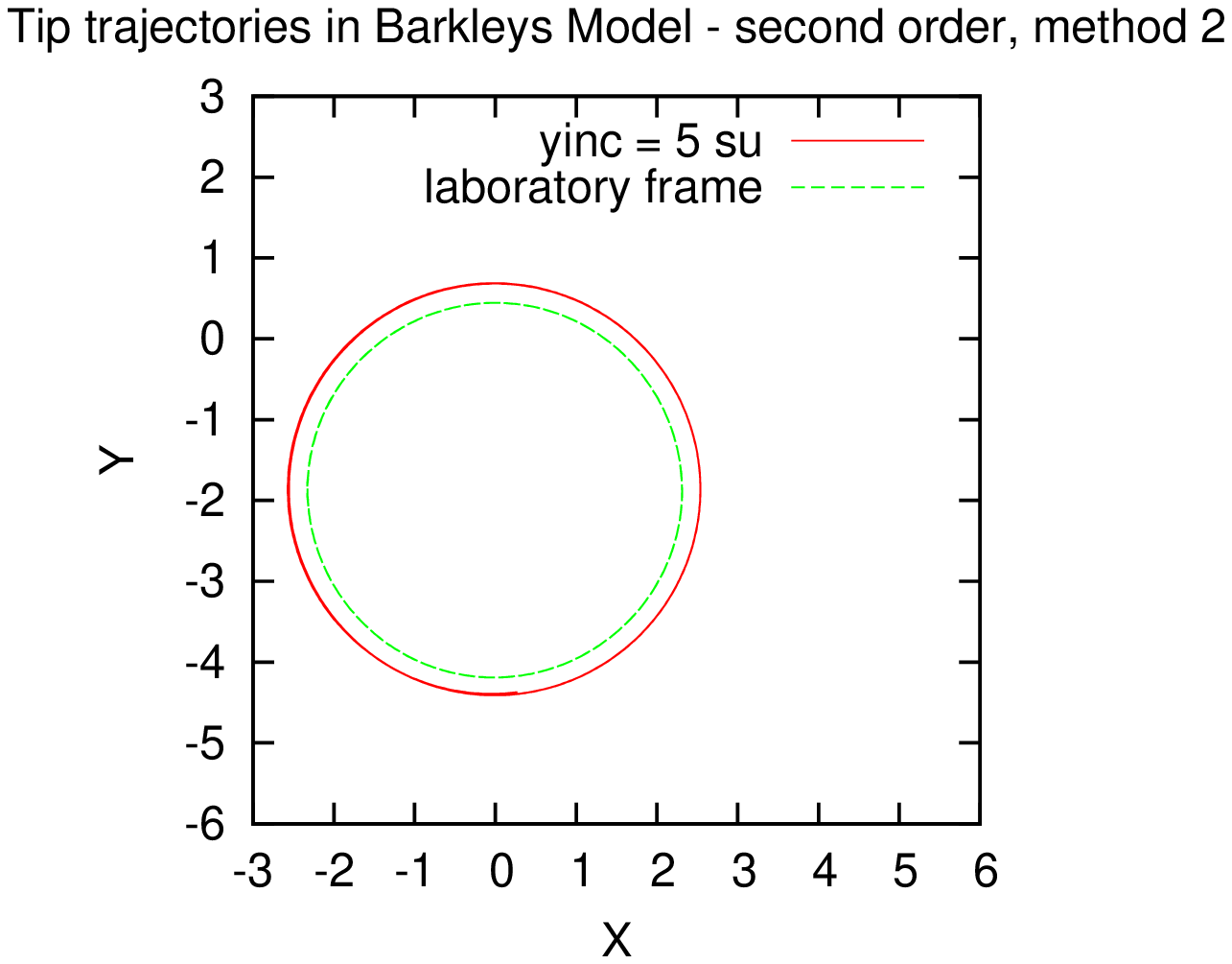}
\end{minipage}
\begin{minipage}{0.4\linewidth}
\centering
\includegraphics[width=0.7\textwidth, angle=-90]{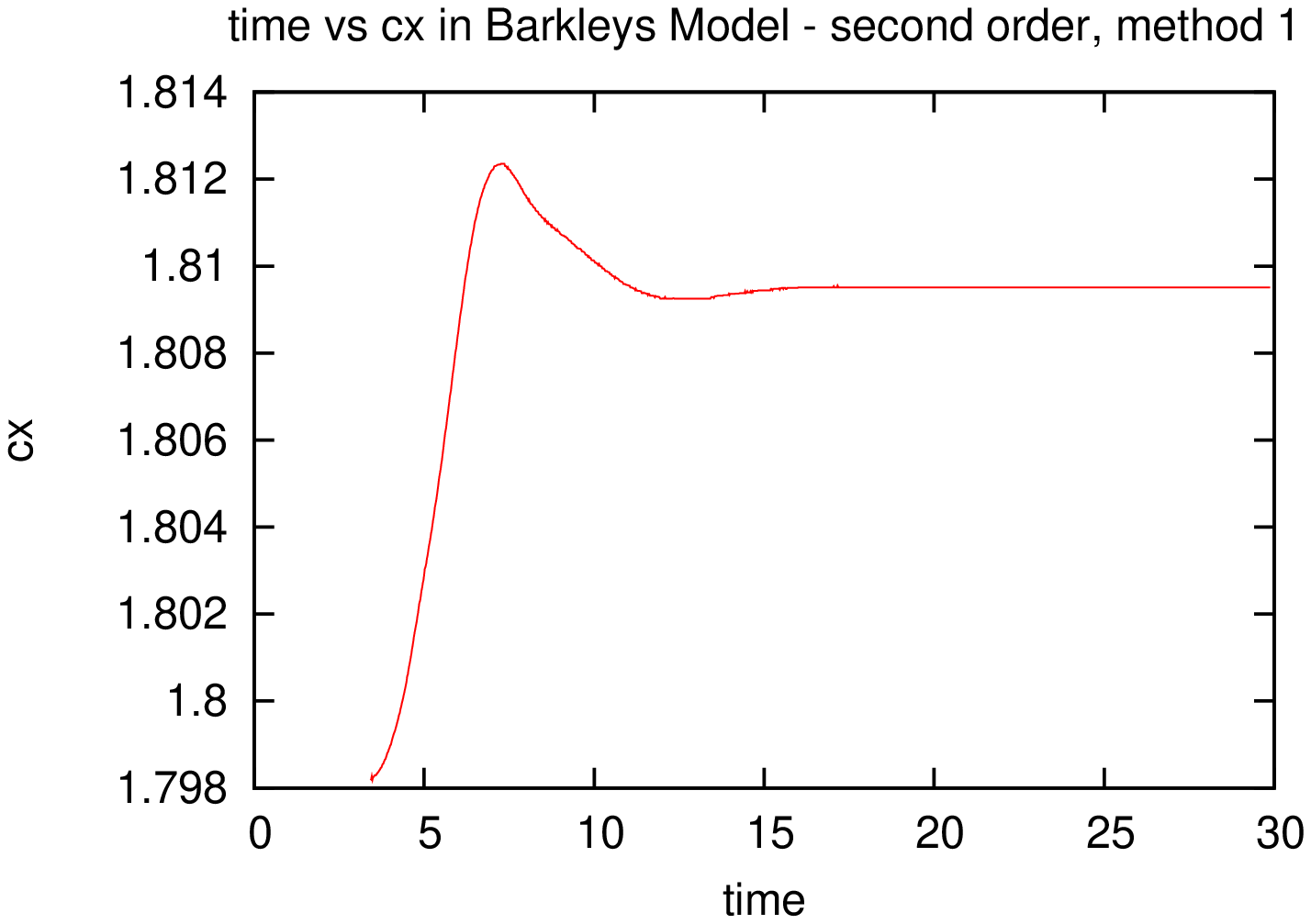}
\end{minipage}
\begin{minipage}{0.4\linewidth}
\centering
\includegraphics[width=0.7\textwidth, angle=-90]{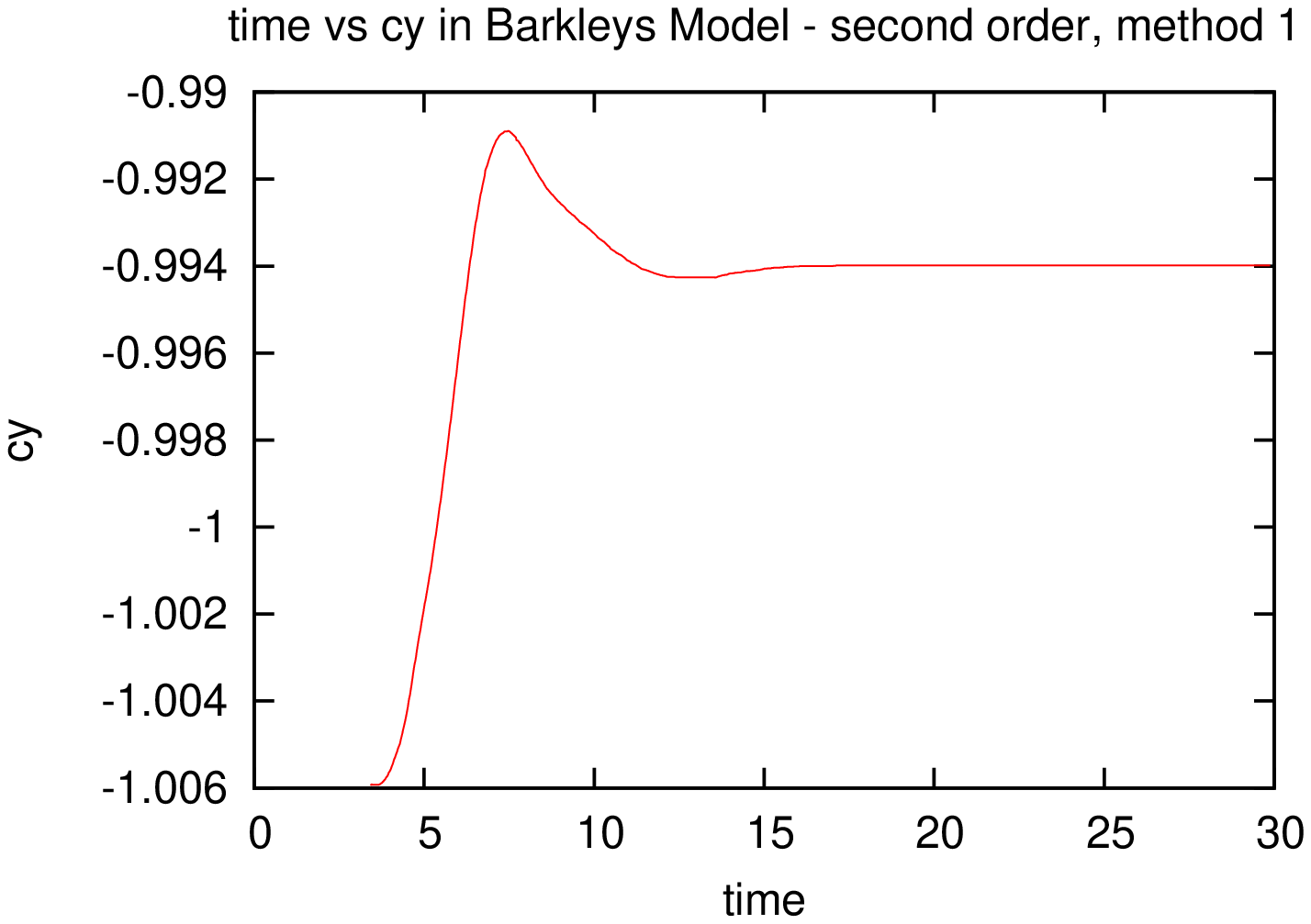}
\end{minipage}
\begin{minipage}{0.4\linewidth}
\centering
\includegraphics[width=0.7\textwidth, angle=-90]{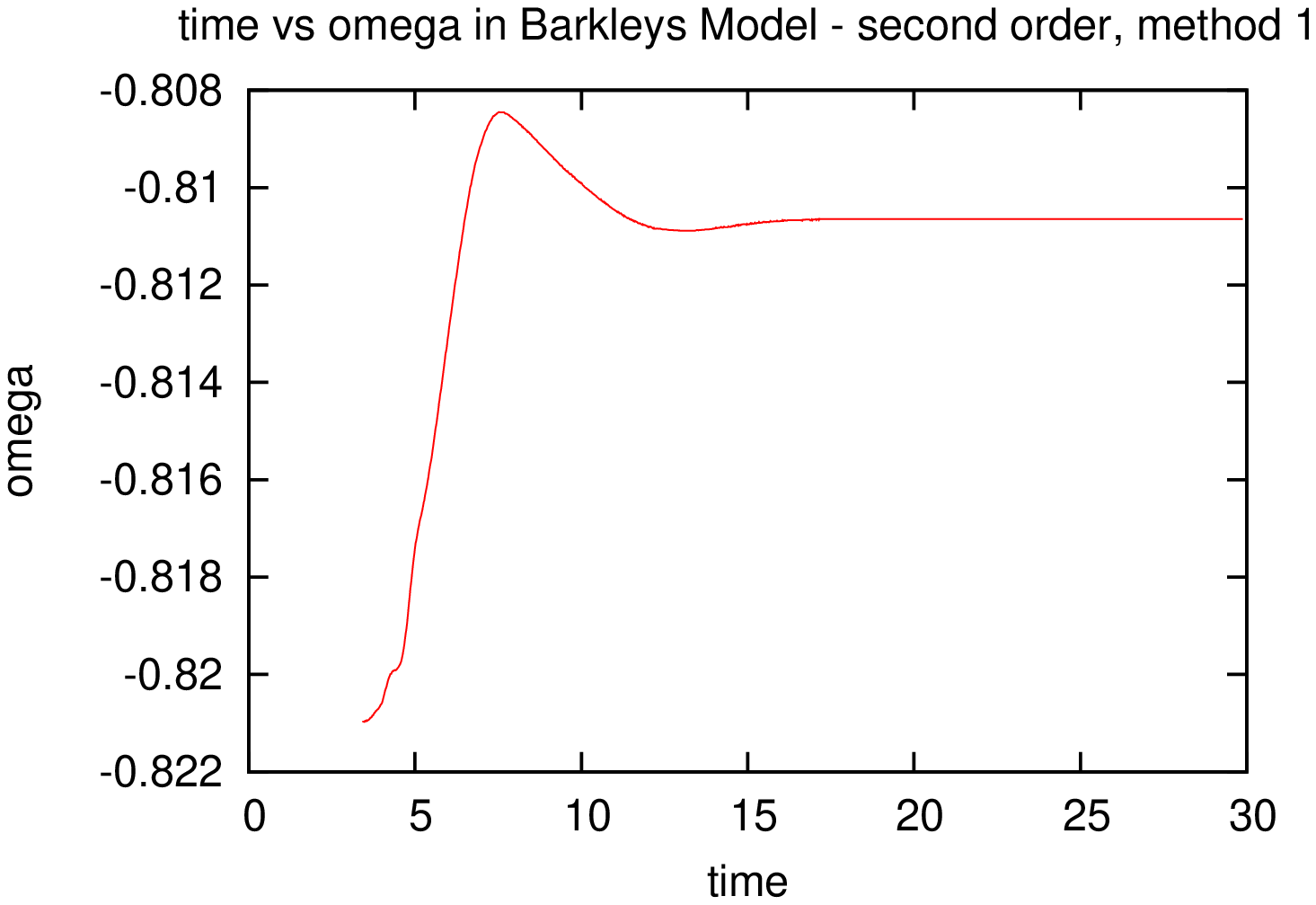}
\end{minipage}
\caption{Rigid rotation: Barkley's model, Second order, Method 2.}
\label{fig:ezf_ex_rw_sec_m2}
\end{center}
\end{figure}

We shown in Figs(\ref{fig:ezf_ex_rw_fir_m1})-(\ref{fig:ezf_ex_rw_sec_m2}) the results from the simulations using first and second order scheme for method 1 and a first and second order for method two. We can see that each of the four simulations produce different results, and we show in table (\ref{tab:ezf_1}) the values of the components of the quotient solution.

\begin{center}
\begin{table}
\begin{tabular}{rccccc}
Simulation &\vline& $c_x$ & $c_y$ & $|c|$ & $\omega$\\
\hline
\hline
First order, method 1  &\vline& 3.848361 & -2.028917 & 4.350446 & -1.426661\\
Second order, method 1 &\vline& 1.586342 & -0.916221 & 1.831922 & -0.815751\\
First order, method 2  &\vline& 4.672928 & -1.217335 & 4.828888 & -1.400537\\
Second order, method 2 &\vline& 1.809512 & -0.993985 & 2.064543 & -0.810646
\end{tabular}
\label{tab:ezf_1}
\caption{Numerical values of the components of the Quotient Solution for each of the four simulations conducted for Rigid Rotation in Barkley's model.}
\end{table}
\end{center}

We note that we expect that the values of $c_x$ and $c_y$ will be different due to the different methods employed, but we expect the absolute value of $c=c_x+ic_y$ to be comparable in each case. Also, we can compare the value of $\omega$ in each simulation since it is a universal solution in each of the simulations.

For the simulation using a first order scheme to calculate the spatial derivatives and Method 1 to calculate the quotient solution (Fig.(\ref{fig:ezf_ex_rw_fir_m1})), we see that first of all the quotient system appears to be unstable. We also see that the value $\omega$ is much different than what we expect (when compared with the value of $\omega$ calculated from the laboratory frame simulations. Also, the reconstructed tip trajectory is not very accurate, with the radius of the reconstructed trajectory being approximately 50\% smaller than the radius of the simulation in the laboratory frame of reference. It is apparent that either the numerical scheme implemented to calculate the spatial derivatives and/or the method of calculating the quotient system need refining.

Therefore, we decided to try out a second order accurate numerical scheme to calculate the spatial derivatives, whilst still using Method 1. The results are shown in Fig.(\ref{fig:ezf_ex_rw_sec_m1}). We can still see that there are still instabilities in the quotient system. However, the values of $|c|$ and $\omega$ are such that the reconstructed tip trajectory is extremely accurate. We can therefore see that the introduction of the second order accurate scheme gives us accurate calculations of the quotient solution. However, \chg[p147gram]{this} does not eliminate the instabilities within the quotient solution.

We therefore tried Method 2 to calculate the quotient solution, whilst using a first order accurate scheme. The results are shown in Fig.(\ref{fig:ezf_ex_rw_fir_m2}). As explained in Sec.(\ref{sec:ezf_numerics_imp}), we require that the second pinning point is specified explicitly. We \chg[p147spell]{therefore} take this to be $(x_{inc},y_{inc})=(0,5)$ in space units. This time, it appears that this method has eliminated the instabilities present within the quotient solution. However, by using the first order scheme we see that the values of $|c|$ and $\omega$ are such that the reconstructed tip trajectory is still not very accurate.

So, we have seen that the use of Method 2 eliminates the instabilities within the quotient solution, and the use of a second order scheme gives us accurate calculation of the quotient solution, so let us consider using both of these together. The results are shown in Fig.(\ref{fig:ezf_ex_rw_sec_m2}). We can see that although the reconstructed trajectory appears not to be as accurate as the use of Method 1 with the second order scheme, we have the advantage of not having instabilities within the solution. We therefore recommend that the second order scheme with method 2 is used in future simulations, since the presence of the instabilities by using Method 1 could potentially give misleading results, should the instabilities be strong enough.

\subsection{Meander}

For completeness, we shall now show an example of a meandering spiral wave. We shall use FHN model to illustrate this using the following model parameters:

\begin{itemize}
 \item $\beta$ = 0.7
 \item $\gamma$ = 0.5
 \item $\varepsilon$ = 0.2
\end{itemize}

The numerical parameters are as per rigid rotation example in Sec.(\ref{sec:ezf_rw}).

Since we are considering meandering spiral waves, we note that $c_x$, $c_y$ and $\omega$ are no longer constant, but are oscillating and time dependent. Therefore, we shall compare the limit cycles and the reconstructed trajectories.

\begin{figure}[bth]
\begin{center}
\begin{minipage}[htbp]{0.49\linewidth}
\centering
\includegraphics[width=0.7\textwidth, angle=-90]{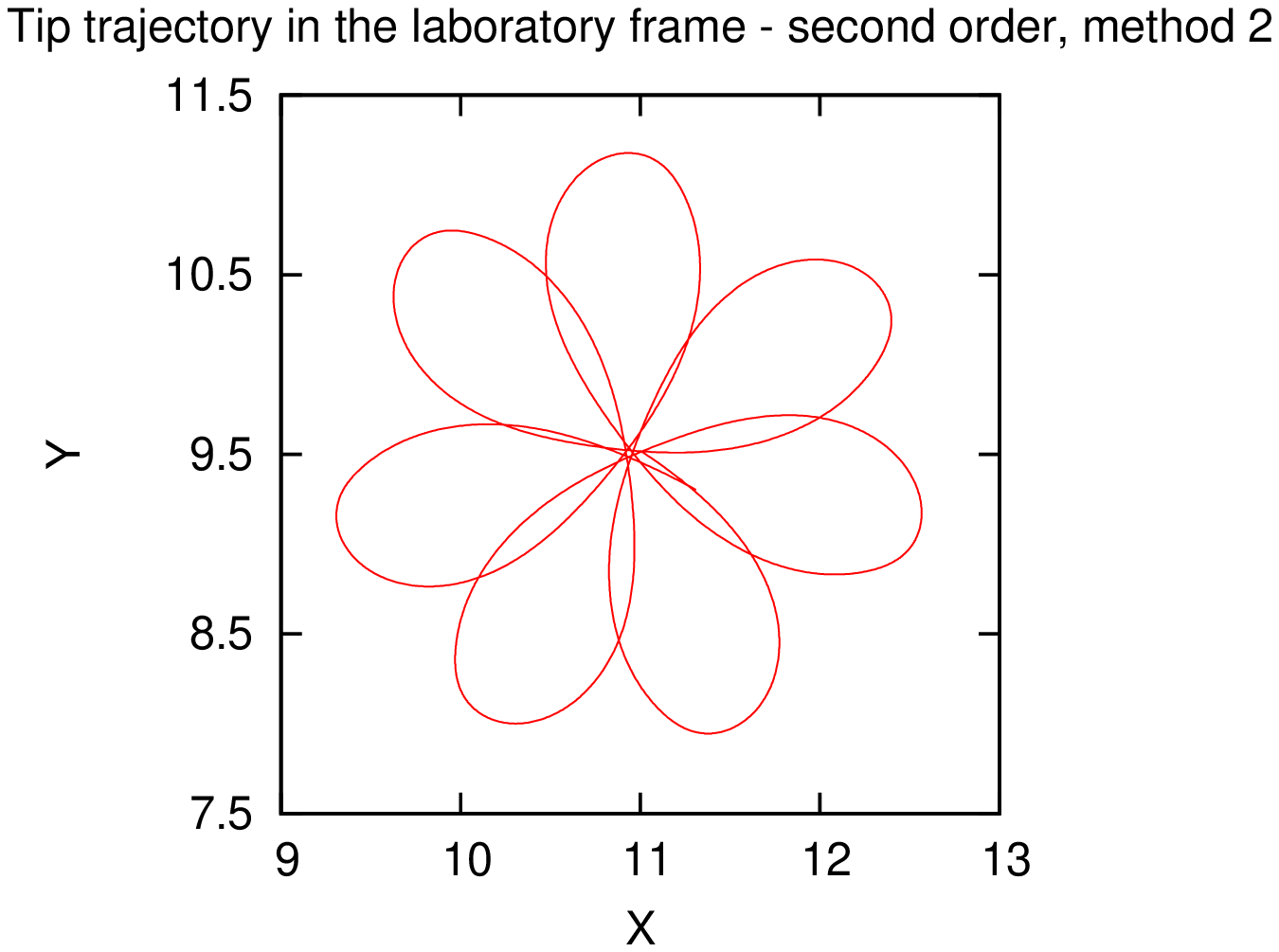}
\end{minipage}
\begin{minipage}[htbp]{0.49\linewidth}
\centering
\includegraphics[width=0.7\textwidth, angle=-90]{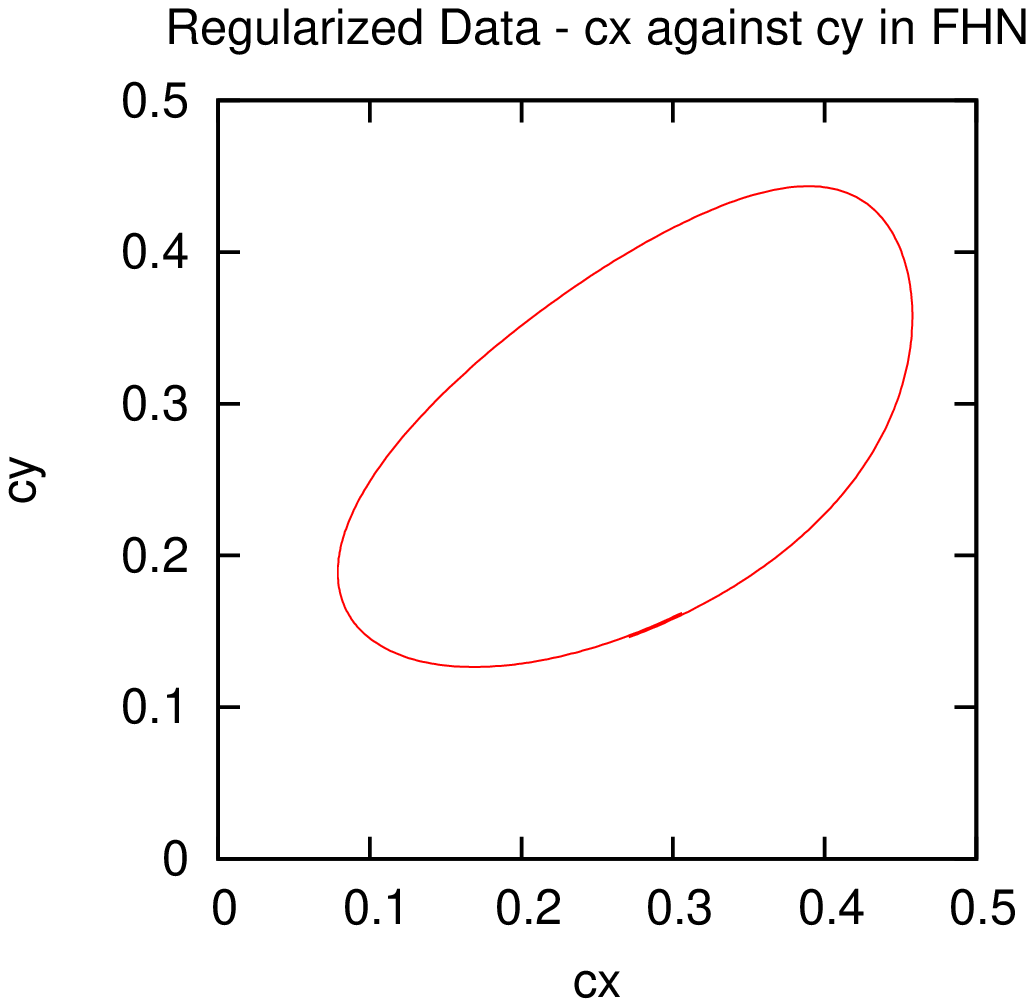}
\end{minipage}
\begin{minipage}[htbp]{0.49\linewidth}
\centering
\includegraphics[width=0.7\textwidth, angle=-90]{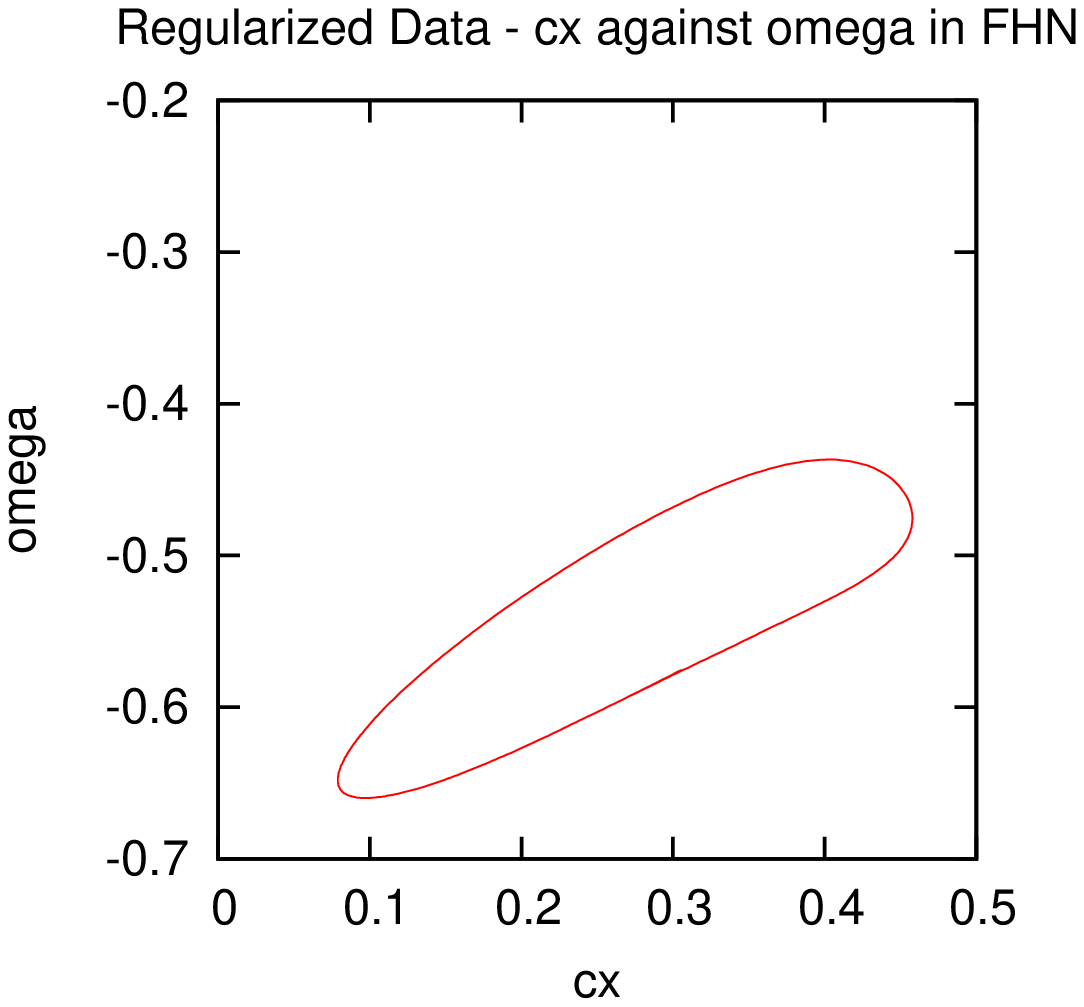}
\end{minipage}
\begin{minipage}[htbp]{0.49\linewidth}
\centering
\includegraphics[width=0.7\textwidth, angle=-90]{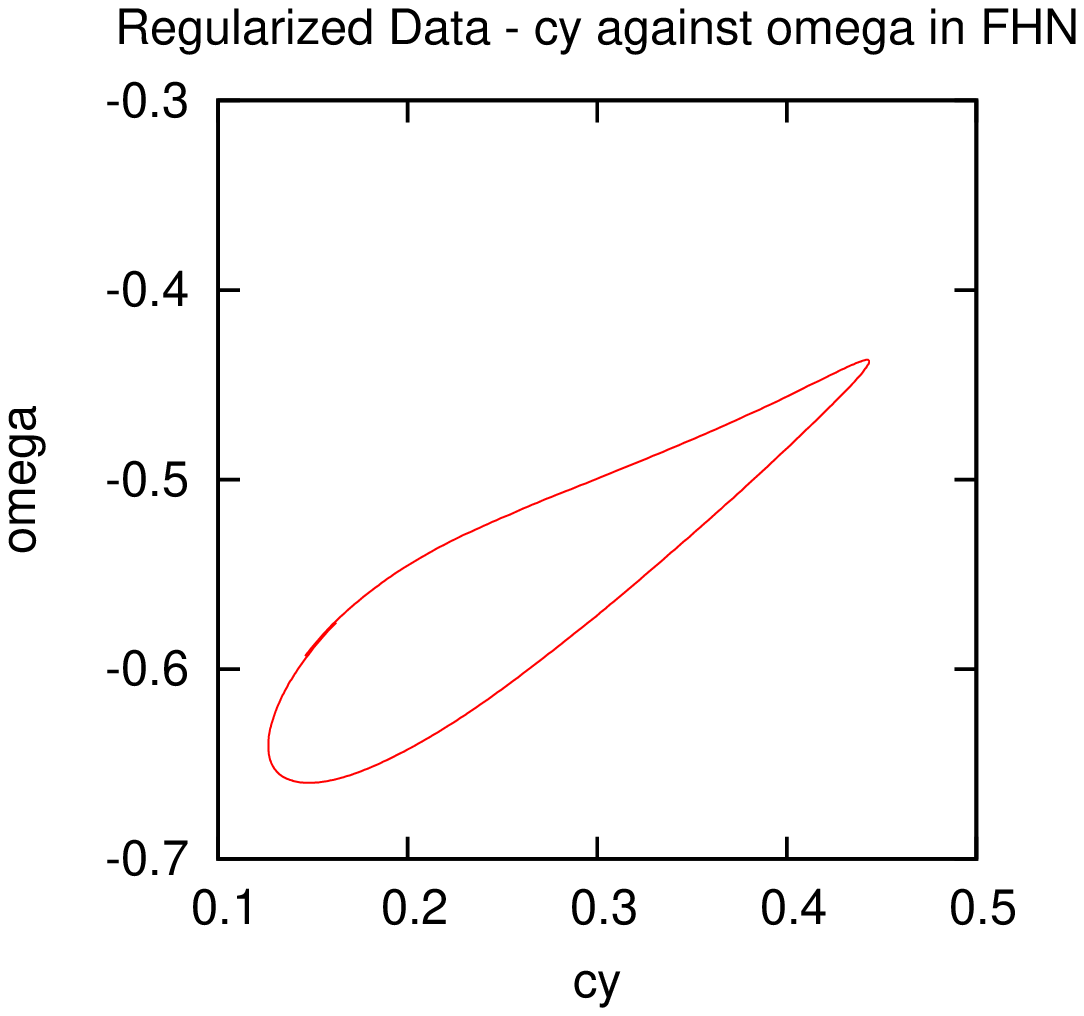}
\end{minipage}
\caption{Meander in the FHN model and in the laboratory frame of reference.}
\label{fig:ezf_ex_mrw_reg}
\end{center}
\end{figure}

Before we proceed to the example, let us consider the data in the laboratory frame of reference. The tip trajectory is given by Fig.(\ref{fig:ezf_ex_mrw_reg}).

As with the example for rigid rotation, we can also determine averaged values of the components of the quotient solution by regularizing the data and using numerical differentiation. The results are also shown in Fig.(\ref{fig:ezf_ex_mrw_reg})\chg[p148gram]{.}

We can clearly see that we should observe limit cycle solutions in the quotient solution. 

We know from the rigid rotation example that the use of Methods 1 and 2 and also the use of first and second order \chg[p148spell]{schemes} give different results. We shall show how similar results will occur in the meandering example

\begin{figure}[p]
\begin{center}
\begin{minipage}{0.4\linewidth}
\centering
\includegraphics[width=0.7\textwidth, angle=-90]{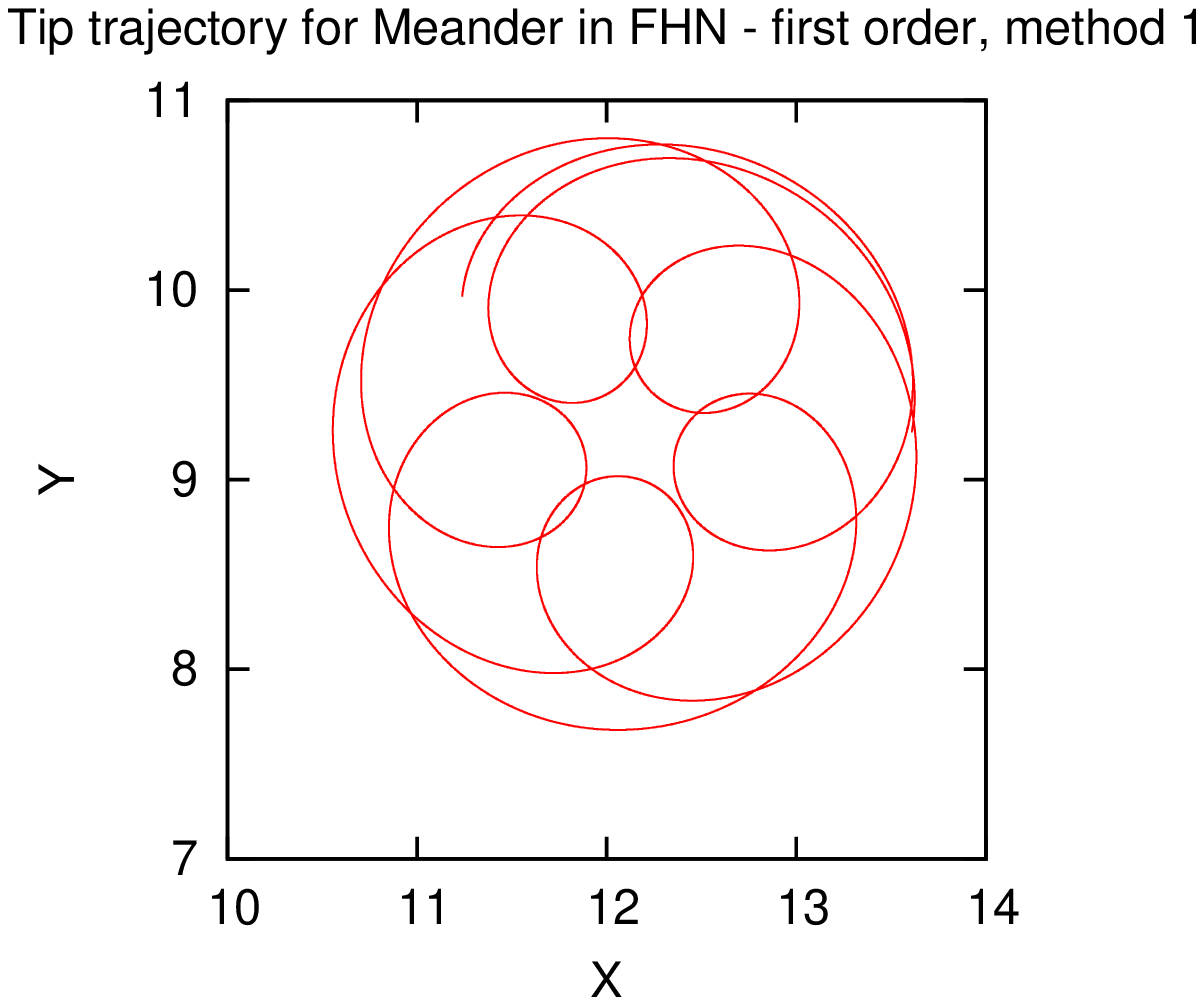}
\end{minipage}
\begin{minipage}{0.4\linewidth}
\centering
\includegraphics[width=0.7\textwidth, angle=-90]{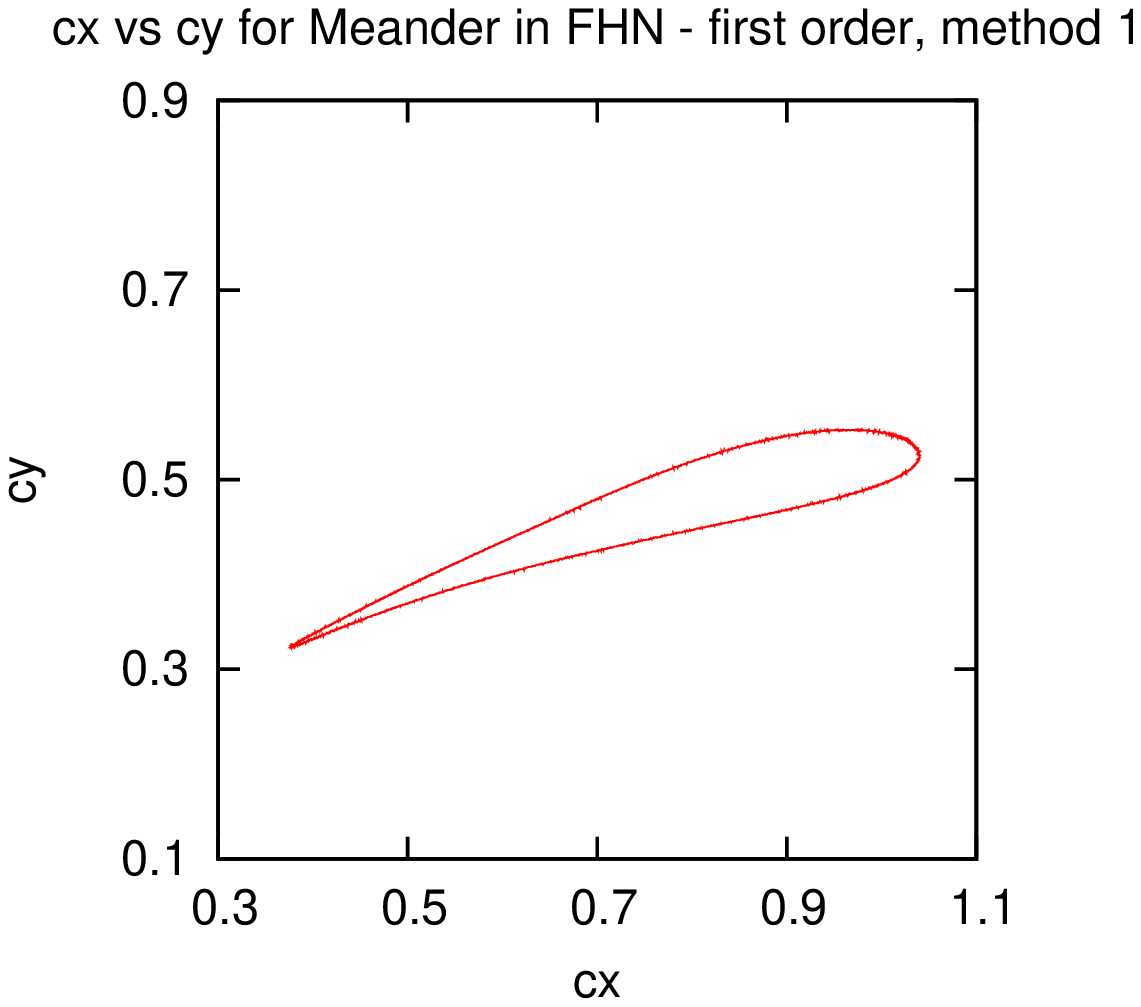}
\end{minipage}
\begin{minipage}{0.4\linewidth}
\centering
\includegraphics[width=0.7\textwidth, angle=-90]{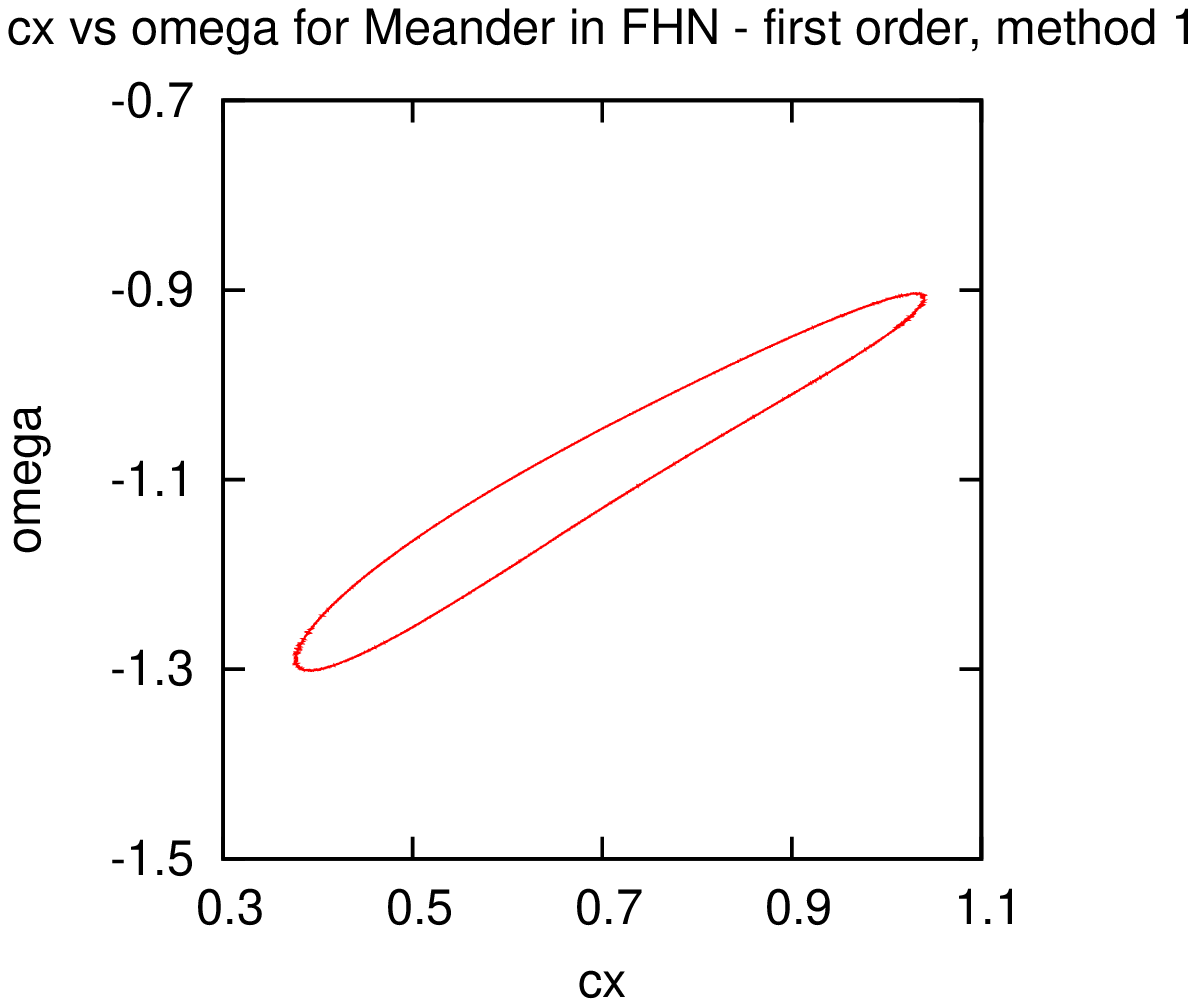}
\end{minipage}
\begin{minipage}{0.4\linewidth}
\centering
\includegraphics[width=0.7\textwidth, angle=-90]{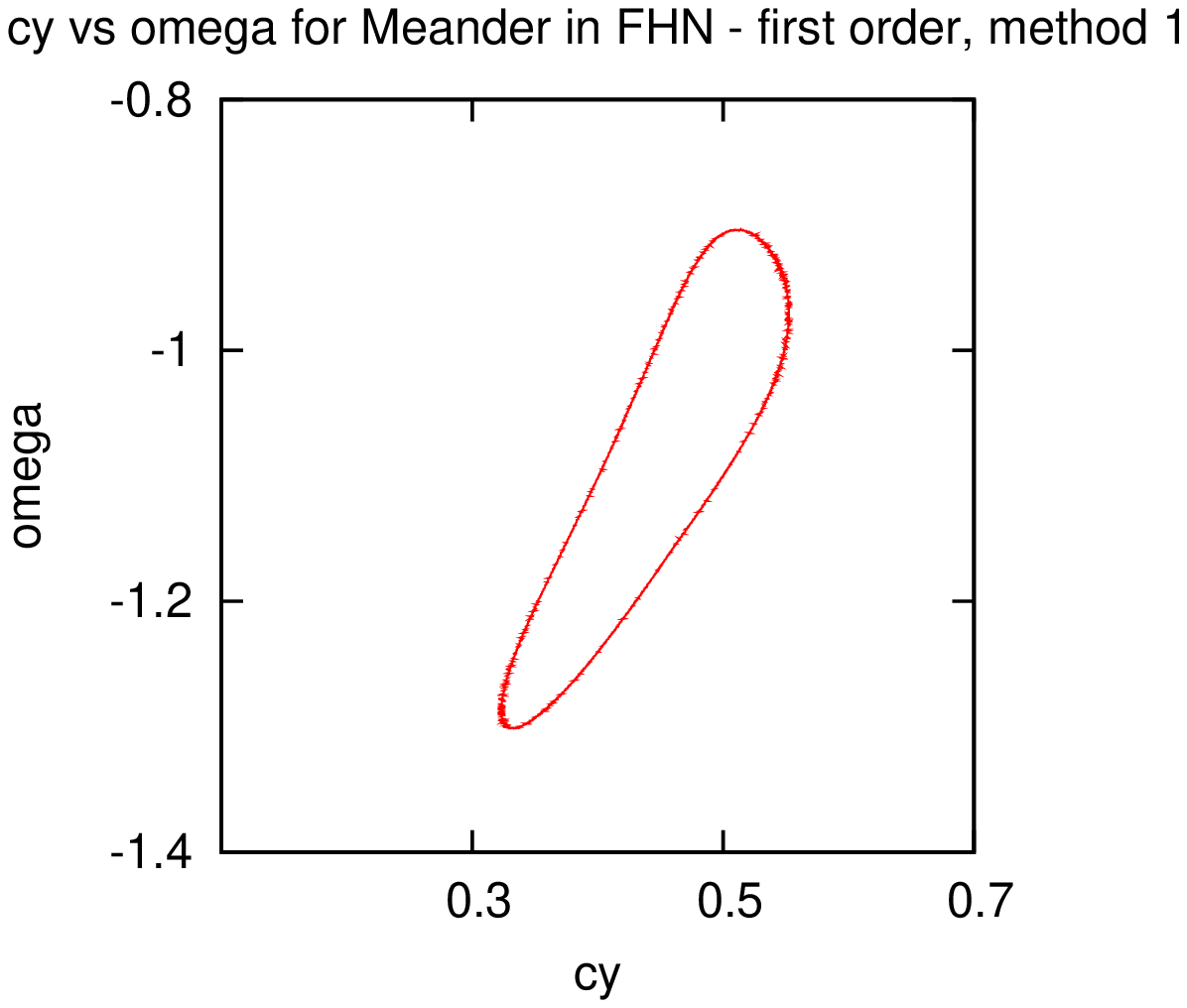}
\end{minipage}
\caption[Meander: FHN model, first order, method 1]{FHN model, First order, Method 1: (top left) Reconstructed Trajectory; (top right) $c_x$ versus $c_y$; (bottom left) $c_x$ versus $\omega$; (bottom right) $c_y$ versus $\omega$}
\label{fig:ezf_ex_mrw_fir_m1}
\end{center}
\end{figure}

Consider the first order with Method 1. The results are shown in Fig.(\ref{fig:ezf_ex_mrw_fir_m1}). We observe that the tip trajectory reconstructed from the quotient solution is nowhere near \chgex[ex]{similar to} what the trajectory should be like. It has five petal not three and also the petal are facing inwards, not outwards as required. So, all the physical characteristics of our spiral in the laboratory frame of reference are not reflected in the comoving frame of reference when we reconstructed the tip. However, what is not quite evident here is that there are ``wobbles'' in the quotient solution. The quotient solution must not be calculated properly and to an acceptable level of accuracy due to the fact that the reconstructed trajectory is nothing \chgex[ex]{similar to} what it should be.

\begin{figure}[p]
\begin{center}
\begin{minipage}{0.4\linewidth}
\centering
\includegraphics[width=0.7\textwidth, angle=-90]{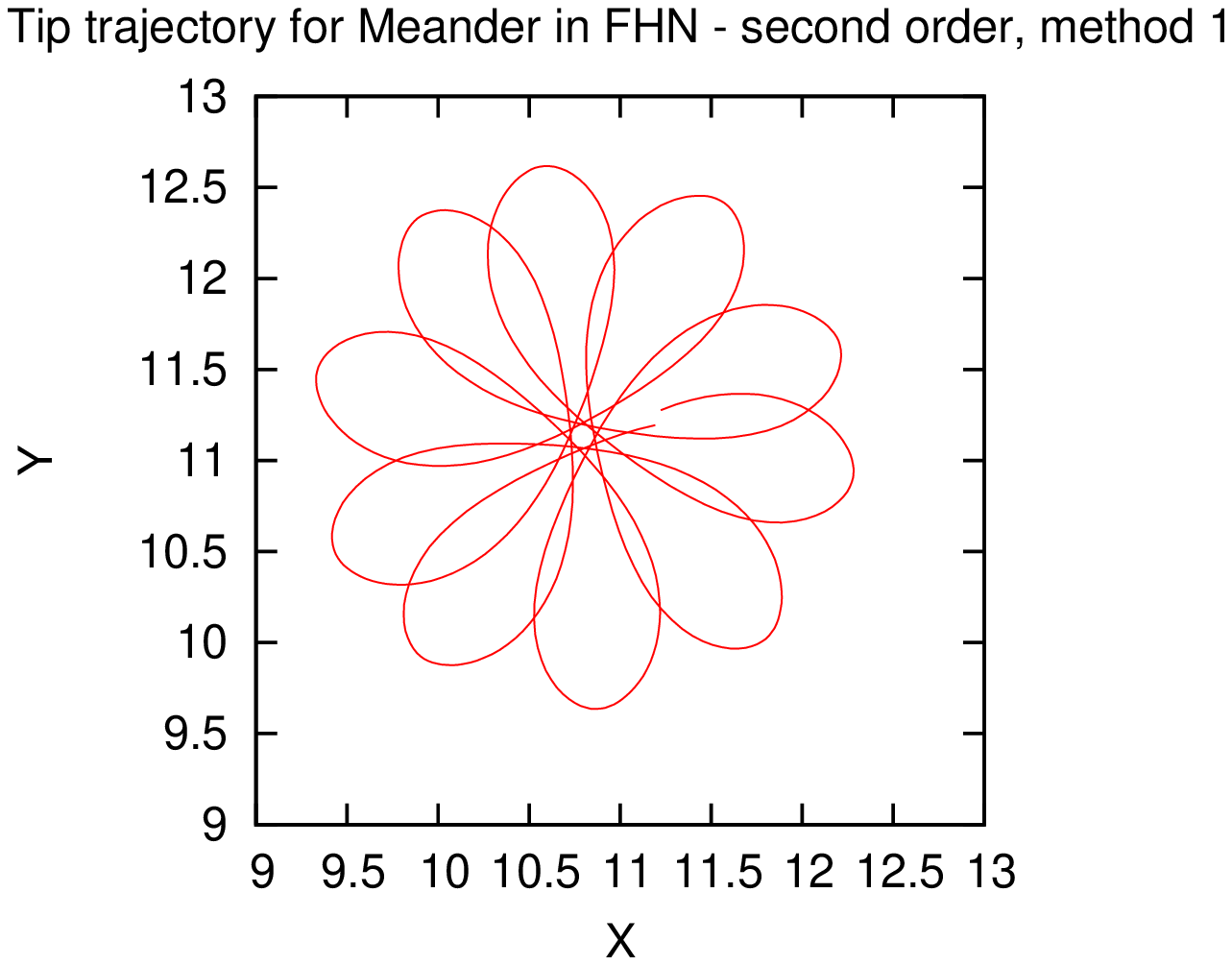}
\end{minipage}
\begin{minipage}{0.4\linewidth}
\centering
\includegraphics[width=0.7\textwidth, angle=-90]{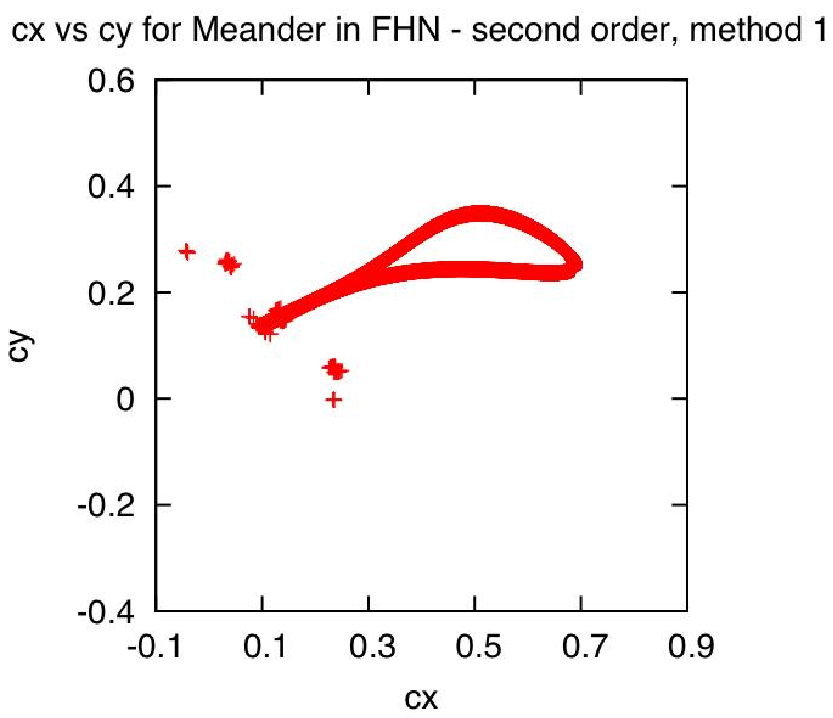}
\end{minipage}
\begin{minipage}{0.4\linewidth}
\centering
\includegraphics[width=0.7\textwidth, angle=-90]{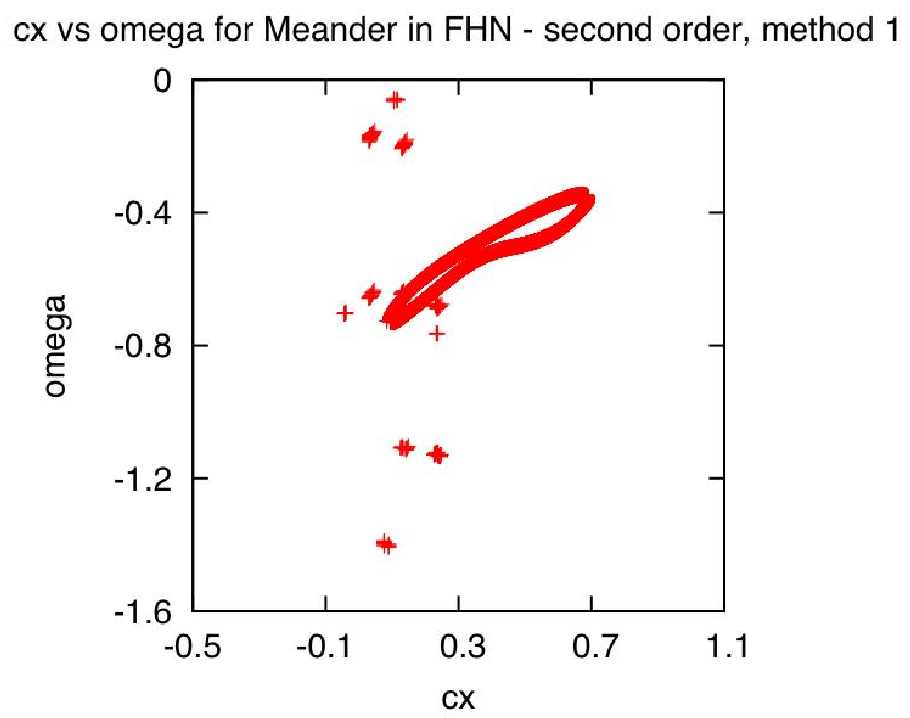}
\end{minipage}
\begin{minipage}{0.4\linewidth}
\centering
\includegraphics[width=0.7\textwidth, angle=-90]{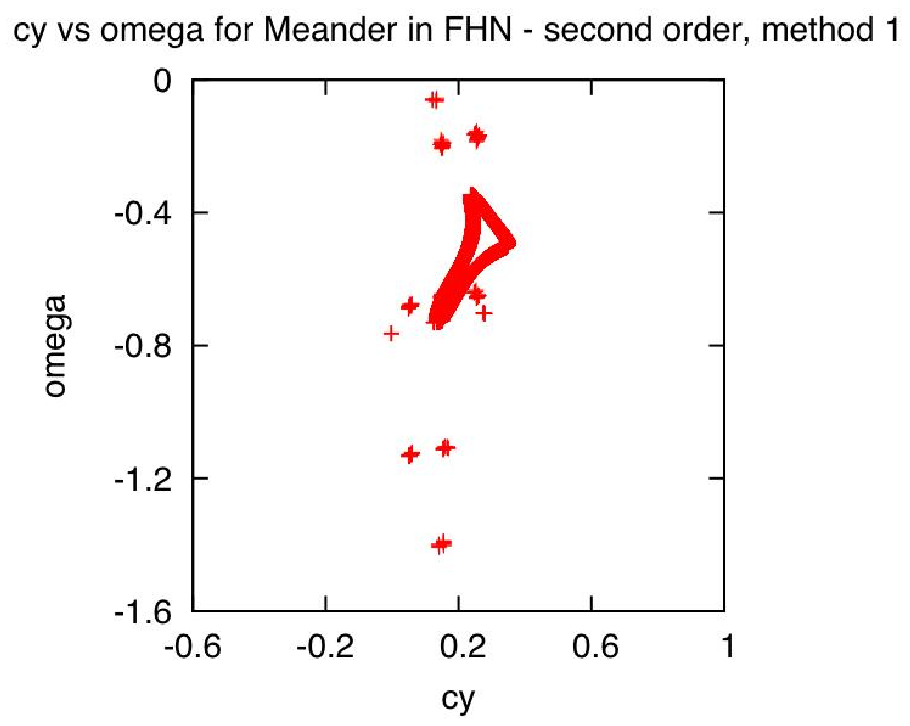}
\end{minipage}
\caption[Meander: FHN model, second order, method 1]{FHN model, Second order, Method 1: (top left) Reconstructed Trajectory; (top right) $c_x$ versus $c_y$; (bottom left) $c_x$ versus $\omega$; (bottom right) $c_y$ versus $\omega$}
\label{fig:ezf_ex_mrw_sec_m1}
\end{center}
\end{figure}

Now consider introducing the second order \chg[p151gram1]{scheme} whilst retaining Method 1, Fig.(\ref{fig:ezf_ex_mrw_sec_m1}). We now see that the anomalies in the quotient are very much evident, so much so that we have plotted the data using just points and not lines, since the plots would be unreadable otherwise. So, we also notice that the trajectory is very much like what we want. As expected, the quotient data has wobbles but we get an accurately calculated tip trajectory.

\begin{figure}[p]
\begin{center}
\begin{minipage}{0.4\linewidth}
\centering
\includegraphics[width=0.7\textwidth, angle=-90]{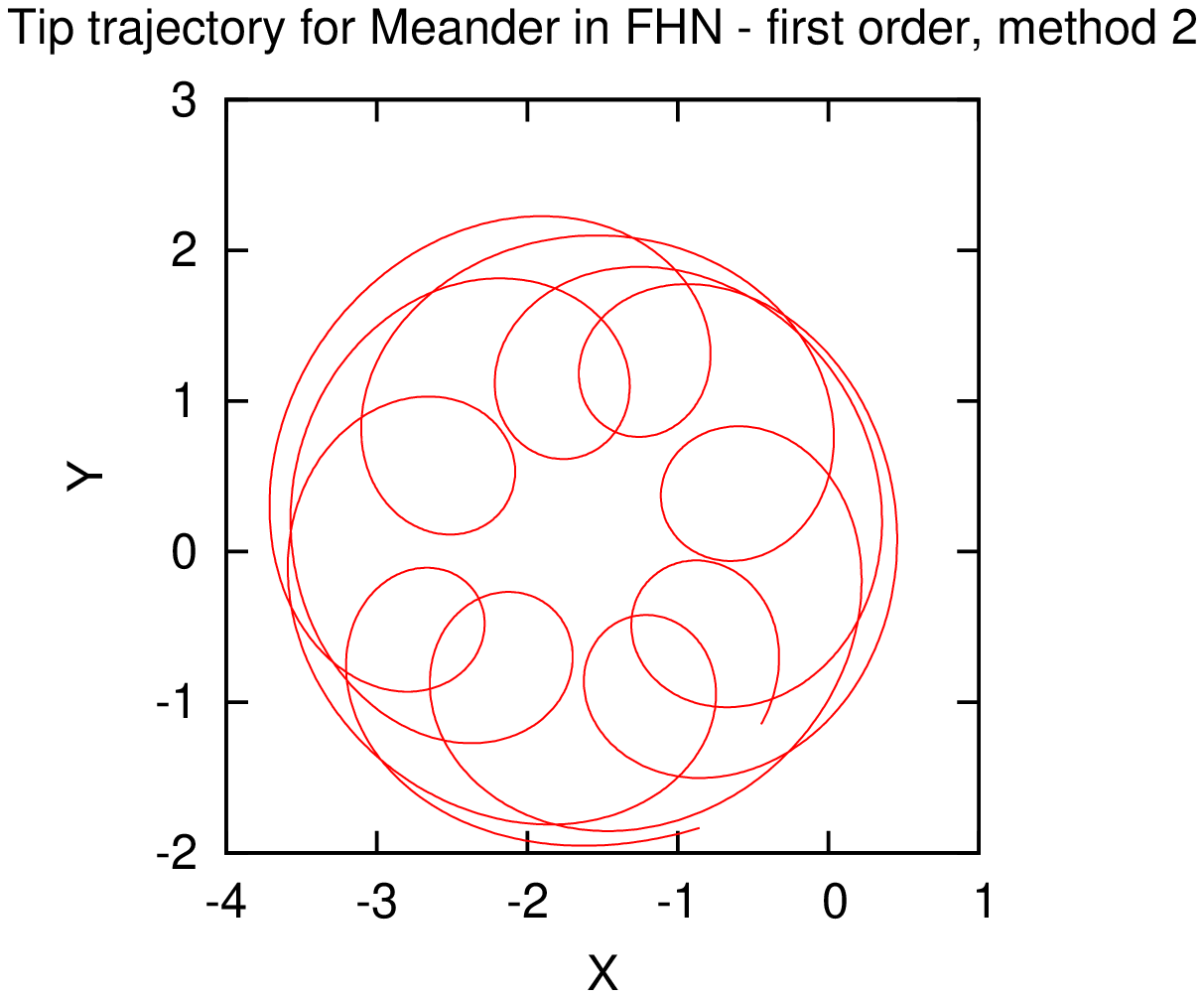}
\end{minipage}
\begin{minipage}{0.4\linewidth}
\centering
\includegraphics[width=0.7\textwidth, angle=-90]{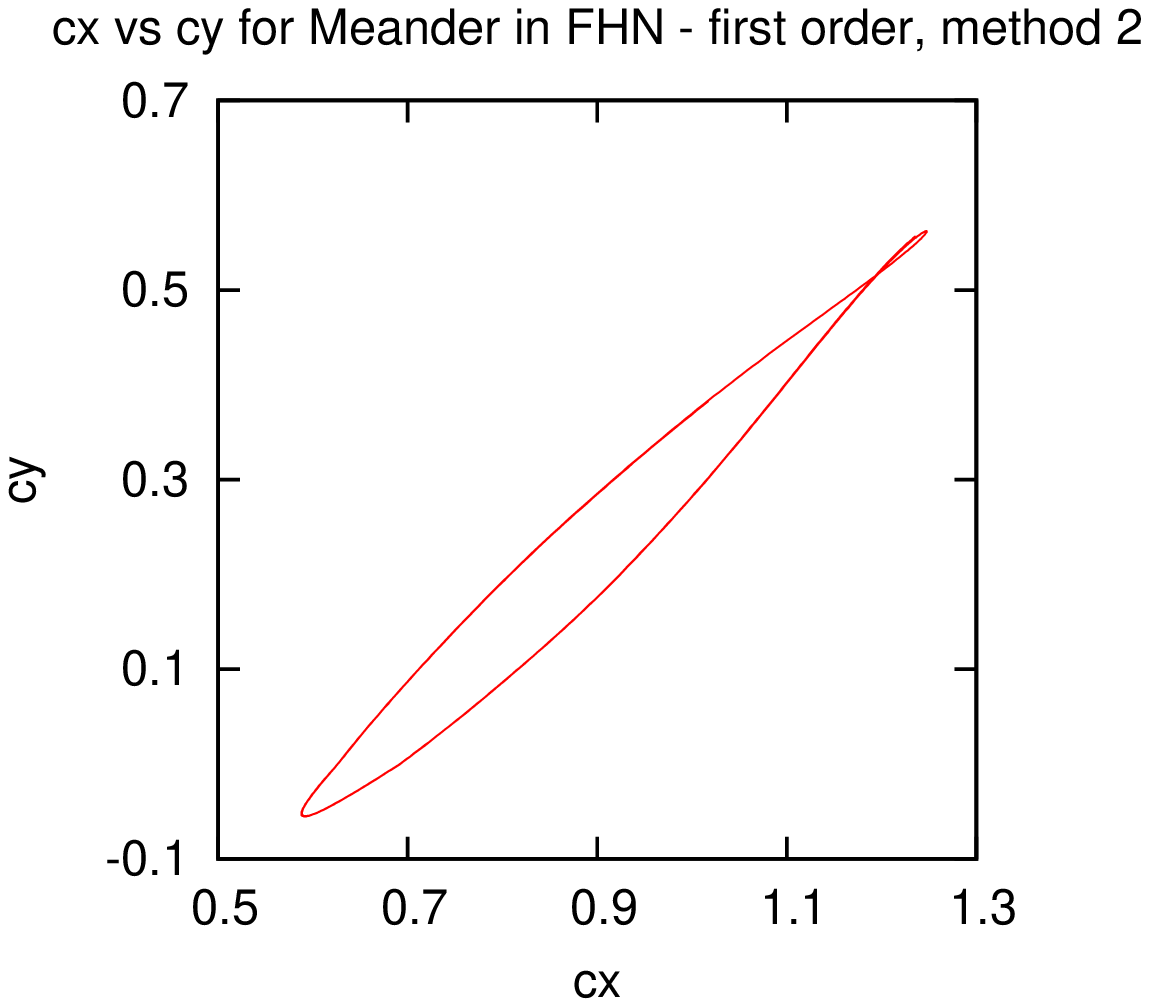}
\end{minipage}
\begin{minipage}{0.4\linewidth}
\centering
\includegraphics[width=0.7\textwidth, angle=-90]{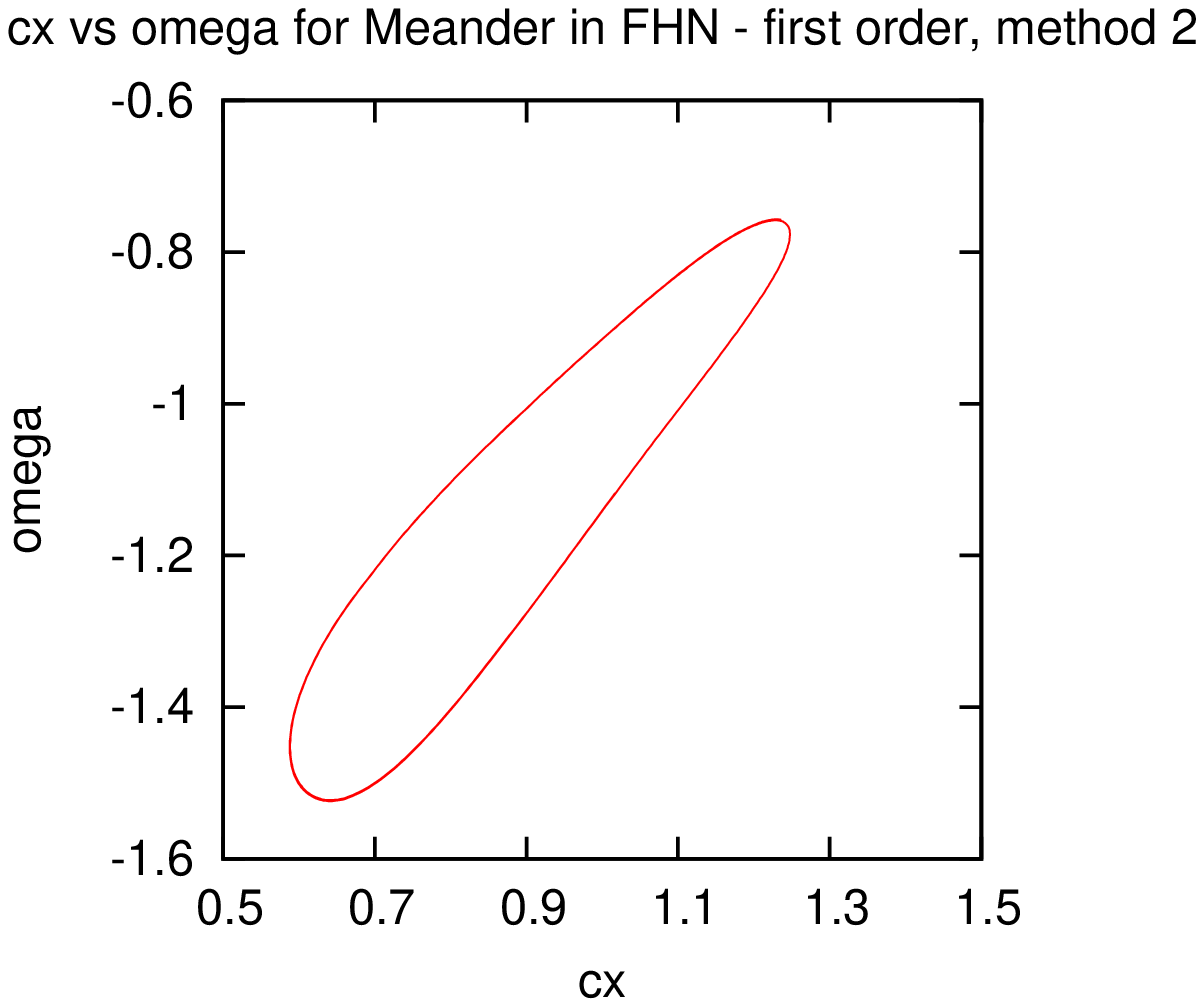}
\end{minipage}
\begin{minipage}{0.4\linewidth}
\centering
\includegraphics[width=0.7\textwidth, angle=-90]{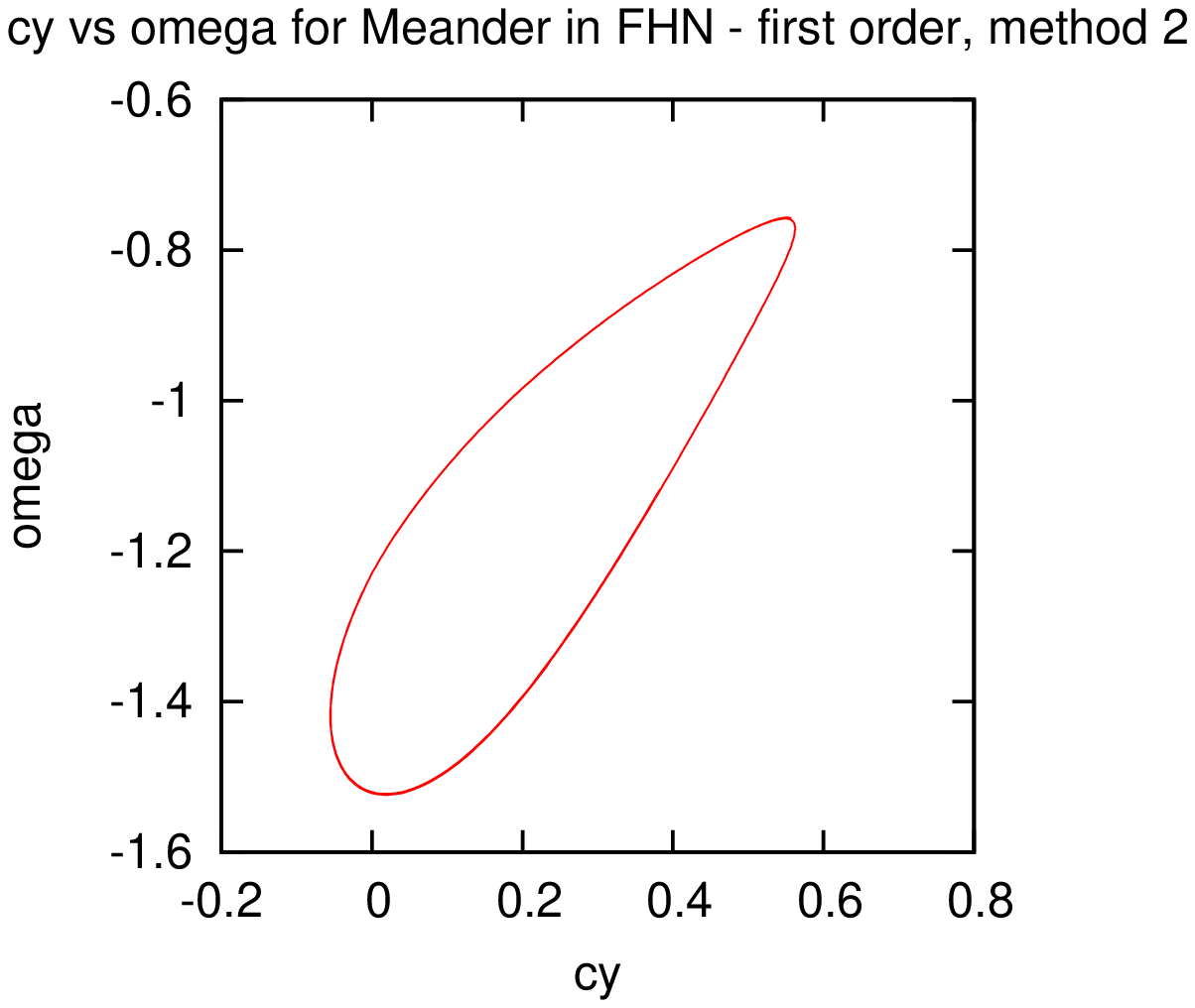}
\end{minipage}
\caption[Meander: FHN model, first order, method 2]{FHN model, First order, Method 2: (top left) Reconstructed Trajectory; (top right) $c_x$ versus $c_y$; (bottom left) $c_x$ versus $\omega$; (bottom right) $c_y$ versus $\omega$}
\label{fig:ezf_ex_mrw_fir_m2}
\end{center}
\end{figure}

Let us now use a first order scheme but this time with Method 2, Fig.(\ref{fig:ezf_ex_mrw_fir_m2}). We observe that although there are now no instabilities in the quotient solution, the quotient \chg[p151gram2]{solution} is not very accurate due to the tip trajectory being nothing like what it should.

\begin{figure}[p]
\begin{center}
\begin{minipage}{0.4\linewidth}
\centering
\includegraphics[width=0.7\textwidth, angle=-90]{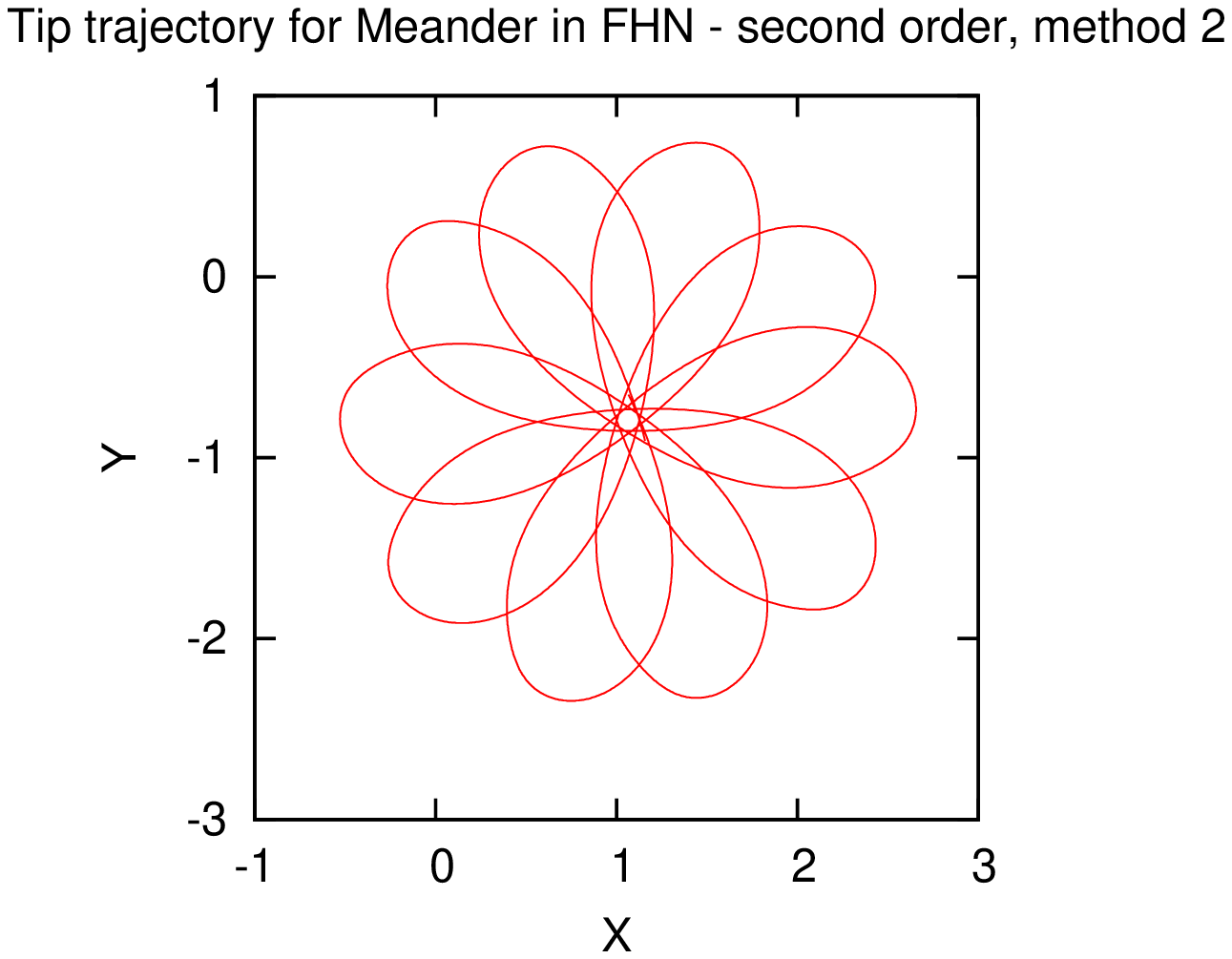}
\end{minipage}
\begin{minipage}{0.4\linewidth}
\centering
\includegraphics[width=0.7\textwidth, angle=-90]{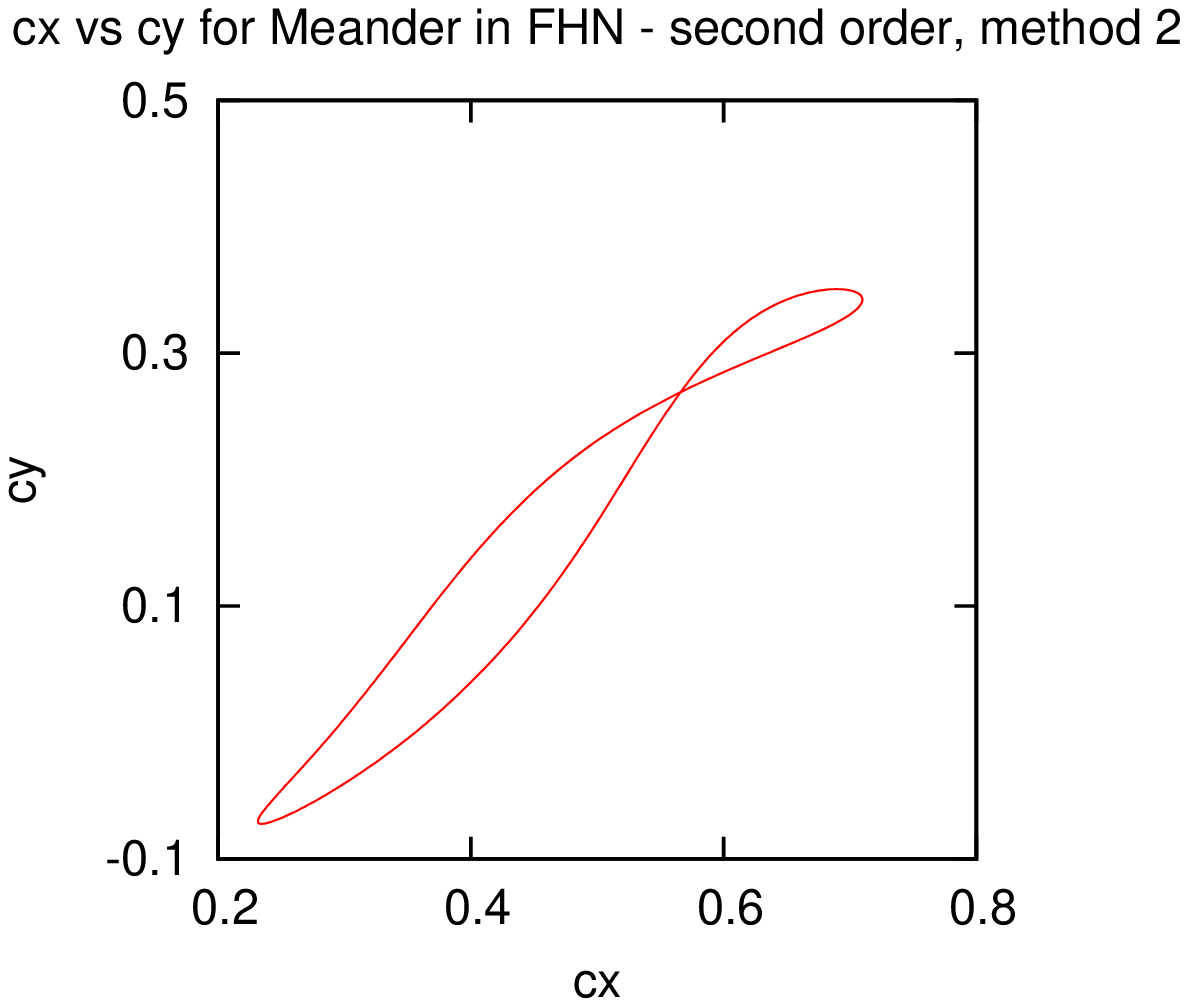}
\end{minipage}
\begin{minipage}{0.4\linewidth}
\centering
\includegraphics[width=0.7\textwidth, angle=-90]{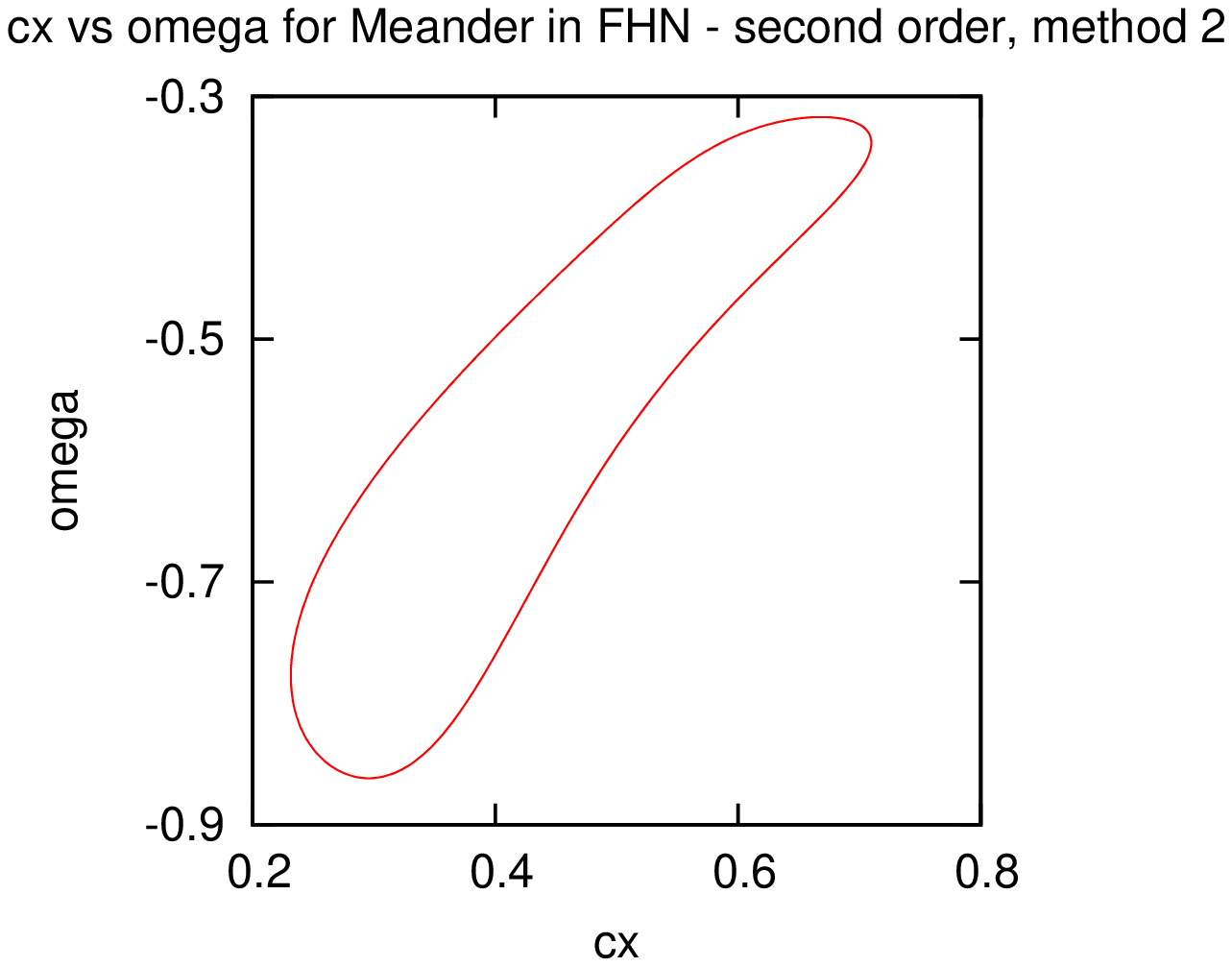}
\end{minipage}
\begin{minipage}{0.4\linewidth}
\centering
\includegraphics[width=0.7\textwidth, angle=-90]{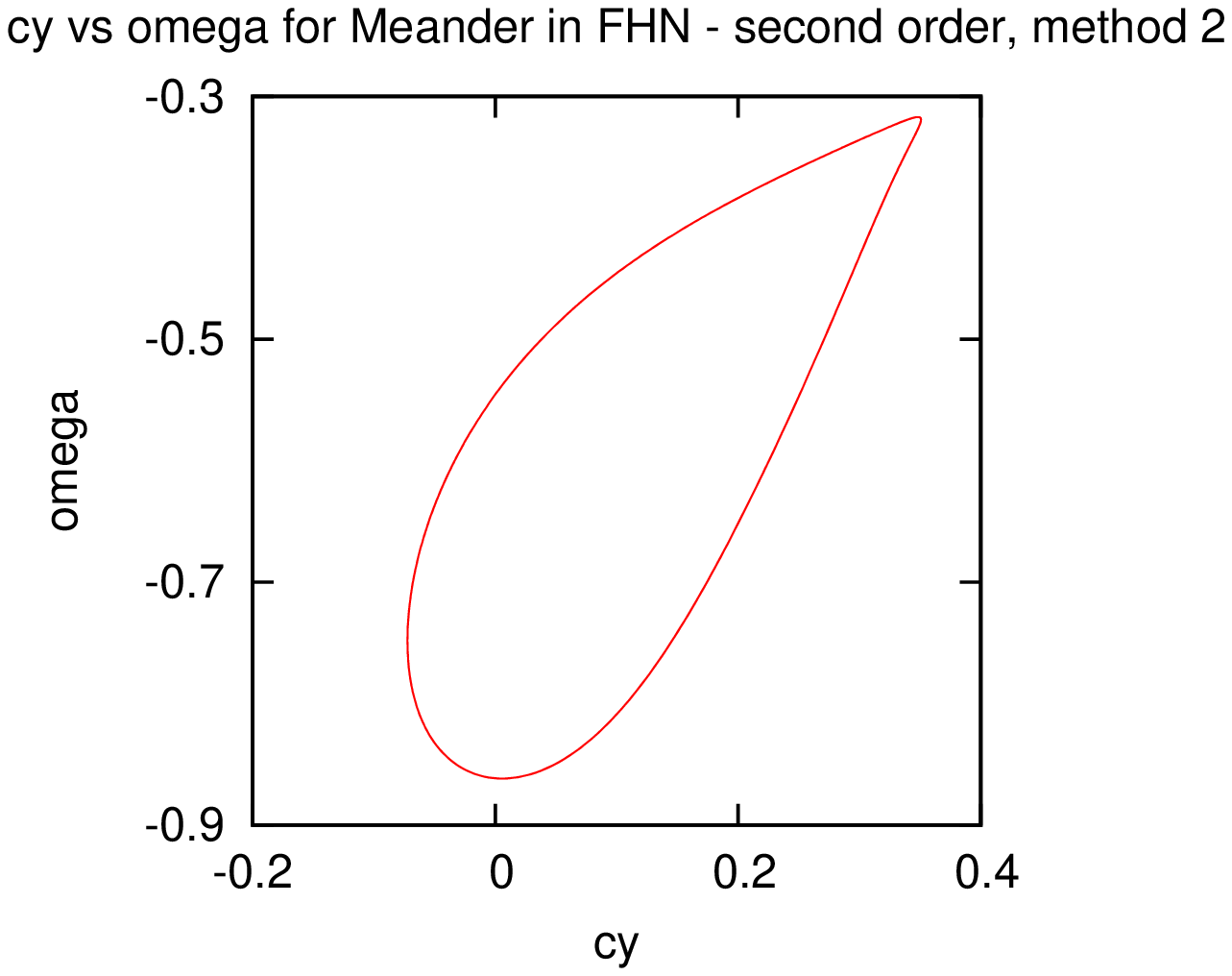}
\end{minipage}
\caption[Meander: FHN model, second order, method 2]{FHN model, Second order, Method 2: (top left) Reconstructed Trajectory; (top right) $c_x$ versus $c_y$; (bottom left) $c_x$ versus $\omega$; (bottom right) $c_y$ versus $\omega$}
\label{fig:ezf_ex_mrw_sec_m2}
\end{center}
\end{figure}

Therefore, we use a second order scheme with Method 2, Fig.(\ref{fig:ezf_ex_mrw_sec_m2}). As expected, this gives an accurately calculated tip trajectory. 

\newpage

\chgex[4]{}\chg[]{
\begin{center}
\begin{table}
\begin{tabular}{rccccc}
$y_{inc}$ &\vline& $|c|$ & $\omega$ & $r$ & \% diff. to lab. frame\\
\hline
\hline
Lab. frame &\vline& - & - & 2.2774975 & -\\
2 s.u. &\vline&  2.104634549 & -0.8715078235 & 2.414934775 & 6.0345741\%\\
3 s.u.  &\vline& 2.089801862 & -0.8805307150 & 2.373343514 & 4.2083916\%\\
4 s.u. &\vline&  2.080863056 & -0.8919446468 & 2.332950888 & 2.4348386\%\\
5 s.u. &\vline&  2.085102207 & -0.8917885423 & 2.338112801 & 2.661487\%
\end{tabular}
\label{tab:chap5_compare_rw}
\caption{Dependence of the accuracy of the solution on the second pinning point.}
\end{table}
\end{center}

\subsubsection{Accuracy by varying the second pinning point ($x_{inc},y_{inc}$)}

We finish this section by showing how the solutions differ when the second pinning is placed at different distances from the origin (first pinning point). We will show two tests: one for a rigidly rotating spiral wave solution to Barkley's model; the second for a meandering spiral wave solution in FHN. We shall use a box size of 20 s.u., with a spacestep of $\Delta_x=0.1$ and timstep of $\Delta_t=2.5\times10^{-3}$ in all the simulations.

\begin{figure}[bth]
\begin{center}
\begin{minipage}[htbp]{0.7\linewidth}
\centering
\includegraphics[width=0.7\textwidth, angle=-90]{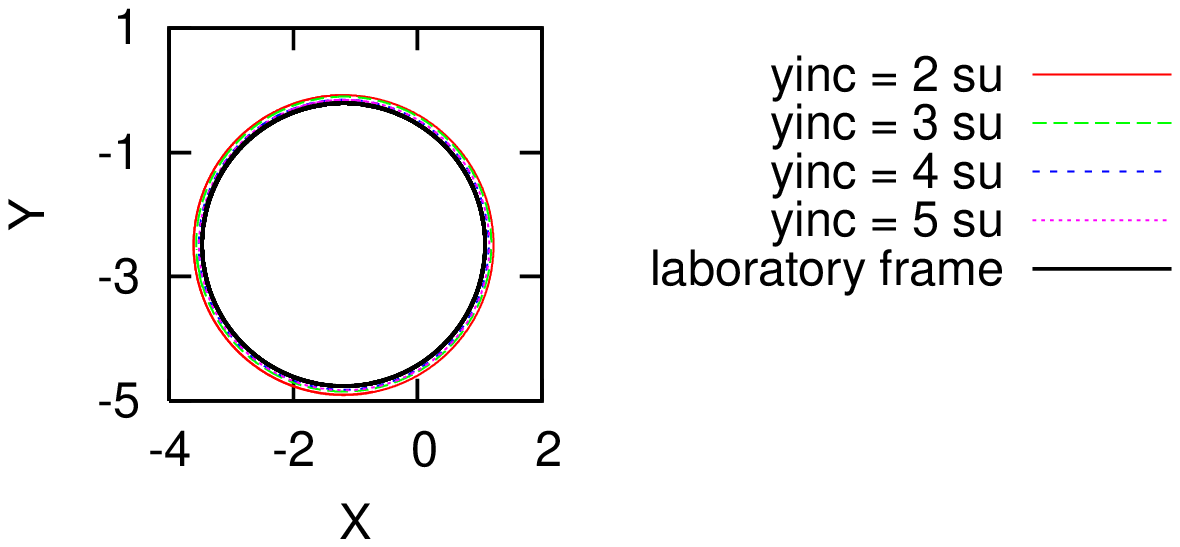}
\end{minipage}
\caption[Rigid rotation: dependency on the second pinning point]{Barkley's model, Second order, method 2: Tip Trajectories for varying $y_{inc}$ illustrating the dependency of the solution on the position of the second pinning point}
\label{fig:ezf_ex_yinc}
\end{center}
\end{figure}

Consider a rigidly rotating spiral wave solution and Fig.(\ref{fig:ezf_ex_yinc}). We have $x_{inc}=0$ throughout the simulations, and then, for each different simulation, we have moved $y_{inc}$ by 1 s.u. starting from 2 s.u. We show in table (\ref{tab:chap5_compare_rw}) the values of the translational speed ($|c|=\sqrt{c_x^2+c_y^2}$) and $\omega$ as determined by EZ-Freeze, together with the radius of the trajectory calculated as:

\begin{equation*}
r = \frac{|c|}{|\omega|}
\end{equation*}
\\
and also how the radius of the reconstructed trajectory compares to the radius in the laboratory frame of reference, as a percentage.

We see that when we have the second pinning point relatively close to the first pinning point, the calculations are not as accurate compared to the pinning point further away. However, as we move the second point further away from the first pinning point, we find that the calculations get more accurate. In fact, for $y_{inc}=4$ s.u. and 5 s.u we see that the solutions are quite similar.

Finally, we show how the trajectories for a meandering solution varied with the position of the second pinning. Again we have numerical parameters as for rigid rotation (box size of 20 s.u., the spacestep of $\Delta_x=0.1$ and the timstep of $\Delta_t=2.5\times10^{-3}$), and we kept $x_{inc}=0$ for all simulations. We show the results in Fig.(\ref{fig:ezf_ex_yinc}) for a range if values of $y_{inc}$.

\begin{figure}[bth]
\begin{center}
\begin{minipage}[htbp]{0.49\linewidth}
\centering
\includegraphics[width=1.0\textwidth, angle=-90]{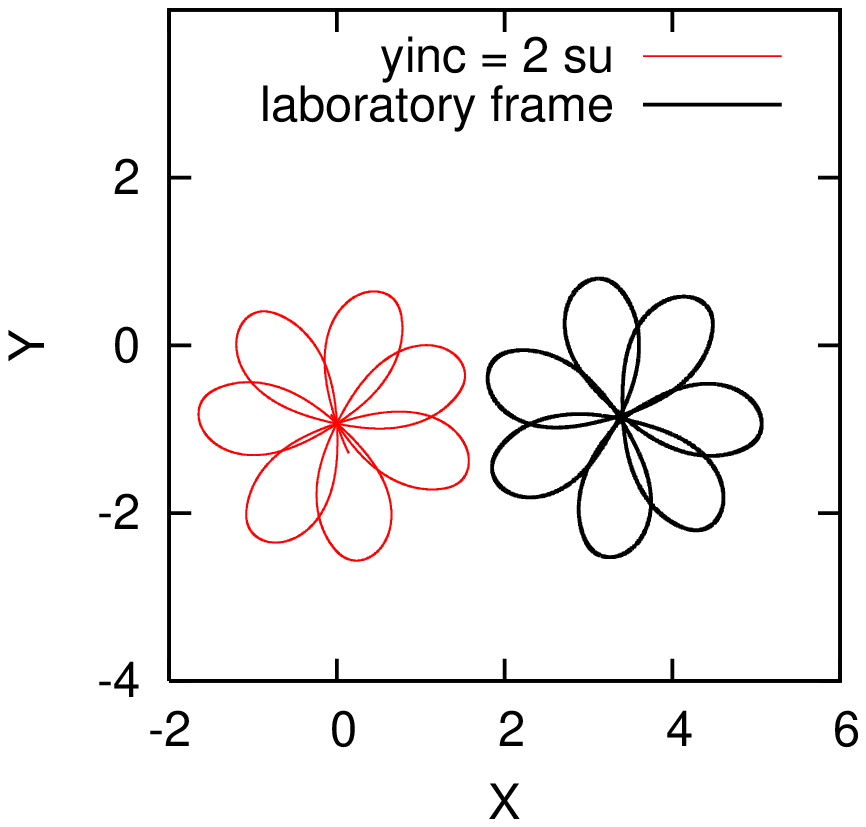}
\end{minipage}
\begin{minipage}[htbp]{0.49\linewidth}
\centering
\includegraphics[width=1.0\textwidth, angle=-90]{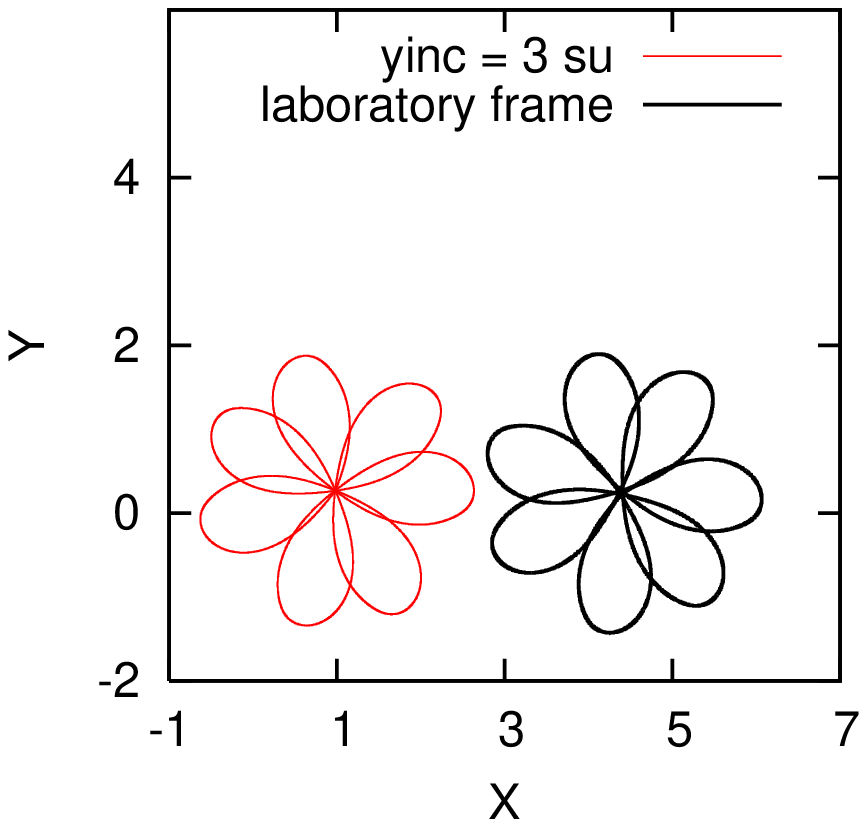}
\end{minipage}
\begin{minipage}[htbp]{0.49\linewidth}
\centering
\includegraphics[width=1.0\textwidth, angle=-90]{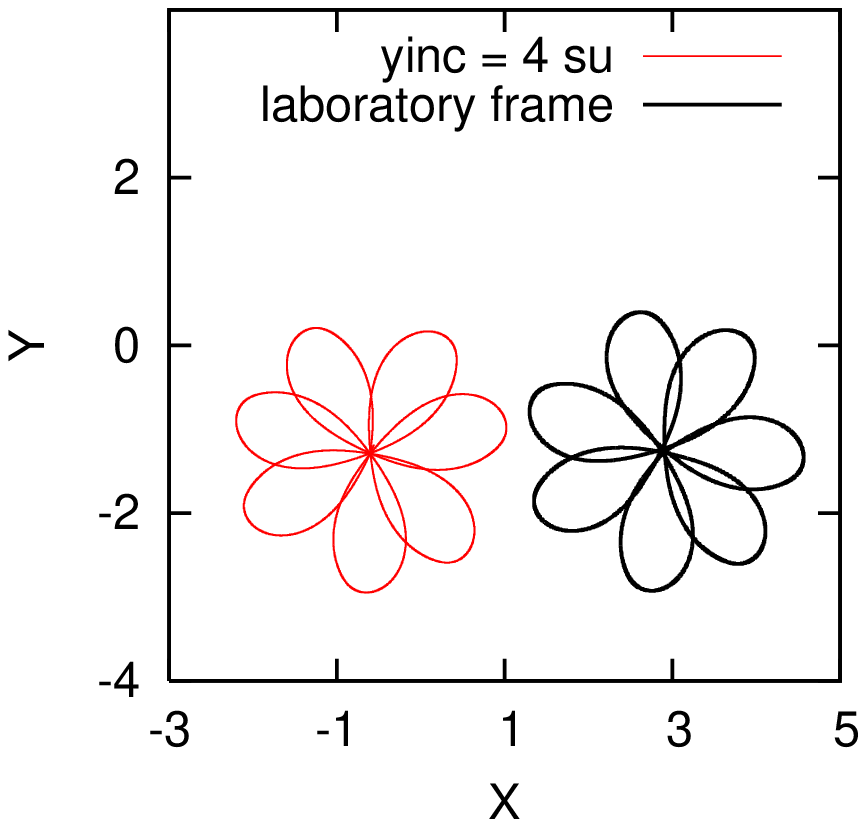}
\end{minipage}
\begin{minipage}[htbp]{0.49\linewidth}
\centering
\includegraphics[width=1.0\textwidth, angle=-90]{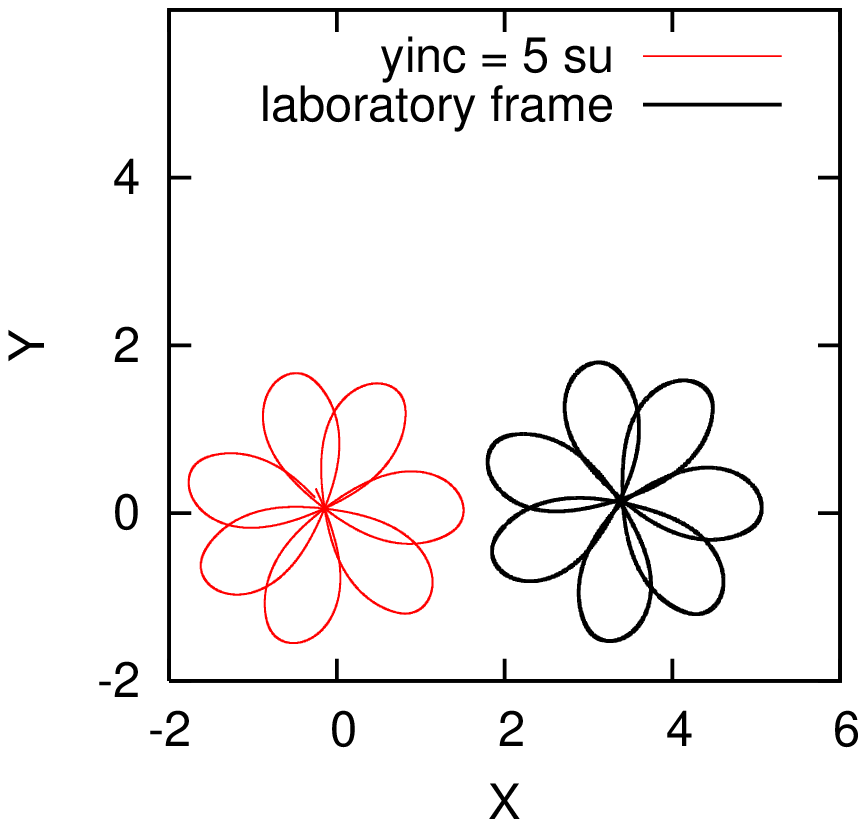}
\end{minipage}
\caption[Meander: dependency on the second pinning point]{Barkley's model, Second order, method 2: reconstructed meander tip trajectory and the original tip trajectory in the laboratory frame of reference for $y_{inc}$=2 s.u. (top left), $y_{inc}$=3 s.u. (top right), $y_{inc}$=4 s.u. (bottom left) and $y_{inc}$=5 s.u. (bottom right).}
\label{fig:ezf_ex_yinc}
\end{center}
\end{figure}

Firstly, we note that the reconstructed trajectory in each case is very close to what the trajectory is like in the laboratory frame of reference. We can also see that for meander, the position of the second pinning point makes only a small difference, if any, to the accuracy of the quotient solution and hence the reconstruction of the original tip trajectory, i.e. the tip trajectory in the laboratory frame of reference. Therefore, the reconstruction of the the tip trajectory for a rigidly rotating spiral wave is much more sensitive to the position of the second pinning than it is for meandering spiral waves.
}

\section{Convergence Testing of EZ-Freeze}
\label{sec:ezf_convergence}
Numerical accuracy is important in any numerical method. So we decided to test out how accurate the numerical methods implemented into EZ-Freeze were. 

We note that we have used a first order accurate forward Euler scheme to calculate the time derivatives. Therefore, one test we decided to conduct was convergence in the timestep. In this case, for varying values of the timestep, we should observe a linear relationship between the solutions generated and the timestep.

We also note that our optimal numerical scheme for the spatial derivatives is second order accurate. Therefore, we decided to look at convergence in spacestep. In this case, we should get a quadratic relationship between the solutions and the spacestep.

Lastly, one of the advantages of EZ-Freeze, is the fact that the simulations can be done in a smaller box compared to conventional techniques (i.e. simulations done in a laboratory frame of reference). So, the last test was convergence in box size. In this case, we should observe that the solutions converge to particular values as the box size increases.


\subsection{Methods}

We shall perform the tests on a rigidly rotating spiral wave in Barkley's model. We should observe the components of the quotient solution should provide us with the results mentioned above.

The model parameters used were:

\begin{eqnarray*}
a &=& 0.52\\
b &=& 0.05\\
\varepsilon &=& 0.02
\end{eqnarray*}

We also state below the ``best'' numerical and physical parameters that we used throughout the tests:

\[\begin{array}{rcl}
\Delta_t &=& 1.11\times10^{-4}\\
\Delta_x &=& \frac{1}{15}\\
\mbox{Box Size}\quad LX &=& 60 \mbox{s.u.}
\end{array}\]

We also note that the timestep is controlled in EZ-Freeze by a parameter called $t_s$, which is the ratio of the timestep to the diffusion stability limit. This optimal timestep corresponds to $t_s=0.1$. The timestep is given by the formula:

\begin{equation}
\Delta_t = \frac{t_s\Delta_x^2}{4}
\end{equation}

Also, we have that we must have a grid size of $NX=901$ for a spacestep of $\Delta_x=\frac{1}{15}$.

In the convergence test for the timestep, $\Delta_t$, we kept $LX$ and $\Delta_x$ at the optimal values, and varied the timestep by adjusting the parameter $t_s$. Starting at the optimal value, we increased $t_s$ in steps of 0.04 each time, until the solution became unstable.

For convergence in the box size size, we kept the $\Delta_t$ and $\Delta_x$ at the optimal values, and varied the box size by starting at the optimal value \chg[p156gram]{}reduced the box size by 5 s.u. each time, until we got to $LX=15$. After this we decreased the box size in steps of 1 s.u. The reason for this will become apparent in the results section.

For the convergence in the spacestep, $\Delta_x$, we must mention at this point the importance of the pinning points. We came across several problems with the testing and one of these problems was tied down to the position of the second tip pinning condition. We noted in Sec.(\ref{sec:ezf_numerics_imp}) that the second pinning point could vastly improve the accuracy of the calculation of the quotient system, in particular, $\omega$. When we first conducted the tests we set the second point at $(x_{inc},y_{inc})=(0,2)$ in space units. As we found out, the position of the second pinning point is too close to the first and therefore the results generated were not very clear. Therefore, we decided to put the second pinning point further away from the first. In fact we tried to put it near one of the boundaries, keeping $x_{inc}=0$ and setting $y_{inc}=\frac{LX}{2}-1$, i.e. 1 s.u. away from the boundary. Unfortunately, this yielded rather strange and, what we conceived, as inaccurate results. One of the reasons for this that is conjected by the author is that the distance between the pinning points is more than the wavelength of the spiral. Therefore, in the tests conducted, we set $y_{inc}$ such that it is just within one full rotation of the spiral. So, if we denote the wavelength as $\Lambda$ then:

\begin{equation}
y_{inc} < \Lambda
\end{equation} 

So, for both convergence in timestep and spacestep we set $y_{inc}=15$ s.u., making sure that, for convergence in the spacestep, the position of the second pinning point is preserved.

Therefore, for convergence in spacestep, we considered setting the spacestep as $\Delta_x=\frac{1}{i}$, where $i\leq15$ and $i\in\mathbb{Z}$. Therefore, we started with $i=15$ and decreased $i$ by 1 for each step.

Finally, we note that all simulations were carried out using both Dirichlet boundary conditions and Neumann boundary conditions.

\subsection{Convergence in Barkley's Model}

We will now show the results of the three different tests in Barkley's model. For each subsection that follows, we shall show how $c_x$, $c_y$ and $\omega$ all depend on the spacestep, timestep, and box size. We shall also show the final spiral wave solutions for each simulation within the convergence tests. These solutions will help the reader to further understand the relationship between $c_x$, $c_y$ and $\omega$ and the numerical \& physical parameters.

Also, we shall show the plots for both Neumann and Dirichlet boundary conditions.


\subsubsection{Convergence in the spacestep}

\begin{figure}[tbh]
\begin{center}
\begin{minipage}{0.6\linewidth}
\centering
\includegraphics[width=0.7\textwidth, angle=-90]{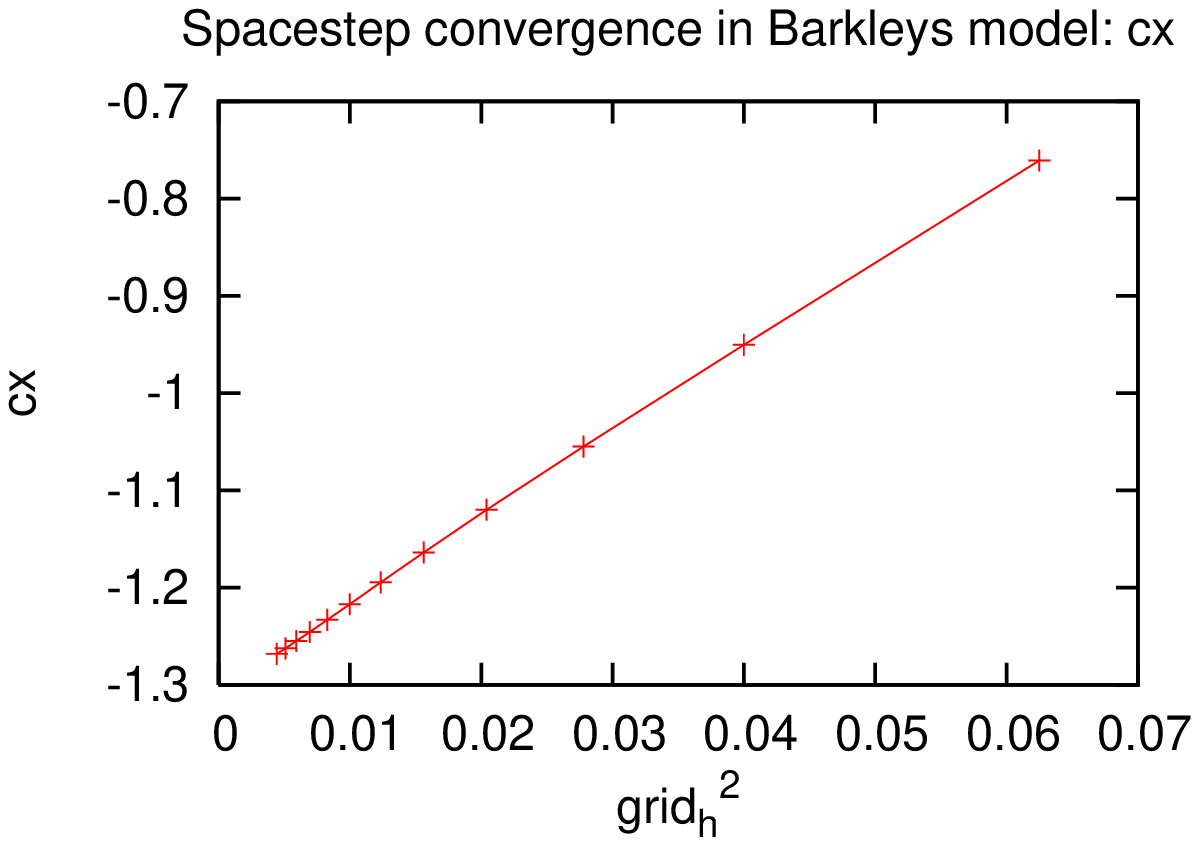}
\end{minipage}
\begin{minipage}{0.6\linewidth}
\centering
\includegraphics[width=0.7\textwidth, angle=-90]{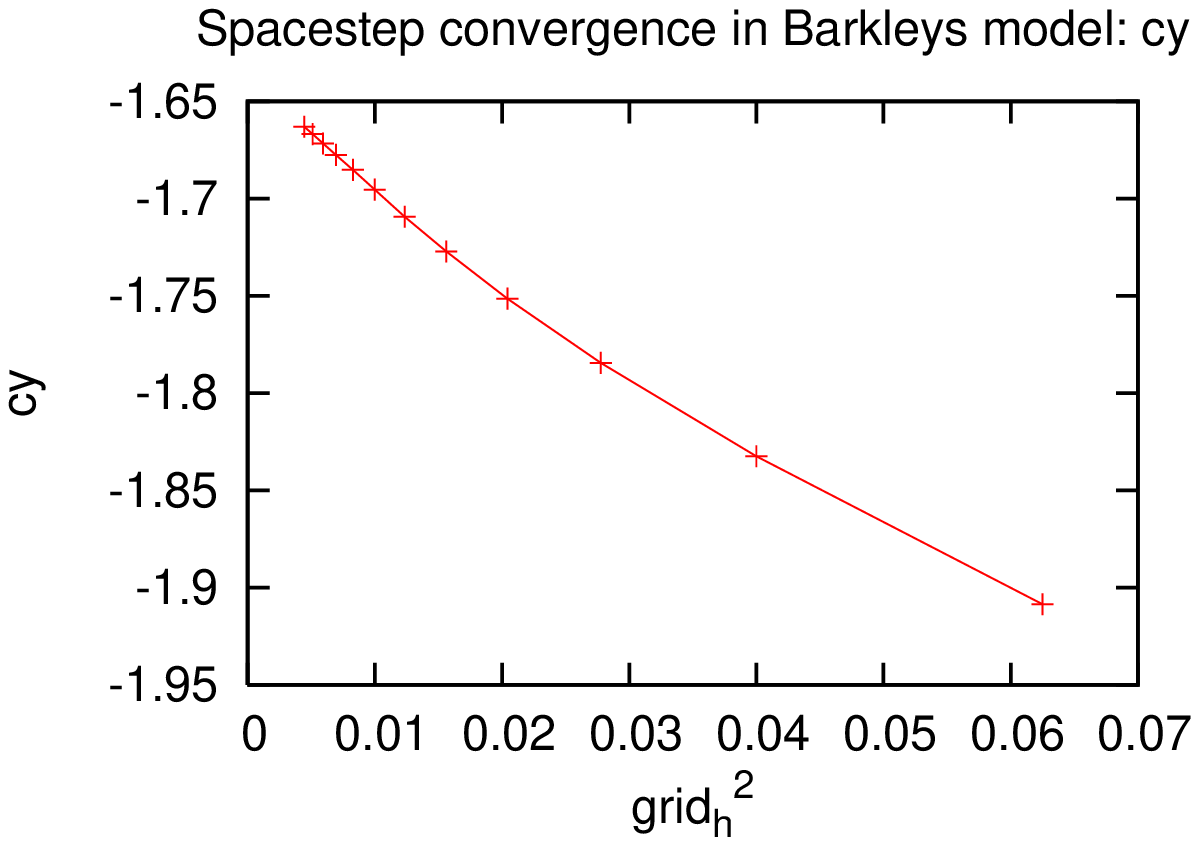}
\end{minipage}
\begin{minipage}{0.6\linewidth}
\centering
\includegraphics[width=0.7\textwidth, angle=-90]{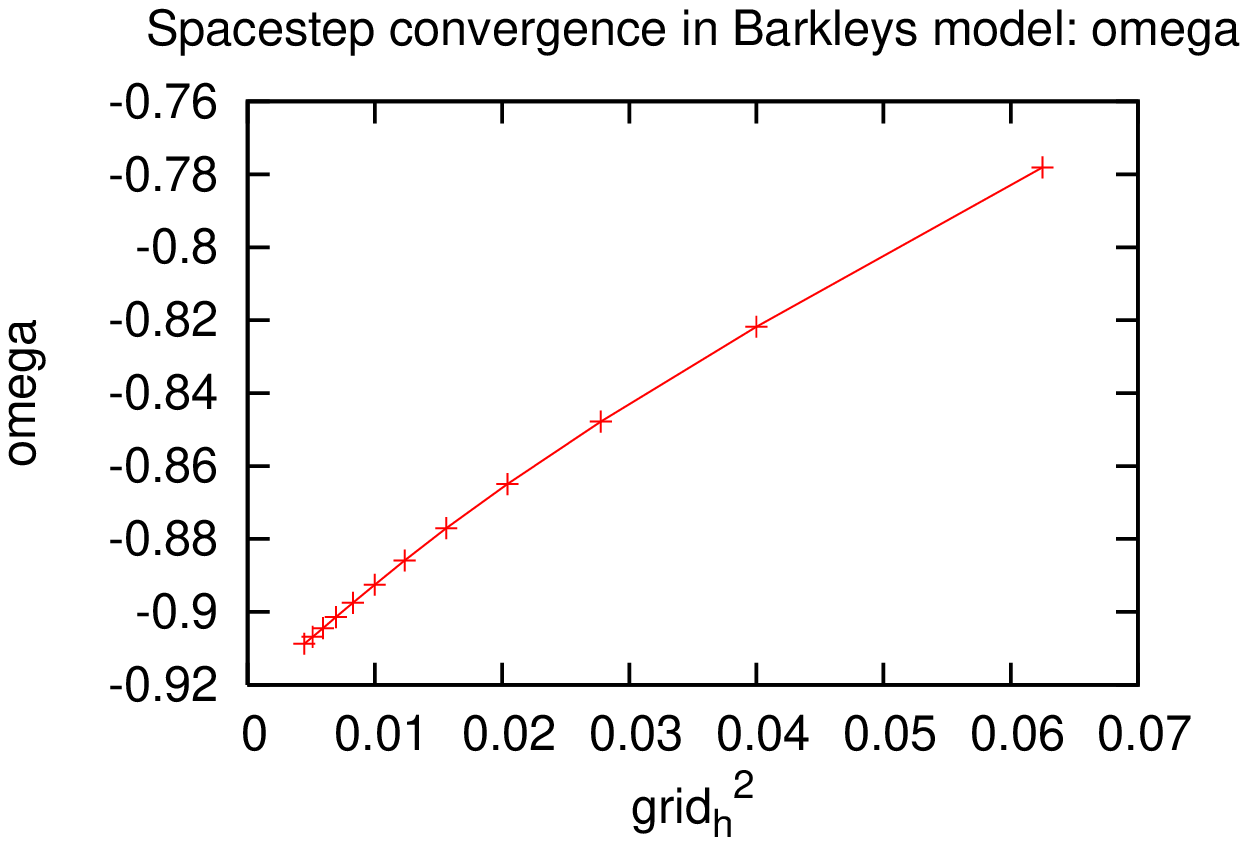}
\end{minipage}
\caption[Spacestep convergence: Neumann boundary condition]{Convergence in spacestep, using Barkley's model and Neumann Boundary conditions with the box size fixed at $L_X=60$, and the timestep fixed at $\Delta_t=1.11\times10^{-4}$.}
\label{fig:ezf_conv_1_nbc}
\end{center}
\end{figure}

In Fig.(\ref{fig:ezf_conv_1_nbc}), we show the plots of $c_x$, $c_y$ and $\omega$ against $\Delta_x^2$ using Neumann boundary conditions. Since we are using a second order scheme, we should find that these plots produce a linear relationship. 

We note that the box size was kept constant throughout the simulations with $L_X$ = 60 s.u. We also kept the timestep constant with $\Delta_t=1.11\times10^{-4}$, by varying $t_s$ accordingly.

From (\ref{fig:ezf_conv_1_nbc}), it is clear that there is a linear relationship between the advection coefficients and $\Delta_x^2$, which is what we expect with a second order accurate scheme. Although the plots involving $c_y$ and $\omega$ are not exactly linear relationships, they are very close to being linear, and therefore we can conclude that they is a linear relationship between all three advection coefficients and $\Delta_x^2$.

Next we show in Fig.(\ref{fig:ezf_conv_1_dbc}) the results from using Dirichlet boundary conditions. These results appear to be almost exactly the same as those for Neumann boundary conditions. So we can conclude that the choice of boundary conditions is irrelevant in these tests since both give the same results.

\begin{figure}[tbh]
\begin{center}
\begin{minipage}{0.32\linewidth}
\centering
\includegraphics[width=0.7\textwidth]{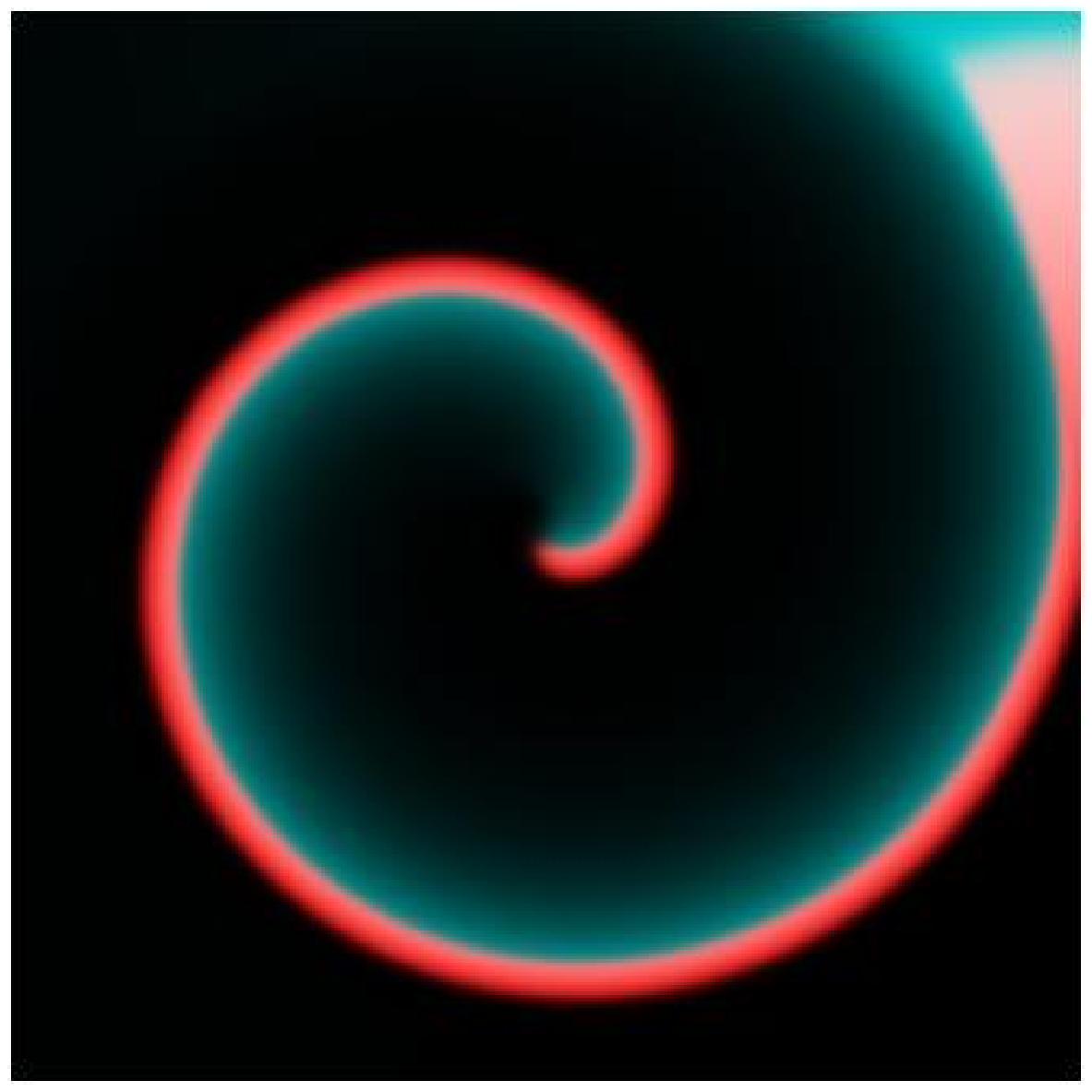}
\end{minipage}
\begin{minipage}{0.32\linewidth}
\centering
\includegraphics[width=0.7\textwidth]{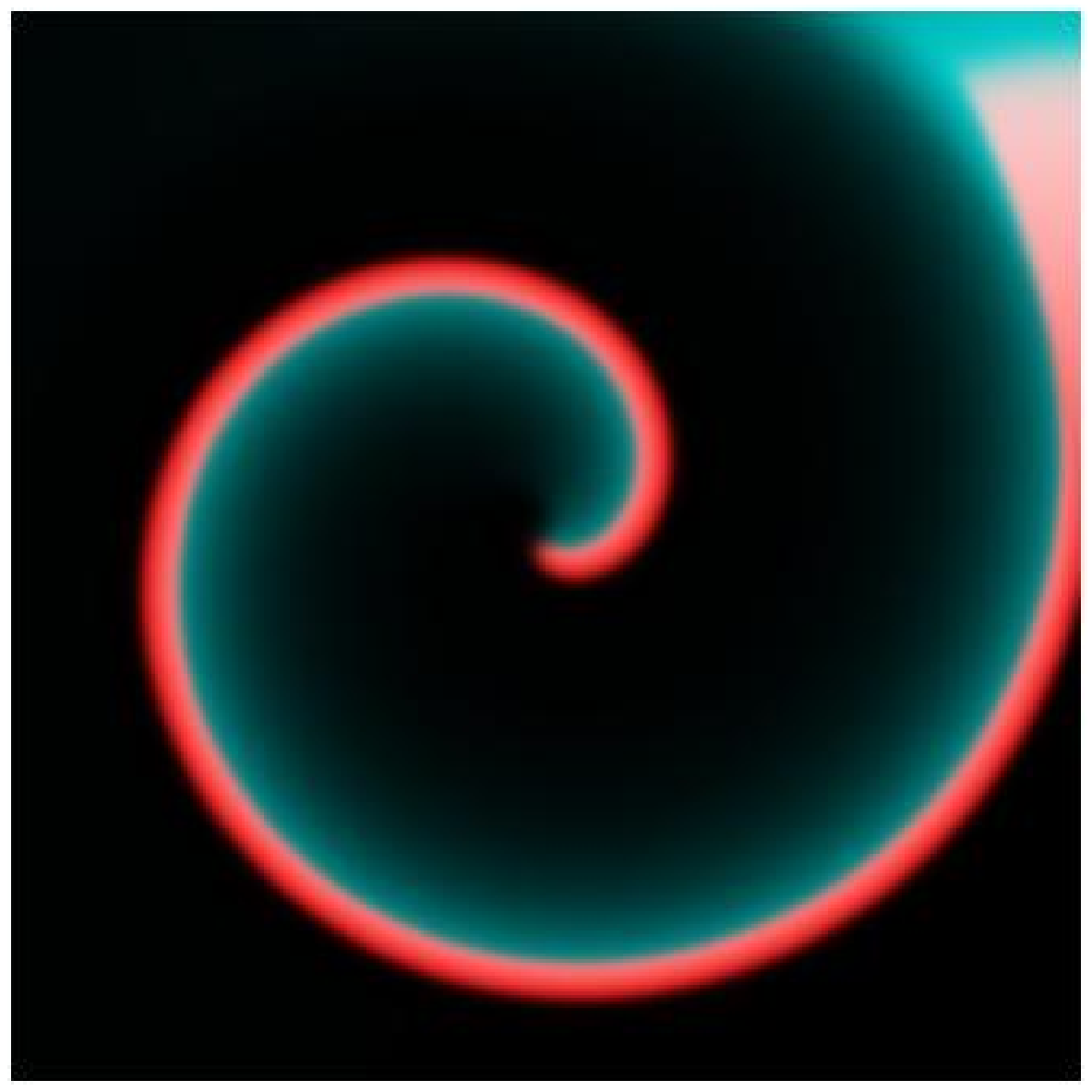}
\end{minipage}
\begin{minipage}{0.32\linewidth}
\centering
\includegraphics[width=0.7\textwidth]{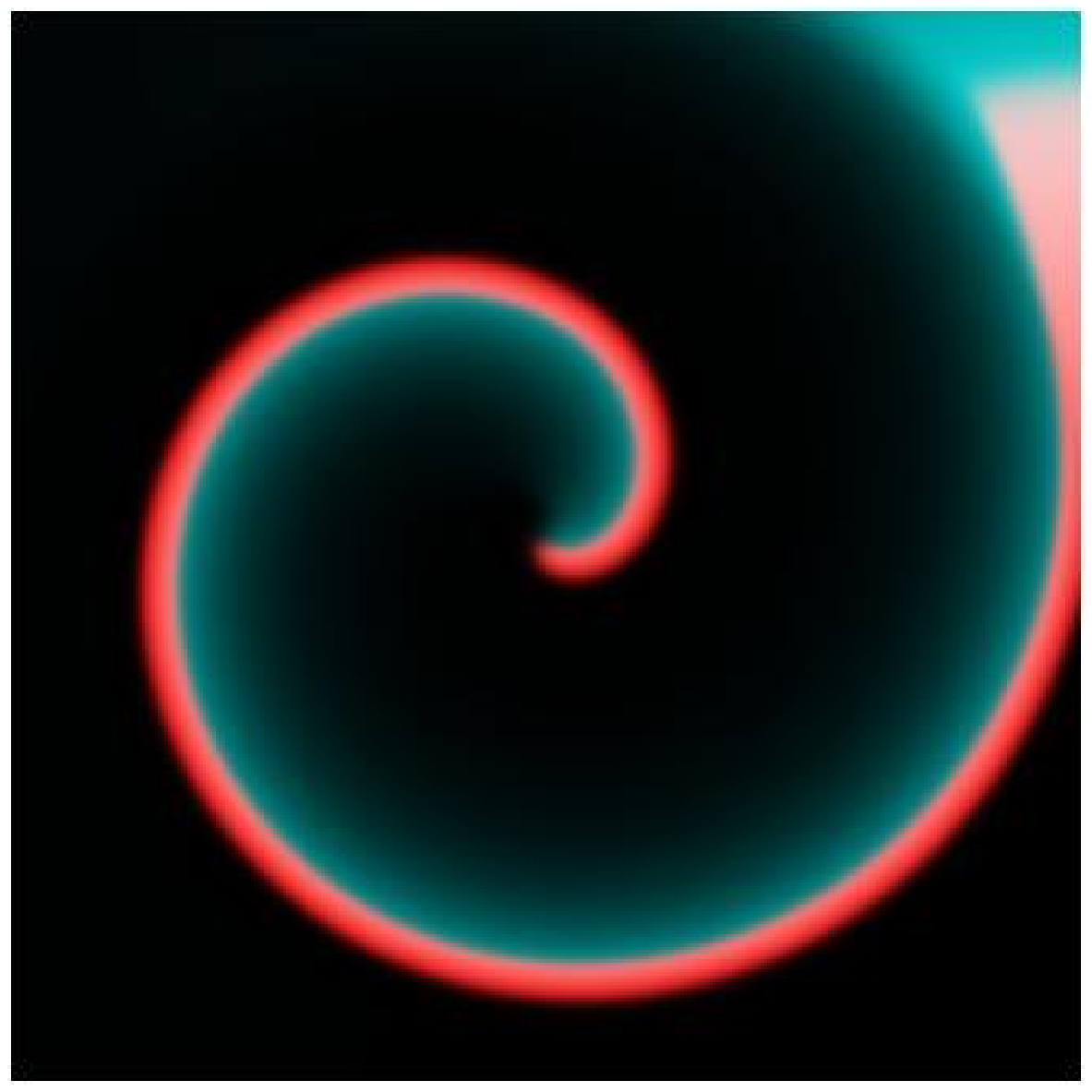}
\end{minipage}
\begin{minipage}{0.32\linewidth}
\centering
\includegraphics[width=0.7\textwidth]{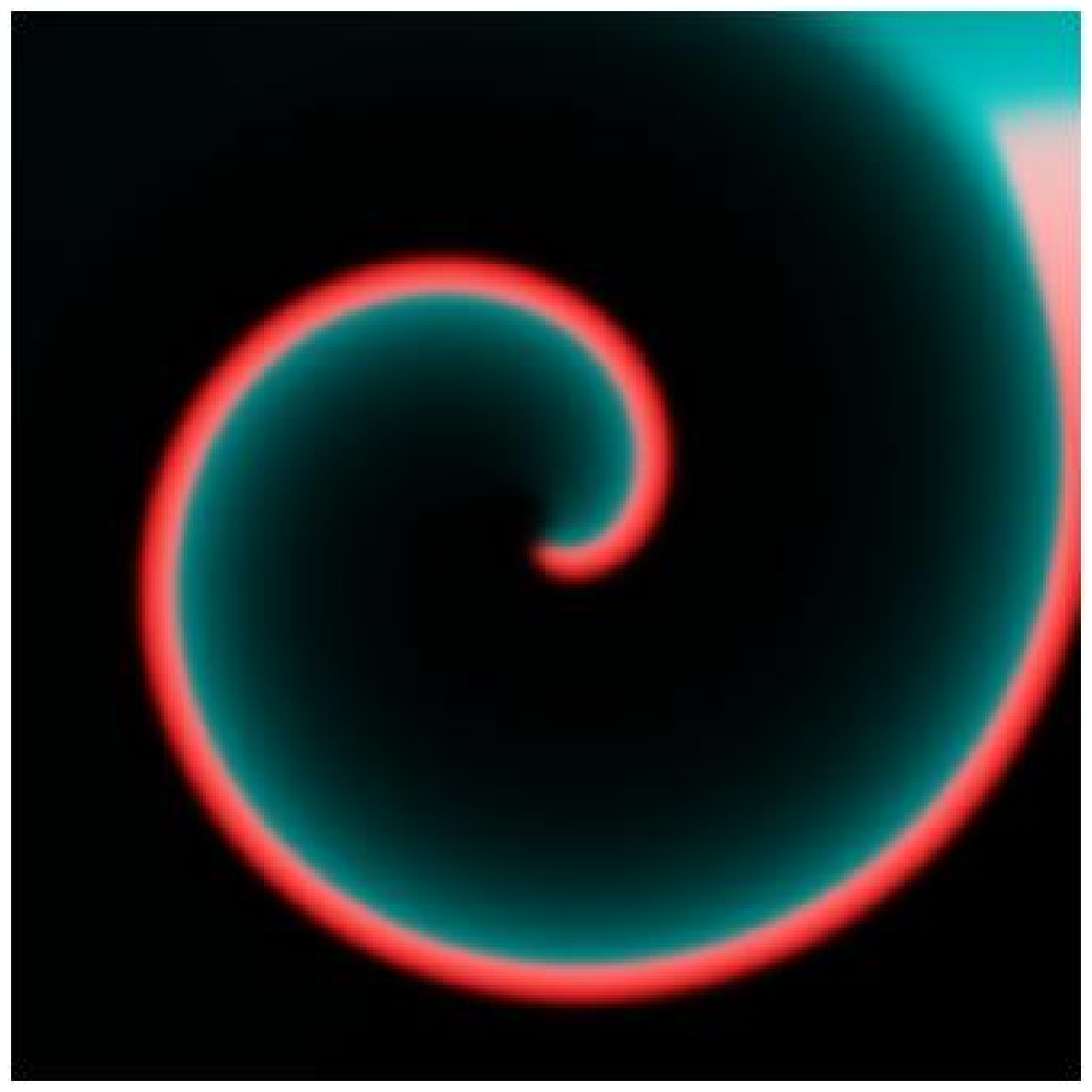}
\end{minipage}
\begin{minipage}{0.32\linewidth}
\centering
\includegraphics[width=0.7\textwidth]{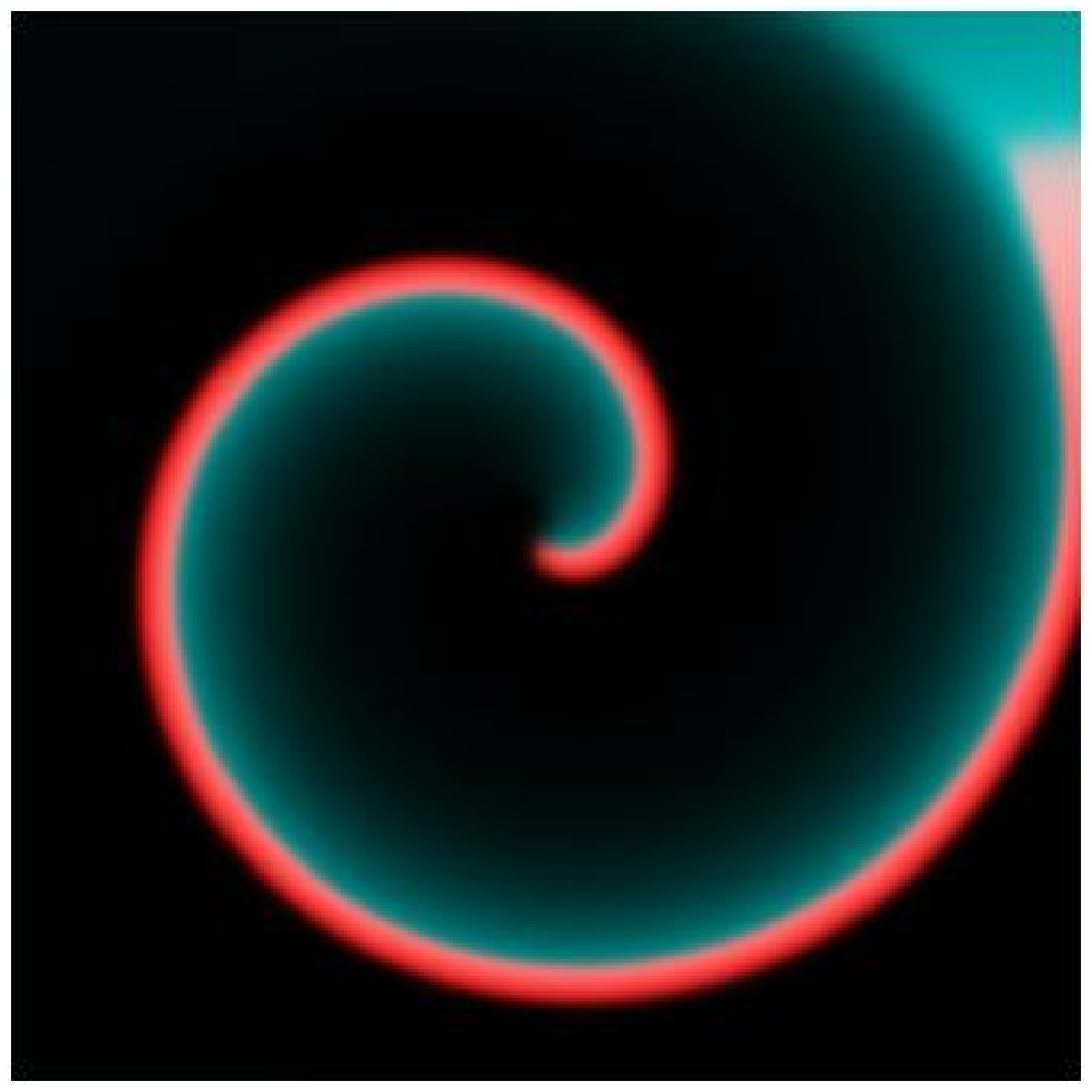}
\end{minipage}
\begin{minipage}{0.32\linewidth}
\centering
\includegraphics[width=0.7\textwidth]{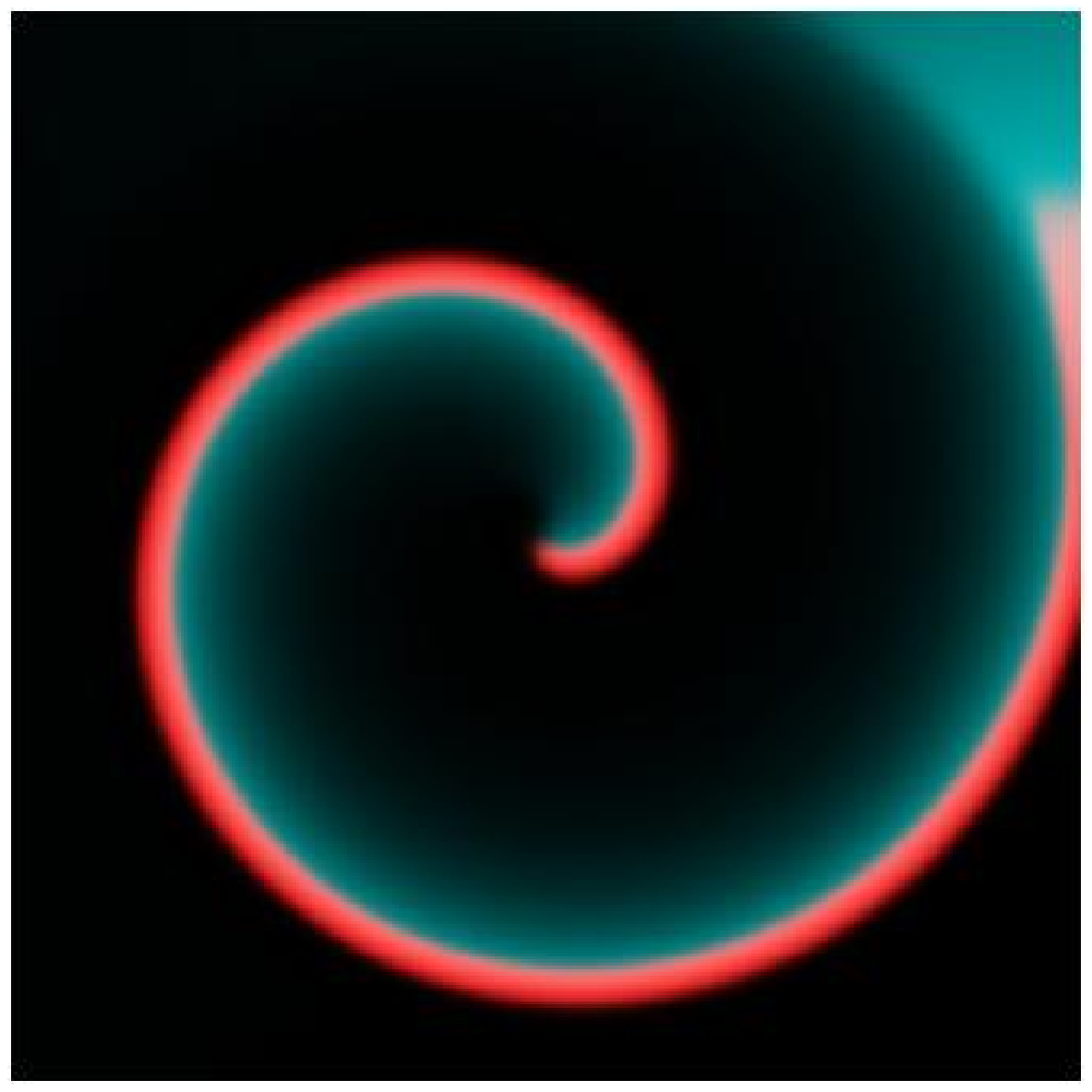}
\end{minipage}
\begin{minipage}{0.32\linewidth}
\centering
\includegraphics[width=0.7\textwidth]{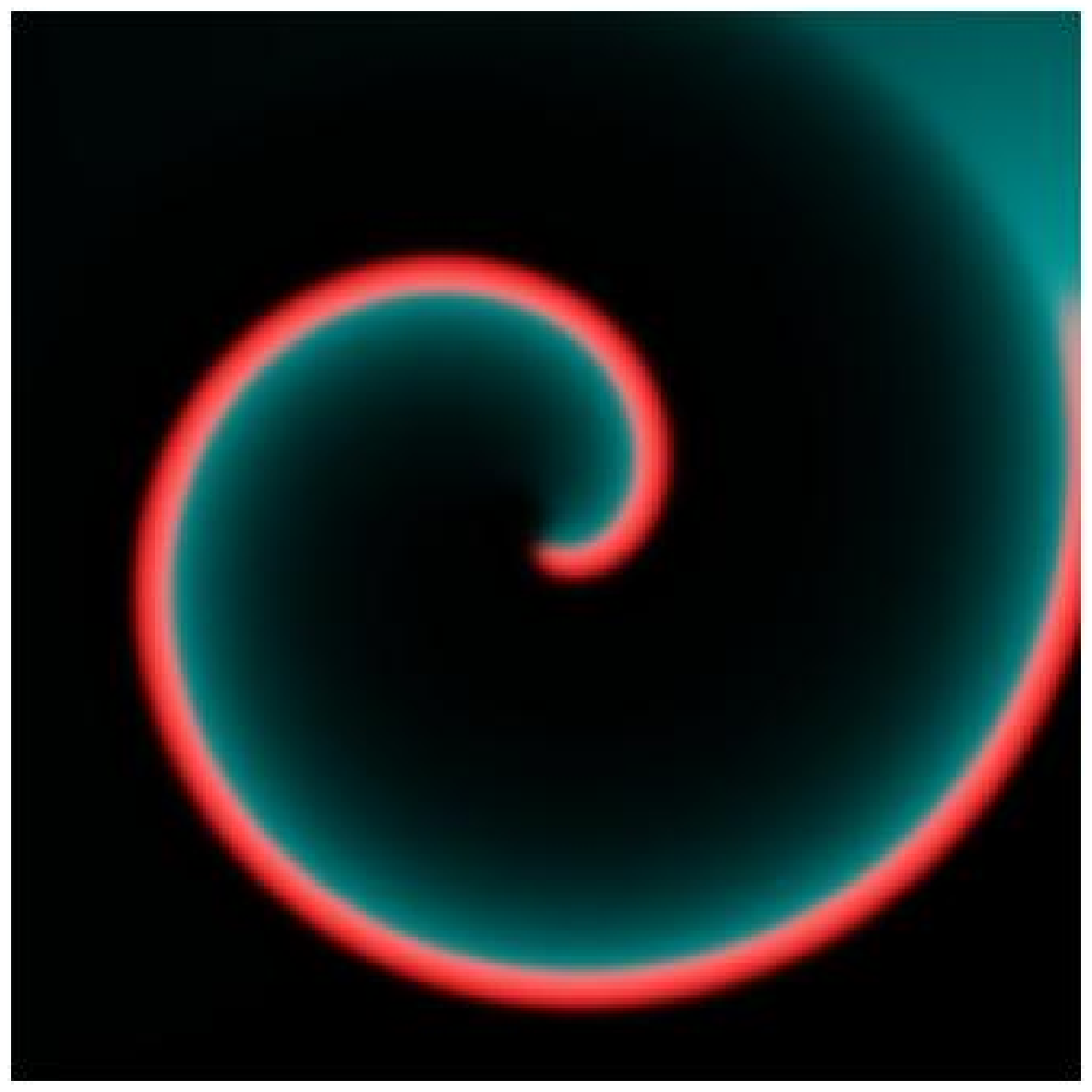}
\end{minipage}
\begin{minipage}{0.32\linewidth}
\centering
\includegraphics[width=0.7\textwidth]{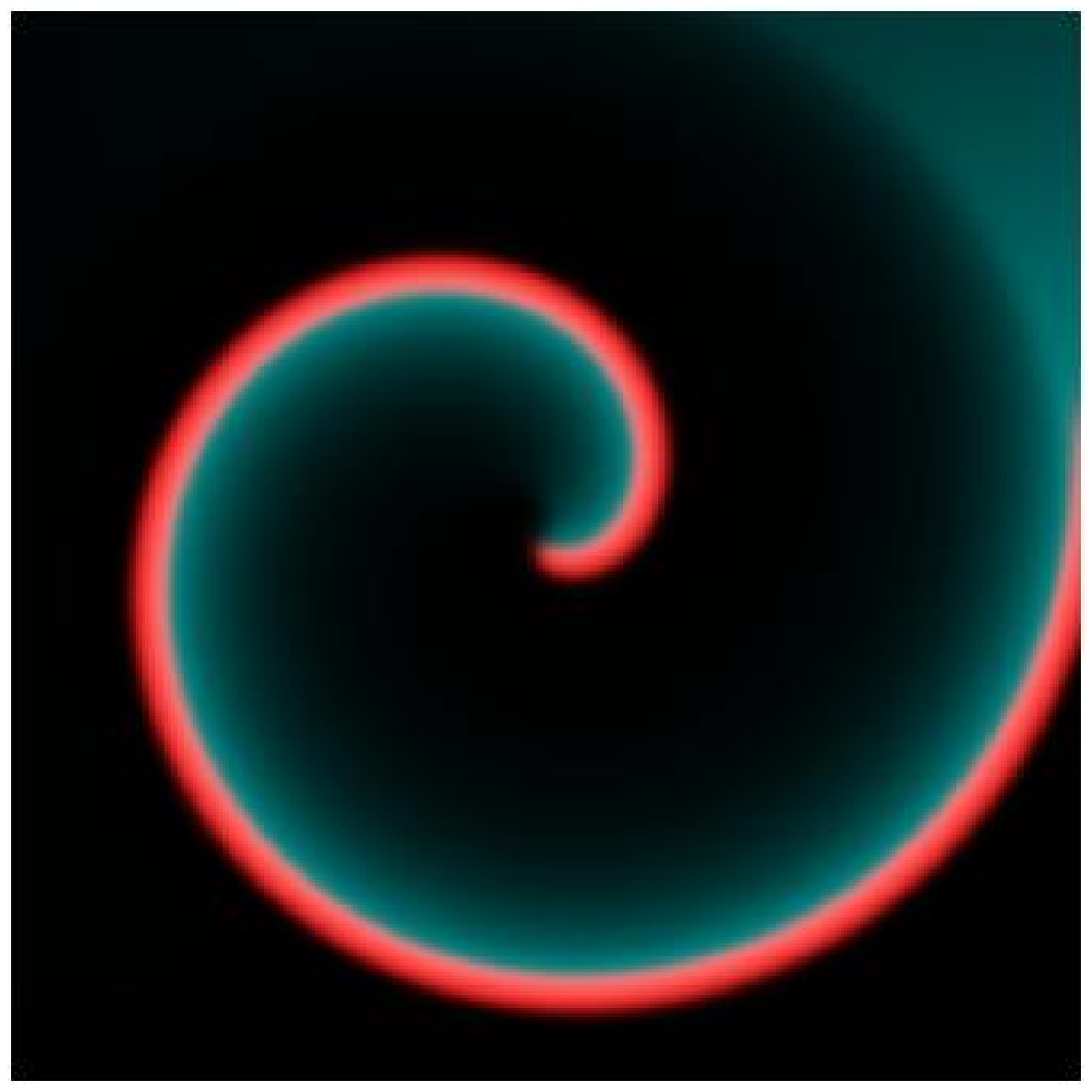}
\end{minipage}
\begin{minipage}{0.32\linewidth}
\centering
\includegraphics[width=0.7\textwidth]{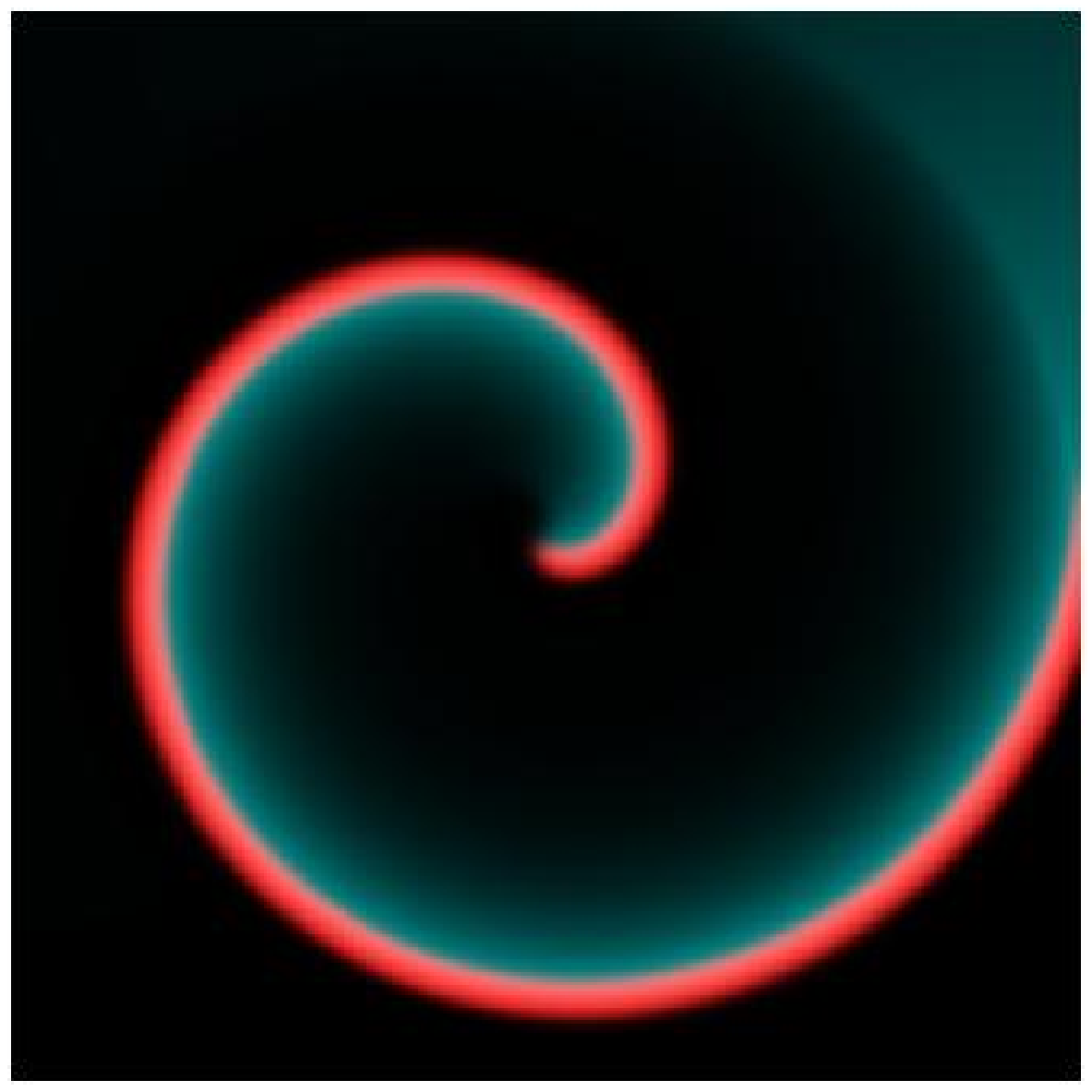}
\end{minipage}
\begin{minipage}{0.32\linewidth}
\centering
\includegraphics[width=0.7\textwidth]{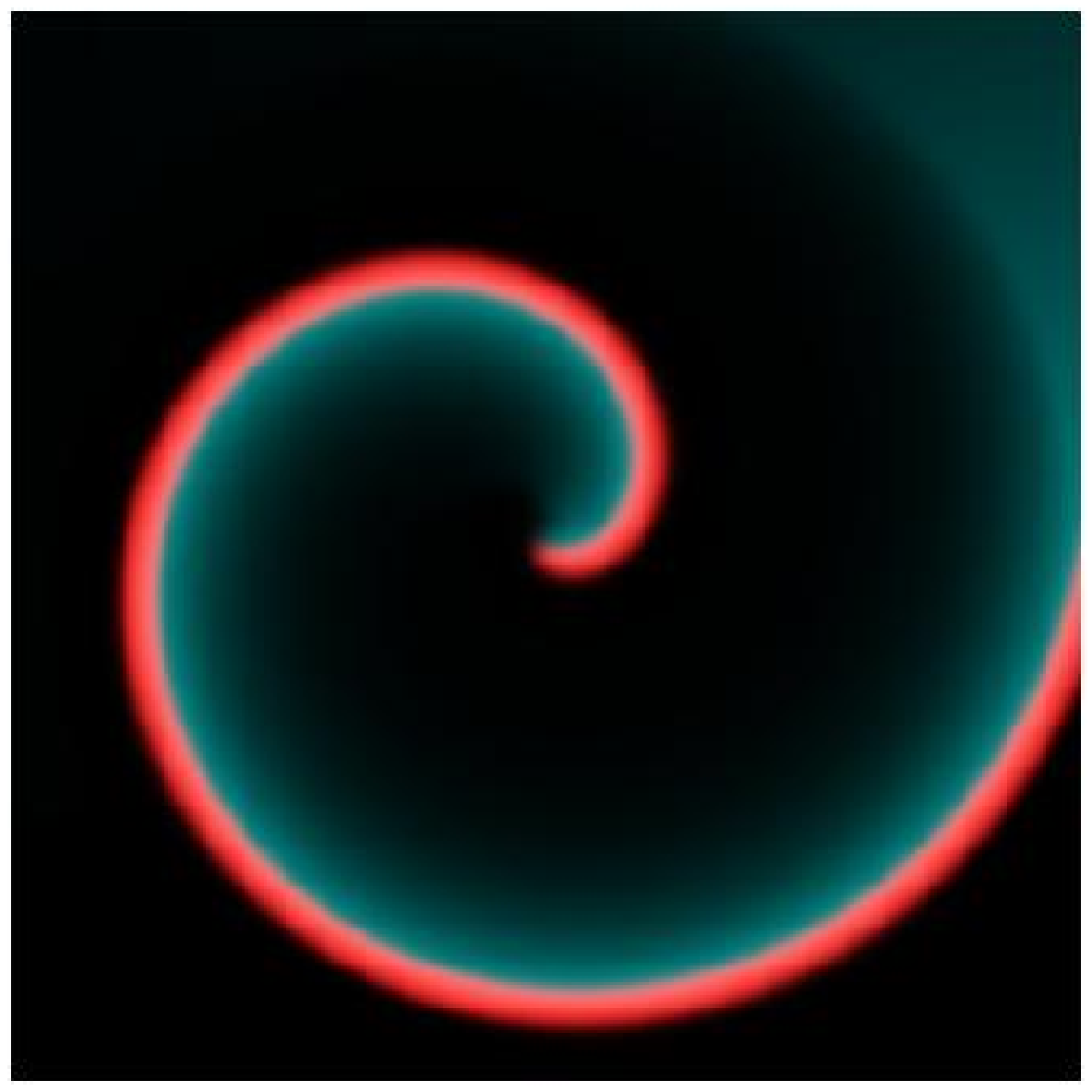}
\end{minipage}
\begin{minipage}{0.32\linewidth}
\centering
\includegraphics[width=0.7\textwidth]{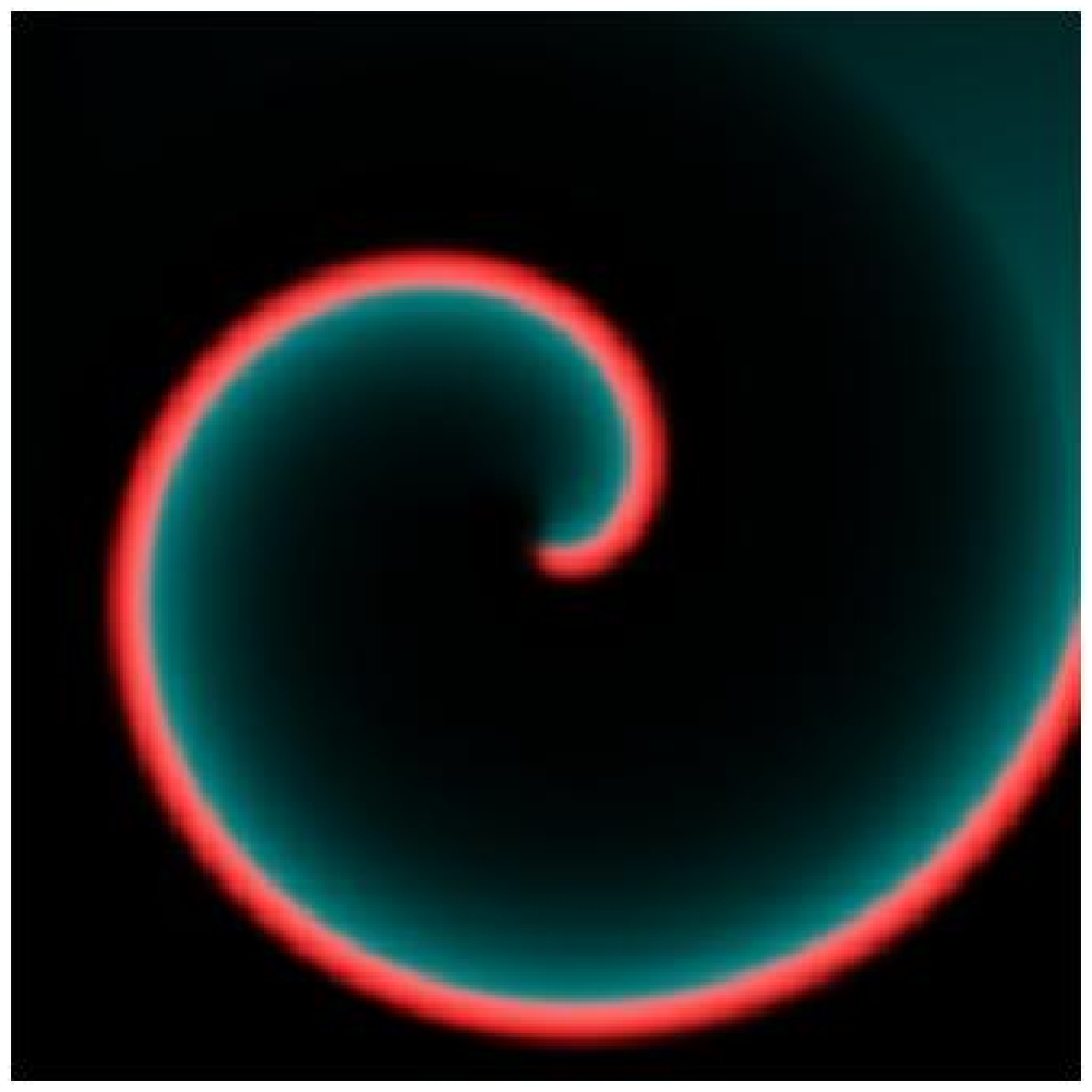}
\end{minipage}
\begin{minipage}{0.32\linewidth}
\centering
\includegraphics[width=0.7\textwidth]{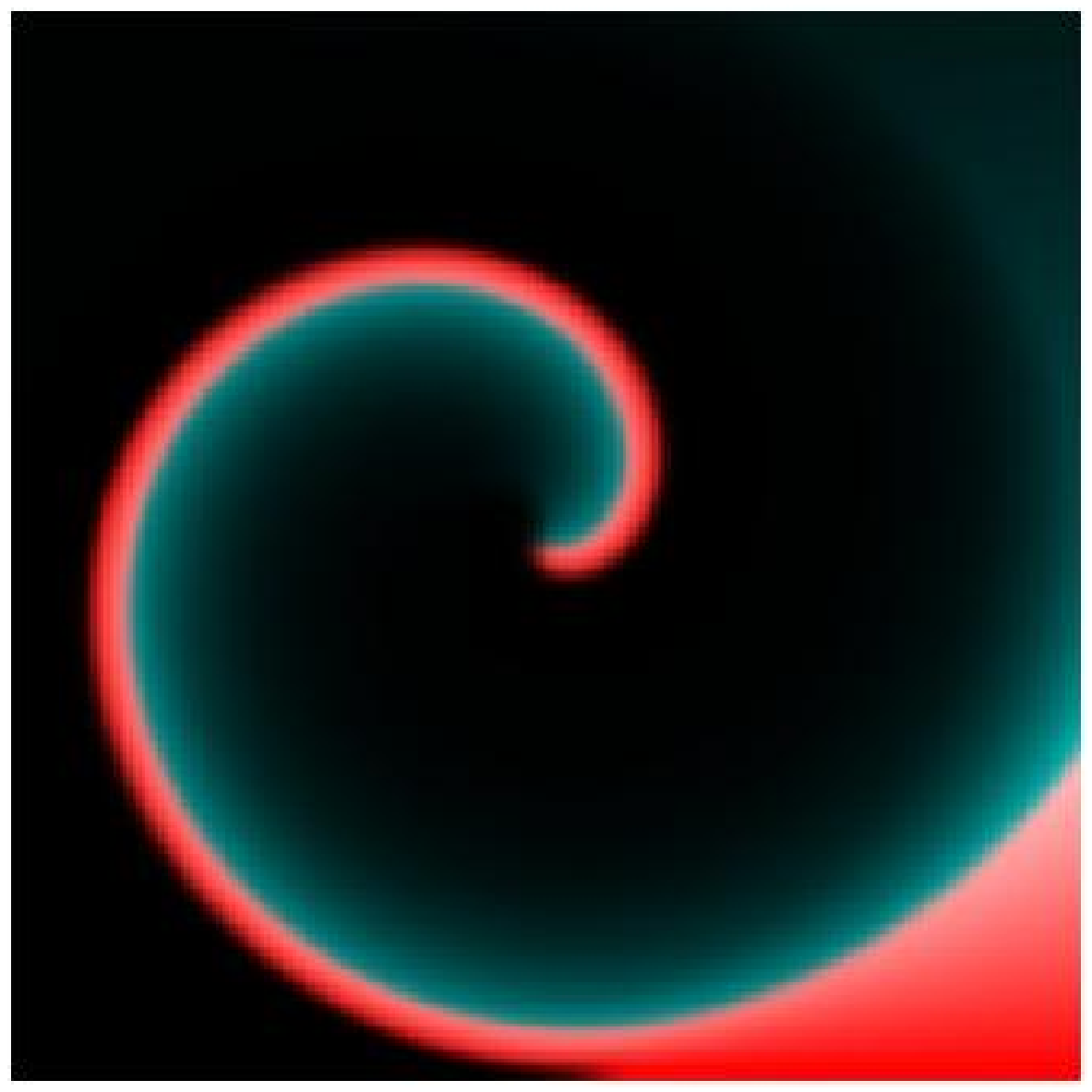}
\end{minipage}
\caption[Spacestep convergence: Neumann boundary conditions, final solutions]{Final Conditions for each run in the convergence testing of the spacestep in Barkley's model using Neumann boundary conditions, starting top left and working right, $\Delta_x=\frac{1}{15}$ (top left) to $\Delta_x=\frac{1}{4}$ (bottom right)}
\label{fig:ezf_conv_1_final_nbc}
\end{center}
\end{figure}

\begin{figure}[tbh]
\begin{center}
\begin{minipage}{0.6\linewidth}
\centering
\includegraphics[width=0.7\textwidth, angle=-90]{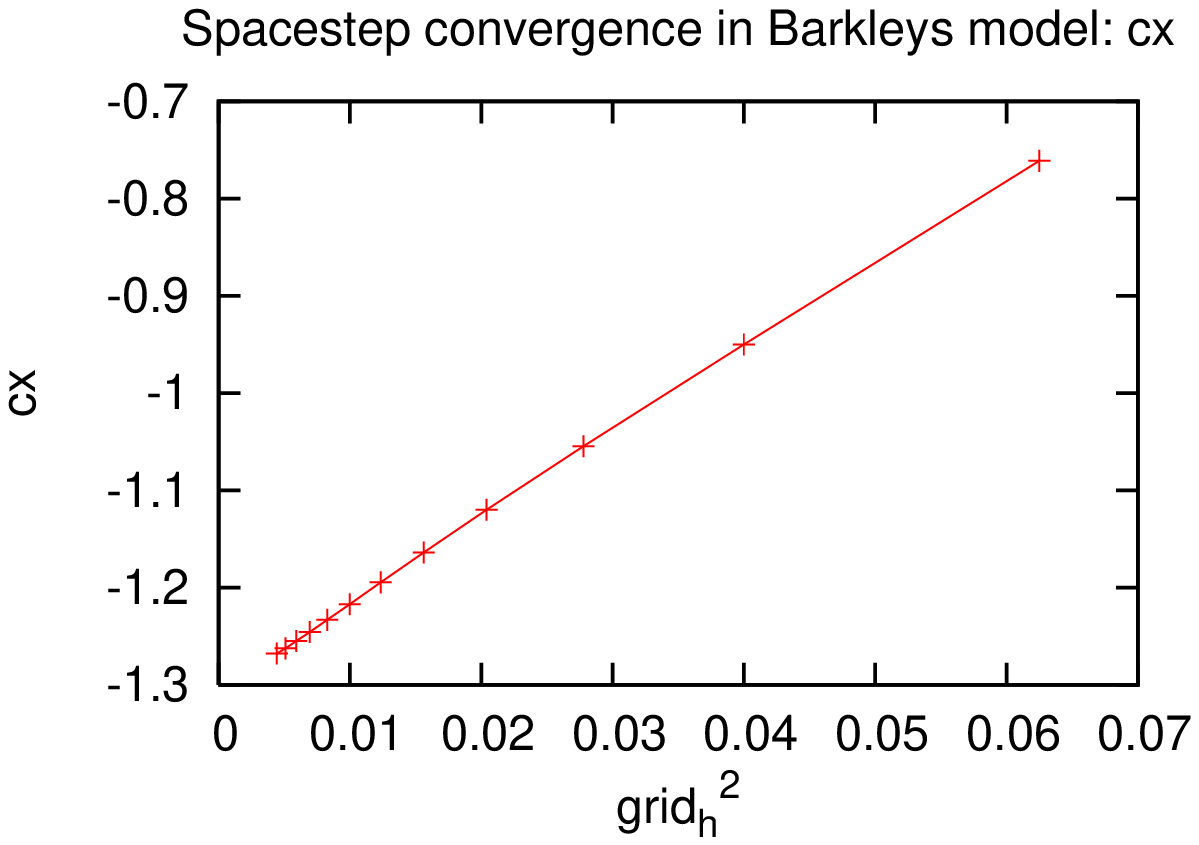}
\end{minipage}
\begin{minipage}{0.6\linewidth}
\centering
\includegraphics[width=0.7\textwidth, angle=-90]{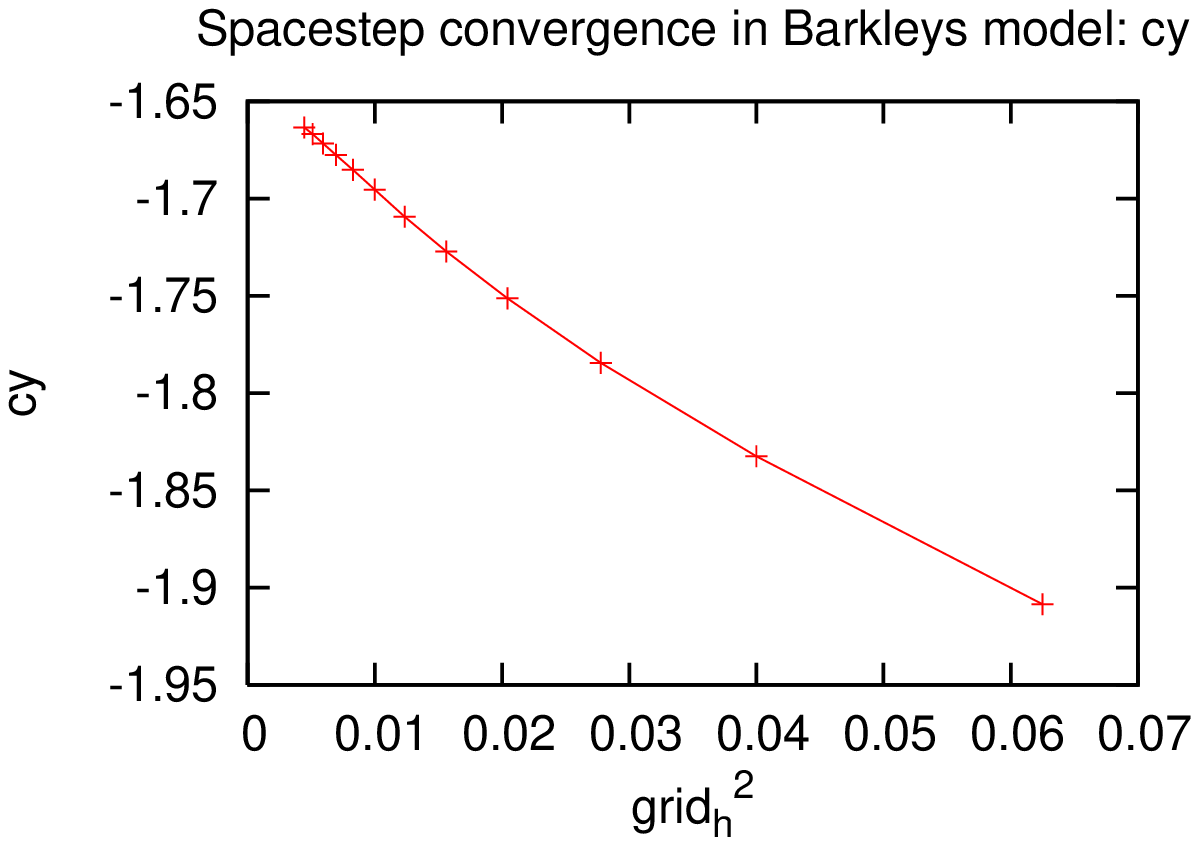}
\end{minipage}
\begin{minipage}{0.6\linewidth}
\centering
\includegraphics[width=0.7\textwidth, angle=-90]{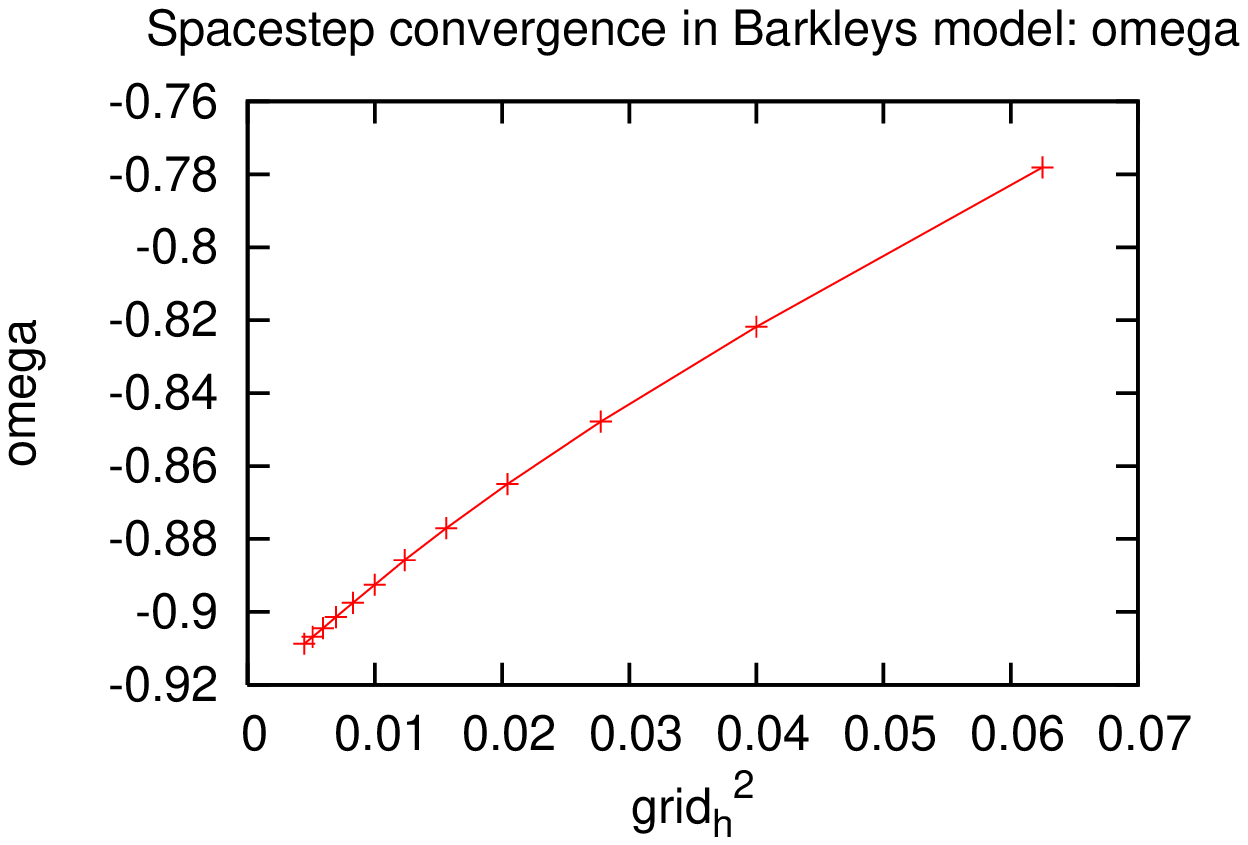}
\end{minipage}
\caption[Spacestep convergence: Dirichlet boundary conditions]{Convergence in spacestep, using Barkley's model and Dirichlet Boundary conditions with the box size fixed at $L_X=60$,and the timestep fixed at $\Delta_t=1.11\times10^{-4}$.}
\label{fig:ezf_conv_1_dbc}
\end{center}
\end{figure}

\begin{figure}[tbh]
\begin{center}
\begin{minipage}{0.32\linewidth}
\centering
\includegraphics[width=0.7\textwidth]{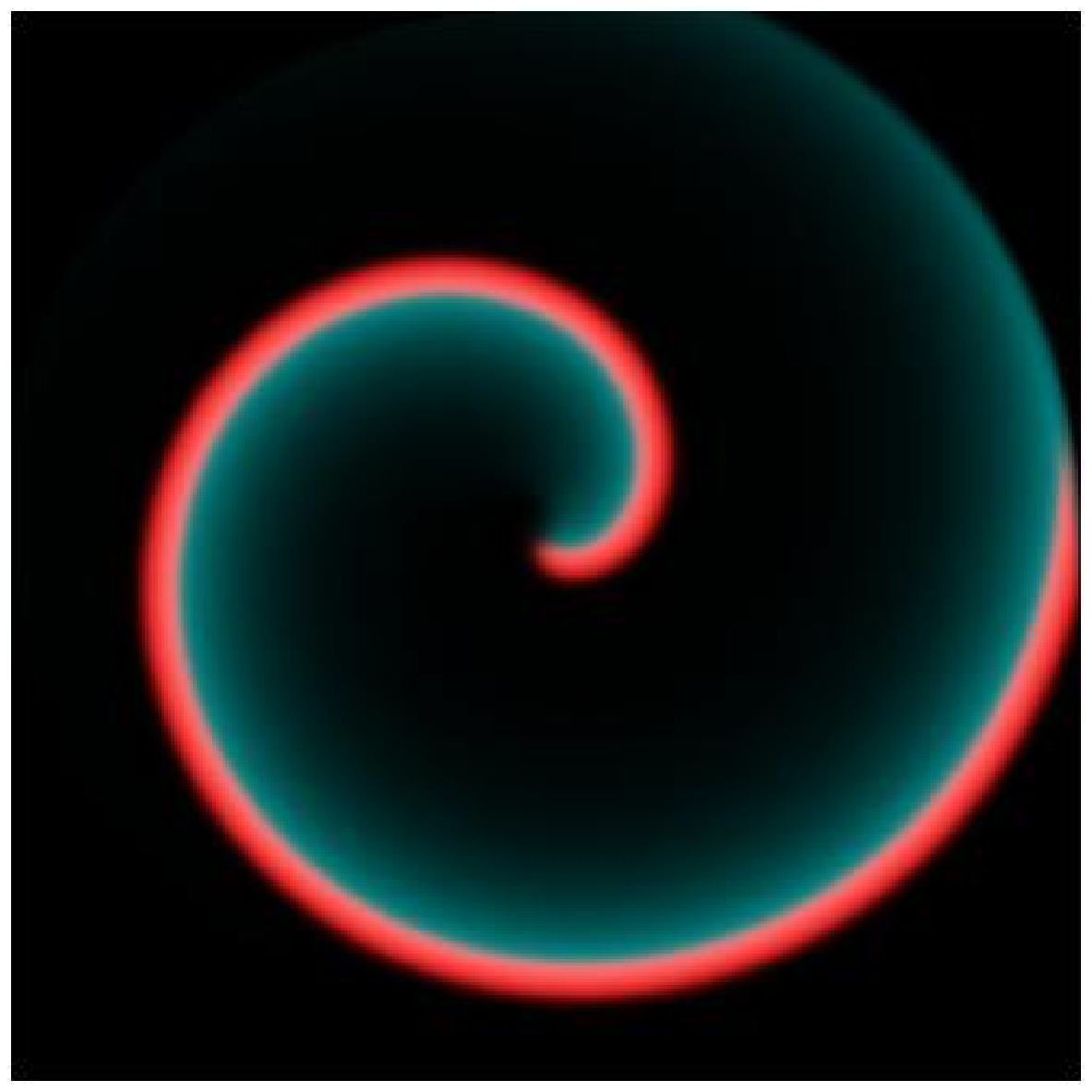}
\end{minipage}
\begin{minipage}{0.32\linewidth}
\centering
\includegraphics[width=0.7\textwidth]{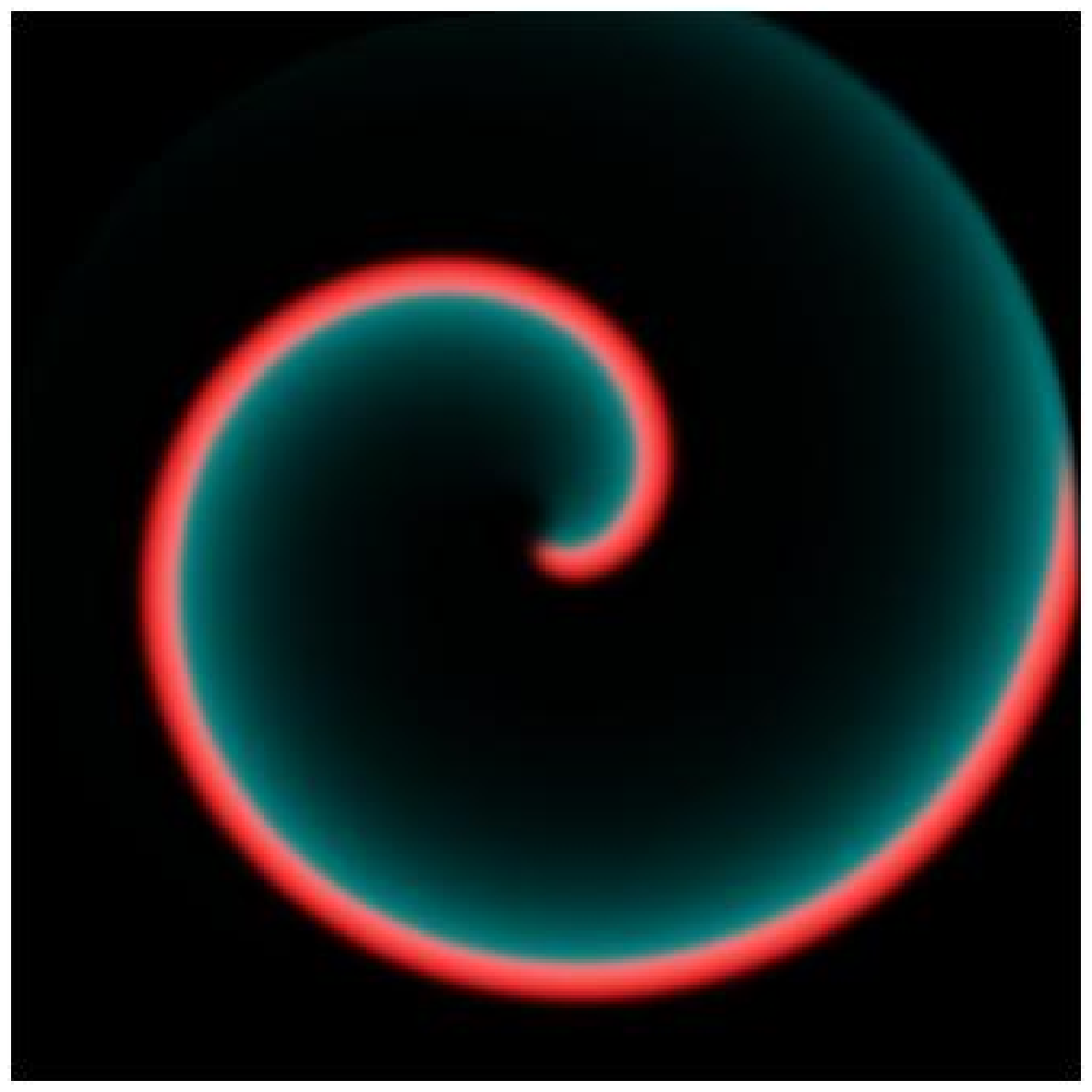}
\end{minipage}
\begin{minipage}{0.32\linewidth}
\centering
\includegraphics[width=0.7\textwidth]{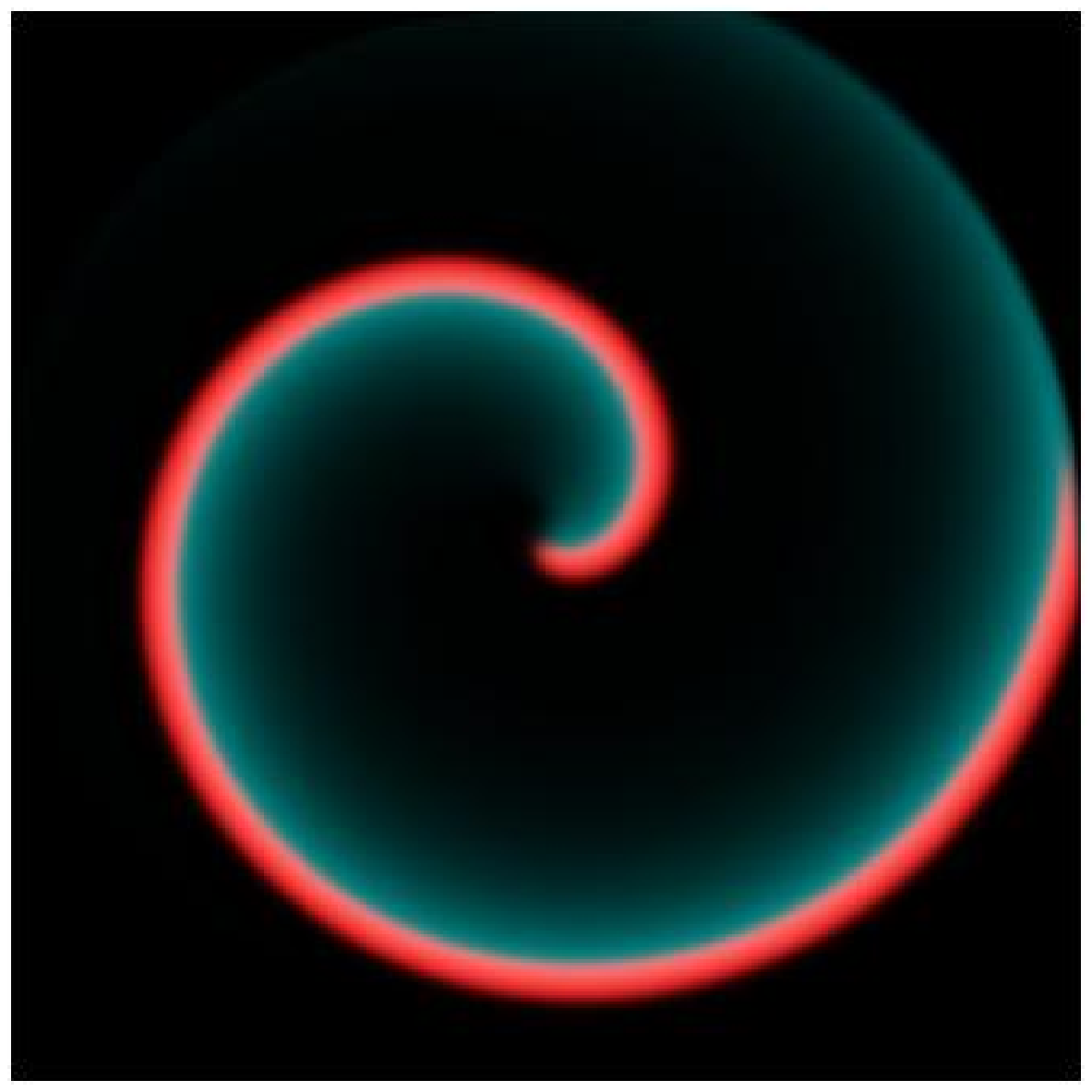}
\end{minipage}
\begin{minipage}{0.32\linewidth}
\centering
\includegraphics[width=0.7\textwidth]{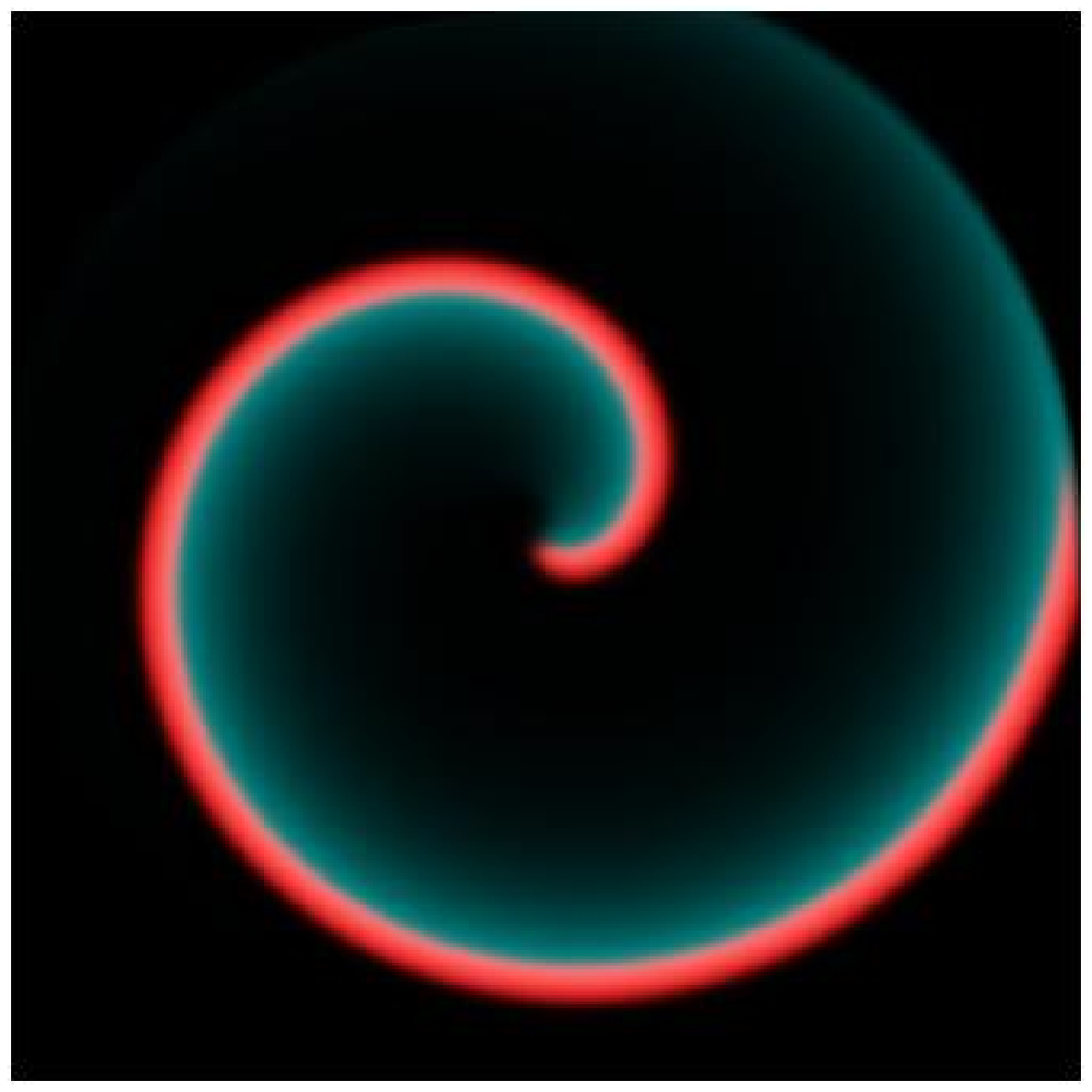}
\end{minipage}
\begin{minipage}{0.32\linewidth}
\centering
\includegraphics[width=0.7\textwidth]{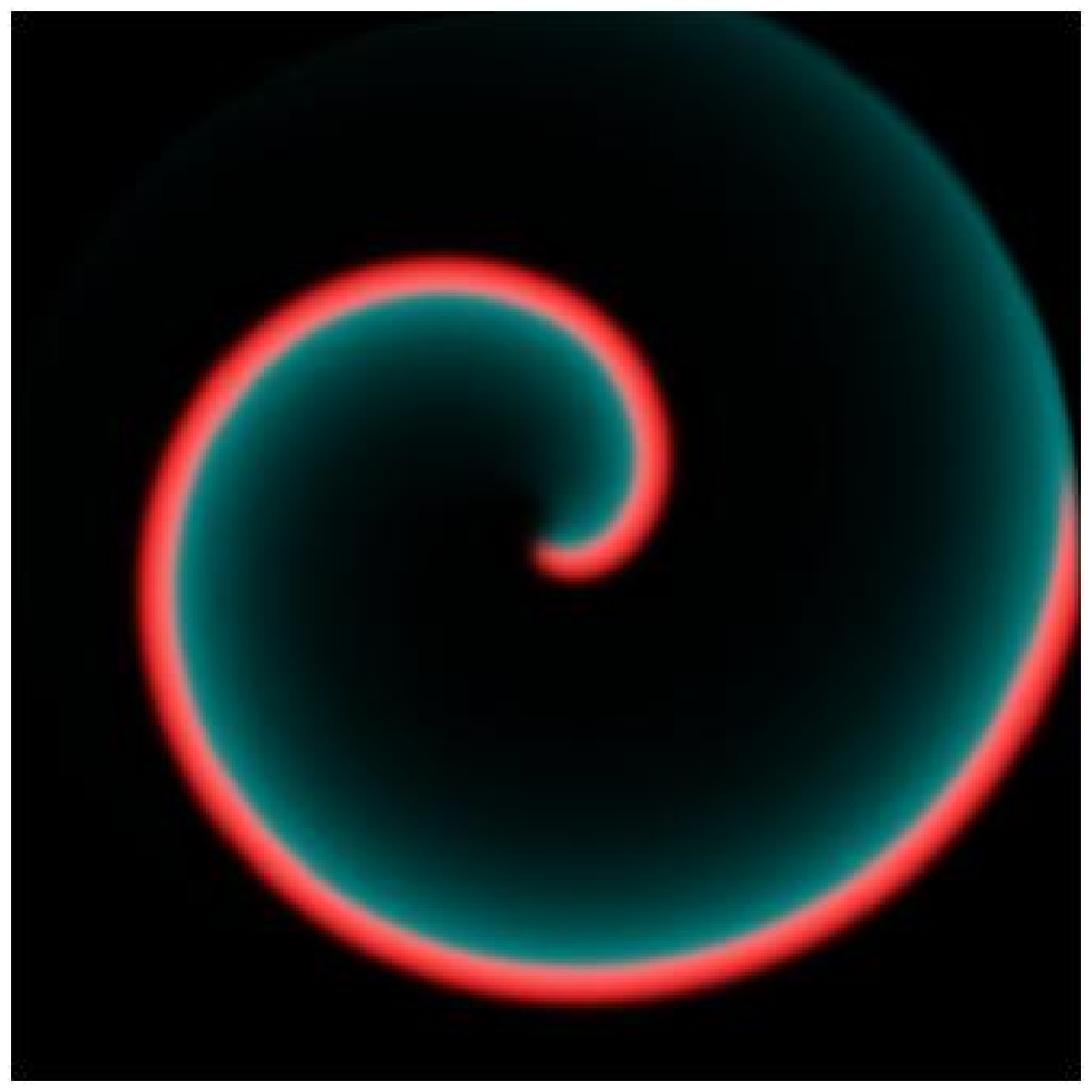}
\end{minipage}
\begin{minipage}{0.32\linewidth}
\centering
\includegraphics[width=0.7\textwidth]{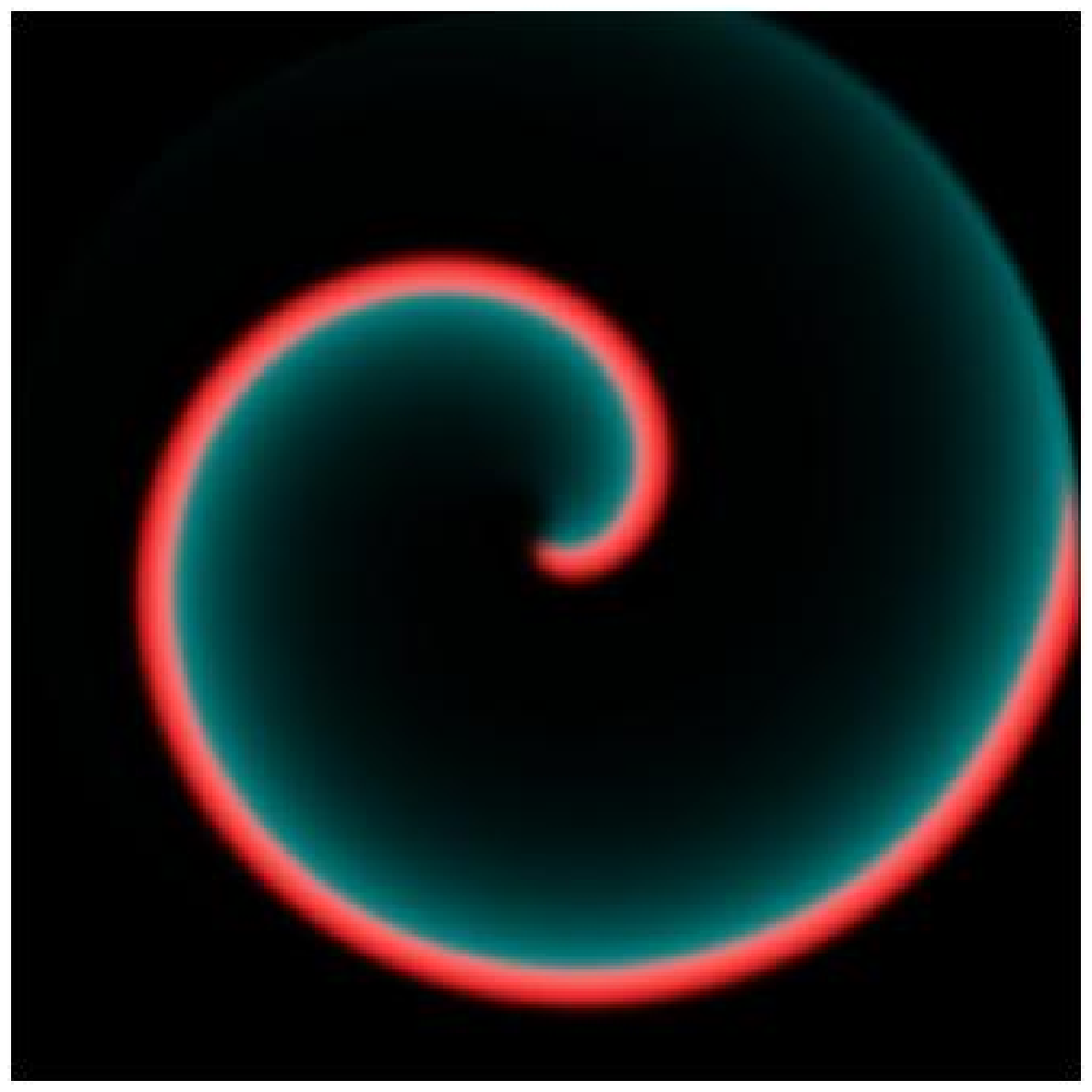}
\end{minipage}
\begin{minipage}{0.32\linewidth}
\centering
\includegraphics[width=0.7\textwidth]{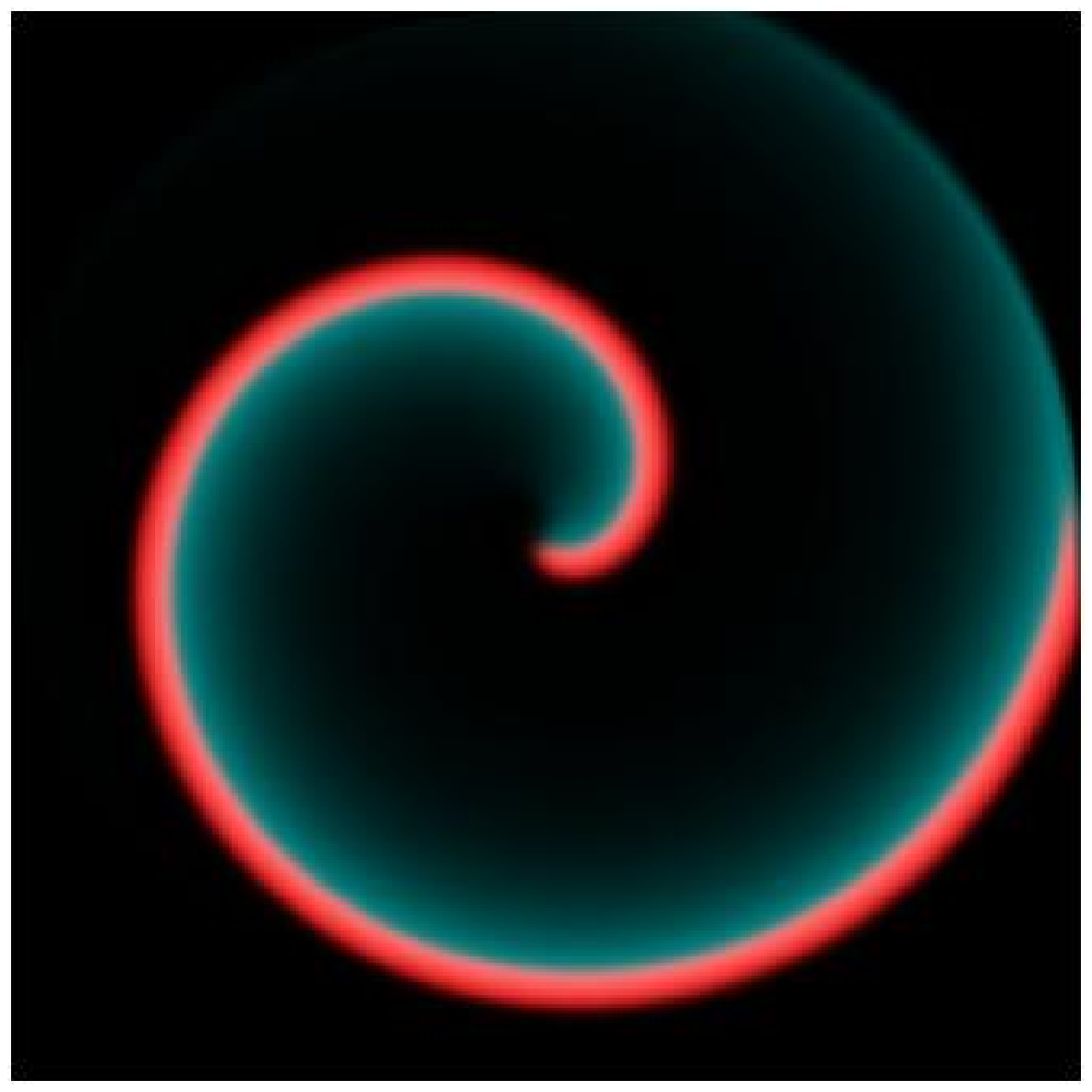}
\end{minipage}
\begin{minipage}{0.32\linewidth}
\centering
\includegraphics[width=0.7\textwidth]{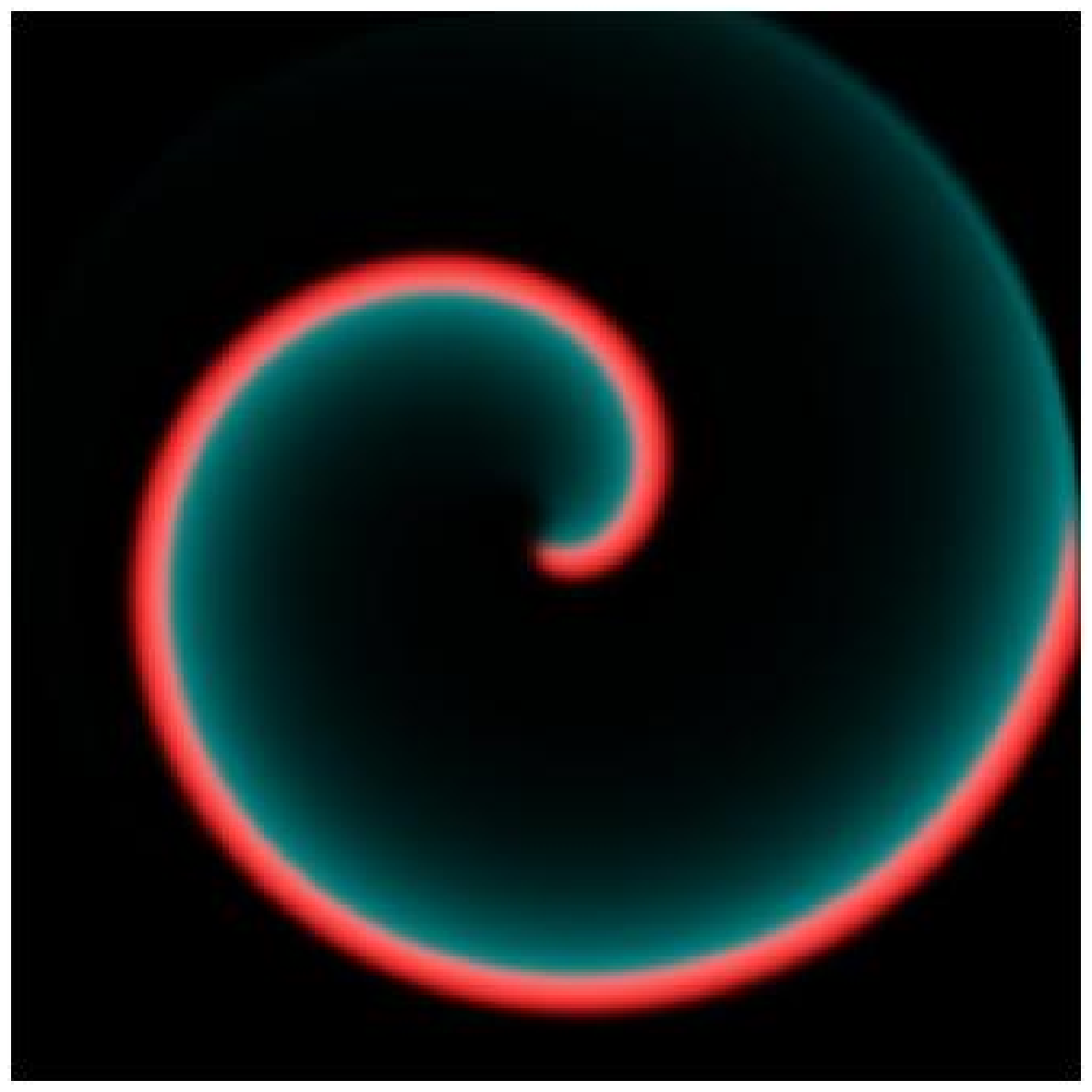}
\end{minipage}
\begin{minipage}{0.32\linewidth}
\centering
\includegraphics[width=0.7\textwidth]{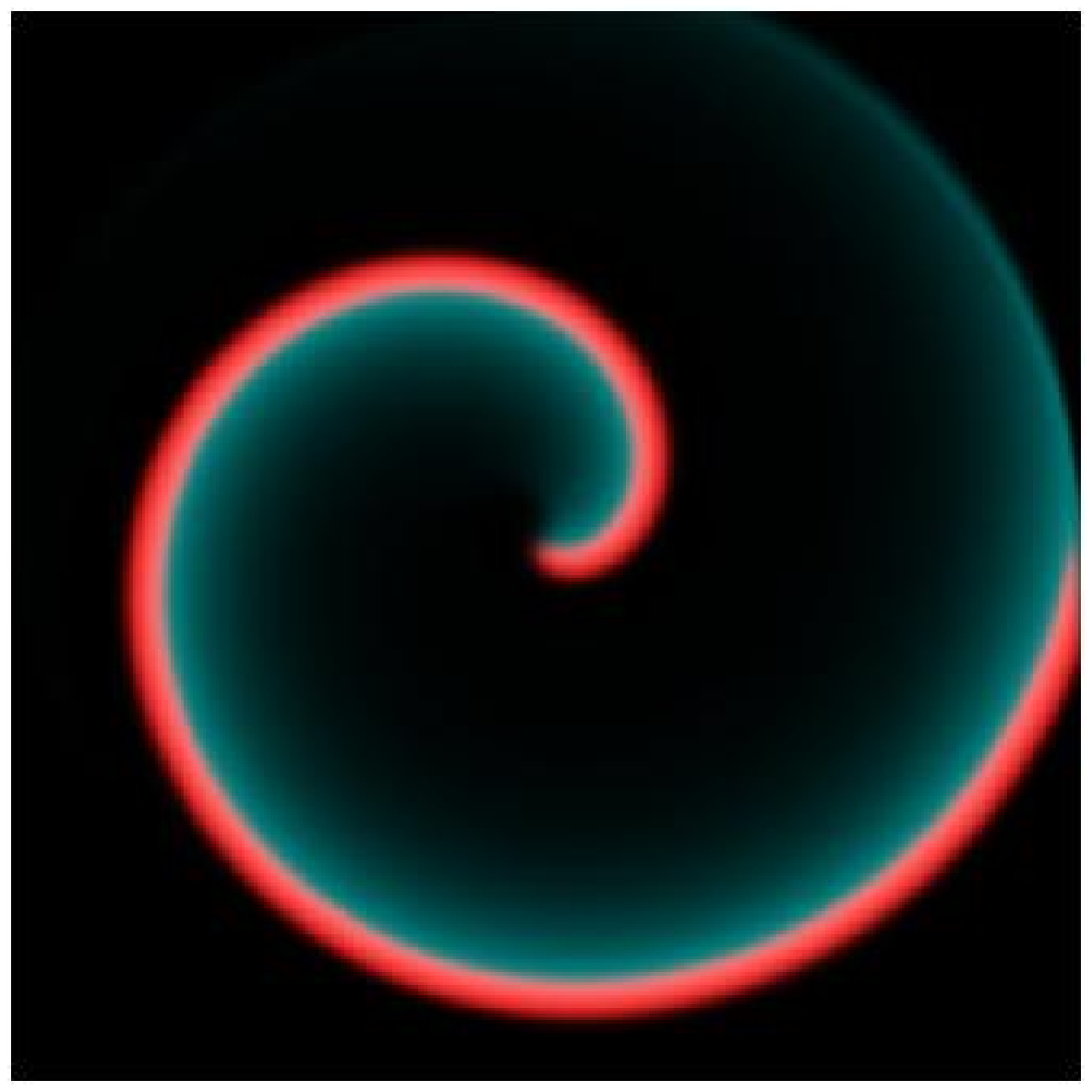}
\end{minipage}
\begin{minipage}{0.32\linewidth}
\centering
\includegraphics[width=0.7\textwidth]{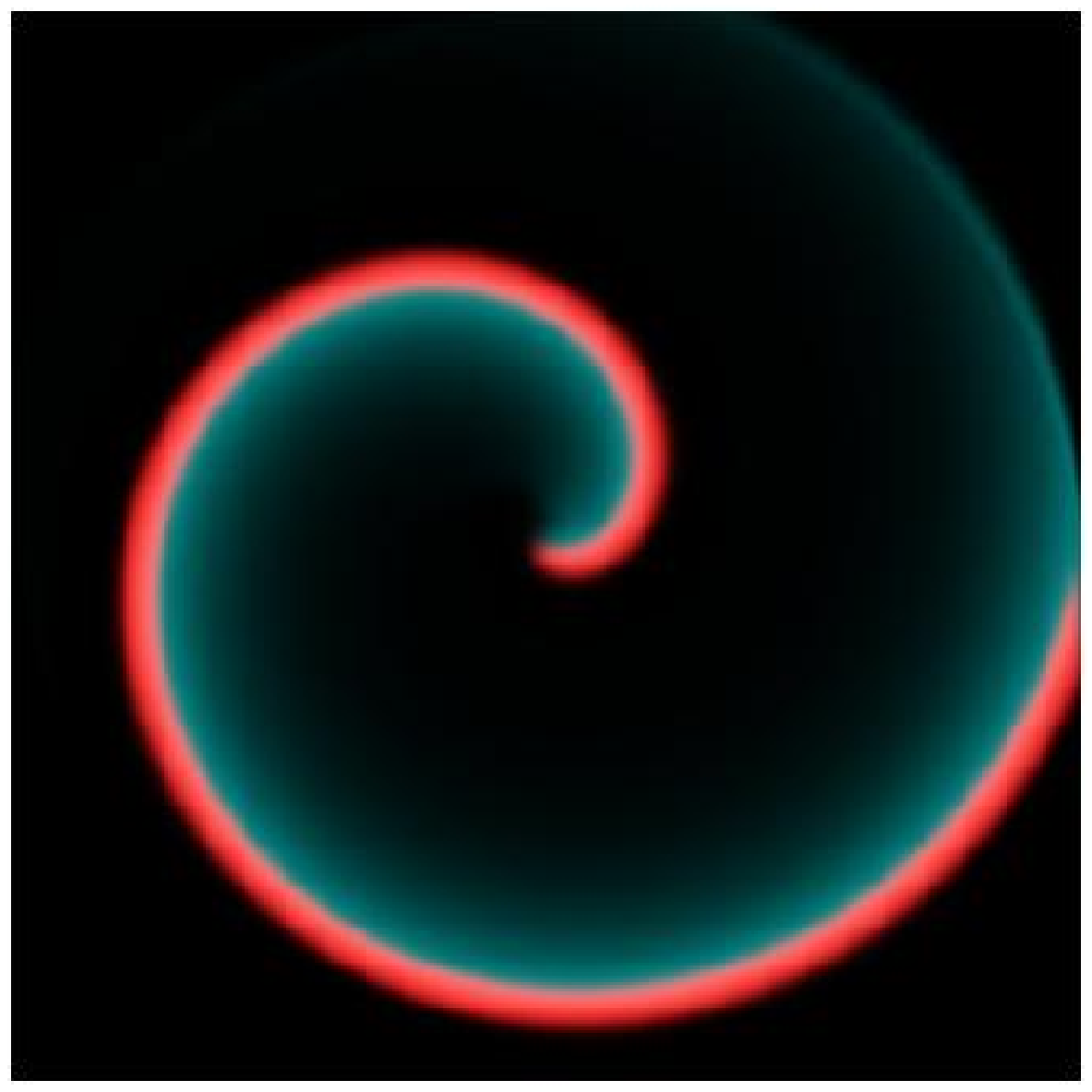}
\end{minipage}
\begin{minipage}{0.32\linewidth}
\centering
\includegraphics[width=0.7\textwidth]{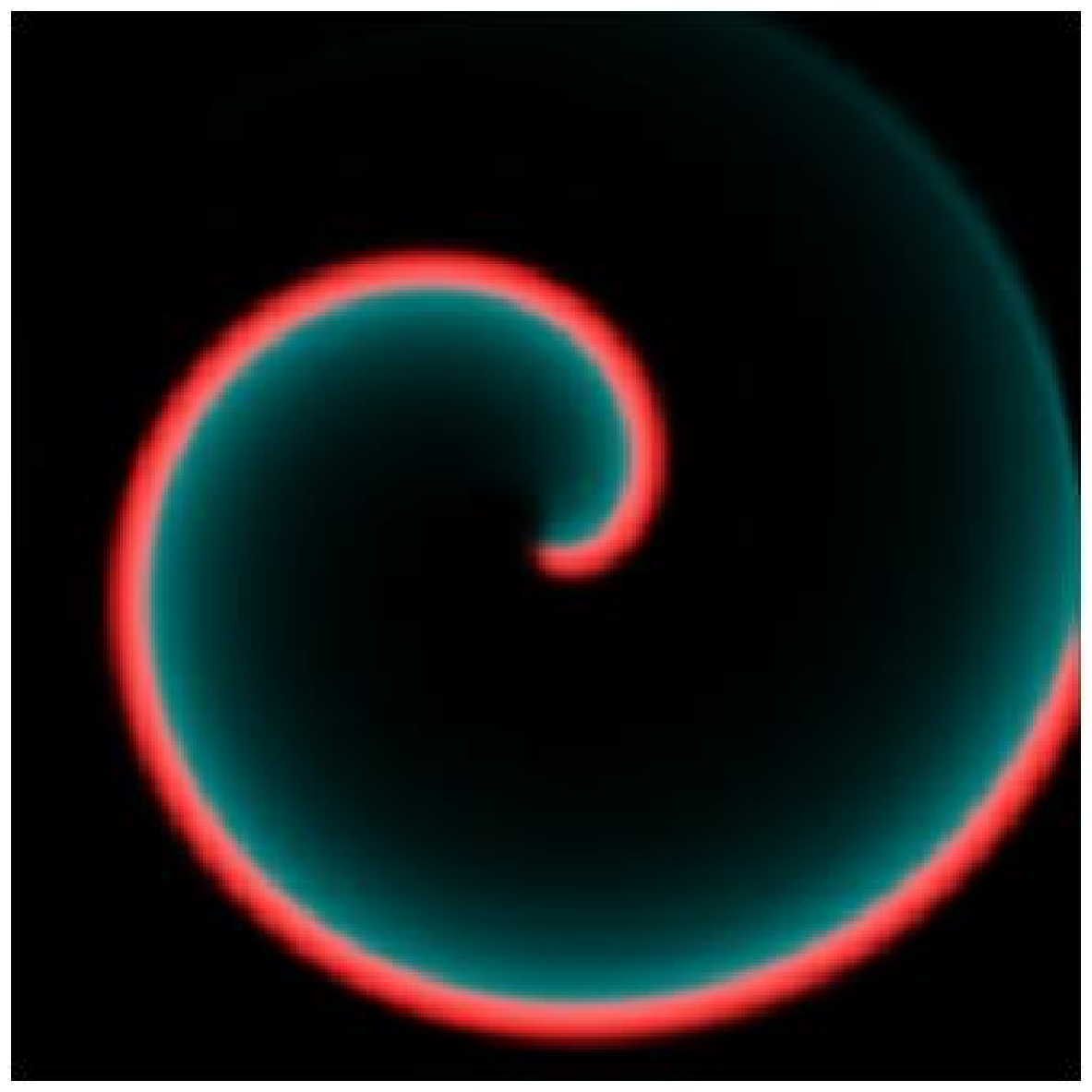}
\end{minipage}
\begin{minipage}{0.32\linewidth}
\centering
\includegraphics[width=0.7\textwidth]{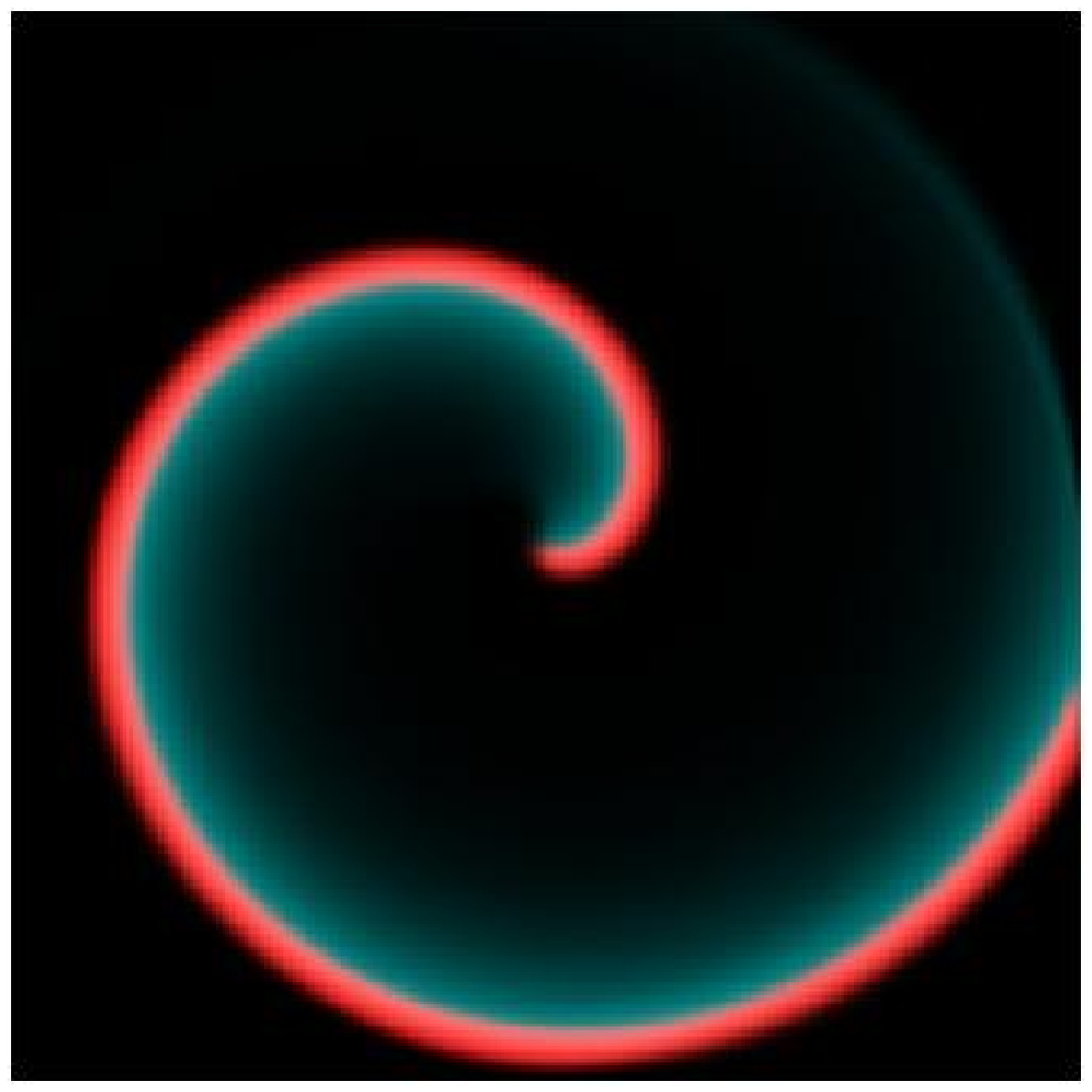}
\end{minipage}
\caption[Spacestep convergence: Dirichlet boundary conditions, final solutions]{Final Conditions for each run in the convergence testing of the spacestep in Barkley's model using Dirichlet boundary conditions, starting top left and working right, $\Delta_x=\frac{1}{15}$ (top left) to $\Delta_x=\frac{1}{4}$ (bottom right)}
\label{fig:ezf_conv_1_final_dbc}
\end{center}
\end{figure}

We also show the final solutions in Figs.(\ref{fig:ezf_conv_1_final_nbc}) and (\ref{fig:ezf_conv_1_final_dbc}).

\clearpage

\clearpage

\clearpage

\clearpage


\subsubsection{Convergence in the box size}

We will now show how $c_x$, $c_y$ and $\omega$ depend on the size of the box. In this instance, we should observe that after a particular size of box, the advection coefficients settle down to a fixed value.

Also, we kept the following numerical parameters constant throughout the simulations:

\begin{itemize}
 \item spacestep, $\Delta_x=\frac{1}{15}$
 \item timestep, $\Delta_t=1.1\times10^{-3}$ (with $t_s=0.1$)
\end{itemize}

Starting from a box size of 60 s.u., we ran a number of simulations with each consecutive simulation have a box size 5 s.u. less than the previous.

\begin{figure}[tbh]
\begin{center}
\begin{minipage}{0.6\linewidth}
\centering
\includegraphics[width=0.7\textwidth, angle=-90]{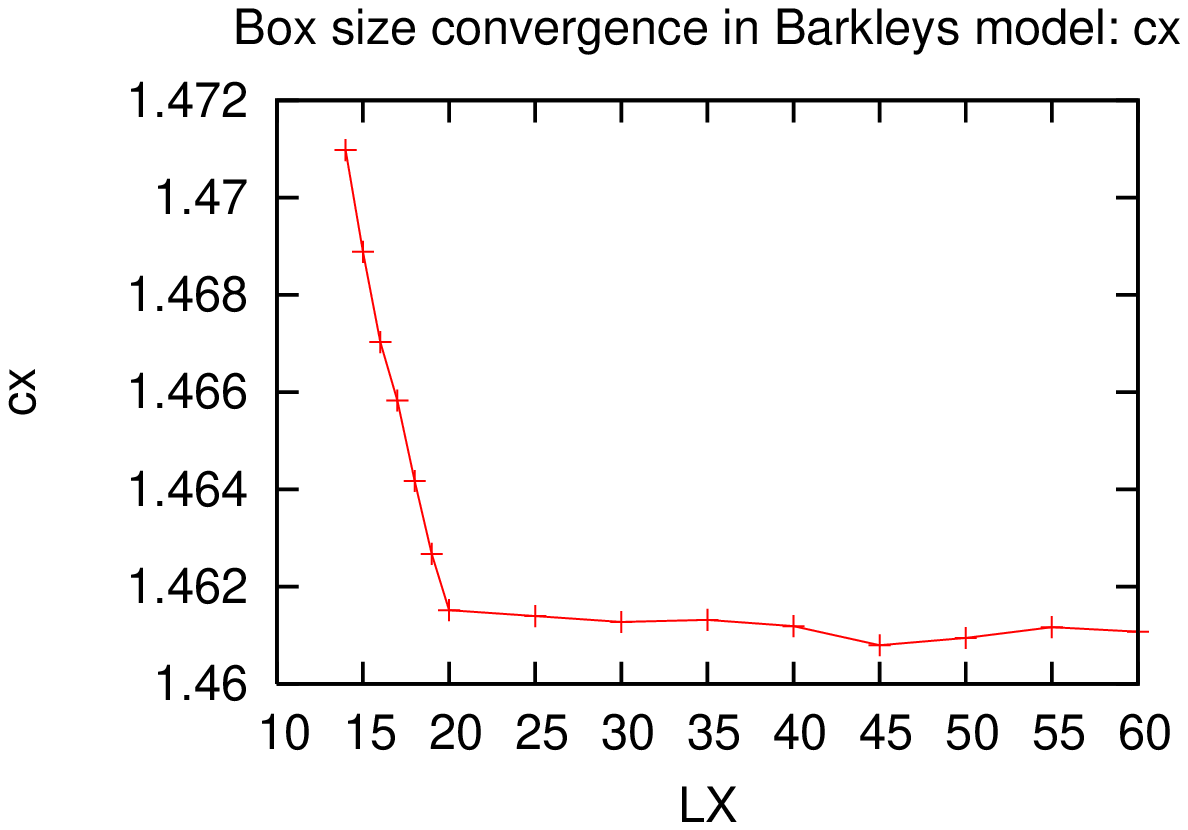}
\end{minipage}
\begin{minipage}{0.6\linewidth}
\centering
\includegraphics[width=0.7\textwidth, angle=-90]{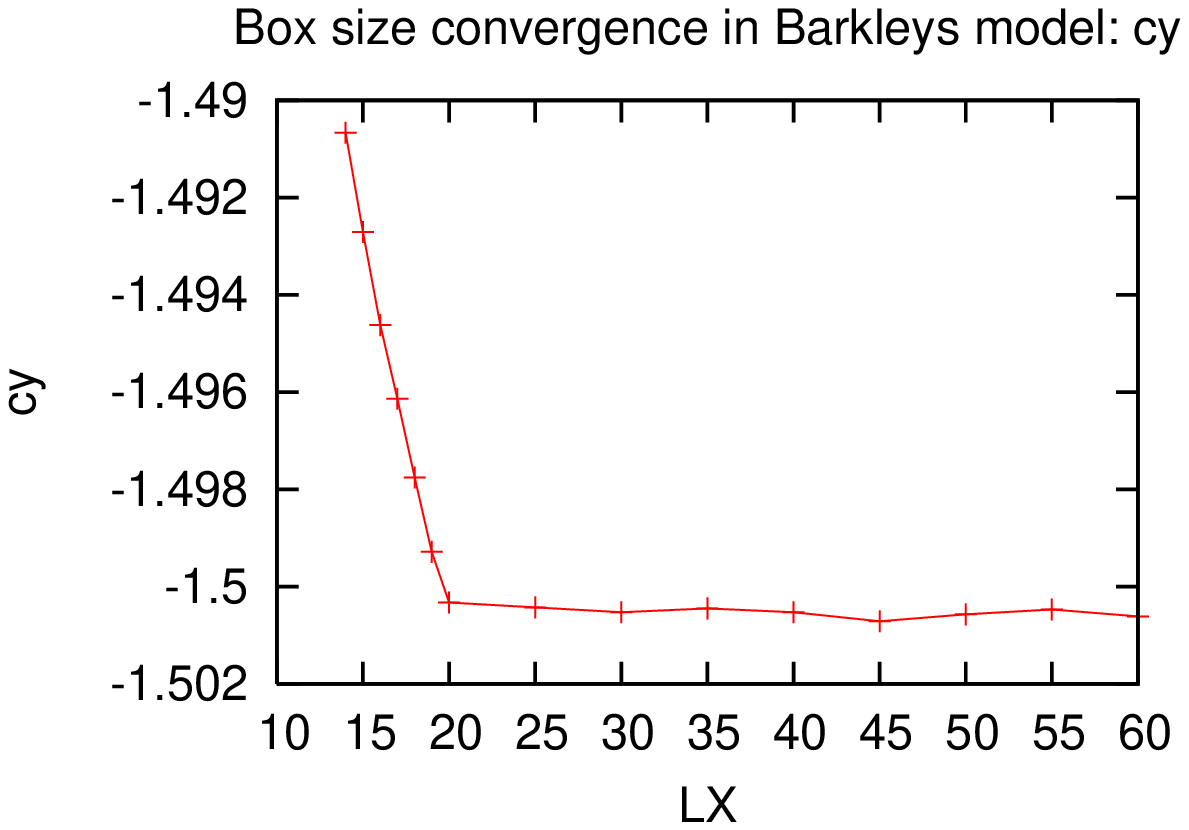}
\end{minipage}
\begin{minipage}{0.6\linewidth}
\centering
\includegraphics[width=0.7\textwidth, angle=-90]{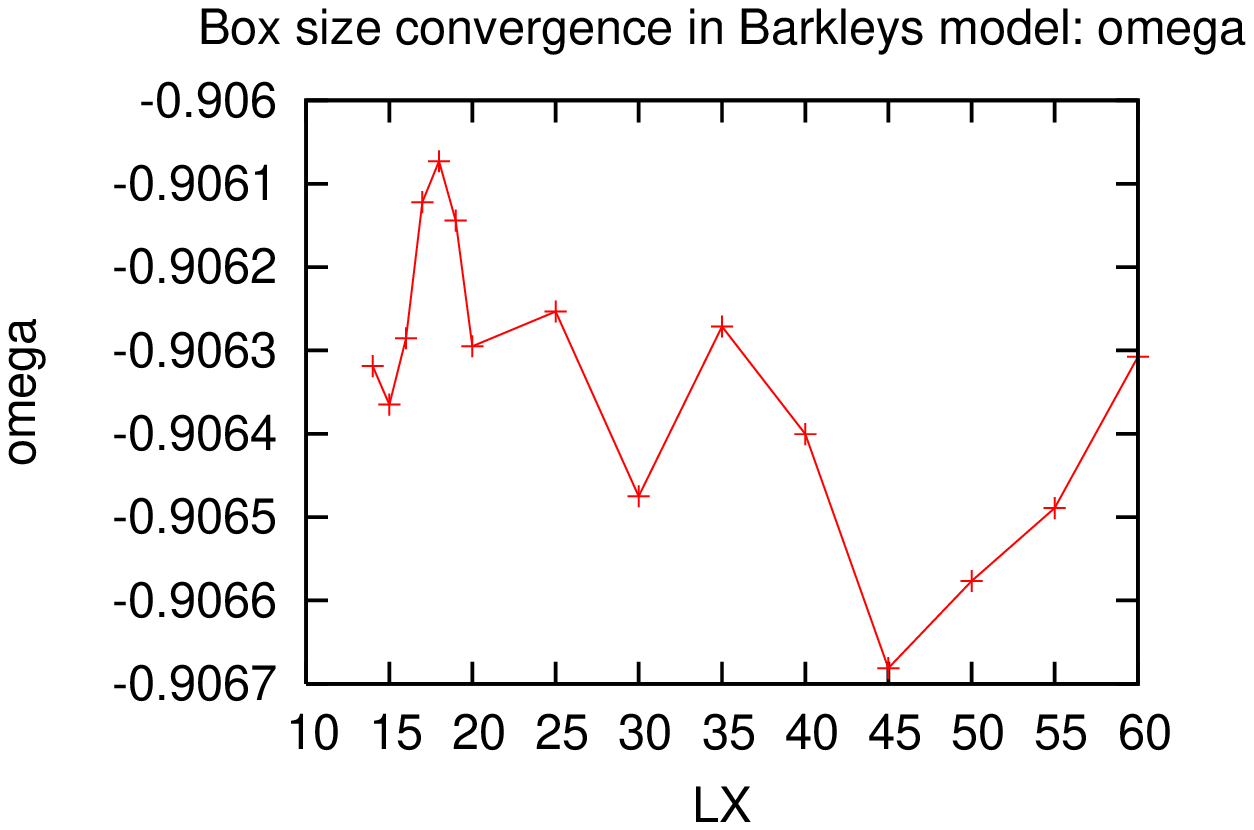}
\end{minipage}
\caption[Box size convergence: Neumann boundary conditions]{Convergence in box size, using Barkley's model and Neumann Boundary conditions with the spacestep fixed at $\Delta_x=\frac{1}{15}$, and the timestep per diffusion stability limit fixed at $t_s=0.1$}
\label{fig:ezf_conv_2_nbc}
\end{center}
\end{figure}

In Fig.(\ref{fig:ezf_conv_2_nbc}) we show the results using Neumann boundary conditions. It is clearly evident that in the plots of $c_x$ and $c_y$ against the box size, we get that the values of $c_x$ and $c_y$ settle down to a fixed value, give or take some very small oscillations of the order of $1.0\times10^{-4}$. In the plot of $\omega$ against the box, it appears that there are some significant oscillations, but again, these oscillations are only of the order $1.0\times10^{-4}$.

We also show in Fig.(\ref{fig:ezf_conv_2_dbc}) the same plots but this time using Dirichlet boundary conditions.  As we can see, we get almost exactly the same results. The values of $c_x$, $c_y$ and $\omega$ all seem to settle down to the same values as those using Neumann boundary conditions.

\begin{figure}[tbh]
\begin{center}
\begin{minipage}{0.32\linewidth}
\centering
\includegraphics[width=0.7\textwidth]{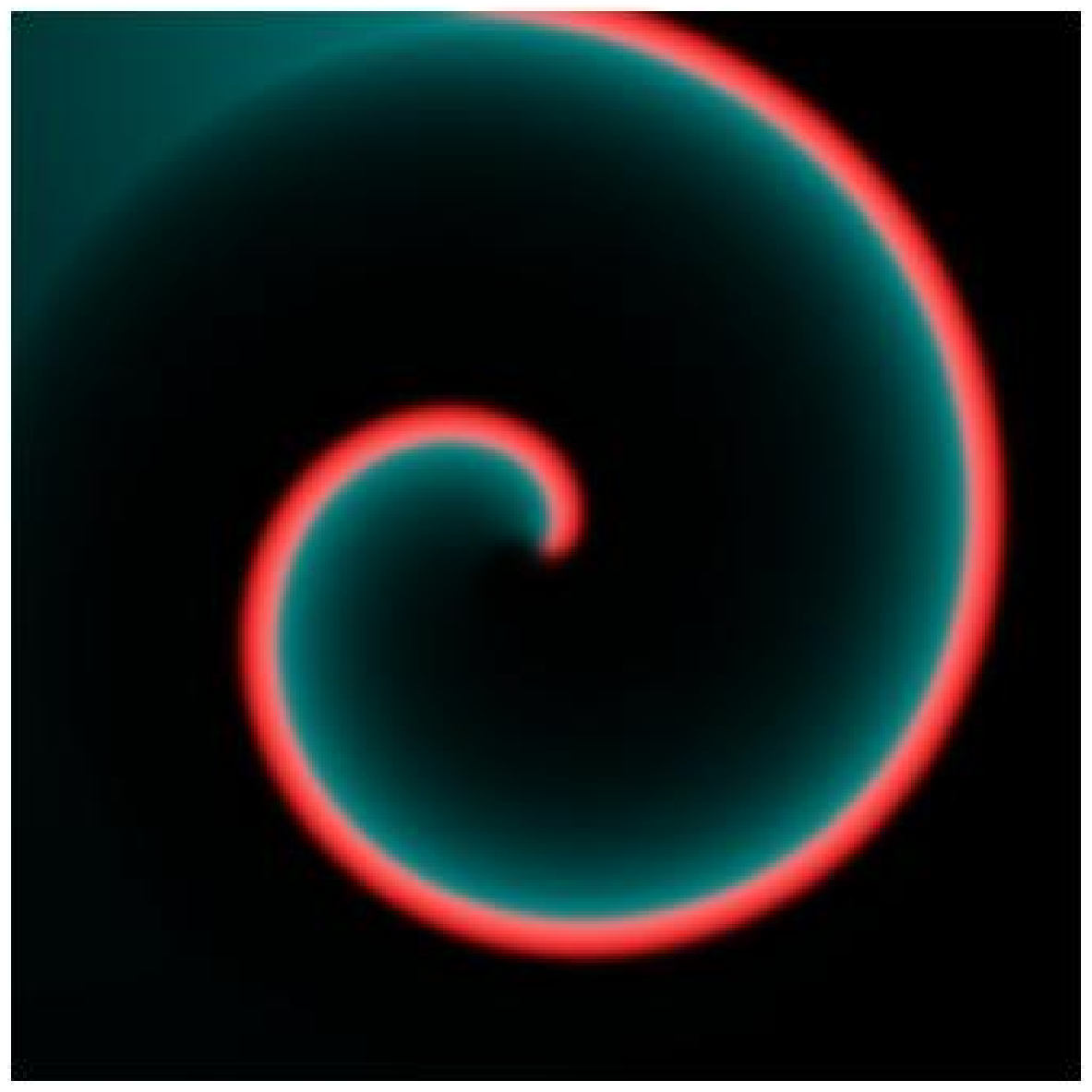}
\end{minipage}
\begin{minipage}{0.32\linewidth}
\centering
\includegraphics[width=0.7\textwidth]{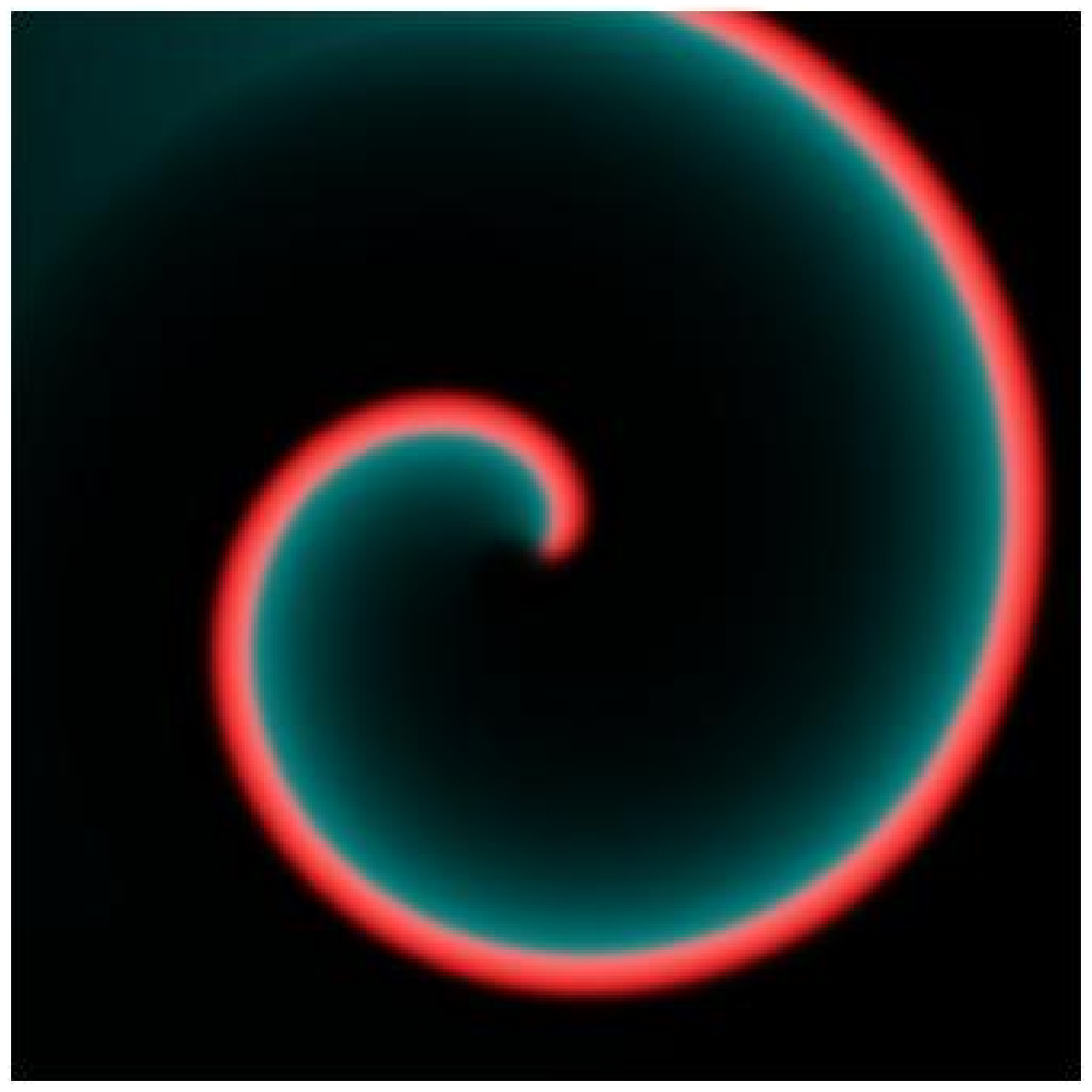}
\end{minipage}
\begin{minipage}{0.32\linewidth}
\centering
\includegraphics[width=0.7\textwidth]{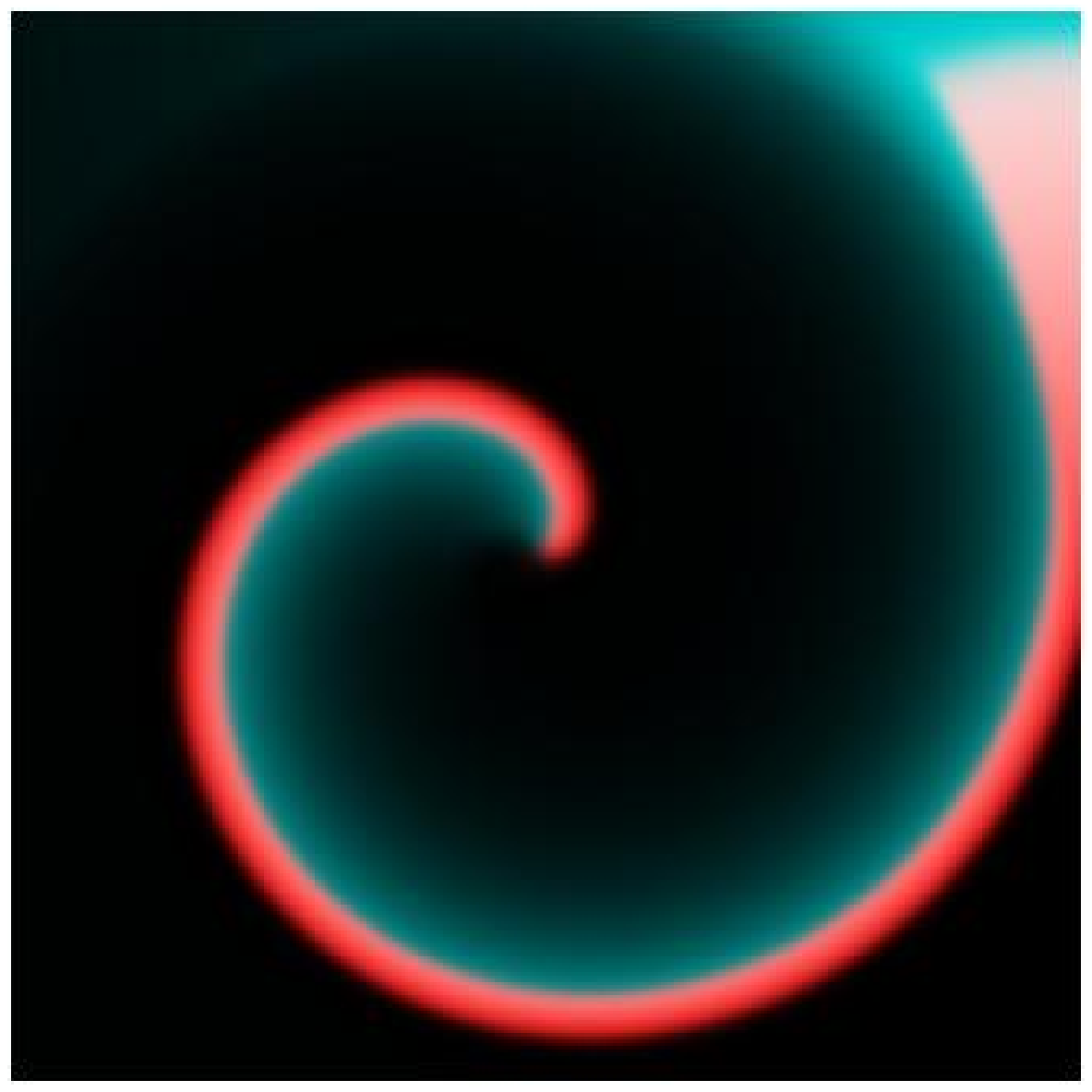}
\end{minipage}
\begin{minipage}{0.32\linewidth}
\centering
\includegraphics[width=0.7\textwidth]{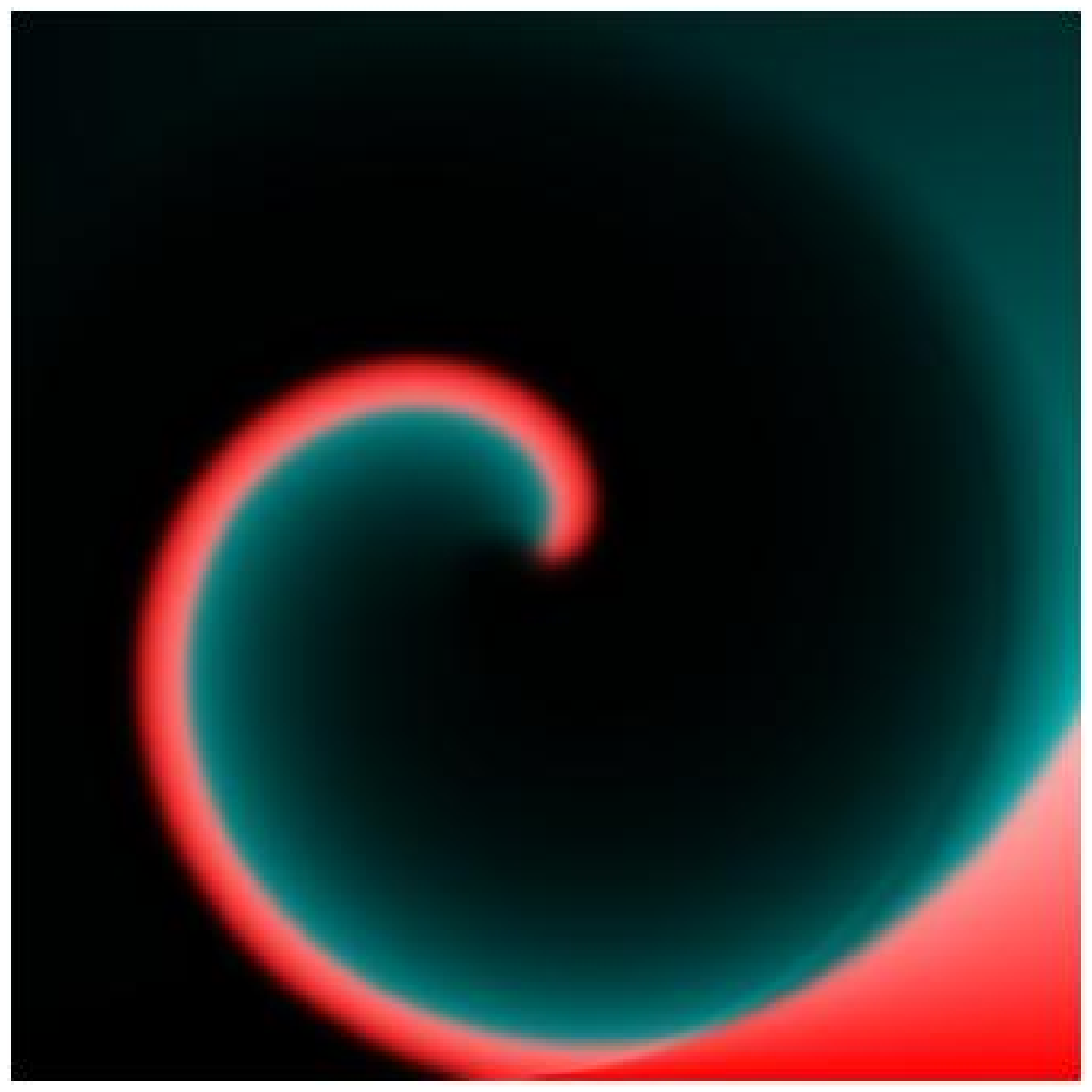}
\end{minipage}
\begin{minipage}{0.32\linewidth}
\centering
\includegraphics[width=0.7\textwidth]{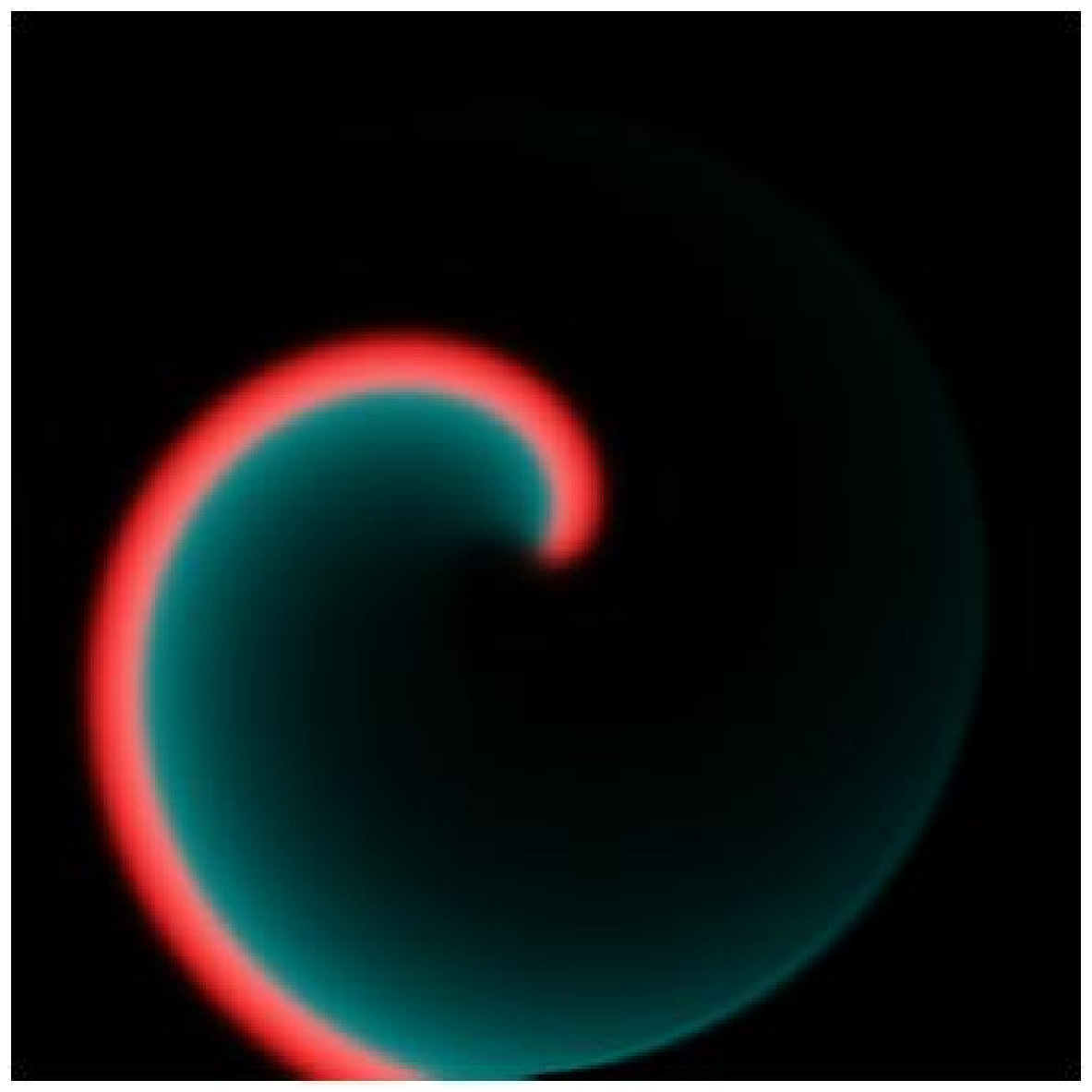}
\end{minipage}
\begin{minipage}{0.32\linewidth}
\centering
\includegraphics[width=0.7\textwidth]{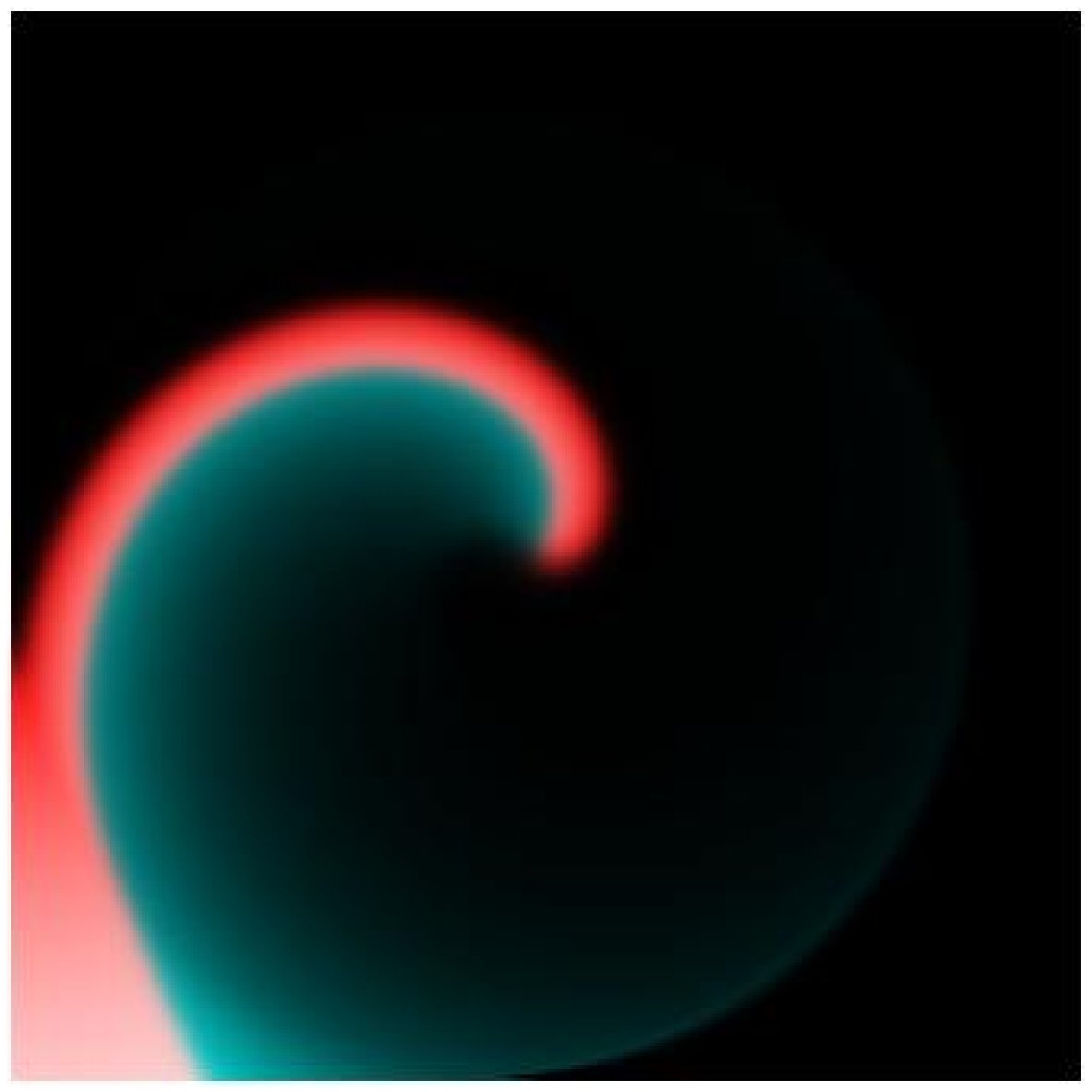}
\end{minipage}
\begin{minipage}{0.32\linewidth}
\centering
\includegraphics[width=0.7\textwidth]{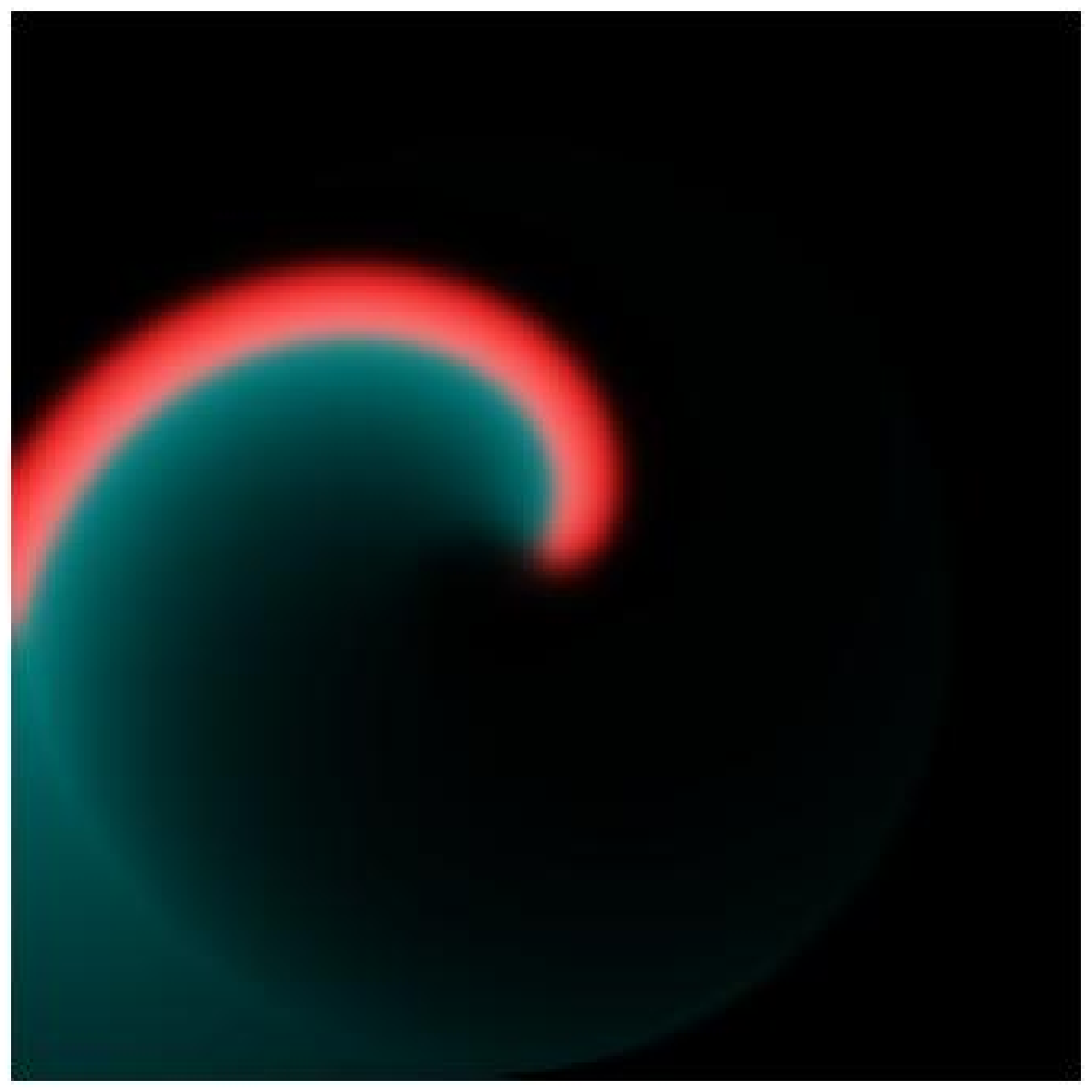}
\end{minipage}
\begin{minipage}{0.32\linewidth}
\centering
\includegraphics[width=0.7\textwidth]{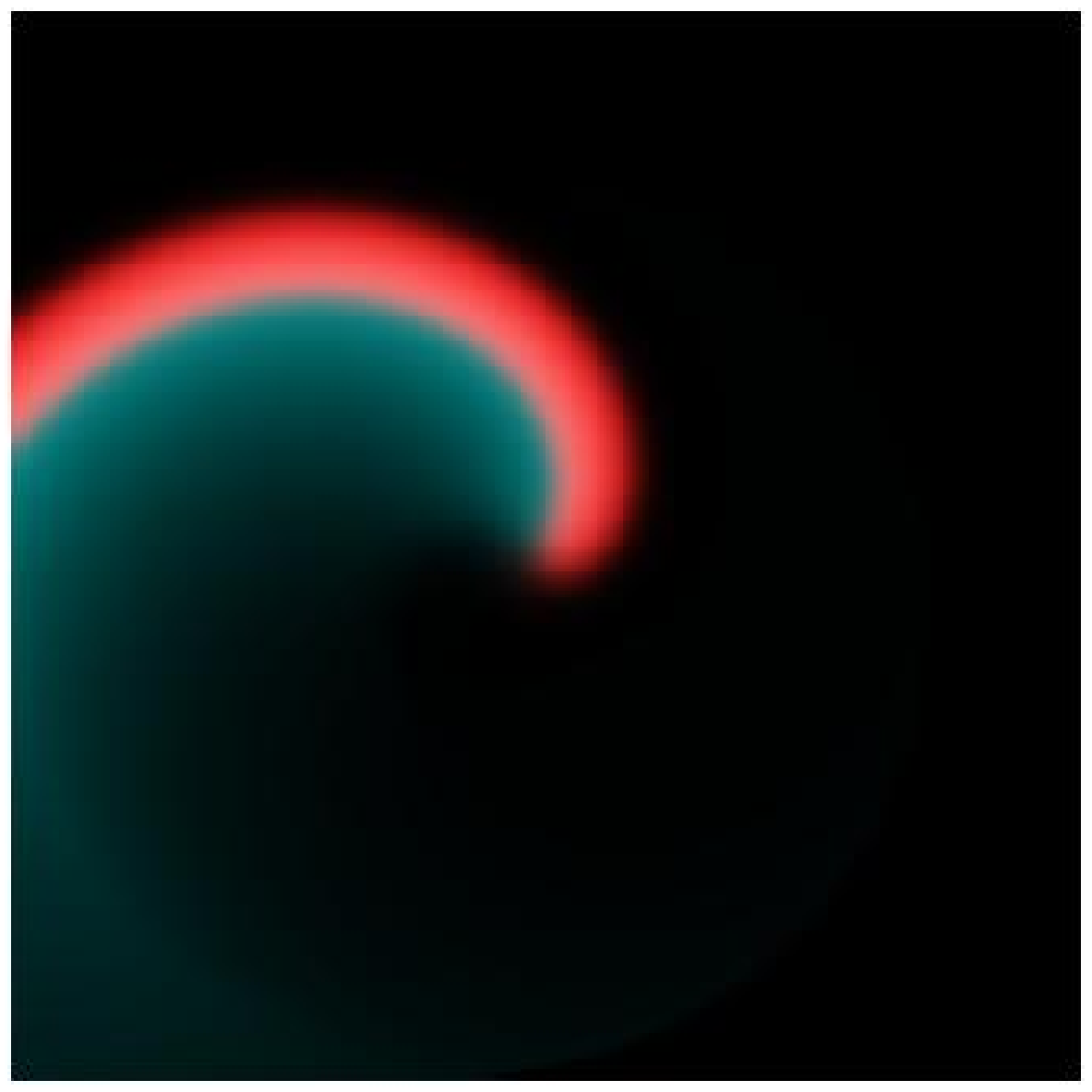}
\end{minipage}
\begin{minipage}{0.32\linewidth}
\centering
\includegraphics[width=0.7\textwidth]{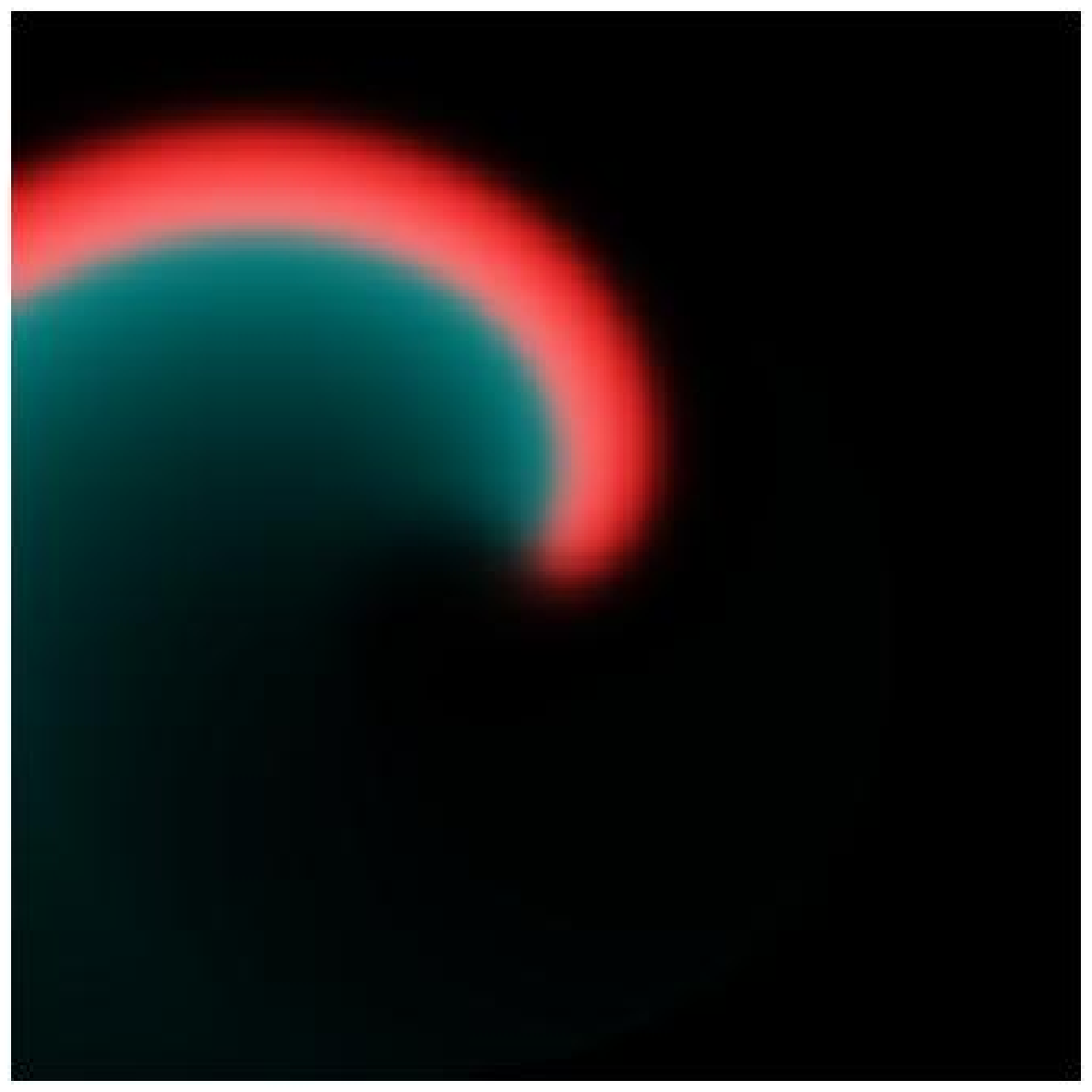}
\end{minipage}
\begin{minipage}{0.32\linewidth}
\centering
\includegraphics[width=0.7\textwidth]{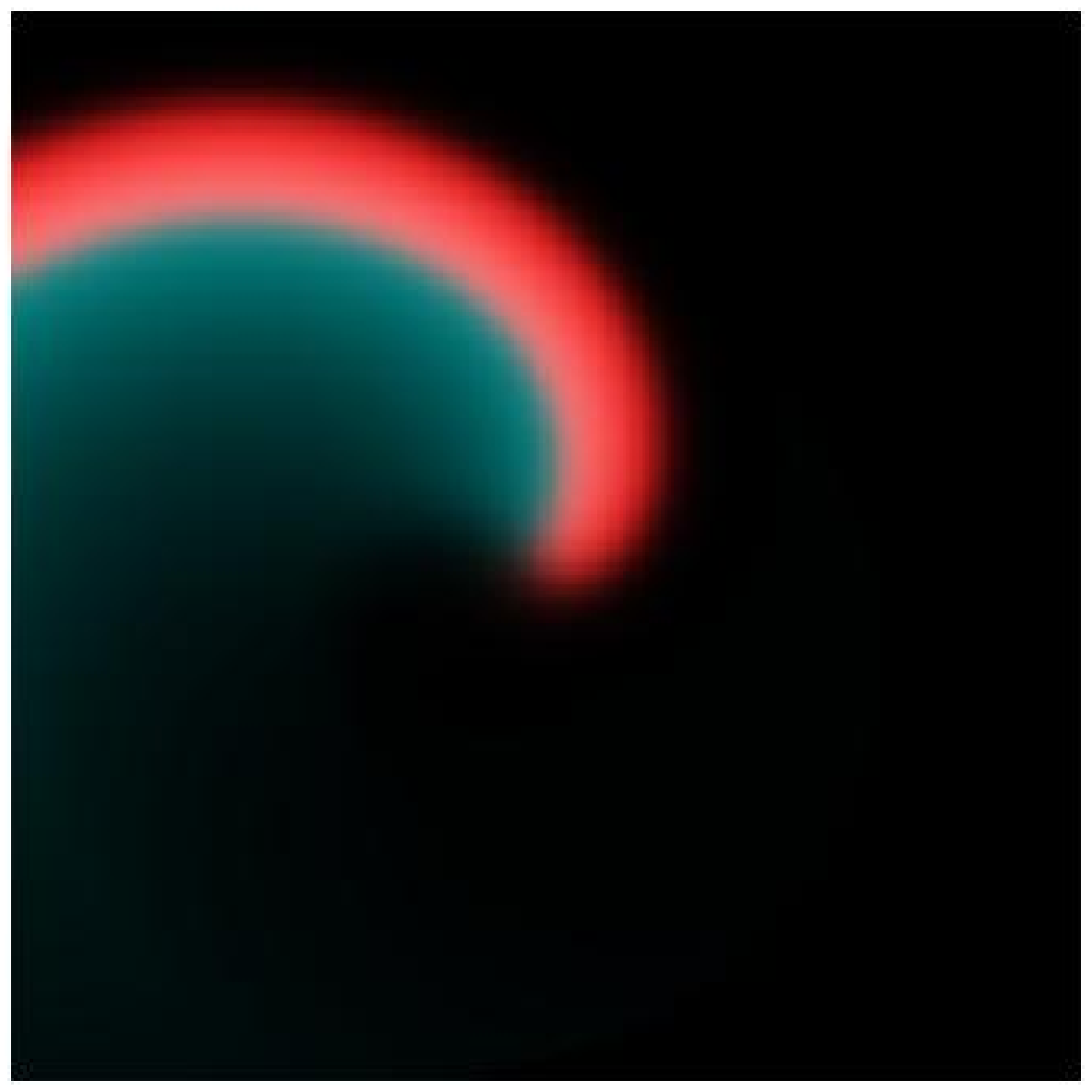}
\end{minipage}
\begin{minipage}{0.32\linewidth}
\centering
\includegraphics[width=0.7\textwidth]{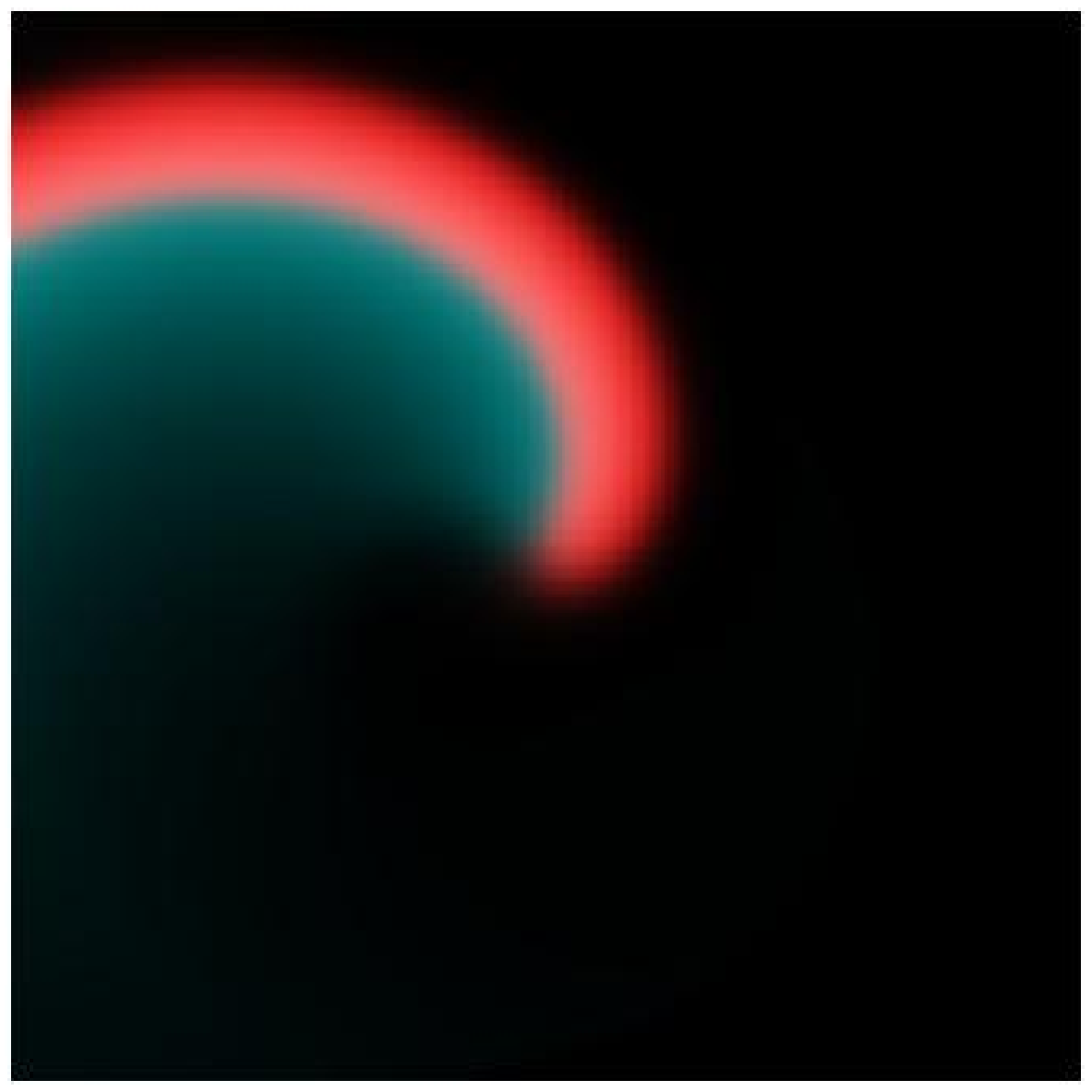}
\end{minipage}
\begin{minipage}{0.32\linewidth}
\centering
\includegraphics[width=0.7\textwidth]{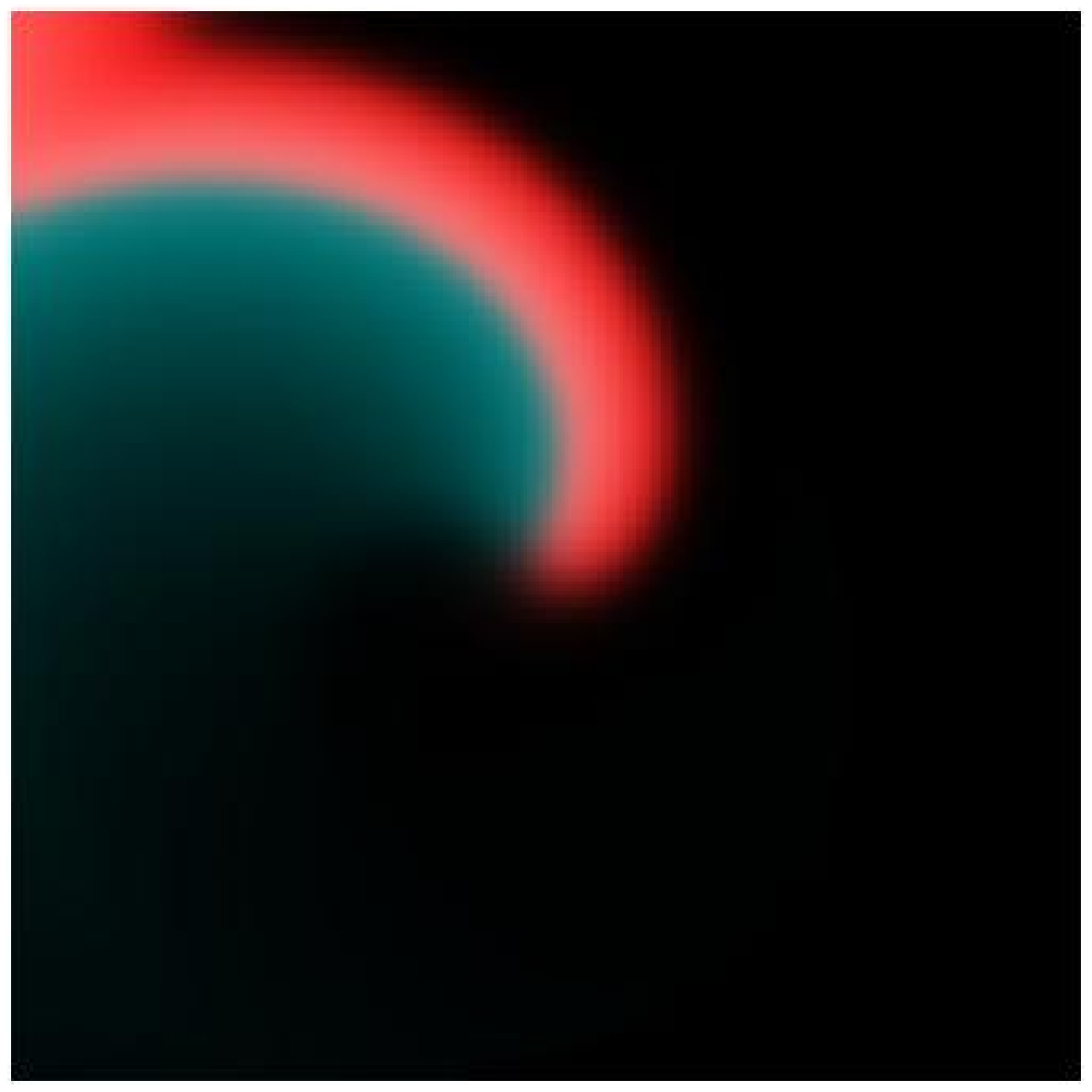}
\end{minipage}
\begin{minipage}{0.32\linewidth}
\centering
\includegraphics[width=0.7\textwidth]{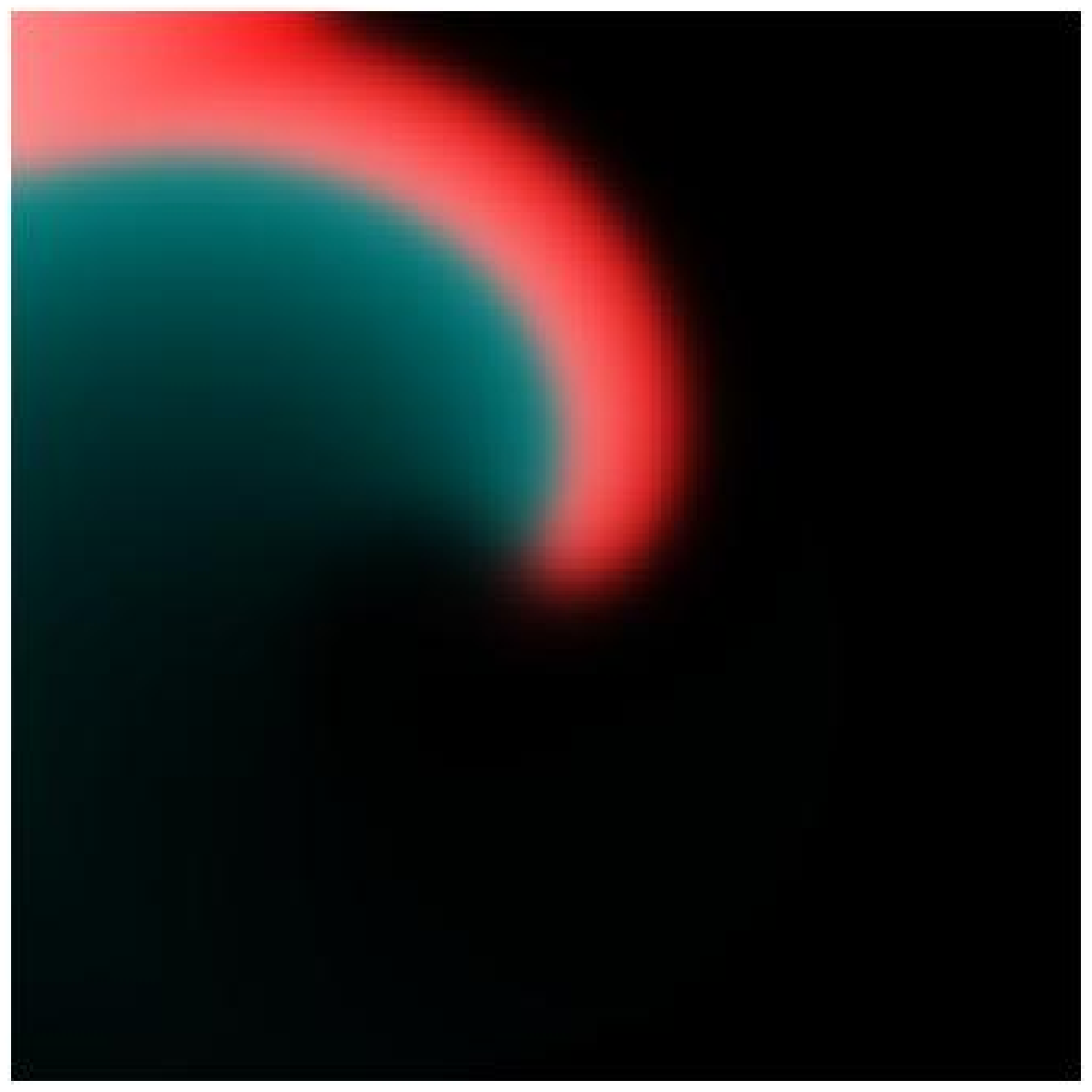}
\end{minipage}
\begin{minipage}{0.32\linewidth}
\centering
\includegraphics[width=0.7\textwidth]{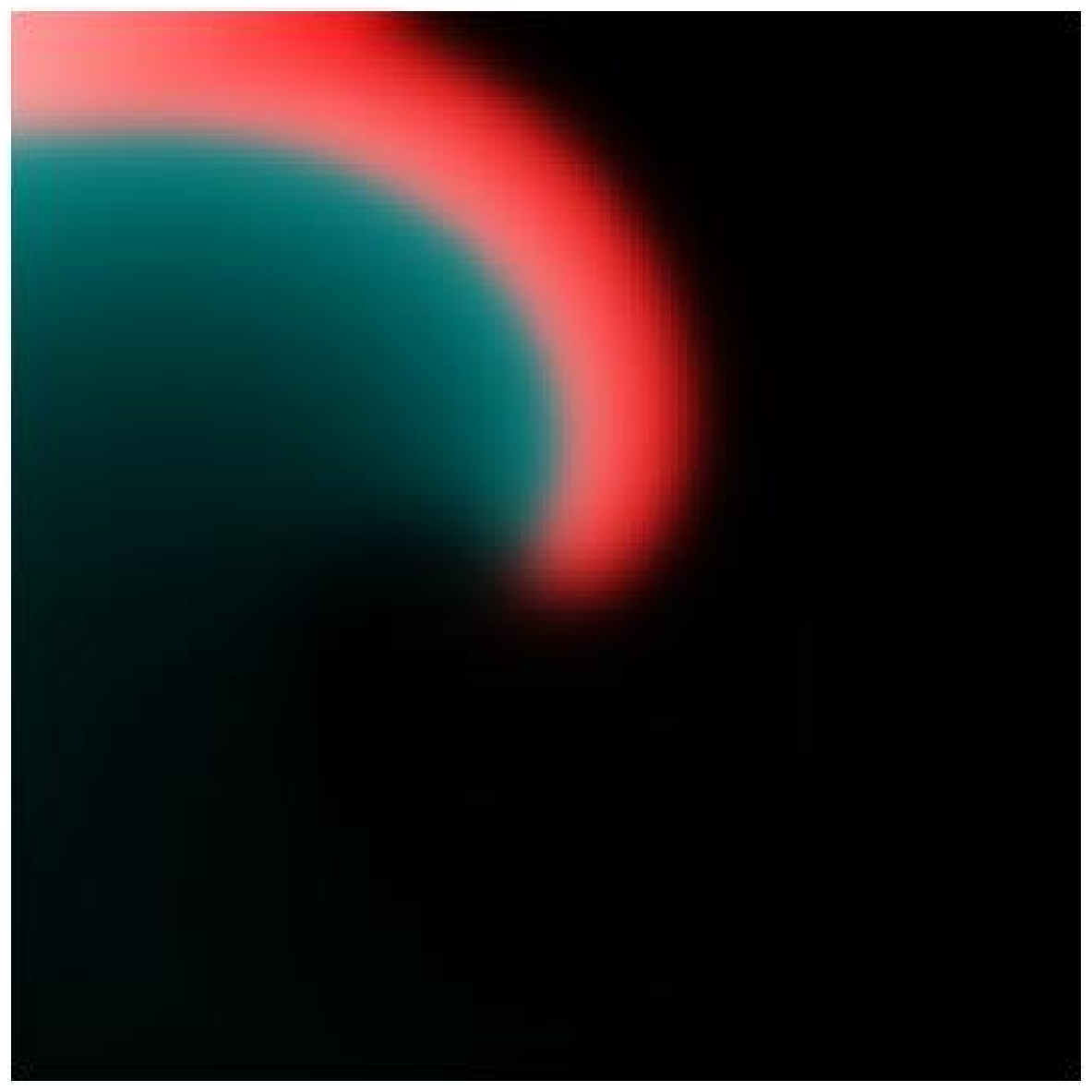}
\end{minipage}
\begin{minipage}{0.32\linewidth}
\centering
\includegraphics[width=0.7\textwidth]{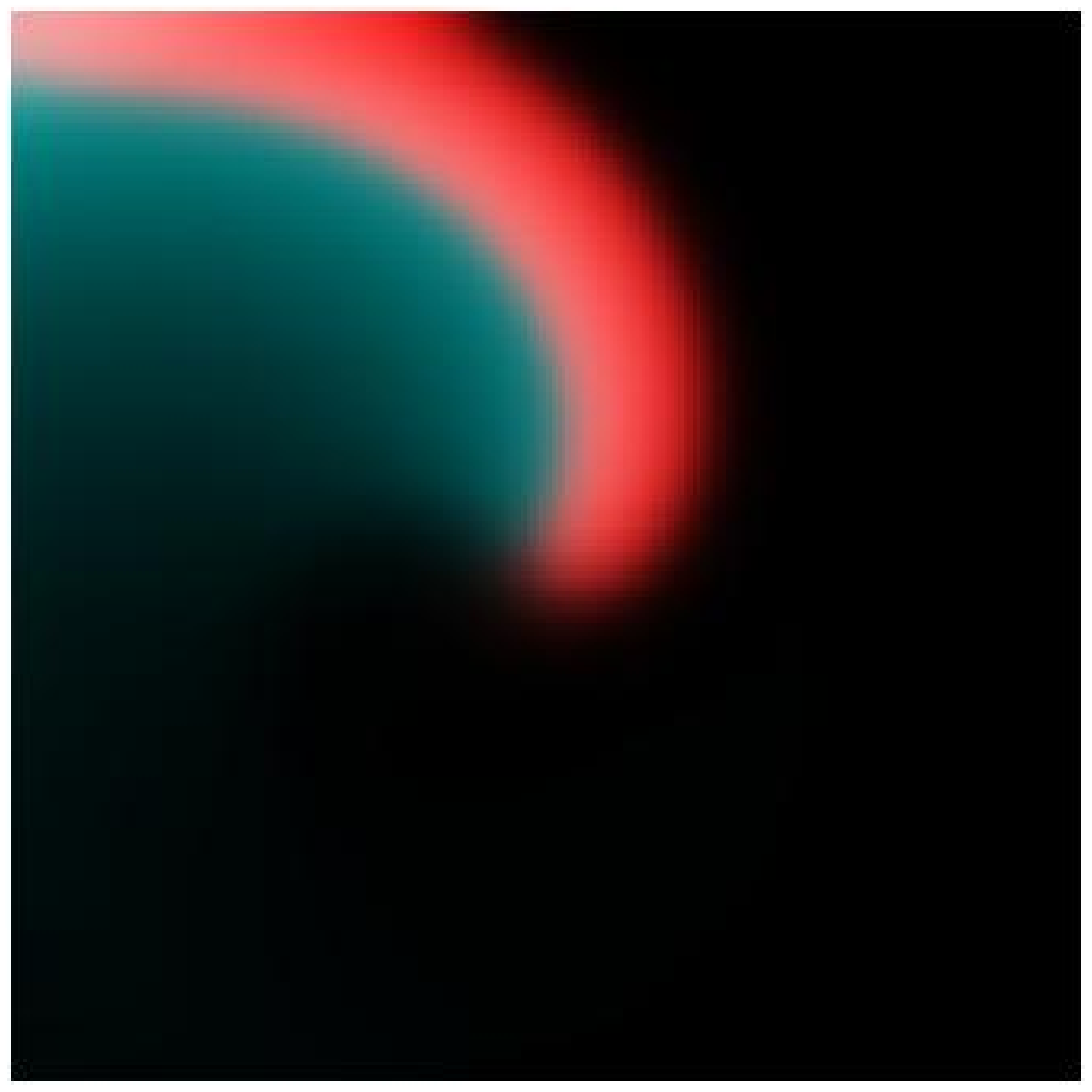}
\end{minipage}
\caption[Box size convergence: Neumann boundary conditions, final solutions]{Final Conditions for each run in the convergence testing of the box size in Barkley's model using Neumann boundary conditions, starting top left and working right, $L_X=60$ (top left) to $L_X=15$ (bottom right)}
\label{fig:ezf_conv_2_final_nbc}
\end{center}
\end{figure}

\begin{figure}[tbh]
\begin{center}
\begin{minipage}{0.6\linewidth}
\centering
\includegraphics[width=0.7\textwidth, angle=-90]{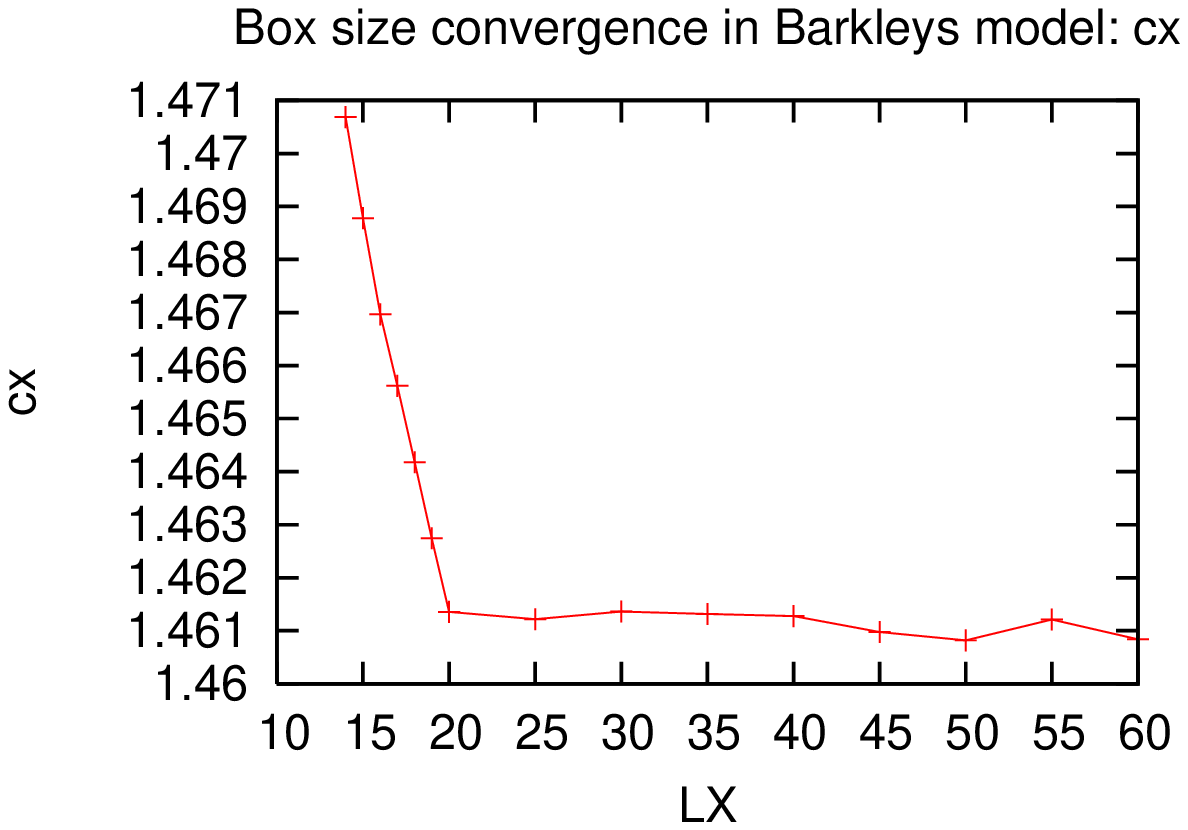}
\end{minipage}
\begin{minipage}{0.6\linewidth}
\centering
\includegraphics[width=0.7\textwidth, angle=-90]{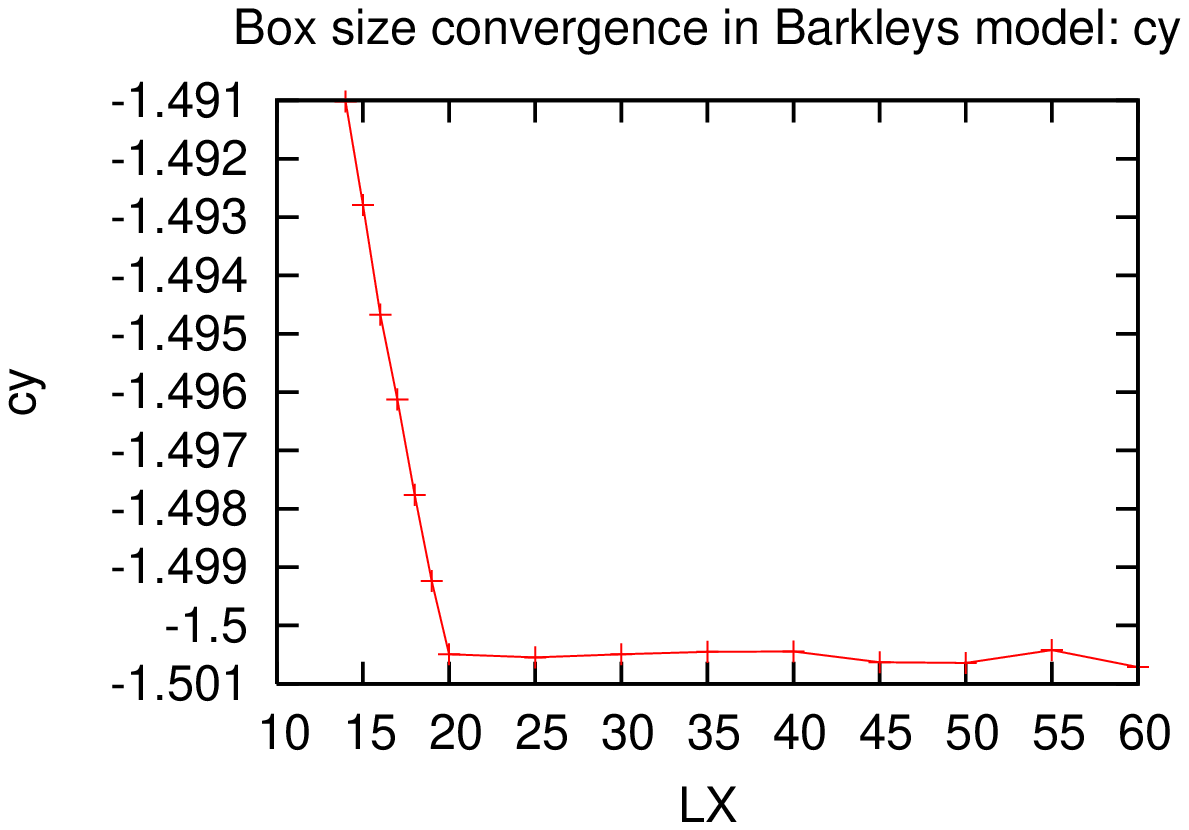}
\end{minipage}
\begin{minipage}{0.6\linewidth}
\centering
\includegraphics[width=0.7\textwidth, angle=-90]{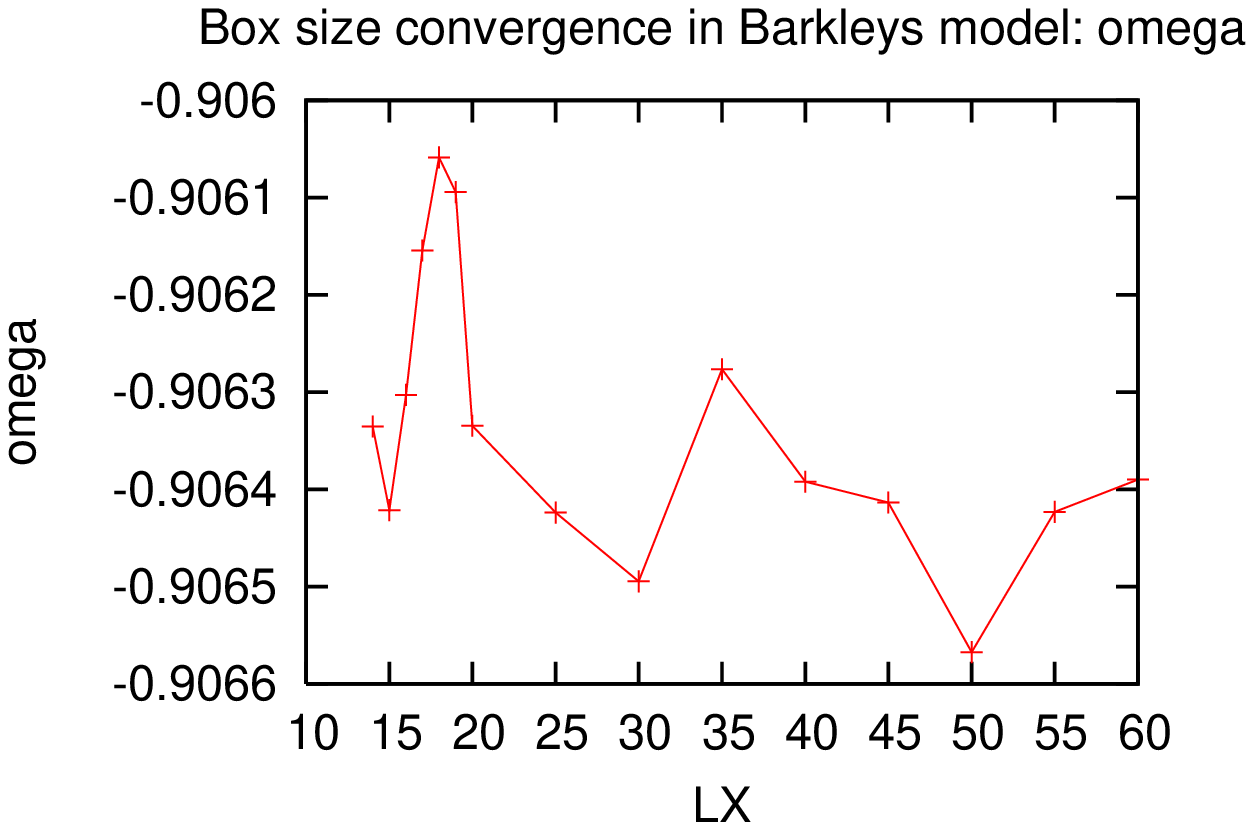}
\end{minipage}
\caption[Box size convergence: Dirichlet boundary conditions]{Convergence in box size, using Barkley's model and Dirichlet Boundary conditions with the spacestep fixed at $\Delta_x=\frac{1}{15}$, and the timestep per diffusion stability limit fixed at $t_s=0.1$}
\label{fig:ezf_conv_2_dbc}
\end{center}
\end{figure}

\begin{figure}[tbh]
\begin{center}
\begin{minipage}{0.32\linewidth}
\centering
\includegraphics[width=0.7\textwidth]{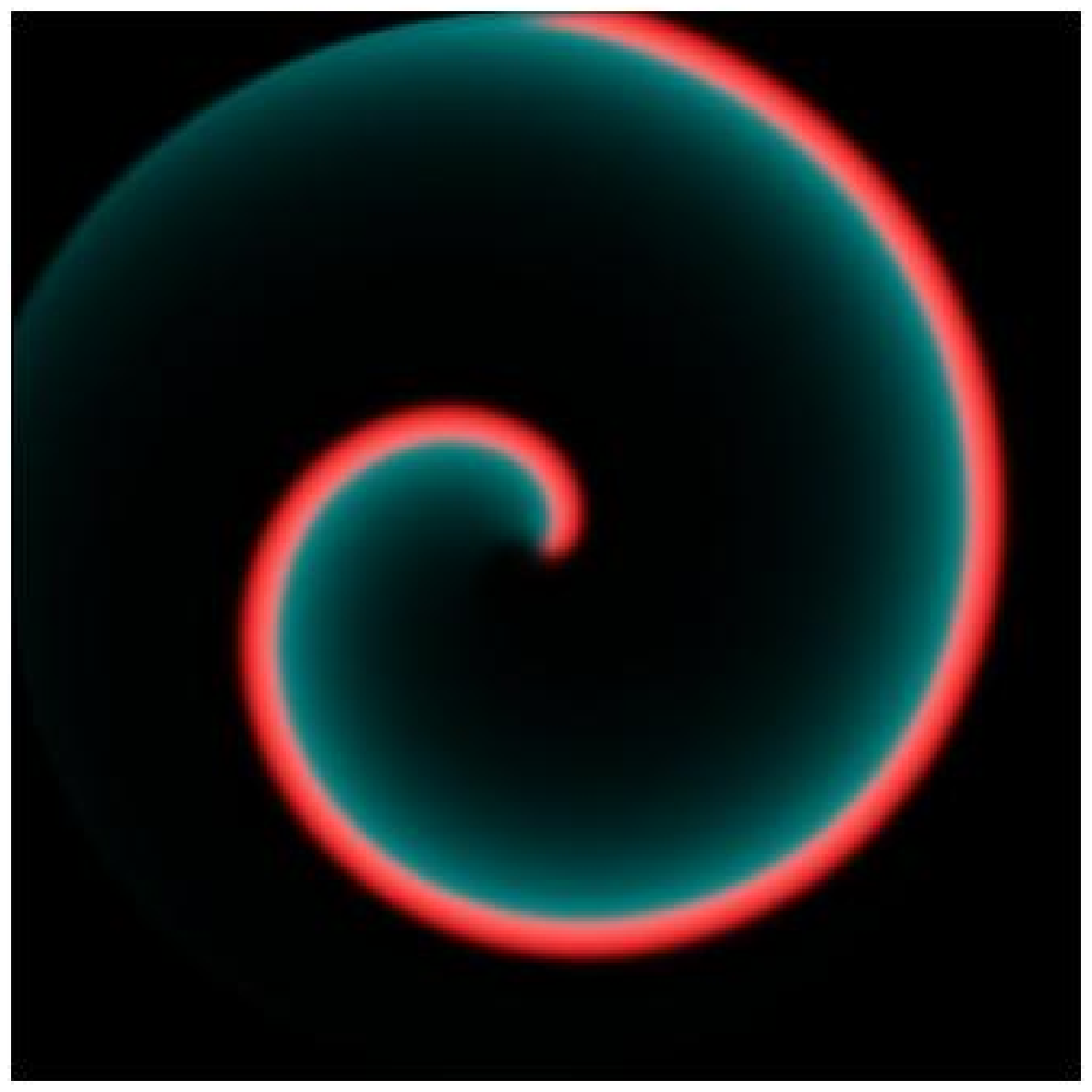}
\end{minipage}
\begin{minipage}{0.32\linewidth}
\centering
\includegraphics[width=0.7\textwidth]{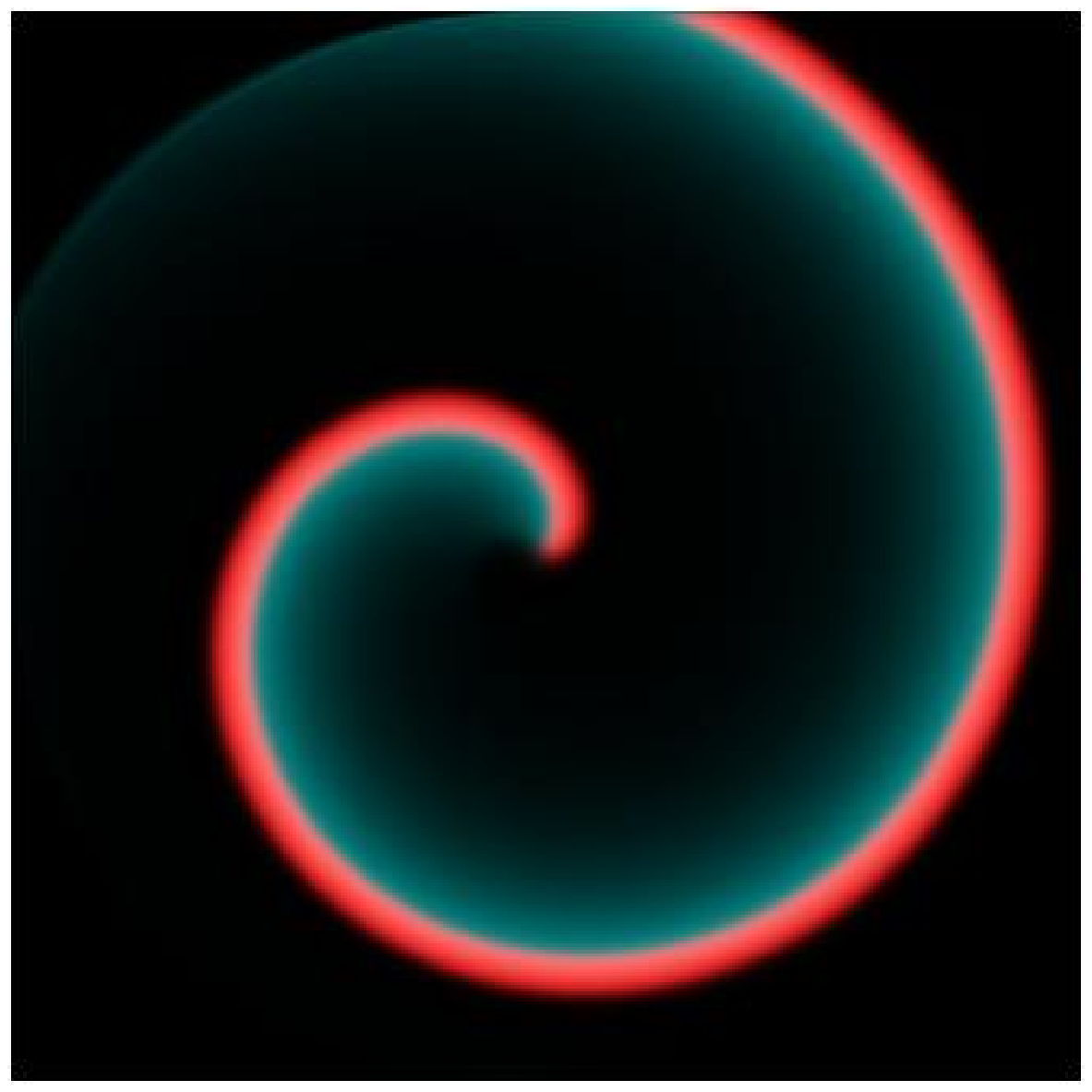}
\end{minipage}
\begin{minipage}{0.32\linewidth}
\centering
\includegraphics[width=0.7\textwidth]{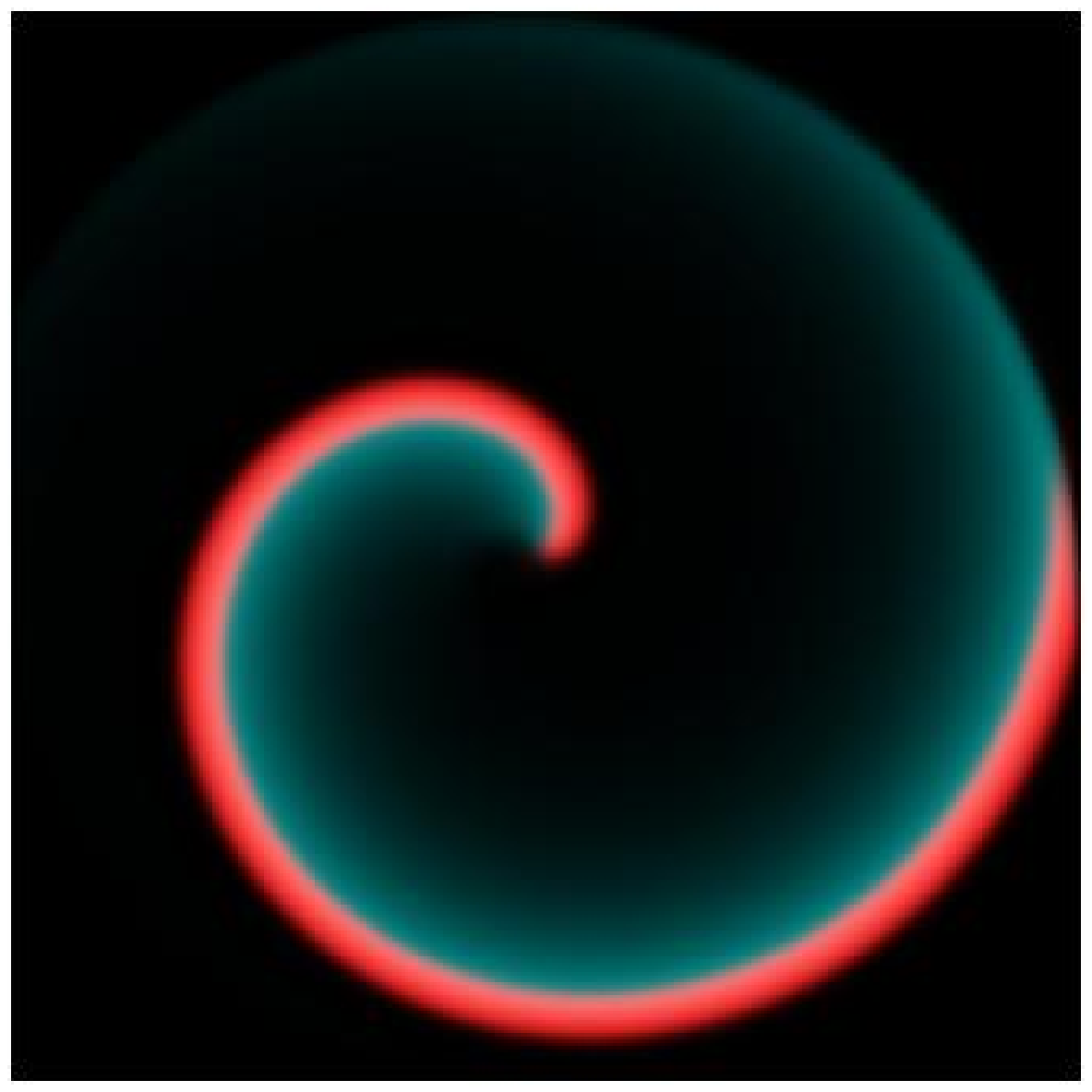}
\end{minipage}
\begin{minipage}{0.32\linewidth}
\centering
\includegraphics[width=0.7\textwidth]{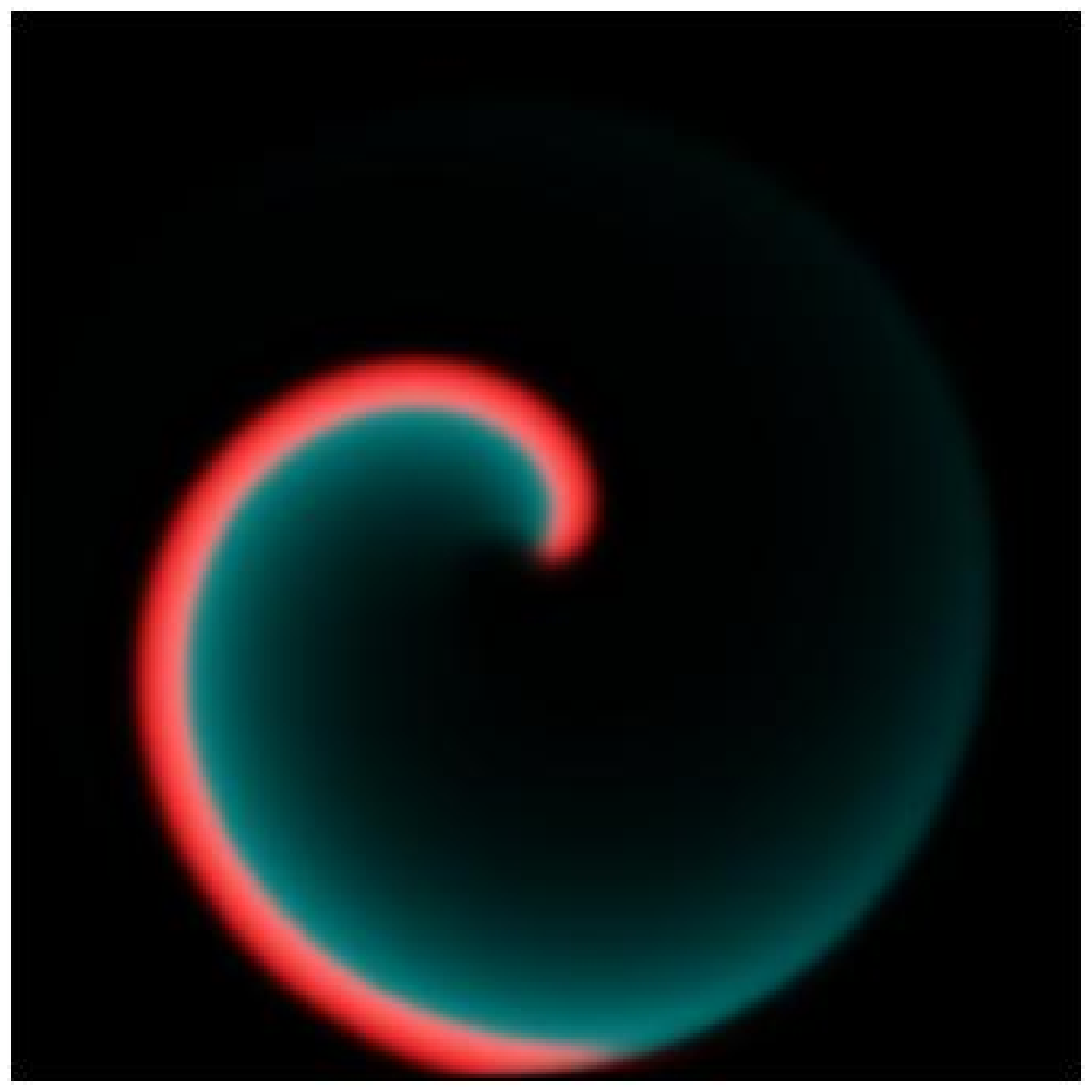}
\end{minipage}
\begin{minipage}{0.32\linewidth}
\centering
\includegraphics[width=0.7\textwidth]{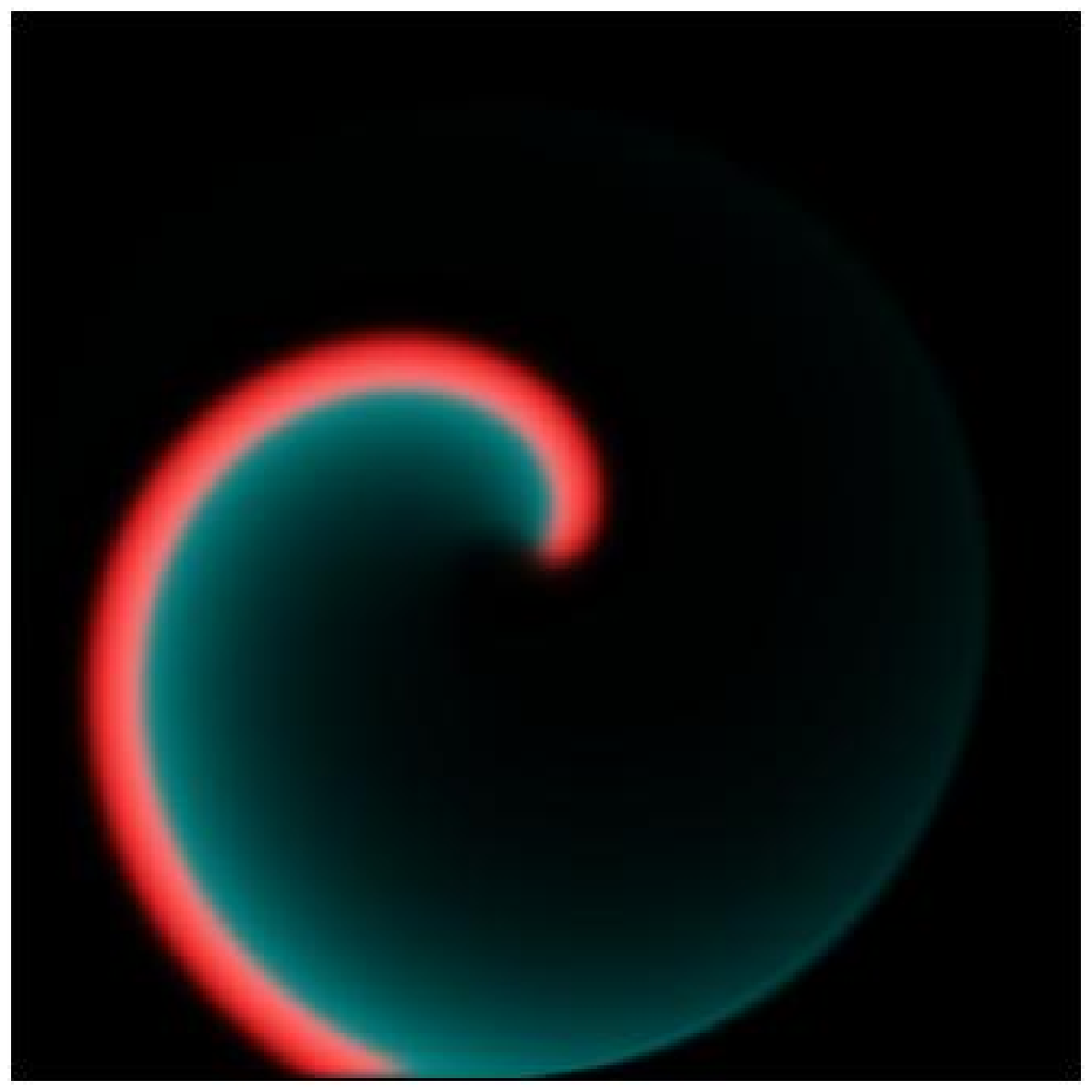}
\end{minipage}
\begin{minipage}{0.32\linewidth}
\centering
\includegraphics[width=0.7\textwidth]{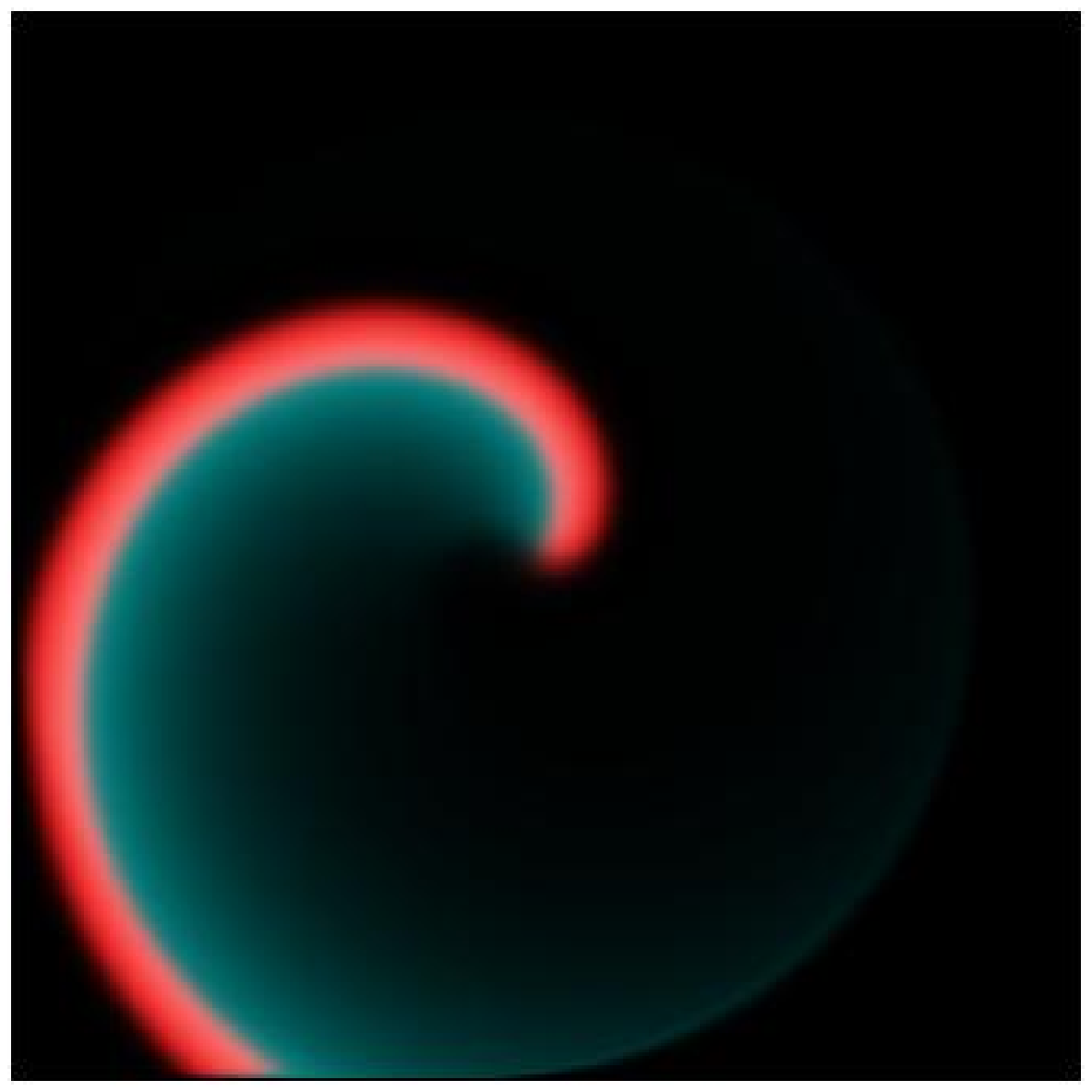}
\end{minipage}
\begin{minipage}{0.32\linewidth}
\centering
\includegraphics[width=0.7\textwidth]{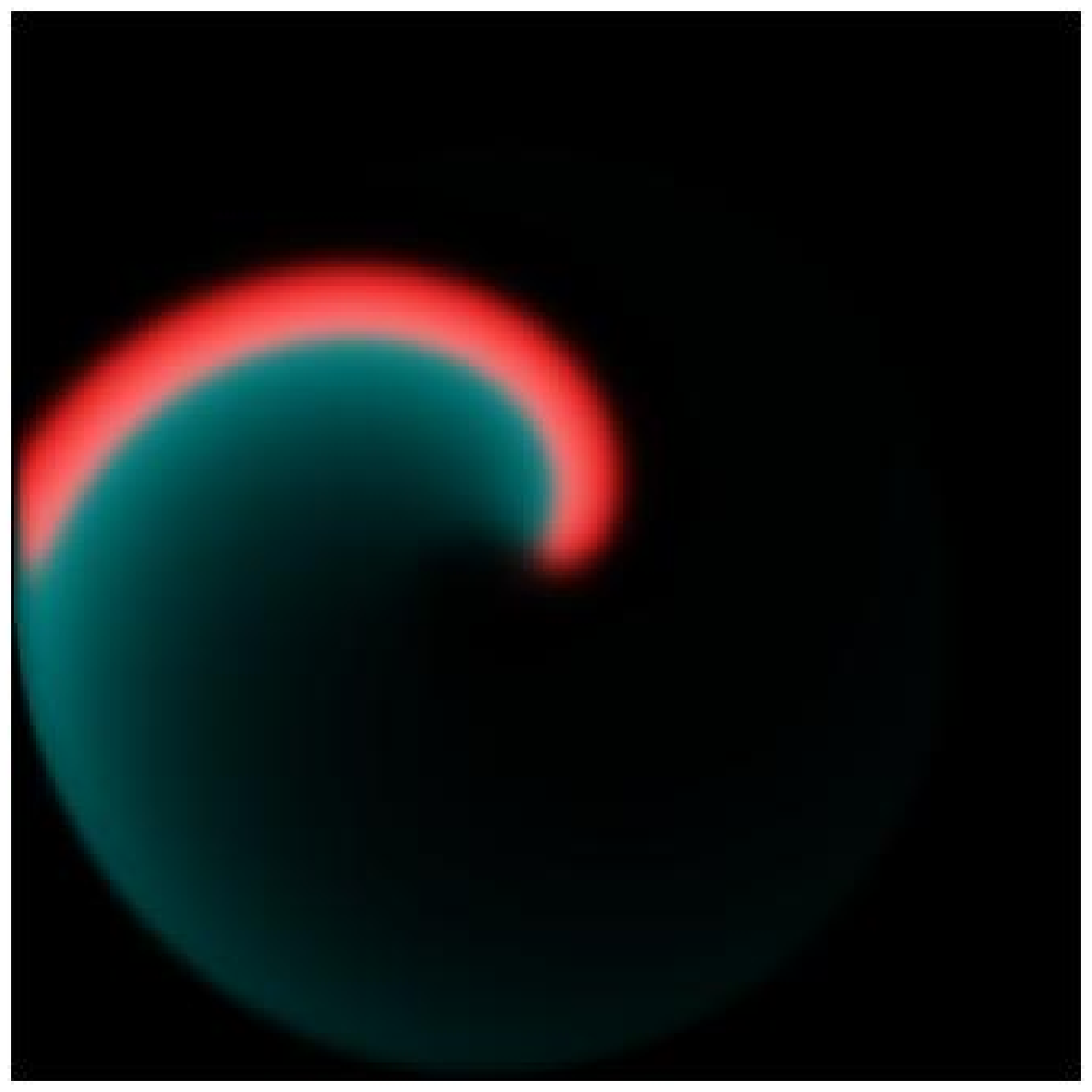}
\end{minipage}
\begin{minipage}{0.32\linewidth}
\centering
\includegraphics[width=0.7\textwidth]{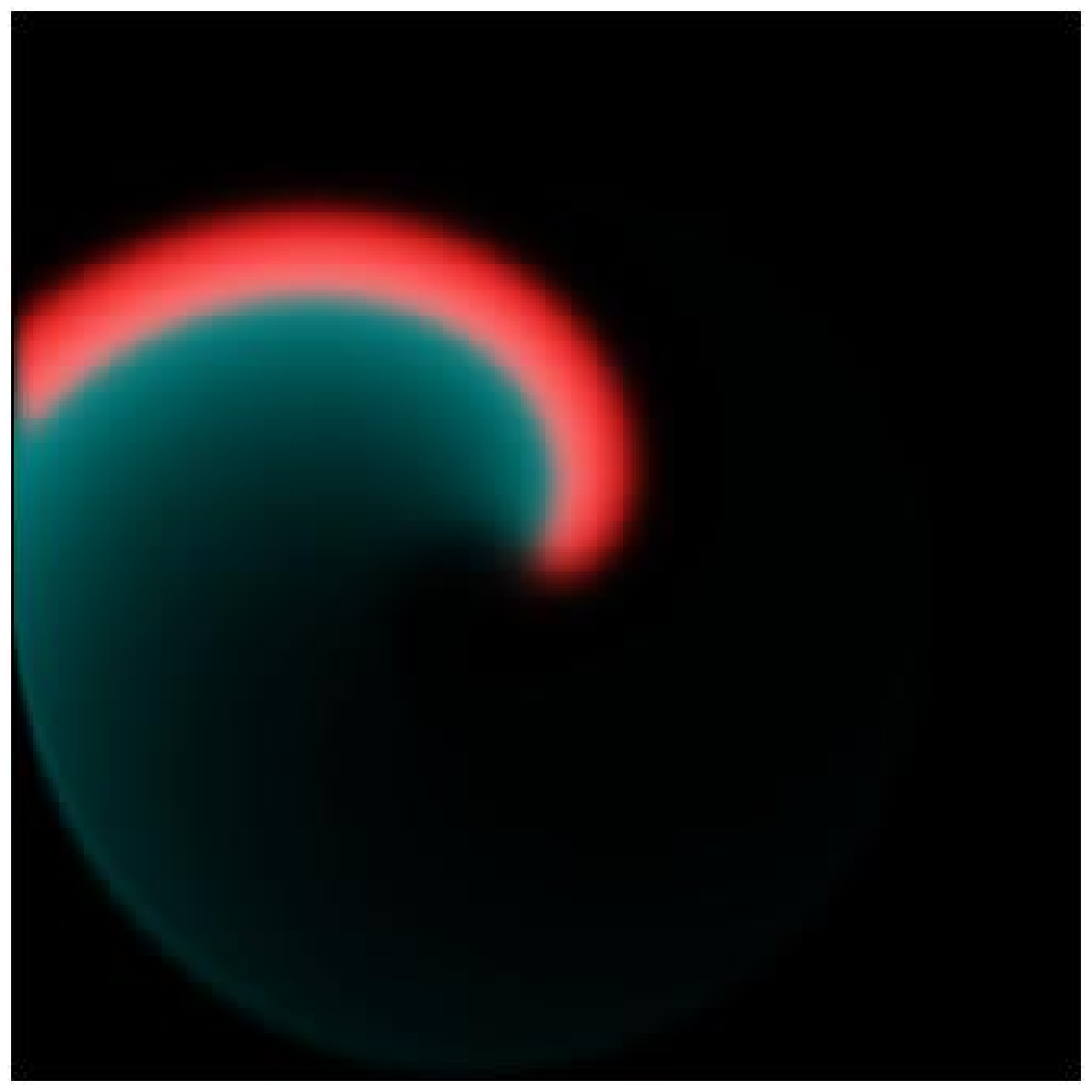}
\end{minipage}
\begin{minipage}{0.32\linewidth}
\centering
\includegraphics[width=0.7\textwidth]{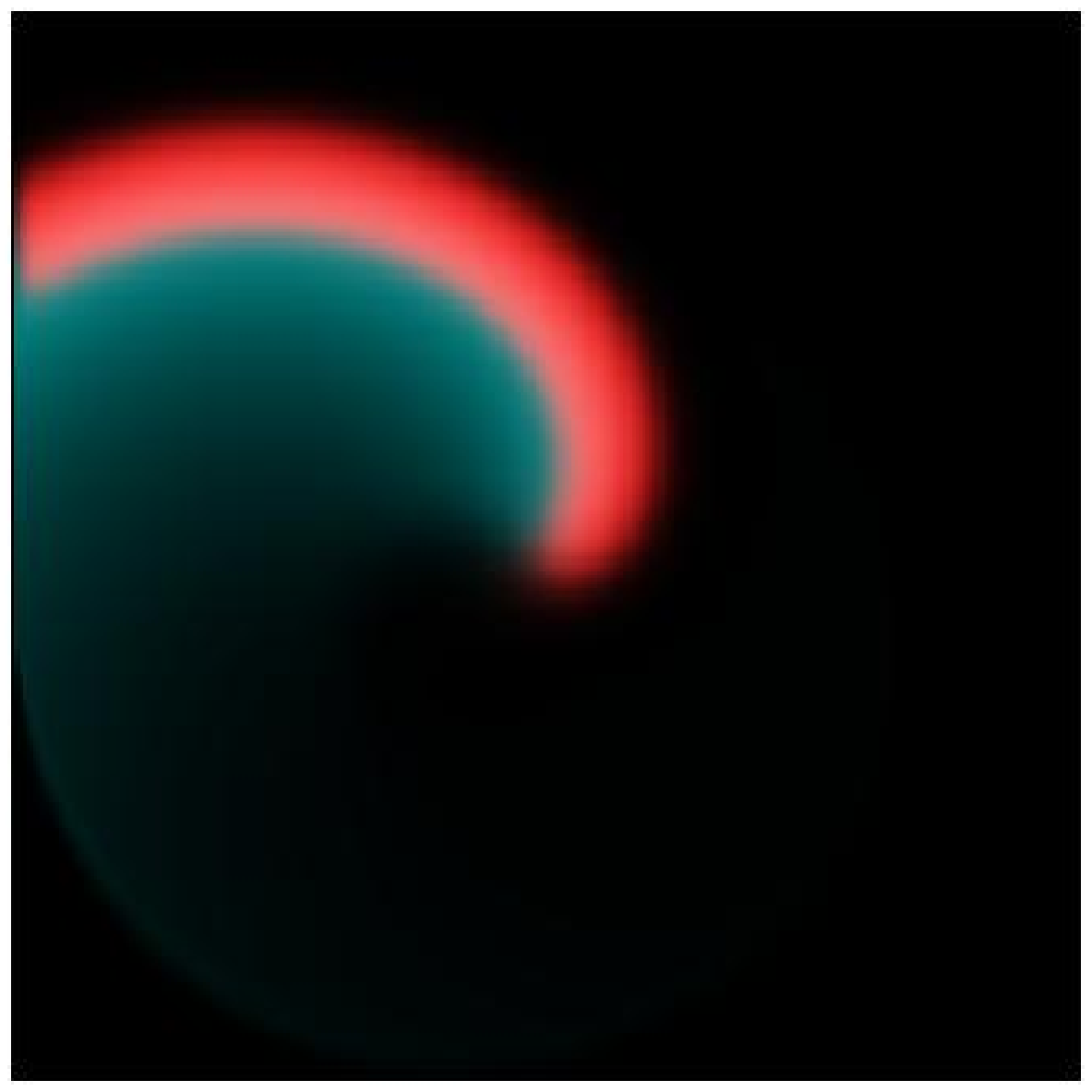}
\end{minipage}
\begin{minipage}{0.32\linewidth}
\centering
\includegraphics[width=0.7\textwidth]{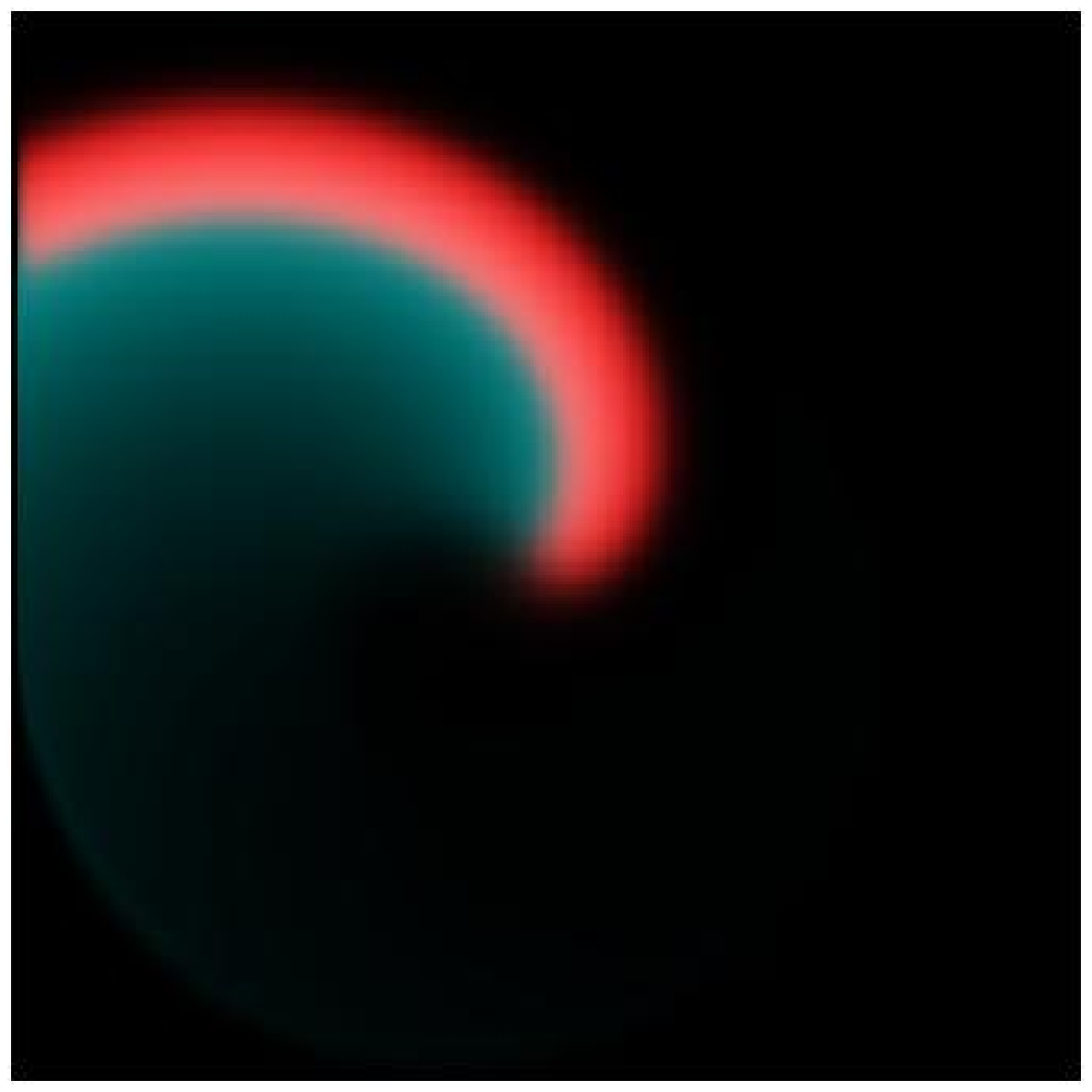}
\end{minipage}
\begin{minipage}{0.32\linewidth}
\centering
\includegraphics[width=0.7\textwidth]{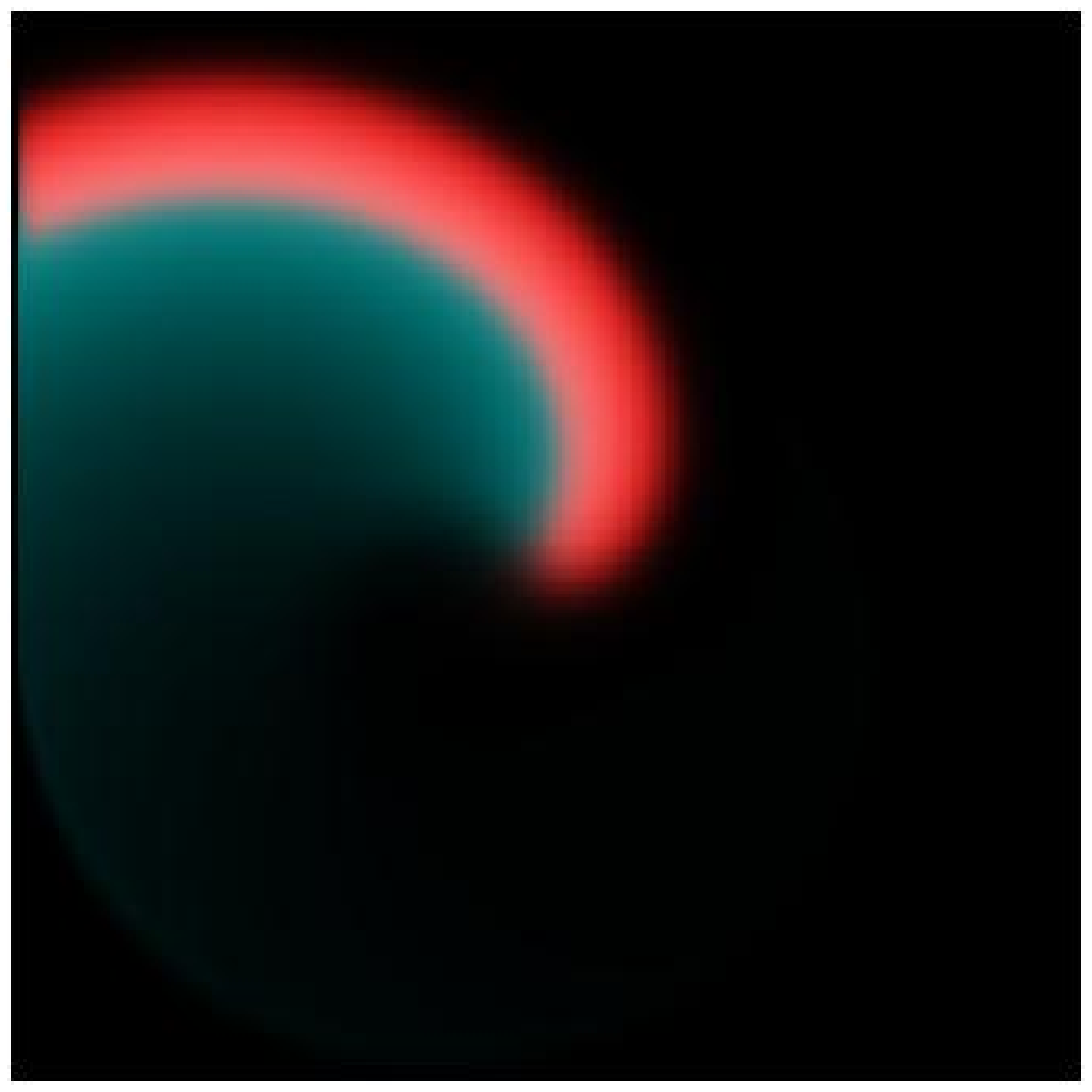}
\end{minipage}
\begin{minipage}{0.32\linewidth}
\centering
\includegraphics[width=0.7\textwidth]{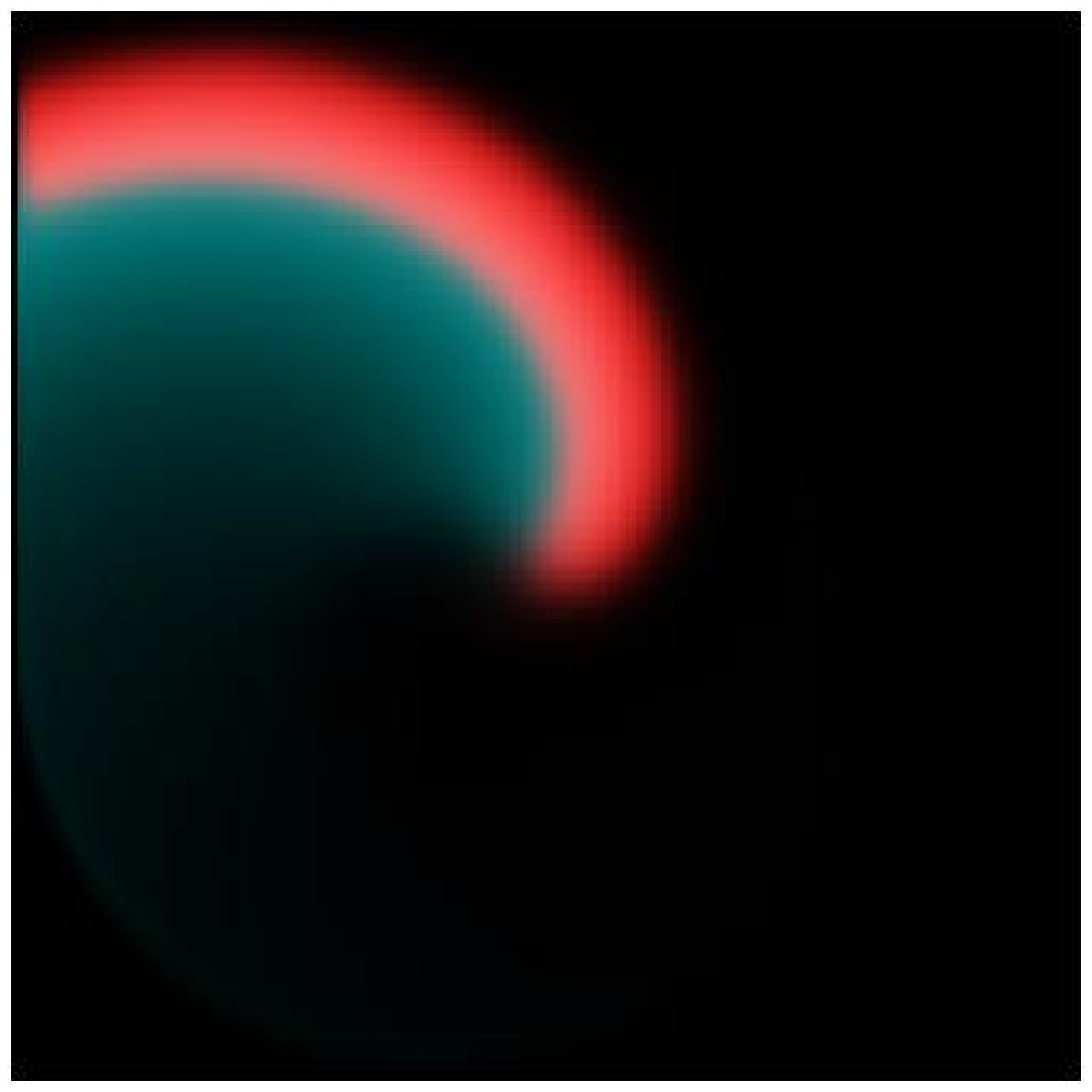}
\end{minipage}
\begin{minipage}{0.32\linewidth}
\centering
\includegraphics[width=0.7\textwidth]{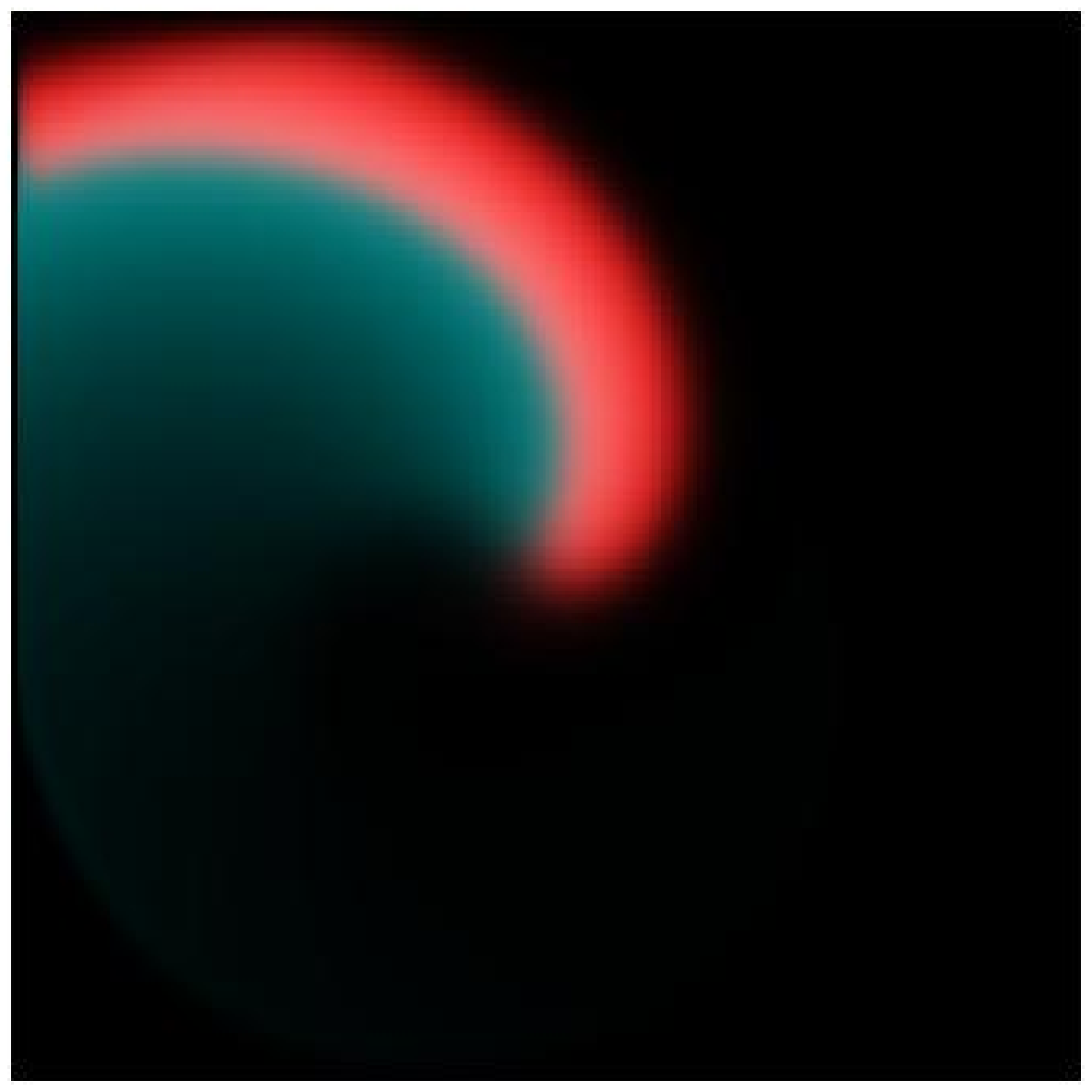}
\end{minipage}
\begin{minipage}{0.32\linewidth}
\centering
\includegraphics[width=0.7\textwidth]{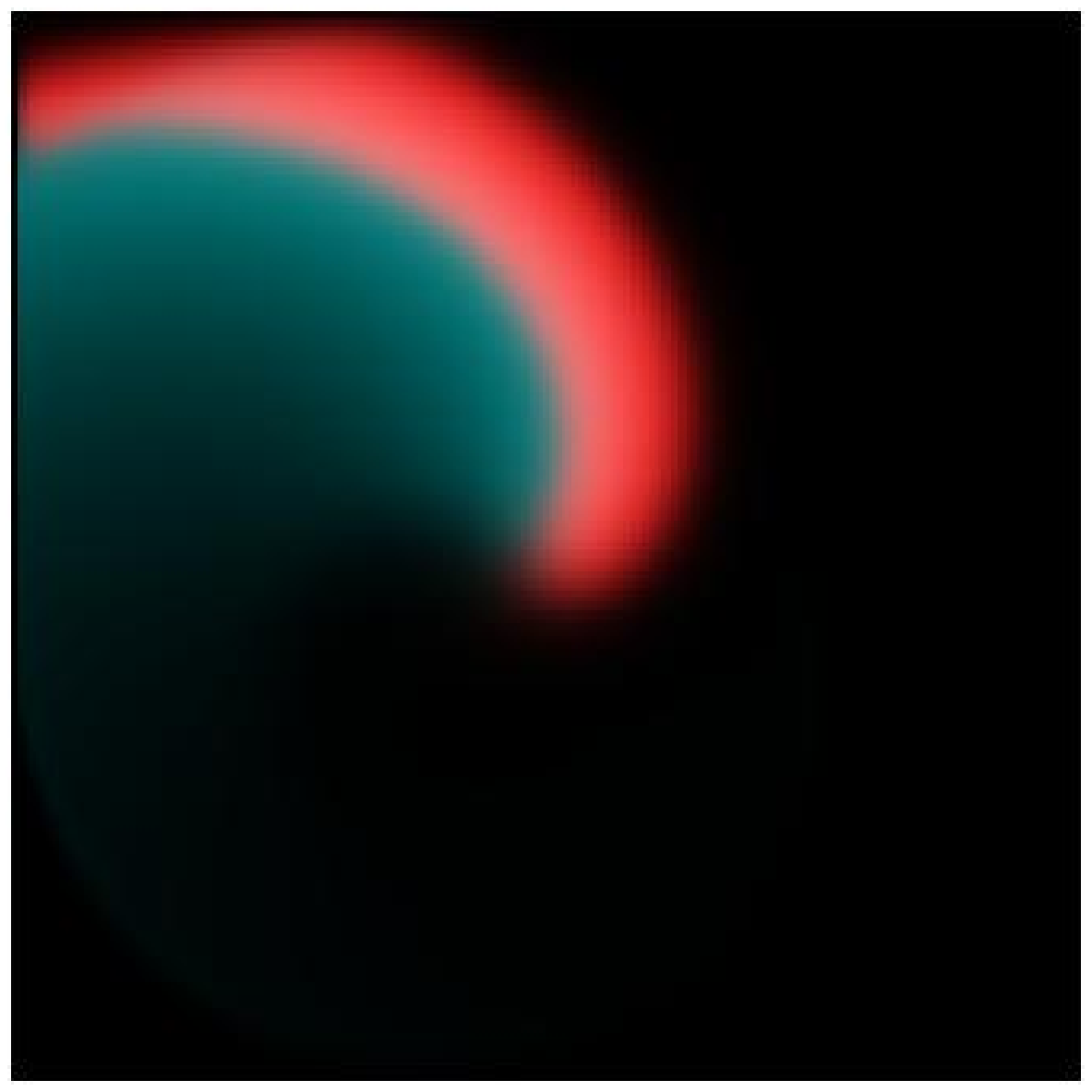}
\end{minipage}
\begin{minipage}{0.32\linewidth}
\centering
\includegraphics[width=0.7\textwidth]{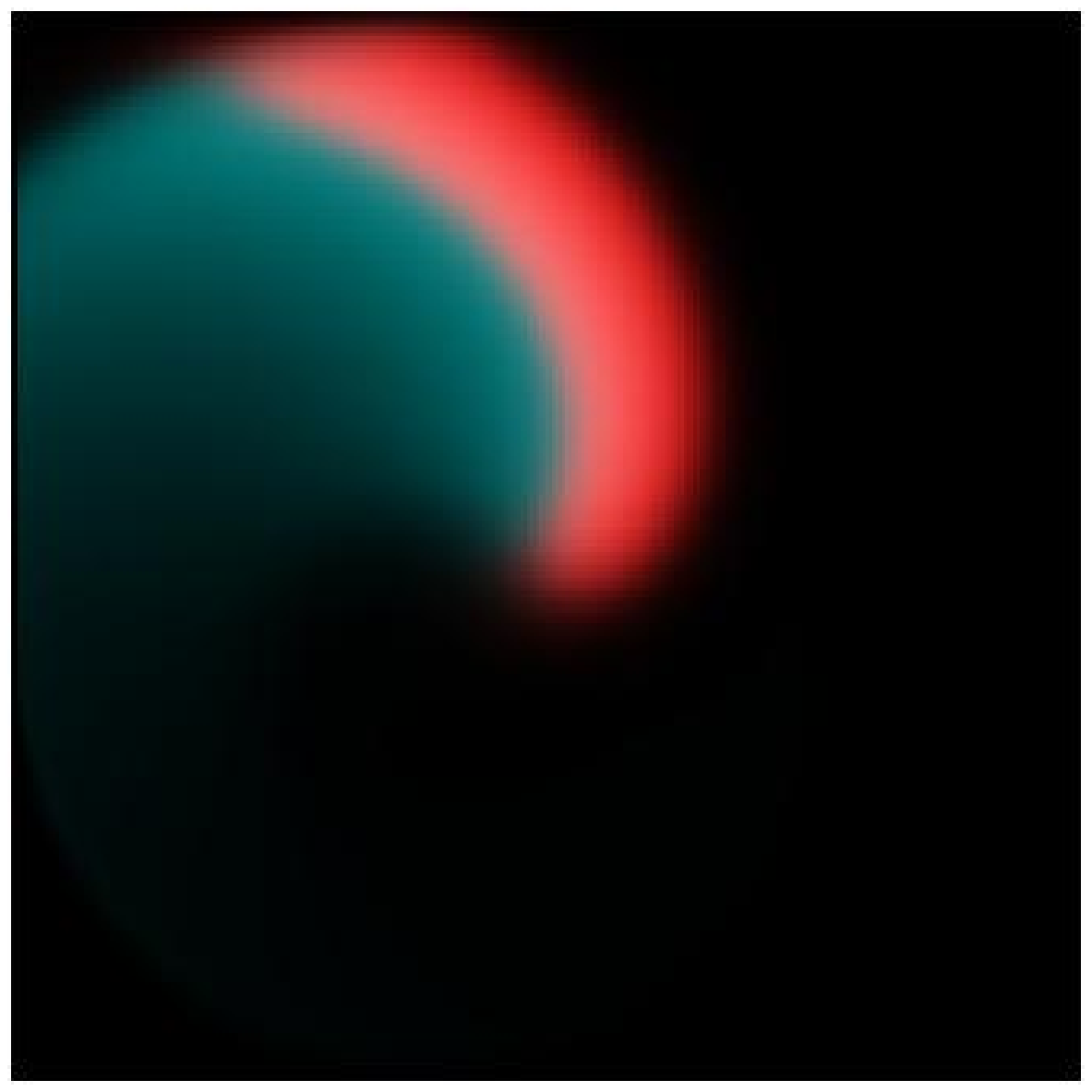}
\end{minipage}
\caption[Box size convergence: Dirichlet boundary conditions, final solutions]{Final Conditions for each run in the convergence testing of the box size in Barkley's model using Dirichlet boundary conditions, starting top left and working right, $L_X=60$ (top left) to $L_X=15$ (bottom right)}
\label{fig:ezf_conv_2_final_dbc}
\end{center}
\end{figure}

We also show the spiral wave solutions for each of the simulations within the tests in Figs.(\ref{fig:ezf_conv_2_final_nbc}) and (\ref{fig:ezf_conv_2_final_dbc}).

\clearpage

\clearpage

\clearpage

\clearpage


\subsubsection{Convergence in the timestep}

\begin{figure}[tbh]
\begin{center}
\begin{minipage}{0.6\linewidth}
\centering
\includegraphics[width=0.7\textwidth, angle=-90]{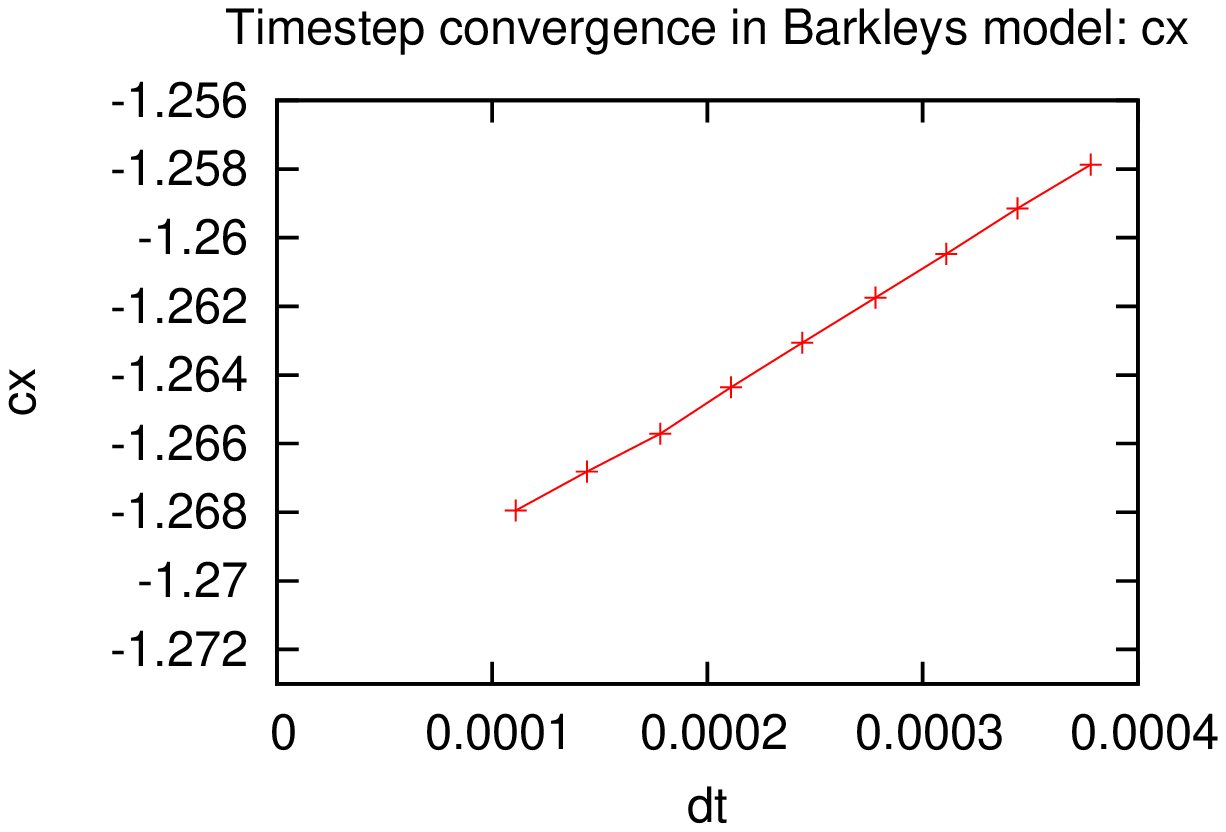}
\end{minipage}
\begin{minipage}{0.6\linewidth}
\centering
\includegraphics[width=0.7\textwidth, angle=-90]{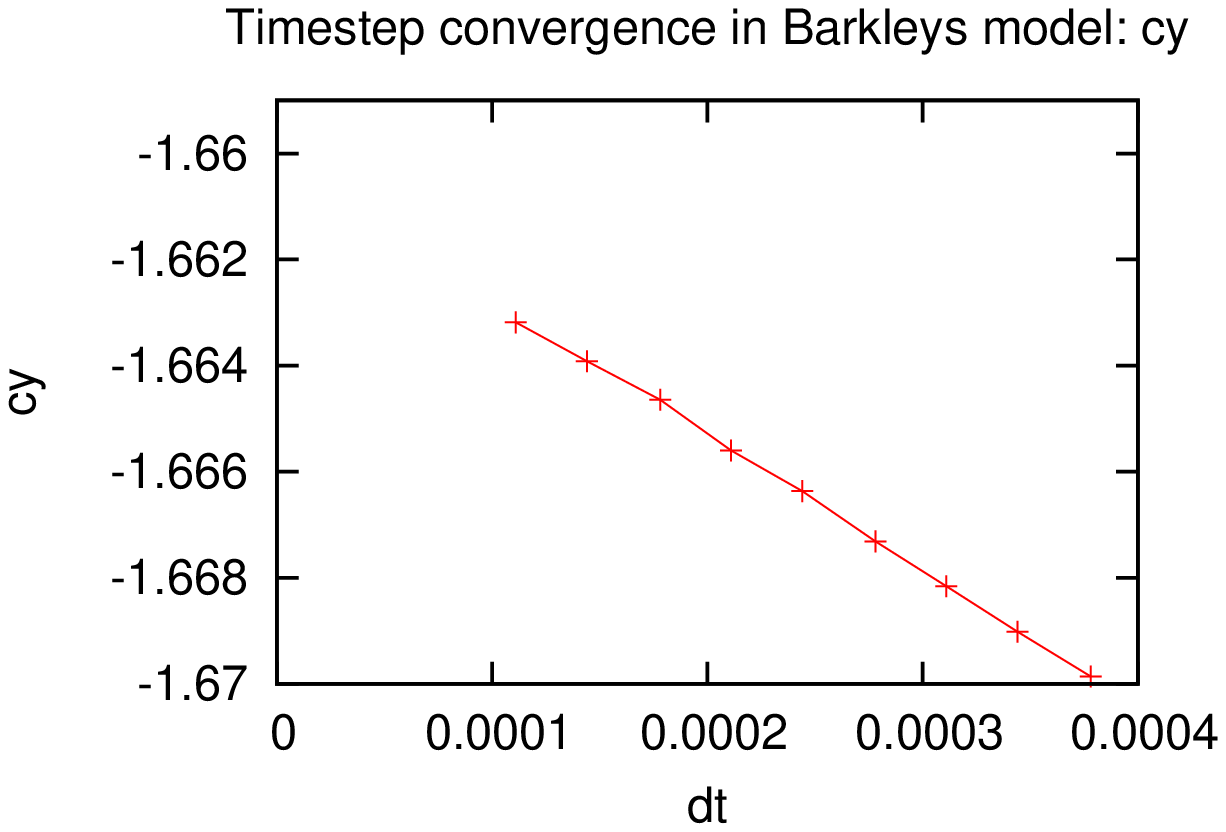}
\end{minipage}
\begin{minipage}{0.6\linewidth}
\centering
\includegraphics[width=0.7\textwidth, angle=-90]{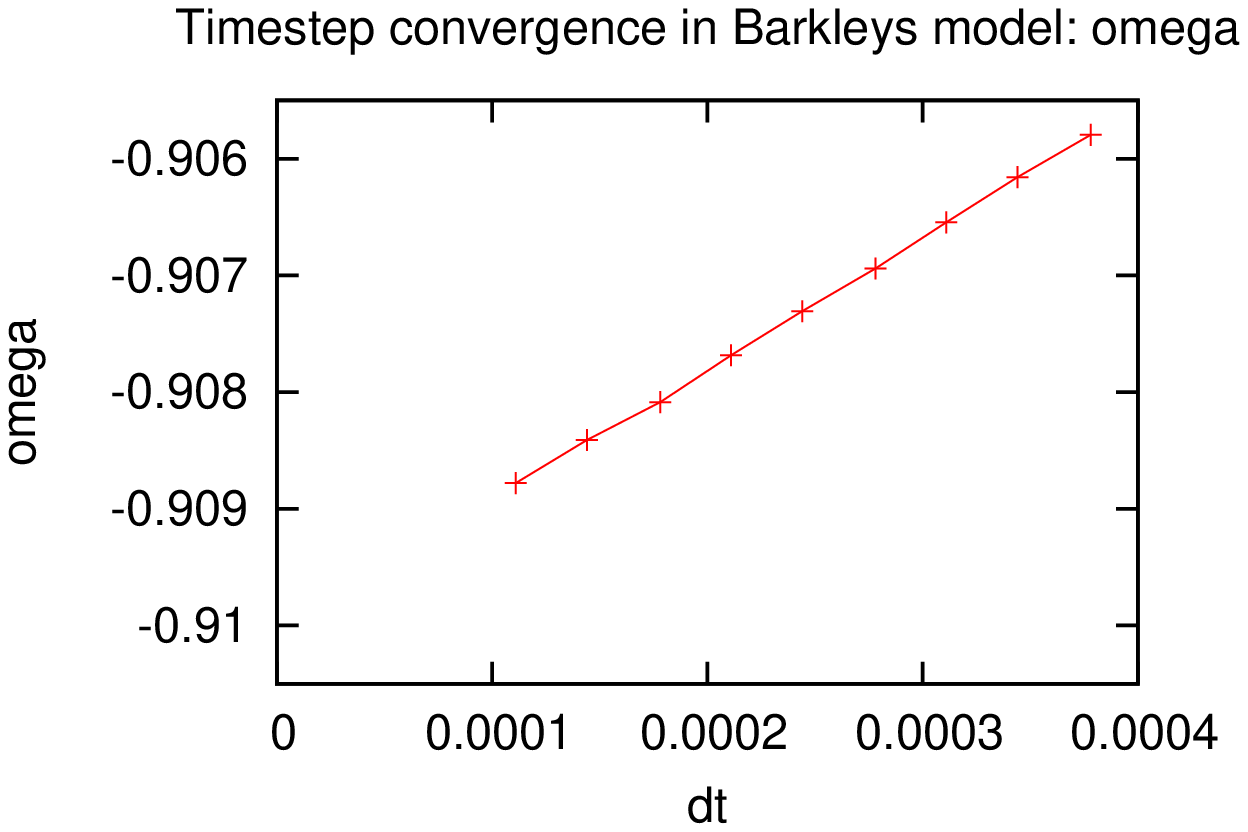}
\end{minipage}
\caption[Timestep convergence: Neumann boundary conditions]{Convergence in timestep, using Barkley's model and Neumann Boundary conditions with the spacestep fixed at $\Delta_x=\frac{1}{15}$, and the box size fixed at $L_X=60$}
\label{fig:ezf_conv_3_nbc}
\end{center}
\end{figure}

Finally for Barkley's model, we consider the convergence in the timestep. The results for Neumann boundary conditions are shown in Fig.(\ref{fig:ezf_conv_3_nbc}). We also show the results for the simulations using Dirichlet boundary conditions in Fig.(\ref{fig:ezf_conv_3_dbc}).

In this case, we kept the following parameters constant:

\begin{itemize}
 \item spacestep, $\Delta_x=\frac{1}{15}$
 \item box size, $L_X=60$ s.u.
\end{itemize}

Starting from $t_s=0.1$, we ran a number of simulations until the instabilities were too great. In each simulation we increased the value of $t_s$ by 0.03 each time.

We can see that there is a linear relationship between the advection coefficients and the timestep, which is expected due to the time derivatives being numerically approximated to first order in the timestep. 

It is also clear again that the choice of boundary conditions is irrelevant, since we get the same results no matter what conditions we use.

\clearpage

\begin{figure}[tbh]
\begin{center}
\begin{minipage}{0.32\linewidth}
\centering
\includegraphics[width=0.7\textwidth]{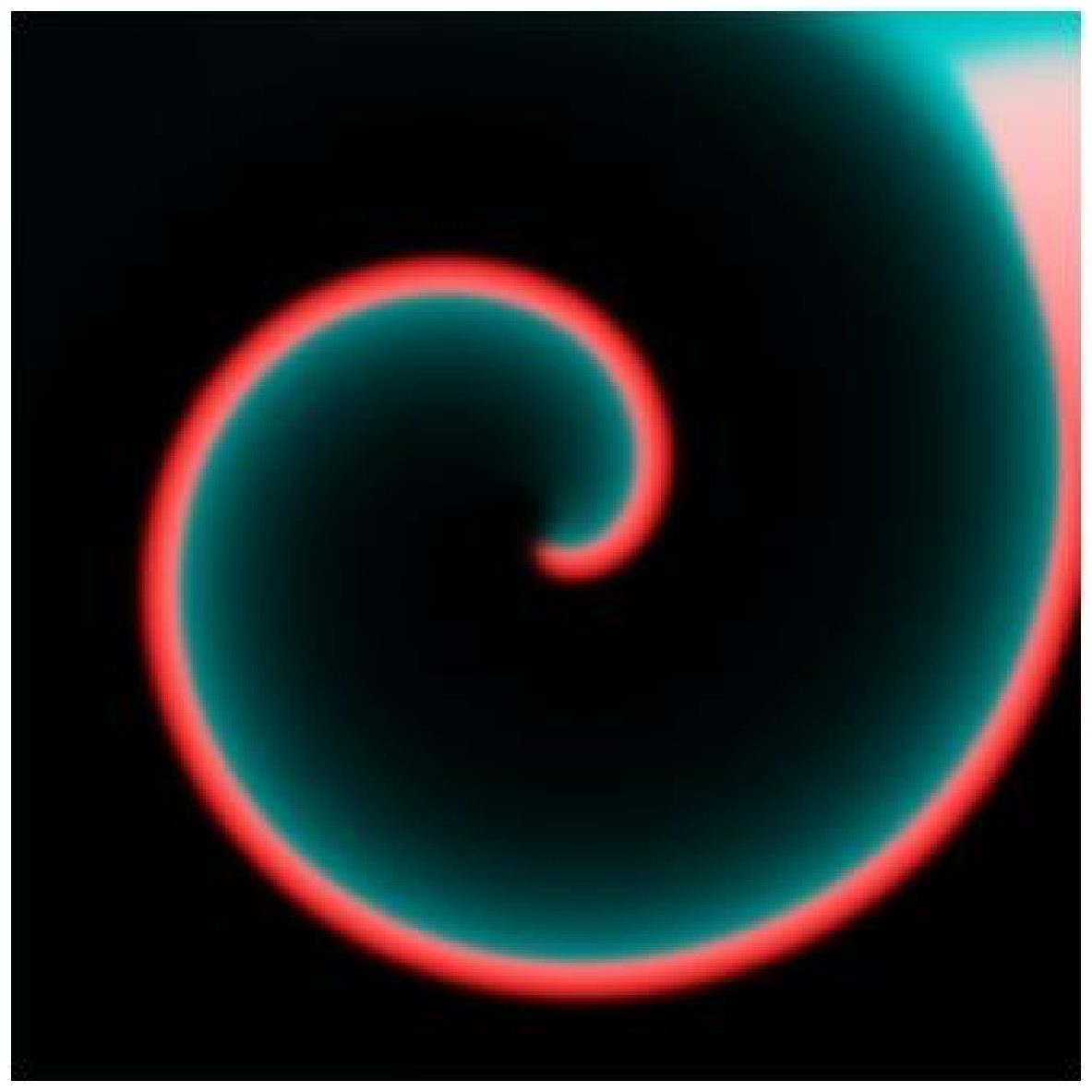}
\end{minipage}
\begin{minipage}{0.32\linewidth}
\centering
\includegraphics[width=0.7\textwidth]{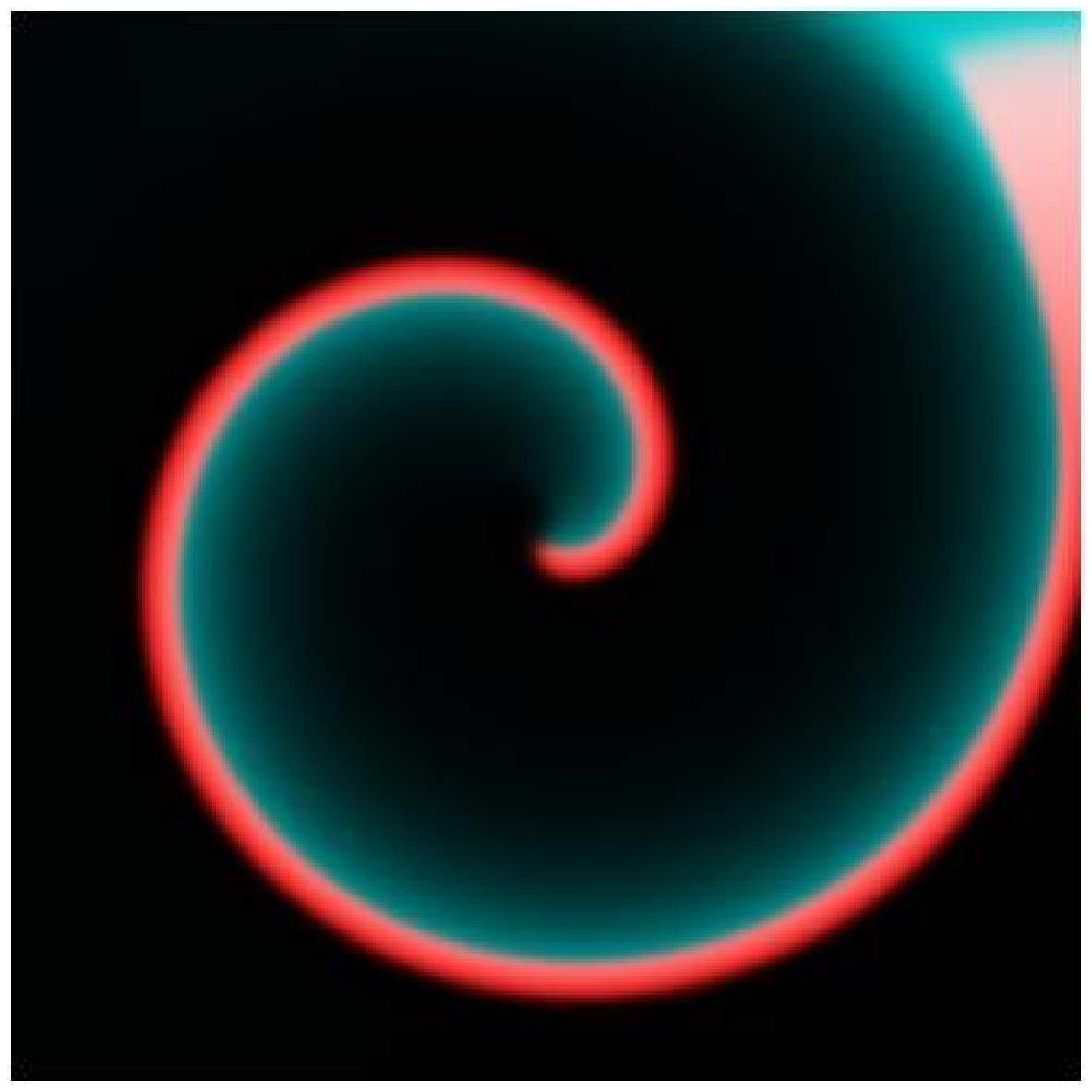}
\end{minipage}
\begin{minipage}{0.32\linewidth}
\centering
\includegraphics[width=0.7\textwidth]{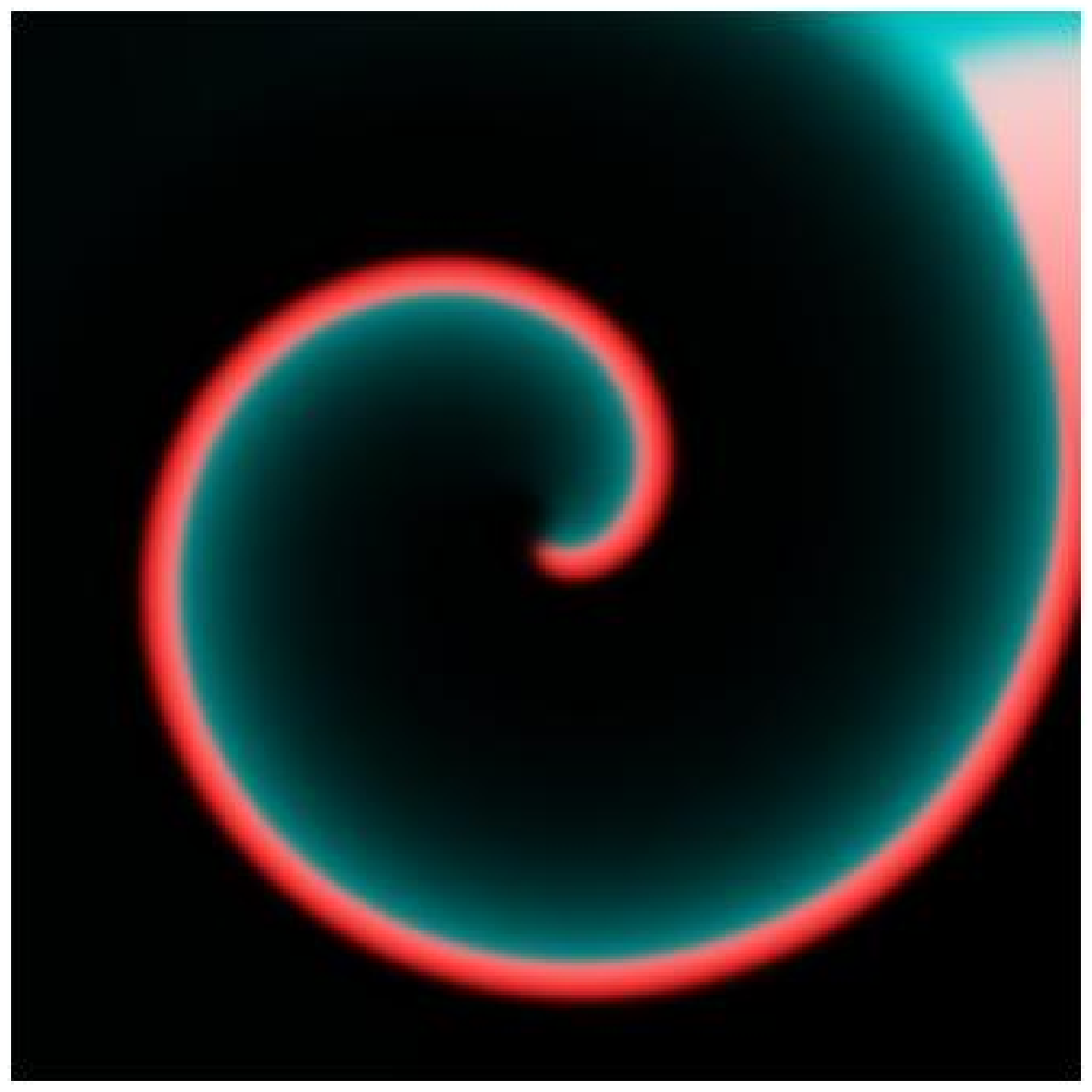}
\end{minipage}
\begin{minipage}{0.32\linewidth}
\centering
\includegraphics[width=0.7\textwidth]{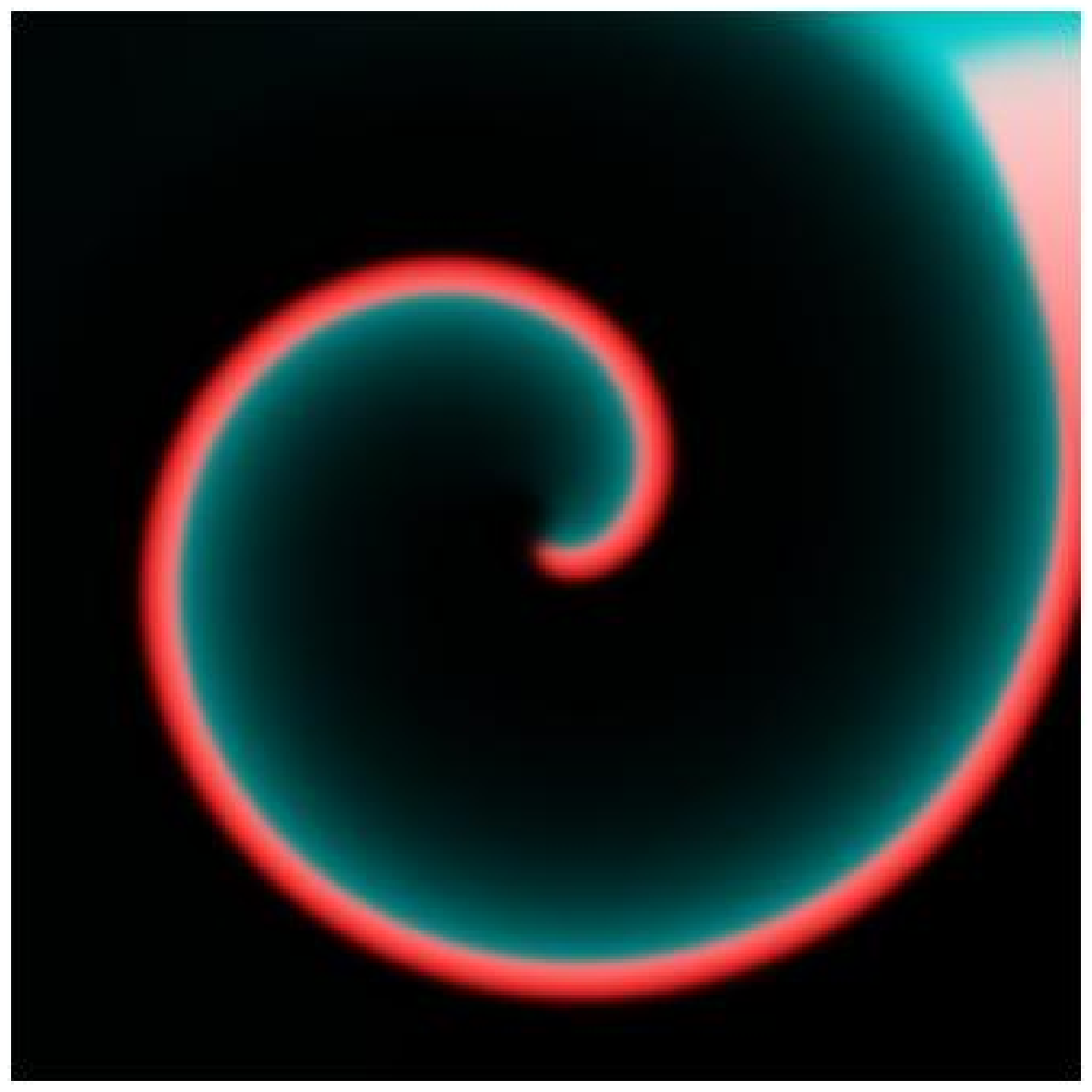}
\end{minipage}
\begin{minipage}{0.32\linewidth}
\centering
\includegraphics[width=0.7\textwidth]{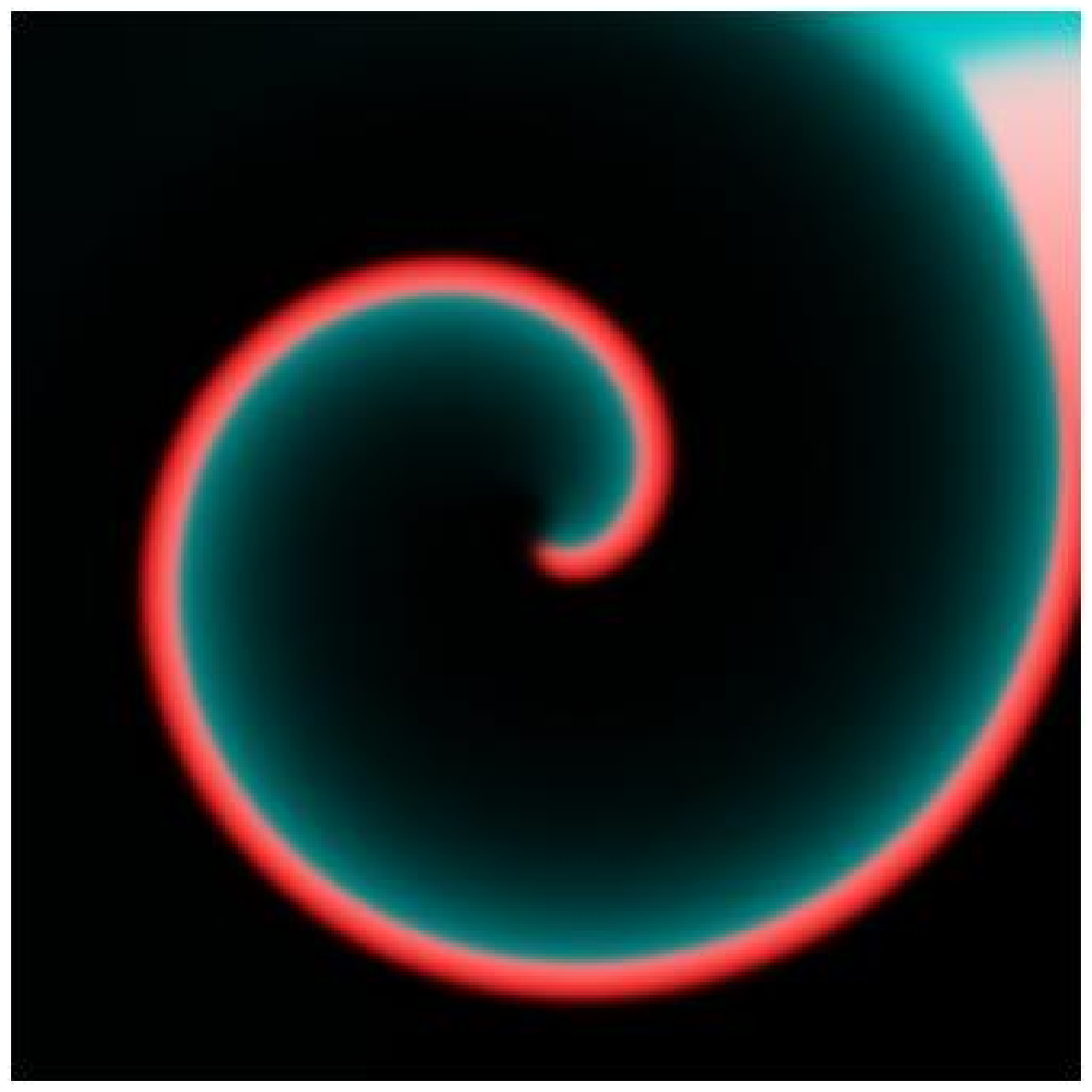}
\end{minipage}
\begin{minipage}{0.32\linewidth}
\centering
\includegraphics[width=0.7\textwidth]{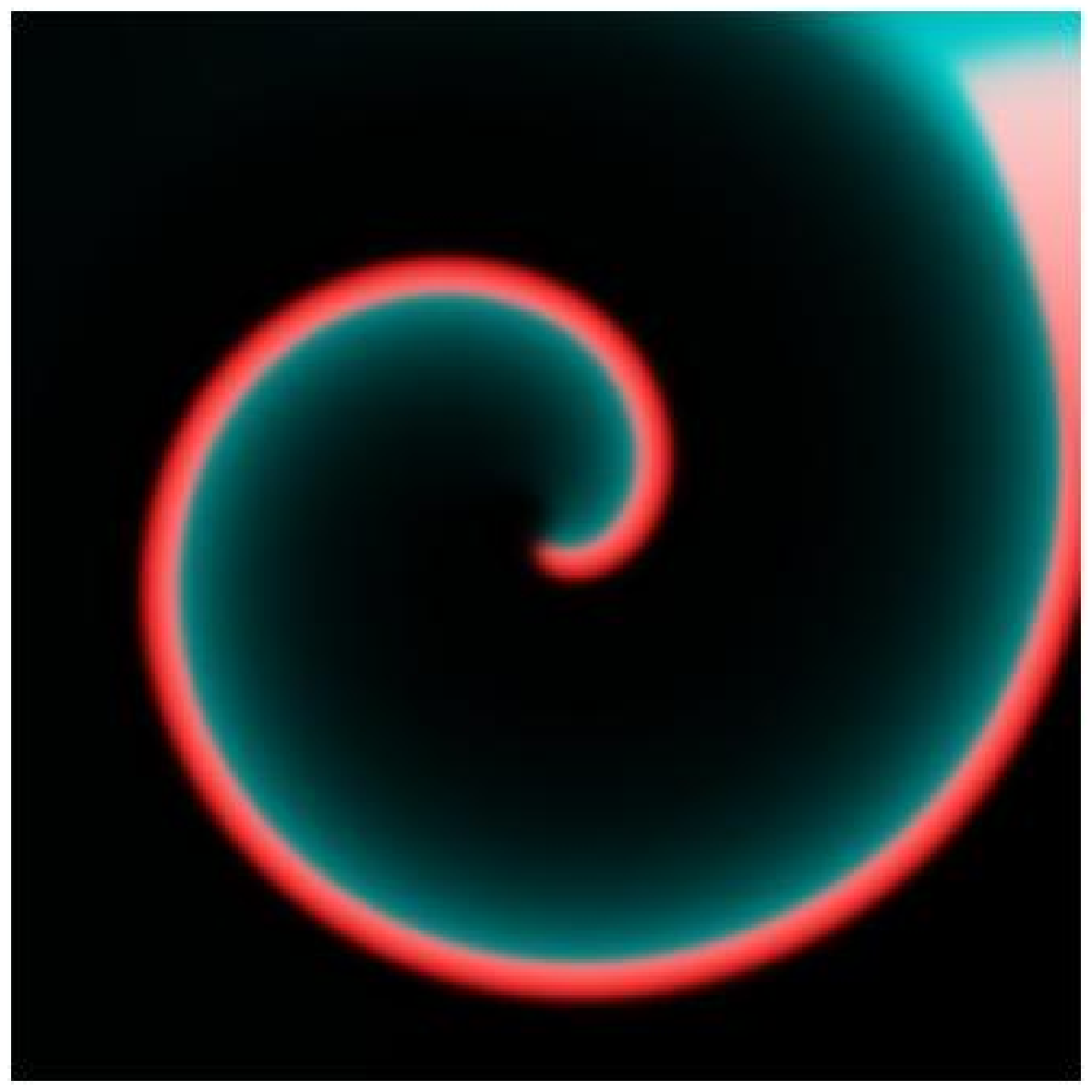}
\end{minipage}
\begin{minipage}{0.32\linewidth}
\centering
\includegraphics[width=0.7\textwidth]{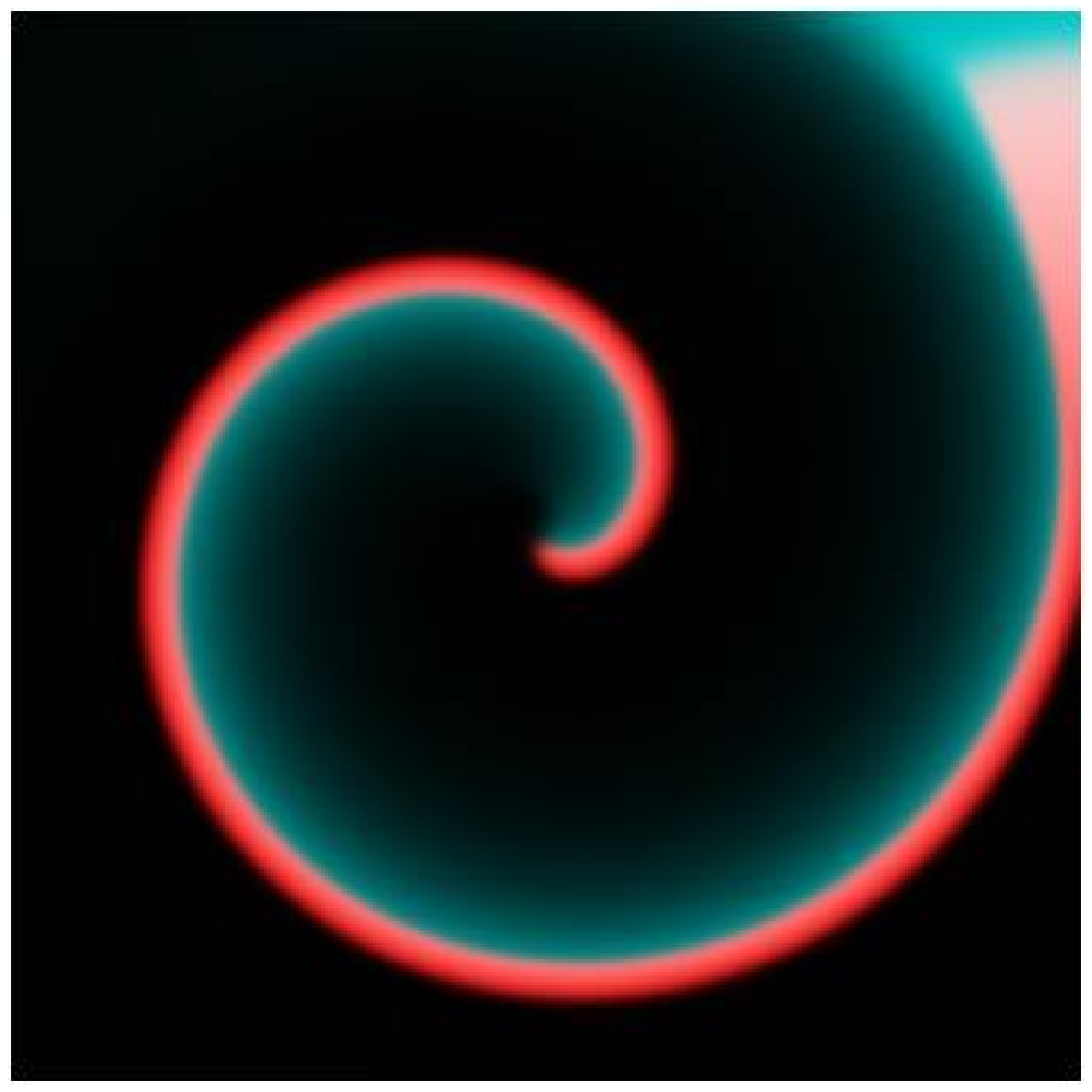}
\end{minipage}
\begin{minipage}{0.32\linewidth}
\centering
\includegraphics[width=0.7\textwidth]{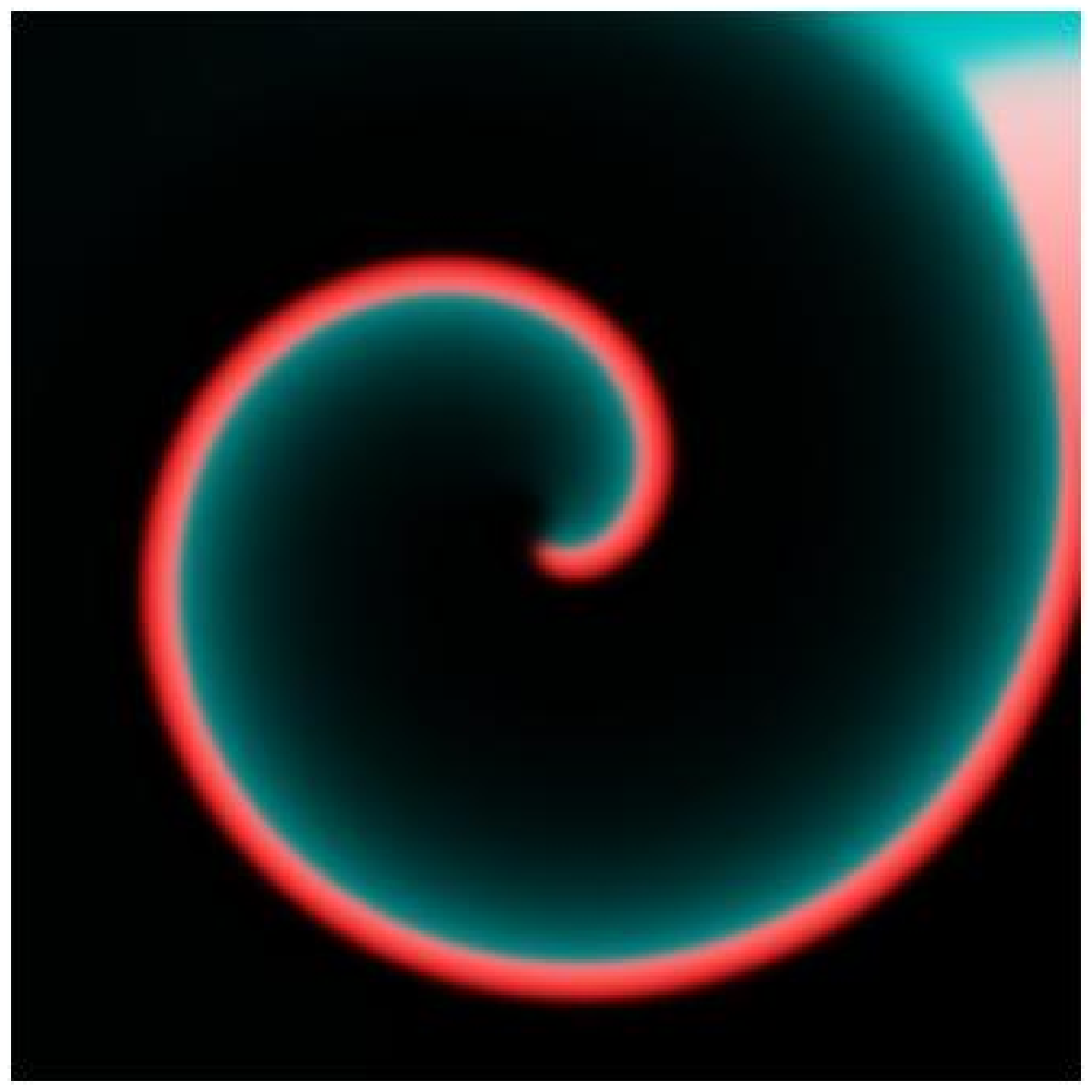}
\end{minipage}
\begin{minipage}{0.32\linewidth}
\centering
\includegraphics[width=0.7\textwidth]{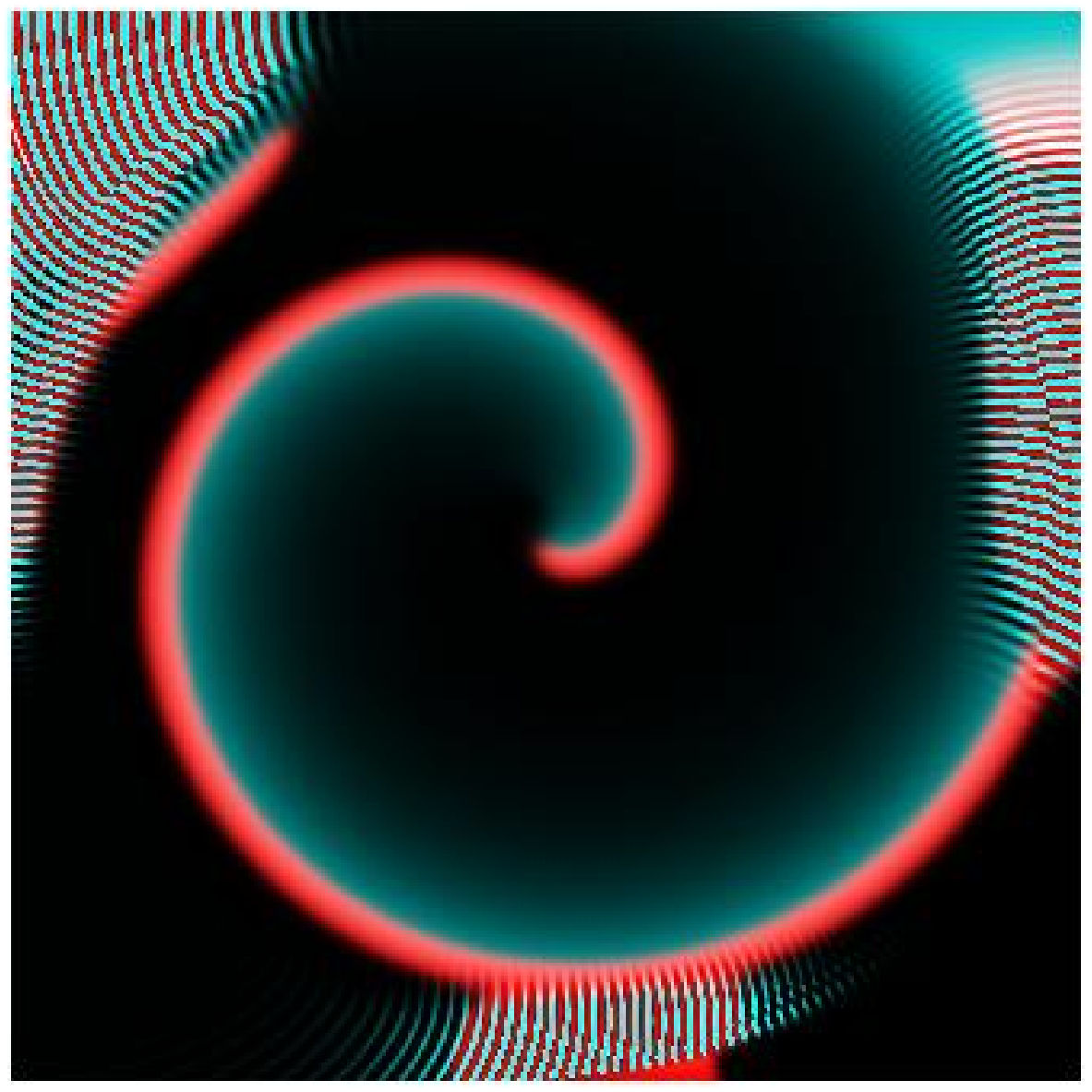}
\end{minipage}
\caption[Timestep convergence: Neumann boundary conditions, final solutions]{Final Conditions for each run in the convergence testing of the timestep in Barkley's model using Neumann boundary conditions, starting top left and working right, $t_s=0.1$ (top left) to $t_s=0.34$ (bottom right)}
\label{fig:ezf_conv_3_final_nbc}
\end{center}
\end{figure}

\clearpage

\begin{figure}[tbh]
\begin{center}
\begin{minipage}{0.6\linewidth}
\centering
\includegraphics[width=0.7\textwidth, angle=-90]{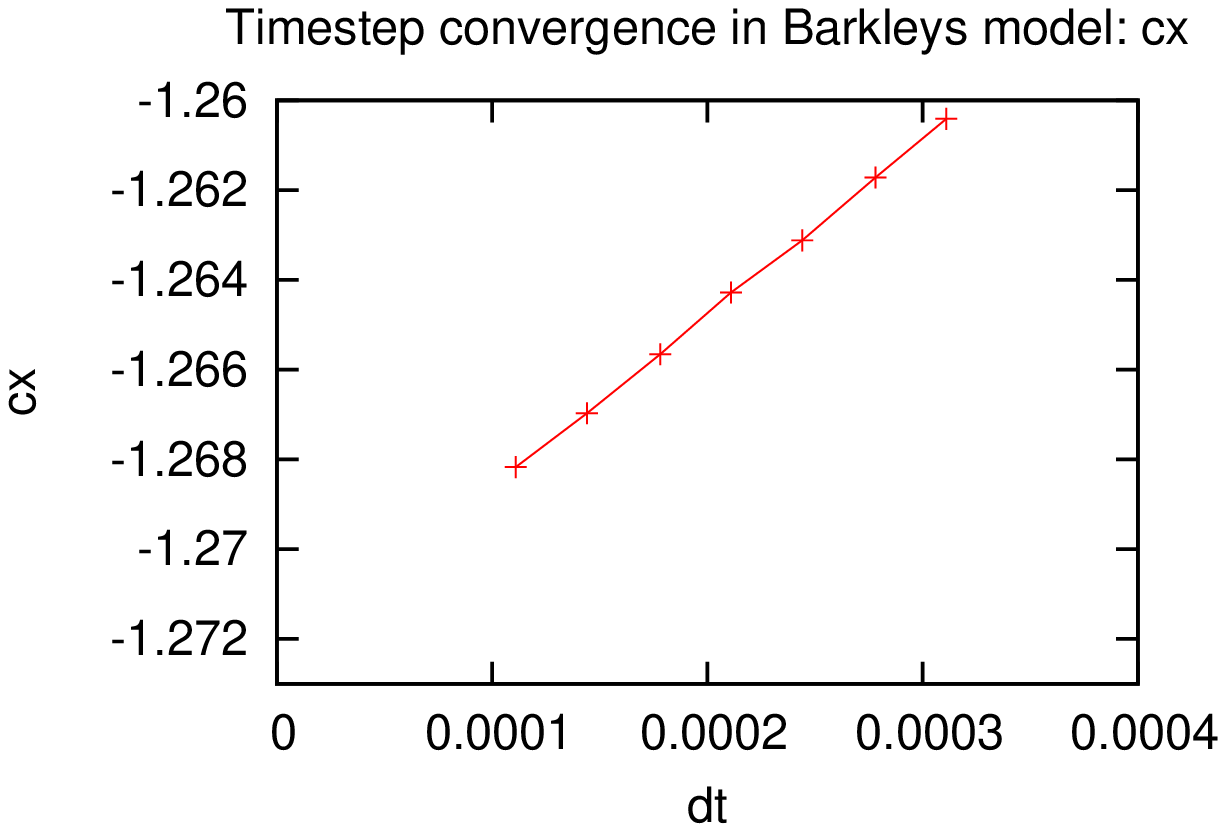}
\end{minipage}
\begin{minipage}{0.6\linewidth}
\centering
\includegraphics[width=0.7\textwidth, angle=-90]{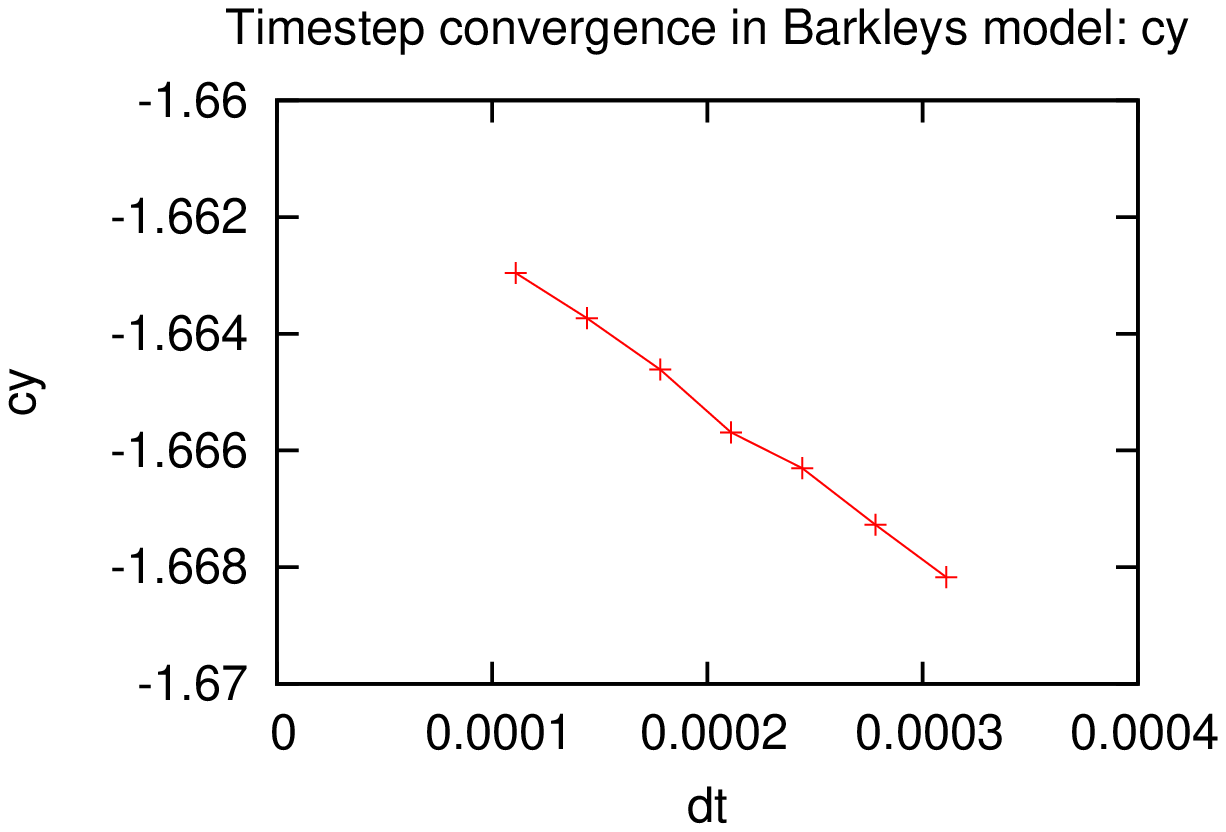}
\end{minipage}
\begin{minipage}{0.6\linewidth}
\centering
\includegraphics[width=0.7\textwidth, angle=-90]{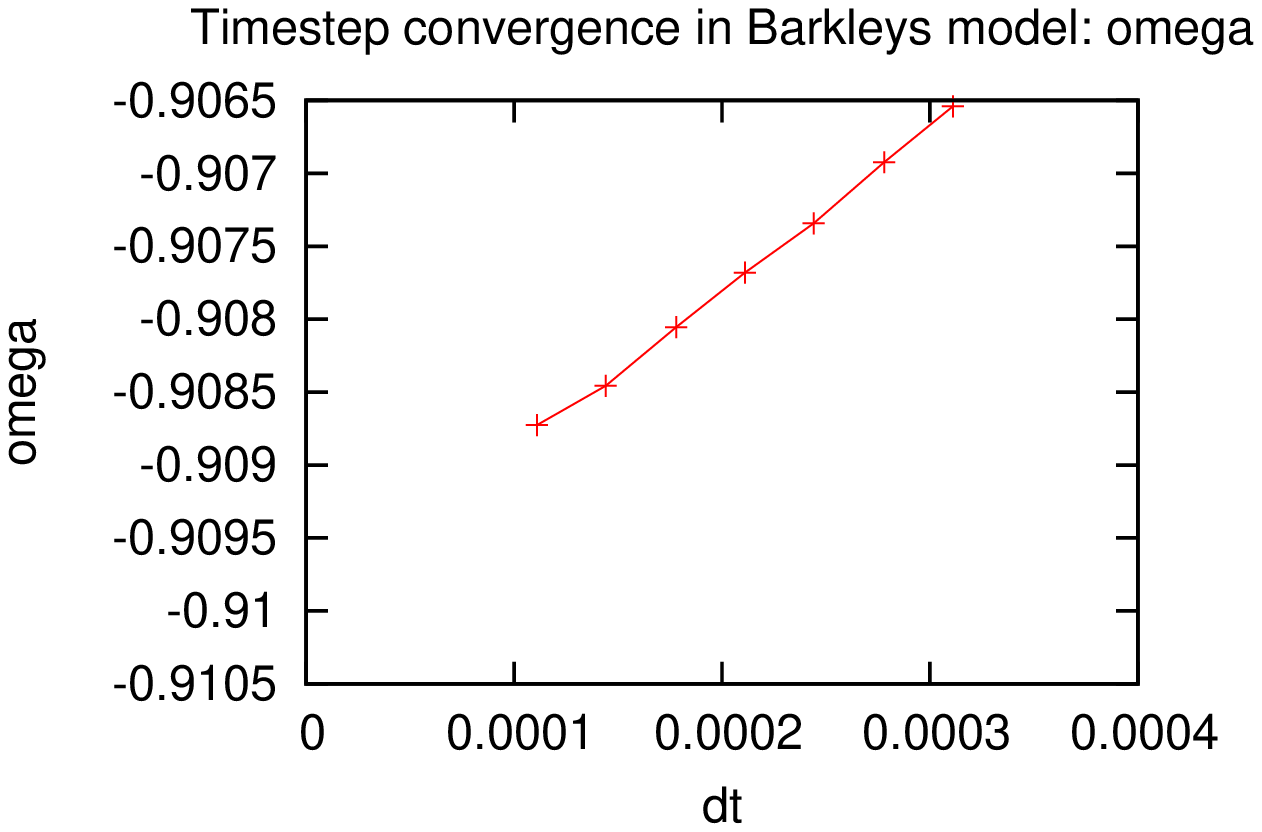}
\end{minipage}
\caption[Timestep convergence: Dirichlet boundary conditions]{Convergence in box size, using Barkley's model and Dirichlet Boundary conditions with the spacestep fixed at $\Delta_x=\frac{1}{15}$, and the timestep per diffusion stability limit fixed at $t_s=0.1$}
\label{fig:ezf_conv_3_dbc}
\end{center}
\end{figure}

\clearpage

\begin{figure}[tbh]
\begin{center}
\begin{minipage}{0.32\linewidth}
\centering
\includegraphics[width=0.7\textwidth]{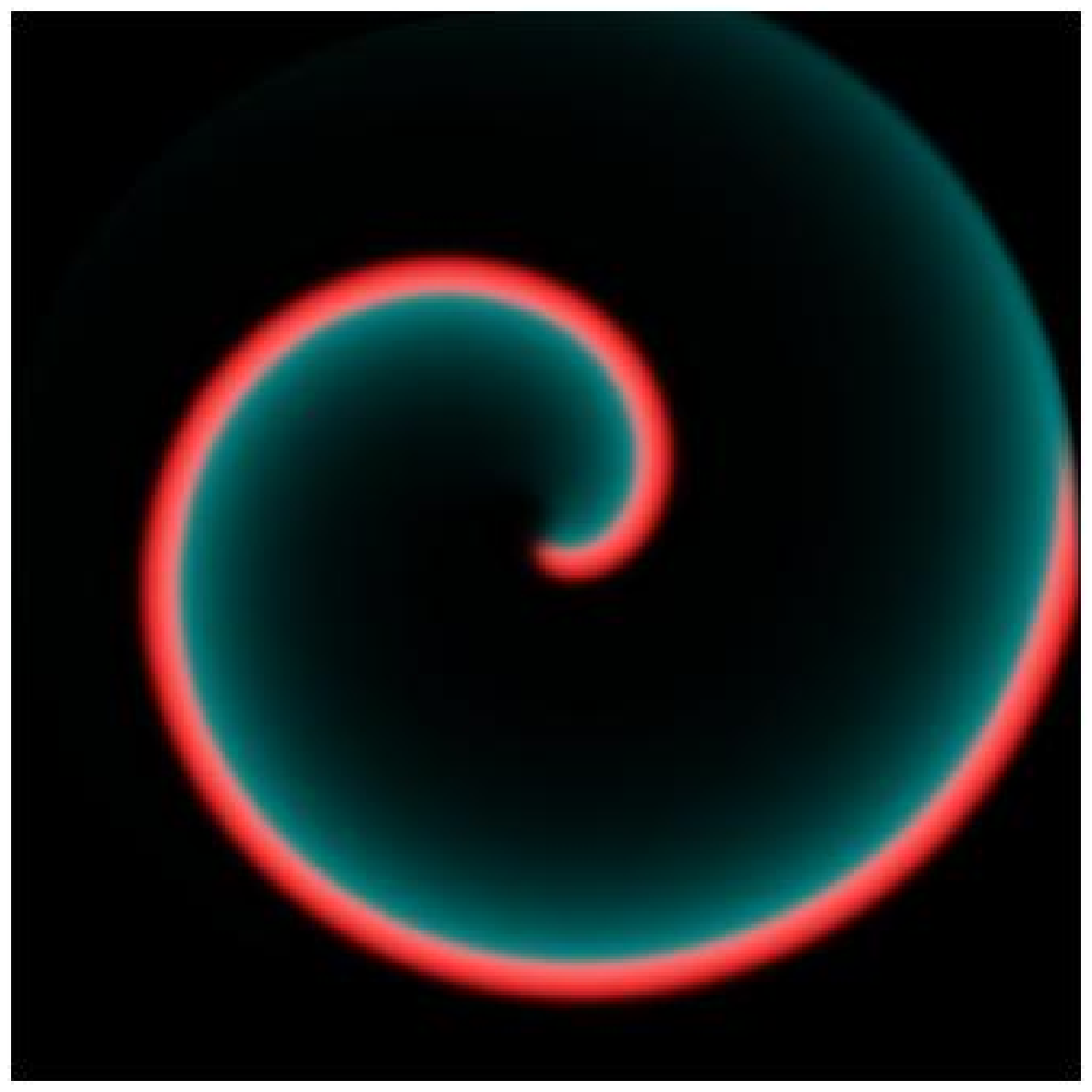}
\end{minipage}
\begin{minipage}{0.32\linewidth}
\centering
\includegraphics[width=0.7\textwidth]{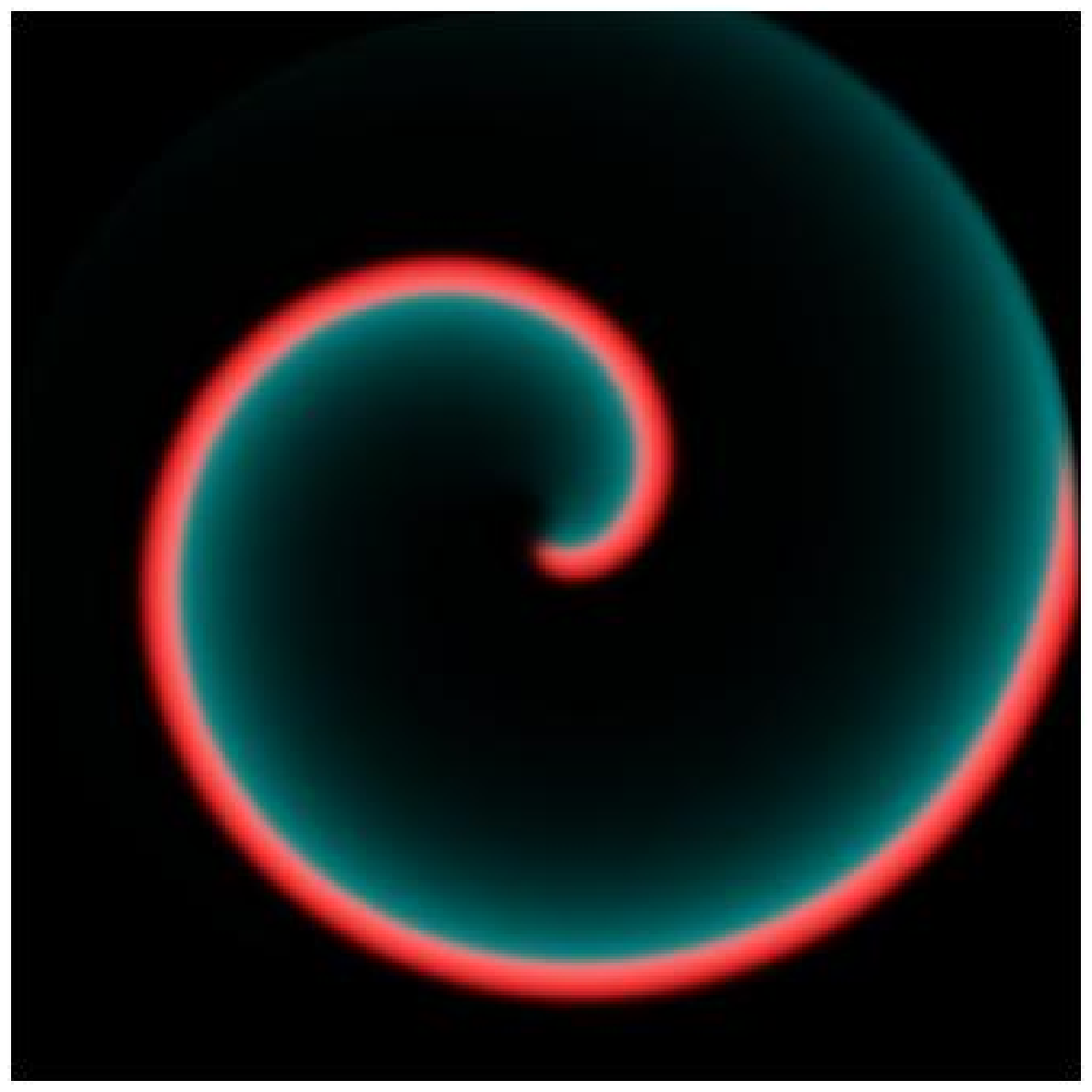}
\end{minipage}
\begin{minipage}{0.32\linewidth}
\centering
\includegraphics[width=0.7\textwidth]{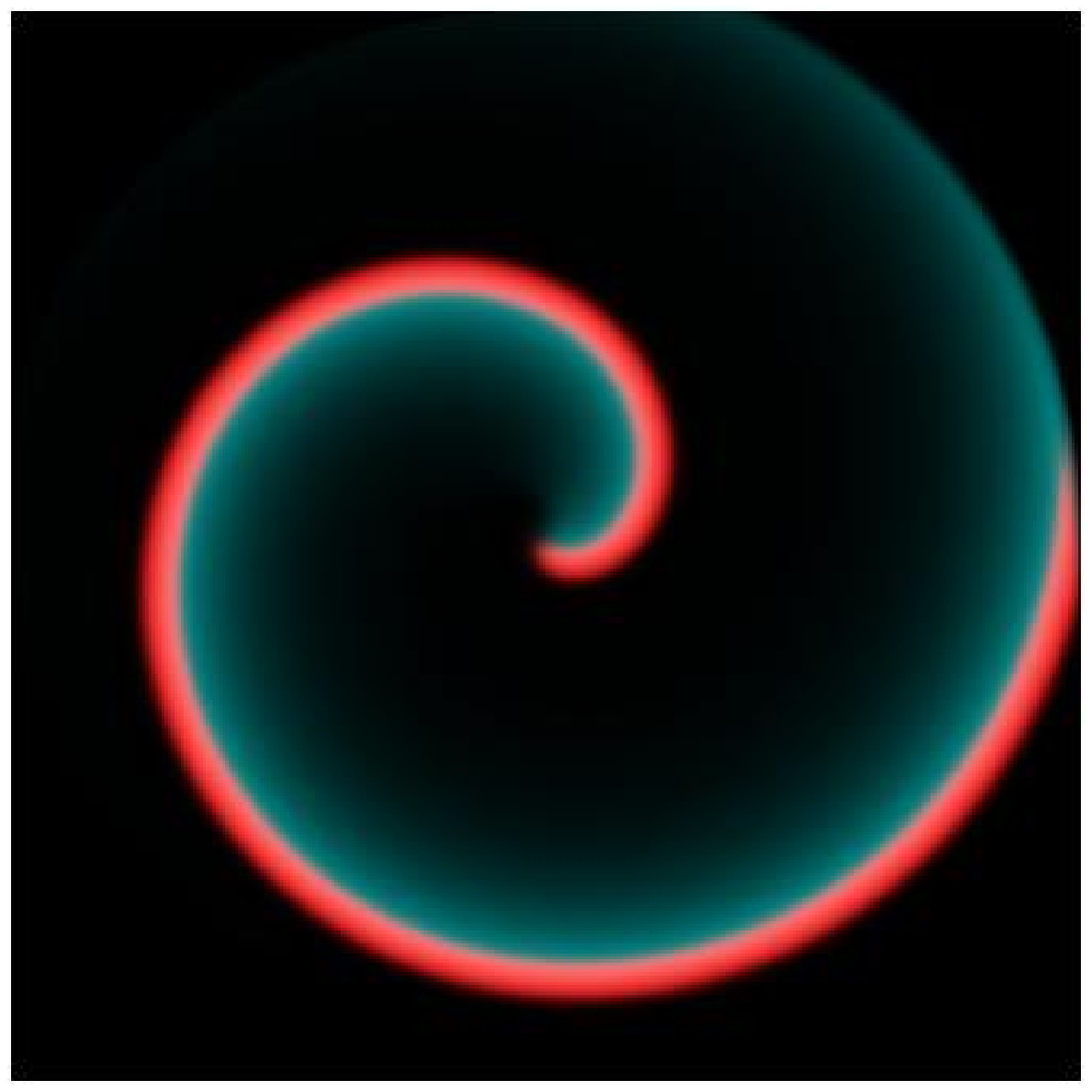}
\end{minipage}
\begin{minipage}{0.32\linewidth}
\centering
\includegraphics[width=0.7\textwidth]{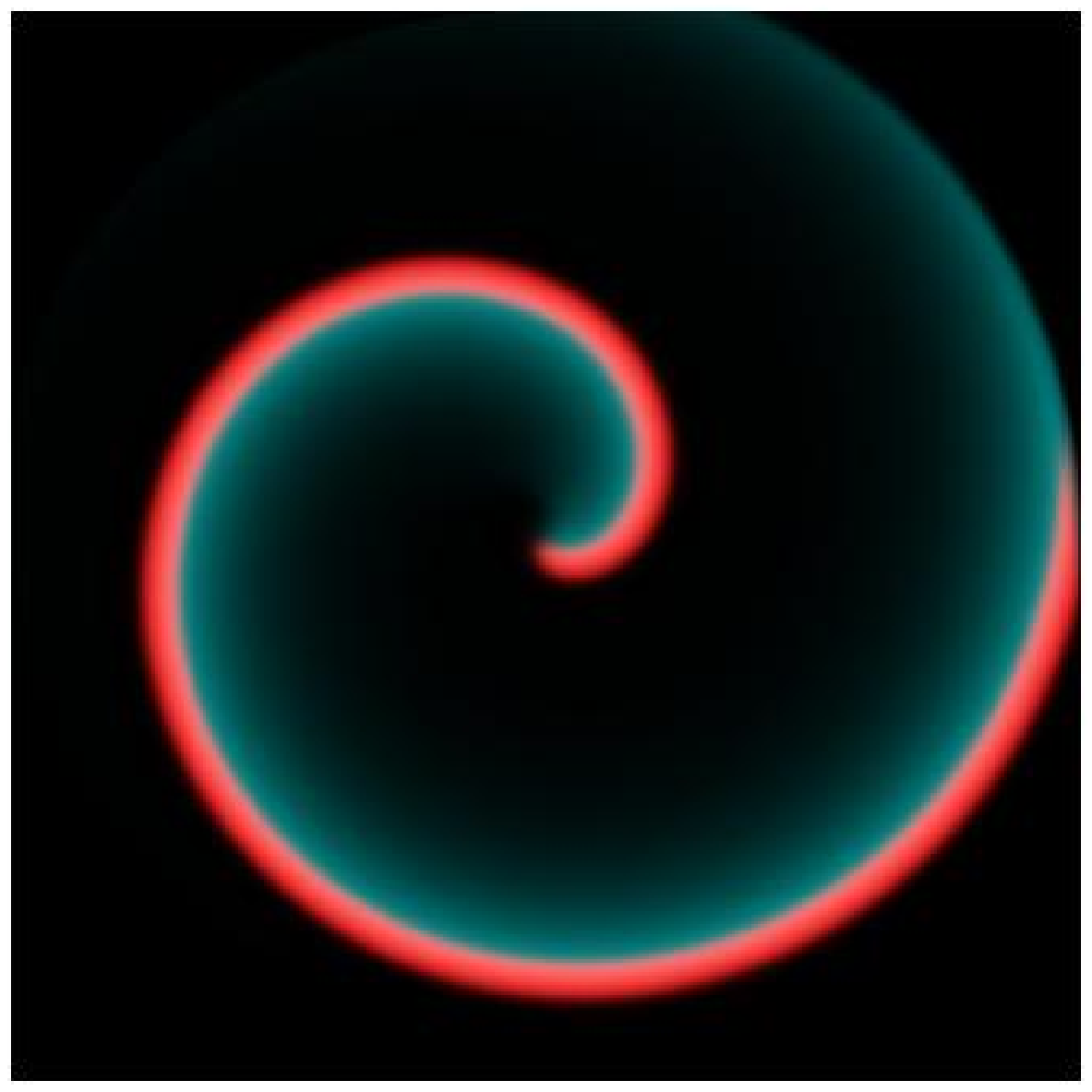}
\end{minipage}
\begin{minipage}{0.32\linewidth}
\centering
\includegraphics[width=0.7\textwidth]{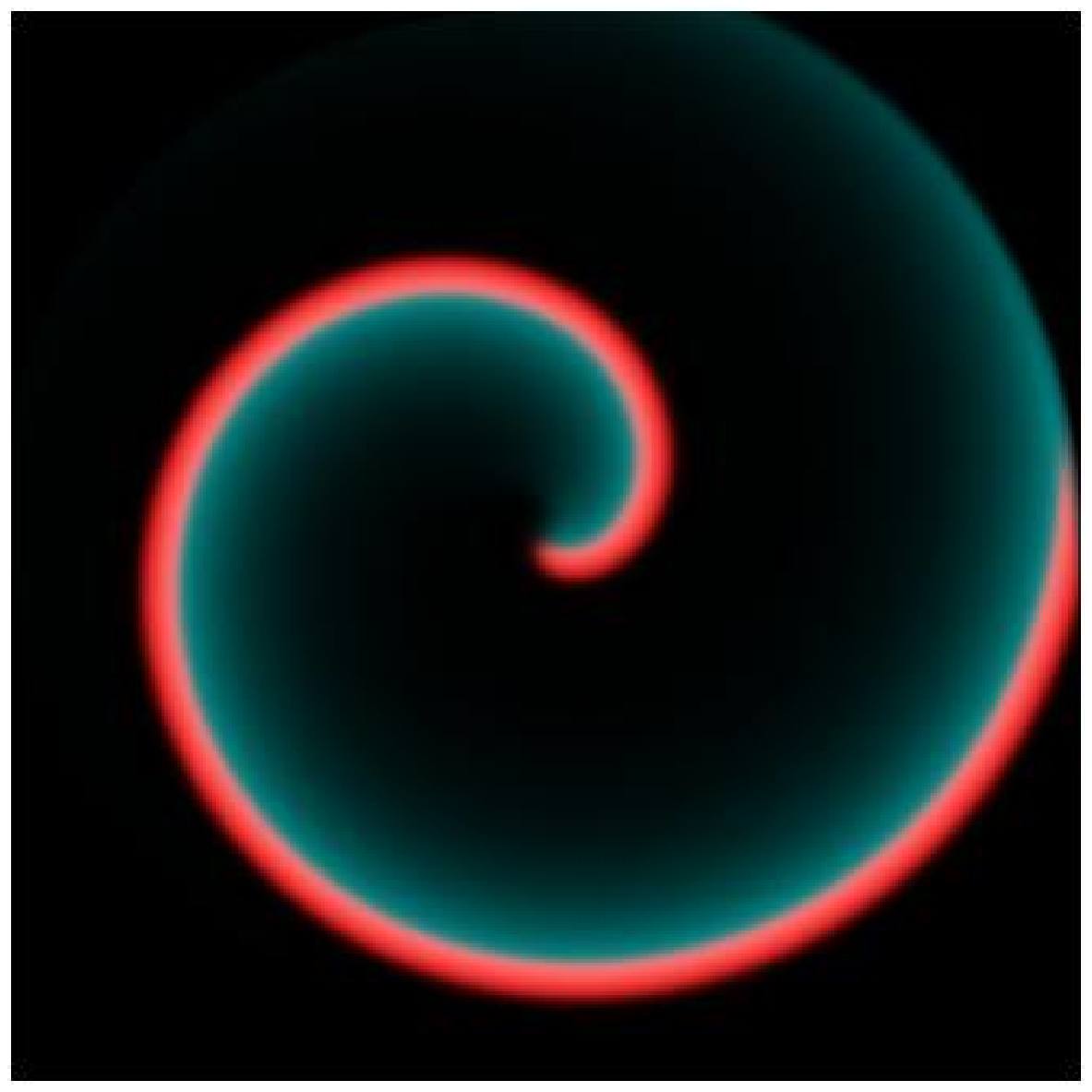}
\end{minipage}
\begin{minipage}{0.32\linewidth}
\centering
\includegraphics[width=0.7\textwidth]{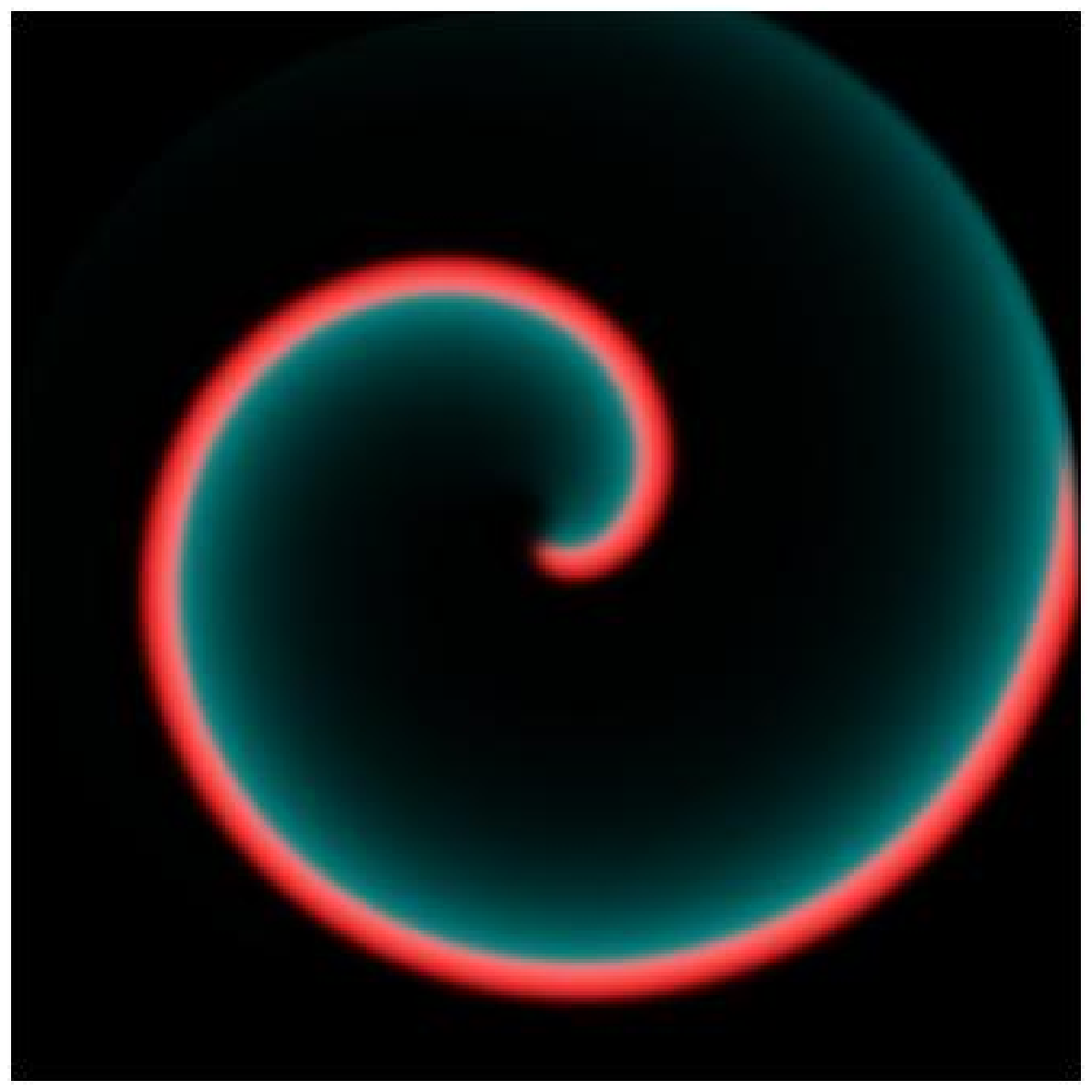}
\end{minipage}
\begin{minipage}{0.32\linewidth}
\centering
\includegraphics[width=0.7\textwidth]{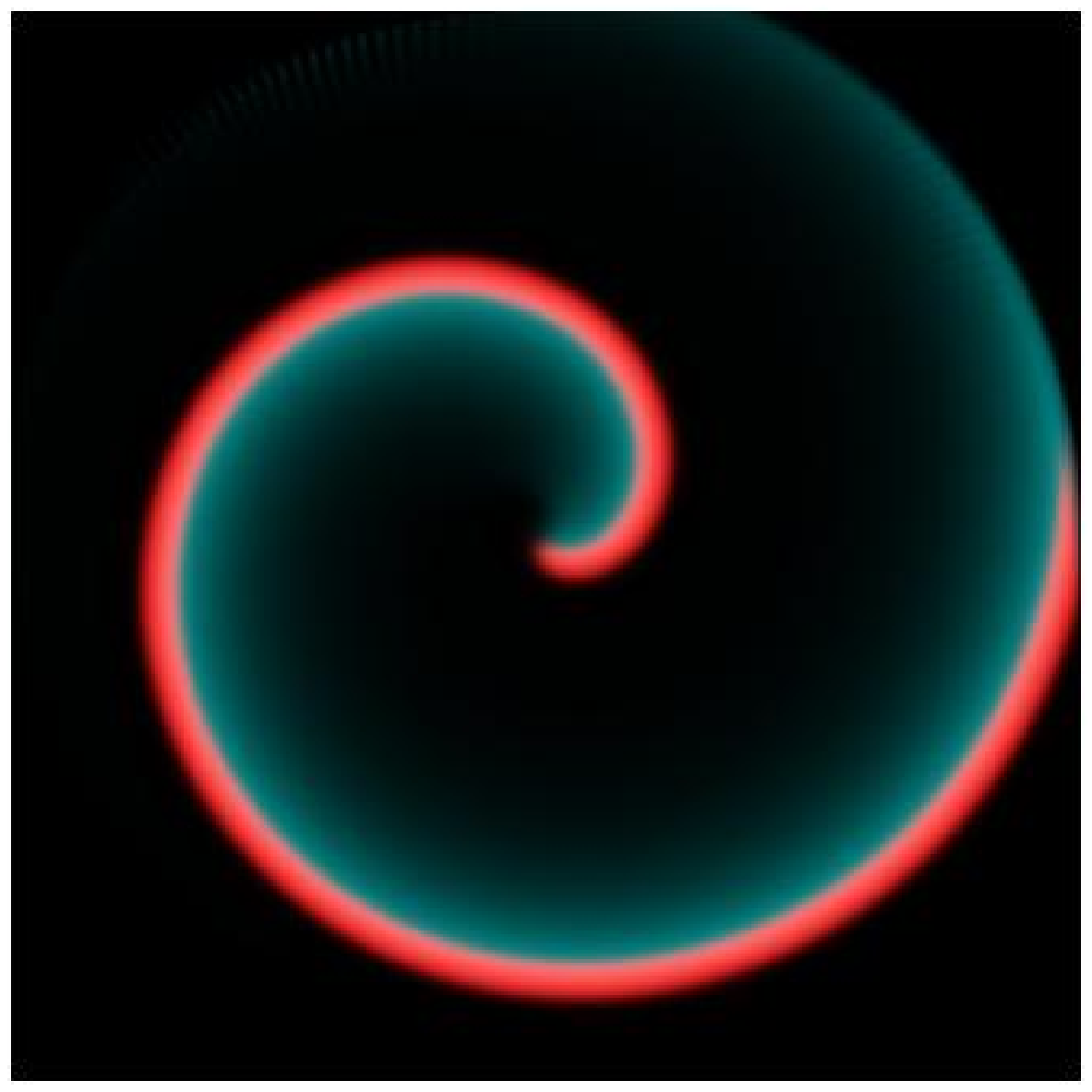}
\end{minipage}
\caption[Timestep convergence: Dirichlet boundary conditions, final solutions]{Final Conditions for each run in the convergence testing of the timestep in Barkley's model using Dirichlet boundary conditions, starting top left and working right, $t_s=0.1$ (top left) to $t_s=0.28$ (bottom right)}
\label{fig:ezf_conv_3_final_dbc}
\end{center}
\end{figure}

\clearpage


\subsection{Results \& Conclusions}

To summarize, we have shown that the convergence testing in Barkley's model using the best numerical methods within EZ-Freeze (i.e. second order accuracy in the spatial derivatives and using method 2) shows that the numerical methods used are working as they should. The convergence in spacestep is a quadratic relationship, as we would expect with a second order scheme. The convergence in the time step produces a linear relationship, which, yet again, is as expected. And finally, the convergence in the box \chg[p172gram]{size} shows that the advection coefficients converge to a value (plus or minus a small numerical error) after a particular box size (about 20 s.u.).

We can also see that the type \chgex[ex]{of} boundary conditions we use is irrelevant, indicating that the range of physical and numerical parameters we have used, are such that the spiral wave solution is not affected by the boundaries.

\section{Application I: 1:1 Resonance in Meandering Spiral Waves}
\label{sec:ezf_121}
In this section, we illustrate one of the many uses of EZ-Freeze to practical situations.

The study of meandering spiral spiral near 1:1 resonance can be hard to study due to the large box sizes needed in the numerical study. We note that 1:1 resonance in meandering spiral waves occurs when the Euclidean frequency, $\omega_0$, is the same as the Hopf frequency, $\omega_H$. If we recall the analytical equations of motion for a meandering spiral wave (no drift due to symmetry breaking perturbations):

\begin{eqnarray}
R &=& R_0+A\left(\begin{array}{c}\sin(\alpha)\\-\cos(\alpha) \end{array}\right)\nonumber\\
\label{eqn:solxy1}
&&+B
\left(\begin{array}{cc}\cos(\alpha) & -\sin(\alpha)\\ \sin(\alpha) & \cos(\alpha) \end{array}\right)\left(\begin{array}{c} m_1\sin(\beta)+n_2\cos(\beta) \\ m_2\sin(\beta)-n_1\cos(\beta) \end{array}\right)
\end{eqnarray}
\\
where:
\chg[p172eqns]{}
\[\begin{array}{rclcrcl}
\alpha &=& \omega_0t+\theta_0 &,& \beta &=& \omega_Ht+\phi\\
A &=& \frac{\bc_0}{\omega_0} &,& B &=& \frac{2r}{\omega_H(\omega_H^2-\omega_0^2)}\\
c_1 &=& c_{11}+ic_{12} &,& \omega_1 &=& \omega_{11}+i\omega_{12}\\
m_1 &=& \omega_H^2c_{11}-\omega_0\bc_0\omega_{11} &,& m_2 &=& \omega_H(c_0\omega_{12}-\omega_0c_{12})\\
n_1 &=& \omega_H(\bc_0\omega_{11}-\omega_0c_{11}) &,& n_2 &=& \omega_H^2c_{12}-\omega_0c_0\omega_{12}
\end{array}\]

We see that if $|\omega_0|=|\omega_H|$ then $B=\infty$. The parameter $B$ determines the core radius of the trajectory. So, when $|\omega_0|=|\omega_H|$ then we have spontaneous drift.

However, we are still in a meandering state, and therefore, we should observe limit cycle solutions in the quotient system. 

The purpose of this investigation, is to uncover the properties of the quotient system at 1:1 resonance and determine whether or not these properties are significantly different than other limit cycles in the vicinity of the 1:1 resonance but not at it.


\subsection{Method}

Throughout this investigation, we shall use the FHN model. The physical and numerical parameters used in the testing were:

\begin{itemize}
\item Box size, $L_X$  = 20 s.u.
\item Spacestep, $\Delta_x$ = 0.125 s.u
\item Timestep, $\Delta_t$  = $3.90625\times10^{-4}$ t.u.
\end{itemize}

We note that the first tests that we did, which we report here, \chgex[ex]{were} done using the second order accurate numerical scheme, together with method 1. We shall report on our finding of these results, together with a run of tests done with a second order scheme and also method 2. The main differences between the results generated using methods 1 and 2 were the instabilities in the quotient system. However, we found that there were no obvious anomalies as such when using method 1 and, when compared with the results from method 2, the results were very similar.

\begin{figure}[tbhp]
\begin{center}
\begin{minipage}[htbp]{0.7\linewidth}
\centering
\includegraphics[width=0.99\textwidth,angle=-178]{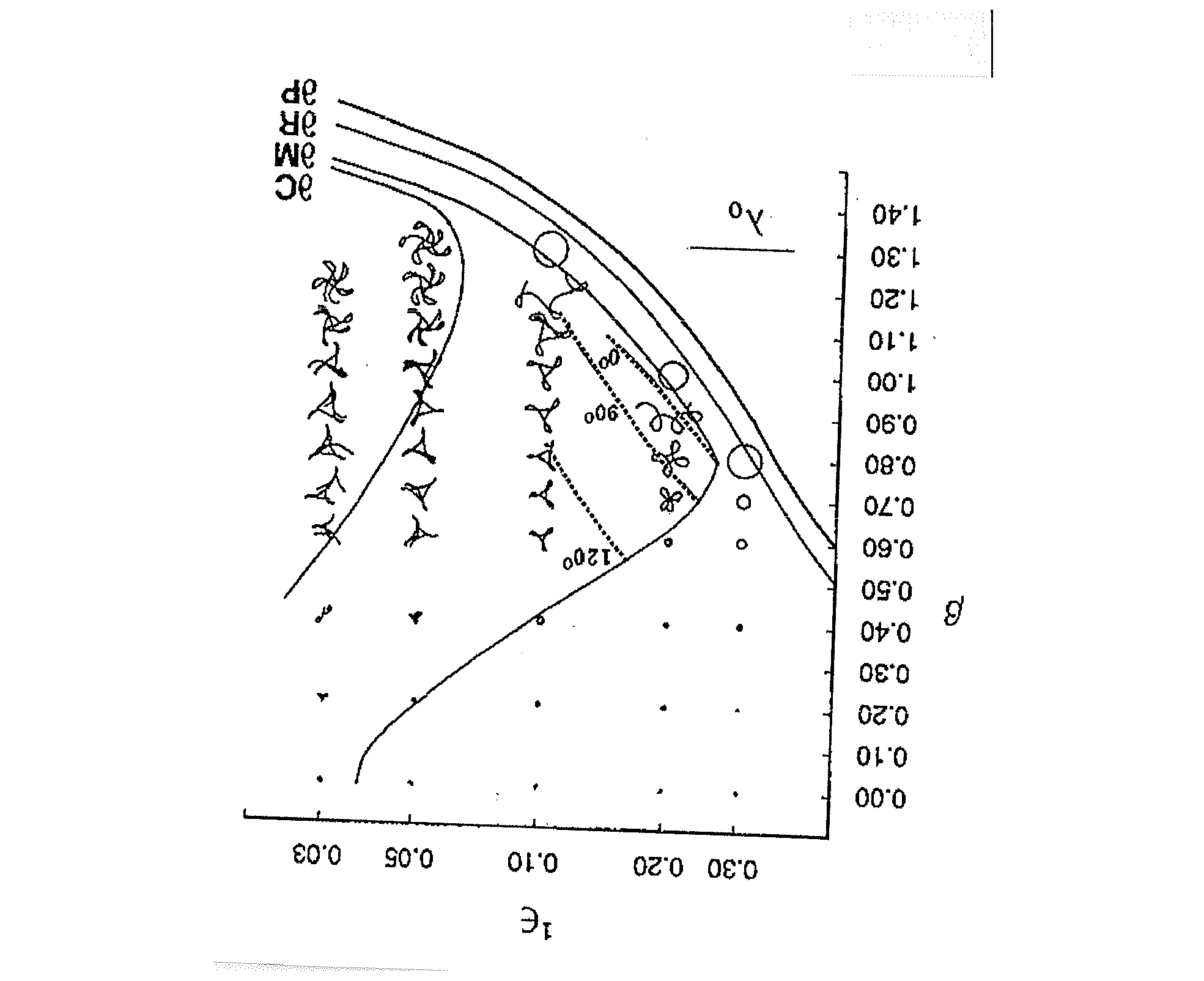}\end{minipage}
\caption{Parametric Portrait for FHN model with $\gamma$=0.5 \cite{Winf91}}
\label{fig:ezf_121_winf}
\end{center}
\end{figure}

So, the first task we had to do was to locate a point of resonance, or get as close as we could to a point of resonance. We do this using the ``Flower Garden'' published by Winfree in his 1991 paper to get a rough estimate of what parameters we should use to find 1:1 resonance \cite{Winf91}. This is shown in Fig.(\ref{fig:ezf_121_winf})

What we did was to keep two of the parameters fixed (in this case we fixed $\gamma$ and $\epsilon$) and estimate the other parameter ($\beta$, which we denote as the free parameter) from the flower garden. Once we have a rough estimate, we rerun EZ-Freeze using these parameters and see what trajectory we get. If we find that we have an outward meandering trajectory (i.e. the petals of the trajectory are facing outwards), then we need to increase the free parameter. If we get an inward meandering trajectory, we need to decrease the value of the free parameter. We do this until we get a value of the parameter which gives us 1:1 resonance or as close to it as we can get.

Once we have found this point, we run EZ-Freeze using the parameters we have just found. We also run it for some parameter values either side of the value of the free parameter.


\subsection{Results}

As mentioned in the above section, we used two methods in this investigation. Therefore, we split this subsection into two; one section for the original results we obtained using a second order scheme and method 1, and another using the more uptodate techniques, these being a second order scheme and method 2.

\subsection{Results using method 1: Run 1}

The first set of runs we conducted used the following parameters which gave us 1:1 resonance:

\begin{itemize}
\item $\beta$ = 0.93535
\item $\gamma$ = 0.5
\item $\varepsilon$ = 0.2
\end{itemize}

The free parameter $\beta$, was found to be $\beta_0=0.93535$ at 1:1 resonance. So we decided to run simulations for this value of $\beta$ together with parameters either side of this value, with these being $\beta_{\pm i}=\beta_0\pm\frac{i}{100}$ for $i=0,1,2$.

\begin{figure}[tbh]
\begin{center}
\begin{minipage}[htbp]{0.49\linewidth}
\centering
\includegraphics[width=0.7\textwidth, angle=-90]{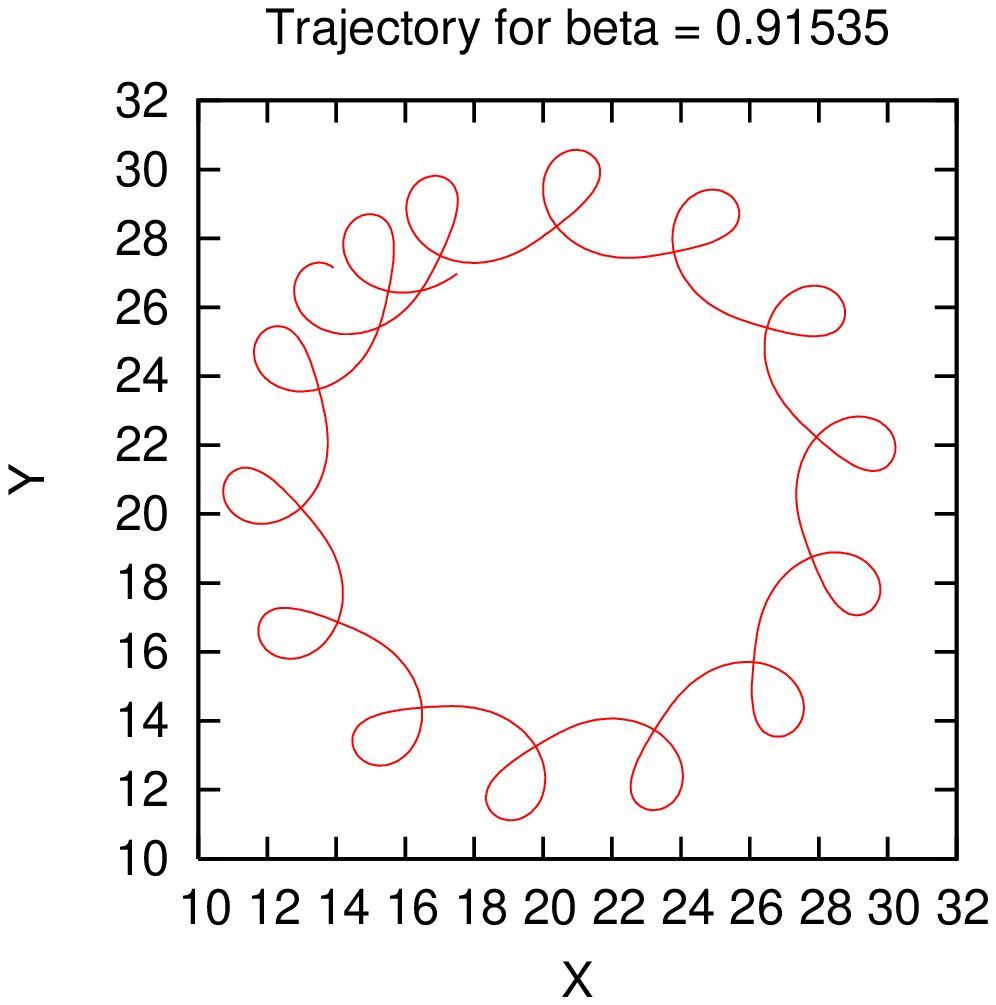}
\end{minipage}
\begin{minipage}[htbp]{0.49\linewidth}
\centering
\includegraphics[width=0.7\textwidth, angle=-90]{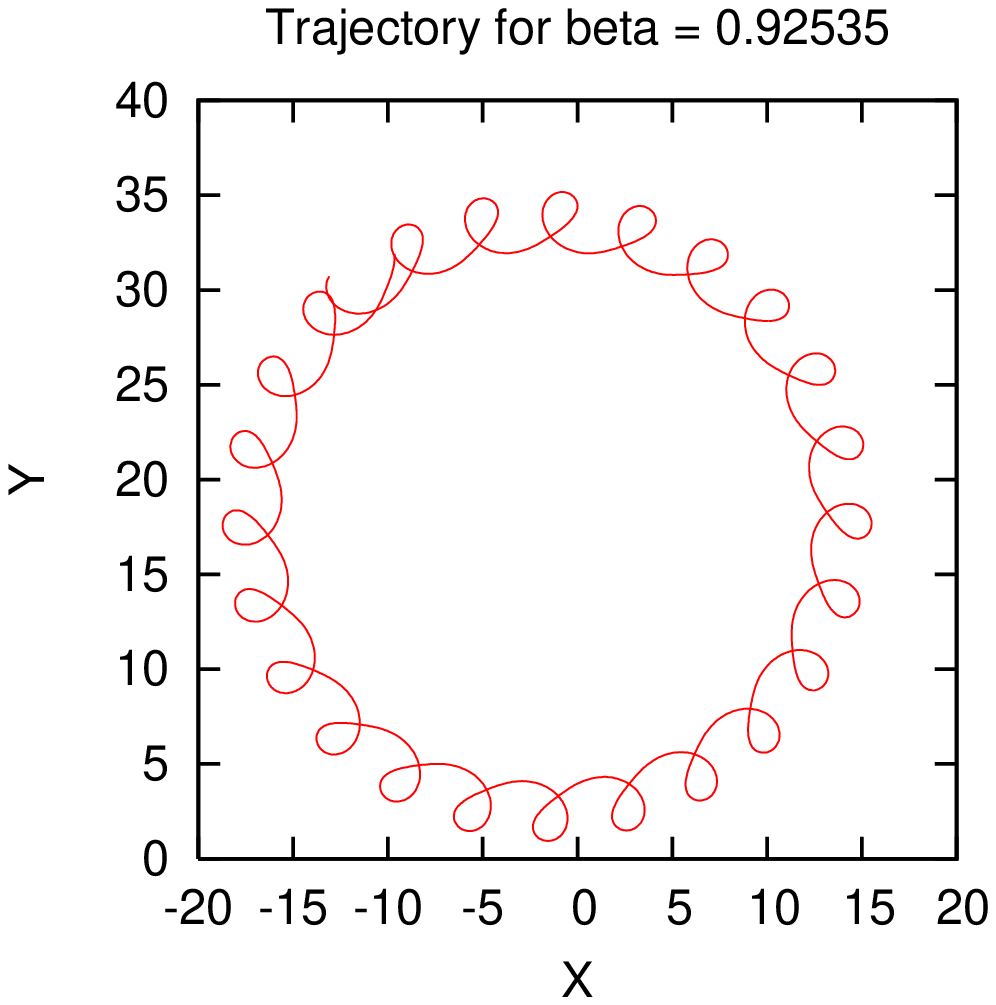}
\end{minipage}
\begin{minipage}[htbp]{0.49\linewidth}
\centering
\includegraphics[width=0.7\textwidth, angle=-90]{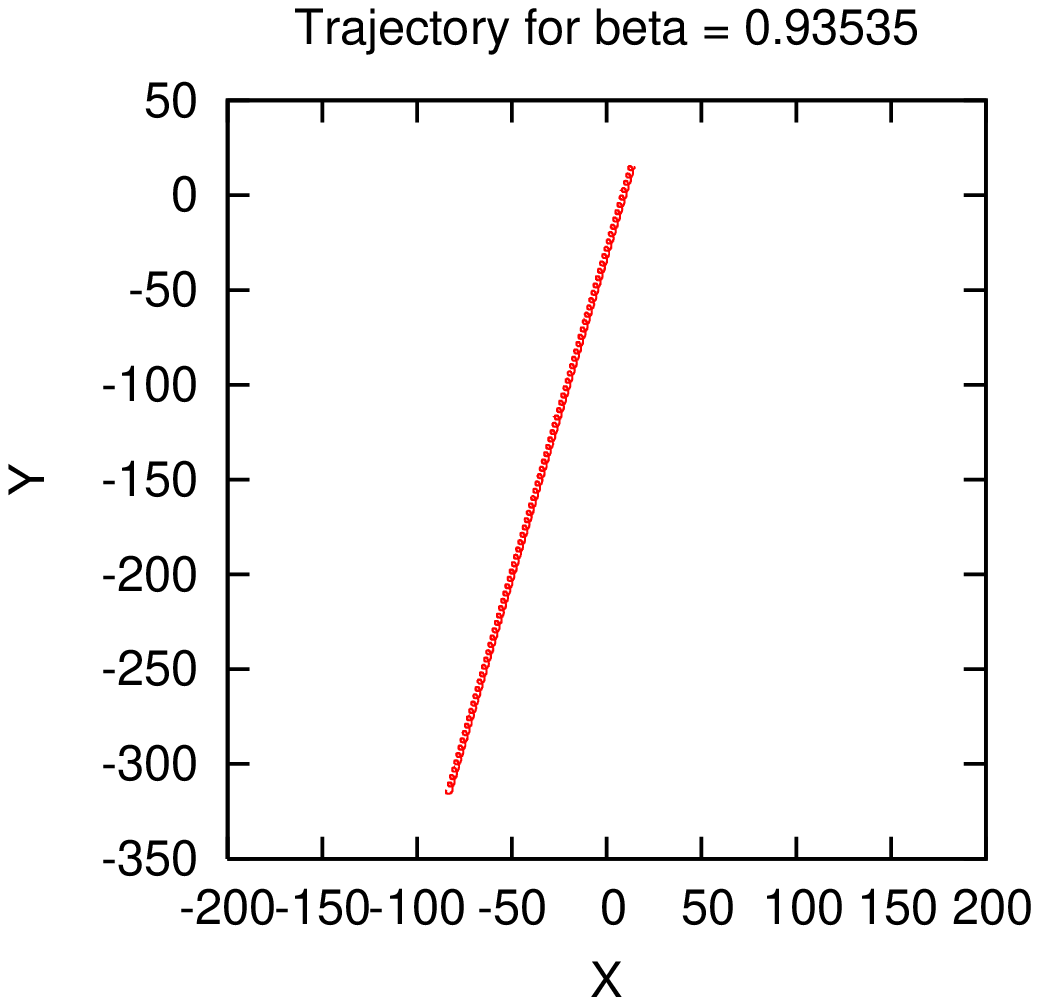}
\end{minipage}
\begin{minipage}[htbp]{0.49\linewidth}
\centering
\includegraphics[width=0.7\textwidth, angle=-90]{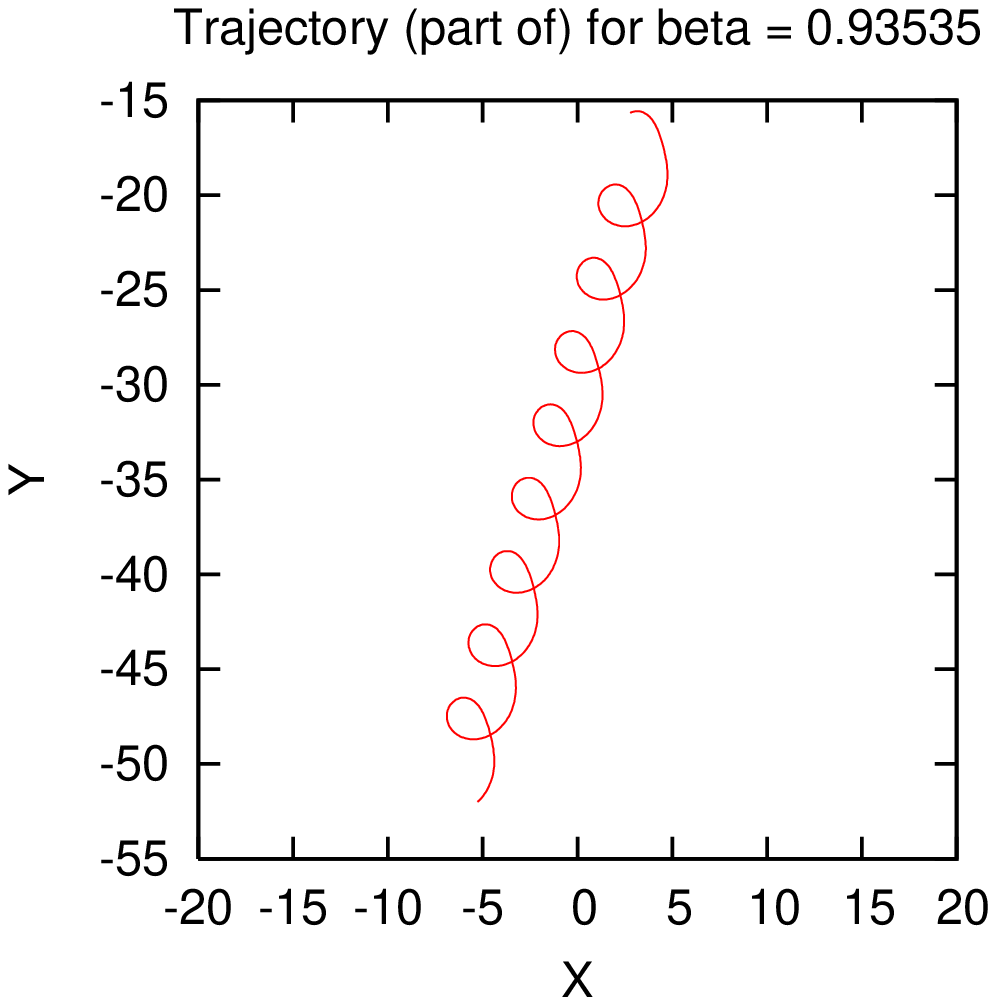}
\end{minipage}
\begin{minipage}[htbp]{0.49\linewidth}
\centering
\includegraphics[width=0.7\textwidth, angle=-90]{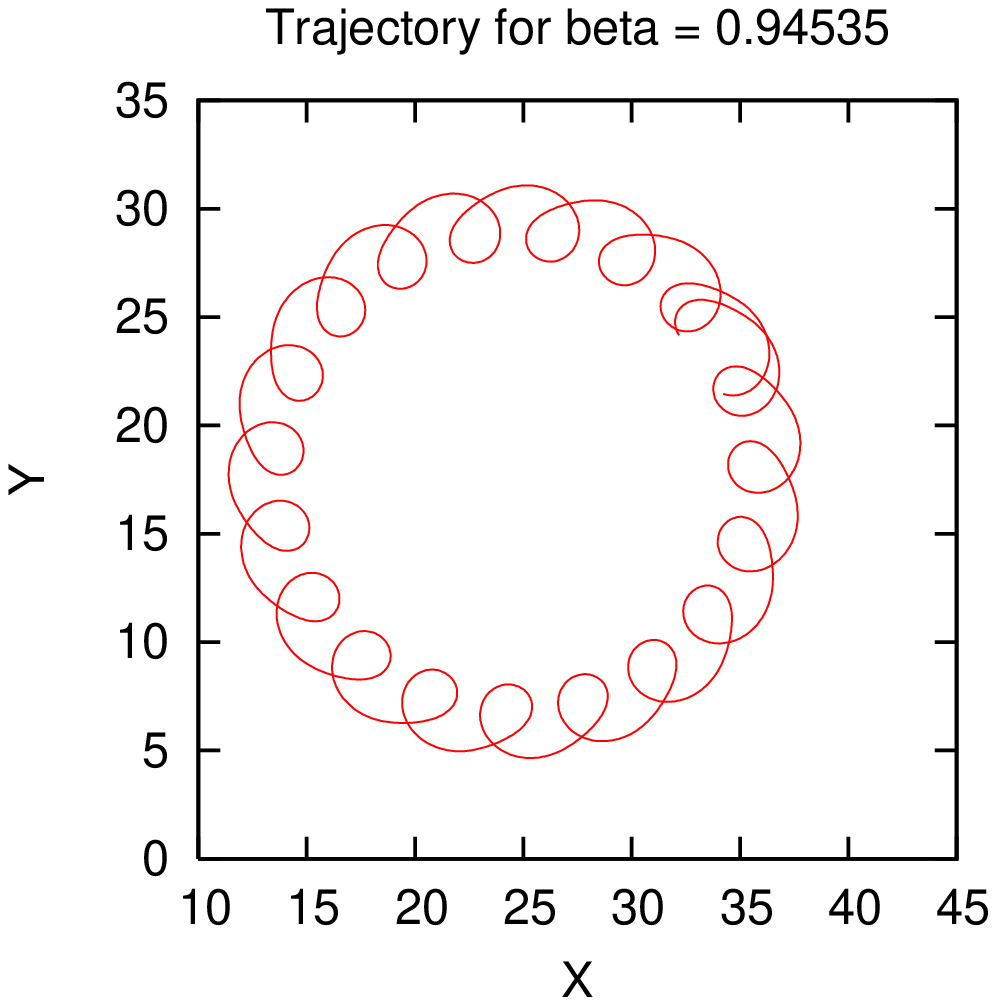}
\end{minipage}
\begin{minipage}[htbp]{0.49\linewidth}
\centering
\includegraphics[width=0.7\textwidth, angle=-90]{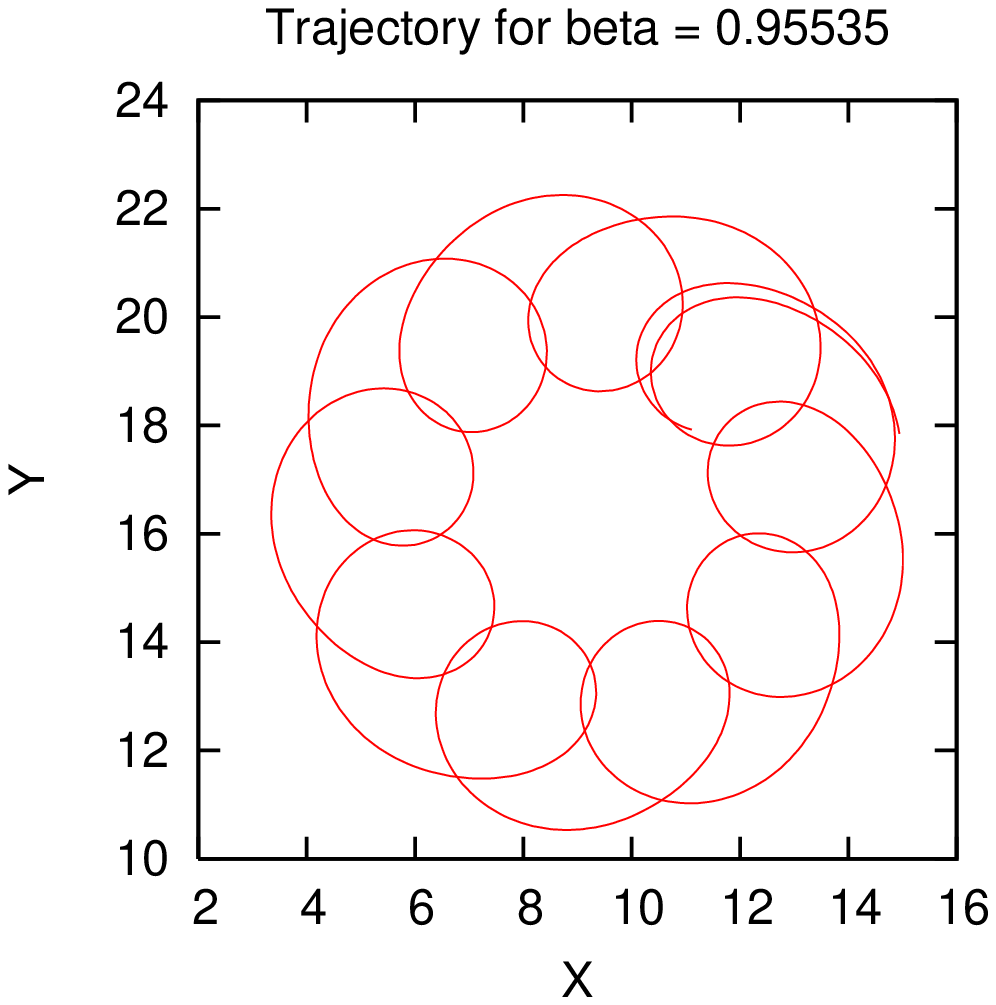}
\end{minipage}
\caption[1:1 resonance testing run 1: trajectories]{FHN, Second Order, Method 1. The trajectories for each of the values of $\beta$ used in the simulation with $\gamma=0.5$ and $\varepsilon=0.2$. The trajectories have been reconstructed by numerically integrating the tip trajectory equations using the numerical values of the advection coefficients. The trajectories are: (Top Left) $\beta=0.91535$; (Top Right) $\beta=0.92535$; (Middle Left) $\beta=0.93535$; (Middle left) $\beta=0.93535$ (part of the trajectory for this value of $\beta$ is shown to give the reader a perspective of what the trajectory is like close up); (Bottom Left) $\beta=0.94535$; (Bottom Right) $\beta=0.95535$}
\label{fig:ezf_121_old1_trajs}
\end{center}
\end{figure}

\begin{figure}[tbh]
\begin{center}
\begin{minipage}{0.49\linewidth}
\centering
\includegraphics[width=1.0\textwidth, angle=-90]{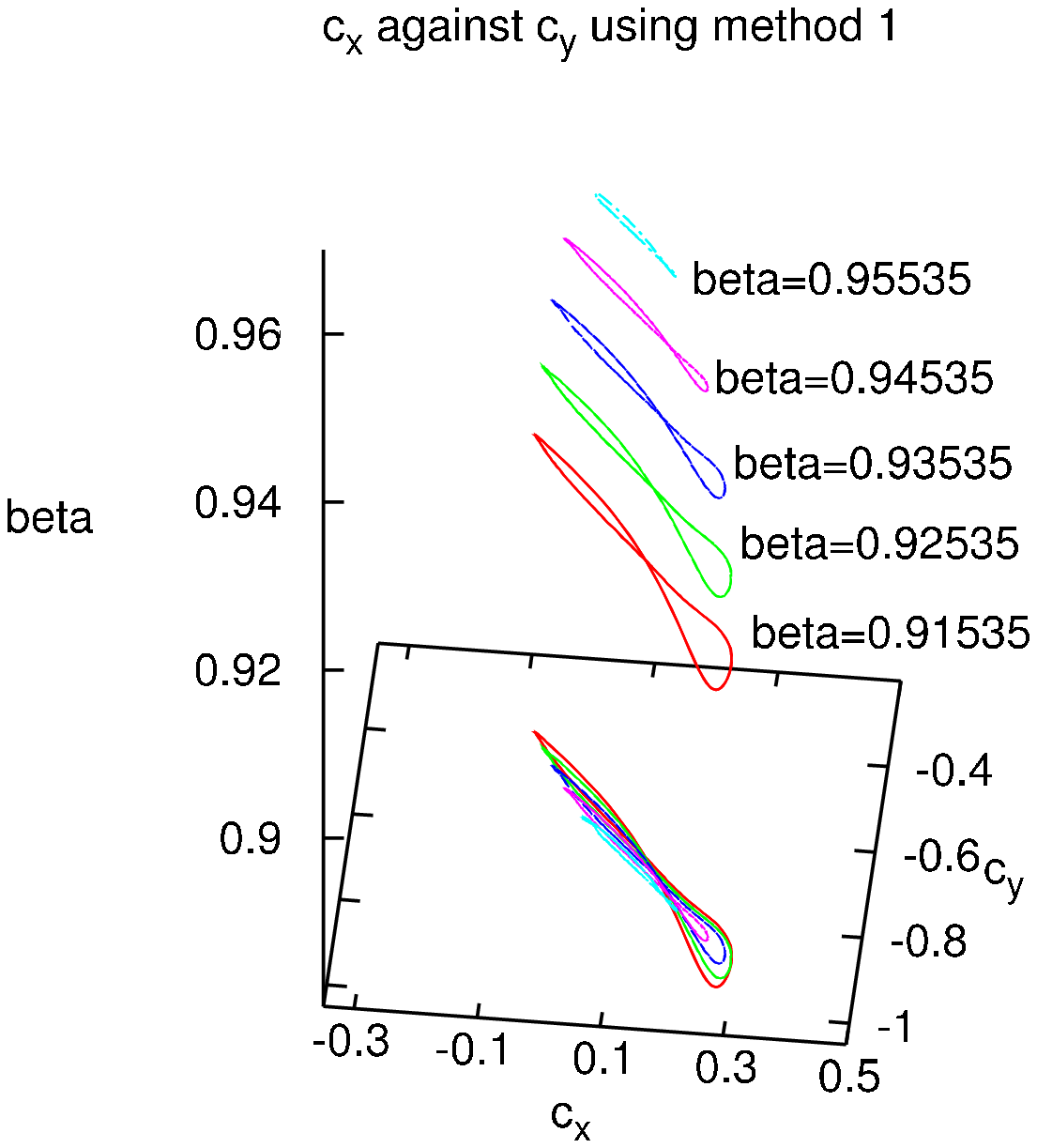}
\end{minipage}
\begin{minipage}{0.49\textwidth}
\centering
\includegraphics[width=1.0\textwidth, angle=-90]{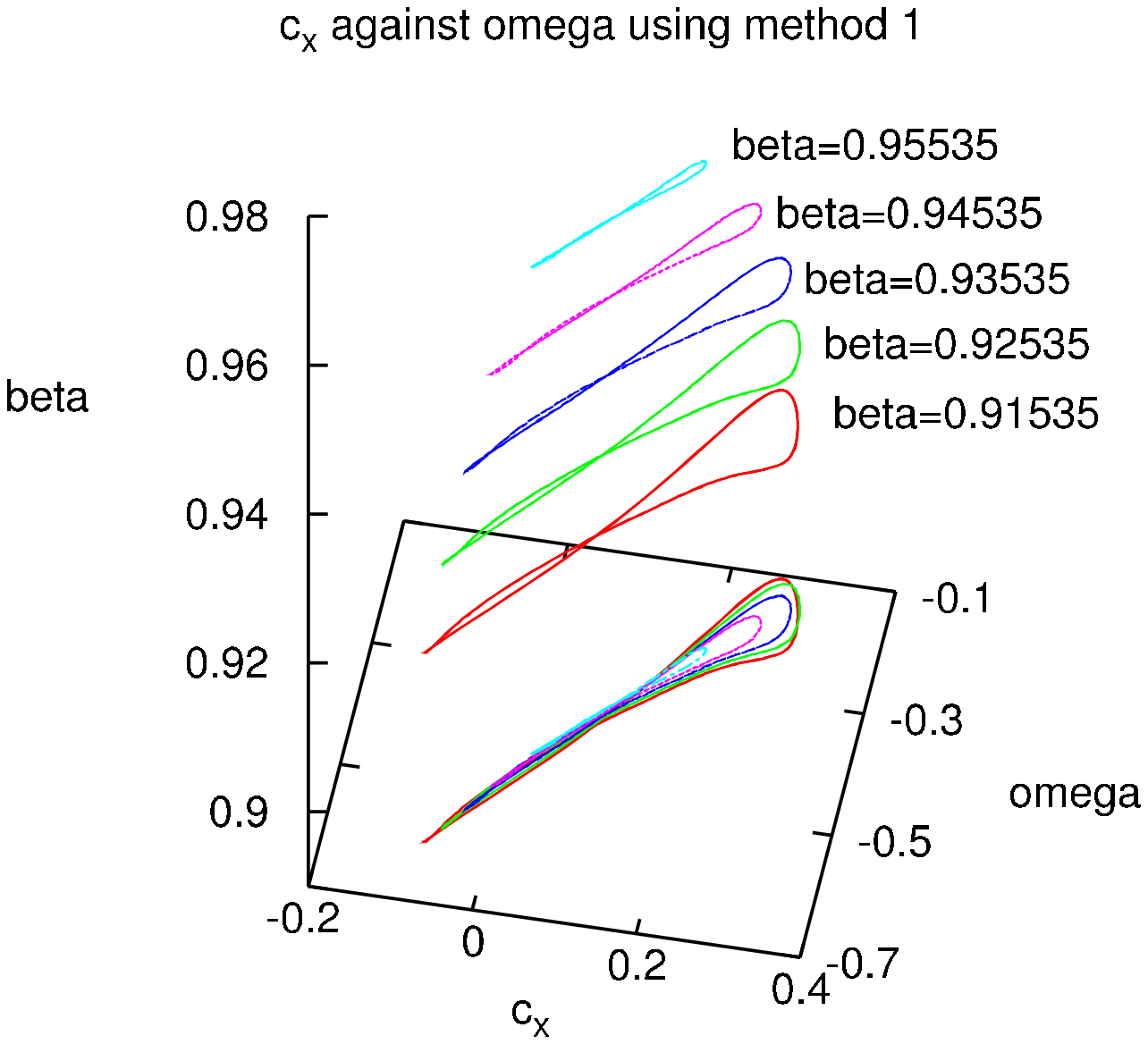}
\end{minipage}
\begin{minipage}{0.49\linewidth}
\centering
\includegraphics[width=1.0\textwidth, angle=-90]{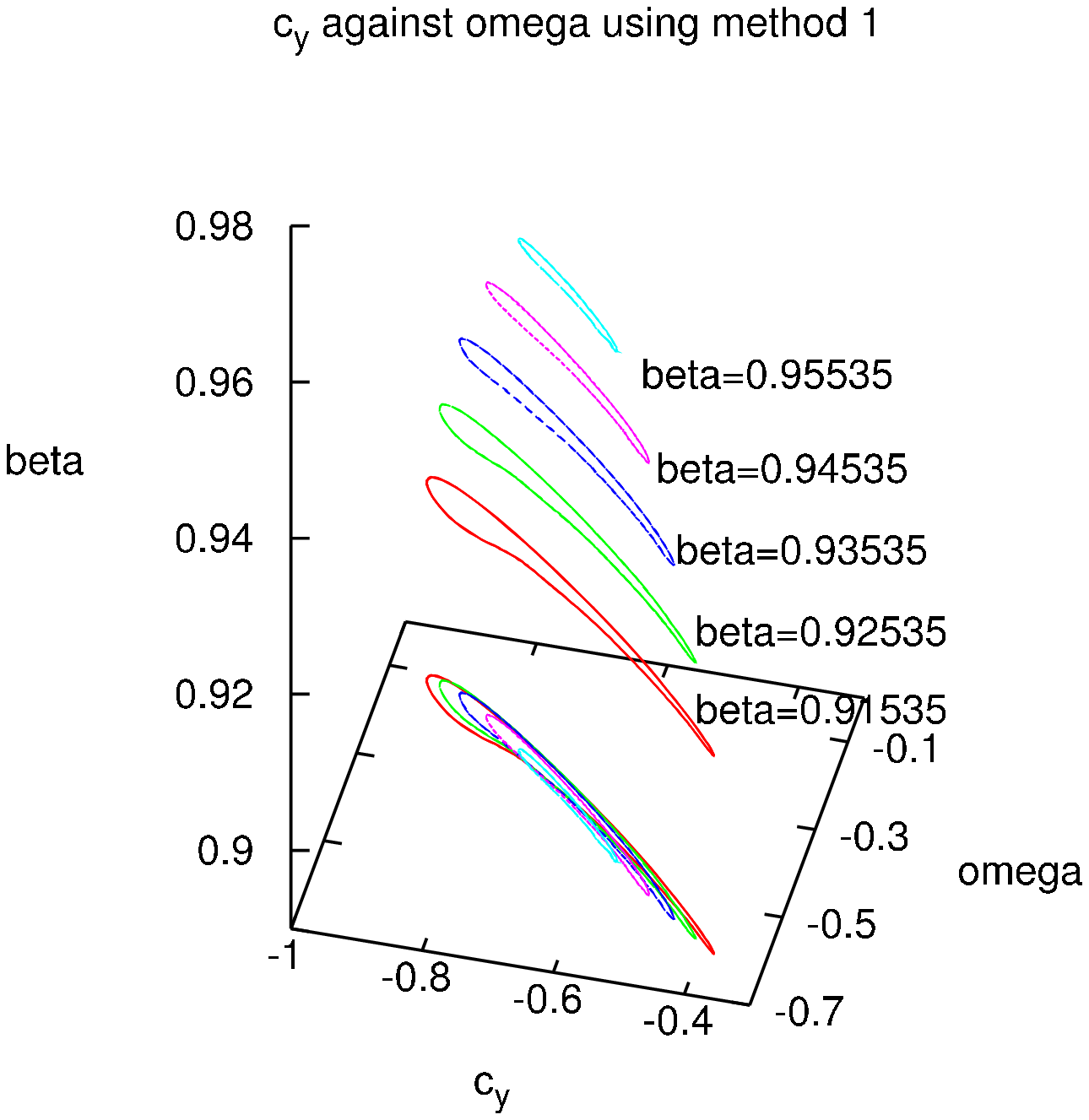}
\end{minipage}
\caption[1:1 resonance testing run 1: quotient solutions]{Quotient solutions. FHN, Second Order, Method 1 with $\gamma=0.5$ and $\varepsilon=0.2$. The parameter $\beta$ is varied and its values are detailed on the plots above.}
\label{fig:ezf_121_old1_quots}
\end{center}
\end{figure}

The results are shown in Figs.(\ref{fig:ezf_121_old1_trajs})-(\ref{fig:ezf_121_old1_quots}).

The trajectories as shown in Fig.(\ref{fig:ezf_121_old1_trajs}) were numerically reconstructed from the advection coefficients, as detailed in Sec.(\ref{sec:ezf_numerics_imp}).

We can see that at $\beta_f$ we get a trajectory that is very close to 1:1 resonance. The parameters either side of $\beta_f$ give either outward meandering trajectories ($\beta<\beta_f$) or inward meandering trajectories ($\beta>\beta_f$).

Also, we note that the length of the trajectory at $\beta_f$ is approximately 360 s.u. If we were to conduct this in the laboratory frame of reference, we would have needed a box length of at least 370 s.u. (the extra to allow for the solution to be near, but not at, the boundary) and a grid size of 2960$\times$2960 to give us a spacestep of 0.125. Computationally, this would have taken a very long time, but we have conducted this simulation in approximately two and a half hours. We estimate that this would have taken about five weeks to perform in a laboratory frame of reference.

If we wanted to, we could let this simulation run for a very long, due to the fact that the spiral wave never reaches the boundaries. We also conducted the simulations knowing that the boundaries would not have a great effect on the solution.

Let us look at the quotient system as shown in Fig.(\ref{fig:ezf_121_old1_quots}). The plots in the figure have been drawn to give the reader a good perspective of how the advection coefficients are related to each other. In each case, we have also included a $\beta$-axis to help the reader compare the plots for each value of $\beta$ used.

One point that we can note is that there is no obvious difference between the quotient system corresponding to the 1:1 resonance (shown in \textcolor{blue}{blue}). We can see that as we get nearer to the $\partial M$ boundary (this is the boundary separating the regions of meandering spirals and rigidly rotating spirals (see Fig.(\ref{fig:ezf_121_winf}))), the amplitudes of the limit cycles decrease. This is as expected, since at the $\partial M$ boundary (also note that this is a Hopf locus), we have that the limit cycles are zero and grow with a $\frac{1}{2}$ power law \cite{Bark90}.

We also note that the limit cycles are quite thin, with one end of the limit cycle being thicker than the other. If we were to continue the parameters so that we were getting closer to the Hopf locus, then we should observe a more ellipsoidal limit cycle.

\subsection{Results using method 1: Run 2}

The next set of runs we conducted, were done using $\beta_f=0.812$, $\gamma=0.5$ and $\varepsilon=0.25$. The other values of $\beta$ are determined as $\beta_{\pm i}=\beta_0\pm\frac{i}{200}$ for $i=0,1,2$.

\begin{figure}[tbh]
\begin{center}
\begin{minipage}[htbp]{0.49\linewidth}
\centering
\includegraphics[width=0.7\textwidth, angle=-90]{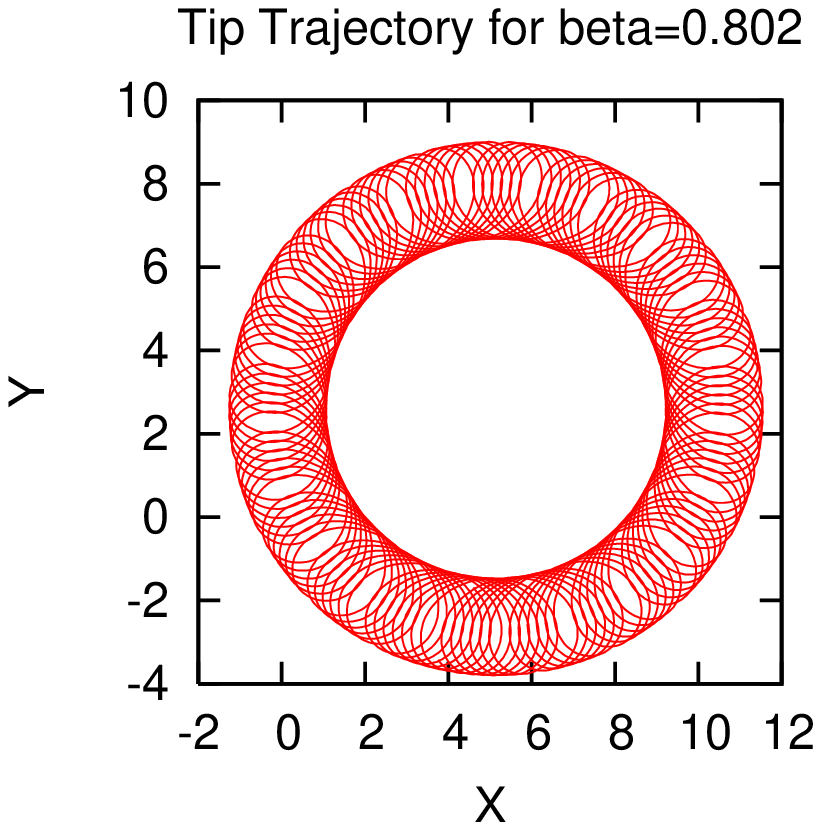}
\end{minipage}
\begin{minipage}[htbp]{0.49\linewidth}
\centering
\includegraphics[width=0.7\textwidth, angle=-90]{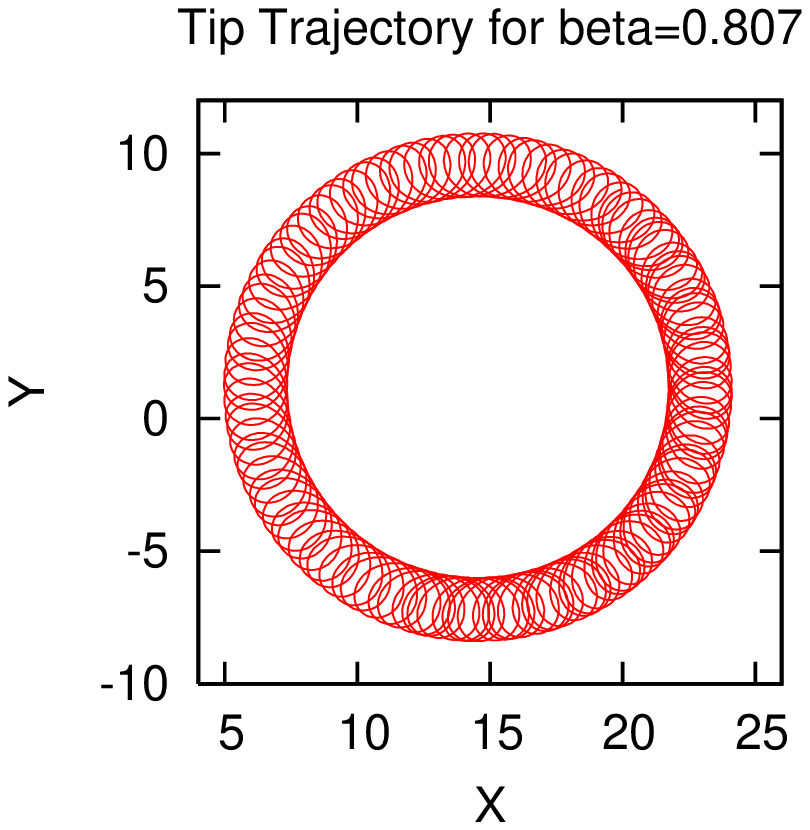}
\end{minipage}
\begin{minipage}[htbp]{0.49\linewidth}
\centering
\includegraphics[width=0.7\textwidth, angle=-90]{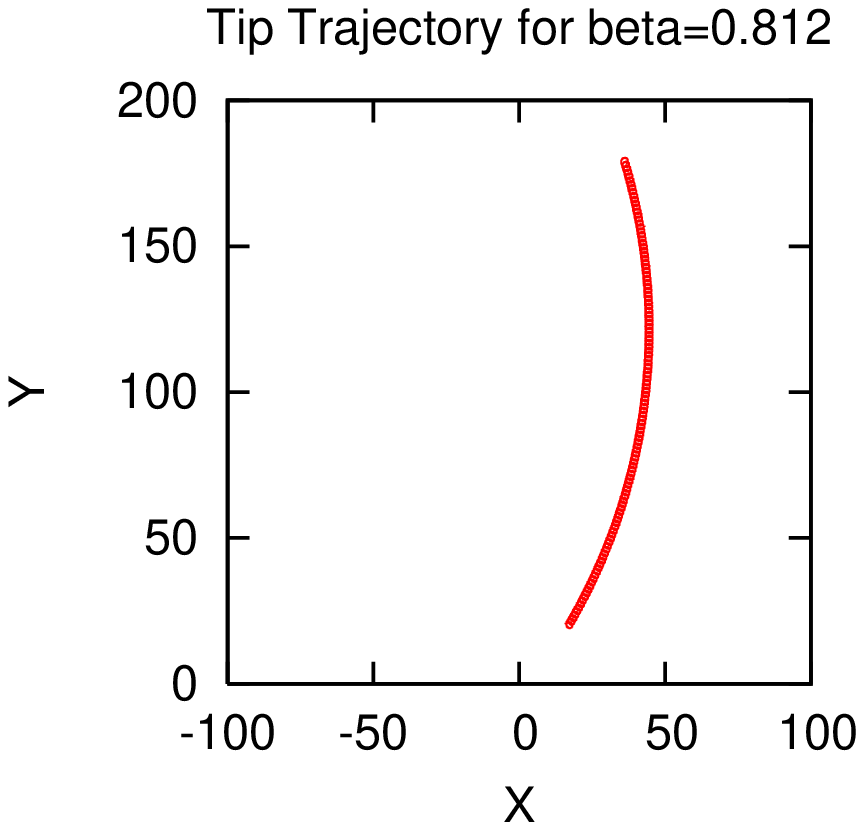}
\end{minipage}
\begin{minipage}[htbp]{0.49\linewidth}
\centering
\includegraphics[width=0.7\textwidth, angle=-90]{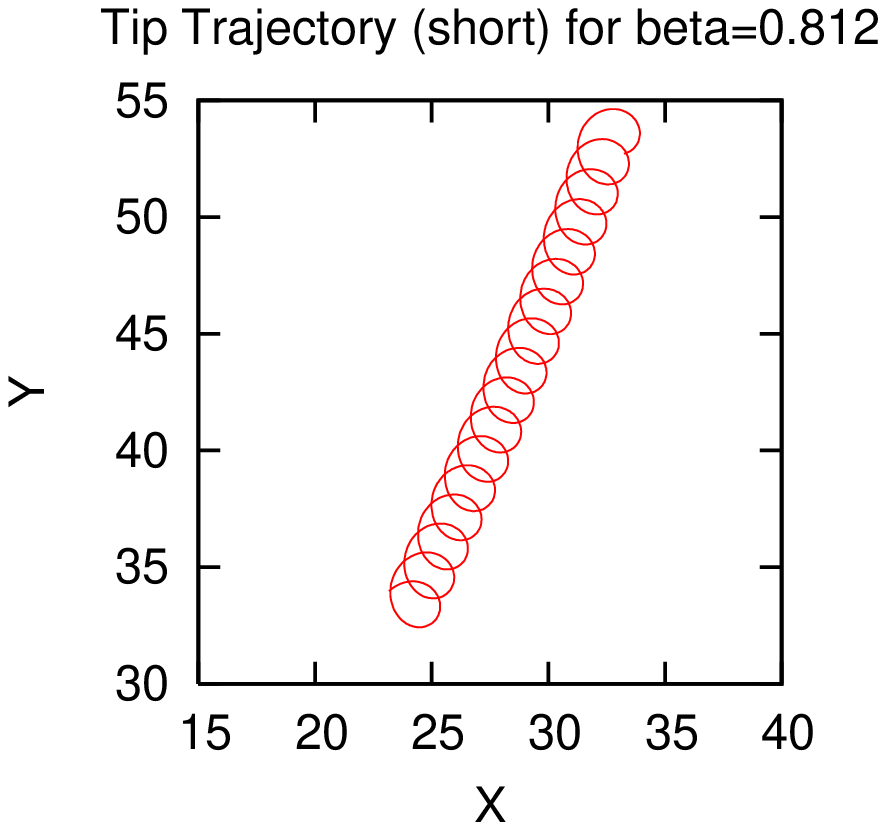}
\end{minipage}
\begin{minipage}[htbp]{0.49\linewidth}
\centering
\includegraphics[width=0.7\textwidth, angle=-90]{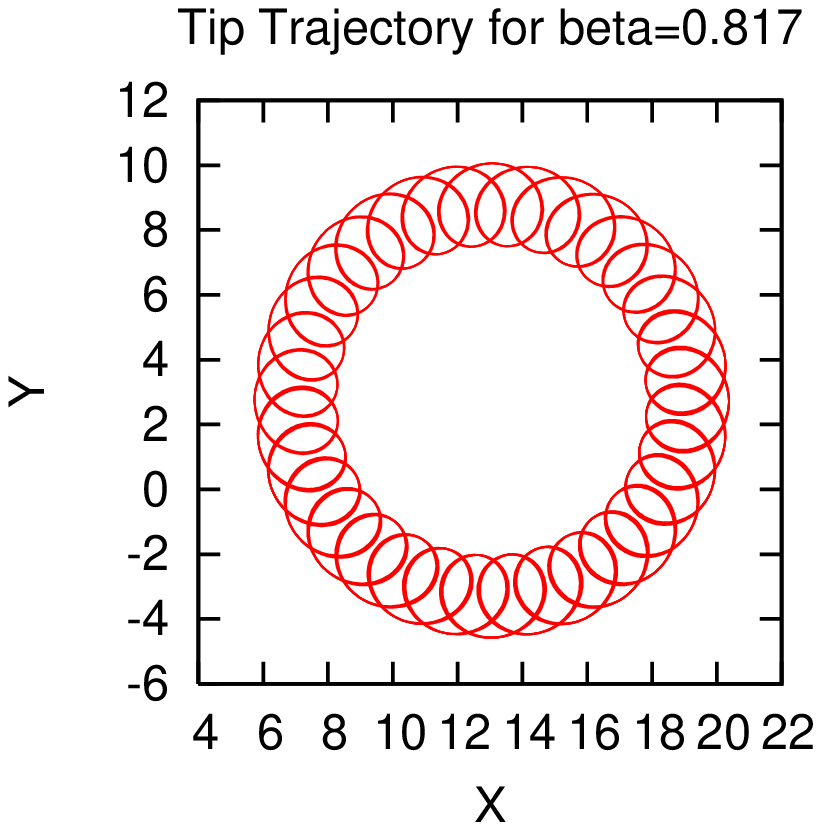}
\end{minipage}
\begin{minipage}[htbp]{0.49\linewidth}
\centering
\includegraphics[width=0.7\textwidth, angle=-90]{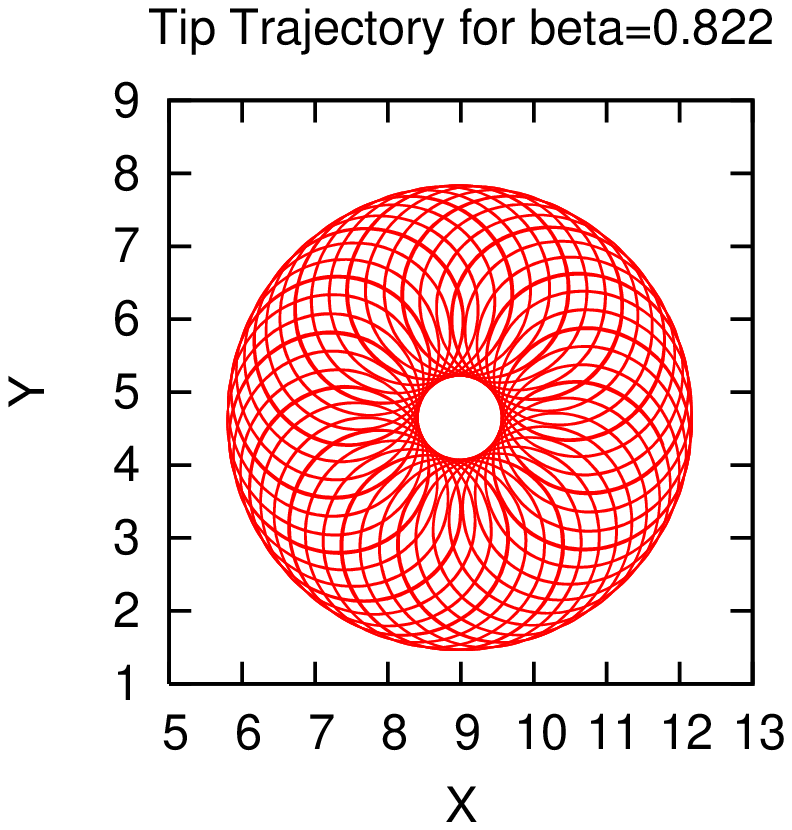}
\end{minipage}
\caption[1:1 resonance testing run 2: trajectories]{FHN, Second Order, Method 1. The trajectories for each of the values of $\beta$ used in the simulation with $\gamma=0.5$ and $\varepsilon=0.25$. The trajectories have been reconstructed by numerically integrating the tip trajectory equations using the numerical values of the advection coefficients. The trajectories are: (Top Left) $\beta=0.802$; (Top Right) $\beta=0.807$; (Middle Left) $\beta=0.812$; (Middle left) $\beta=0.812$ (part of the trajectory for this value of $\beta$ is shown to give the reader a perspective of what the trajectory is like close up); (Bottom Left) $\beta=0.817$; (Bottom Right) $\beta=0.822$}
\label{fig:ezf_121_old2_trajs}
\end{center}
\end{figure}

\begin{figure}[tbh]
\begin{center}
\begin{minipage}{0.49\linewidth}
\centering
\includegraphics[width=1.0\textwidth, angle=-90]{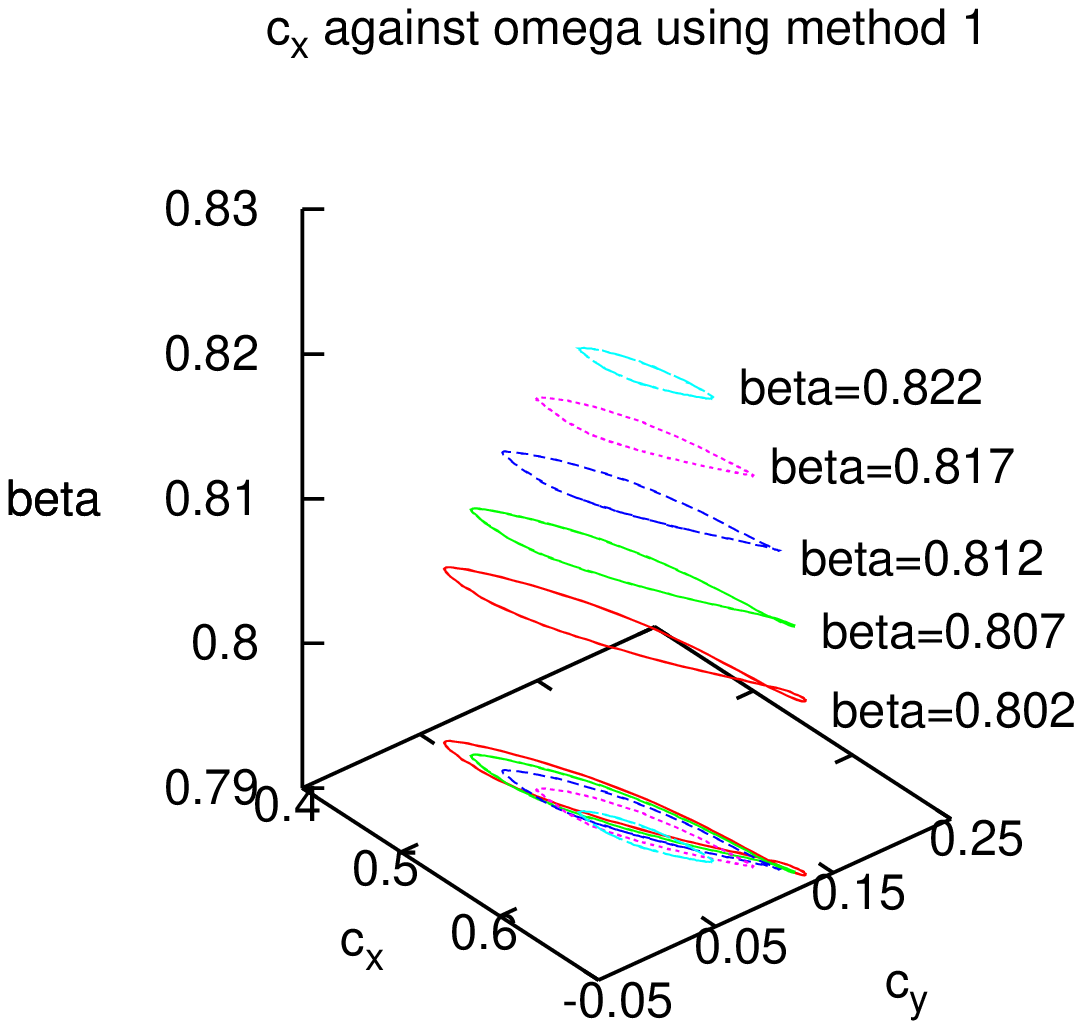}
\end{minipage}
\begin{minipage}{0.49\linewidth}
\centering
\includegraphics[width=1.0\textwidth, angle=-90]{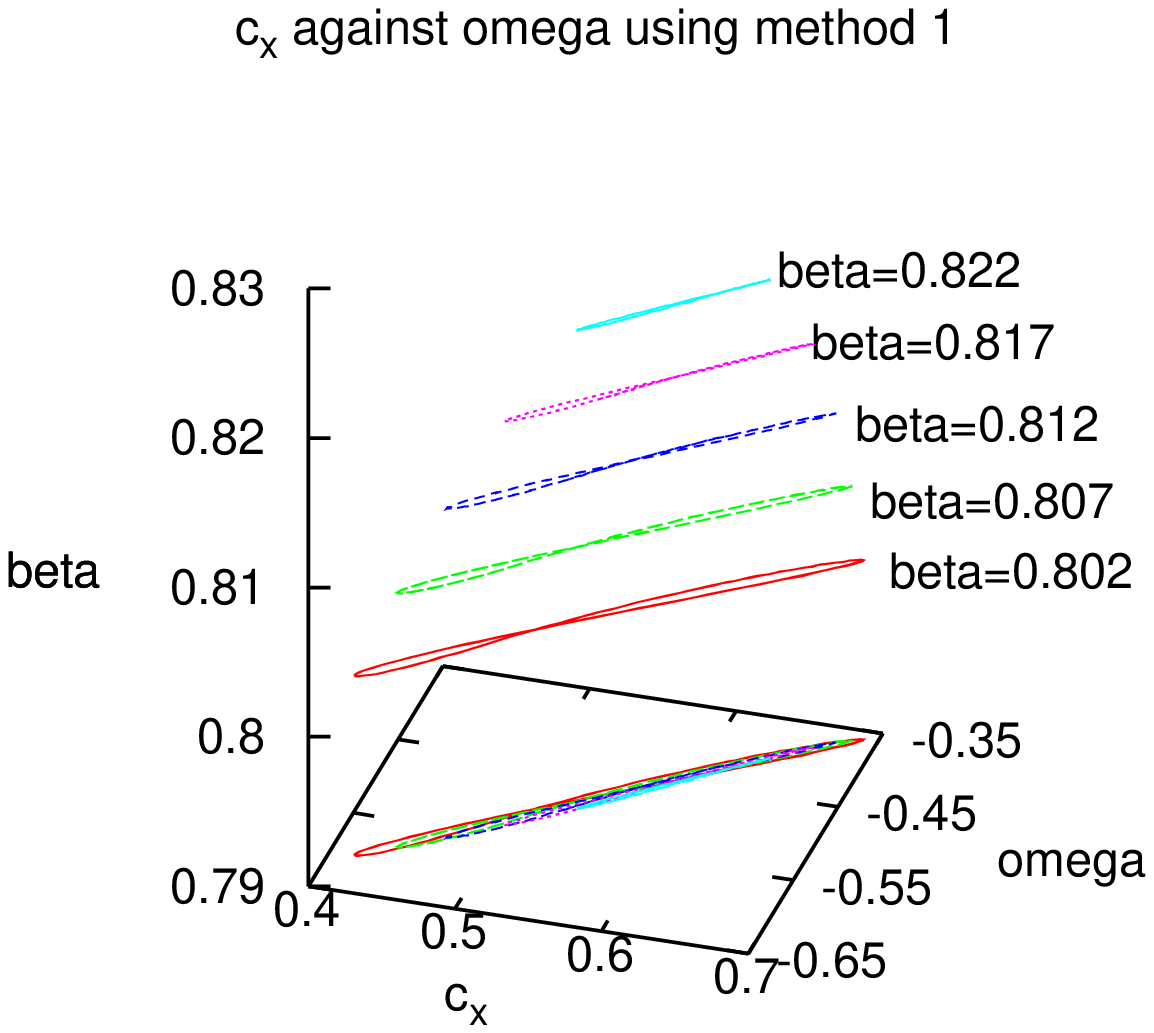}
\end{minipage}
\begin{minipage}{0.49\linewidth}
\centering
\includegraphics[width=1.0\textwidth, angle=-90]{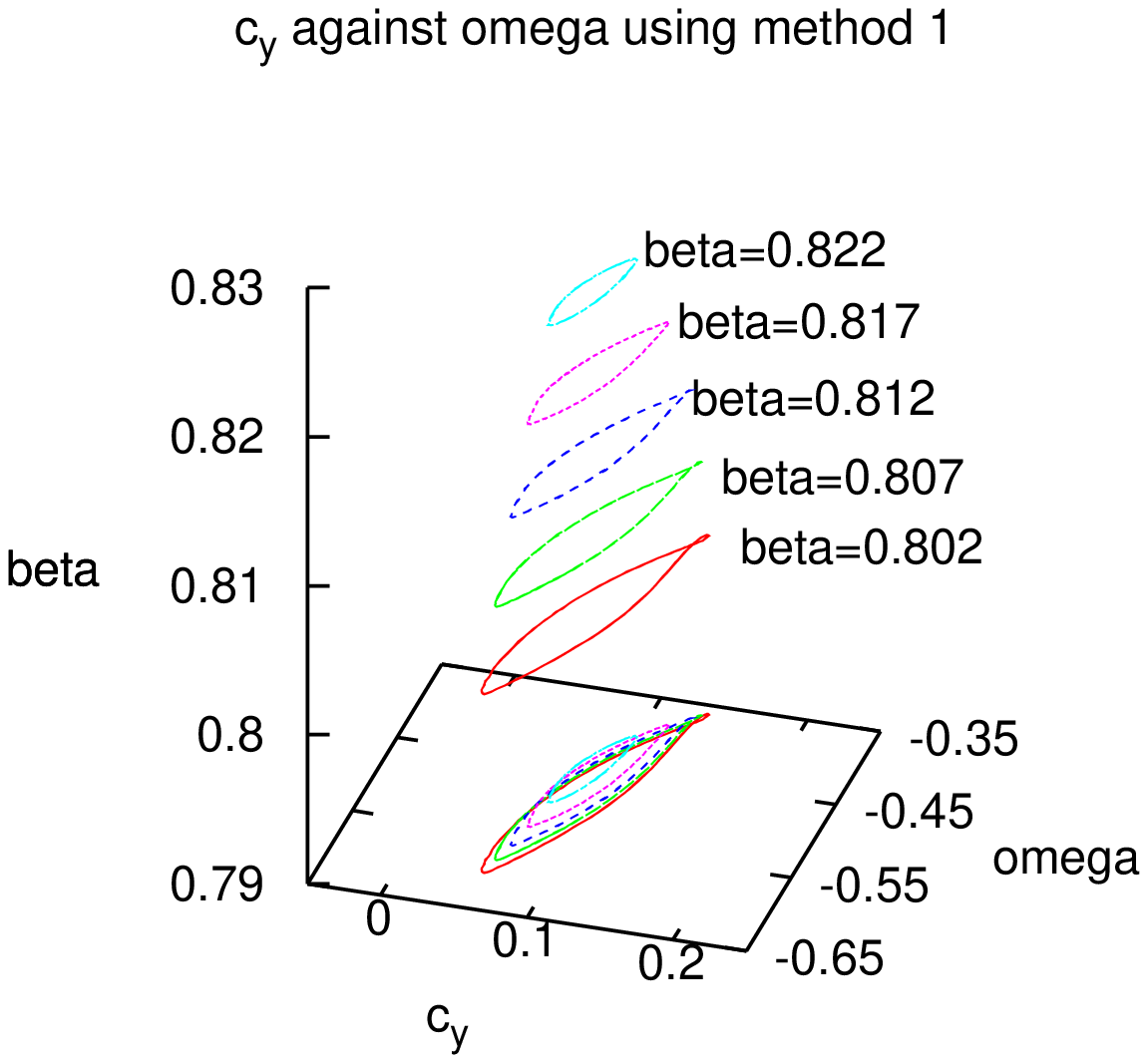}
\end{minipage}
\caption[1:1 resonance testing run 2: quotient solutions]{Quotient solutions. FHN, Second Order, Method 1 with $\gamma=0.5$ and $\varepsilon=0.25$. The parameter $\beta$ is varied and its values are detailed on the plots above.}
\label{fig:ezf_121_old2_quots}
\end{center}
\end{figure}

The results are shown in Figs.(\ref{fig:ezf_121_old2_trajs})-(\ref{fig:ezf_121_old2_quots}). As we can see, the results are qualitatively very similar. We have quite narrow limit cycles. We also note that we are quite close to the codim-2 point (the point at which the locus of 1:1 resonance meets the Hopf locus), we get more ellipsoidal limit cycles.

Again, we see that there is no distinct difference between the limit cycles at the 1:1 resonance points and its environs.

\clearpage


\subsection{Results using method 2}

\begin{figure}[p]
\begin{center}
\begin{minipage}{0.49\linewidth}
\centering
\includegraphics[width=0.7\textwidth, angle=-90]{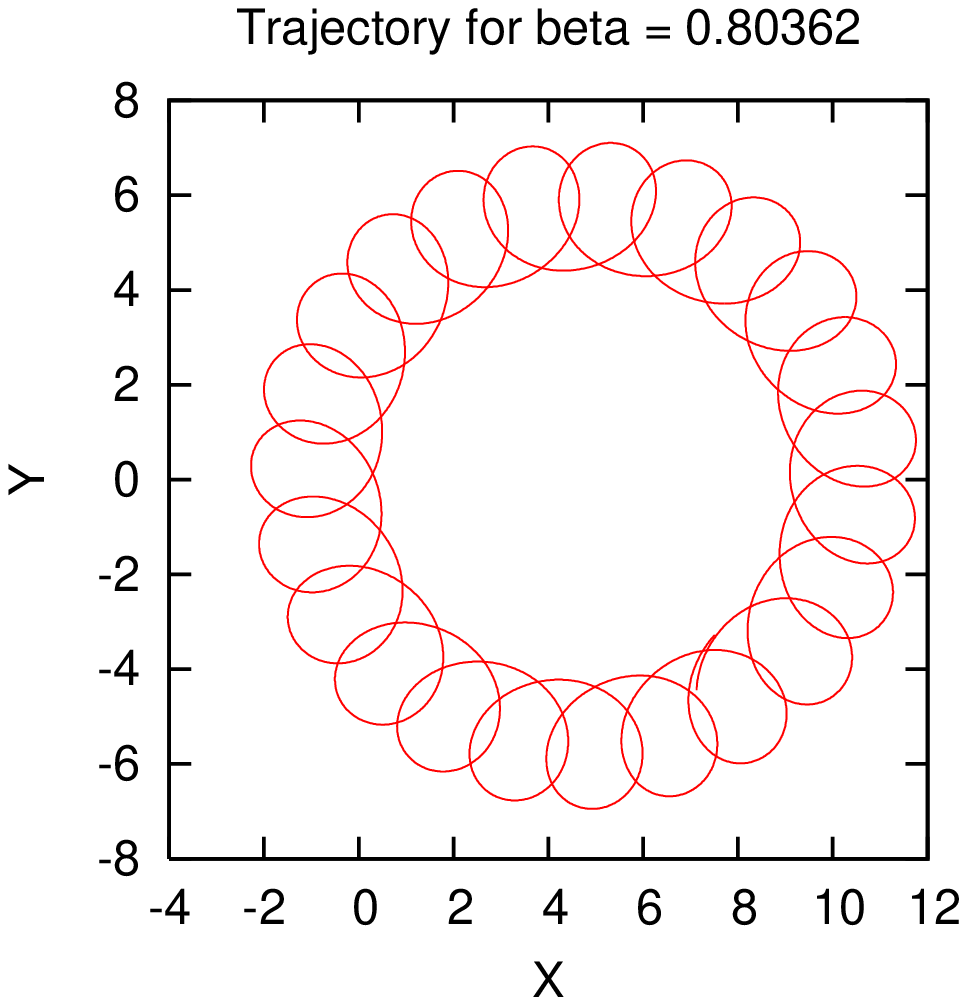}
\end{minipage}
\begin{minipage}{0.49\linewidth}
\centering
\includegraphics[width=0.7\textwidth, angle=-90]{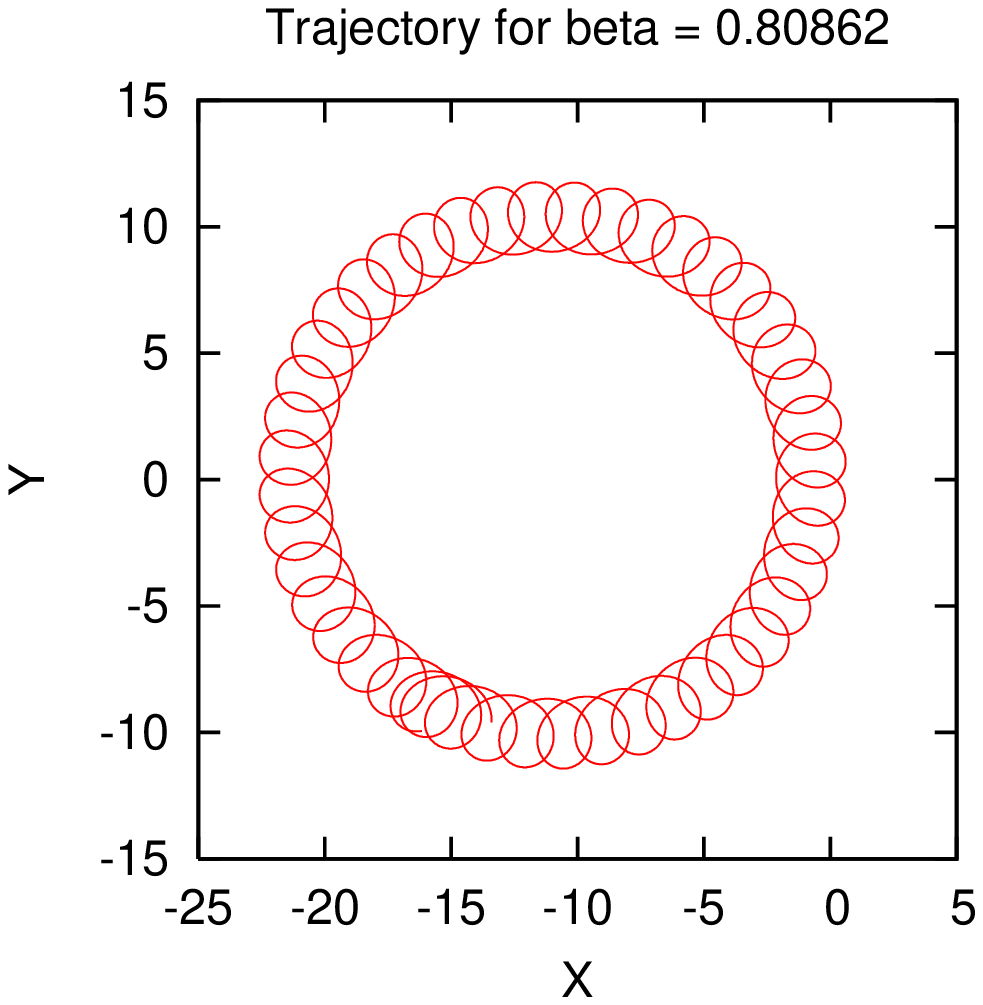}
\end{minipage}
\begin{minipage}{0.49\linewidth}
\centering
\includegraphics[width=0.7\textwidth, angle=-90]{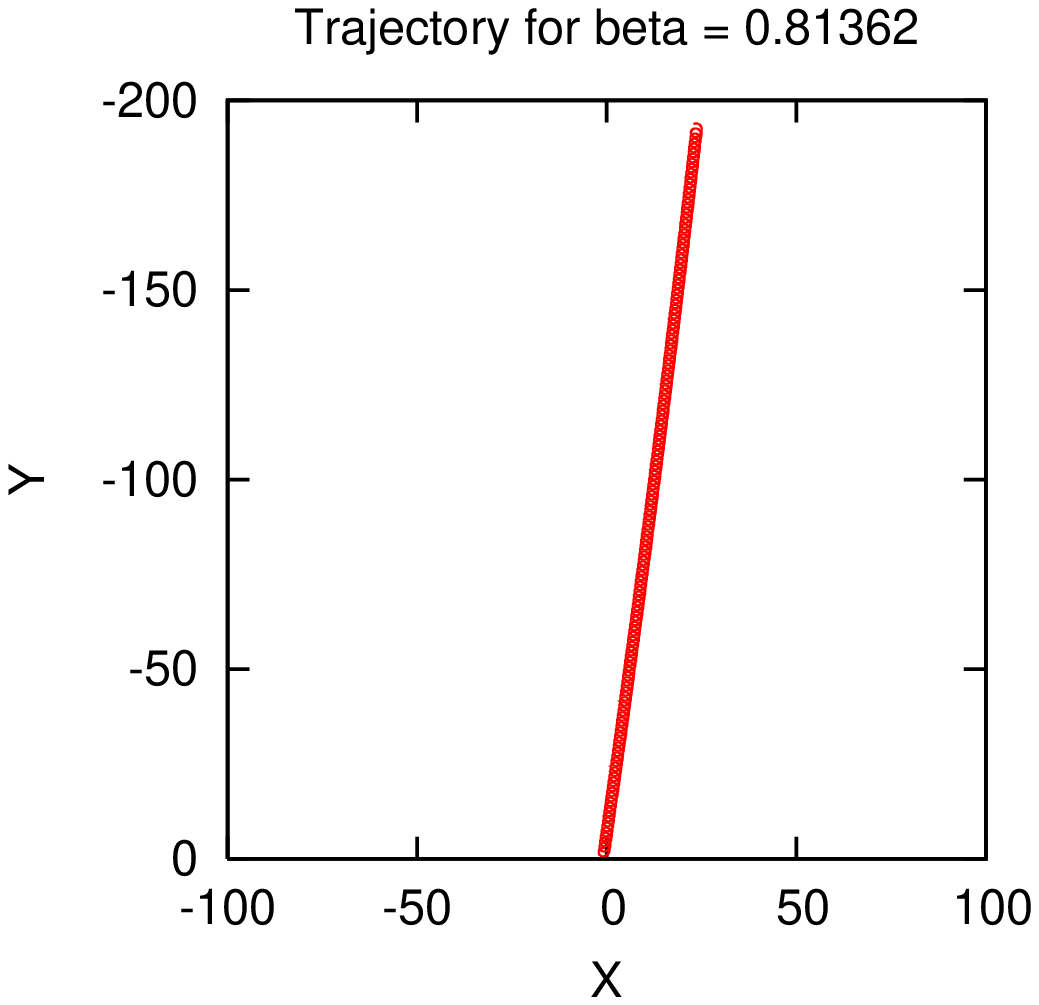}
\end{minipage}
\begin{minipage}{0.49\linewidth}
\centering
\includegraphics[width=0.7\textwidth, angle=-90]{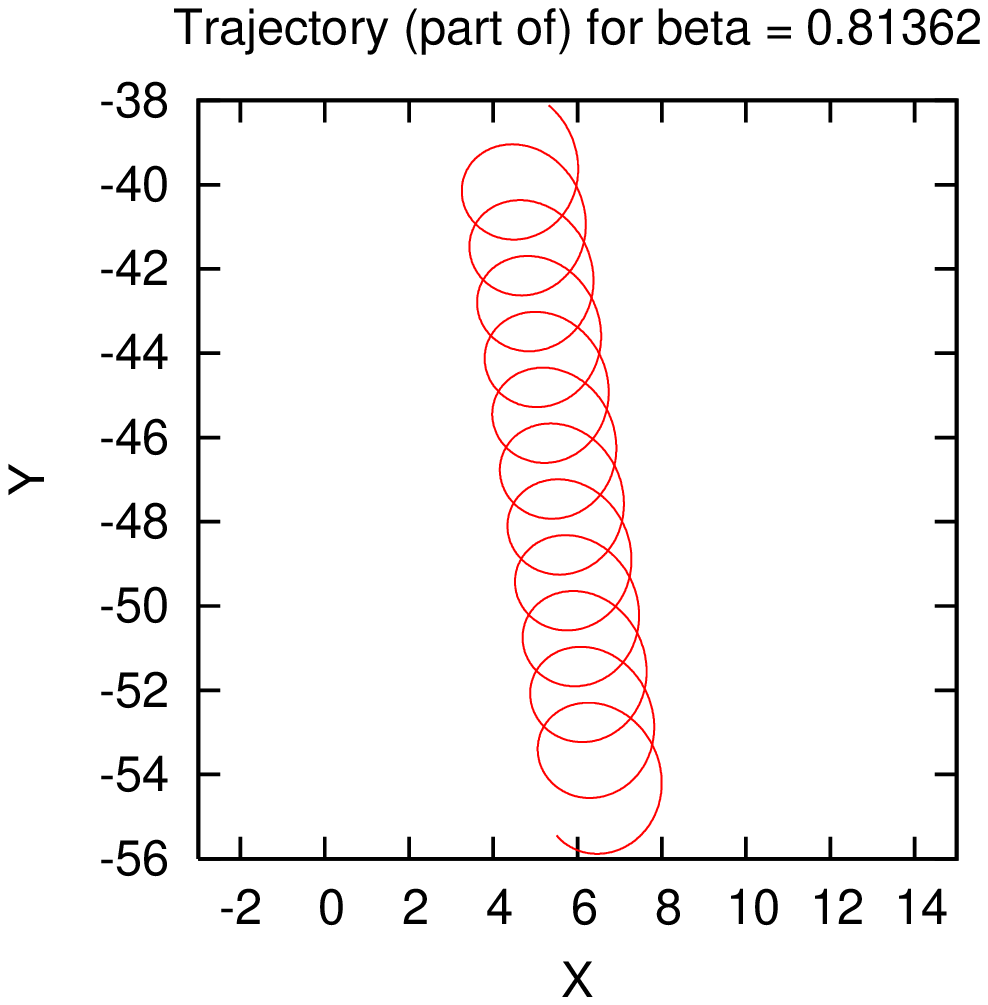}
\end{minipage}
\begin{minipage}{0.49\linewidth}
\centering
\includegraphics[width=0.7\textwidth, angle=-90]{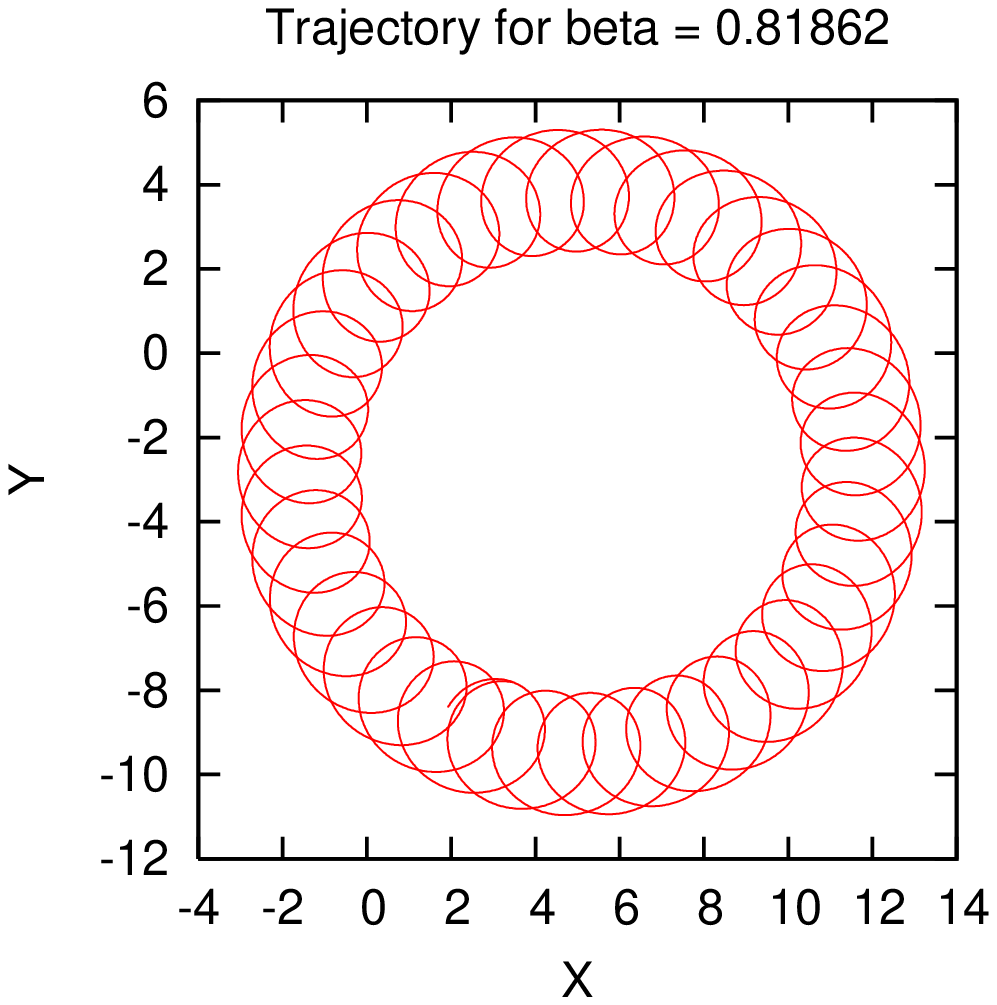}
\end{minipage}
\begin{minipage}{0.49\linewidth}
\centering
\includegraphics[width=0.7\textwidth, angle=-90]{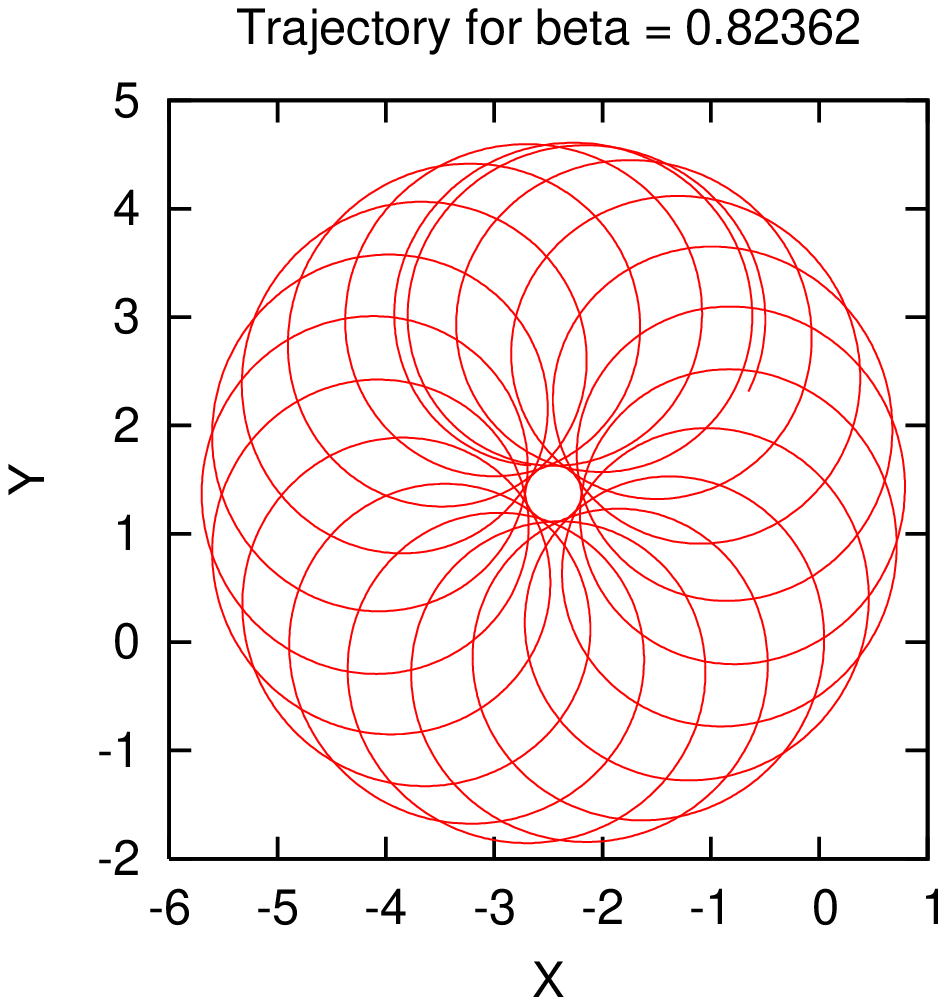}
\end{minipage}
\caption[1:1 resonance testing run 3: trajectories]{FHN, Second Order, Method 2. The trajectories for each of the values of $\beta$ used in the simulation with $\gamma=0.5$ and $\varepsilon=0.25$. The trajectories have been reconstructed by numerically integrating the tip trajectory equations using the numerical values of the advection coefficients. The trajectories are: (Top Left) $\beta=0.80362$; (Top Right) $\beta=0.80862$; (Middle Left) $\beta=0.81362$; (Middle left) $\beta=0.81362$ (part of the trajectory for this value of $\beta$ is shown to give the reader a perspective of what the trajectory is like close up); (Bottom Left) $\beta=0.81862$; (Bottom Right) $\beta=0.82622$}
\label{fig:ezf_121_new_trajs}
\end{center}
\end{figure}

\begin{figure}[p]
\begin{center}
\begin{minipage}{0.49\linewidth}
\centering
\includegraphics[width=1.0\textwidth, angle=-90]{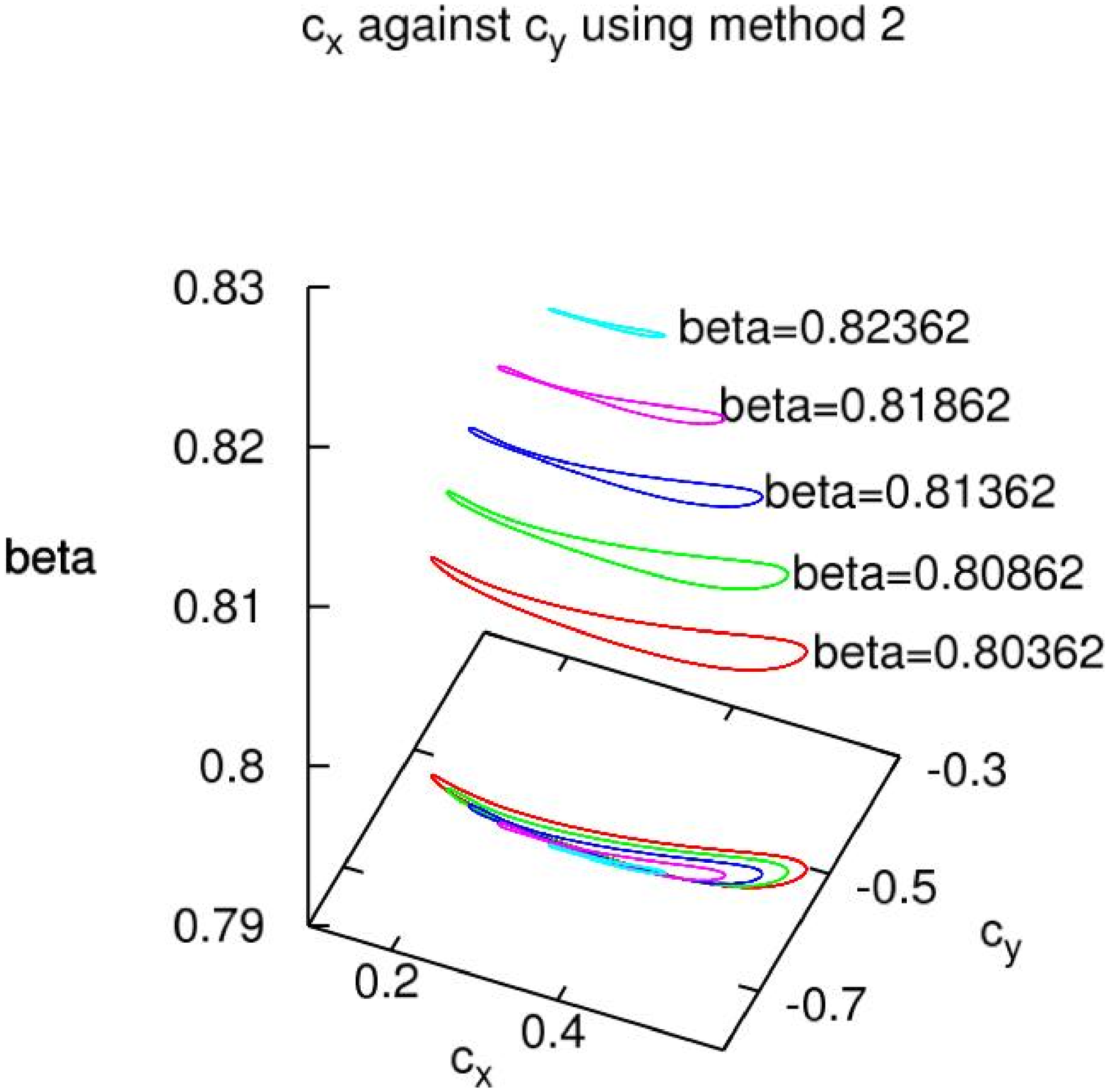}
\end{minipage}
\begin{minipage}{0.49\linewidth}
\centering
\includegraphics[width=1.0\textwidth, angle=-90]{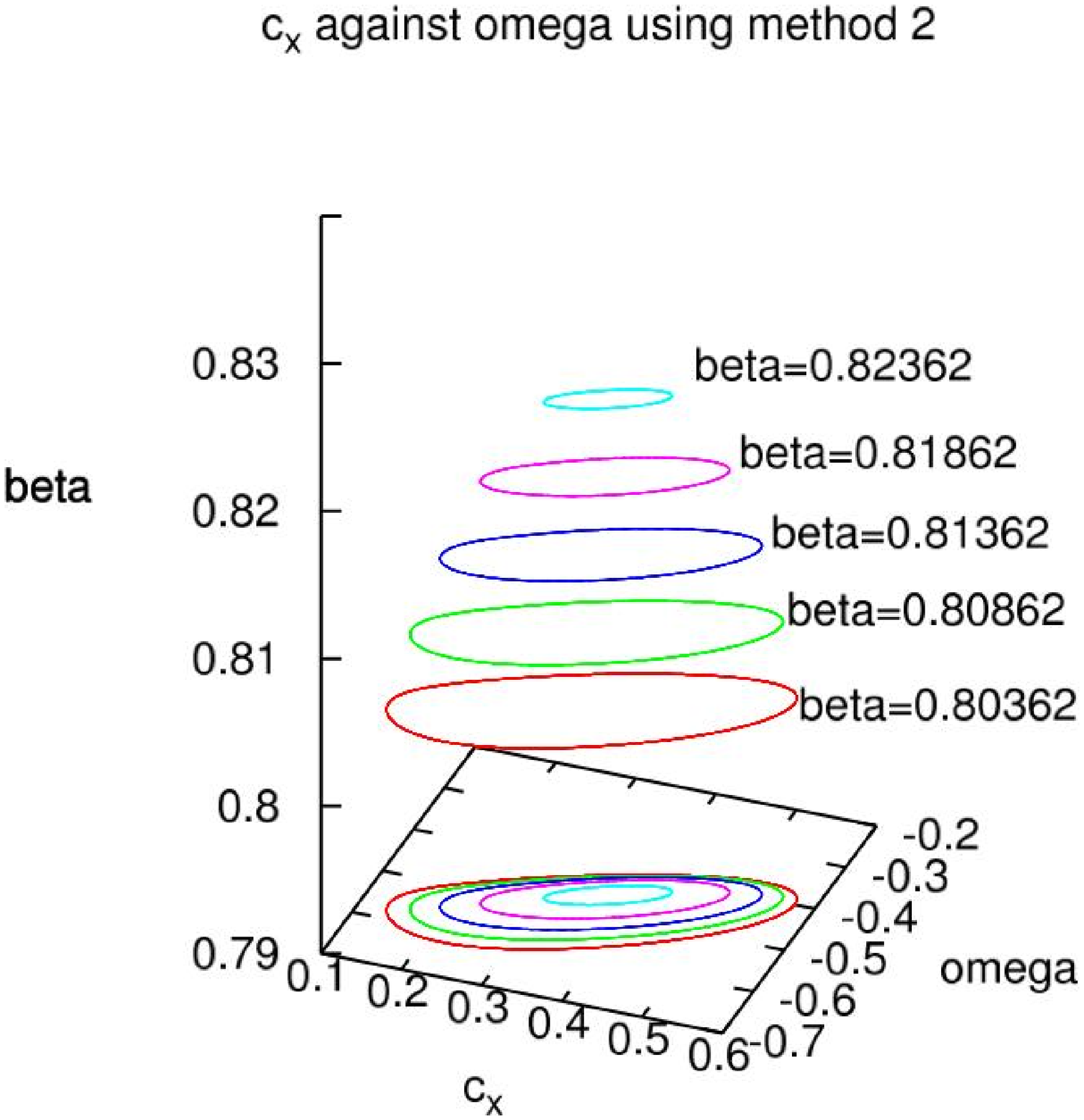}
\end{minipage}
\begin{minipage}{0.49\linewidth}
\centering
\includegraphics[width=1.0\textwidth, angle=-90]{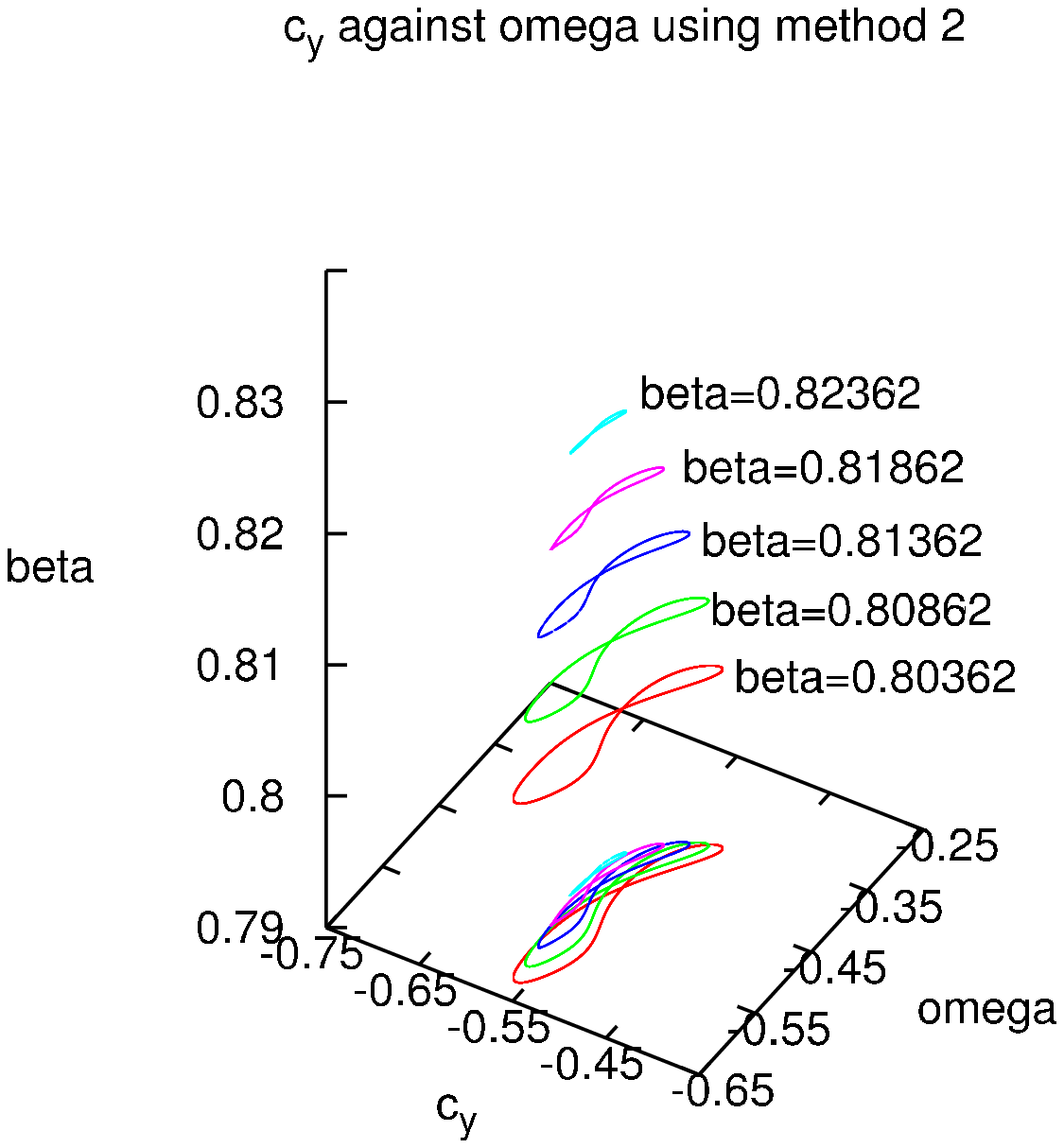}
\end{minipage}
\caption[1:1 resonance testing run 2: quotient solutions]{Quotient solutions. FHN, Second Order, Method 2 with $\gamma=0.5$ and $\varepsilon=0.25$. The parameter $\beta$ is varied and its values are detailed on the plots above:  Quotient solutions: (top left) $c_x$ vs $c_y$; (top right) $c_x$ vs $\omega$; (bottom) $c_y$ vs $\omega$}
\label{fig:ezf_121_new_quots}
\end{center}
\end{figure}

We shall now show the results for when we used not only a second order scheme but also method 2 to calculate the quotient system. The results are shown in Figs.(\ref{fig:ezf_121_new_trajs})-(\ref{fig:ezf_121_new_quots}).

Yet again, we see that the results are qualitatively the same. The limit cycles are generally long and thin, with the limit cycle at the 1:1 resonance point very similar to those generated from the parameters surrounding it.

\subsection{Summary}

To summarize our findings in this investigation, we can say that there is no obvious difference in our numerical studies between the limit cycles at the 1:1 resonance point and those in its environs. In all cases studied, the limit cycles are generally long and narrow, except in a couple of cases. Also, we see that as we move nearer to the $\partial M$ boundary, the limit cycles, as expected decrease in amplitude. With further studies, it could be possible to calculate the amplitudes and possibly show that the increase in the amplitude from the Hopf locus is a $\frac{1}{2}$ power law.

\section{Application II: Large Core Spirals}
\label{sec:ezf_lc}

\subsection{Introduction}

We will now use EZ-Freeze to investigate the behaviour of spirals near the boundary with propagating spirals (in FHN model- denoted by Winfree as $\partial R$ \cite{Winf91}), or with no spirals (Barkley's model). In the investigation of the dynamics of spiral waves near these boundaries, it was noted by two different groups that there was a power law relationship between the radius/period of the spiral and the parameters of the system when studying the asymptotics of the system. However, these powers laws were different.

Hakim and Karma were the first to publish their results on the asymptotic study of spirals near the $\partial R$ boundary \cite{Hakim97}. They concluded that the relationship between one of the model parameters (they assumed by the way that the model used was a general Reaction-Diffusion model, which is parameter dependent) was:

\begin{equation}
R_{tip} \propto(p-p_*)^{-\frac{3}{2}}
\end{equation}
\\
where $R_{tip}$ is the radius of the tip trajectory and $p$ is a parameter of the model, and $p_*$ is the starting parameter.

Elkin et el then noted that the asymptotic approximation of the angular velocity of a rigidly rotating spiral near this boundary was accompanied by a power growth in the period and radius as \cite{bik98a}:

\begin{equation}
R_{tip} \propto(p-p_*)^{-1}
\end{equation}

Hakim and Karma then came back with further analytical evidence of a $-\frac{3}{2}$ power law and extended their studies to further applications such as multiarmed spirals \cite{Hakim99}.

Therefore, we used EZ-Freeze as a tool to investigate these contradictions and show our results below.

As Elkin et al noted, the numerical investigation into these claims can prove to be numerically expensive, in terms of time and system restrictions (memory etc), purely due to the size of the numerical grid we need. For large core spirals, the core radius of the spiral wave can be very large and in the numerical simulations which we performed to generate the following results, we found that the largest radius was approximately 260 space units. For a spacestep of say 0.2, we would need a grid size of at the very least $1300\times1300=1.69\times10^6$, for instance. For a standard desktop computer, the numerical simulations on a grid this size would prove very impractical, if not impossible, to perform.

So, by using EZ-Freeze (i.e. in a frame of reference comoving with the tip of the spiral wave), we can afford a smaller box size, thereby eliminating costly computations, but still get the results as required.

\subsection{Methods}

The advantage of using EZ-Freeze is that is can very quickly determine the values of the advection coefficients, i.e. the values of $c_x$, $c_y$ and $\omega$. From these values we can calculate the radius of the trajectory by using the formula:

\begin{equation}
R_{tip} = \frac{|\bc|}{|\omega|} = \frac{\sqrt{c_x^2+c_y^2}}{|\omega|}
\end{equation}

So, once we have the values of the quotient \chg[p183gram]{solution} for a given set of parameters, we can calculate the relationship between the parameters and the $\omega$ and $R_{tip}$. Consider the following relationship between between a parameter, $a$ and the period, $T$, of the spiral using Hakim \& Karma's results:

\begin{eqnarray}
T &\propto& (p-p_*)^{-\frac{3}{2}}\nonumber\\
\Rightarrow \frac{1}{T} &\propto& (p-p_*)^{\frac{3}{2}}\nonumber\\
\label{eqn:ezf_lc_om_a_hk}
\Rightarrow \omega &\propto& (p-p_*)^{\frac{3}{2}}
\end{eqnarray}

Using the results of Elkin et al, we have, by a similar argument that:

\begin{eqnarray}
T &\propto& (p-p_*)^{-1}\nonumber\\
\label{eqn:ezf_lc_om_a_elk}
\omega &\propto& p-p_*
\end{eqnarray}

Therefore, we investigated which of Eqns.(\ref{eqn:ezf_lc_om_a_hk}) and (\ref{eqn:ezf_lc_om_a_elk}) are indeed correct, if either, from a numerical point of view.

We used Barkley's model with the initial model parameters that were needed to get us as close to the boundary as we could, being found using Barkley's so called ``Flower Garden'' \cite{Bark94a}. These were found to be:

\begin{itemize}
\item $a$ = 0.48
\item $b$ = 0.05
\item $\varepsilon$ = 0.02
\end{itemize}

We also used the following numerical parameters:

\begin{itemize}
\item $L_X$ = $L_Y$ = 30 s.u.
\item $\Delta_x$ = 0.125
\item $\Delta_t$ = $3.9\times10^{-4}$
\item $(x_{inc},y_{inc})$ = (0,14) (in spaceunits)
\end{itemize}

We then ran EZ-Freeze several times, each time the value of $a$ was decreased by 0.001. This was repeated until no spiral wave solution was observed.

It was then possible to observe the relationship between the angular frequency $\omega$ and the parameter $a$.

\subsection{Observations}

\begin{figure}[btp]
\begin{center}
\begin{minipage}[htbp]{0.7\linewidth}
\centering
\psfrag{X}[l]{$a$}
\psfrag{Y}[l]{$\omega$}
\includegraphics[width=0.9\textwidth]{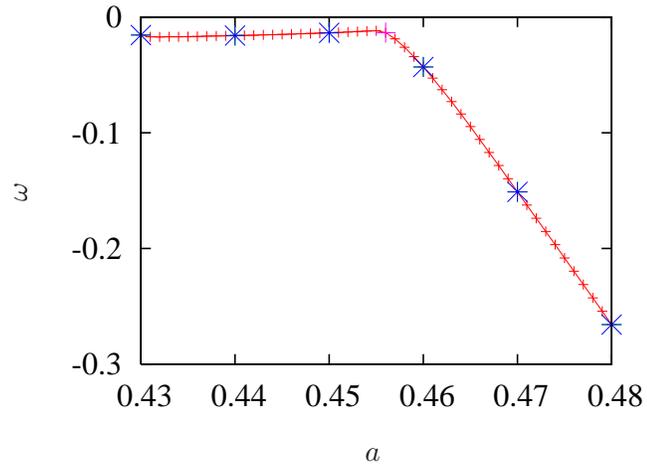}
\end{minipage}
\caption[Large core: $a$ versus $\omega$]{$a$ versus $\omega$}
\label{fig:ezf_lc_omega}
\end{center}
\end{figure}

Before we show the results of the investigation into which asymptotic method is numerical verified, we show in Fig.(\ref{fig:ezf_lc_omega}) the full range of how omega is varying with the parameter as get very close to the boundary $\partial R$.

We can see straightaway that there are two distinct regions of this plot. With the parameters in the range $0.456\leq a\leq0.48$, we have what appears at first sight to be a linear growth. We then reach a point, which we shall denote as the \emph{Critical Point}, at which the growth changes, and then for parameters in the range $0.43\leq a<0.456$ we have a much slower growth (in fact this is decay) and what \chg[p184gram]{}appears to be a stabilizing in $\omega$.

\begin{figure}[btp]
\begin{center}
\begin{minipage}[htbp]{0.49\linewidth}
\centering
\psfrag{X}[l]{$a$}
\psfrag{Y}[l]{$\omega$}
\includegraphics[width=0.9\textwidth]{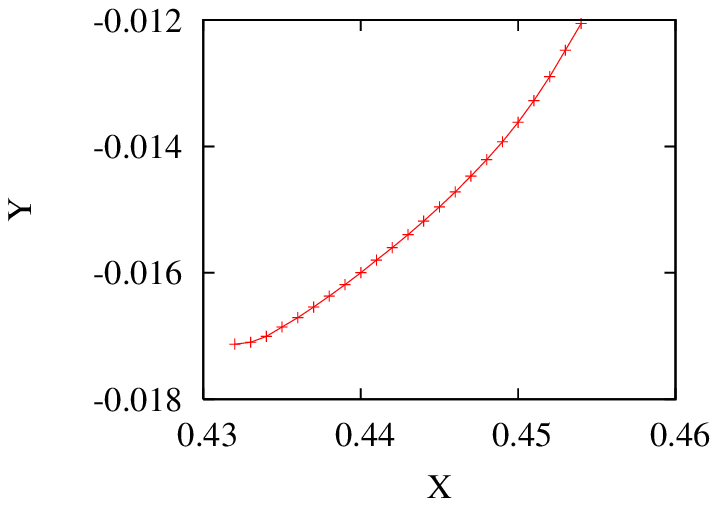}
\end{minipage}
\begin{minipage}[htbp]{0.49\linewidth}
\centering
\psfrag{X}[l]{$a$}
\includegraphics[width=0.9\textwidth]{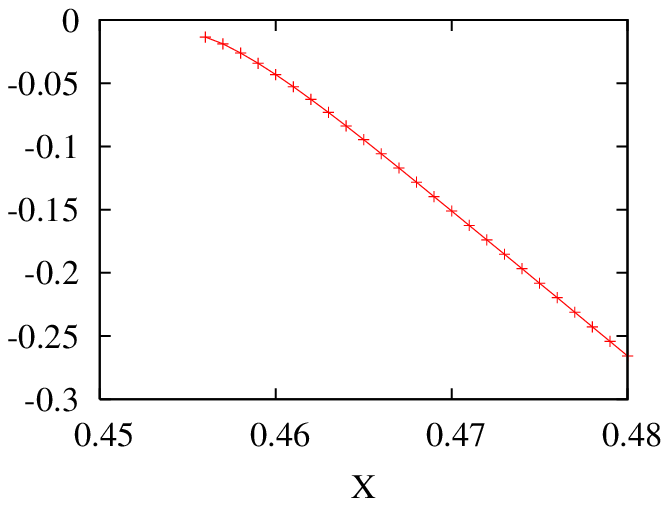}
\end{minipage}
\caption[Large core: $a$ versus $\omega$ split either side of critical point]{(left) $a$ versus $\omega$ ($0.43\leq a<0.456$), and (right) $a$ versus $\omega$ ($0.456\leq a\leq0.48$)}
\label{fig:ezf_lc_omega_split}
\end{center}
\end{figure}

We also show in Fig.(\ref{fig:ezf_lc_omega_split}) the plot in Fig.(\ref{fig:ezf_lc_omega}) split over the two noted halves.

\begin{figure}[btp]
\begin{center}
\begin{minipage}[htbp]{0.45\linewidth}
\centering
\includegraphics[width=0.7\textwidth]{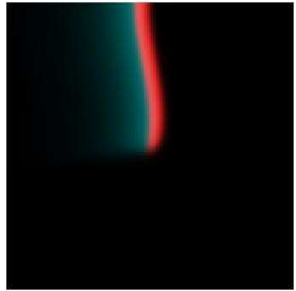}
\end{minipage}
\caption[Spiral wave solution at the critical point]{Spiral wave solution at the critical point, $a=0.456$.}
\label{fig:ezf_lc_crit_fingers_crit_point1}
\end{center}
\end{figure}

\clearpage
The Critical Point, $a=0.456$, is a crucial numerical observation. At this point the so called \emph{Critical Finger} \cite{Hakim99} is formed. We show in Fig.(\ref{fig:ezf_lc_crit_fingers_crit_point1}) the critical finger.

\begin{figure}[btp]
\begin{center}
\begin{minipage}[htbp]{0.45\linewidth}
\centering
\includegraphics[width=0.7\textwidth]{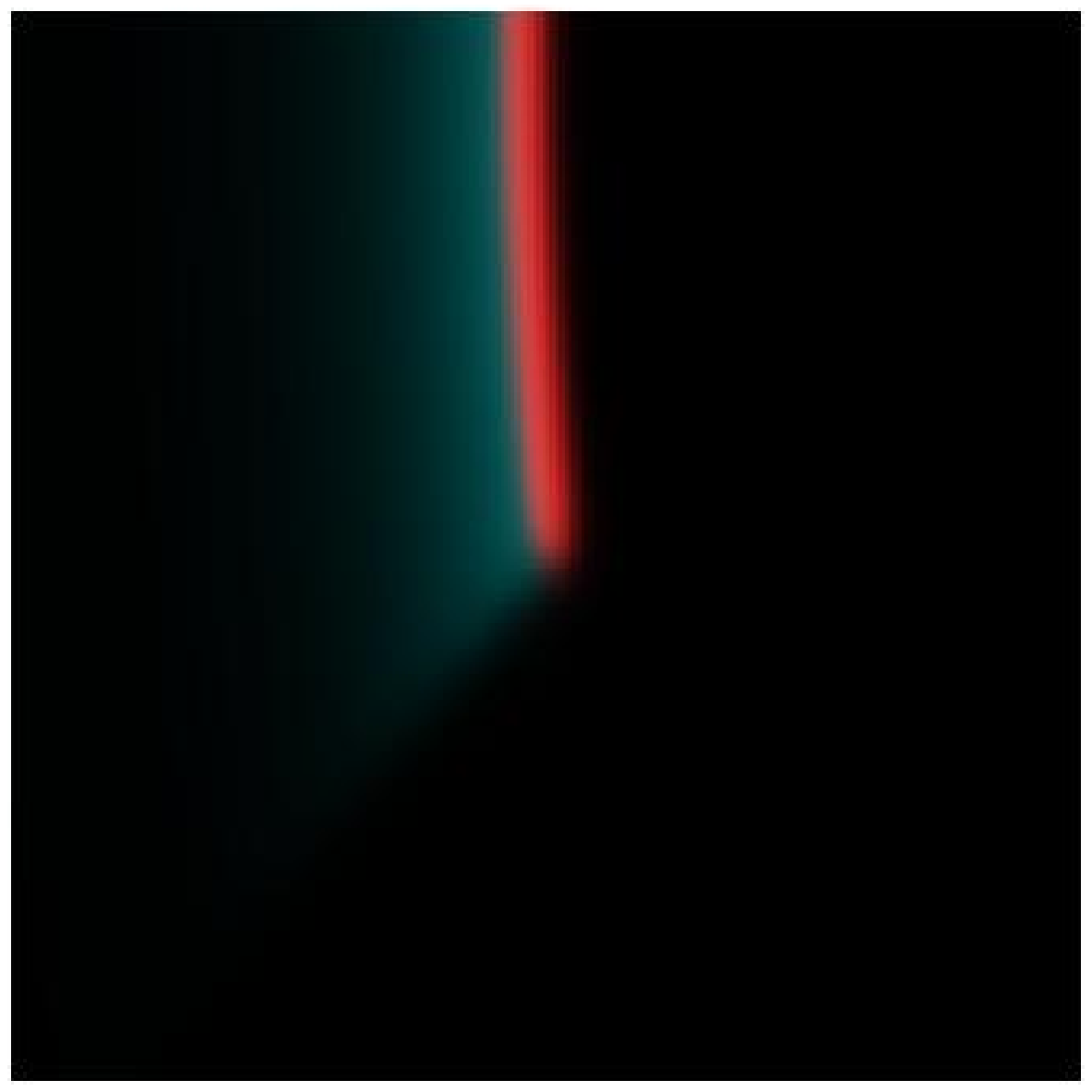}
\end{minipage}
\begin{minipage}[htbp]{0.45\linewidth}
\centering
\includegraphics[width=0.7\textwidth]{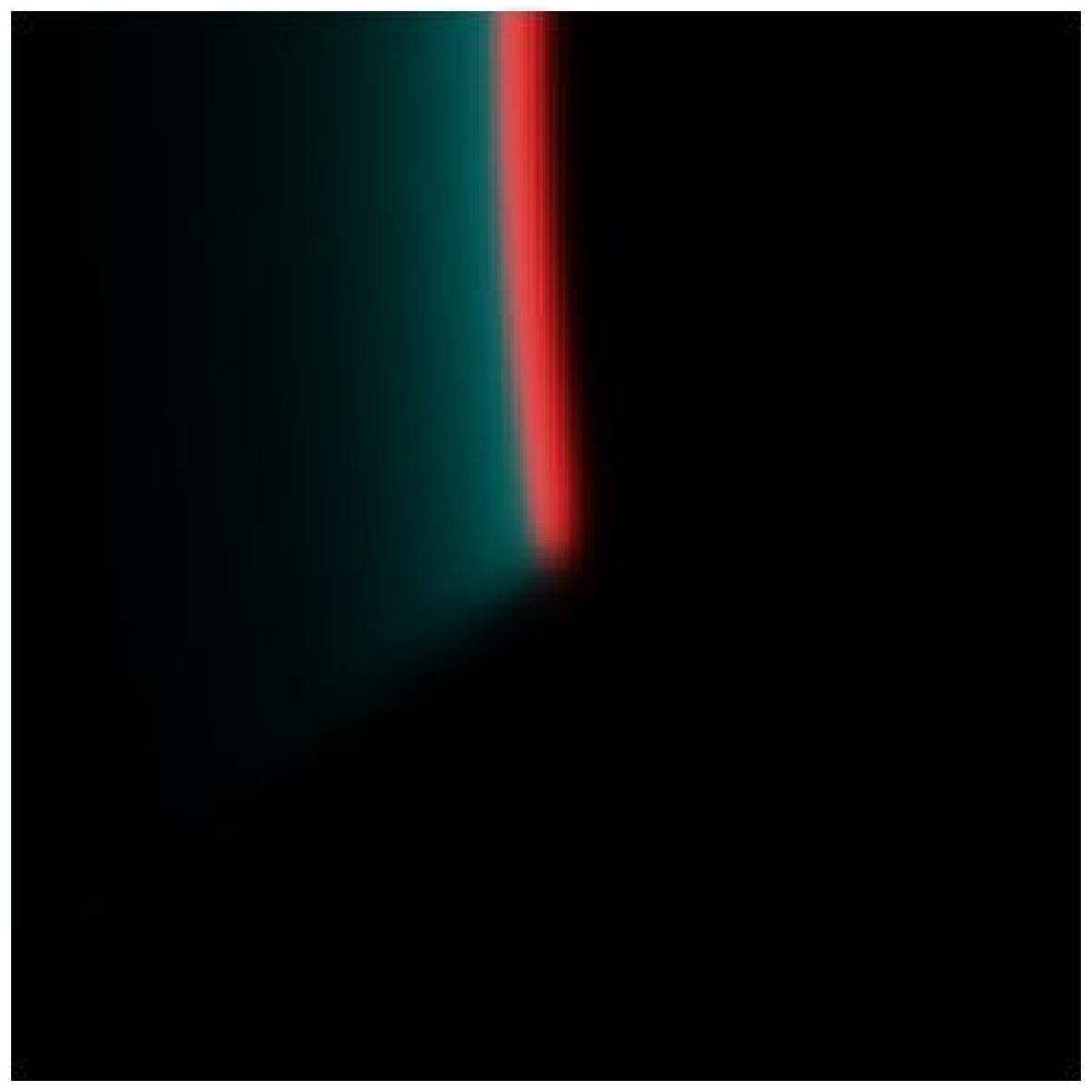}
\end{minipage}
\begin{minipage}[htbp]{0.45\linewidth}
\centering
\includegraphics[width=0.7\textwidth]{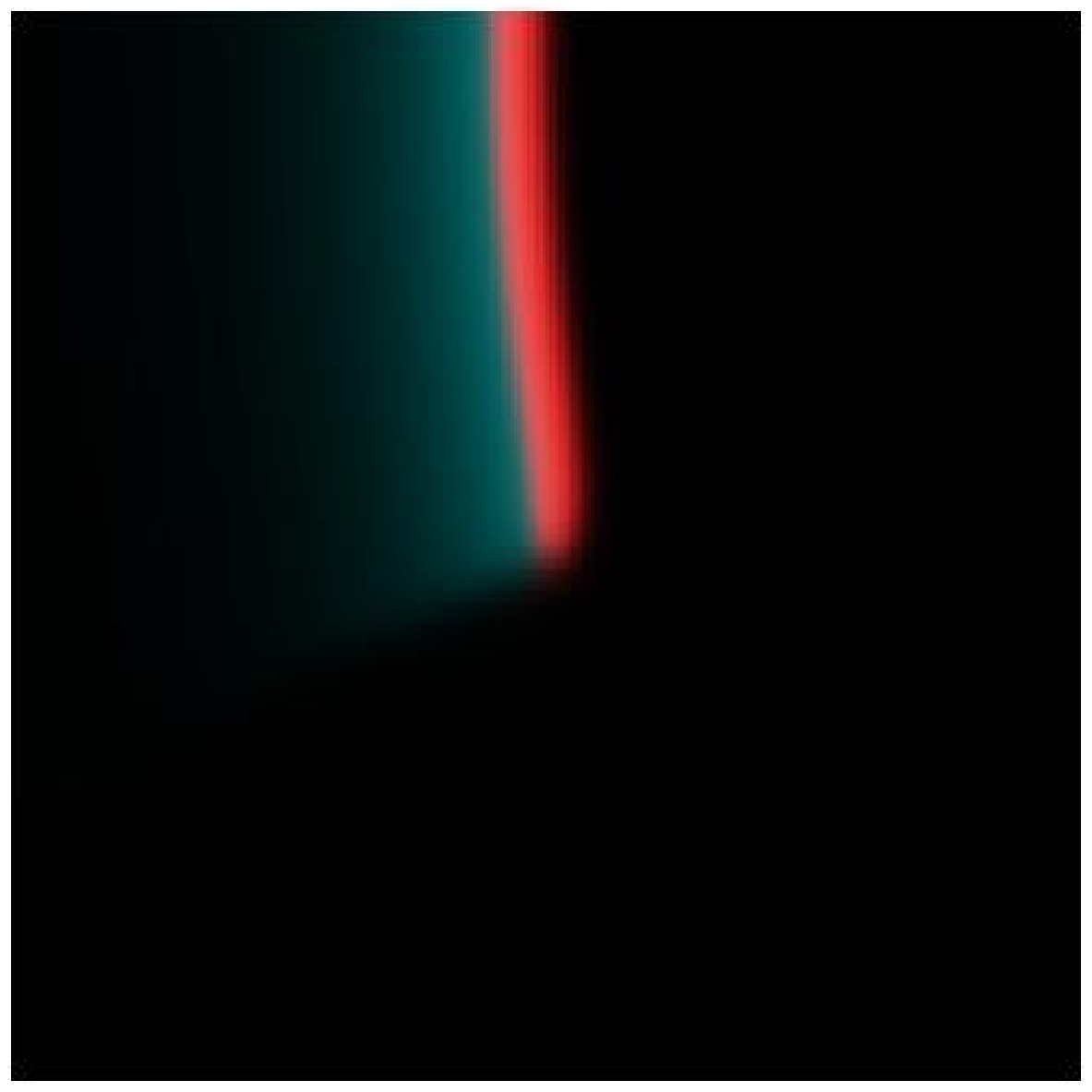}
\end{minipage}
\begin{minipage}[htbp]{0.45\linewidth}
\centering
\includegraphics[width=0.7\textwidth]{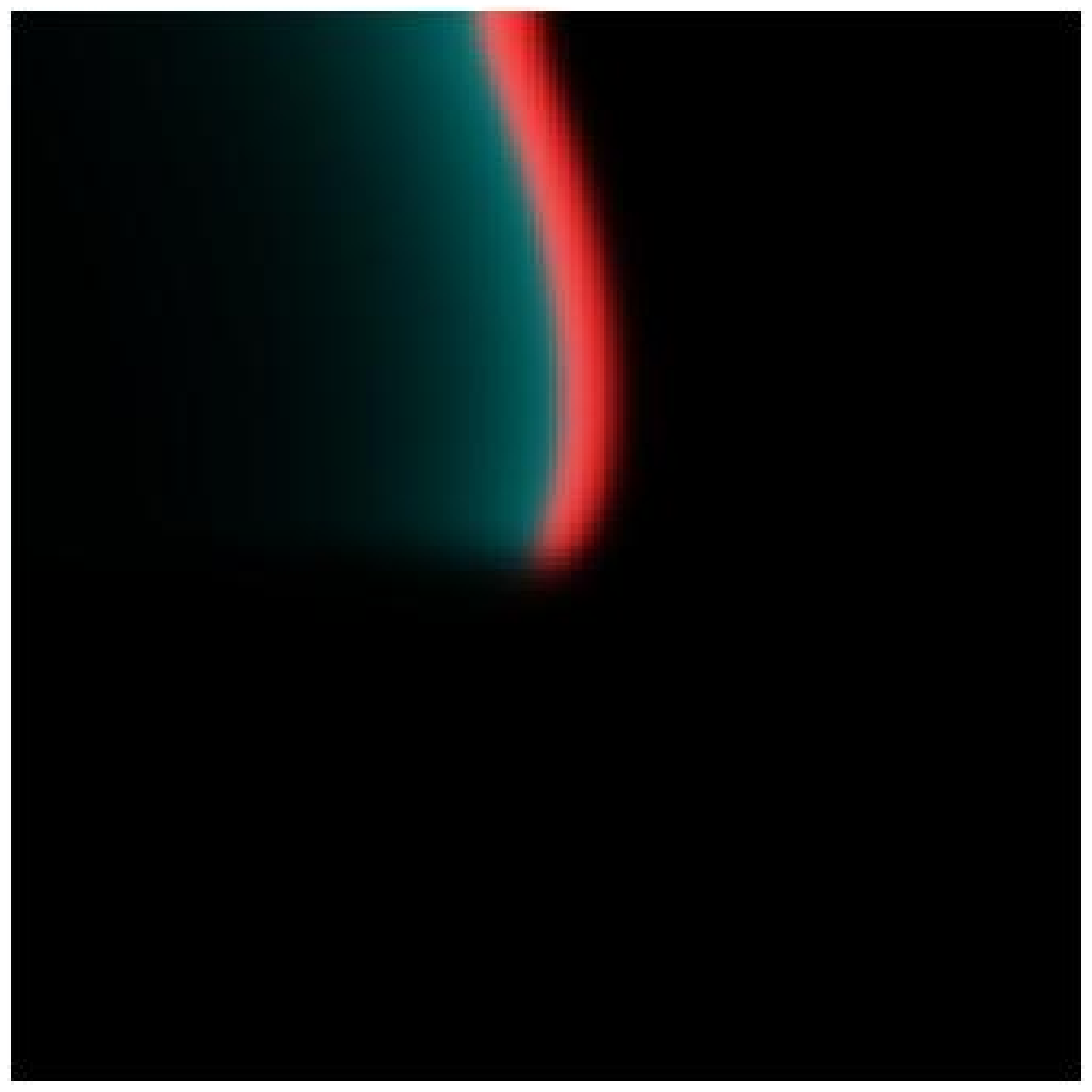}
\end{minipage}
\begin{minipage}[htbp]{0.45\linewidth}
\centering
\includegraphics[width=0.7\textwidth]{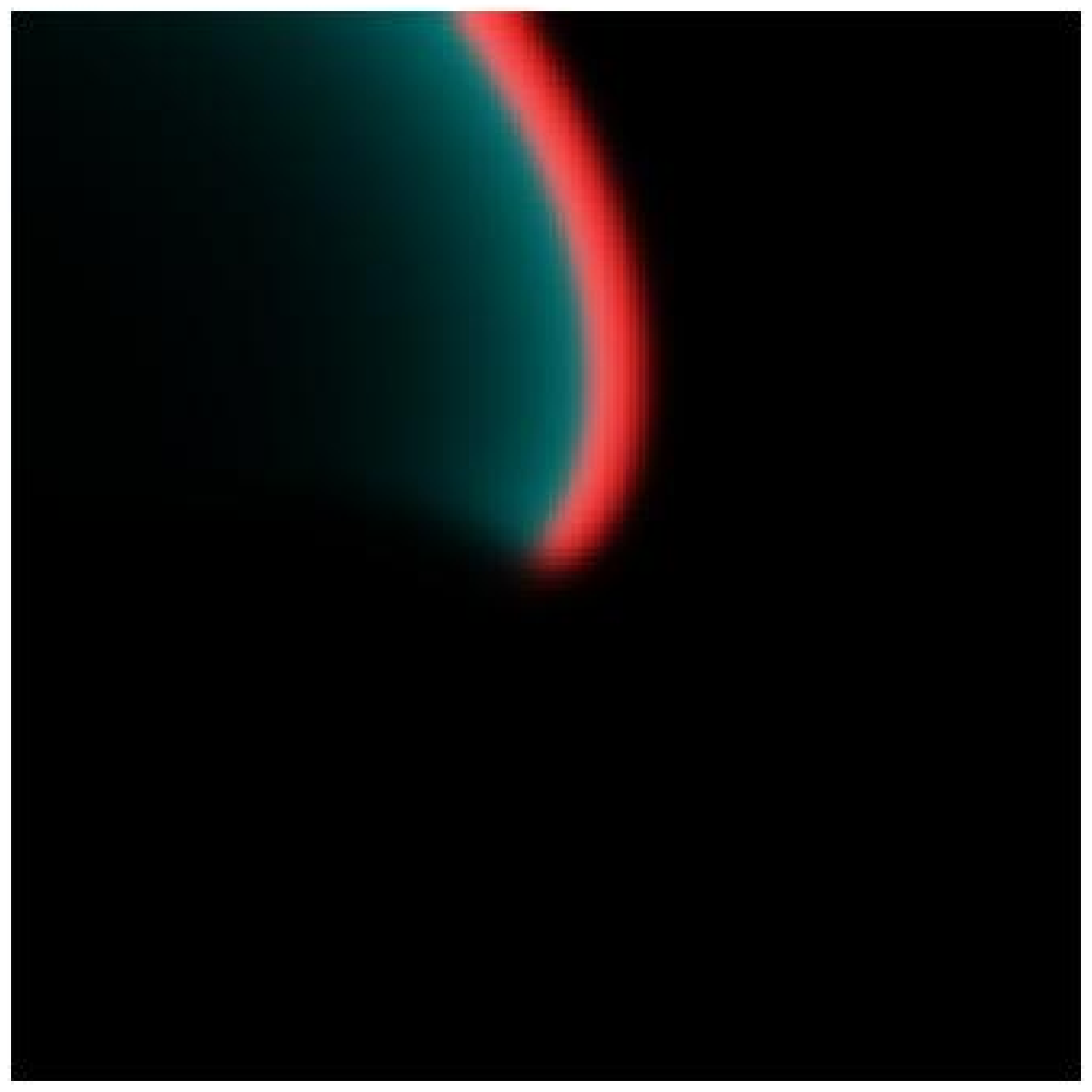}
\end{minipage}
\begin{minipage}[htbp]{0.45\linewidth}
\centering
\includegraphics[width=0.7\textwidth]{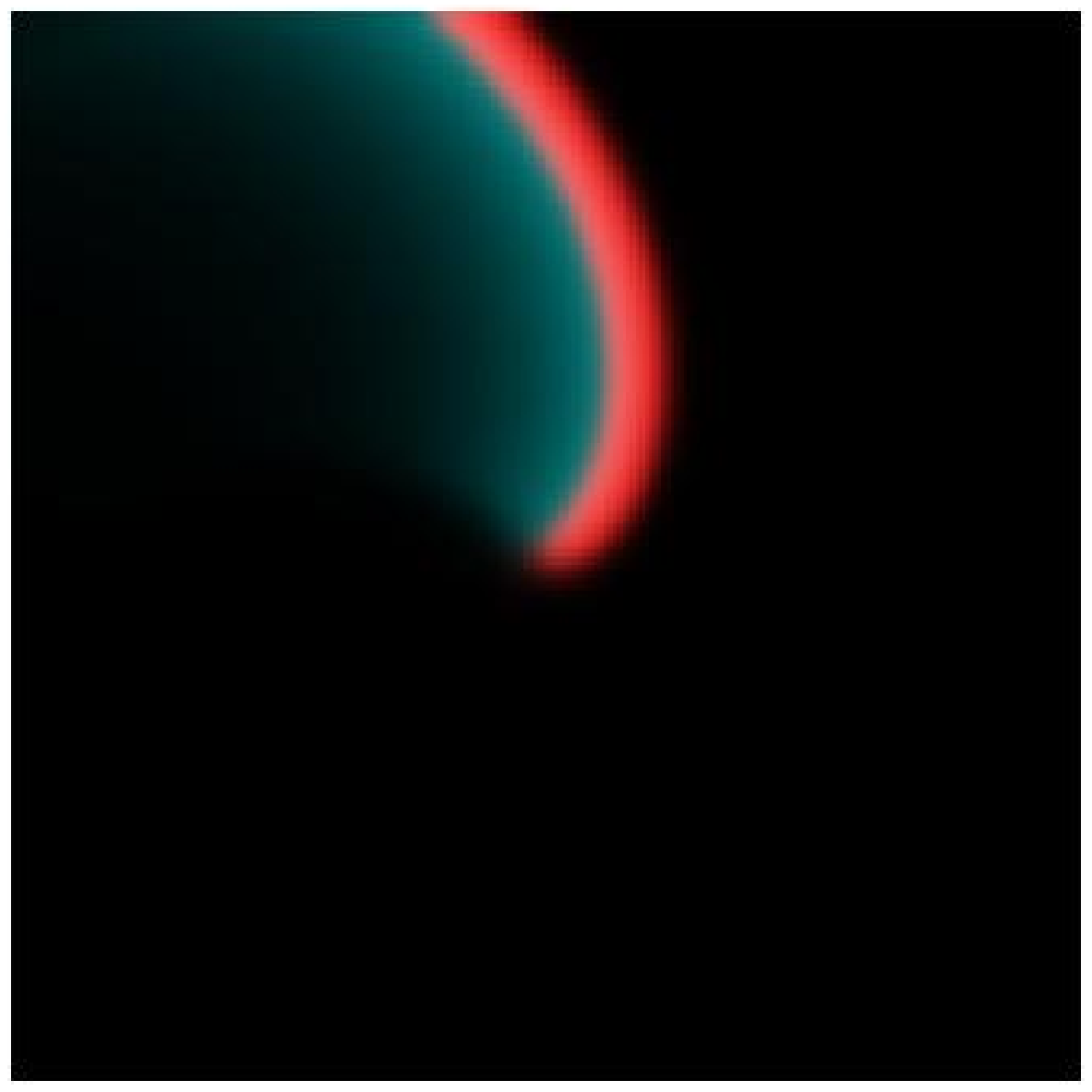}
\end{minipage}
\caption[Large core: solutions]{Plots of the solutions in the comoving frame for various values of $a$, showing critical fingers for $a$=0.43, 0.44 \& 0.45; (top left) $a=0.43$; (top right) $a=0.44$; (middle left) $a=0.45$; (middle right) $a=0.46$; (bottom left) $a=0.47$; (bottom right) $a=0.48$;}
\label{fig:ezf_lc_crit_fingers}
\end{center}
\end{figure}

We also show in Fig.(\ref{fig:ezf_lc_crit_fingers}) the solutions on either side of this critical point, which are also indicated on Fig.(\ref{fig:ezf_lc_omega}) by a {\color{blue}{blue}} cross, with  the critical point shown with a {\color{pink}{pink}} cross.

\begin{figure}[btp]
\begin{center}
\begin{minipage}[htbp]{0.45\linewidth}
\centering
\includegraphics[width=0.7\textwidth]{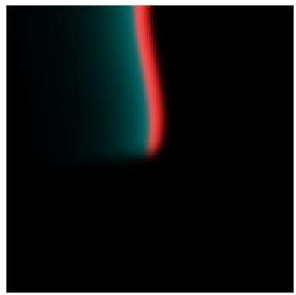}
\end{minipage}
\caption[Spiral wave solution at the critical point]{Spiral wave solution at the critical point, $a=0.456$.}
\label{fig:ezf_lc_crit_fingers_crit_point}
\end{center}
\end{figure}


\subsubsection{Results: Before the Critical Point ($a\geq0.456$)}

\begin{figure}[btp]
\begin{center}
\begin{minipage}[htbp]{0.9\linewidth}
\centering
\psfrag{X}[l]{$a$}
\psfrag{Y}[l]{$|\omega|$}
\includegraphics[width=0.78\textwidth]{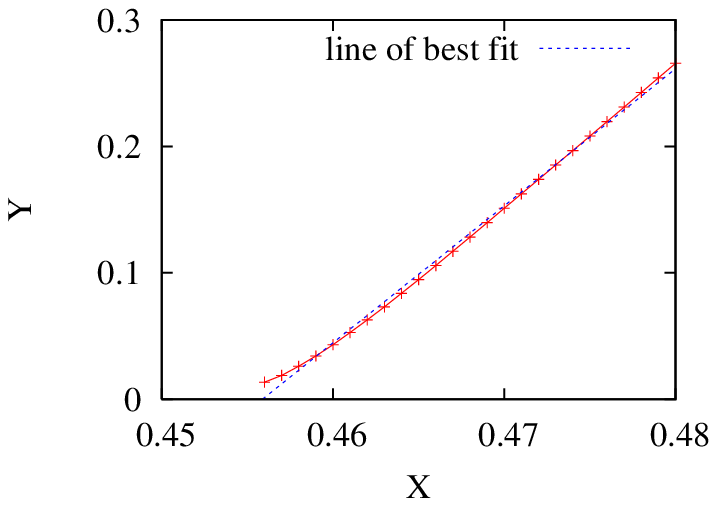}
\end{minipage}
\begin{minipage}[htbp]{0.9\linewidth}
\centering
\psfrag{X}[l]{$a$}
\psfrag{Y}[l]{$|\omega|^{\frac{2}{3}}$}
\includegraphics[width=0.8\textwidth]{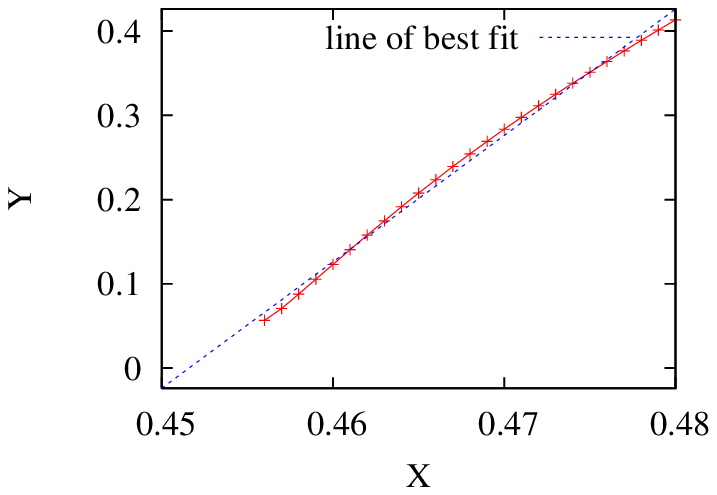}
\end{minipage}
\caption[Large core: right of critical point]{(top) $a$ versus $|\omega|$; (bottom) $a$ versus $|\omega|^{\frac{2}{3}}$}
\label{fig:ezf_lc_front}
\end{center}
\end{figure}

We now show the results for when $a\geq0.456$, which is just before the critical finger is formed. The results are shown in Fig.(\ref{fig:ezf_lc_front}).

It is clear that the relationship is better when we have $a$ against $|\omega|$. Therefore, before we reach the critical point, we can say that the results by Elkin et al is better suited, when we numerically investigate large core spiral waves in Barkley's model.


\subsubsection{Results: After the Critical Point ($a<0.456$)}

\begin{figure}[tb]
\begin{center}
\begin{minipage}[tb]{0.7\linewidth}
\centering
\psfrag{X}[l]{$a$}
\psfrag{Y}[l]{$|\omega|^{4.3}$}
\includegraphics[width=1.0\textwidth]{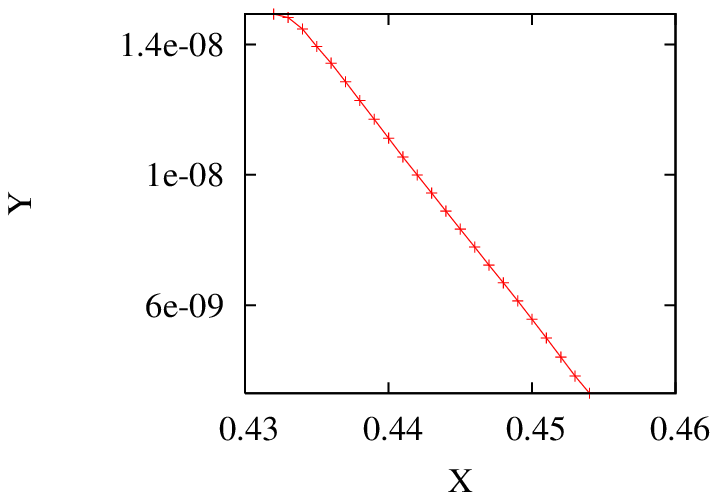}
\end{minipage}
\begin{minipage}[tb]{0.7\linewidth}
\centering
\psfrag{X}[l]{$a$}
\psfrag{Y}[l]{$|\omega|^{4.3}$}
\includegraphics[width=1.0\textwidth]{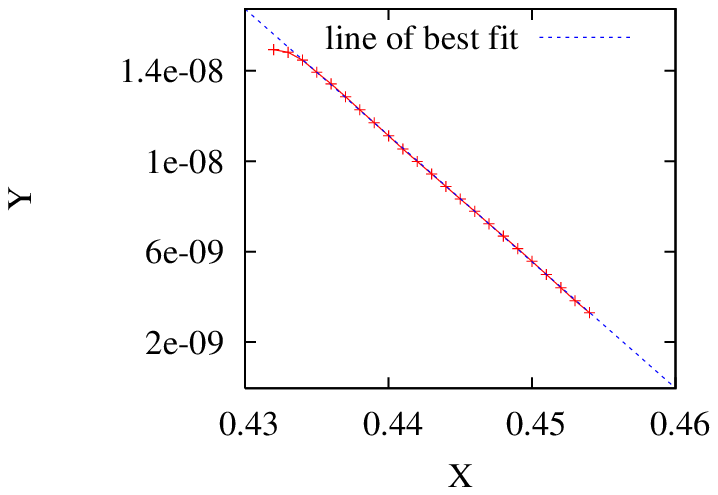}
\end{minipage}
\caption[Large core: left of critical point]{(top) $a$ versus $|\omega|^{4.3}$ with no fitted line; (bottom) $a$ versus $|\omega|^{4.3}$}
\label{fig:ezf_lc_back}
\end{center}
\end{figure}

Let us now consider when $a\geq0.456$, which is just after the critical finger is formed. The results are shown in Fig.(\ref{fig:ezf_lc_back}).

There is, at this present time, no developed asymptotic theory relating to the rate of growth of the period or radius of the spiral wave once we have crossed the critical point. 

What we have shown here is that for Barkley's model, we can see that the rate of growth of the angular velocity is a power law $a\propto\omega^{4.3}$ implying that the relationship between the period, $T$, and the parameter $a$ is:

\begin{equation}
T \propto a^\frac{10}{43}
\end{equation}

This can hopefully be used as a marker for future asymptotic theory into this phenomena.


\subsection{Summary}

To summarize, we can see we have a Critical Point at which the Critical Finger is formed. Numerically, we have shown that the results of Elkin et al are better suited for numerical considerations in Barkley's model. We have shown that, in a region not yet investigated in any literature that we have researched, the relationship between the period and the parameters of the system is again a power law, and is in fact a $\frac{10}{43}$ power law.

\section{Conculsion}
\label{sec:ezf_conclusion}
To conclude, we have seen that we have shown a numerical method of solving a system of equations which provides spiral wave solutions in a frame of reference which is moving with the tip of the spiral wave. This work has resulted in a program called EZ-Freeze.

We have shown that the numerical methods implemented withing EZ-Freeze provide accurate solutions, and in some instances, the solutions are produced with remarkable speed.

We also showed that this method can be used for numerical simulations that have would have not been possible otherwise, or which would have been numerically expensive in terms of CPU time and memory. The classic case was for the investigation into large core spirals. There were two conflicting theories which were developed in the late 1990's, and it was admitted back then that the numerical investigation into such methods would have been computationally expensive. We have shown that EZ-Freeze can be used to investigate such solutions with very little effort. We also confirmed that from our calculations, the results of Elkin et al proved to be better suited to Barkley's model, using the parameters stated. We also showed some initial work into what occurs after the critical point and what direction any researcher who is thinking of studying this phenomena should proceed in.

We also showed some results into the study of 1:1 resonance, which we believe have not yet been shown before. It was interesting to \chg[p190]{note} there that the limit cycles relating to the 1:1 resonance solutions are not qualitatively different form those of other meandering solutions. We have also seen that the shapes of the limit cycles are quite unique and can be studied using EZ-Freeze.

As for future work, we have made the code for EZ-Freeze available online for others to download and use. There are many uses of EZ-Freeze including the generation of the value of $\omega$. This is vital information for any researcher of spiral waves, since we can then determine many things about the spiral wave including the form of the response function (see Chap.\ref{chap:5} for more details).

\chapter{Numerical Calculation of Response Functions}
\label{chap:5}

\section{Introduction}
\label{sec:rf_intro}
This chapter is concerned with the numerical calculation of the response functions to the adjoint linearised Reaction-Diffusion system in the rotating frame of reference. We shall be considering only the FHN system of equations.

We begin the chapter with a review of what a Response Function is and why we need to study them. We shall also review briefly the literature that has been published on this subject and how the works presented in the literature can be implemented into our problem.

We shall then describe the numerical methods used to calculate such functions and will also introduce a program called \verb|evcospi|, which numerically calculates the critical Response Functions. This program is still under development.

We shall then show some examples of how \verb|evcospi| works, and present some results from convergence testing that we conducted on the program.

One of the most important factors in generating accurate numerical solutions from \verb|evcospi| is the use of accurate initial conditions within the program. We shall present a review of the way the initial conditions were originally generated, and then present a more refined and faster approach to generating accurate initial conditions.

This chapter is based on results which are currently in preparation to be published \cite{bik08}.

\section{Response Functions \& Their Importance}
\label{sec:rf_rf}
\subsection{The Response Function Problem}

We shall be considering rigidly rotating spiral wave solutions in the Reaction-Diffusion system of equations. We note from Chap.\ref{chap:4} that in a frame of reference that is comoving with the spiral wave with the same angular velocity $\omega$, the rigidly rotating spiral wave solutions are independent of time. Therefore, the system of equations which we now consider are:

\begin{eqnarray}
\label{eqn:rf_rda}
0 &=& \bDD\nabla^2\bv+\boof(\bv)+\omega\pderiv{\bv}{\theta},
\end{eqnarray}
\begin{eqnarray}
\label{eqn:rf_pin}
v^{(i)}(x_c,y_c,t) &=& v_*.
\end{eqnarray}
\\
where $\bv,\boof\in\mathbb{R}^n,\bDD\in\mathbb{R}^{n\times n}$. Also, we have that $\bv=(v^{(1)},v^{(2)},\hdots,v^{(n)})$, $1\leq i\leq n$ and $v_*$ is a constant (note that the value of $v_*$ depends on the excitable boundaries of the medium as determined by the local kinetics). In this case, $\omega$ is the angular velocity of the spiral wave and is actually an unknown constant in this equation. Therefore, we have $n+1$ unknowns, and since Eqn.(\ref{eqn:rf_rda}) \chg[af]{consists of} $n$ equations, we require an additional equation (Eqn.(\ref{eqn:rf_pin})) to make it a closed system. This system of equations gives us spiral wave solutions in a rotating frame of reference. 

We recall from Chap.\ref{chap:3} that we can formulate an eigenvalue problem, where the linear operator $L$ in this instance is given by:

\begin{equation}
\label{eqn:rf_L}
L\balp = \bDD\nabla^2\balp+\booF(\bv_0)\balp+\omega\pderiv{\balp}{\theta}.
\end{equation}
\\
where $\booF(\bv_0)=\pderiv{\boof(\bv_0)}{\bv}$. \chg[af]{We note that in this chapter, we are concerned with just the rotating frame of reference, not the comoving as used in Chap.\ref{chap:3}.} The eigenvalue problem is therefore:

\begin{equation}
L\bphi_i = \lambda_i\bphi_i,
\end{equation}
\\
where the eigenfunctions $\bphi_i$ corresponding to the eigenvalues $\lambda_i$ are time independent functions.

Also, we can construct the adjoint eigenvalue problem:

\begin{equation}
\label{eqn:rf_evalue_prob_rf}
L^+\bpsi_j = \bar{\lambda}_j\bpsi_j,
\end{equation}
\\
where the adjoint operator $L^+$ is given by:
\chgex[ex]{}
\begin{equation}
\label{eqn:rf_adj}
\chg[]{L^+\bbet = \bDD\nabla^2\bbet+\booF^+(\bv_0)\bbet-\omega\pderiv{\bbet}{\theta}},
\end{equation}
\\
and $\booF(\bv_0)=\pderiv{\boof(\bv_0)}{\bv}$.

The eigenfunctions to this adjoint linear operator are what are termed as {\bf Response Functions}. So, to calculate them, we need to solve Eqn.(\ref{eqn:rf_evalue_prob_rf}).

Now, as we have seen from Chap.\ref{chap:3}, when we have a rigidly rotating spiral wave solution, there are three critical eigenvalues that are located on the imaginary axis. These eigenvalues are $\lambda_{0,\pm1}=0,\pm i\omega$, and relate to the symmetry of Reaction-Diffusion-Advection system. 

Therefore, we have that the eigenvalues corresponding to the Response Functions are $\bar{\lambda}_{0,\pm1}=0,\mp i\omega$. 

Also, we have that the Response Functions $\bpsi_j$ and the eigenfunctions $\bphi_i$ to the operator $L$ satisfy the biorthogonality condition:

\begin{equation}
(\bpsi_j,\bphi_i) = \delta_{ji}
\end{equation}
\\
where $\delta_{ji}$ is the Kronecker delta function.


\subsection{Importance of Response Functions}

Before we move on, we shall briefly review the relevant literature regarding the calculation of response functions and their uses.

It has been said that Response Functions are as important to spiral waves, as mass is to matter \cite{bik03}. This is a very strong and critical opinion. But the reasoning behind such a statement is that you just can't seem to get away from Response Functions when studying spiral waves.

Biktasheva et al \cite{bik03}, showed that the velocities of a drifting spiral wave can be formulated as:

\begin{eqnarray*}
\partial_t\Phi &=& \epsilon F_0(R,\omega t-\Phi)\\
\partial_tR    &=& \epsilon F_1(R,\omega t-\Phi)
\end{eqnarray*}
\\
where the functionals $F_n$ are given by:

\begin{eqnarray*}
\bar{F}_n(t) &=& e^{in\Phi}\int_{t-\frac{\pi}{\omega}}^{t+\frac{\pi}{\omega}}\frac{\omega\dd\tau}{2\pi}\int\int_{\mathbb{R}^2}\dd^2re^{-in\omega\tau}\times\\
                &&  \langle\bpsi_n\left(\rho(r-R),\theta(r-R)+\omega\tau-\Phi\right),\bh\rangle,\quad n=0,\pm1\nonumber
\end{eqnarray*}
\\
and $\bh$ is a symmetry breaking perturbation.

Clearly, the Response Functions $\bpsi_n$ feature prominently in the prediction of spiral wave drift. 

Similarly, in our research (Chap.3), we see that Response Functions play a vital role in determining the equations of motion for the tip of a drifting spiral wave:
\chg[p194eqn]{}
\begin{eqnarray*}
\odRt  &=&   \left[c_0-\epsilon(2(\bar{\bpsi}_1,\bht(\bv_0,\br,t))+\frac{\bc_0}{\omega_0}(\bpsi_0,\bht(\bv_0,\br,t)))\right]e^{i\Theta}\\
\odTht &=& \omega_0-\epsilon (\bpsi_0,\bht(\bv_0,\br,t))
\end{eqnarray*}
\\
where $\bht$ is a transformed symmetry breaking perturbation.

We have also seen that we can have Floquet Response Functions when we are considering the dynamics of meandering spiral waves that are drifting. The system of equations that determine the solution in the space of group orbits, is given by:

\begin{eqnarray*}
\deriv{\bV_0}{\tau}  &=& \bg(\bV_0)\\
\deriv{\bV_1}{\tau}  &=& L\bV_1+\bS+O(\epsilon)\\
\deriv{\theta}{\tau} &=& \epsilon(\bpsi_*,\bk)
\end{eqnarray*}
\\
where 
\begin{eqnarray*}
\bg(\bV_0) &=& \booF(\bV_0)+(\bc_0,\Bh_\br)\bV_0+\omega_0\Bh_\theta\bV_0\\
L\bV_1     &=& \deriv{\booF(\bV_0)}{\bV}+(\bc_0,\Bh_\br)\bV_1+\omega_0\Bh_\theta\bV_1\\
\bS        &=& (\bc_1,\Bh_\br)\bV_0+\omega_1\Bh_\theta\bV_0+\bHt-(\bpsi_0,\til{\bH})\deriv{\bV_0}{\tau}
\end{eqnarray*}

This system is solved using a solvability condition:

\begin{eqnarray*}
(\bpsi_n,\bV_0) &=& 0\\
(\bpsi_n,\bS)   &=& 0
\end{eqnarray*}
 
Clearly, we have that the Response Functions, or in this case, the Floquet Response Functions featuring prominently in the equations.

Therefore, it is evident that the \chg[af]{response} functions are extremely important in the study of spiral waves.

\section{Numerical Implementation}
\label{sec:rf_numerics_imp}
We shall now review the methods that are employed into \verb|evcospi| which numerically solve the adjoint eigenvalue problem to give us the Response Function solutions:

\begin{equation}
\label{eqn:rf_adj_prob}
L^+\bpsi_j = \bar{\lambda}_j\bpsi_j
\end{equation}
\\
where $L^+$ is given by \chgex[ex]{}(\ref{eqn:rf_adj}). We note that in the definition of $L$, we have two unknown variables built into the operator - $\bv_0$ and $\omega$.

Therefore, the first problem that we need to solve is to find what $\bv_0$ and $\omega$ are. Once we have found these, we can substitute them into (\ref{eqn:rf_adj_prob}) and find the Response Functions.

So, in the next couple of sections we shall describe the general methods that we will be using to find the initial solutions and then the eigenvalues and eigenfunctions for a general eigenvalue problem.

The techniques implemented into \verb|evcospi| are based on methods documented by Wheeler and Barkley \cite{Bark06}.


\subsection{Calculation of the initial conditions}

Firstly, we note that we are now going to consider just a general eigenvalue problem of the form:

\begin{equation*}
\label{eqn:rf_prob}
L\bphi_i = \lambda_i\bphi_i
\end{equation*}
\\
where $L$ is given by \chg[af]{}(\ref{eqn:rf_L}). As mentioned, we need to find an initial steady state solution $\bv_0$ and also the rotational velocity $\omega$. The first step we take is to get an initial approximation to $\bv_0$ and $\omega$. This means solving the following \chg[af]{equations}:

\begin{eqnarray}
\label{eqn:rf_steady_eqn}
0 &=& \bDD\nabla^2\bv_0+\boof(\bv_0)+\omega\pderiv{\bv_0}{\theta}
\end{eqnarray}
\begin{eqnarray}
v_0^{(i)}(x_c,y_c,t) &=& v_*\nonumber
\end{eqnarray}
\\
where $\bv_0,\boof\in\mathbb{R}^n$, $\bDD\in\mathbb{R}^{n\times n}$, $\bv_0=(v_0^{(1)},v_0^{(2)},\hdots,v_0^{(n)})$, $1\leq i \leq n$, and $v_*$ is a carefully chosen value depending on the local kinetics of the model being used and which is chosen within the region of excitability of the model.

There are several ways to generate an initial approximation to $\bv_0$, and these methods are described in Sec.(\ref{sec:rf_ic}).

So, once we get an initial approximation to $\bv_0$ and $\omega$, we then need to find just how good this initial approximation is. We do this by calculating the norm of the RHS of Eqn.(\ref{eqn:rf_steady_eqn}), and we call upon the subroutine set out in LAPACK (this is a standard numerical analysis package used to solve a large number of linear equations)\cite{LAPACK}. The routine called upon is \verb|dnrm2_|, which calculates the residual norm. 

We then apply Newton Iterations to the nonlinear problem (\ref{eqn:rf_steady_eqn}) and repeat these iterations until the solution converges to a specified order. After each iteration, we perform a check of how good the solution is after that iteration by calling the routine \verb|dnrm2_|. 


\subsection{Numerical Calculation of the eigenvalues and eigenfunctions}

Once we have a refined steady state solution and value of $\omega$ we then need to solve the following problem:

\begin{eqnarray*}
L^+\bpsi_j &=& \bar{\lambda}_j\bpsi_j
\end{eqnarray*}

For the rest of this section, we shall consider a general eigenvalue problem and describe (briefly) the methods which we used to solve this, which by the way, are implemented into \verb|evcospi|.

Consider a general eigenvalue problem:

\begin{eqnarray}
\label{eqn:rf_num_evalprob}
 \bA\bx &=& \mu\bx
\end{eqnarray}
\\
where $\bx\in\mathbb{R}^n$ is the eigenvector corresponding to the eigenvalue $\mu$, and $\bA\in\mathbb{R}^{n\times n}$.

We would like to solve system (\ref{eqn:rf_num_evalprob}) numerically to find both the eigenvalues and corresponding eigenvectors, in the knowledge that we know what the eigenvalues are from theoretical considerations.

Let us assume that we \chg[af]{can} approximate the eigenvalues $\mu$ by $\nu$, i.e. $\mu\approx\nu$, where $\nu$ are eigenvalues of matrix $\bB$. We also assume that the eigenvectors corresponding to $\nu$ are the same as those for $\mu$:

\begin{eqnarray}
\label{eqn:rf_num_evalprob1}
\bB\bx &=& \nu\bx
\end{eqnarray}

Using a Cayley transform, we can establish a relationship between the matrices $\bA$ and $\bB$:

\begin{eqnarray*}
\bA &=& (\zeta I+\bB)^{-1}(\xi I+\bB)
\end{eqnarray*}
\\
where $\xi,\zeta\in\mathbb{R}$. Similarly, we have a relationship between the eigenvalues:

\begin{eqnarray*}
 \mu &=& \frac{\xi+\nu}{\zeta+\nu}
\end{eqnarray*}

By doing this transformation, we can transform the eigenvalues that we want to the eigenvalues with the greatest magnitude, simply by varying the two parameters $\xi$ and $\zeta$.

The next step is to use \emph{Arnoldi Iterations} to find the dominant eigenvalues. This is a more stable form of the \emph{Power Iteration}. A Power Iteration to solve the eigenvalue problem (\ref{eqn:rf_num_evalprob1}) is when we take an arbitrary vector, $\til{\bx}$, which we assume to be normalized (i.e. $||\til{\bx}||=1$), and apply matrix $B$ to it. Each iteration consists of normalizing the previously found vector and applying $B$ to it.

For the eigenvalue problem (\ref{eqn:rf_num_evalprob1}), let us assume that the eigenvector has the form:

\begin{equation*}
\bx = \sum_i a_i\bx_i
\end{equation*}

Applying Power Iterations we get:

\begin{eqnarray*}
B\bx &=& \sum_i\lambda_ia_i\bx_i\\
B^2\bx &=& \sum_i\lambda^2_ia_i\bx_i\\
\hdots && \hdots\\
B^n\bx &=& \sum_i\lambda^n_ia_i\bx_i
\end{eqnarray*}

Now, as $n\rightarrow\infty$, we find that the largest eigenvalue dominates and that:

\begin{eqnarray*}
B^n\bx &=& \lambda^n_*a_*\bx_*(1+o(1))
\end{eqnarray*}

Hence we can establish the eigenvector $\bx_*$ corresponding to the eigenvalue with the largest absolute value, $\lambda_*$. An estimate of this eigenvalue is given by \cite{kumar}:

\begin{equation*}
\lambda_* = \frac{\bx^T_*B\bx_*}{\bx^T_*\bx_*}
\end{equation*}

The Arnoldi Method is very similar to the Power Iteration. However, instead of iterating just one vector $\bx$, it iterates a finite dimensional ``Krylov'' subspace, the basis of which is orthonormalized after each iteration by Gram-Schmidt orthogonalization.

The program \verb|evcospi| takes advantage of the numerical package \emph{ARPACK}, which includes all the routines necessary to use the Arnoldi method \cite{ARPACK}. Also, to use some of the techniques used in the Newton Iterations, it utilises the routines found in \verb|LAPACK| \cite{LAPACK}.

\section{Generation of Initial Conditions}
\label{sec:rf_ic}
In this section we shall describe the methods that are used to firstly generate the initial conditions (IC's) and then how they are implemented into \verb|evcospi|. We will then describe a method which we have devised to generate initial conditions quicker then the other methods but still retaining high orders of accuracy. 

The accuracy of initial conditions within \verb|evcospi| is critical, since the program is sensitive to initial conditions. If we had a set of initial conditions that are not very accurate, the performing Newton Iterations may result in the initial solution  not being refined, and therefore the calculation of the Response Functions not being made.


\subsection{Generation of IC's using EZ-Spiral}

There are three methods by which \verb|evcospi| IC's are generated. The first is by using the program EZ-Spiral; the next is using a program called AUTO; and the last method is using the final conditions generated from a run of \verb|evcospi| as the initial conditions for the next run. We shall describe two of these methods below (EZ-Spiral and EZ-Freeze). We will not go into any sort of detail regarding AUTO, since this method is due to be made redundant within \verb|evcospi|.

As mentioned earlier on in this thesis, EZ-Spiral is a program that is available to download free of charge and which generates spiral wave solutions using fast numerical calculations \cite{bark91,barkweb}. It originally incorporated Barkley's model but it has been adjusted to include FHN kinetics.

The idea is to generate a rigidly rotating spiral wave solution whose center of rotation is as close to the center of the box as possible. When a wave is initiated, the user can plot the tip path of the spiral wave. It was advised to move the spiral wave using the arrow keys so that the spiral is rotating around a point which is as close to the center of the box as possible. There are some major disadvantages to this. Firstly, how do you tell when it is rotating around the center of the box? Do you get a ruler out and measure, or do you just go ``by eye''? This is a very crude method and one which actually threw up problems when trying to implement it into \verb|evcospi|.

The next task that needs to be done is to find what the rotational velocity, $\omega$, is. To do this the user was advised to plot the trajectory of the $u$-field against time as recorded in the \verb|history.dat| file produced by EZ-Spiral. From this plot, the user could estimate the period, $T$, of the spiral (by measuring the time period between successive peaks for instance) and then calculate by hand the angular velocity using:

\begin{equation*}
\omega = \frac{2\pi}{T}
\end{equation*}

The final conditions generated by EZ-Spiral could then be used for the IC's in \verb|evcospi|. 

How the IC's are read into \verb|evcospi| is of vital importance. The IC's must also be as accurate as possible in order to generate accurate Response Functions. For EZ-Spiral the major problem was the fact that the final conditions are generated on a cartesian grid, but \verb|evcospi| requires the IC's in polar form. So, we must transform the data from a cartesian grid to a polar grid.

The program \verb|evcospi|, does this by taking the \chg[p199]{angular step} as given by the user \chg[p199]{in the command line} and the EZ-Spiral final conditions file,\chg[p199]{} working out what the cartesian equivalent is for each polar grid point, and \chg[af]{calculating} the nearest cartesian grid point to each polar point. The $u$ and $v$ values for each polar point are taken as the values of $u$ and $v$ fields at the nearest cartesian grid point to each polar point.

The main and obvious disadvantage to this is that the allocation of the $u$ and $v$ values to each grid point is very crude. The method just picks out the nearest $u$ and $v$ values and allocates them to the polar grid coordinates. This therefore, provides inaccuracies within the solution and therefore the performance of Newton Iterations on that solution greatly increases the chances of the solution not converging.


\subsection{Generation of IC's using EZ-Freeze}

We shall now describe the method with which EZ-Freeze can be used to generate IC's for evcospi.

We know from Chap.\ref{chap:4} that EZ-Freeze has several uses. One of these uses is to calculate $\omega$, the rotational velocity of a spiral wave. We have also shown that this value is very good in terms of numerical accuracy.

So, there are two processes which we use within EZ-Freeze which we will utilise to give use accurate IC's for \verb|evcospi|. 

The first is to calculate $\omega$ for a rigidly rotating spiral wave. This is achieved by the following algorithm:

\begin{enumerate}
\item Initiate a spiral wave with the desired physical and numerical parameters. Note that the numerical parameters must be sufficient to give us an accurate spiral wave solution (see convergence testing);
\item Once the wave has settled from its transient period to a rigidly rotation spiral wave (\chg[af]{done} by observing the tip trajectory), switch on the advection terms by pressing \verb|z|;
\item The wave will oscillate for a brief moment and then the advection coefficients ($c_x$, $c_y$ and $\omega$) will converge to their final values;
\item Either terminate the program by pressing \verb|q| or allow the program to run through it full number of timesteps as determined in the \verb|task.dat| file (denoted by \verb|nsteps|). 
\end{enumerate}

A final conditions file \verb|fc_ecx.dat| will be generated, amongst other files, which is written in the ``own'' conditions usable by \verb|evcospi|. We shall now describe the algorithm incorporated into EZ-Freeze which generates these conditions.

\subsubsection{Algorithm for generating the file fc\_ecx.dat}

EZ-Freeze can be used to calculate not only the angular velocity, $\omega$, but also the translational velocities, $\bc=(c_x,c_y)$, of the spiral. We note from Sec.(\ref{sec:rev_swdyn}) that the radius of the trajectory is given by:

\begin{equation}
\label{eqn:fc_ic_rs}
R_c = \frac{|\bc|}{|\omega|} = \frac{\sqrt{c_x^2+c_y^2}}{|\omega|}
\end{equation}

\begin{figure}[btp]
\begin{center}
\begin{minipage}[htbp]{0.49\linewidth}
\centering
\psfrag{a}{$(X,Y)$}
\psfrag{b}{$\omega$}
\psfrag{c}{$\bc$}
\includegraphics[width=0.7\textwidth]{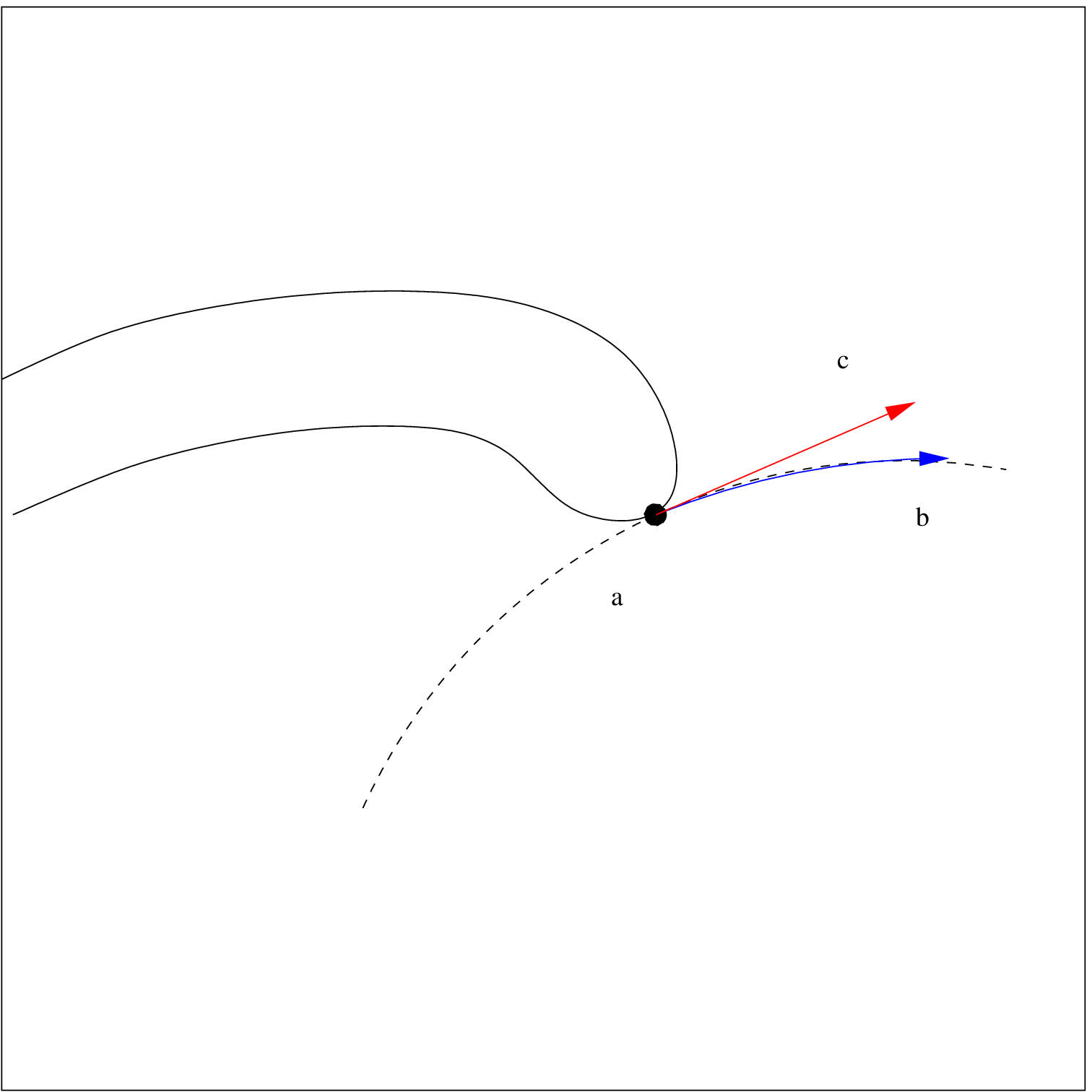}
\end{minipage}
\begin{minipage}[htbp]{0.49\linewidth}
\centering
\psfrag{a}{$(X,Y)$}
\psfrag{b}{$R_c$}
\psfrag{c}{$(x_c,y_c)$}
\includegraphics[width=0.7\textwidth]{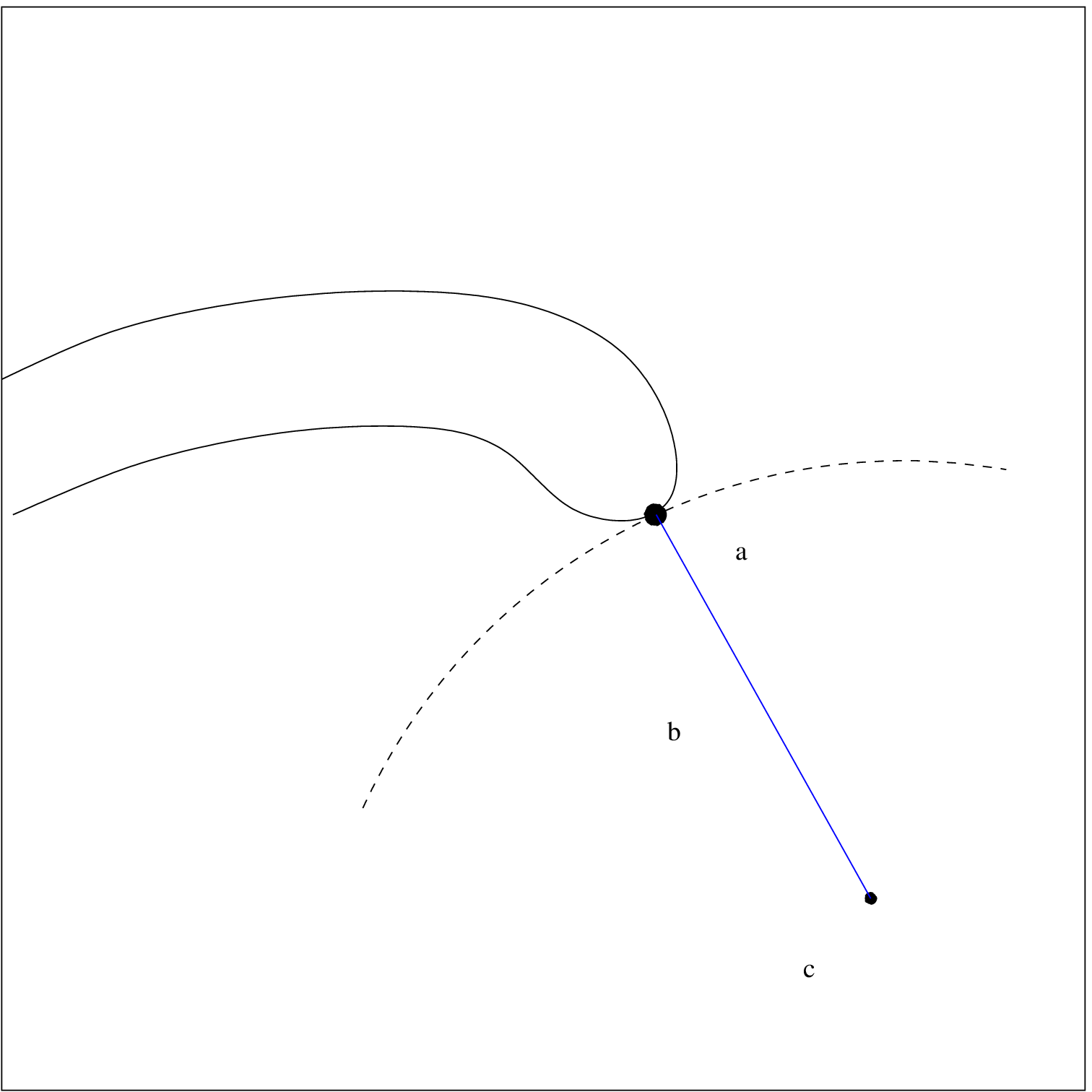}
\end{minipage}
\caption[Determining the radius of the trajectory and the center of rotation]{Illustration of the translational and angular velocities used in determining the radius of the trajectory (left), and determination of the center of rotation (right).}
\label{fig:rf_ic_radius}
\end{center}
\end{figure}

Our next task is to determine the center of rotation of spiral. This is a straightforward task and, by considering Fig.(\ref{fig:rf_ic_radius}), we can see that the center of rotation is given by:

\begin{eqnarray*}
(x_c,y_c) &=& (X,Y)-\hat{\bn}R_c\text{sgn}(\omega)
\end{eqnarray*}
\\
where $\hat{\bn}$ is the unit normal along the radius $R_c$. This can further be reduced by using Eqn.(\ref{eqn:fc_ic_rs}):

\begin{eqnarray*}
(x_c,y_c) &=& (X,Y)-\hat{\bn}R_c\text{sgn}(\omega)\nonumber\\
&=& (X,Y)-\frac{(-c_y,c_x)}{|\bc|}\frac{|\bc|}{|\omega|}\text{sgn}(\omega)\nonumber\\
&=& (X,Y)-\frac{(-c_y,c_x)}{|\omega|}\text{sgn}(\omega)\nonumber\\
&=& (X,Y)-\frac{(-c_y,c_x)}{\omega}
\end{eqnarray*}

Having determined the center of rotation, we can determine the radius of the inscribed circle which will form the maximum radius of the polar grid we will be using.

\begin{figure}[btp]
\begin{center}
\begin{minipage}[htbp]{0.7\linewidth}
\centering
\psfrag{a}{$R_{IC}$}
\psfrag{b}{$R_c$}
\includegraphics[width=0.7\textwidth]{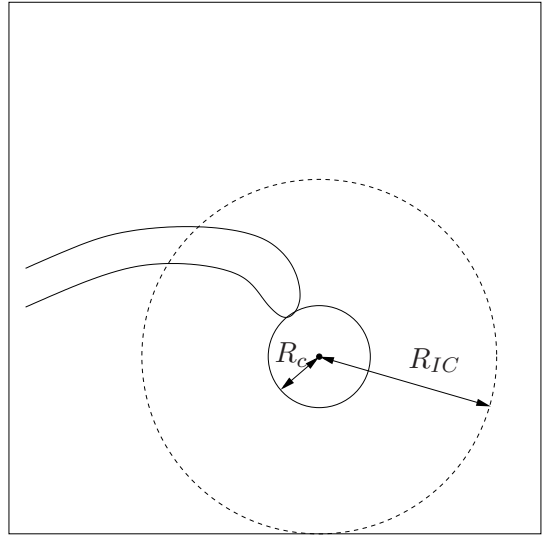}
\end{minipage}
\caption{Determination of the inscribed circle.}
\label{fig:rf_ic_inscribed}
\end{center}
\end{figure}

This is, yet again, another \chgex[ex]{straightforward} task, made even easier by the fact that we are using a square (or \chg[af]{rectangular)} domain. All we need to do is determine the minimum distance from the center of rotation to the boundaries of the domain. This is illustrated in Fig.(\ref{fig:rf_ic_inscribed}).

Therefore, the radius of the inscribed circle is:

\begin{equation*}
r_{IC} = \min\left\{x_c, L_X-x_c, y_c, L_Y-y_c\right\}
\end{equation*}

Next, we must determine the numerical parameters on the polar grid. We know that the maximum radius of the Inscribed Circle is $r_{IC}$. The actual radius of the inscribed circle which we will define in several paragraphs time, is denoted by $R_{IC}$. In order to get an accurate transformation from cartesian coordinates to polar coordinates, we must have that the radial step, \chg[p202]{$\Delta_r$,} and the angular step, $\Delta_\theta$, must satisfy:

\begin{eqnarray*}
\Delta_r      &\leq& \Delta_x\\
\Delta_\theta &\leq& \frac{\Delta_x}{R_{IC}}
\end{eqnarray*}
\\
where $\Delta_x$ is the space step.

Hence, we can say that the number of radial grid points (circles on the polar grid) must be:

\begin{equation*}
n_r = \frac{r_{IC}}{\Delta_r}
\end{equation*}

Since we must have that $N_r\in\mathbb{Z}^+$, then we must round $n_r$ down to the nearest integer, $N_r$. The actual radius of the inscribed circle is therefore:

\begin{equation*}
R_{IC} = \Delta_r\,N_r
\end{equation*}

Also, if the angular step is $\Delta_\theta\leq\frac{\Delta_x}{R_{IC}}$, then we have that the number of angular grid points is:

\begin{eqnarray*}
n_\theta &=& \frac{2\pi}{\Delta_\theta}\nonumber\\
\Rightarrow n_\theta &\geq& \frac{2\pi R_{IC}}{\Delta_x}
\end{eqnarray*}

Now, we must take into account the way that the information is used in \verb|evcospi|. The derivatives for the diffusion are calculated using a 5-point Laplacian. This must mean that the number of angular points is divisible by four. Therefore, our final number of angular grid points is:

\begin{eqnarray*}
N_\theta &=& 4\left(\left[\frac{n_\theta}{4}\right]+1\right)
\end{eqnarray*}
\\
where $[x]$ is the integer part of $x$. 

Finally, our angular step is now refined to:

\begin{equation*}
\Delta_\theta = \frac{2\pi}{N_\theta}
\end{equation*}

We have now determined the values of our numerical parameters ($\Delta_\theta$, $\Delta_r$, $N_\theta$, $N_r$, and $R_{IC}$) for the polar grid.

\begin{figure}[btp]
\begin{center}
\begin{minipage}[htbp]{0.7\linewidth}
\centering
\psfrag{a}[r]{$(x_0,y_1)$}
\psfrag{b}[r]{$(x_0,y_0)$}
\psfrag{c}{$(x_1,y_1)$}
\psfrag{d}{$(x_1,y_0)$}
\psfrag{e}{$(x,y)$}
\includegraphics[width=0.7\textwidth]{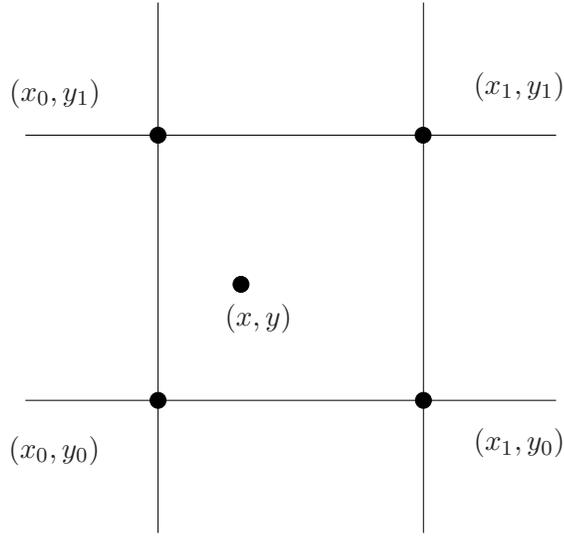}
\end{minipage}
\caption{Linear Interpolation.}
\label{fig:rf_ic_bilinear}
\end{center}
\end{figure}

We next need to use bilinear interpolation to get an estimate of the $u$ and $v$-field values at each polar grid point. Consider, Fig.(\ref{fig:rf_ic_bilinear}).

Take a point on the polar grid $(i,j)$, where $i$ is the radial point and $j$ is the angular point. We first need to convert the polar coordinates, $(i,j)$, to cartesian coordinates, $(x,y)$. This is done using:

\begin{eqnarray*}
x &=& Xc+(i\Delta_r\cos(j\Delta_\theta))\\
y &=& Yc+(i\Delta_r\sin(j\Delta_\theta))
\end{eqnarray*}
\\
where $(X_c,Y_c)$ are the coordinates for the center of the spiral. We note that $x,y\in\mathbb{R}$ and can be \chg[af]{represented} as:

\begin{eqnarray*}
x &=& x_0+p\\
y &=& y_0+q
\end{eqnarray*}
\\
where $x_0,y_0\in\mathbb{Z}$, and $0<p<1$, $0<q<1$.

The value of the point \chg[p204gram]{$u(x,y)$} using bilinear interpolation is:
\chg[p204eqn]{}
\begin{eqnarray*}
\chg[]{u(x,y)} &=& u(x_0,y_0)(1-p)(1-q)+u(x_1,y_0)p(1-q)\nonumber\\
&& +u(x_0,y_1)(1-p)q+u(x_1,y_1)pq
\end{eqnarray*}
\\
where $x_1=x_0+1$ and $y_1=y_0+1$. This is done for all points on the polar grid.

The next task is to make sure that data calculated above which is written to the file \verb|fc_ecx.dat|, is in an order which can be read using \verb|evcospi|'s ``own'' format.

Once this is done, the file can be copied to a directory which is \chg[af]{accessible} by \verb|evcospi| and then can be used as the the user sees fit.

\section{Examples: ec.x}
\label{sec:rf_examples}
We shall now consider an example using the FHN model and we shall generate the IC's for a rigidly rotating spiral wave using EZ-Freeze. We refer the reader to Chap.\ref{chap:4} for details \chg[af]{of} the FHN model and how this model is implemented into the code.

Using, a second order accurate scheme, the parameters chosen in the cartesian grid were:
\chg[chap6]{}
\begin{itemize}
\item Model Parameters: $a$=0.68, $b$=0.5, $\varepsilon=0.3$
\chg[]{\item Numerical Parameters: $L_x=L_y$=30 s.u., $\Delta_x=\frac{1}{3}$, $ts$=0.05 $\Rightarrow$ $\Delta_t=1.389\times10^{-3}$.}
\end{itemize}
\chg[]{We set the second pinning condition at $(x_{inc},y_{inc})=(0,\frac{20}{3})$ s.u

We found that the values of the advection coeffecients were $c_x=0.5203459859$, $c_y=-0.2124501765$ and $\omega=-0.5724548697$. The radius of the tip trajectory was therefore $R_c=0.981816$ s.u., which means that the center of rotation was at $(x_c,y_c)=(14.8174,15.2814)$ and the radius of the inscribed circle was $R_{IC}=14.7186$ s.u.

Therefore, the numerical \& physical parameters on the polar grid were found to be:}
\chg[]{
\begin{itemize}
\item $\Delta_r=\frac{1}{3}$, $N_r$=43, $\Delta_\theta=2.27652\times10^{-2}$, $N_\theta$=276.
\end{itemize}}

\begin{figure}[p]
\begin{tabular}[t]{cccc}
    \panel{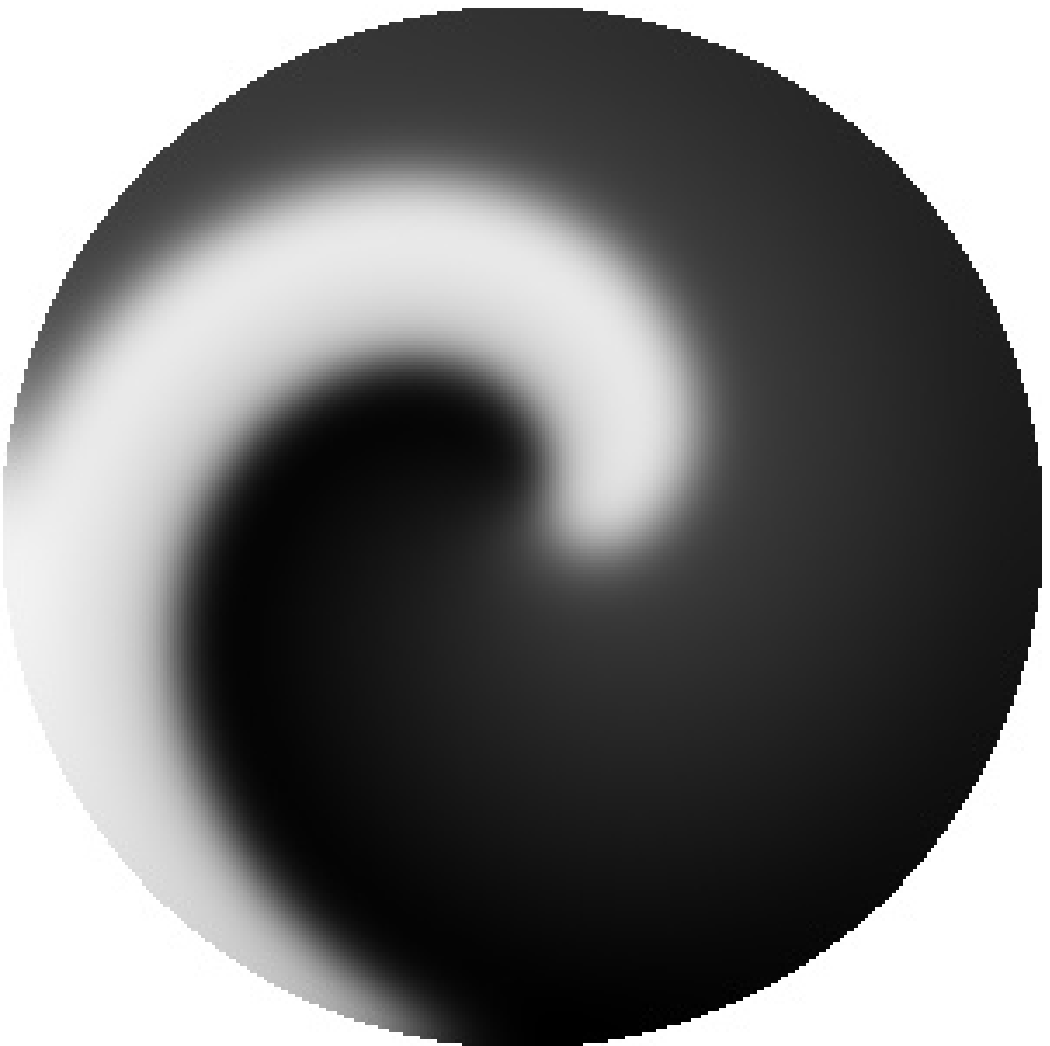} & 
    \panel{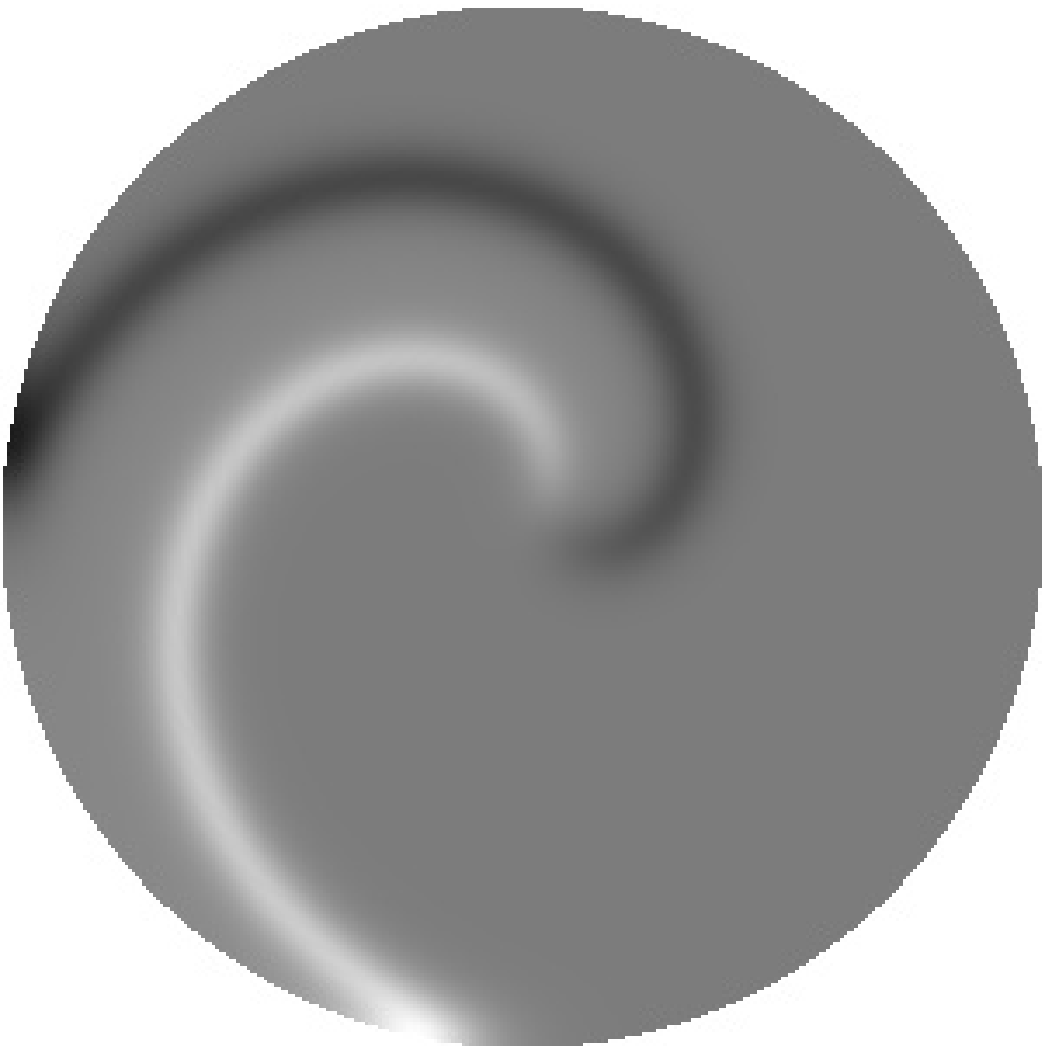} &
    \panel{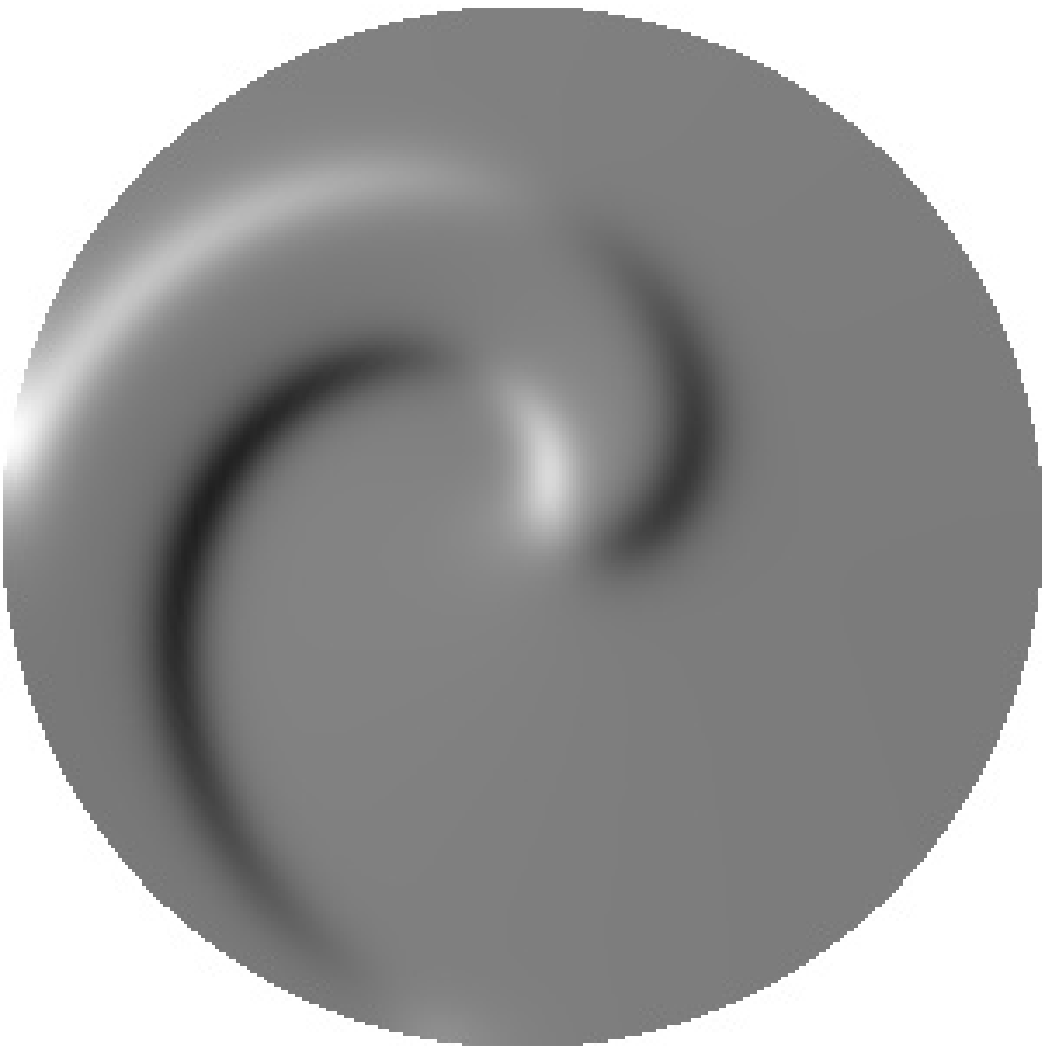} &
    \panel{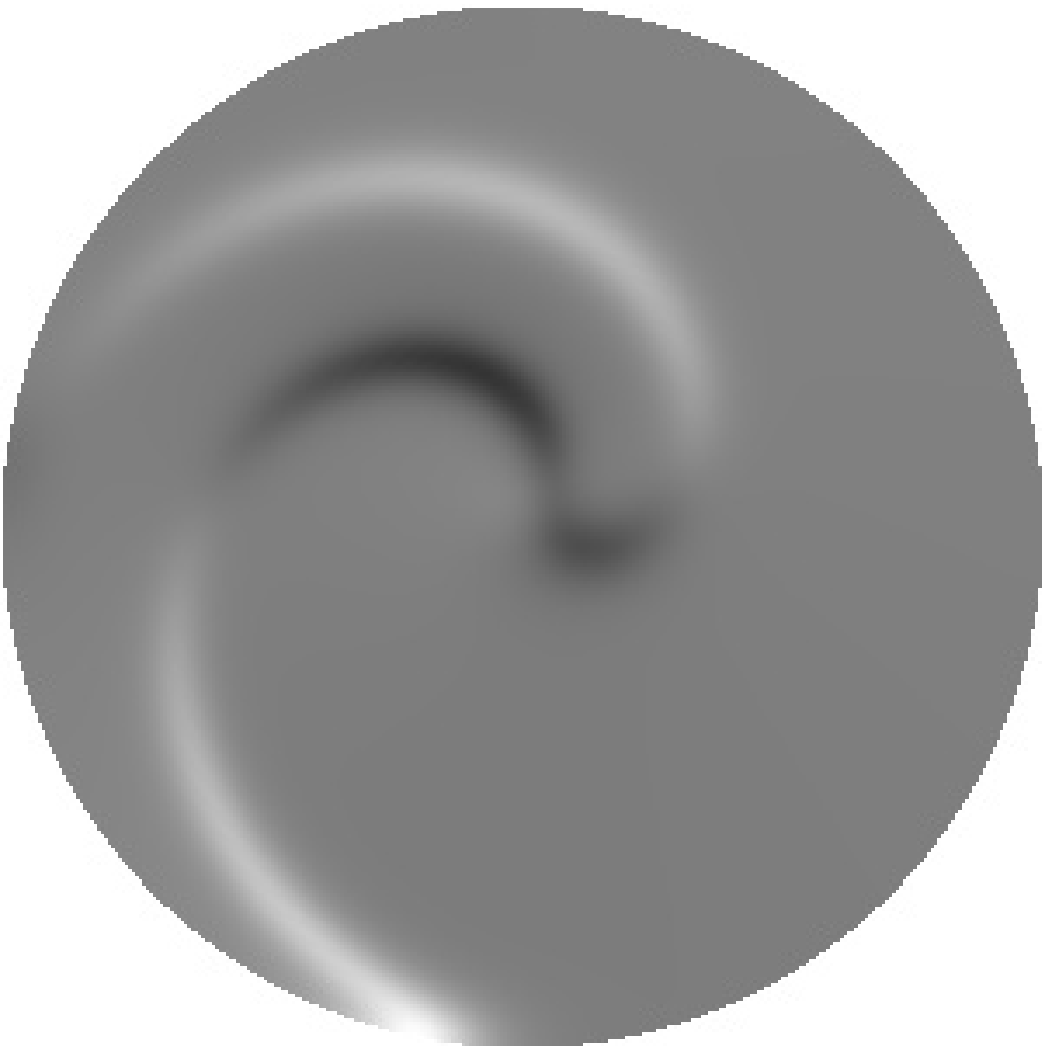}\\
    \panel{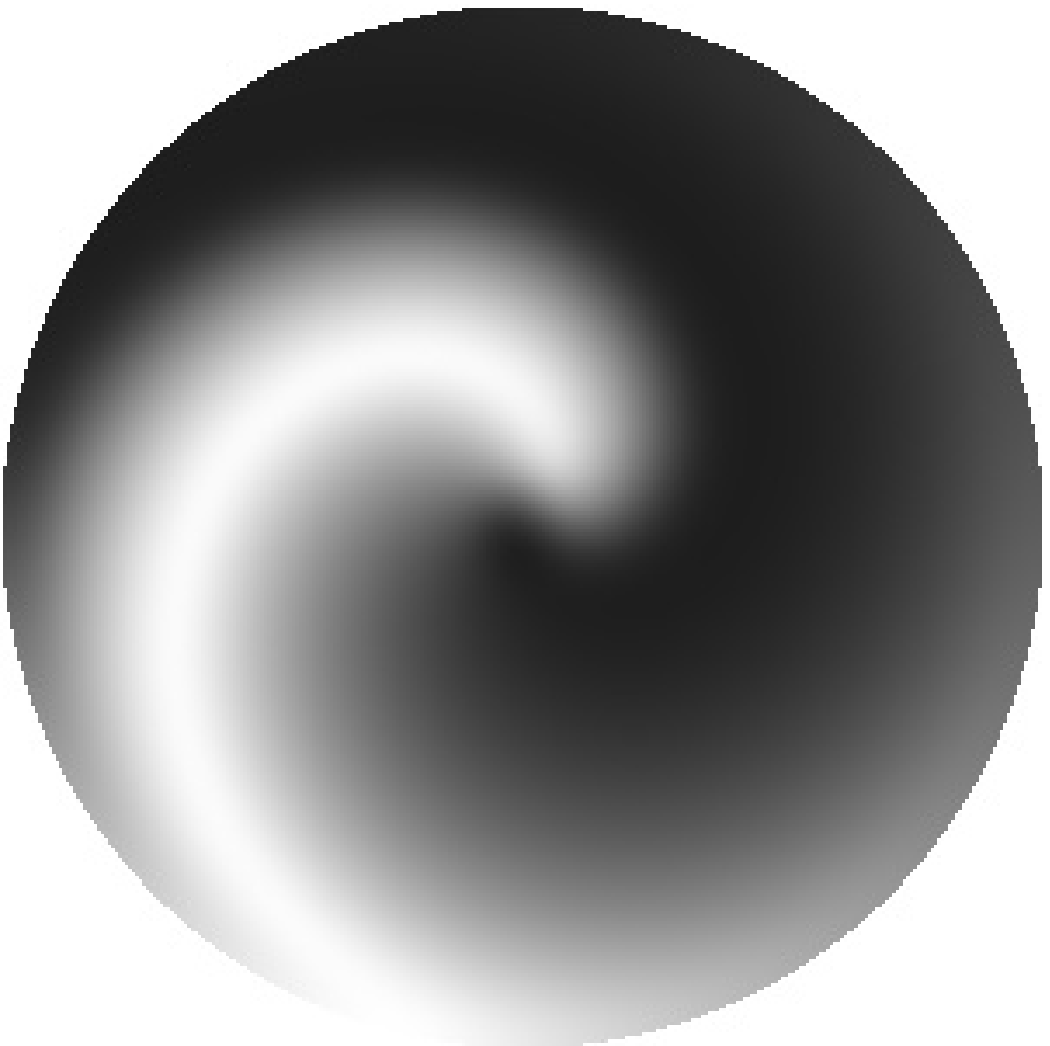} & 
    \panel{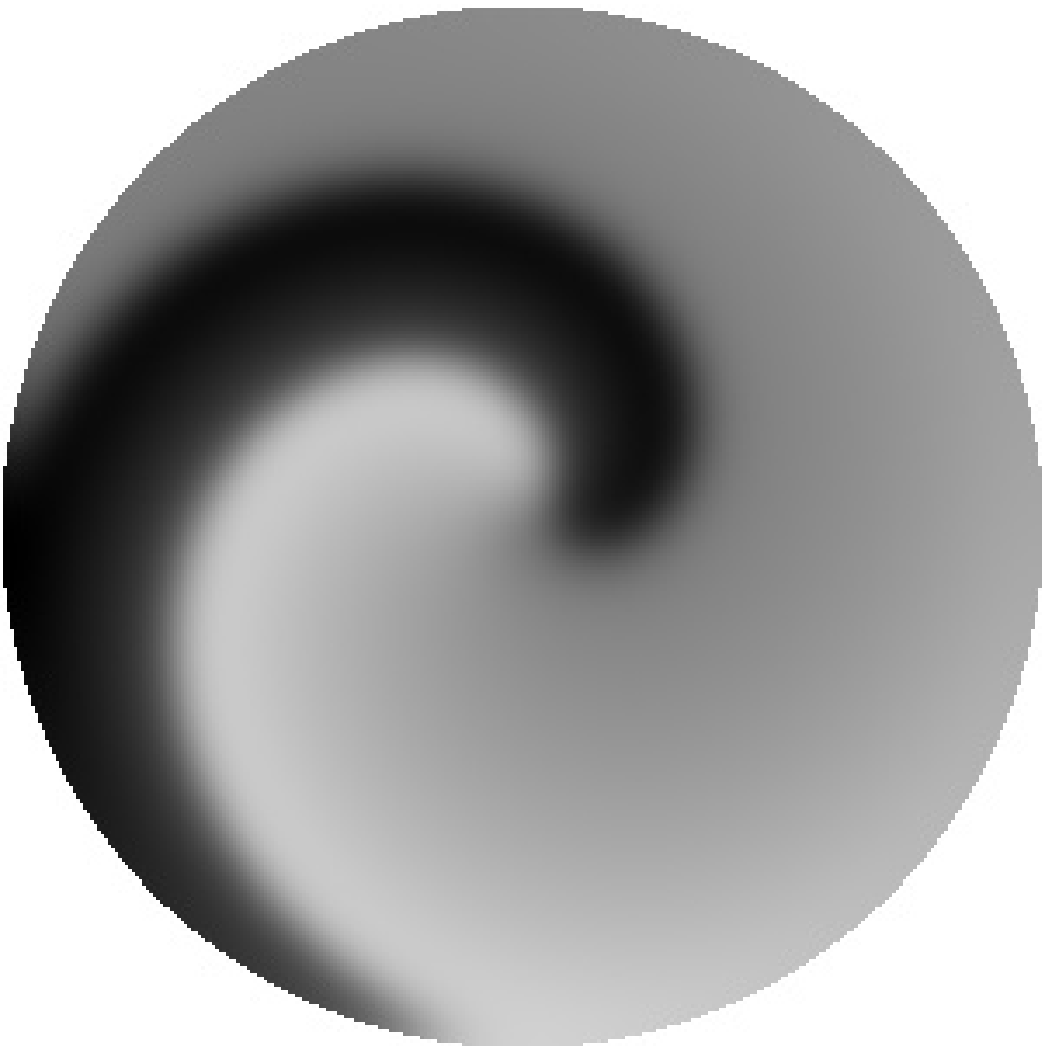} &
    \panel{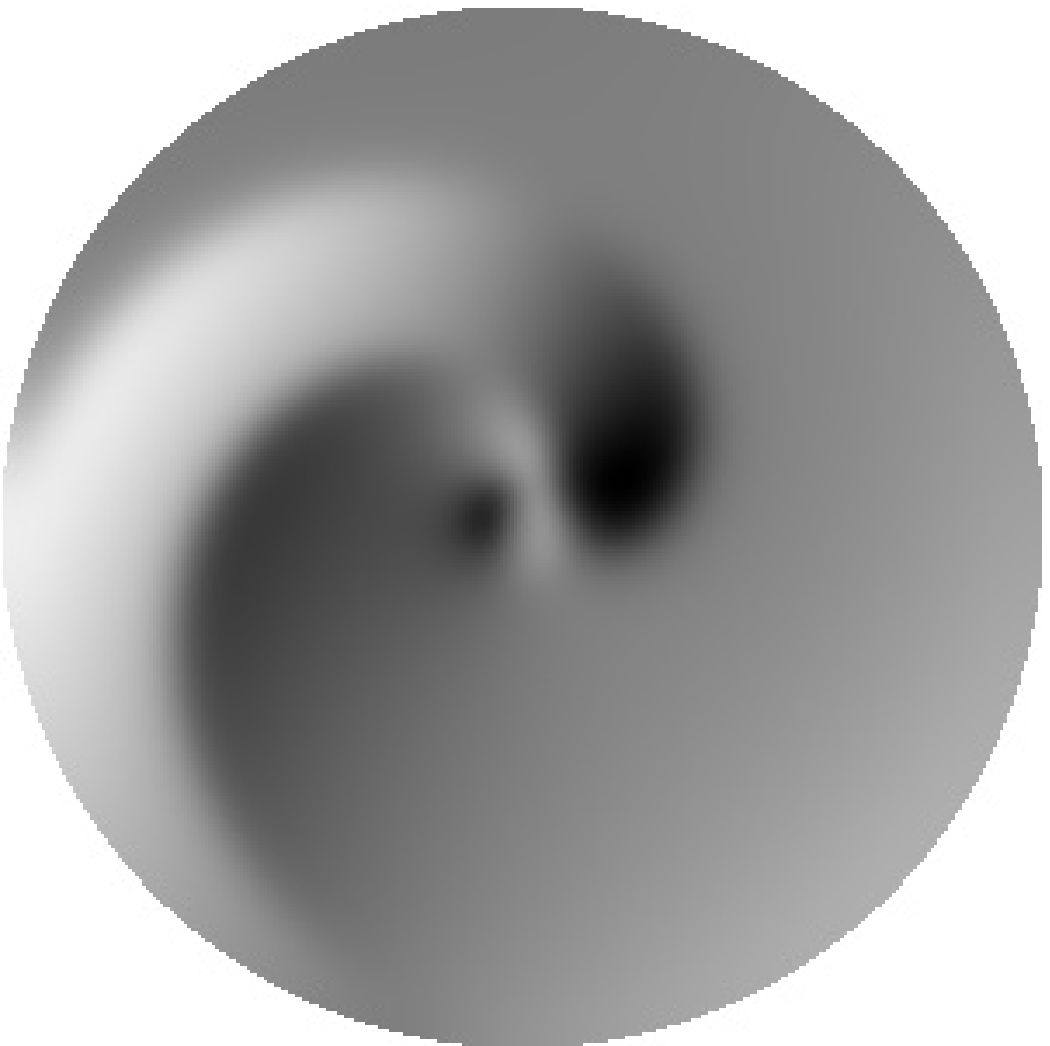} &
    \panel{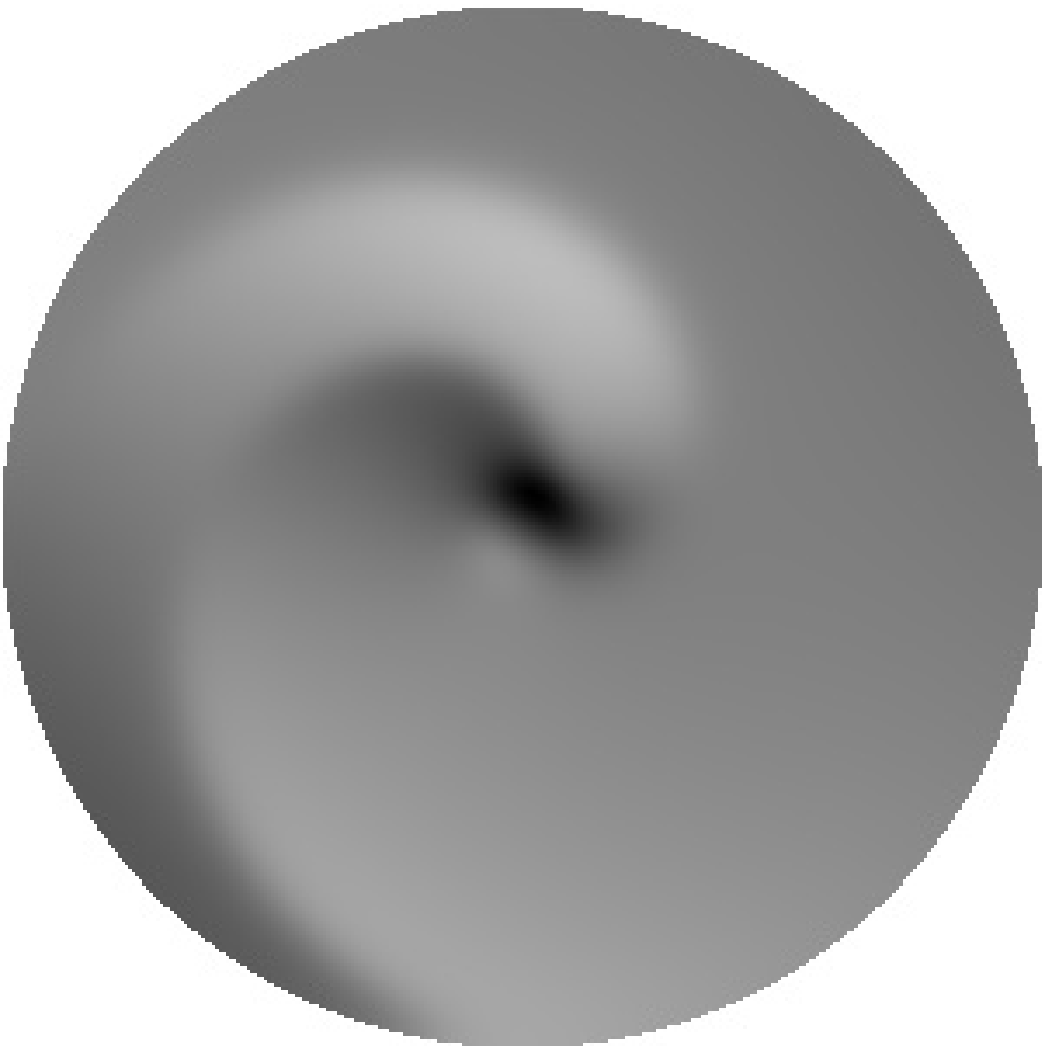}\\
    spiral &
    $\bphi_0$ &
    {\rm Re}$\{\bphi_1\}$ &
    {\rm Im}$\{\bphi_1\}$ 
\end{tabular}
\caption[Numerical solution of the Goldstone modes]{The $u$-field is shown at the top, with the $v$-field at the bottom.The refined spiral solution is shown on the left, with the particular Goldstone modes, $\bphi_i$ shown as detailed.}
\label{fig:chap6_ex1}
\end{figure}

\begin{figure}[p]
\begin{tabular}[t]{ccccc}
    \panel{chapter5/examples/spiral-u.eps} & 
    \panel{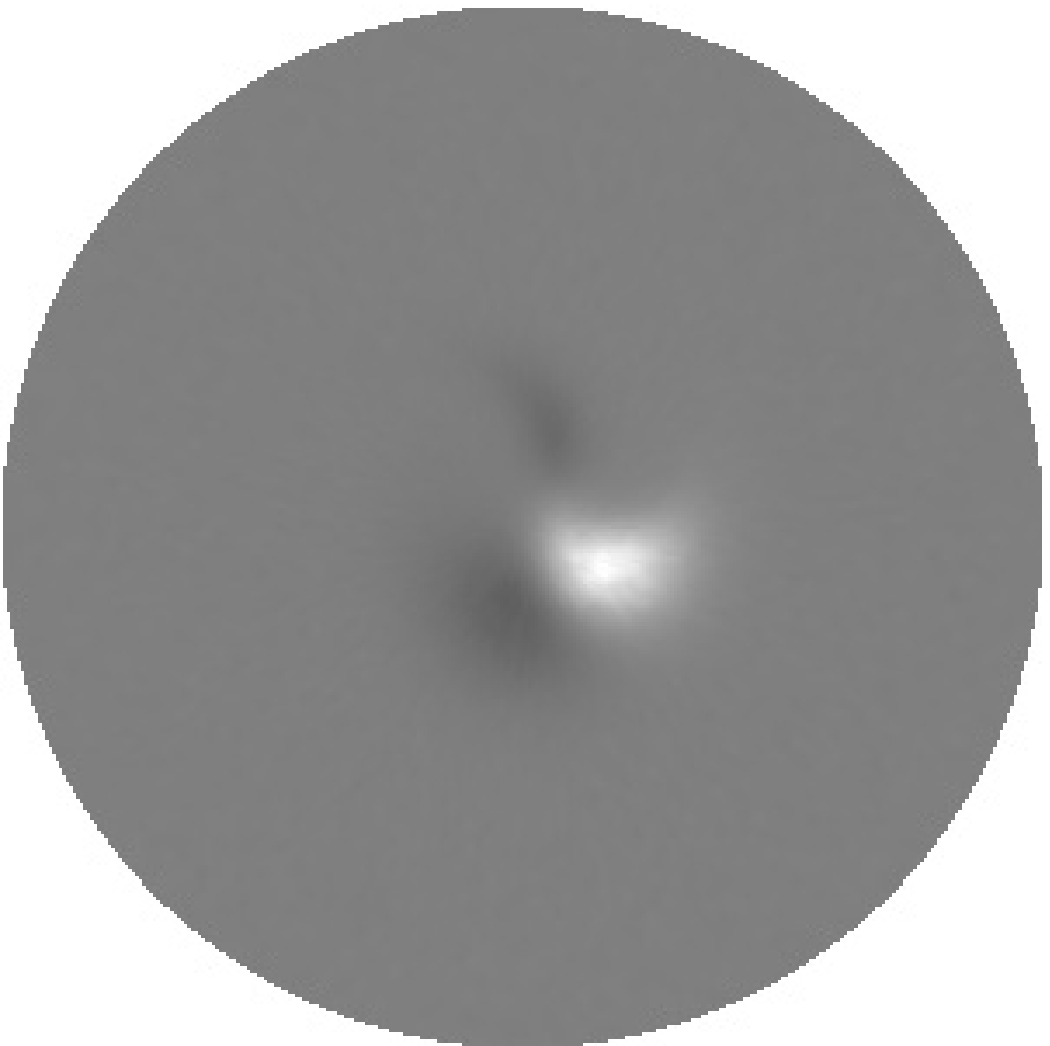} &
    \panel{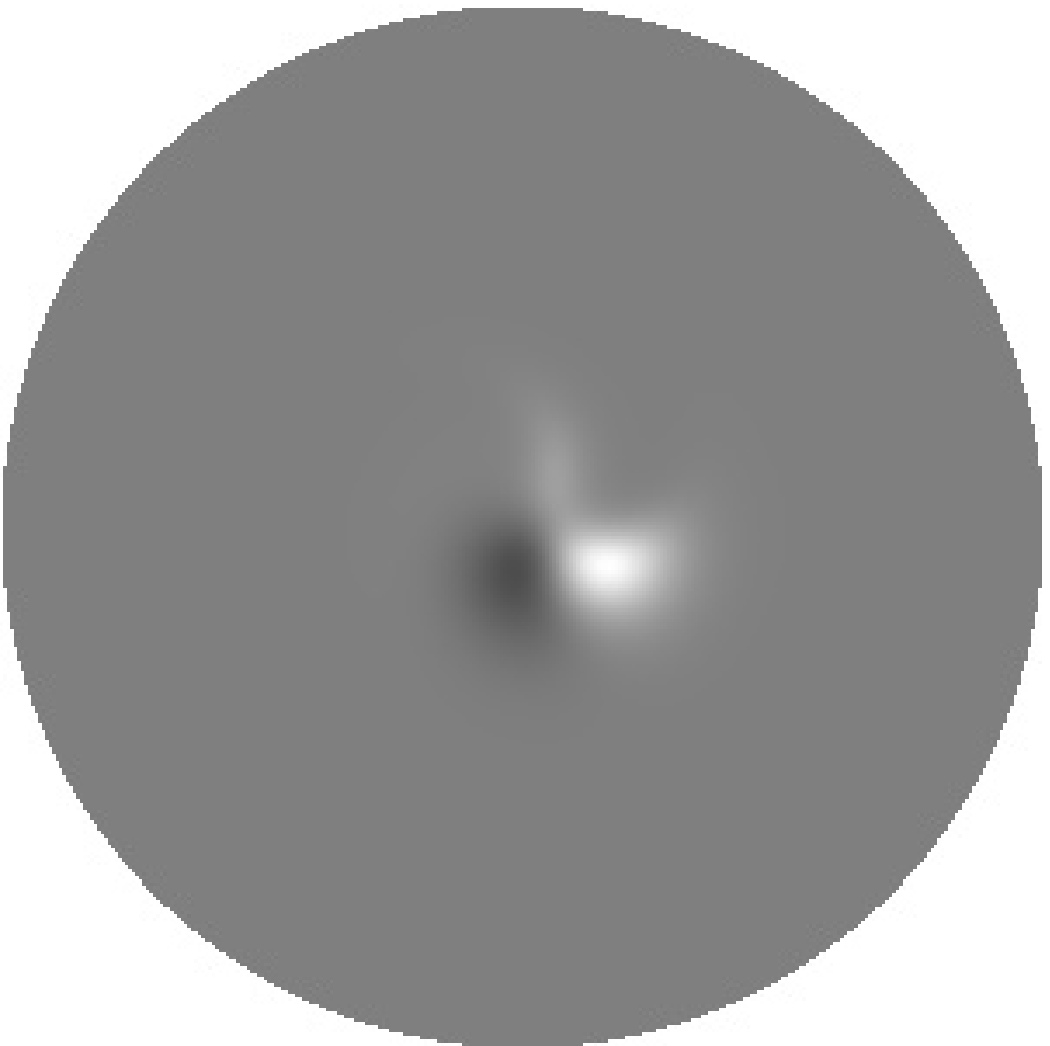} &
    \panel{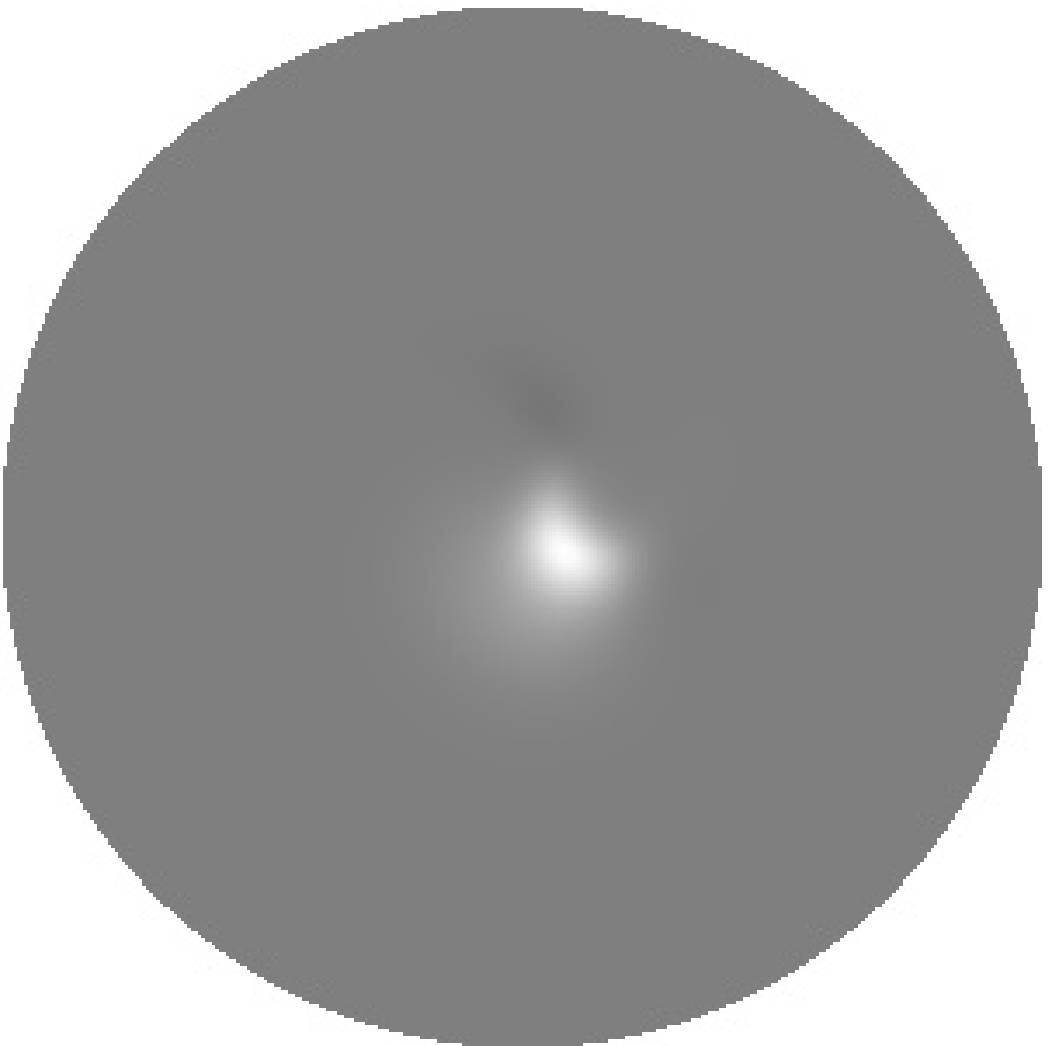}\\
    \panel{chapter5/examples/spiral-v.eps} & 
    \panel{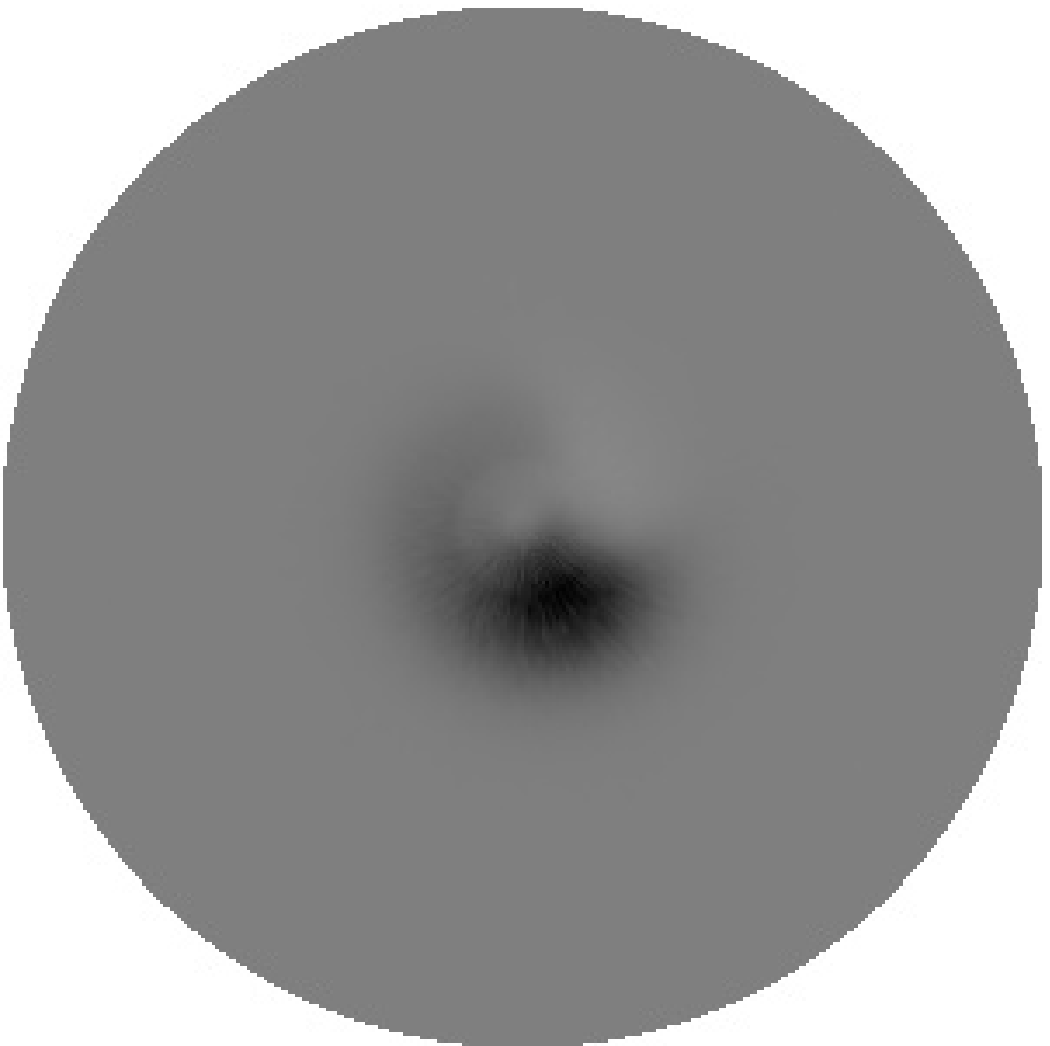} &
    \panel{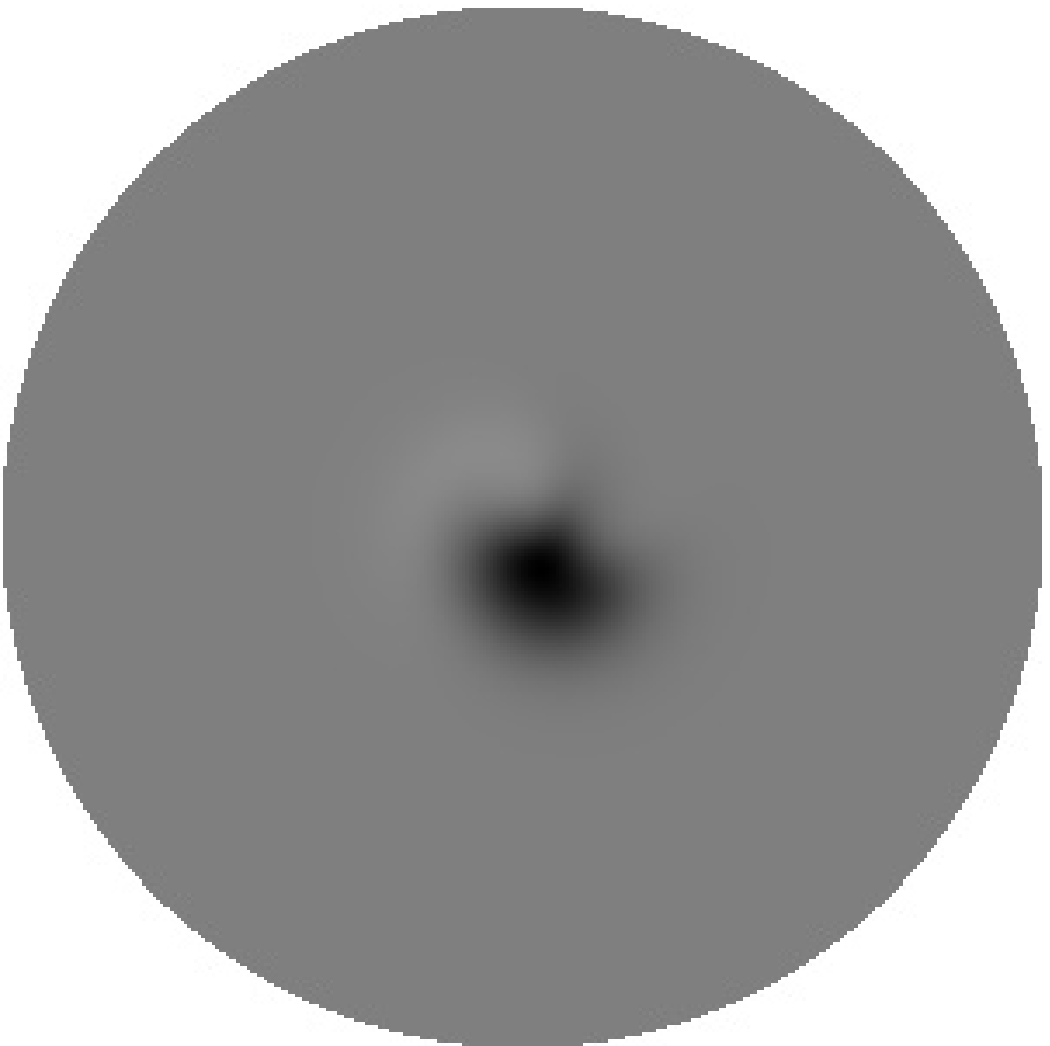} &
    \panel{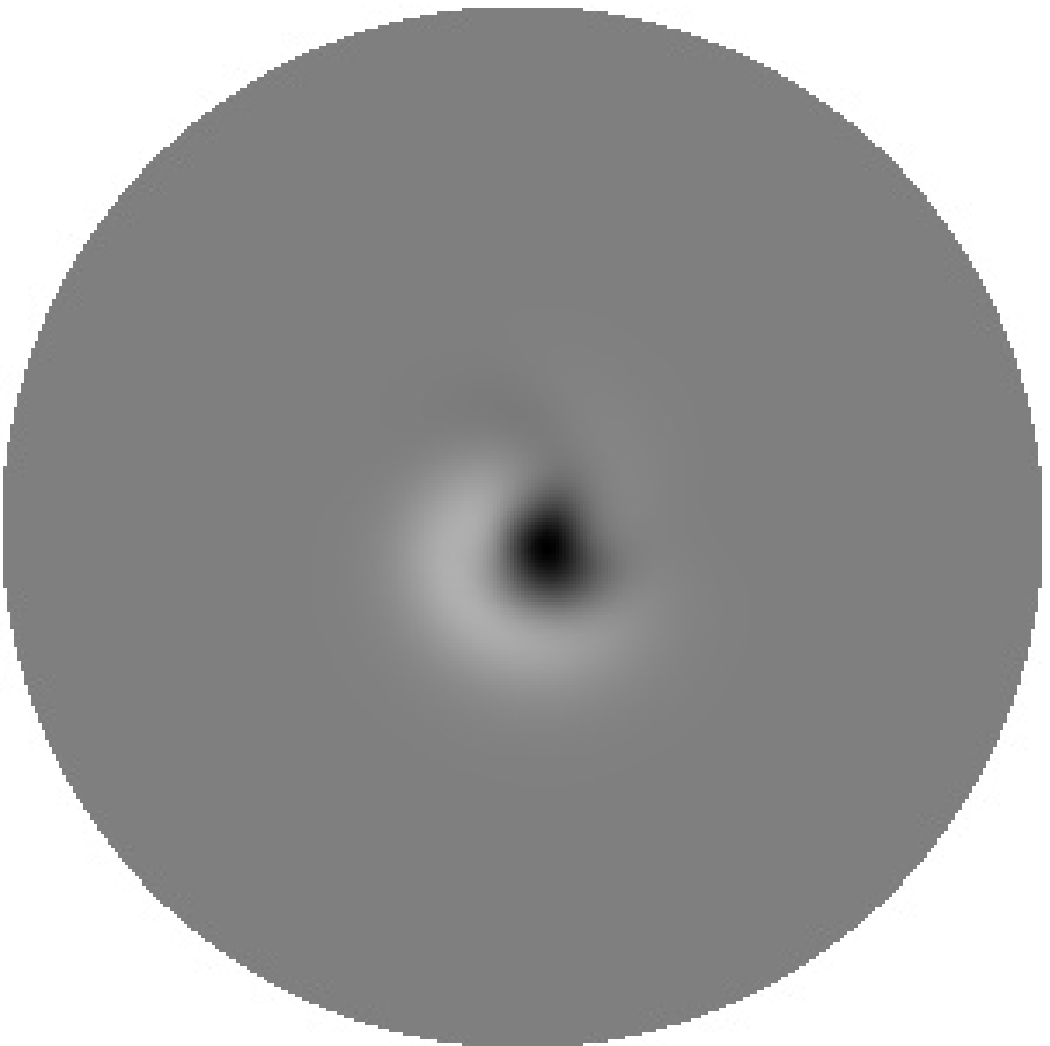}\\
     spiral &
    $\bpsi_0$ &
    {\rm Re}$\{\bpsi_1\}$  &
    {\rm Re}$\{\bpsi_1\}$  
\end{tabular}
\caption[Numerical solution of the response functions]{The $u$-field is shown at the top, with the $v$-field at the bottom. This time the response functions, $\bpsi_j$ are plotted.}
\label{fig:chap6_ex2}
\end{figure}

\chg[]{We then ran evcospi using the IC's generated above, and got the results as shown in Figs.(\ref{fig:chap6_ex1})-(\ref{fig:chap6_ex2})}

The main observation is the localisation of the response functions near the core of the spiral. This localisation was conjectured by Biktashev [ref], shown numerically by [refs] for the Complex Ginzburg Landau Equation, but first calculated accurately by Biktasheva et al \cite{bik08}. This localisation is a surprising feature of response functions and can be taken advantage numerically by, for example, calculating the drift of the spiral wave considering only the core of the spiral. Indeed, it can be said that the speed of a drifting spiral wave is determined only by what is happening at the core of the spiral.


\section{Convergence Testing}
\label{sec:rf_convergence}
We will detail the work we did in initiating the convergence testing for \verb|evcospi|. The purpose of the program is to see how \chg[p205gram]{accurate the solutions were by considering the variation of the numerical parameters.} 

We conducted tests for convergence in the radial step, angular step, and disk radius\chg[af]{, with the idea being} to see how accurately the eigenvalues were calculated. We recall that from analytical considerations there are three critical \chg[af]{eigenvalues} - $\lambda_{0,\pm1}=0,\pm i\omega$. Numerical \chg[af]{calculations} always carry an error and therefore, by comparing the numerically calculated eigenvalues to the values we \chg[p205spell]{know} they should be, then we can see how good our numerical approximation is.

\subsection{Methods}

We started with initial conditions generated using \chg[p208spell1]{EZ-Spiral} and continued the parameter to the following set of ``ideal'' parameters.

\begin{itemize}
 \item Numerical Parameters
  \begin{itemize}
   \item disk radius = 40
   \item Number of angular points = 76
   \item number of radial points = 250
  \end{itemize}
 \item Model Parameters
  \begin{itemize}
   \item a = 0.5
   \item b = 0.68
   \item $\varepsilon$ = 0.3
  \end{itemize}
\end{itemize}

Once these IC's are generated, then we decrease one of the parameters from it's ideal value using continuation by a suitably chosen step in that parameter. We repeat this procedure, \chg[p208spell2]{decreasing} the parameter each time by the fixed step, until the solution no longer exists. 

We repeat this for all the numerical parameters.


\subsection{Disk Size}

\begin{figure}[p]
\begin{center}
\begin{minipage}[tbp]{0.49\linewidth}
\centering
\includegraphics[width=0.7\textwidth, angle=-90]{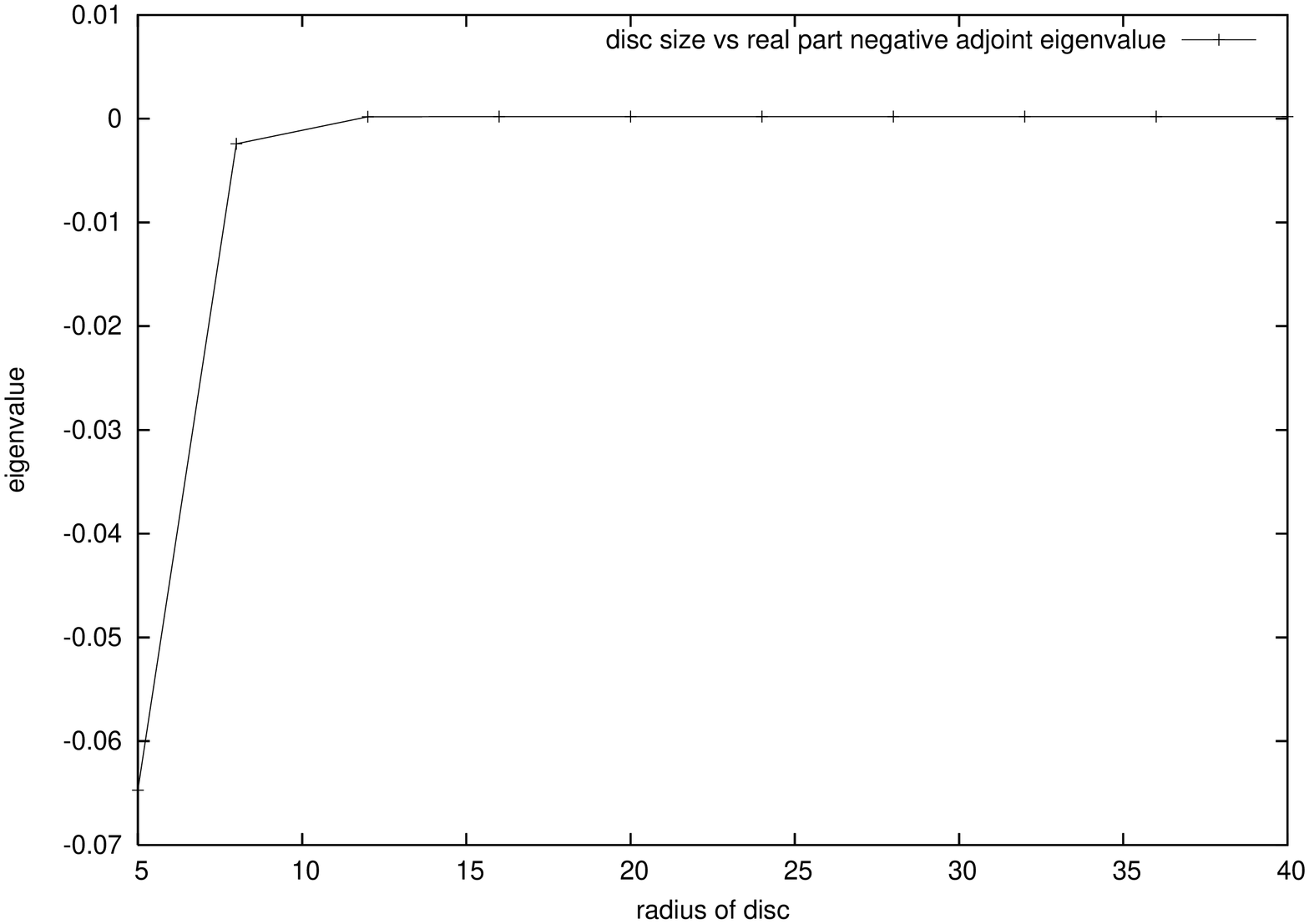}
\end{minipage}
\begin{minipage}[tbp]{0.49\linewidth}
\centering
\includegraphics[width=0.7\textwidth, angle=-90]{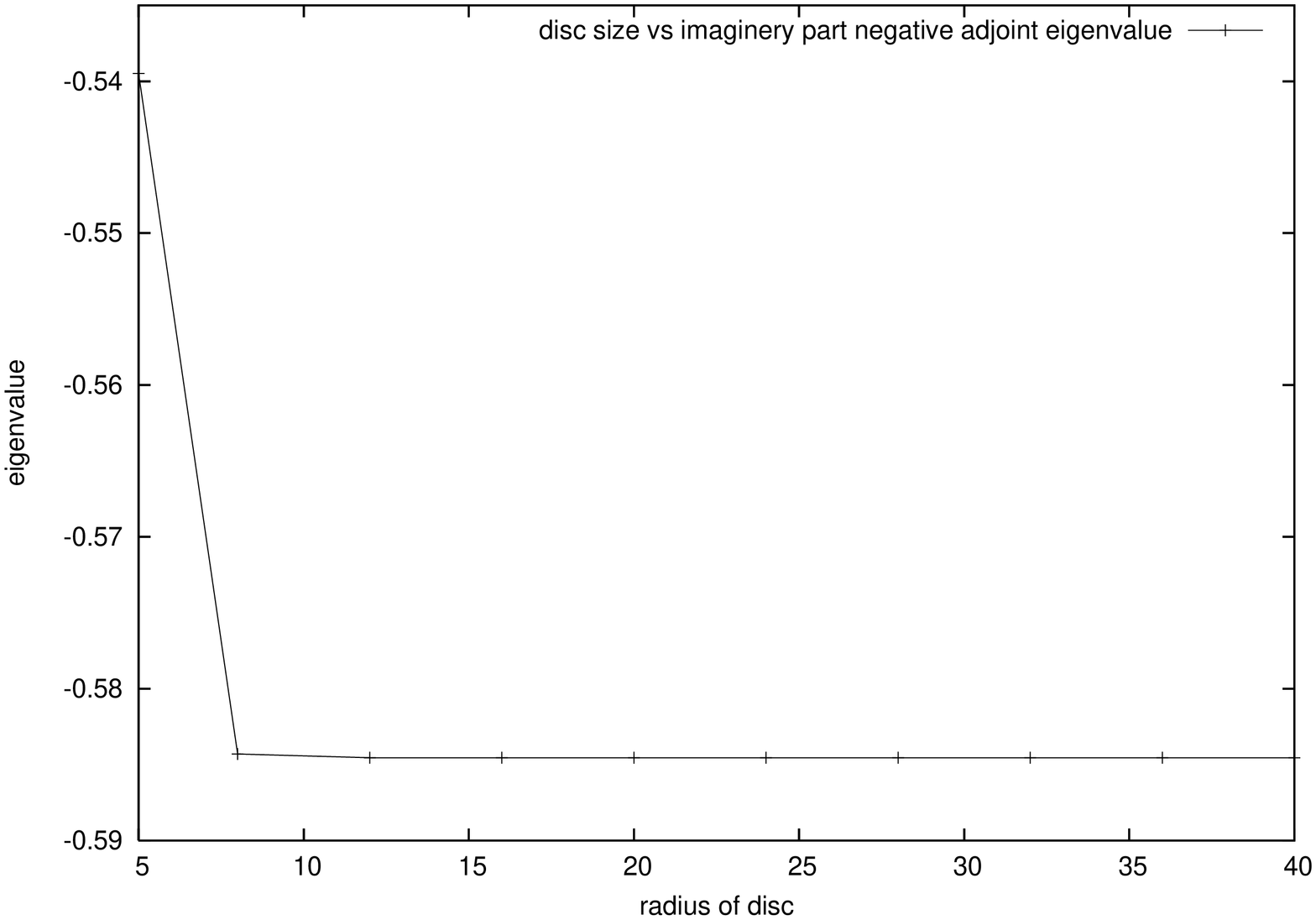}
\end{minipage}
\begin{minipage}[tbp]{0.49\linewidth}
\centering
\includegraphics[width=0.7\textwidth, angle=-90]{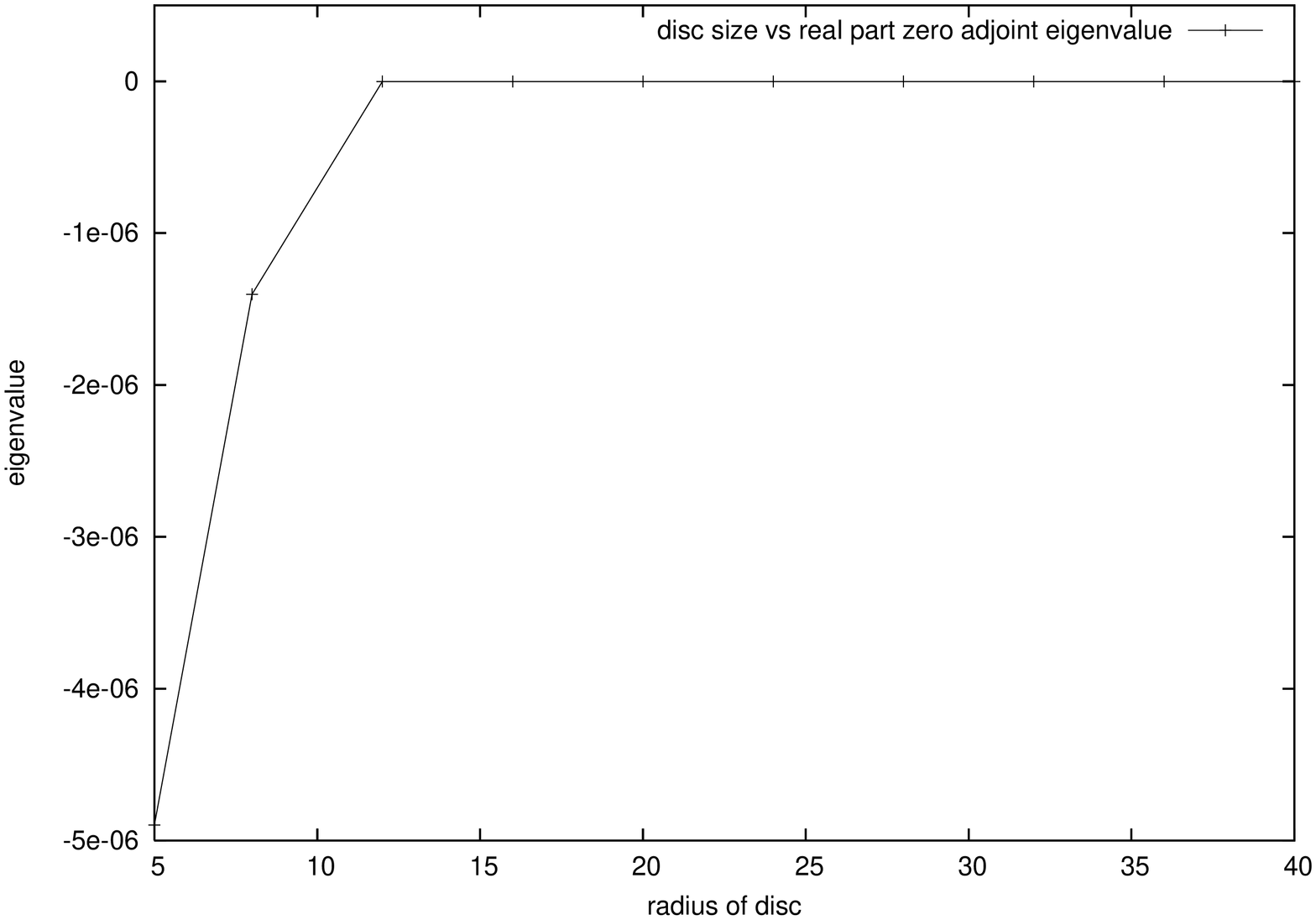}
\end{minipage}
\begin{minipage}[tbp]{0.49\linewidth}
\centering
\includegraphics[width=0.7\textwidth, angle=-90]{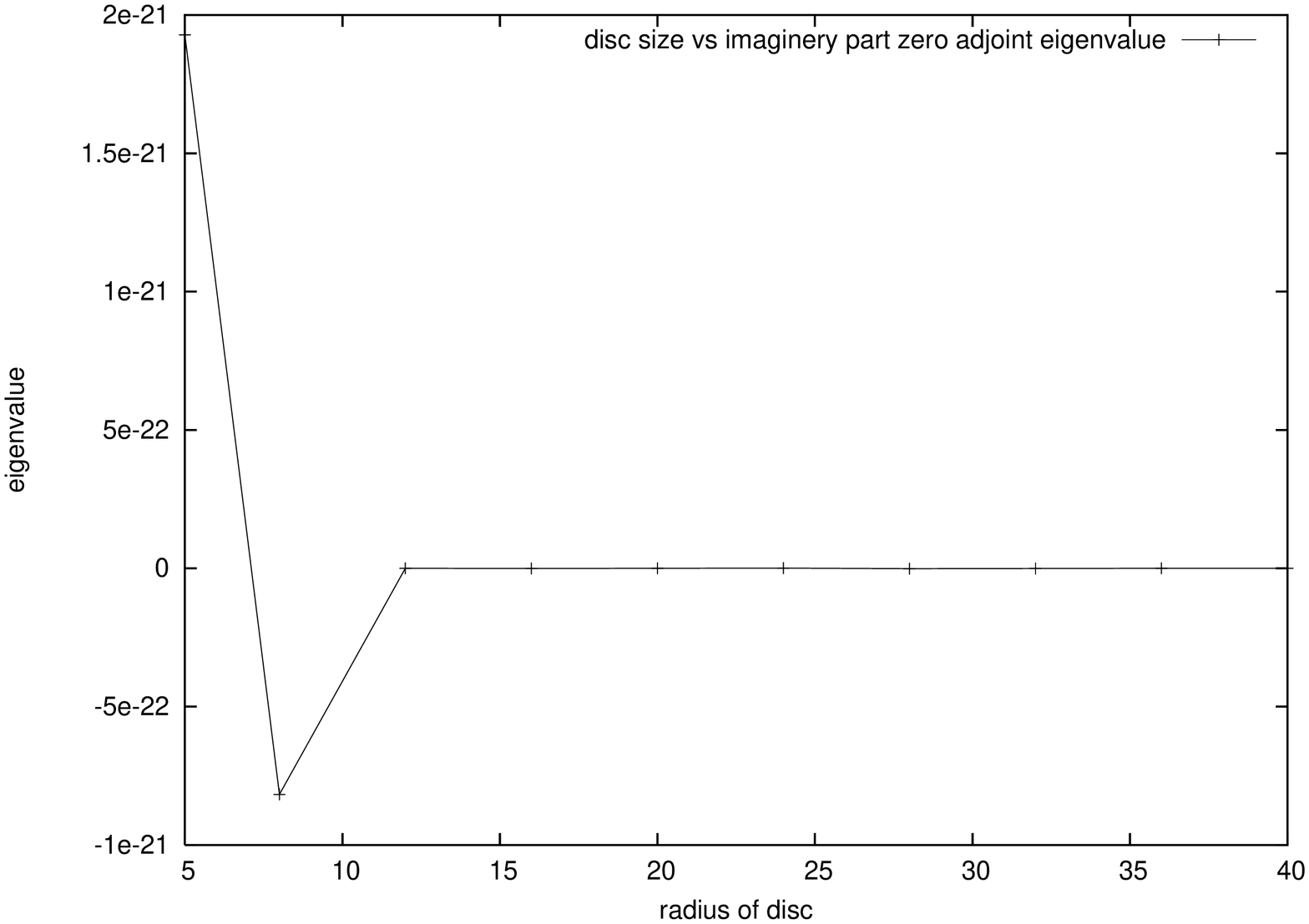}
\end{minipage}
\begin{minipage}[tbp]{0.49\linewidth}
\centering
\includegraphics[width=0.7\textwidth, angle=-90]{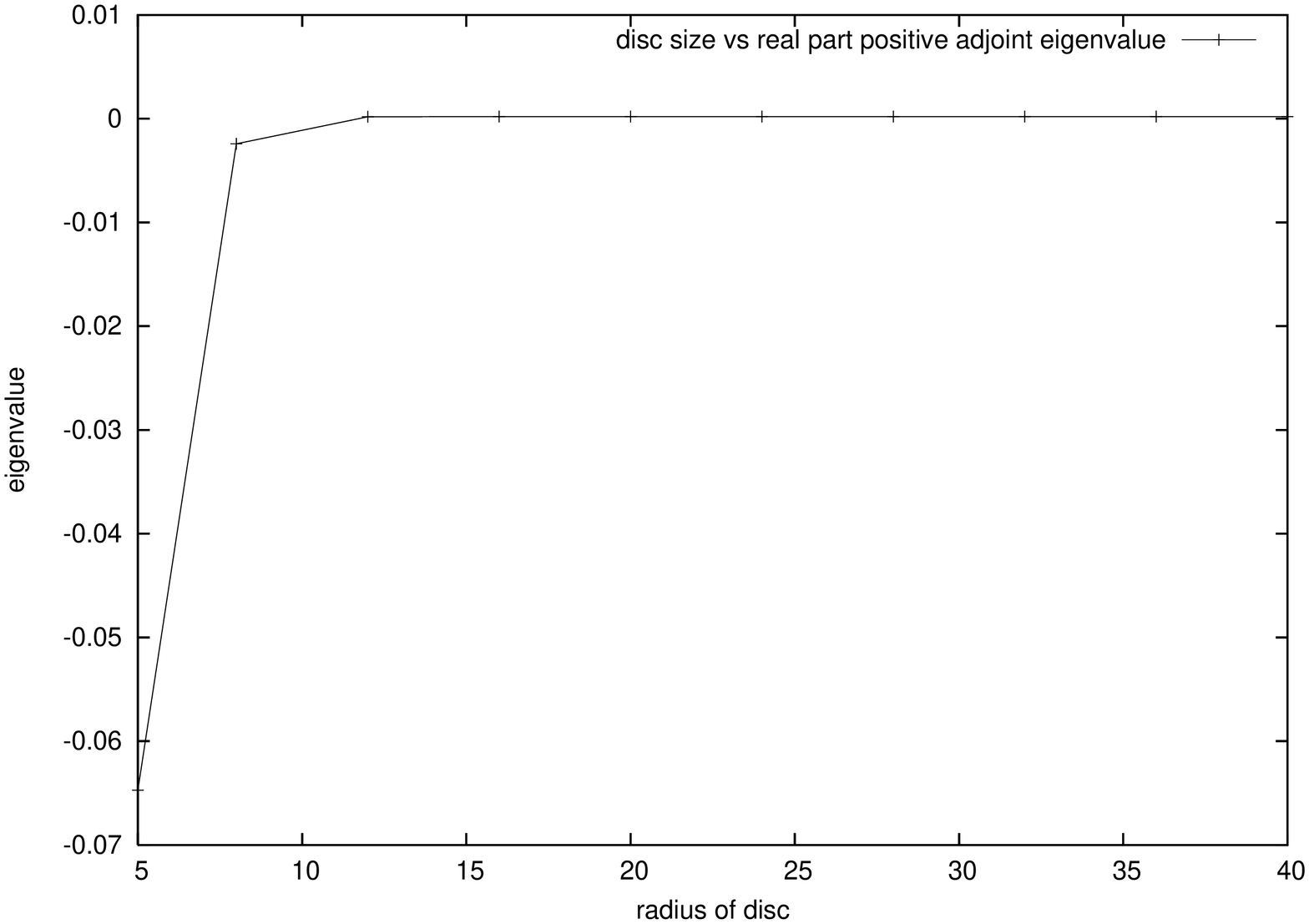}
\end{minipage}
\begin{minipage}[tbp]{0.49\linewidth}
\centering
\includegraphics[width=0.7\textwidth, angle=-90]{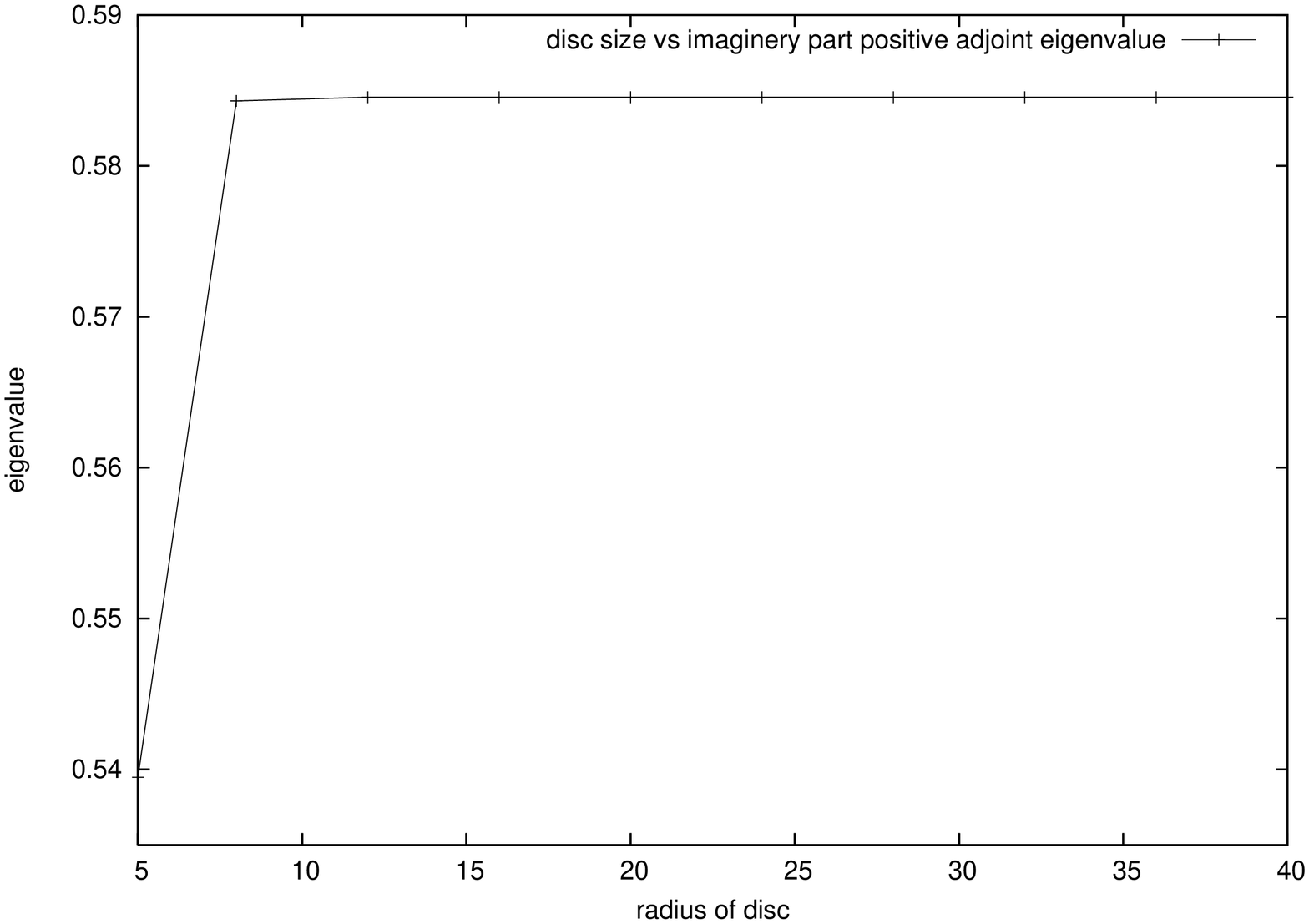}
\end{minipage}
\caption[Disk Size Convergence]{Disk Size Convergence: Real (left column) and \chg[af]{imaginary} (right column) parts of the eigenvalues, corresponding to the analytical eigenvalues $-i\lambda$ (top), 0 (middle) and $i\lambda$ (bottom).}
\label{fig:rf_ex_conv_disc1}
\end{center}
\end{figure}

We show the results of the convergence testing of the disk size in Fig.(\ref{fig:rf_ex_conv_disc1}). We started with a disk radius size of 40 s.u. and worked our way back to 4 s.u. in steps of 4 s.u.

We can see that the eigenvalues converge to their absolute values between disk sizes 8 and 12 s.u. 

Let us represent the eigenvalues as:
\chg[p208spell3]{}
\begin{equation*}
\lambda_n = \Lambda_n+\chg[]{\delta_n}
\end{equation*}
\\
where $\lambda_n$ is the actual numerically calculated eigenvalue, $\Lambda_n=in\omega$, $\omega$ is the numerically found value, and $\delta_n$ is the error. We call $\Lambda_n$ the converged eigenvalue. We also note that $n=0,\pm1$.

For disk sizes greater than 12 s.u. we note that the value of the angular velocity and the converged eigenvalues are:

\begin{itemize}
 \item Linear operator $L$:
  \begin{itemize}
   \item $\omega$       = -0.582022542082
   \item $\Lambda_{1}$  = 0.00020000998-i0.584547959642
   \item $\Lambda_{0}$  = 1.0$\times10^{-13}$-i1.0$\times10^{-27}$
   \item $\Lambda_{-1}$ = 0.00020000998+i0.584547959642
  \end{itemize}
 \item Adjoint linear operator $L^+$:
  \begin{itemize}
   \item $\omega$             = -0.582022542082
   \item $\bar{\Lambda}_{1}$  = 0.00020001193+i0.584547960827
   \item $\bar{\Lambda}_{0}$  = 1.0$\times10^{-9}$-i1.0$\times10^{-24}$
   \item $\bar{\Lambda}_{-1}$ = 0.00020001193-i0.584547960827
  \end{itemize}
\end{itemize}

So we can see that the eigenvalue have converged to a very accurately calculated figures. We see that the real parts must be zero but that they are of the orders $1.0\times10^{-13}$ and $1.0\times10^{-9}$ for the zero eigenvalue and adjoint eigenvalue respectively, and of the order $1.0\times10^{-4}$ for the other \chg[af]{eigenvalues.} The \chg[af]{imaginary} parts are of the orders $1.0\times10^{-27}$ and $1.0\times10^{-24}$ for the zero eigenvalue and adjoint eigenvalue respectively, and $1.0\times10^{-3}$ for the others.

So a high order of accuracy is achieved.

\clearpage

\subsection{Angular Step}

We conduct a similar analysis for the convergence in the angular step. We started with $N_\theta=76$ and worked backwards in steps of 4. This ensures that we still have $N_\theta \text{mod} 4=0$.

For $N_\theta>44$, we have that the value of the angular velocity and the converged eigenvalues are:

\begin{itemize}
 \item Linear operator $L$:
  \begin{itemize}
   \item $\omega$       = -0.5820225
   \item $\Lambda_{1}$  = 0.000200-i0.584547
   \item $\Lambda_{0}$  = 1.0$\times10^{-11}$-i1.0$\times10^{-27}$
   \item $\Lambda_{-1}$ = 0.000200+i0.584547
  \end{itemize}
 \item Adjoint linear operator $L^+$:
  \begin{itemize}
   \item $\omega$             = -0.5820225
   \item $\bar{\Lambda}_{1}$  = 0.000200+i0.584547
   \item $\bar{\Lambda}_{0}$  = 1.0$\times10^{-7}$-i1.0$\times10^{-23}$
   \item $\bar{\Lambda}_{-1}$ = 0.000200-i0.584547
  \end{itemize}
\end{itemize}

Again, the accuracy is quite good with the real parts of the orders $1.0\times10^{-11}$ and $1.0\times10^{-7}$ for the zero eigenvalue and adjoint eigenvalue respectively, and of the order $1.0\times10^{-4}$ for the other \chg[af]{eigenvalues.} The \chg[af]{imaginary} parts are of the orders $1.0\times10^{-27}$ and $1.0\times10^{-23}$ for the zero eigenvalue and adjoint eigenvalue respectively, and $1.0\times10^{-3}$ for the others.

\begin{figure}[p]
\begin{center}
\begin{minipage}[tbp]{0.49\linewidth}
\centering
\includegraphics[width=0.7\textwidth, angle=-90]{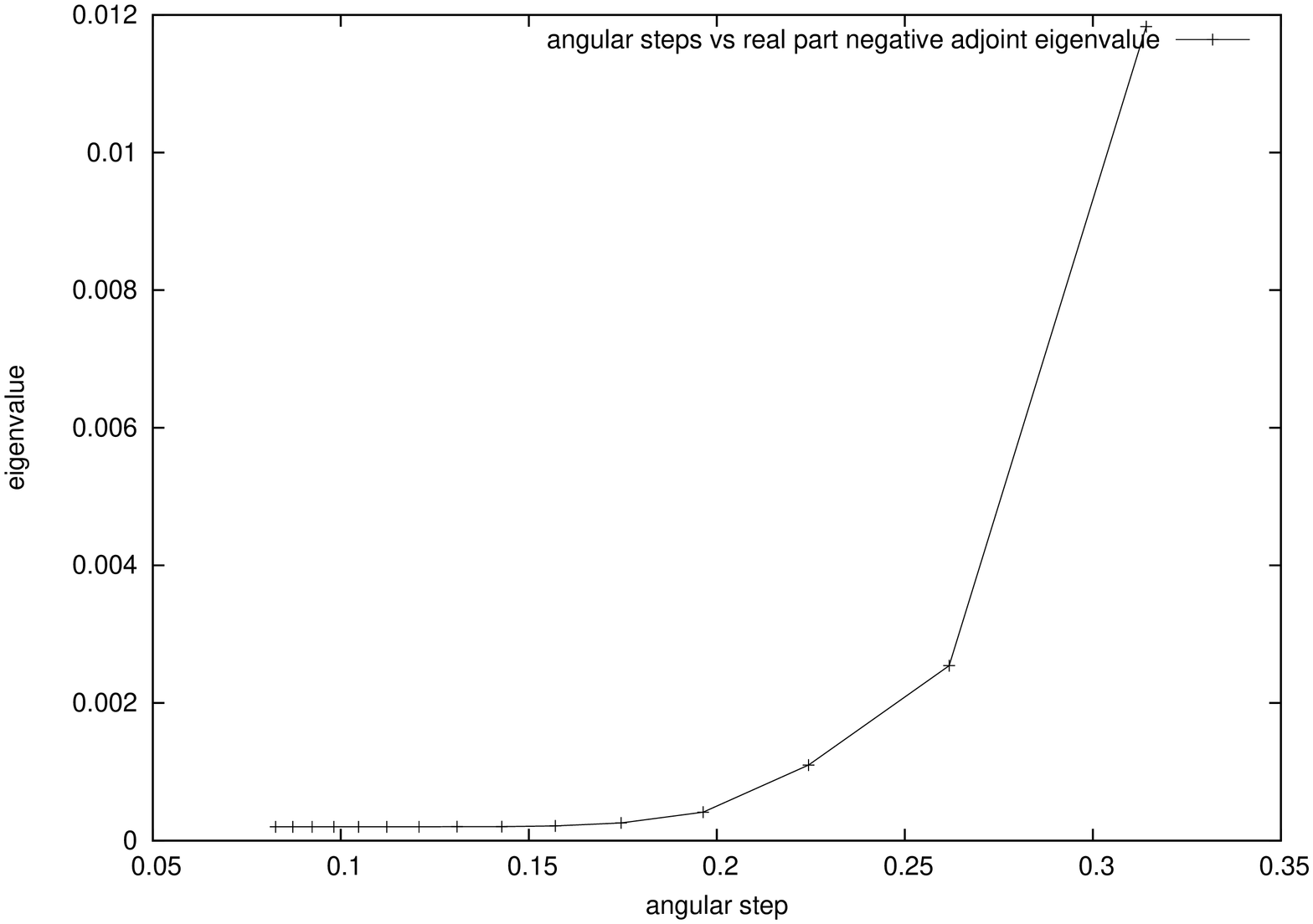}
\end{minipage}
\begin{minipage}[tbp]{0.49\linewidth}
\centering
\includegraphics[width=0.7\textwidth, angle=-90]{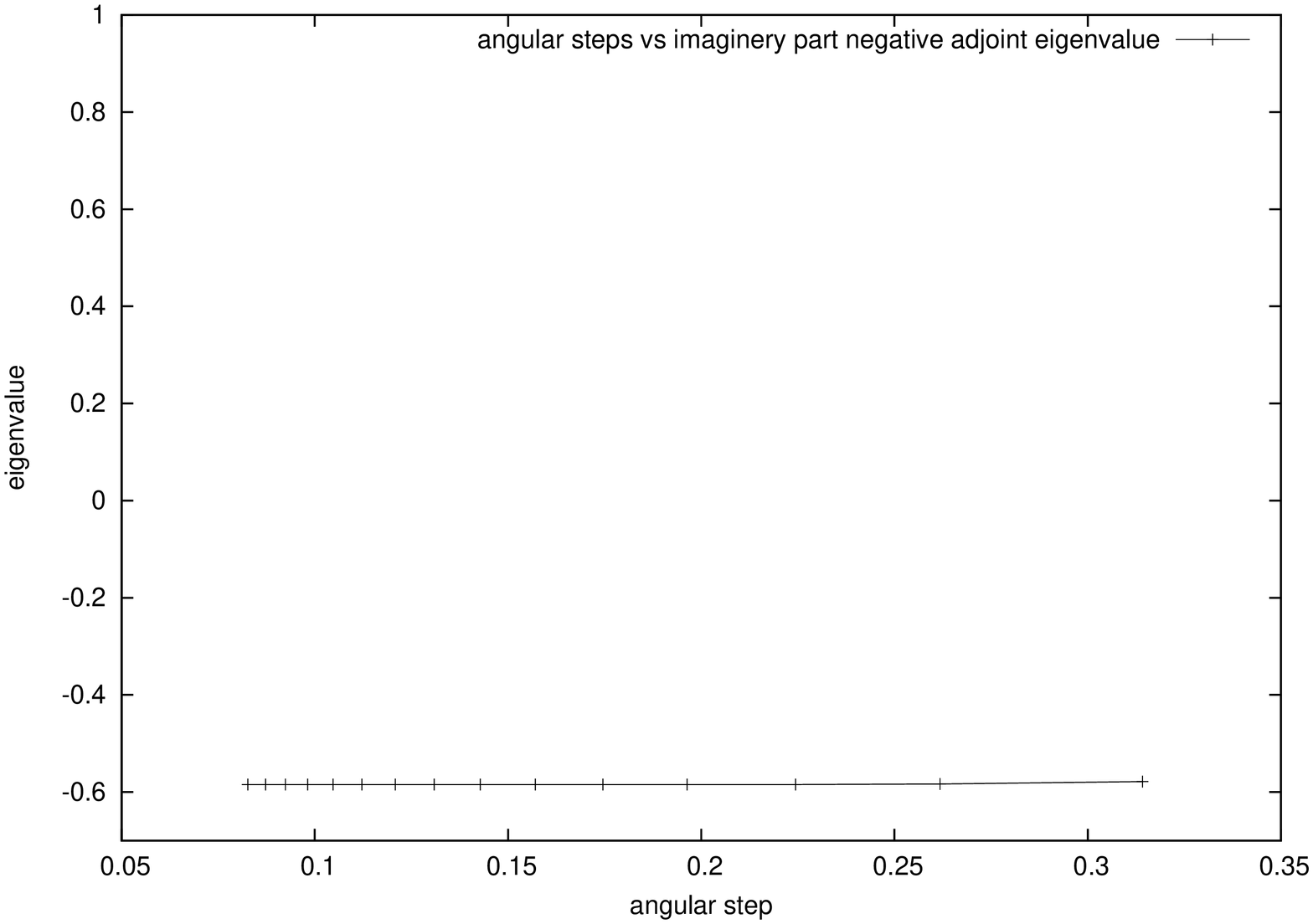}
\end{minipage}
\begin{minipage}[tbp]{0.49\linewidth}
\centering
\includegraphics[width=0.7\textwidth, angle=-90]{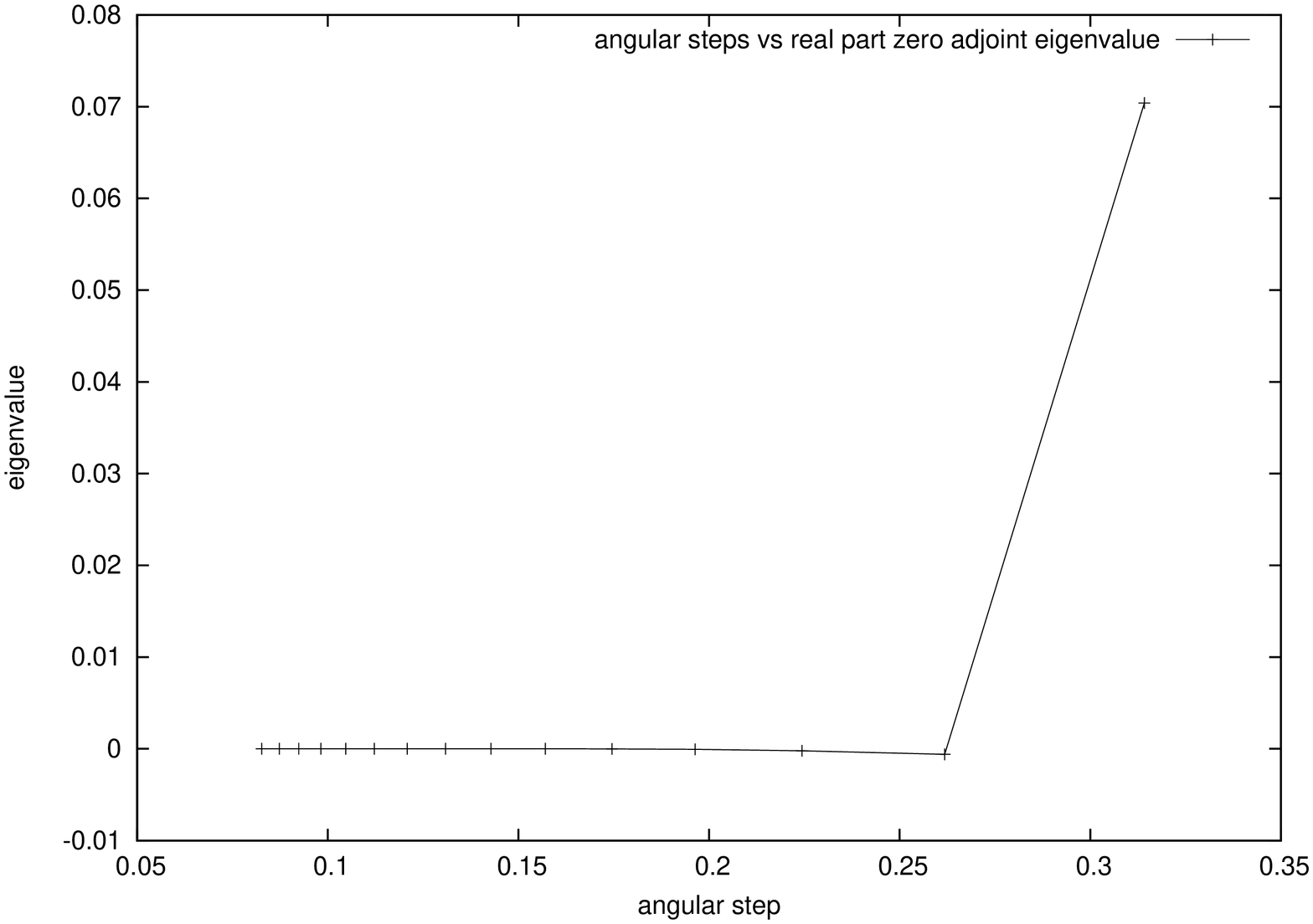}
\end{minipage}
\begin{minipage}[tbp]{0.49\linewidth}
\centering
\includegraphics[width=0.7\textwidth, angle=-90]{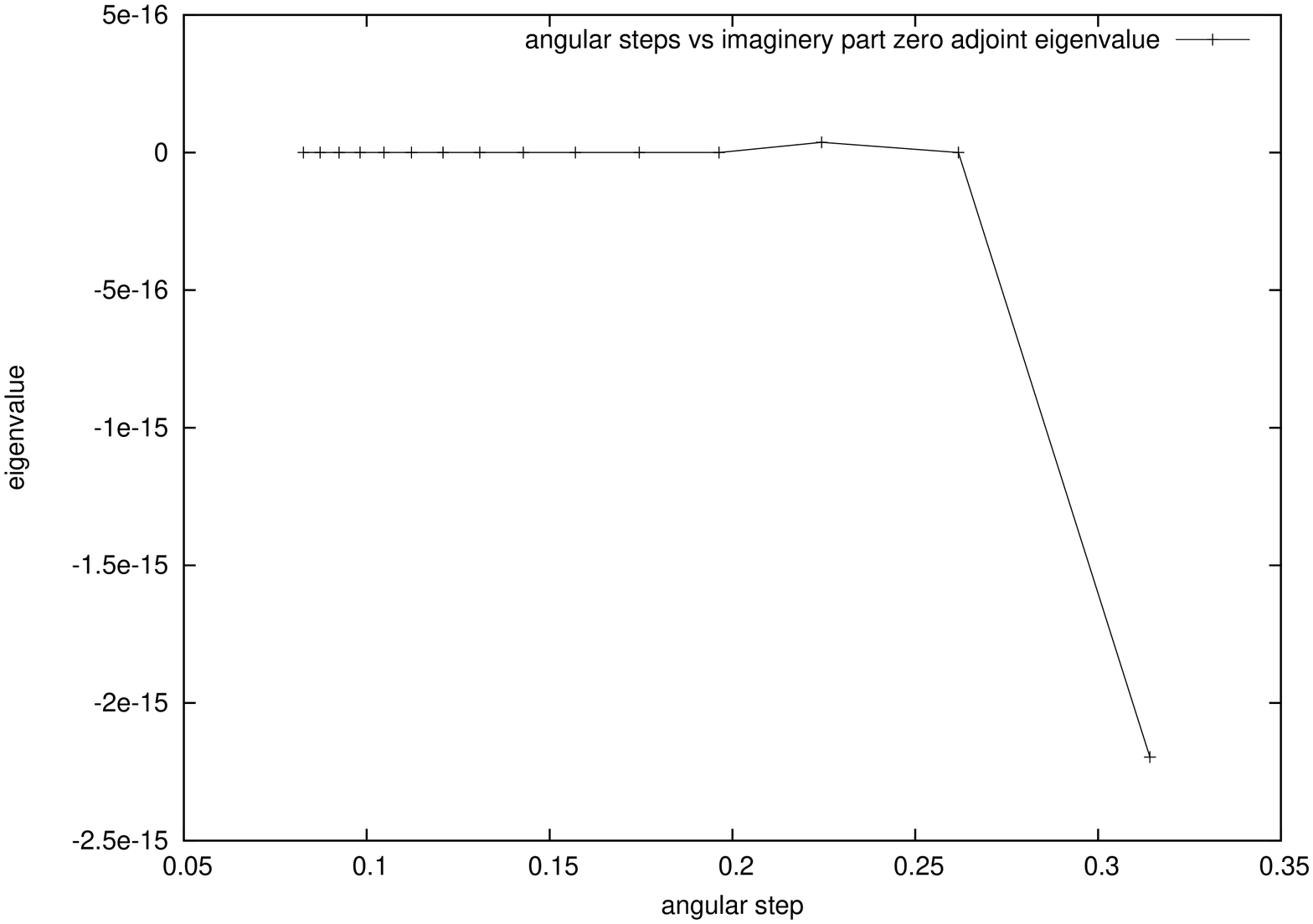}
\end{minipage}
\begin{minipage}[tbp]{0.49\linewidth}
\centering
\includegraphics[width=0.7\textwidth, angle=-90]{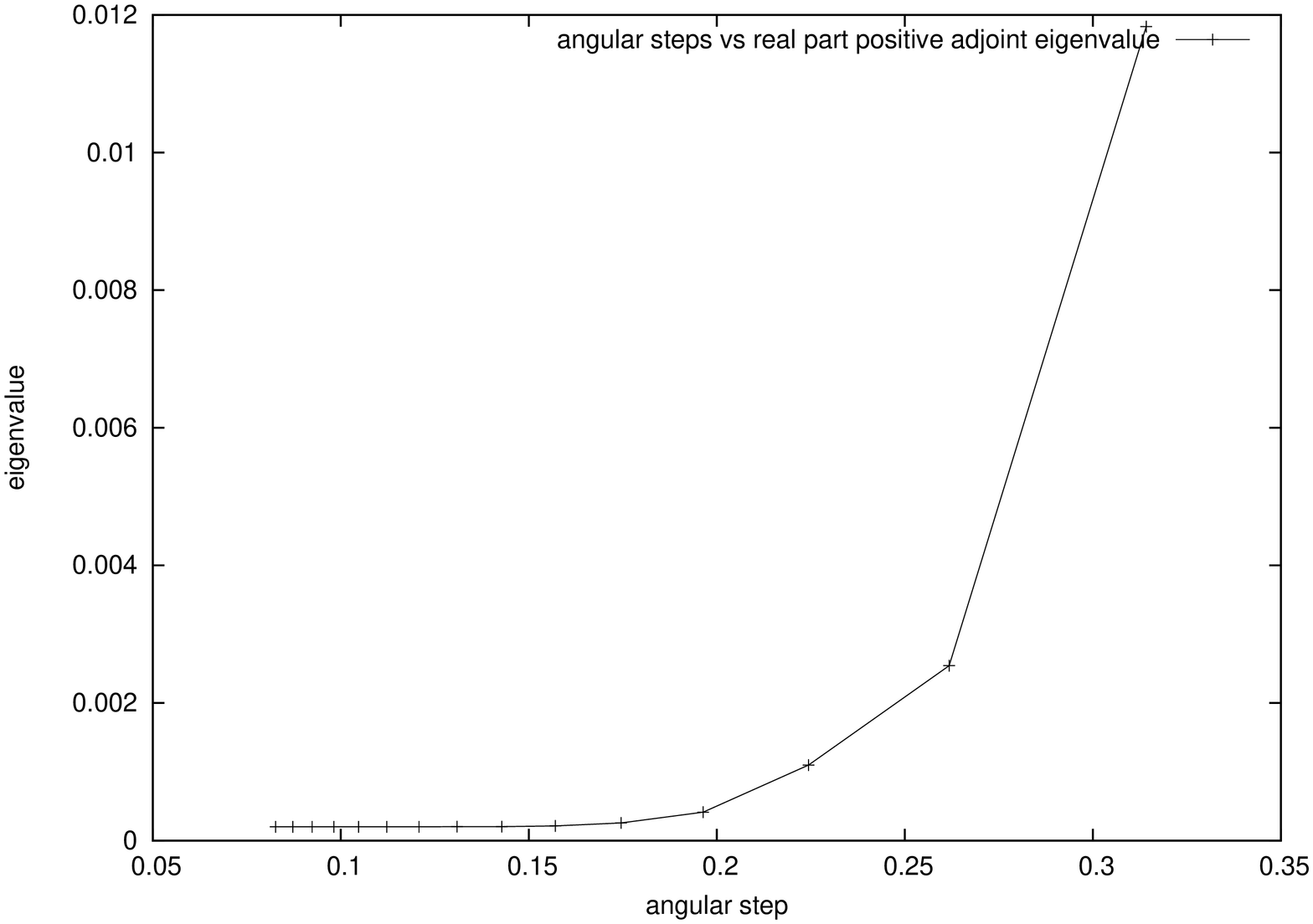}
\end{minipage}
\begin{minipage}[tbp]{0.49\linewidth}
\centering
\includegraphics[width=0.7\textwidth, angle=-90]{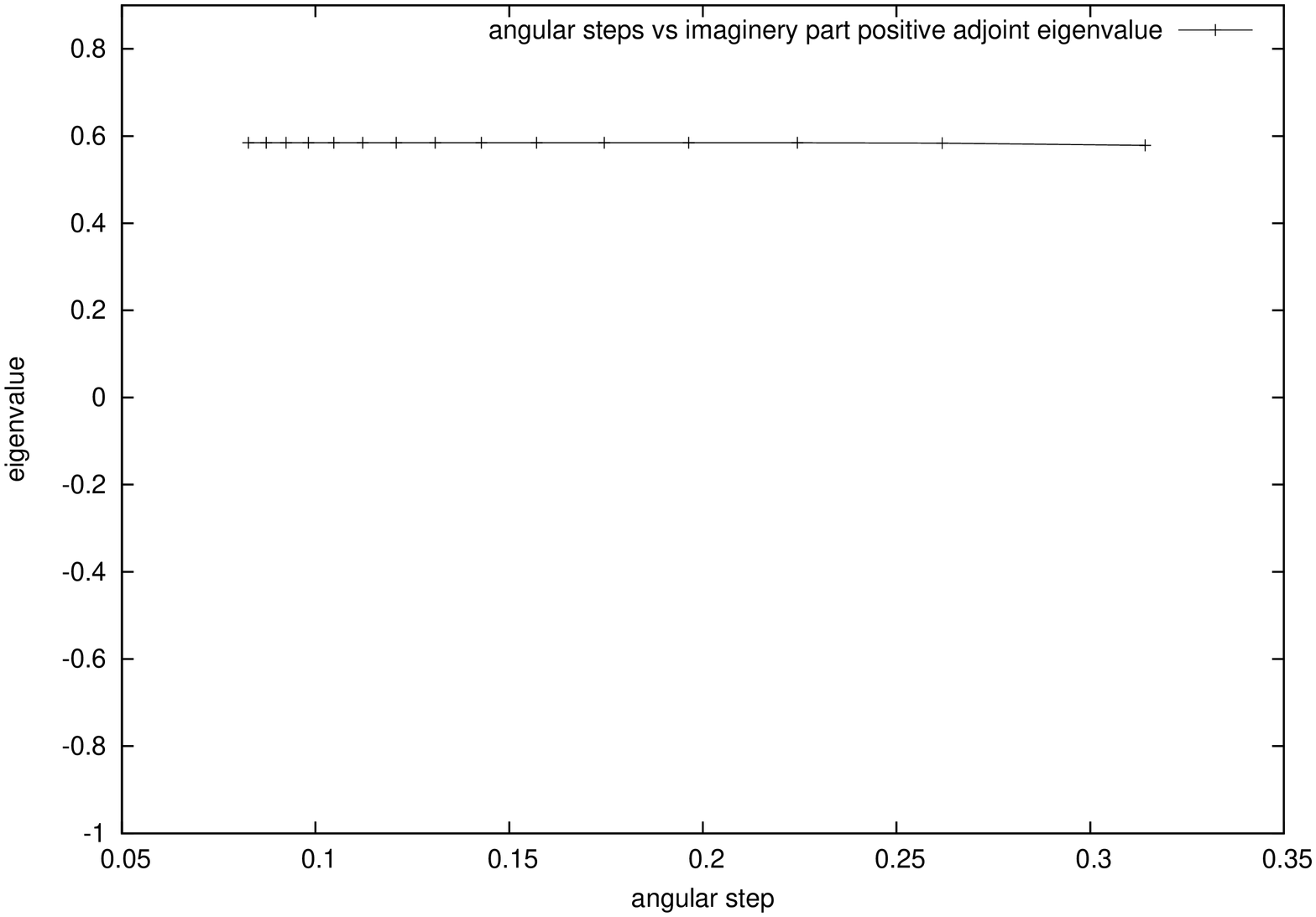}
\end{minipage}
\caption[Angular Convergence]{Angular Convergence: Real (left column) and \chg[af]{imaginary} (right column) parts of the eigenvalues, corresponding to the analytical eigenvalues $-i\lambda$ (top), 0 (middle) and $i\lambda$ (bottom).}
\label{fig:rf_ex_conv_ang1}
\end{center}
\end{figure}
\clearpage

\subsection{Radial Step}

We now consider the plots for convergence in the radial step. We kept the box size fixed at 10 s.u. and varied the number of radial grid points, therefore giving us convergence in the Radial step. Since the numerical radial derivatives are to second order, we plot the results with the square of the radial step.

We find that for the real parts of the eigenvalues, convergence happens for $N_r>100$. We also note there are instabilities in the plot for real part of the zero eigenvalue. These instabilities are of the order $1.0\times10^{-7}$ and is therefore relatively small. We also feel that this did not affect the result generated from this test.

The \chg[af]{imaginary} parts take a bit longer to converge. The plot for \chg[af]{imaginary} part of the zero eigenvalue may look very \chg[af]{chaotic} but in fact the \chg[af]{amplitude} of the \chg[af]{oscillations} here are of the order $1.0\times10^{-22}$, i.e. extremely small. The other eigenvalues appear to converge at $N_r>150$.

\begin{figure}[p]
\begin{center}
\begin{minipage}[tbp]{0.49\linewidth}
\centering
\includegraphics[width=0.7\textwidth, angle=-90]{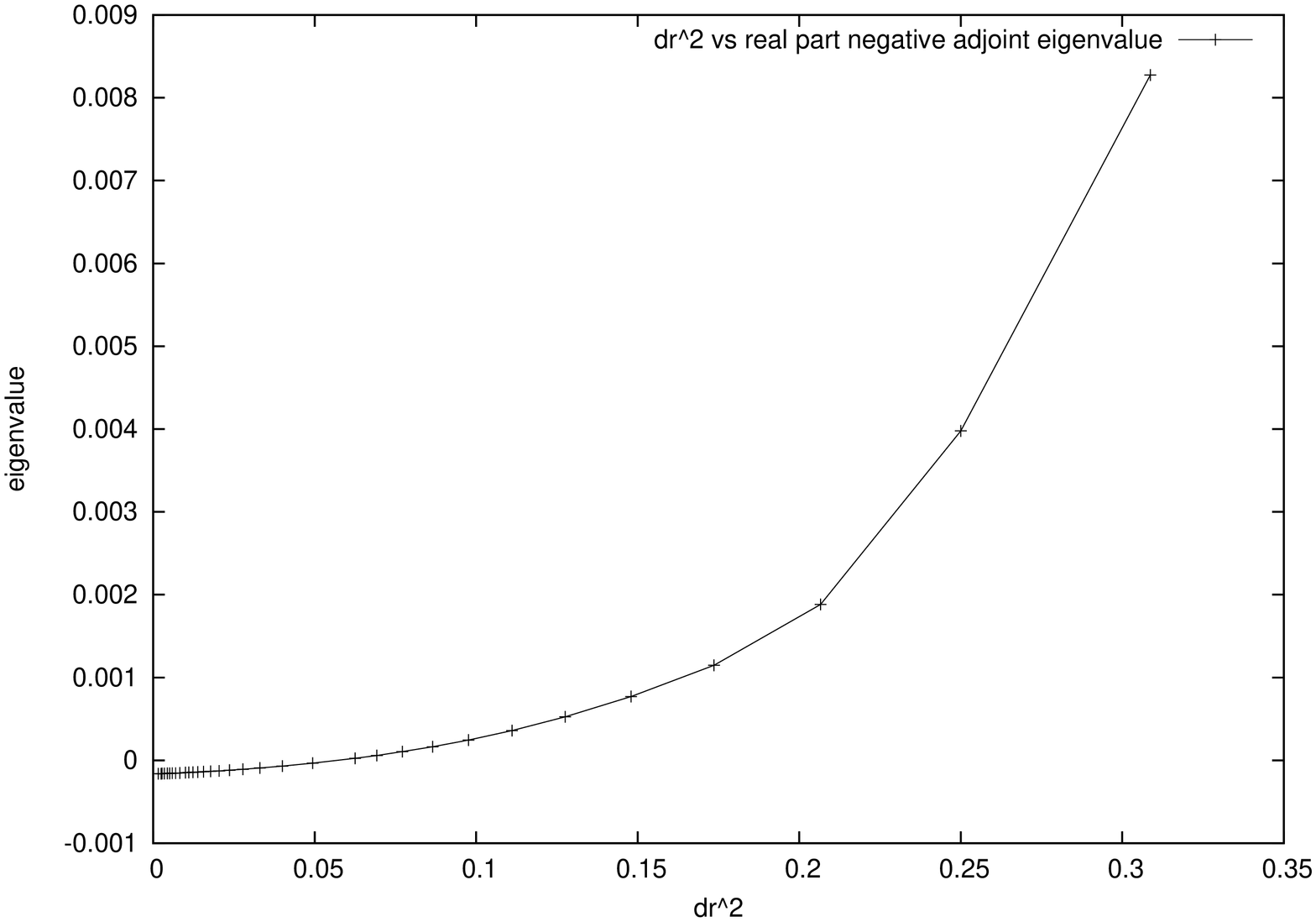}
\end{minipage}
\begin{minipage}[tbp]{0.49\linewidth}
\centering
\includegraphics[width=0.7\textwidth, angle=-90]{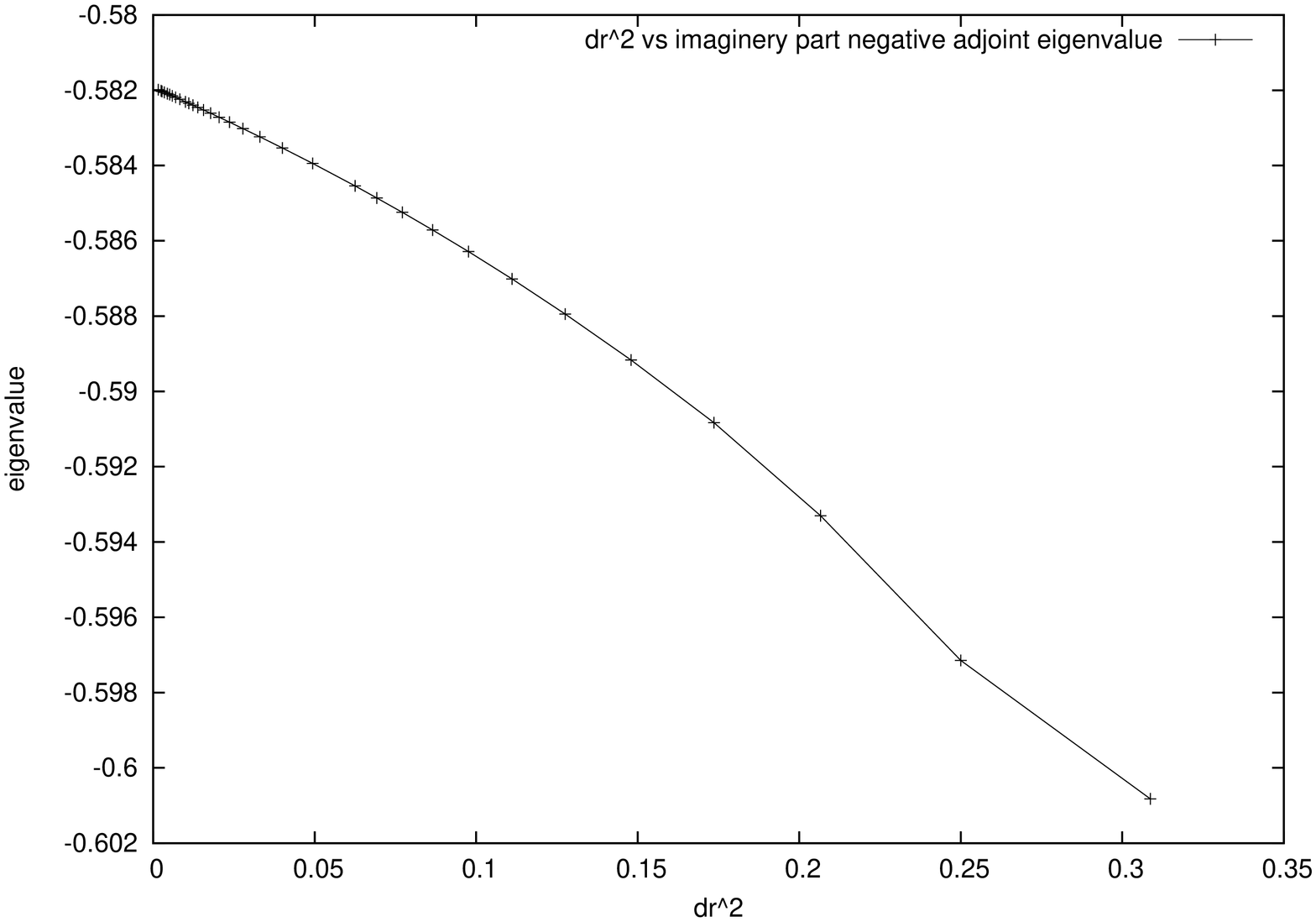}
\end{minipage}
\begin{minipage}[tbp]{0.49\linewidth}
\centering
\includegraphics[width=0.7\textwidth, angle=-90]{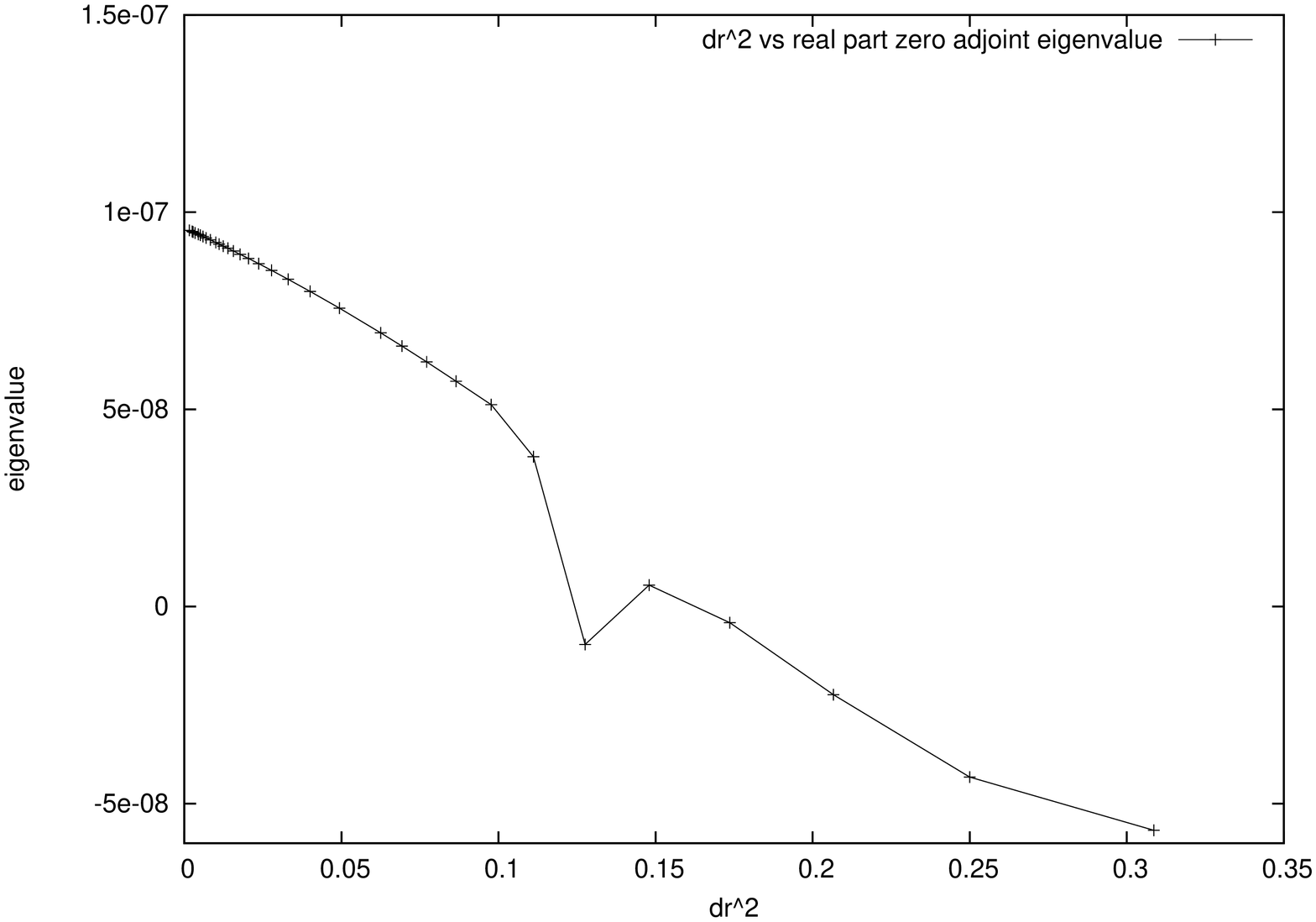}
\end{minipage}
\begin{minipage}[tbp]{0.49\linewidth}
\centering
\includegraphics[width=0.7\textwidth, angle=-90]{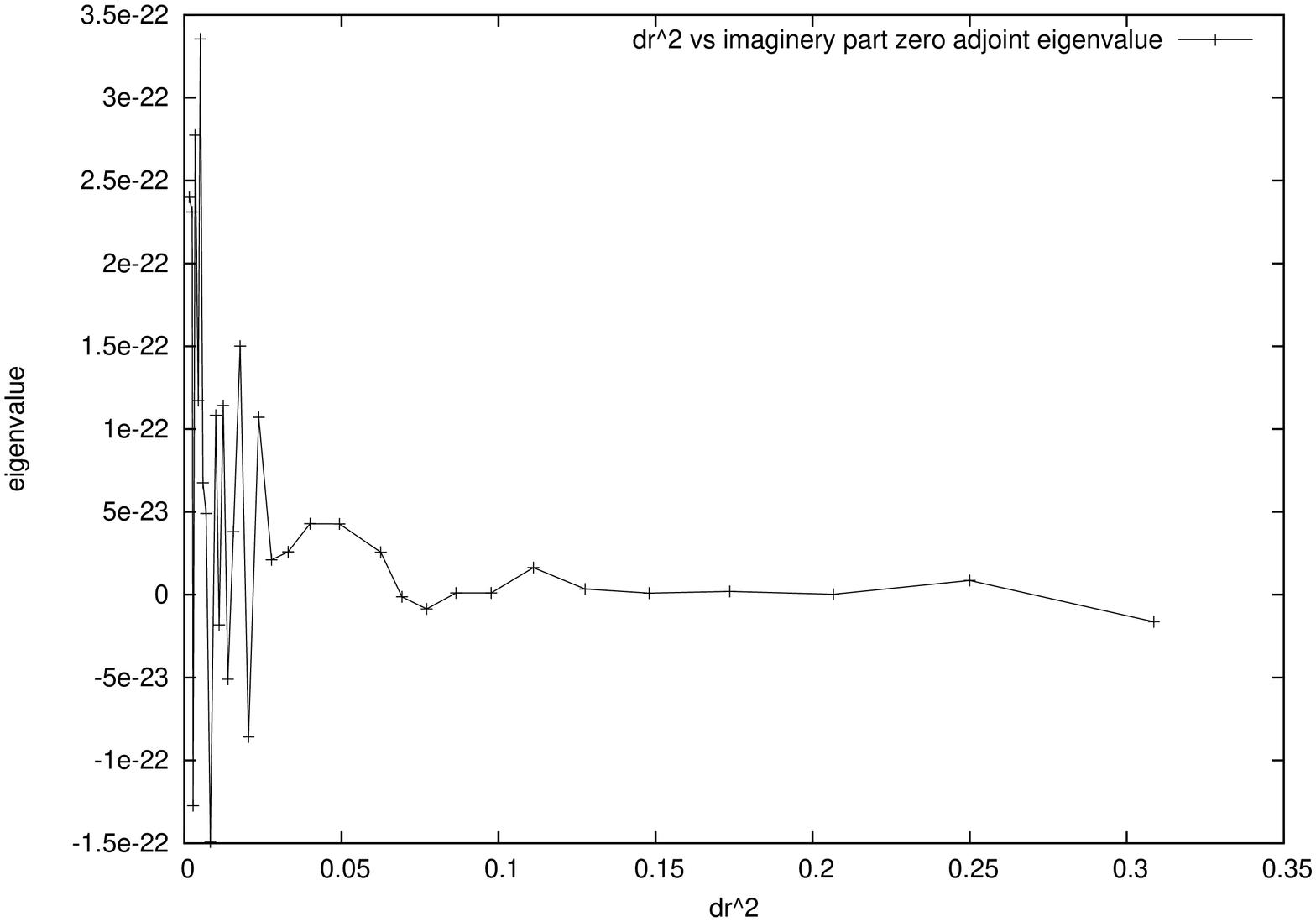}
\end{minipage}
\begin{minipage}[tbp]{0.49\linewidth}
\centering
\includegraphics[width=0.7\textwidth, angle=-90]{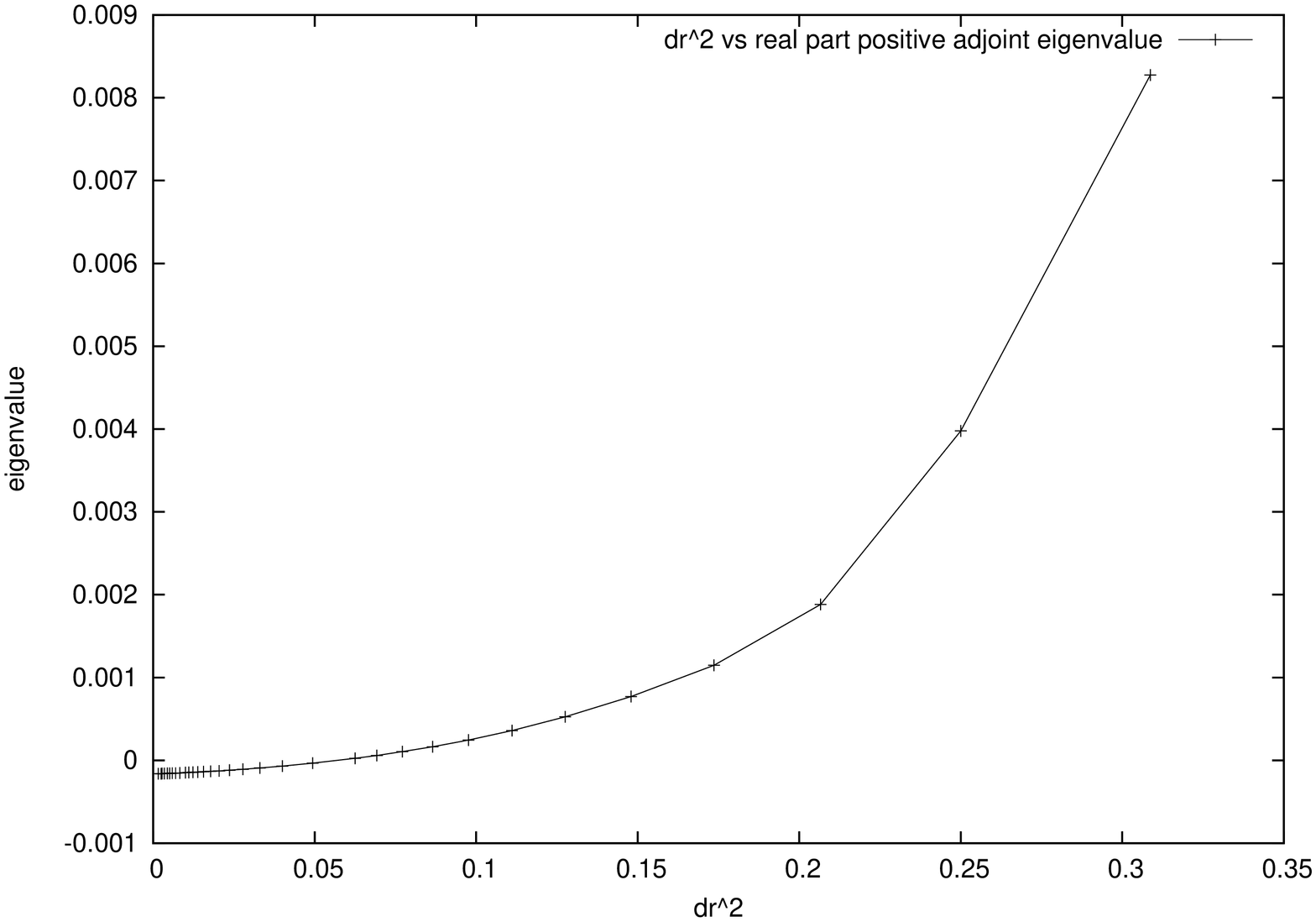}
\end{minipage}
\begin{minipage}[tbp]{0.49\linewidth}
\centering
\includegraphics[width=0.7\textwidth, angle=-90]{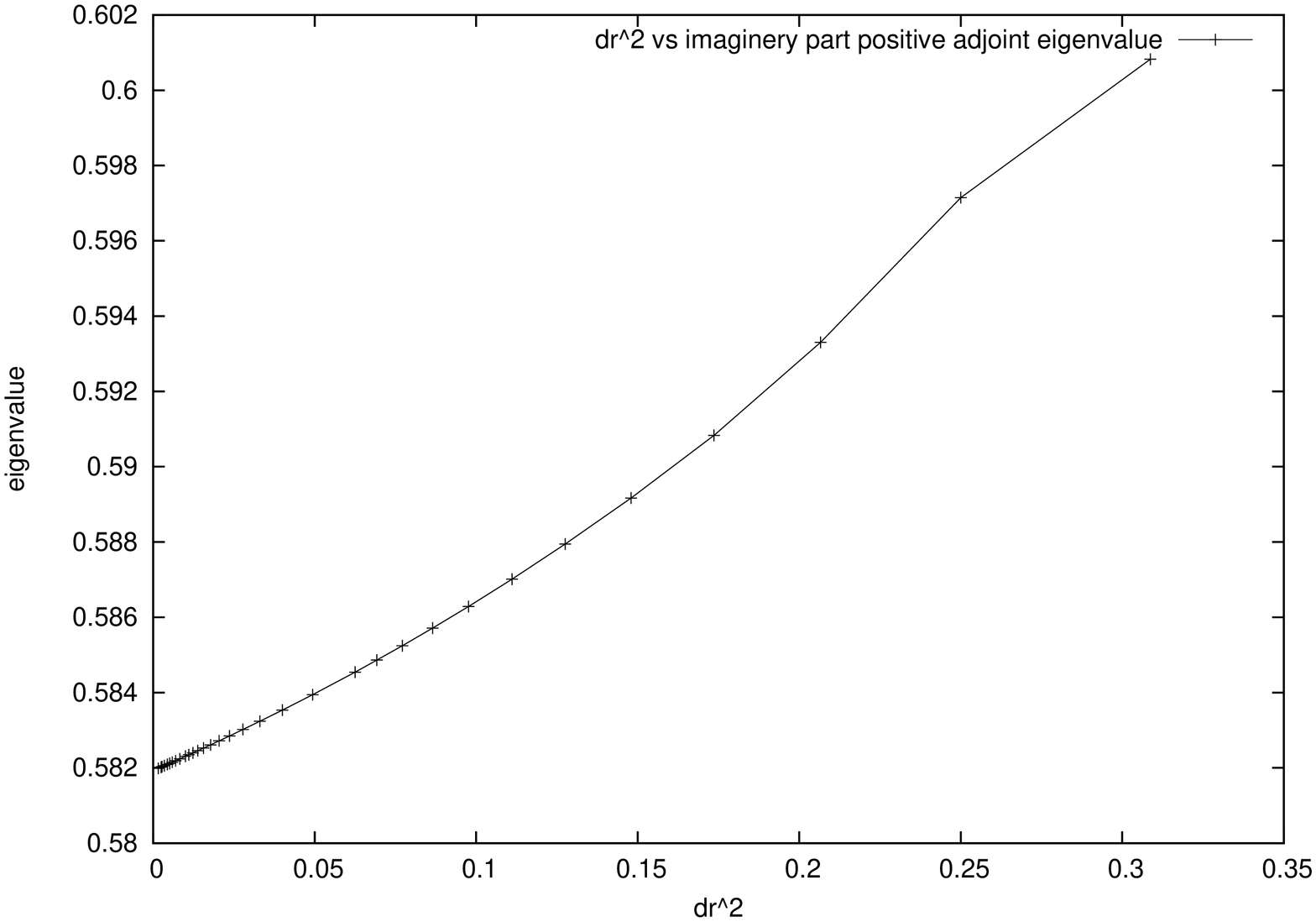}
\end{minipage}
\caption[Radial Convergence]{Radial Convergence plotting $dr^2$ against the Real and \chg[af]{imaginary} parts of the eigenvalues: Real (left column) and \chg[af]{imaginary} (right column) parts of the eigenvalues, corresponding to the analytical eigenvalues $-i\lambda$ (top), 0 (middle) and $i\lambda$ (bottom).}
\label{fig:rf_ex_conv_rad2}
\end{center}
\end{figure}
\clearpage

\section{Conculsion}
\label{sec:rf_conclusion}
To conclude, we have seen that \verb|evcospi| is a very versatile and useful tool in the study of the \chg[af]{response} functions of spiral waves. We have seen that the convergence testing of the program indicates that the program is running as it should and that the numerical methods used have been implemented correctly into the code.

We have also seen that EZ-Freeze can be used to generate initial conditions for \verb|evcospi|. These are produced very quickly and also very accurately. They are generated in a format which is in the \verb|evcospi| ``own'' format, meaning that the \verb|evcospi|.

This work is currently ongoing and we hope that the program will be made available very soon for other researchers to use. The program will also be extended to look at the response functions for scroll waves.

\chapter{Conclusions \& Further Work}
\label{chap:6}

The main results from this work are:

\begin{itemize}
\item A theory of drift of spiral waves using the method of quotient
system by the special Euclidean group is developed.
\item Our new theory has been applied to the drift of rigidly rotating
spirals and the results are consistent with those produce by earlier
theories.
\item Our new theory has been applied to the drift of meandering spirals
for which there \chg[p216]{has} been no complete theory earlier. The theory
predicts the drift of meandering spirals and also a possibility of phase
locking between external stimulation and meandering, which has been
observed experimentally.
\item The method of quotient system by the special Euclidean group has
been implemented numerically (EZ-Freeze program). The numerical
convergence of this method has been demonstrated.
\item The numerical implementation of the method of quotient system can
be used to study drift of spiral waves on indefinitely long time intervals.
\item The numerical implementation of the method of quotient system has
been used to investigate the behaviour of the spiral waves near the
``large core" boundaries, which are difficult to study by standard
methods. It has been demonstrated that the new method can distinguish
between different theoretical asymptotics.
\item The numerical implementation of the method of quotient system has
been used to investigate the behaviour of the spiral waves near the ``l:1
resonance" parametric line. It has been demonstrated that the behaviour
of the quotient system does not demonstrate any peculiarities across
this line.
\item The numerical implementation of the method of quotient system has
been used to generate rigidly rotating spiral wave solutions in the
format suitable for the use in \verb|evcospi|, a program for calculation of the
response functions of spiral waves.
\item Numerical convergence of the solutions obtained by \chg[p217spell]{}\verb|evcospi| has
been demonstrated.
\end{itemize}

Future directions for this work include:

\begin{itemize}
\item Extend our theory to include at least $O(\epsilon^2$) terms to see if frequency locking can be detected. Also, studying locking in FHN model.
\item Use EZ-Freeze to investigate frequency locking.
\item Asymptotic investigation of the behaviour of the solutions after the critical point when studying large core spirals.
\item Investigation into the shape of the limit cycles of the drifting and meandering spiral spiral waves and how the behaviour of the spiral wave is dependent on the shape of the limit cycles.
\item Investigating further the numerical evidence of a Hopf bifurcation from rigid rotation to meander by using EZ-Freeze to study the amplitude of the limit cycles.
\item Another interesting line of research would be to see how our theory could be adapted to three dimensional scroll waves.
\end{itemize}

\appendix
\chapter{Definitions}

Throughout this \chg[p218gram1]{thesis}, we will be using particular words and phrases. It is 
assumed that the reader will have no specific knowledge of these terms and 
therefore we introduce some basic definitions for the reader's perusal.


\section{Dynamical System}
Although not specifically mentioned in this \chg[p218gram2]{thesis}, the definition of a 
Dynamical System should be known as the systems studied throughout this 
report are indeed Dynamical Systems. A Dynamical System is a pair 
$\{X,\phi^t\}$, where X is a state space and $\phi^t : X \to X $ is a family 
of evolution operators satisfying the conditions $\phi^0=id$ and 
$\phi^{t+s}=\phi^t\circ\phi^s$. A State Space is a set containing all 
possible states of the system [6].

\chg[p218para]{The most common way to represent a continuous-time dynamical system is 
with a set of differential equations.}


\section{Quasiperiodicity}

Meandering waves display quasiperiodic motion. Quasiperiodic motion is a 
regular but \chg[p218gram3]{non-chaotic} motion which consists of a combination of period 
motions with a trajectory that after a sufficient period of time pass 
arbitrarily close to an earlier value\chg[p218gram4]{, but never on it.}

A dynamical system is called \emph{k-quasi-periodic} if it can be written in 
the form:

\[x(t)=f(\omega_1t, \omega_2t, .. , \omega_kt)\]
\\
where $\omega_i$ for $i=1,..,k$ are such that they are not rationally 
related, i.e. the \chg[p219spell]{relationship} between say $\omega_i$ and $\omega_j$ is such 
that $\frac{\omega_i}{\omega_j}\ne\frac{p}{q}, p\in\mathbb{Z}, 
q\in\mathbb{Z}$. $\omega_i$ are said to be the frequencies of the 
system.

A system which has just 2 frequencies, $\omega_1$ and $\omega_2$, is known 
just as quasi-periodic. The motion of a quasi-periodic system can be 
described as the motion on the \chg[p219gram]{2-torus} (see below). Since the 
frequencies, $\omega_1$ and $\omega_2$, are not rationally related, the 
trajectory of the system on the torus never crosses, although one point on 
the torus, which the trajectory passes through, may be arbitrarily close to 
another point on the trajectory, which is also a point on the Torus.

As a final note, when the relationship between the frequencies becomes 
rational somehow, then we say that the frequencies are locked. 


\section{Torus}

A \emph{n-Torus}, $\mathbb{T}^n$, is the product of $n$-circles, i.e:

\[\mathbb{T}^n=S^1\times  S^1\times .... \times S^1\]

The 2-Torus is a 3 dimensional shape resembling a "doughnut". A typical 
2-Torus is \chg[af]{shown in Fig.(\ref{torus})}:

\begin{figure}[h]
\begin{center}
\begin{minipage}[b]{0.5\linewidth}
\centering
\includegraphics[width=\textwidth, angle=-90]{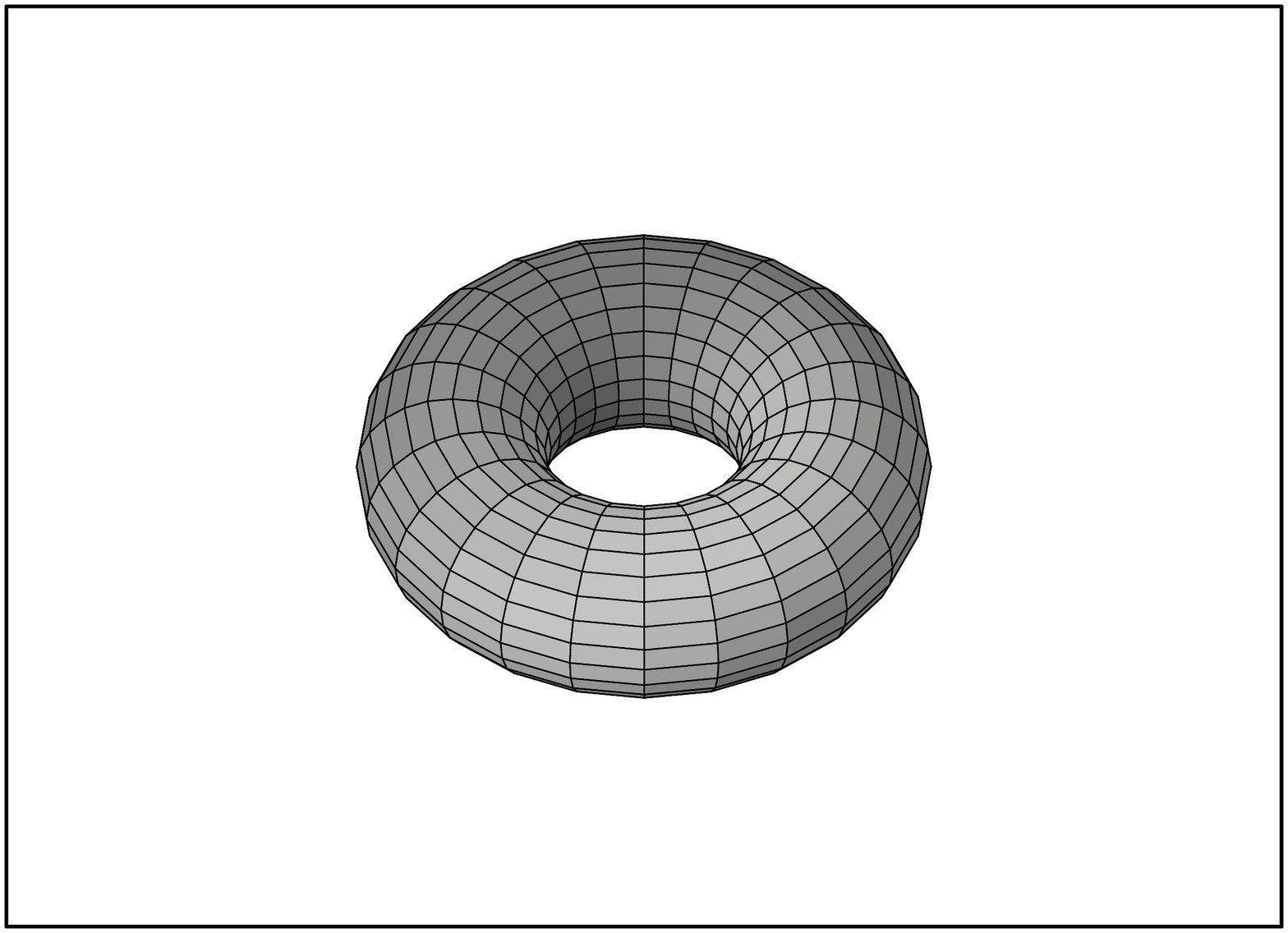}
\caption{A typical 2-Torus.}
\label{torus}
\end{minipage}
\end{center}
\end{figure}

The 2-Torus can be created by taking a rectangle and fixing the pairs of 
opposite corners together ensuring that there are no twists.


\section{Bifurcations}

In laymans terms, a bifurcation point separates regions of stability in a particular system. Consider for example the following simple system of ODE's:

\begin{eqnarray*}
\dot{x} & = & -x+y \\
\dot{y} & = & ax-y
\end{eqnarray*}

The system has a fixed point at the origin, (0,0), and the stability matrix at this point, J(0,0), is given by:

\begin{eqnarray*}
J(0,0) & = & \left(\begin{array}{cc} -1 & 1\\ a & -1 \end{array}\right) 
\end{eqnarray*}

Eigenvalues are therefore the solution to the equation $\lambda^2+2\lambda+(1+a)=0 \ \Rightarrow \ \lambda_{\pm}=-1 \pm \sqrt{-a}$. Therefore, we can come to the following conclusions about the stability of the fixed point (0,0):

\[\begin{array}{rcccc}
\mbox{if} \ a<-1 & \Rightarrow & \lambda_{\pm} \ \mbox{have opposite signs} & \Rightarrow & \mbox{Saddle Point - Unstable}\\
\mbox{if} \ -1<a\le0 & \Rightarrow & \lambda_{\pm} \ \mbox{are both negative} & \Rightarrow & \mbox{Stable Node}\\
\mbox{if} \ a>0 & \Rightarrow & \lambda_{\pm} \ \mbox{are complex with a positive real part} & \Rightarrow & \mbox{Stable Focus Point}
\end{array}\]

We therefore say that there is a Bifurcation when a=-1, at which point the system goes from unstable to stable.

This is a very simple example of a system possessing a Bifurcation point. There are many different types of bifurcations (Fold, Pitchfork, Neimark-Sacker, Hopf, Hopf-Hopf) but the ones that we are most interested in are the Hopf Bifurcations and also the Neimark-Sacker Bifurcations (sometimes referred to as a Secondary Hopf Bifurcation).


\subsection{Hopf Bifurcation}

A Hopf Bifurcation occurs when a fixed point of the system in question goes from being stable to being unstable and is surrounded by a \chg[p220gram]{stable} limit cycle. The Hopf Bifurcation point is the point at which the complex eigenvalues of the system at the fixed point in question cross the imaginary axis. Hence, the eigenvalues of the system are purely imaginary at the Hopf Bifurcation Point. The eigenvalue equation may be written in the form:

\begin{eqnarray*}
\lambda^2+Tr(J)\lambda+Det(J) & = & 0
\end{eqnarray*}

where $Tr(J)$ is the trace of the Jacobian and $Det(J)$ is the determinant of the Jacobian. Hence, sufficient conditions for the presence of a Hopf bifurcation is $Tr(J)=0$ and $Det(J)>0$, giving the eigenvalues to be $\lambda_{\pm}=\pm i\sqrt{Det(J)}$.


\subsection{Supercritical and Subcritical Hopf Bifurcations}

A Supercritcial Hopf Bifurcation is one in which the limit cycle is attracting, i.e it is a stable limit cycle. Therefore, a Subcritical Hopf Bifurcation is one in which the limit is repelling, i.e. unstable.


\subsection{Hopf Bifurcation Normal Form}
\label{sec:HB}

In this section we provide a brief but thorough review of the Hopf Bifurcation and how it applies to our system and spiral wave solution.

Consider the following dynamical system:

\begin{equation}
\label{eqn:hbxsys}
\dot{\textbf{x}} = \textbf{f}(\textbf{x},p),\quad \textbf{x}\in\mathbb{R}^m,\quad \textbf{x}=(x_1,x_2,\ldots,x_m)
\end{equation}

This system has an equilibrium at $\textbf{x}_*$, satisfying:

\begin{equation*}
\textbf{0} = \textbf{f}(\textbf{x}_*,p)
\end{equation*}

Now the right hand side of (\ref{eqn:hbxsys}), has both linear and non-linear parts. In order to study the full system inclusive of the non-linear parts, we need to use the \emph{Center Manifold Theorem} (this will be introduced later on). Therefore, we need our equilibrium to located at the origin. Let us introduce a change of coordinates:

\begin{equation*}
\textbf{y} = \textbf{x}-\textbf{x}_*
\end{equation*}

Therefore, our original system now becomes:

\begin{equation*}
\label{eqn:hbysys1}
\dot{\textbf{y}} = \textbf{g}(\textbf{y},p),\quad \textbf{y}\in\mathbb{R}^m,\quad \textbf{y}=(y_1,y_2,\ldots,y_m)
\end{equation*}

The equilibrium of this new system is now located at the origin, since if at the equilibrium $\textbf{x}=\textbf{x}_*$, then $\textbf{y}=\textbf{0}$ at the equilibrium:

\begin{equation*}
\textbf{0} = \textbf{g}(\textbf{0},p)
\end{equation*}

The Hopf Bifurcation Theory (see \cite{kuznetsov}) tells us that if a system possesses a Hopf Bifurcation, then there are 2 complex conjugate eigenvalues that lie on the imaginary axis when the Hopf Bifurcation takes place. Also, the real parts of these eigenvalues are monotonic functions of the system parameter. If the real parts are monotonically increasing, then we get a \emph{Supercritical Hopf Bifurcation}. If the real parts are monotonically decreasing, then we get a \emph{Subcritical Hopf Bifurcation}. This then leads to limit cycle solutions being observed in the system.

Let us consider the linear part of the system and assume that the system has eigenvalues in the following form:

\begin{equation*}
\lambda_{1,2} = \epsilon(p)\pm i\omega(p),\quad Re\{\lambda_{i\geq 3}\}<0
\end{equation*}

Now, $\epsilon(p)$ is a monotonic function. Let us assume that it is monotonically increasing. This means that $\epsilon(p)$ has an inverse. In particular this inverse is $p=p(\epsilon)$. Therefore, we have:

\begin{eqnarray}
\label{eqn:hbysys}
\dot{\textbf{y}} &=& \textbf{g}(\textbf{y},\epsilon),\quad \textbf{y}\in\mathbb{R}^m,\quad \textbf{y}=(y_1,y_2,\ldots,y_m)\\
\lambda_{1,2} &=& \epsilon\pm i\omega(\epsilon)\nonumber
\end{eqnarray}

Another restriction that we have with the Center Manifold Theorem is that the system studied must be independent of parameters. System (\ref{eqn:hbysys}) is clearly dependent on parameter $\epsilon$. So how can be make this system independent of parameters? One way is to make $\epsilon$ a dynamical variable of the system. We know that the parameters of our system stay constant for all time and therefore we can introduce the following conditions:

\begin{eqnarray*}
\dot{\textbf{y}} &=& \textbf{g}(\textbf{y},\epsilon)\\
\dot{\epsilon} &=& 0
\end{eqnarray*}

If we let $\textbf{u}=(\textbf{y},\epsilon)^T$, then we get a new parameter independent system:

\begin{equation}
\label{eqn:hbusys}
\dot{\textbf{u}} = \textbf{h}(\textbf{u}),\quad \textbf{u}\in\mathbb{R}^{m+1},\quad \textbf{u}=(u_1,u_2,\ldots,u_m)
\end{equation}

Let us think about an equilibrium point of this system. Let us suppose this equilibrium point is at $\textbf{u}_*$:

\begin{eqnarray*}
\textbf{u}_* &=& \left( \begin{array}{c} \textbf{y}_*\\ \epsilon \end{array}\right)\\
&=& \left( \begin{array}{c} \textbf{0}\\ \epsilon \end{array}\right)
\end{eqnarray*}

This means that we have an infinite amount of equilibrium points. However, as we stated earlier, the Center Manifold Theorem insists that the equilibrium is at the origin. Hence, we must have $\epsilon=0$. Also, this means that in the linear system, we have that $\epsilon=0$.

Let us now split the right hand side of (\ref{eqn:hbusys}) into its linear and non-linear parts:

\begin{equation}
\label{eqn:hbusplit}
\dot{\textbf{u}} = J\textbf{u}+\textbf{N}(\textbf{u}), \quad \textbf{N}(\textbf{u})=O(\textbf{u}^2)
\end{equation}
\\
where $J$ is the Jacobian of the system. Firstly, let us neglect the non-linear terms and look at the linear system:

\begin{equation*}
\label{eqn:hbulin}
\dot{\textbf{u}} = J\textbf{u}
\end{equation*}

The Jacobian is an $(m+1)\times(m+1)$ matrix and is given by:

\begin{equation*}
\label{eqn:jacobian}
J = \left(\begin{array}{cc} \frac{\partial{\textbf{g}}}{\partial{\textbf{y}}} & 0\\ 0 & 0 \end{array}\right)
\end{equation*}

The eigenvalues are exactly the same as those in the system (\ref{eqn:hbysys}) but with an extra eigenvalue $\lambda_0=0$. Therefore the eigenvalues are:

\begin{equation*}
\lambda_0 = 0,\quad \lambda_{1,2} = \pm i\omega_0,\quad Re\{\lambda_{i\geq 3}\}<0
\end{equation*}
\\
remembering that $\epsilon=0$ and also denoting $\omega(0)=\omega_0$. If we denote the corresponding eigenvectors as $v_i$ for $i=0,\ldots,m$, then the solution in the linearised system is:

\begin{eqnarray*}
\label{eqn:usollinear}
\textbf{u} &=& \sum_{i=0}^m {a_i\textbf{v}_ie^{\lambda_it}}\\
&=& a_0\textbf{v}_0e^{\lambda_0t}+a_1\textbf{v}_1e^{\lambda_1t}+a_2\textbf{v}_2e^{\lambda_2t}+\sum_{i=3}^m {a_i\textbf{v}_ie^{\lambda_it}}\\
&=& a_0\textbf{v}_0+a_1\textbf{v}_1e^{i\omega t}+\bar{a}_1\bar{\textbf{v}}_1e^{-i\omega t}+\sum_{i=3}^m {a_i\textbf{v}_ie^{\lambda_it}}
\end{eqnarray*}
\\
noting that because $\lambda_2=\bar{\lambda}_1\Rightarrow \textbf{v}_2=\bar{\textbf{v}}_1$, then $a_2=\bar{a}_1$ for our solution to produce real results. What happens as $t\rightarrow\infty$? We can see that since $Re\{\lambda_{i\geq 3}\}<0$, then $\sum_{i=3}^m {a_i\textbf{v}_ie^{\lambda_it}}\rightarrow 0$. Therefore, over the long term, our solution becomes:

\begin{equation*}
\label{eqn:usollin}
\textbf{u} = a_0\textbf{v}_0+a_1\textbf{v}_1e^{i\omega t}+\bar{a}_1\bar{\textbf{v}}_1e^{-i\omega t}
\end{equation*}

Now, all solutions to the linear system lie on a \emph{Center Subspace}. The Center Subspace is a subspace of space of solutions that is spanned by the eigenvectors of the linearised system. In the following subsection we show an example this statement. Also, the solutions to the linear system can be expressed as a linear combination of the eigenvectors - this is known as the span of the system:

\begin{equation}
\label{eqn:span}
\textbf{u} = \{\alpha_0\textbf{v}_0+\alpha_1\textbf{v}_1+\bar{\alpha}_1\bar{\textbf{v}}_1|\alpha=(\alpha_0,\alpha_1,\bar{\alpha}_1)\}
\end{equation}

So, any solution on this Center Subspace can be expressed in terms of $\alpha$, i.e. $\alpha$ is the coordinate of a solution lying on the Center Subspace. Also, the Center Subspace is such that any solution that does not lie on it is attracted to it over time. This is due to the ``discarded'' terms $\sum_{i=3}^m {a_i\textbf{v}_ie^{\lambda_it}}$, which actually do not actually vanish but are extremely small. We also note that if the solution is on the Center Subspace, then it will stay in it indefinitely. It can also be seen that as we move through time, we get different values of $\alpha$. Therefore, we can say that $\alpha$ is dependent on time.

We now bring back the non-linear terms. In order for us to study the full system inclusive of the non-linear terms, we must use the Center Manifold Theorem.

\begin{thm}[Center Manifold Theorem (CMT)]
\label{thm:cmt}
Given the parameter independent system (\ref{eqn:hbusplit}), then there exists a Center Manifold which is:
\begin{enumerate}
\item tangential at the origin,
\item invariant,
\item attracting.
\end{enumerate}
\end{thm}

In our system we have $\textbf{u}\in\mathbb{R}^{m+1}$. If we want to study our system in the region of the equilibrium, then the CMT allows us to study our full system in a reduced system consisting of $k$ ODE's, where k is the number of eigenvalues with zero real part. In our case, we have 3 eigenvalues with zero real part and hence the CMT implies that we will have 3 ODE's describing the motion of the system near the equilibrium point.

So how does the CMT allow this? Well, let us consider the CMT. Firstly, what does point 3 in Theorem (\ref{thm:cmt}) \chg[p236gram1]{tells us that any solution to our system which does not start} on the Center Manifold will be attracted to the Center Manifold, but the trajectory will not be part of the manifold. This is evident in our system since $Re\{\lambda_{1,2}\}<0$, and therefore over time these will decay to zero. Therefore, we can model the motion using just 3 equations in our.

The second point in the theorem states that any trajectory starting on the Center Manifold will remain on the manifold indefinitely. Hence, we model the system using a system of closed ODE's.

Finally, the first point in the theorem states that in region sufficiently close to the origin, i.e. the equilibrium, there is a one to one correspondence between points on the Center Manifold and points on the Center Subspace. Therefore, in the region of the equilibrium, we can model the motion of the system as motion on the Center Subspace. Therefore, each point on the trajectory of a solution to our system will have the coordinates $\alpha=(\alpha_0,\alpha_1,\bar{\alpha}_1)$. Accordingly, the 3 ODE's that will model this motion will be:

\begin{eqnarray*}
\dot{\alpha}_0 &=& k_0(\alpha)\\
\dot{\alpha}_1&=& k_1(\alpha)\\
\dot{\bar{\alpha}}_1 &=& k_2(\alpha)
\end{eqnarray*}

Now let us now return to our problem. We saw that we could express the solution in the linear system as the span of the eigenvectors (\ref{eqn:span}). The CMT implies that the solution in the full system may also be expressed in this form. Let us now substitute (\ref{eqn:span}) into \chg[p236gram2]{}({\ref{eqn:hbusplit}):

\begin{eqnarray*}
\dot{\textbf{u}} &=& J\textbf{u}+\textbf{N}(\textbf{u})\\
\dot{\alpha}_0\textbf{v}_0+\dot{\alpha}_1\textbf{v}_1+\dot{\bar{\alpha}}_1\bar{\textbf{v}}_1 &=& \alpha_0J\textbf{v}_0+\alpha_1J\textbf{v}_1+\bar{\alpha}_1J\bar{\textbf{v}}_1+\textbf{N}(\alpha))
\end{eqnarray*}

Considering just the linear terms and using $J\textbf{v}_i=\lambda_i\textbf{v}_i$, we get:

\begin{eqnarray*}
\dot{\alpha}_0\textbf{v}_0+\dot{\alpha}_1\textbf{v}_1+\dot{\bar{\alpha}}_1\bar{\textbf{v}}_1 &=& \alpha_0\lambda_0\textbf{v}_0+\alpha_1\lambda_1\textbf{v}_1+\bar{\alpha}_1\bar{\lambda}_1\bar{\textbf{v}}_1
\end{eqnarray*}

Equating coefficients of the eigenvalues, we get:

\begin{eqnarray*}
\dot{\alpha}_0 &=& \alpha_0\lambda_0\\
\dot{\alpha}_1 &=& \alpha_1\lambda_1\\
\dot{\bar{\alpha}}_1 &=& \bar{\alpha}_1\bar{\lambda}_1
\end{eqnarray*}
\\
not adding in the non-linear terms and substituting in the expressions for the eigenvalues we get:
\begin{eqnarray}
\label{eqn:alp0dot}
\dot{\alpha}_0 &=& 0+N_0(\alpha)\\
\label{eqn:alp1dot}
\dot{\alpha}_1 &=& i\omega_0\alpha_1+N_1(\alpha)\\
\label{eqn:alp2dot}
\dot{\bar{\alpha}}_1 &=& -i\omega_0\bar{\alpha}_1+N_2(\alpha)
\end{eqnarray}

Let us consider (\ref{eqn:alp0dot}). We know that $\textbf{u}$ is of the form:

\begin{eqnarray*}
\textbf{u} &=& \alpha_0\textbf{v}_0+\alpha_1\textbf{v}_1+\bar{\alpha}_1\bar{\textbf{v}}_1\\
\Rightarrow \left(\begin{array}{c} \textbf{y}\\ \epsilon \end{array}\right) &=& \alpha_0\left(\begin{array}{c} \textbf{0}\\ 1 \end{array}\right)+\alpha_1\left(\begin{array}{c} \vdots\\ 0 \end{array}\right)+\bar{\alpha}_1\left(\begin{array}{c} \vdots \\ 0 \end{array}\right)
\end{eqnarray*}

Therefore, $\alpha_0=\epsilon\Rightarrow\dot{\alpha}_0=\dot{\epsilon}=0$. Therefore, we have $0=N_0(\textbf{u})$.

Let us now consider (\ref{eqn:alp1dot}). We can also consider (\ref{eqn:alp2dot}) at the same time since it is just the conjugate of (\ref{eqn:alp1dot}). Using Taylor's Expansion, we can expand the non-linear terms and get:

\begin{equation*}
\dot{\alpha}_1 = i\omega_0\alpha_1+A\alpha_0^2+B\alpha_1^2+C\bar{\alpha}_1^2+D\alpha_0\alpha_1+E\alpha_0\bar{\alpha}_1+F\alpha_1\bar{\alpha}_1+O(\alpha^3)
\end{equation*}

Firstly, note that $\alpha_0=\epsilon$ and $i\omega_0=\lambda_1$. Also, in our system, we are denoting the motion of the solutions as lying on the Center Subspace. Therefore we have that $\epsilon=0$. This means that:

\begin{eqnarray*}
\dot{\alpha}_1 &=& (D\epsilon+i\omega_0)\alpha_1+E\epsilon\bar{\alpha}_1+A\epsilon^2+B\alpha_1^2+C\bar{\alpha}_1^2+F\alpha_1\bar{\alpha}_1+O(\alpha^3)\\
&=& 
\lambda_1\alpha_1+B\alpha_1^2+C\bar{\alpha}_1^2+F\alpha_1\bar{\alpha}_1+O(\alpha^3)
\end{eqnarray*}

We now make a Poincar\'e change of variables using:

\begin{equation}
\label{eqn:alpha1}
\alpha_1 = w+p_1w^2+p_2w\bar{w}+p_3\bar{w}^2+O(|w|^3)
\end{equation}

Therefore, we can say that $w$ is given by the inverse of $\alpha_1$:

\begin{equation}
\label{eqn:w}
w = \alpha_1-p_1\alpha_1^2-p_2\alpha_1\bar{\alpha}_1-p_3\bar{\alpha}_1^2+O(|\alpha_1|^3)
\end{equation}

This can be seen by substituting (\ref{eqn:w}) into (\ref{eqn:alpha1}) which will give, to $O(|\alpha_1|^3)$, $\alpha_1=\alpha_1+O(|\alpha_1|^3)$. Differentiating with respect to time:

\begin{eqnarray*}
\dot{w} &=& \dot{\alpha}_1-2p_1\alpha_1\dot{\alpha}_1-p_2(\dot{\alpha}_1\bar{\alpha}_1+\alpha_1\dot{\bar{\alpha}}_1)-2p_3\bar{\alpha}_1\dot{\bar{\alpha}}_1\\
&=& (\lambda_1\alpha_1+B\alpha_1^2+C\bar{\alpha}_1^2+F\alpha_1\bar{\alpha}_1)-2\lambda_1p_1\alpha_1^2-p_2(\lambda_1\alpha_1\bar{\alpha}_1+\bar{\lambda}_1\alpha_1\bar{\alpha}_1)-2\bar{\lambda}_1p_3\bar{\alpha}_1^2\\
&=& \lambda_1\alpha_1+\alpha_1^2(B-2p_1\lambda_1)+\alpha_1\bar{\alpha}_1(F-p_2(\lambda_1+\bar{\lambda}_1))+\bar{\alpha}_1^2(C-2p_3\bar{\lambda}_1)\\
&=& \lambda_1(w+p_1w^2+p_2w\bar{w}+p_3\bar{w}^2)+w^2(B-2p_1\lambda_1)+w\bar{w}(F-p_2(\lambda_1+\bar{\lambda}_1))\\
&& +\bar{w}^2(C-2p_3\bar{\lambda}_1)\\
&=& \lambda_1w+w^2(B-p_1\lambda_1)+w\bar{w}(F-p_2\bar{\lambda}_1)+\bar{w}^2(C-p_3(2\bar{\lambda}_1-\lambda_1))
\end{eqnarray*}
\\
omitting all higher order terms. We can see that we can eliminate all quadratic terms by setting:

\begin{eqnarray*}
p_1 &=& \frac{B}{\lambda_1}\\
p_2 &=& \frac{F}{\bar{\lambda}_1}\\
p_3 &=& \frac{C}{2\bar{\lambda}_1-\lambda_1}
\end{eqnarray*}

We also see that the denominators in the above expressions for $p_1$ to $p_3$ are never zero since $\omega_0>0$. We therefore have:

\begin{equation}
\label{eqn:hbnoquad}
\dot{w} = \lambda_1w+O(|z|^3)
\end{equation}

Let us now look at the cubic terms and express (\ref{eqn:hbnoquad}) as follows:

\begin{equation*}
\dot{w} = \lambda_1w+Gw^3+Hw^2\bar{w}+Kw\bar{w}^2+L\bar{w}^3+O(|w|^4)
\end{equation*}

We now make another Poincar\'e Change of Variables:

\begin{eqnarray}
w &=& z+q_1z^3+q_2z^2\bar{z}+q_3z\bar{z}^2+q_4\bar{z}^3\\
\label{eqn:z}
\Rightarrow z &=& w-q_1w^3-q_2w^2\bar{w}-q_3w\bar{w}^2-q_4\bar{w}^3+\mbox{h.o.t.}
\end{eqnarray}

Differentiating (\ref{eqn:z}) with respect to time gives us:
\chg[p238eqn]{}
\begin{eqnarray*}
\dot{z} &=& \dot{w}-3q_1w^2\dot{w}-q_2(2w\dot{w}\bar{w}+w^2\dot{\bar{w}})-q_3(\dot{w}\bar{w}^2+2w\bar{w}\dot{\bar{w}})-3q_4\bar{w}^2\dot{\bar{w}}\\
&=& (\lambda_1w+Gw^3+Hw^2\bar{w}+Kw\bar{w}^2+L\bar{w}^3)-3q_1\lambda_1w^3-q_2(2\lambda_1w^2\bar{w}\\
&& +\bar{\lambda}_1w^2\bar{w})-q_3(\lambda_1w\bar{w}^2+2\bar{\lambda}_1w\bar{w}^2)-3q_4\bar{\lambda}_1\bar{w}^3\\
&=& \lambda_1w+w^3(G-3q_1\lambda_1)+w^2\bar{w}(H-q_2(2\lambda_1+\bar{\lambda}_1))\\
&& +w\bar{w}^2(K-2q_3\bar{\lambda}_1)+\bar{w}^3(L-q_4(3\bar{\lambda}_1+\lambda))\\
&=& \lambda_1(z+q_1z^3+q_2z^2\bar{z}+q_3z\bar{z}^2+q_4\bar{z}^3)+z^3(G-3q_1\lambda_1)\\
&& +z^2\bar{z}(H-q_2(2\lambda_1+\bar{\lambda}_1))+z\bar{z}^2(K-2q_3\bar{\lambda}_1)+\bar{z}^3(L-q_4(3\bar{\lambda}_1+\lambda))+O(|z|^4)\\
&=& \lambda_1z+z^3(G-2q_1\lambda_1)+z^2\bar{z}(H-q_2(\lambda_1+\bar{\lambda}_1))\\
&& +z\bar{z}^2(K-q_3(2\bar{\lambda}_1-\lambda_1))+\bar{z}^3(L-3q_4\bar{\lambda}_1)
\end{eqnarray*}

In order to eliminate the cubic terms we require the following to be true:

\begin{eqnarray*}
q_1 &=& \frac{G}{2\lambda_1}\\
q_2 &=& \frac{H}{\lambda_1+\bar{\lambda}_1}\\
q_3 &=& \frac{K}{2\bar{\lambda}_1-\lambda_1}\\
q_4 &=& \frac{L}{3\bar{\lambda}_1}
\end{eqnarray*}

We see that the expression for $q_2$ cannot be true since if $\lambda_1=i\omega_0$ then $\lambda_1+\bar{\lambda}_1=0$. Therefore, the denominator for $q_2$ is zero. Hence, we can eliminate all cubic terms apart from the $z^2\bar{z}=z|z|^2$ term. Further analysis shows that for higher order terms (i.e. $O(|z|^4)$), all terms can actually cancel out. Therefore, we are left with the following, which is the Hopf Normal Form:

\begin{equation*}
\label{eqn:normalform}
\dot{z} = \lambda_1z+q_2z|z|^2
\end{equation*}
\\
where $z$ is, in fact, a limit cycle solution. Let us now work our way back to the original system. We know that the solution to (\ref{eqn:hbusys}) is:

\begin{equation*}
\textbf{u} = \alpha_0\textbf{v}_0+\alpha_1\textbf{v}_1+\bar{\alpha}_1\bar{\textbf{v}}_1+O(|\alpha_1|^3)
\end{equation*}

In the system given by (\ref{eqn:hbysys}), the solution will then be:

\begin{equation*}
\textbf{y} = \alpha_1\textbf{w}_1+\bar{\alpha}_1\bar{\textbf{w}}_1+O(|\alpha_1|^3)
\end{equation*}
\\
and in the first system:

\begin{equation}
\label{eqn:xsolx1}
\textbf{x} = \textbf{x}_*+\alpha_1\textbf{w}_1+\bar{\alpha}_1\bar{\textbf{w}}_1+O(|\alpha_1|^3)
\end{equation}

Finally, we note that $\alpha_1=z+O(|z|^3)$ and therefore we get that (\ref{eqn:xsolx1}) can be expressed as:

\begin{equation}
\label{eqn:xsolx}
\textbf{x} = \textbf{x}_*+z\textbf{w}_1+\bar{z}\bar{\textbf{w}}_1+O(|z|^3)
\end{equation} 

All we now need to do is apply this to our spiral wave solutions. We know that in our system (\ref{eqn:pderot}) the solutions generated are of the form $\{\textbf{v},\textbf{c},\omega\}$. This is infact our $x$ in (\ref{eqn:xsolx}).

\begin{equation*}
\left(\begin{array}{c} \textbf{v}\\ \textbf{c} \\ \omega\end{array}\right) = \left(\begin{array}{c} \textbf{v}_*\\ \textbf{c}_* \\ \omega_*\end{array}\right)+z\left(\begin{array}{c} \textbf{w}_v\\ \textbf{w}_c \\ w_\omega\end{array}\right)+\bar{z}\left(\begin{array}{c} \bar{\textbf{w}}_v\\ \bar{\textbf{w}}_c \\ \bar{w}_\omega\end{array}\right)+\mbox{h.o.t.}
\end{equation*}

Therefore we now have expressions for $\textbf{c}=(c_x,c_y)$ and $\omega$:

\begin{eqnarray*}
c_x &=& c_{x*}+zc_{x1}+\bar{z}\bar{c}_{x1}+O(|z|^3)\\
c_x &=& c_{y*}+zc_{y1}+\bar{z}\bar{c}_{y1}+O(|z|^3)\\
\omega &=& \omega_*+z\omega_1+\bar{z}\bar{\omega}_1+O(|z|^3)
\end{eqnarray*}


\section{Codimension (Codim)}

$Codimesion$, or $Codim$ as it is sometimes referred to as, is basically the difference in the dimensions of 2 comparing spaces or manifolds. Let $X$ be a vector space of dimension $a$, and $Y$ a sub space of $X$, i.e. $Y\subset X$, of dimension $b$. Then $X$ is said to have Codim = $a-b$.

As an example, take the Poincare Cross Section of a 2-Torus. The 2-Torus has Dim=3 and the Poincare Cross Section has Dim=2. Hence, the Codim of the Torus is 1. Another way to look at this is that we have to introduce just one restriction to the 2-Torus in order to reduce its dimension to 2 - hence the Codim is 1.


\section{Excitability}

A system is classed as \emph{Excitable} and displays \emph{Excitability} if 
it is capable of generating a wave of some description \chg[p221gram1]{and supporting its propagation.} \chg[p221gram1]{Cardiac} tissue is classed as an excitable media  since it is capable of \chg[p221gram1]{supporting the propagation of waves}. Consider \chg[p221gram1]{Fig.(\ref{wave1})}:

\begin{figure}[h]
\begin{center}
\begin{minipage}[b]{1.0\linewidth}
\centering
\includegraphics[width=0.5\textwidth]{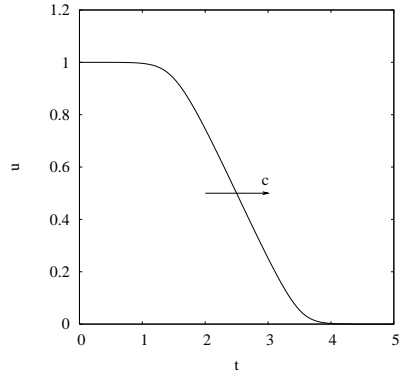}
\caption{A \chg[af]{traveling} wave displaying propagation.}
\label{wave1}
\end{minipage}
\end{center}
\end{figure}

Now, if the wave front is moving in the direction shown for increasing time, 
i.e. in the positive x-direction, then this shows that the wave is 
propagating. If the wave is moving in the opposite direction for increasing 
time then the wave is not propagating but is decaying. Hence, we require 
that for a system to display excitability, then the wave front must move in 
the positive x-direction as shown.


\section{Euclidean Symmetry}

A system is said to be invariant under under \emph{Euclidean Symmetry} if it 
possesses the properties of a Symmetry Group and obeys Euclidean Symmetry 
laws. We will \chg[af]{throughout} this report consider only the 2 dimensional 
Euclidean Group, \chg[p221gram2]{$SE(2)$,} possessing the following properties:

\begin{itemize}
\item Identity element.
\item Invariance under rotations.
\item Invariance under translations.
\end{itemize}

To show how these properties work, consider the following example.


\subsection{An Example of Euclidean Symmetry: The Laplacian}

Take a point (x,y) $\in\mathbb{R}^2$. Let us now rotate this point by angle $\phi_1$ and translate it by (x',y'), calling the new set of coordinates (x$_1$,y$_1$). In order to determine what the new coordinates are in terms of the old coordinates, consider Fig.(\ref{fig:example}) showing how the new coordinates relate to the original ones:

\begin{figure}[htbp]
\begin{center}
\psfrag{0}[tl]{$y_1$}
\psfrag{1}[tl]{$y$}
\psfrag{2}[tl]{$x_1$}
\psfrag{3}[tl]{$x$}
\psfrag{4}[tl]{$\theta$}
\psfrag{5}[tl]{$\theta$}
\includegraphics[width=0.65\textwidth]{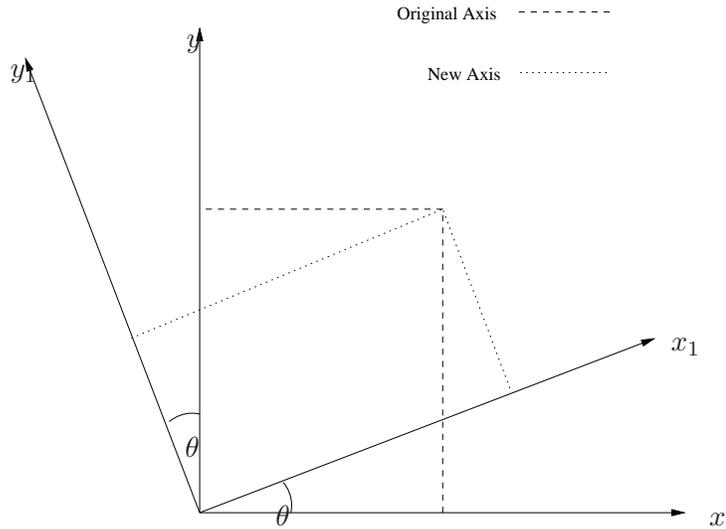}
\end{center}
\caption{Rotation of the \chg[af]{original} axis}
\label{fig:example}
\end{figure}

Now, using some elementary geometry theories, we can observe that the new coordinates are given by:

\[\left(\begin{array}{c} x_1\\ y_1 \end{array}\right) = \left(\begin{array}{cc} cos\theta & sin\theta\\ -sin\theta & cos\theta \end{array}\right)\left(\begin{array}{c} x\\y \end{array}\right)\]

Therefore, using the above notations, we can see that a rotation by angle $\phi_1$ followed by a translation by (x',y') gives us new coordinates of:

\[\left(\begin{array}{c} x_1\\ y_1 \end{array}\right) = \left(\begin{array}{c} x'\\y' \end{array}\right) + \left(\begin{array}{cc} cos\phi_1 & sin\phi_1\\ -sin\phi_1 & cos\phi_1 \end{array}\right)\left(\begin{array}{c} x\\y \end{array}\right)\]

Let us now do a similar rotation to the new coordinates, this time rotating by an angle of $\phi_2$ and then translating by $(x'',y'')$:

\begin{eqnarray*}
\left(\begin{array}{c} x_2\\ y_2 \end{array}\right) &=& \left(\begin{array}{c} x''\\y'' \end{array}\right) + \left(\begin{array}{cc} cos\phi_2 & sin\phi_2\\ -sin\phi_2 & cos\phi_2 \end{array}\right)\left(\begin{array}{c} x'\\y' \end{array}\right)\\
&& +\left(\begin{array}{cc} cos\phi_2 & sin\phi_2\\ -sin\phi_2 & cos\phi_2 \end{array}\right)\left(\begin{array}{cc} cos\phi_1 & sin\phi_1\\ -sin\phi_1 & cos\phi_1 \end{array}\right)\left(\begin{array}{c} x\\y \end{array}\right)
\end{eqnarray*}
\\
which, letting $\phi_{1+2}=\phi_1+\phi_2$, can be reduced to:

\[\left(\begin{array}{c} x_2\\ y_2 \end{array}\right) = \left(\begin{array}{c} x''\\y'' \end{array}\right) + \left(\begin{array}{cc} cos\phi_2 & sin\phi_2\\ -sin\phi_2 & cos\phi_2 \end{array}\right)\left(\begin{array}{c} x'\\y' \end{array}\right)+\left(\begin{array}{cc} cos\phi_{1+2} & sin\phi_{1+2}\\ -sin\phi_{1+2} & cos\phi_{1+2} \end{array}\right)\left(\begin{array}{c} x\\y \end{array}\right)\]

Therefore, we can represent these transformations using {\boldmath{x}}=(x,y) and represent the matrices as:

\[\mbox{A}(\phi)=\left(\begin{array}{cc} cos\phi & sin\phi\\ -sin\phi & cos\phi \end{array}\right)\]

Hence:

\[\mathbf{x_2}=\mathbf{x''}+\mbox{A}(\phi_{2})\mathbf{x'}+\mbox{A}(\phi_{1+2})\mathbf{x}\]

Now, we have our new set of coordinates, $(x_2, y_2)$ in terms of the old set of coordinates, $(x, y)$, where $x', x'', \phi_2, \phi_{1+2}$ are all constants. To see whether the Laplacian is invariant under these transformations. Firstly, consider the $u$ field:

\begin{eqnarray*}
\frac{\partial{u}}{\partial{x}} & = & \frac{\partial{u}}{\partial{x_2}}\frac{\partial{x_2}}{\partial{x}} + \frac{\partial{u}}{\partial{y_2}}\frac{\partial{y_2}}{\partial{x}}\\
\Rightarrow \frac{\partial{u}}{\partial{x}} & = & cos(\phi_{1+2})\frac{\partial{u}}{\partial{x_2}} - sin(\phi_{1+2})\frac{\partial{u}}{\partial{y_2}}\\
\Rightarrow \frac{\partial^2{u}}{\partial{x^2}} & = & cos^2(\phi_{1+2})\frac{\partial^2{u}}{\partial{x_2^2}} + sin^2(\phi_{1+2})\frac{\partial^2{u}}{\partial{y_2^2}}
\end{eqnarray*}

Similarly, for the second derivative of $u$ with respect to y, we get:

\begin{eqnarray*}
\frac{\partial^2{u}}{\partial{y^2}} & = & sin^2(\phi_{1+2})\frac{\partial^2{u}}{\partial{x_2^2}} + cos^2(\phi_{1+2})\frac{\partial^2{u}}{\partial{y_2^2}}
\end{eqnarray*}

Therefore, the Laplacian is given by:

\begin{eqnarray*}
\Delta u & = & \frac{\partial^2{u}}{\partial{x^2}} + \frac{\partial^2{u}}{\partial{y^2}}\\
\Rightarrow \Delta u & = & (cos^2(\phi_{1+2})\frac{\partial^2{u}}{\partial{x_2^2}} + sin^2(\phi_{1+2})\frac{\partial^2{u}}{\partial{y_2^2}}) + (sin^2(\phi_{1+2})\frac{\partial^2{u}}{\partial{x_2^2}} + cos^2(\phi_{1+2})\frac{\partial^2{u}}{\partial{y_2^2}})\\
\Rightarrow \Delta u & = & \frac{\partial^2{u}}{\partial{x_2^2}} + \frac{\partial^2{u}}{\partial{y_2^2}}\\
\Rightarrow \Delta u & = & \Delta u_2
\end{eqnarray*}

where $u_2 = u(x_2,y_2)$. Hence we can conclude that the Laplacian is invariant under the actions of an \chg[p224gram]{SE(2)} group.

\section{A Bit of Group Theory}

The information that follows has been taken from various references - \cite{GroupLed}, \cite{GroupHum}, and \cite{Patterns}


\subsection{Definition}

A group is defined to be a set of elements, $G$, together with an operation, $\circ$, which obey the following axioms:

\begin{enumerate}
\item \emph{Closure} - $\forall g,h\in G$ then $g\circ h \in G$.
\item \emph{Associativity} - $\forall f,g,h,\in G$ then $(f\circ g)\circ h = f\circ (g\circ h)$
\item \emph{Identity} - $\forall g\in G\quad \exists\quad e\in G$ such that $g\circ e=g$
\item \emph{Inverse} - $\forall g\in G\quad \exists\quad h\in G$ such that $g\circ h=e$
\end{enumerate}

A further axiom is \emph{Commutativity} which states:

\[\forall g,h \in G, \quad g\circ h = h\circ g\]

A group obeying this axion is said to be an \emph{Abelian Group}.


\subsection{Invariance and Equivariance}

If $g\in G$ and $x\in X$ is a set of solutions to a particular equation (in our case we would be looking at the set of all solutions $u=u(\textbf{r},t)=(u_1,u_2,..)\in \mathbb{R}^l$ to the Reaction-Diffusion system of equations, then we say that a solution $x'\in X$ is invariant if:

\begin{equation*}
\label{eqn:invariant}
x' = g\cdot x
\end{equation*}
\\
i.e. $x$ is a solution and $x'$ is a solution which is derived by applying the action of a group element $g\in G$ to the original solution, $x$. We say that ``solutions $x\in X$ are invariant under action of $g\in G$".

Let $Y,Z\subset G$ and define a map $f:Y\mapsto Z$. Then we say that $f$ is equivariant if:

\begin{equation*}
\label{eqn:equivariant}
f(gy) = gf(y)\quad y\in Y
\end{equation*}


\subsection{Direct Product}

Let $(G,\circ_G)$ and $(H,\circ_H)$ be groups. The the direct product between these groups are:

\begin{equation*}
\label{eqn:directproduct}
G\times H = \{(g,h)|g\in G, h\in H\}
\end{equation*}


\subsection{Orbit}
\label{sec:orbit}

The Orbit of an element $x\in X$, is the set of all possible transformations of $x$ under the action of group elements which belong to the set $X$:

\begin{equation*}
\label{eqn:orbit}
G_x = \{g\cdot x|x\in X,g\in G\}
\end{equation*}


\subsection{Stabiliser}

The Stabiliser of $x\in X$ is defined as:

\begin{equation*}
\label{eqn:stab}
G_{stab} = \{g\cdot x=x|x\in X,g\in G\}
\end{equation*}

The stabiliser is sometimes often referred to as the \emph{Isotropy Subgroup}.


\subsection{Centre}

The center of a group $G$ is defined to be:

\begin{equation*}
\label{eqn:center}
C(G) = \{t\in G|tg=gt\quad \forall g\in G\}
\end{equation*} 

Therefore, in this case, we are looking at an Abelian Group (see above).


\subsection{Normaliser}

The Normaliser of a subgroup $H\subset G$ is defined as:

\begin{equation*}
\label{eqn:normaliser}
N_G(H) = \{g\in G:gHg^{-1}=H\}
\end{equation*}


\subsection{Normal and Quotient Groups}

We say that a subgroup $H\subset G$ is a normal subgroup if it is invariant under action of elements of $G$:

\begin{equation*}
\label{eqn:normal}
N\lhd G = \{\forall g\in G:gHg^{-1}=H\}
\end{equation*}

If $N$ is a normal subgroup then we define the Quotient group to be the set of all \emph{left cosets} of $N$ in $G$ and is denoted by $G/N$.


\subsection{Euclidean Groups - $E(2)$, $SE(2)$, $SO(2)$ and $\mathbb{Z}(2)$}
\label{sec:euclid}

The Euclidean Group, $E(2)$, if the 2 dimensional group consisting of all rotations, translations and reflections. A subgroup $E(2)$ is the group of rotations and translation, $SE(2)$, which has 2 further subgroups - the group of rotations, $SO(2)$, and the group of translations, \chg[p226gram]{$\mathbb{R}^2$}. Note that $E(n)$ is the Euclidean group in $n$ dimensions. 
\chg[p226eqn]{}
\begin{equation*}
(SO(2)\cup \chg[]{\mathbb{R}^2})\subset SE(2) \subset E(2)
\end{equation*}


\subsubsection{Group action on a solution $u(\textbf{r},t)$}

In this subsection, we show that the following is true:

\begin{equation*}
T(g)u(\textbf{r},t) = u(g^{-1}\textbf{r},t)
\end{equation*}

Firstly, consider the 1-dimensional case $\textbf{r}=x$. In Fig.(\ref{fig:shifted})), have $u(x)$ being moved parallel to the x-axis in the positive direction, i.e. we start at $x_1$ and shift it to $x_1+X=x_2$, where $X=x_2-x_1$. We have a solution $u(x)$, denoted in Fig.(\ref{fig:shifted}) by the solid line, and this solution is shifted in the positive $x$ direction via a group action, i.e. $T(g)u(x)=\tilde{u}(x)$.

\begin{figure}[htbp]
\begin{center}
\begin{minipage}[b]{0.6\linewidth}
\centering
\psfrag{b}[l]{$x_1$}
\psfrag{c}[l]{$x_2$}
\psfrag{u}[t]{$u$}
\psfrag{x}[l]{$x$}
\psfrag{a}[l]{$u_1$}
\includegraphics[width=0.9\textwidth]{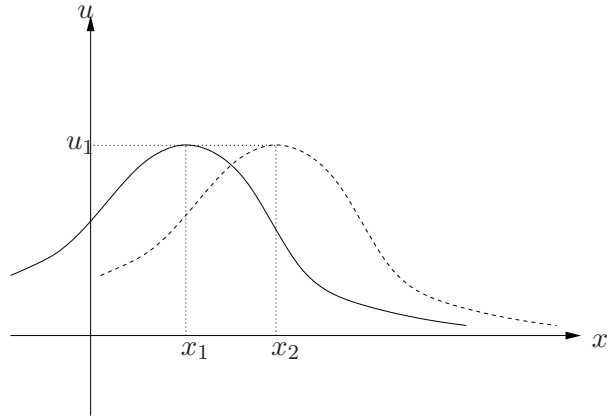}
\end{minipage}
\caption[Translational shift of a wave train]{$u(x)$ shifted along the x-axis by R. The original function, $u$ is shown as a solid line, while the new shifted function, $\tilde{u}$, is shown as a dashed line.}
\label{fig:shifted}
\end{center}
\end{figure}

Now, we have that:

\begin{equation*}
u(x_1) = u_1
\end{equation*}
\\
Also, we have that:

\begin{eqnarray*}
\tilde{u}(x_2) &=& u_1\\
\Rightarrow \tilde{u}(x_2) &=& u(x_1)\\
\Rightarrow T(g)u(x_2) &=& u(x_1)
\end{eqnarray*}

The last equation above comes from the fact that $\tilde{u}=T(g)u$. Also, since the transformation affects the spatial coordinate then let $T(g)u(x_2)=u(\tilde{x}_2)$, where $\tilde{u}$ is some transformed version of $u$. We then have:

\begin{eqnarray*}
u(\tilde{x}_2) &=& u(x_1)\\
\Rightarrow \tilde{x}_2 &=& x_1\\
\Rightarrow \tilde{x}_2 &=& x_2-X\\
\Rightarrow \tilde{x}_2 &=& g^{-1}x_2
\end{eqnarray*}

Therefore, we finally get that:

\begin{eqnarray*}
T(g)u(x_2) &=& u(\tilde{x}_2)\\
\Rightarrow T(g)u(x_2) &=& u(g^{-1}{x}_2)
\end{eqnarray*}

Similar arguments can be arranged for the 2 dimensional case and it can be shown that the following is true for $g\in SE(2)$:

\begin{equation*}
T(g)u(\textbf{r},t) = u(g^{-1}\textbf{r},t)
\end{equation*}


\subsubsection{Invariance Properties}

In their 1996 paper, Biktashev et al considered the symmetry with respect to the Special Euclidian Group, $SE(2)$, i.e. the group of rotations and translations. It can be shown that spiral wave solutions are actually invariant under actions from $SE(2)$. That is, if we take a spiral, rotate and move it to a different position within the domain in which it is rotating, then we will still have a spiral wave solution only in a different position and phase. We prove this in the following paragraphs.

Let assume that we have a Reaction Diffusion Equation:

\begin{equation}
\label{eqn:rdeinv}
\partial_t{u} = \textbf{D}\nabla^2u+f(u)
\end{equation}
\\
with $u=u(x,y,t)$ and $u\in \mathbb{R}^2$. Let us assume that the solution, $u$, to this equation, now undergoes an action of the the group element $g\in SE(2)$:

\begin{equation*}
\label{eqn:utild}
\tilde{u} = T(g)u(x,y,t) = u(g^{-1}x,g^{-1}y,t) = u(\tilde{x},\tilde{y},t)
\end{equation*}

If we let the group element be a rotation followed by a translation, i.e. $g=\{R,\theta\}$ where $\theta$ is the angle of rotation and $R=(X,Y)$ is the translation vector, then we have that $g^{-1}=\{-\theta,-R\}$, i.e. a translation (in the opposite direction, followed by a rotation:

\begin{eqnarray*}
\left(\begin{array}{c} \tilde{x}\\ \tilde{y} \end{array}\right) &=& \left(\begin{array}{cc} \cos\theta & \sin\theta\\ -\sin\theta & \cos\theta \end{array}\right)\left(\begin{array}{c} x-X\\ y-Y \end{array}\right)\\
\Rightarrow \tilde{x} &=& (x-X)\cos\theta+(y-Y)\sin\theta\\
\mbox{and}\quad \tilde{y} &=& -(x-X)\sin\theta+(y-Y)\cos\theta
\end{eqnarray*}

Let us now assume that $\tilde{u}$ is a solution to (\ref{eqn:rdeinv}):

\begin{equation*}
\label{eqn:rdeinvtil}
\partial_t{\tilde{u}} = \textbf{D}\nabla^2\tilde{u}+f(\tilde{u})
\end{equation*}

Consider the right hand side of (\ref{eqn:rdeinvtil}). Eqn.(\ref{eqn:utild}) gives us:
\begin{eqnarray*}
\frac{\partial{\tilde{u}}}{\partial{t}} &=& \frac{\partial{u}}{\partial{t}} 
\end{eqnarray*}
\\
since the transformation applied only affects the spatial variables and not the temporal ones. Consider now the Laplacian terms:

\begin{eqnarray*}
\nabla^2\tilde{u} &=& \frac{\partial^2{\tilde{u}}}{\partial{\tilde{x}}^2}+\frac{\partial^2{\tilde{u}}}{\partial{\tilde{y}}^2}
\end{eqnarray*}

Now, consider just the differentiation with respect to $x$:

\begin{eqnarray*}
\frac{\partial{\tilde{u}}}{\partial{\tilde{x}}} &=& \frac{\partial{u}}{\partial{x}}\frac{\partial{x}}{\partial{\tilde{x}}}+\frac{\partial{u}}{\partial{y}}\frac{\partial{y}}{\partial{\tilde{x}}}\\
&=& \cos\theta\frac{\partial{u}}{\partial{x}}+\sin\theta\frac{\partial{u}}{\partial{y}}
\end{eqnarray*}

Differentiating again with respect to $\tilde{x}$:

\begin{equation*}
\frac{\partial^2{u}}{\partial{\tilde{x}}^2} = \cos^2\theta\frac{\partial^2{u}}{\partial{x}^2}+\sin^2\theta\frac{\partial^2{u}}{\partial{y}^2}
\end{equation*}

Similar calculations yield:

\begin{equation*}
\frac{\partial^2{u}}{\partial{\tilde{y}}^2} = \sin^2\theta\frac{\partial^2{u}}{\partial{x}^2}+\cos^2\theta\frac{\partial^2{u}}{\partial{y}^2}
\end{equation*}

Hence the Laplacian for $\tilde{u}$ is given by:

\begin{eqnarray*}
\nabla^2\tilde{u} &=& (\cos^2\theta\frac{\partial^2{u}}{\partial{x}^2}+\sin^2\theta\frac{\partial^2{u}}{\partial{y}^2})+(\sin^2\theta\frac{\partial^2{u}}{\partial{x}^2}+\cos^2\theta\frac{\partial^2{u}}{\partial{y}^2})\\
&=&(\cos^2(\theta)+\sin^2(\theta))\frac{\partial^2{u}}{\partial{x}^2}+(\cos^2(\theta)+\sin^2(\theta))\frac{\partial^2{u}}{\partial{y}^2}\\
&=& \frac{\partial^2{u}}{\partial{x}^2}+\frac{\partial^2{u}}{\partial{y}^2}\\
&=& \nabla^2u
\end{eqnarray*}

Finally, consider the function $f(\tilde{u})$:

\begin{eqnarray*}
f(\tilde{u}) &=& f(u((\tilde{x},\tilde{y},t)))\\
&=& f(u((x-X)\cos\theta+(y-Y)\sin\theta,-(x-X)\sin\theta+(y-Y)\cos\theta,t))\\
&=& f(u(x,y,t))\\
&=& f(u)
\end{eqnarray*}

So, putting all our results together gives us:

\begin{eqnarray*}
\partial_t{\tilde{u}} &=& \textbf{D}\nabla^2\tilde{u}+f(\tilde{u})\\
\Rightarrow \partial_t{u} &=& \textbf{D}\nabla^2u+f(u)
\end{eqnarray*}

Hence, we can conclude that the Reaction Diffusion Equation is invariant under Euclidean Symmetry.


\section{Manifolds}

\subsection{Definition}

A \emph{Manifold} is an Euclidean Space of solutions which has codimension k and also members of a Topological Space (the original space) are mapped into this Euclidean Space. It is in affect, a represention of the original solutions but in a space of lower dimension.

In group theoretical terms, a Manifold can be represented as a Quotient Group (see above):

\begin{equation*}
\mathcal{M} = \{G_x|x\in X\} = X/G
\end{equation*}


\subsection{Simple Examples}

\subsubsection{The circle in $\mathbb{R}^2$}

Consider the circle as shown in Fig.(\ref{fig:circle}). A small arc on this circle can be represent as a line provided that the arc (neighbourhood) is small enough. Therefore, we can represent a neighbourhood in $\mathbb{R}^2$ as a straight line in $\mathbb{R}$. Therefore our manifold, the straight line, has codim-1, reducing our original space from 2 dimensions to 1 dimension.

\begin{figure}[htbp]
\begin{center}
\begin{minipage}[b]{0.6\linewidth}
\centering
\includegraphics[width=0.9\textwidth]{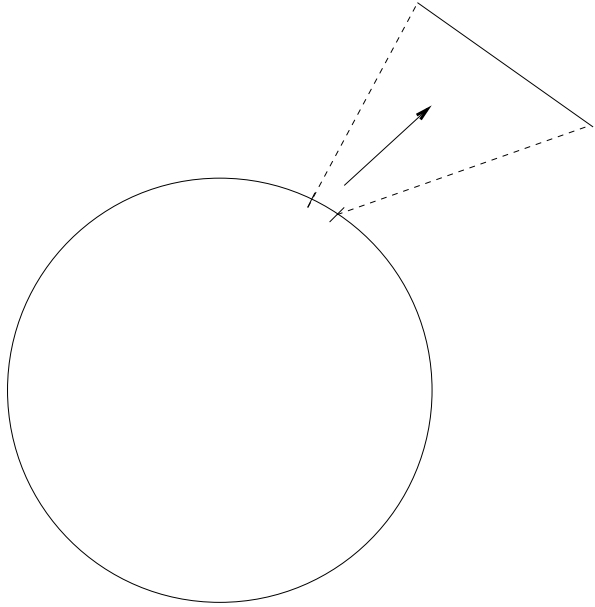}
\end{minipage}
\caption[Manifold illustration 1]{A circle with an small arc that has been enlarged.}
\label{fig:circle}
\end{center}
\end{figure}


\subsubsection{Group of rotations about the origin}

Consider all possible rotations about the origin in $\mathbb{R}^2$. If we take a point $(x,y)\in\mathbb{R}^2$, then this would belong to one of the circles as shown in Fig.(\ref{fig:circle1}). Therefore, every point in $\mathbb{R}^2$ belongs to one of these circles.

\begin{figure}[htbp]
\begin{center}
\begin{minipage}[b]{0.6\linewidth}
\centering
\psfrag{a}[l]{$y$}
\psfrag{b}[l]{$x$}
\includegraphics[width=0.9\textwidth]{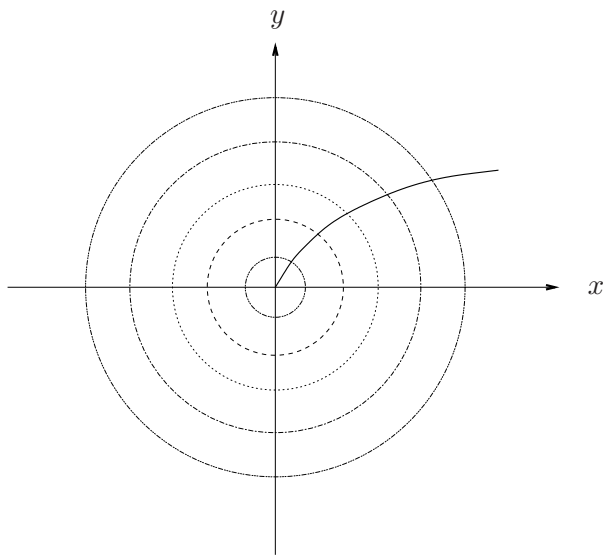}
\end{minipage}
\caption[Manifold illustration 2]{A family of circles in the $(x,y)$ plane with line starting at the origin passing through each circle exactly once.}
\label{fig:circle1}
\end{center}
\end{figure}

Therefore, if we know a point on a particular circle we can determine all other points on that circle by applying all rotations to that point. So, all we need to determine all possible points on a circle is one single point. We can do this for all other circles and therefore we get a line of solutions as shown in Fig.(\ref{fig:circle1}). This line is a manifold of codim-1.


\section{Banach Spaces}
\label{sec:banach}

\subsection{Definition}

A \emph{Banach Space} is a \emph{Complete Normed Vector Space} \chg[p230]{with a converging Cauchy sequence.} This means that a Banach Space is a Vector Space (Function Space) over the real or complex numbers with a norm $||\cdot||$. This definition is best viewed through examples. 


\subsection{Example}

Consider a function of space and time, $f(x,t)$. Let us for a moment fix time and vary $x$. We get the picture shown in Fig.(\ref{fig:ban3}). If we now consider our function at the next few timesteps, we get the picture as shown in Fig.(\ref{fig:ban4}).

\begin{figure}[htbp]
\begin{center}
\begin{minipage}[b]{0.48\linewidth}
\centering
\psfrag{a}[l]{$f(\textbf{x})$}
\psfrag{b}[l]{$\textbf{x}$}
\includegraphics[width=1.0\textwidth]{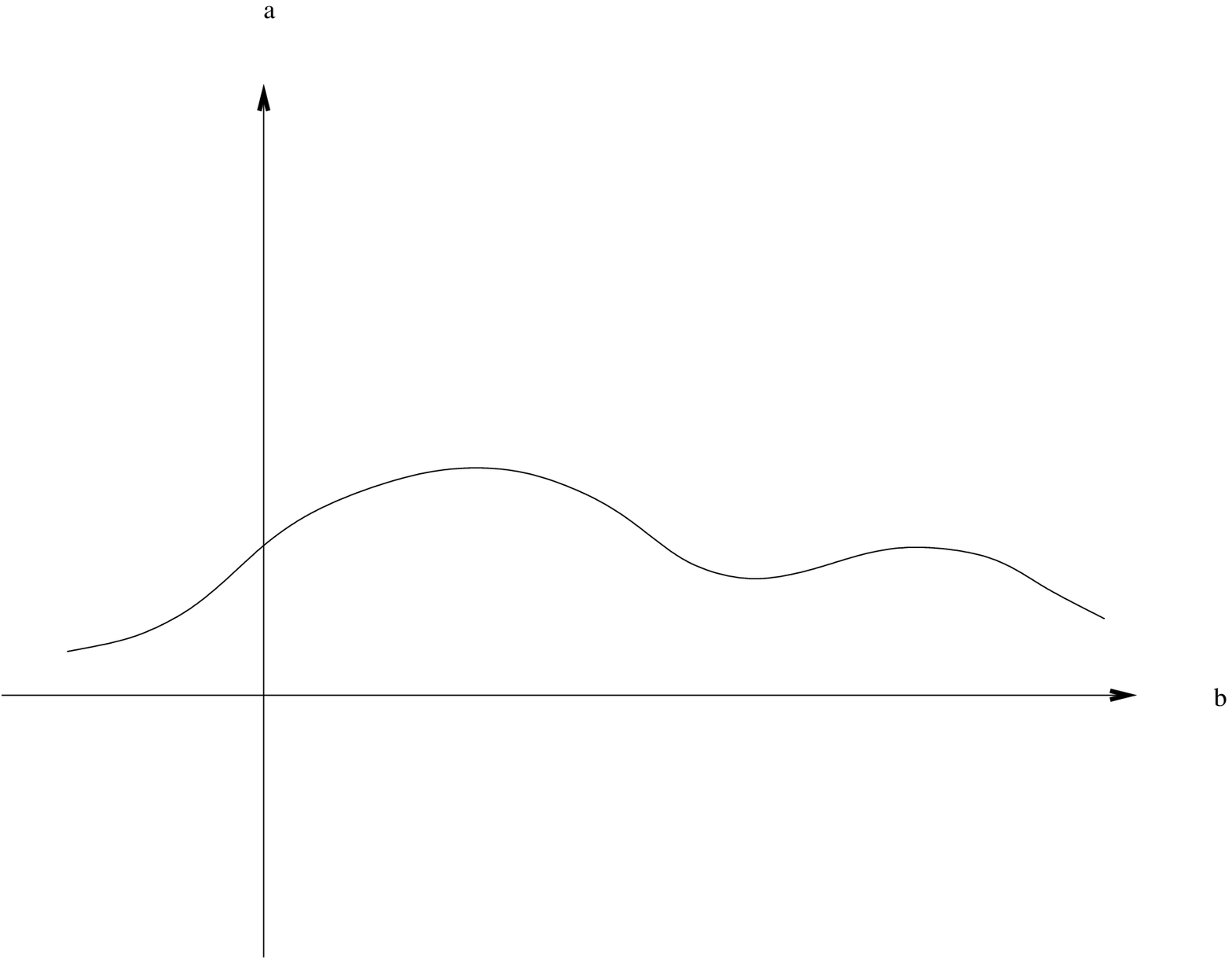}
\caption{The graph of $f(\textbf{x},t)$ with $t$ fixed.}
\label{fig:ban3}
\end{minipage}
\begin{minipage}[b]{0.48\linewidth}
\centering
\psfrag{a}[l]{$f(\textbf{x})$}
\psfrag{b}[l]{$\textbf{x}$}
\psfrag{c}[l]{$t_1$}
\psfrag{d}[l]{$t_2$}
\psfrag{e}[l]{$t_3$}
\psfrag{f}[l]{$t_4$}
\includegraphics[width=1.0\textwidth]{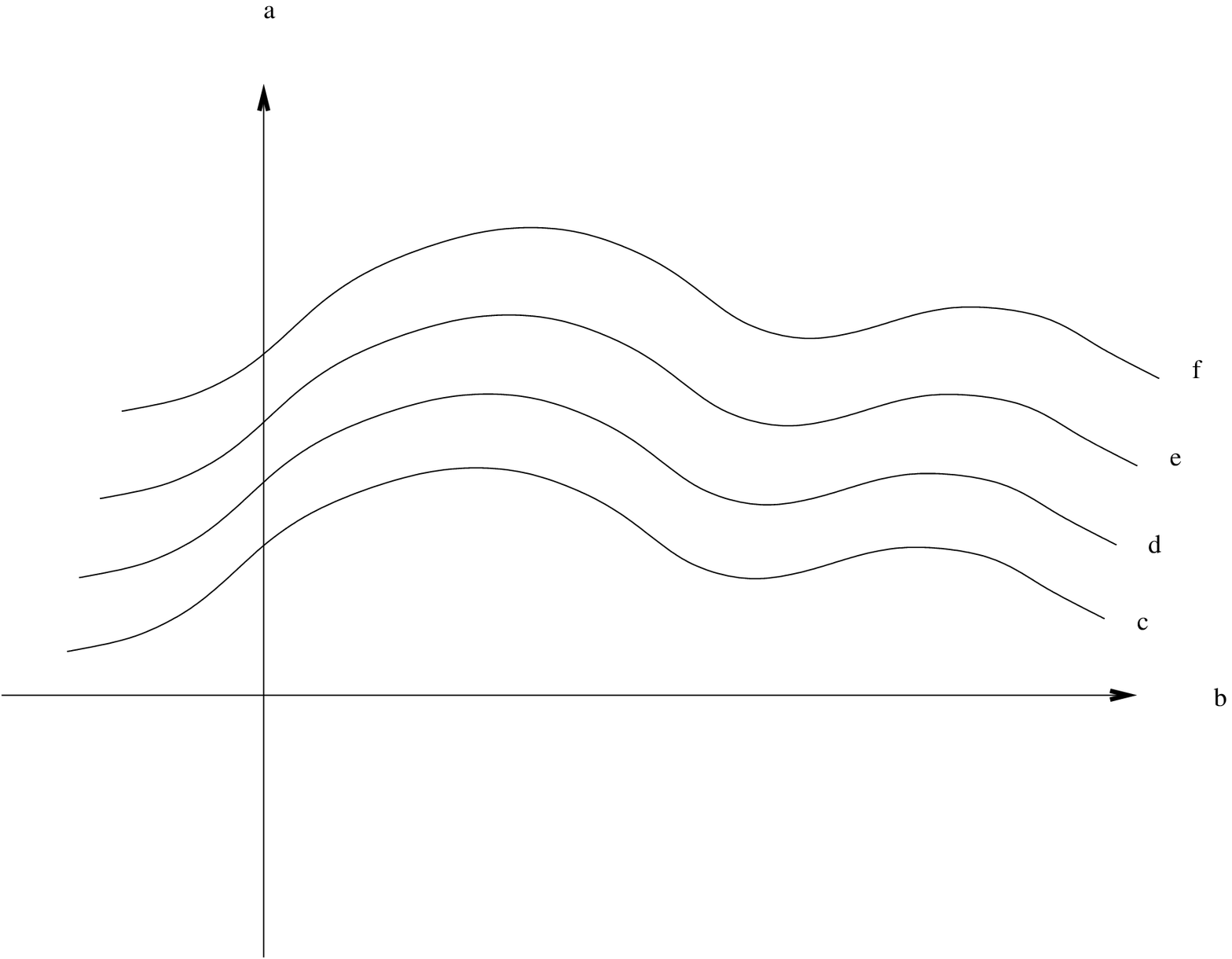}
\caption[Banch space: a function in the original space]{The graph of $f(\textbf{x},t)$ for various fixed values of time.}
\label{fig:ban4}
\end{minipage}
\end{center}
\end{figure}

Now, at each moment in time, we can represent each curve as a point in an n-dimensional vector space, where n in the number of space steps we have used to determine the curve (i.e. we have $x=\{x_1, \ldots ,x_n\}$). For ease of example and illustration, let us assume that $\textbf{x}=(x_1,x_2,x_3,x_4,x_5)$. This will give us 5 values of $f(\textbf{x})$ which can be represented as a point in a Banach Space with 5 dimensions (note: this is artificial as a Banach Space generally has an infinite number of dimensions, but this 5 dimensional Banach Space will show the particular properties found in a normal Banach Space). Figs.(\ref{fig:ban5}) and (\ref{fig:ban6}) show how this is illustrated.

\begin{figure}[htbp]
\begin{center}
\begin{minipage}[b]{0.48\linewidth}
\centering
\psfrag{a}[l]{$f(\textbf{x})$}
\psfrag{b}[l]{$\textbf{x}$}
\psfrag{c}[l]{$t_1$}
\psfrag{d}[l]{$t_2$}
\psfrag{e}[l]{$t_3$}
\psfrag{f}[l]{$t_4$}
\psfrag{g}[l]{$x_1$}
\psfrag{h}[l]{$x_2$}
\psfrag{i}[l]{$x_3$}
\psfrag{j}[l]{$x_4$}
\psfrag{k}[l]{$x_5$}
\includegraphics[width=1.0\textwidth]{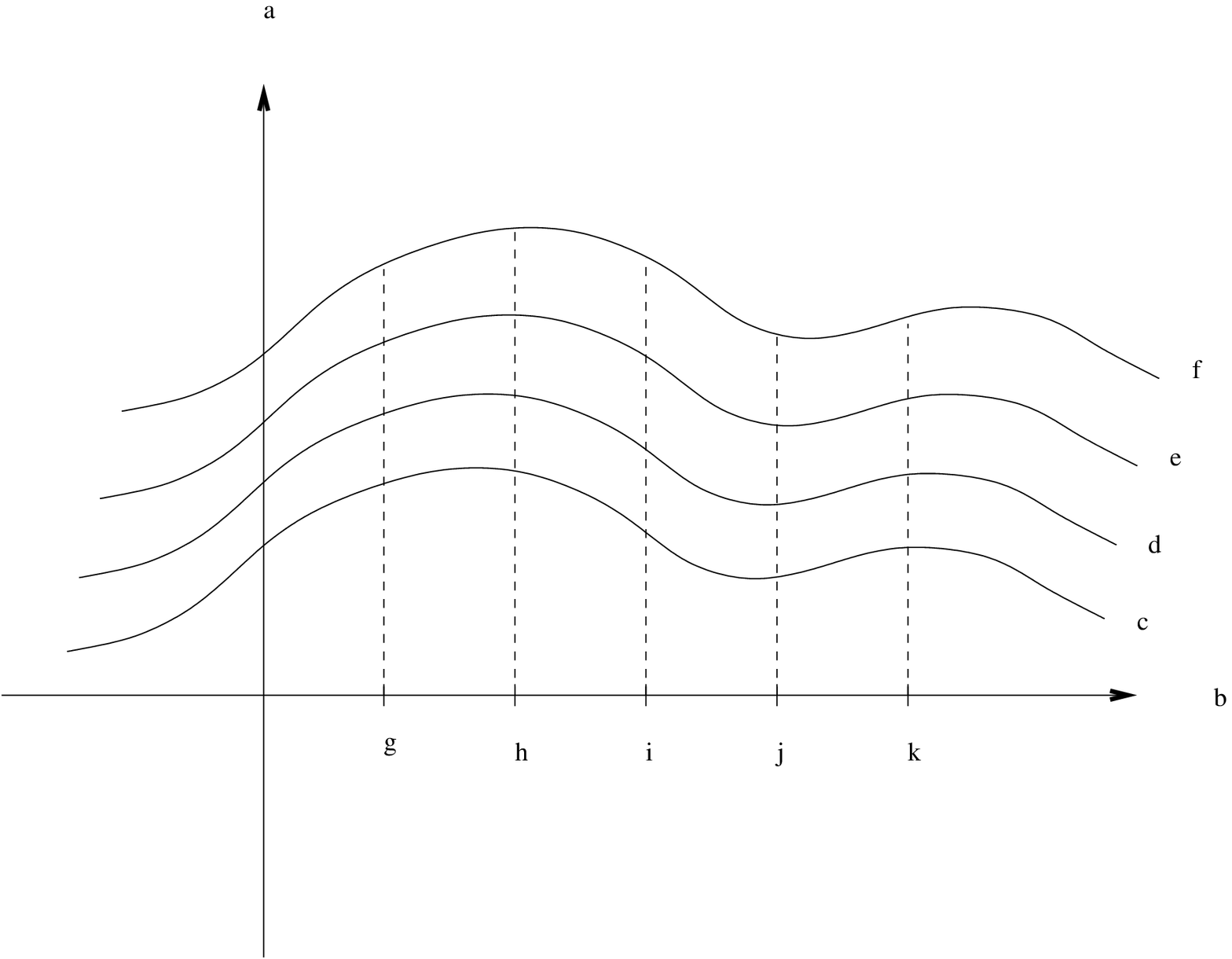}
\caption{The graph of $f(\textbf{x},t)$ for various fixed values of time.}
\label{fig:ban5}
\end{minipage}
\begin{minipage}[b]{0.48\linewidth}
\centering
\psfrag{a}[l]{$f(x_1)$}
\psfrag{b}[l]{$f{x_2}$}
\psfrag{c}[l]{$f(x_3)$}
\psfrag{d}[l]{$f(x_4)$}
\psfrag{e}[l]{$f(x_5)$}
\includegraphics[width=1.0\textwidth]{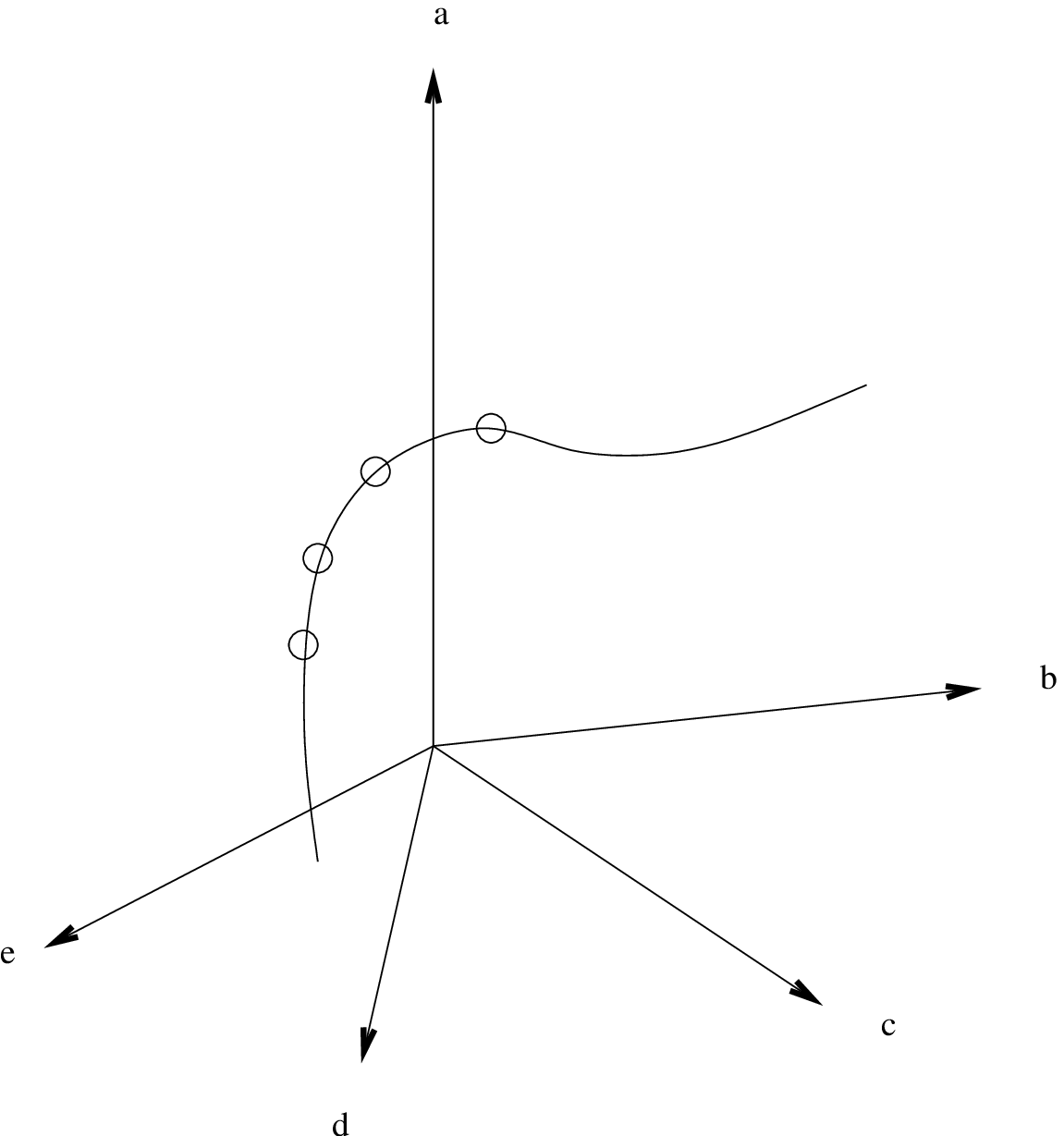}
\caption[Banch space: a function in the Banach space]{The trajectory in the Banach Space with each circle representing the function $f(\textbf{x},t)$ for a particular value of $t$.}
\label{fig:ban6}
\end{minipage}
\end{center}
\end{figure}

Therefore, in the vector space, we get a series of points that are dependent on time, which form a curve in this particular space. This vector space is a Banach Space, provided that the distance between each successive point is finite. So, we have represented our function $f(x,t)$ as a function in the Banach Space that is dependent only on time and not space, $F(t)$ say.
\chapter{EZ-Freeze}

\ez is a program which simulates spiral waves in a frame of reference which is moving with the tip of the wave. The motivation behind the program is within \cite{bik96}, in which the a Reaction-Diffusion-Advection type dynamical system was derived. The program is based on the very popular program EZ-Spiral by Dwight Barkley \cite{barkweb}. We refer to \cite{Foul08} for a more complete mathematical analysis of solving spiral waves in the comoving frame of reference.

\ez uses both Barkley's and FitzHugh-Nagumo's (FHN) models, together with a choice of boundary conditions (Neumann or Dirichlet) and whether interactive graphics are to be used or not.

The following set of intructions are organised as follows:

\begin{description}
\item[Section 2] {\bf Getting Started}: This will talk you through how to get the program working by following two examples.
\item[Section 3] {\bf Users Manual}: This will decribe the important uses of {\ez}and how the user can do particular tasks.
\item[Section 4] {\bf Programmers Manual}: This section will describe how the program works, detailing the differences between \ez and EZ-Spiral, the function of each file, and descriptions of the main functions used.
\item[Section 5] {\bf Mathematics}: We will describe the mathematical problem from which this program was first conceived, including the numerical methods used within \ez.
\end{description}


\section{Getting Started: A Quick Guide}

We will talk you through getting the program started by following a couple of examples - one example for a rigidly rotating spiral wave and another example for a meandering spiral wave. We will initial choose the parameters such that fast numerical calculations are illustrated. However, we will see that the tip trajectories recontructed from the quotient data (the quotient data is the name given to the coefficients to the advection terms - see \cite{bik96} for details) are not very accurate. So, we provide details at the end of this section of changes to be made to the numerical parameters in order to get a more accurate reconstruction of the tip trajectory (note that the refined parameters still provide for fast numerical calculations, but obviously not as fast as first illustrated).

At each step, issue the command to the left of the table (you will obviously need to have a terminal screen open and you will also need to be in the directory which contains the program files). The commentary to the right of the table describes what each command does.

Before we start, you will need to adjust \verb|Makefile| to reflect the type of C compiler you have. This is shown by the flag \verb|CC| on line 42, which is initially set to \verb|CC=cc|. Adjust this as you need to.

We are now ready to work through the examples. At each line in the tables below, issue the command shown to the left.

\begin{center}
    \begin{tabular}{ | c | p{9.5cm} |}
    \hline
    Command & Commentary \\ \hline\hline
    \verb|make ezfreeze| & This makes the program \verb|ezfreeze| using Barkley kinetics and interactive graphics (these are the default settings in the \verb|Makefile| - see Sec.(\ref{sec:ezf_Makefile}) for further details).\\ \hline
    \verb|./ezfreeze| & Executes the program \verb|ezfreeze|. An X-Window should appear.\\ \hline
    Bring X-Window to top & In order to run the simulation, the X-Window displaying the simulation needs to be the Window on the system which is on top of all others. This can be achieved by clicking on the window with the mouse, placing the mouse arrow within the window or tabbing until the window comes to the top.\\ \hline
    Press \verb|SPACE| bar & This will initiate the simulation.\\ \hline
    Press \verb|t| key & This will display the tip path from time of pressing the key. At the beginning of the simulation, you may notice several tips. Press \verb|t| twice to erase the tips and start the drawing process again. You should notice a smooth tip pattern being plotted.\\ \hline
    Press \verb|z| key & Switches on advection terms (i.e. moves to a comoving frame of reference). Do this once the tip has traced out about three full circles.\\
    \hline
    \end{tabular}
\end{center}

Now, leave the simulation to run until it self-terminates (initially, the number of time steps have been set at 50,000 and so the program will self terminate when the number of time steps has been exhausted). You should notice that the spiral wave within the comoving frame becomes stationary. This means that we have a Rigidly Rotating Spiral Wave. Fig.(\ref{fig:ezf_rw}) shows snap shots of the evolution of the spiral.

\begin{figure}[btp]
\begin{center}
\begin{minipage}[htbp]{0.24\linewidth}
\centering
\includegraphics[width=0.7\textwidth]{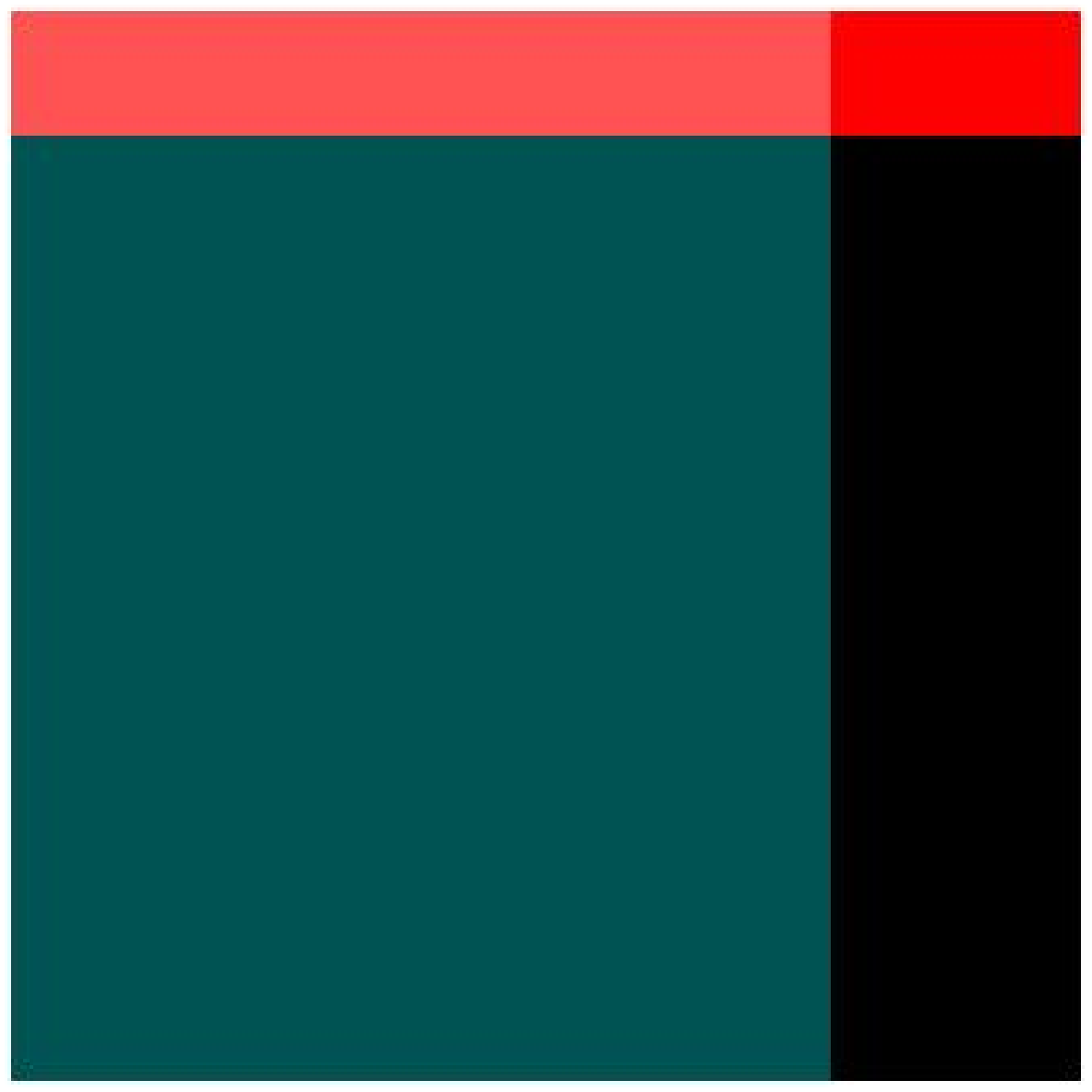}
\end{minipage}
\begin{minipage}[htbp]{0.24\linewidth}
\centering
\includegraphics[width=0.7\textwidth]{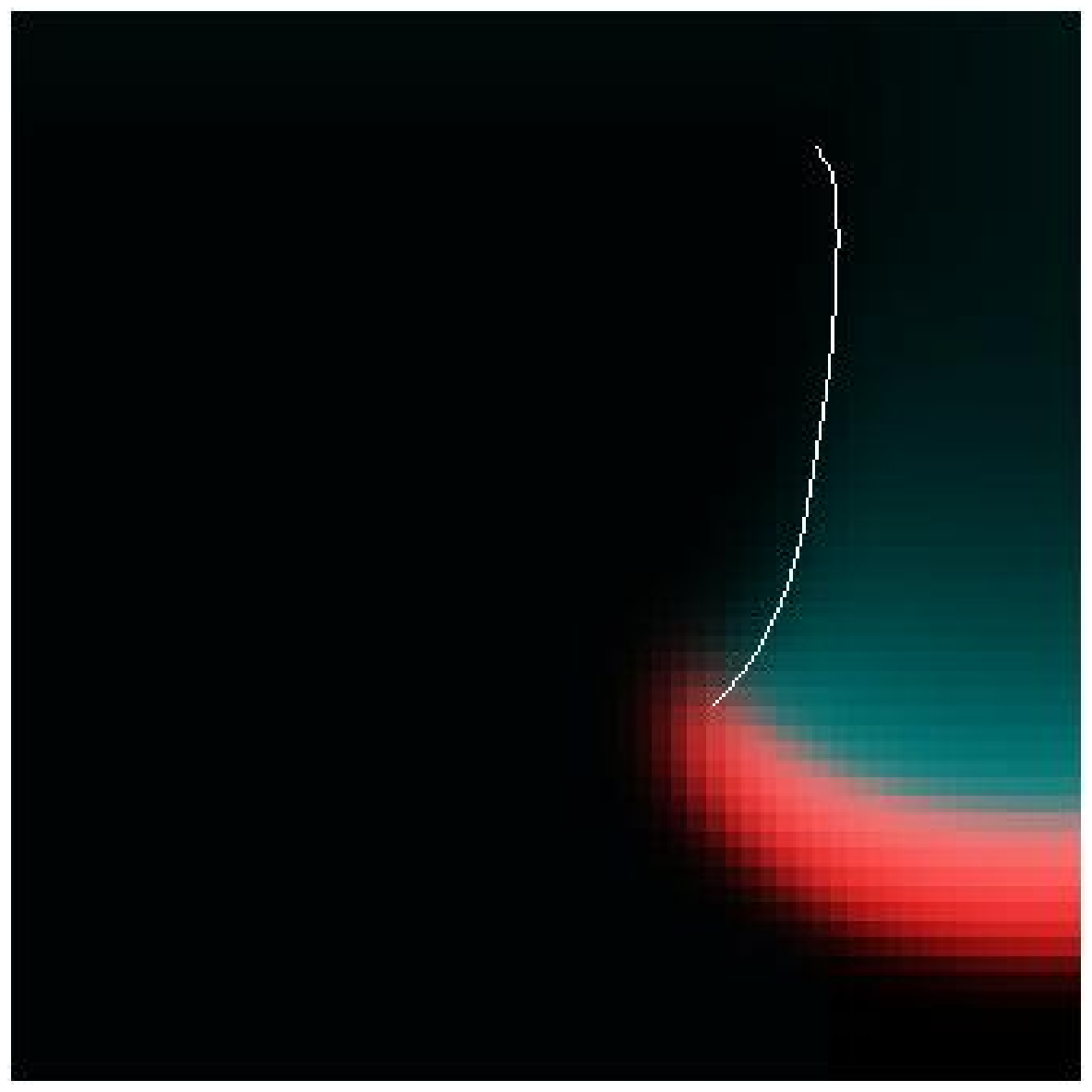}
\end{minipage}
\begin{minipage}[htbp]{0.24\linewidth}
\centering
\includegraphics[width=0.7\textwidth]{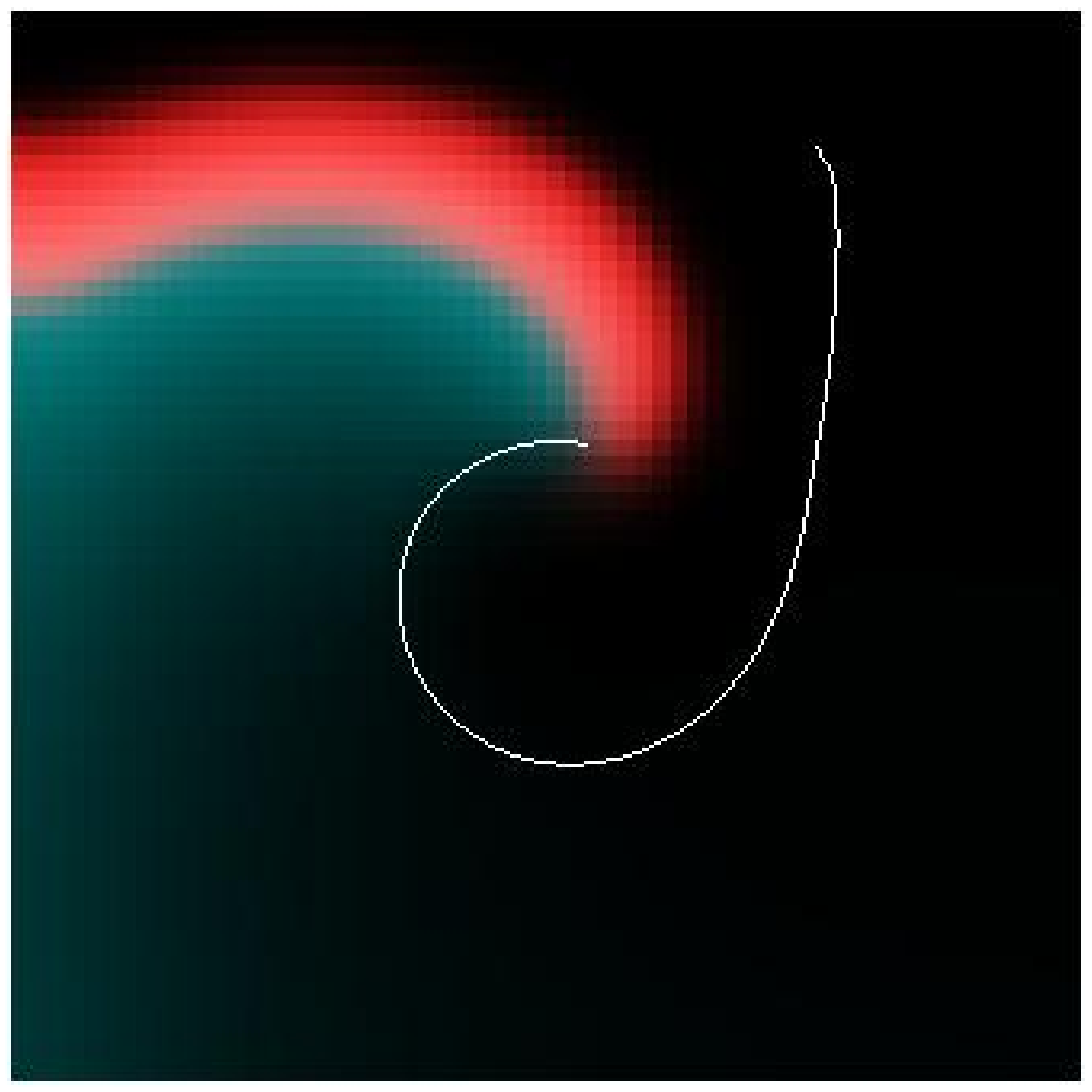}
\end{minipage}
\begin{minipage}[htbp]{0.24\linewidth}
\centering
\includegraphics[width=0.7\textwidth]{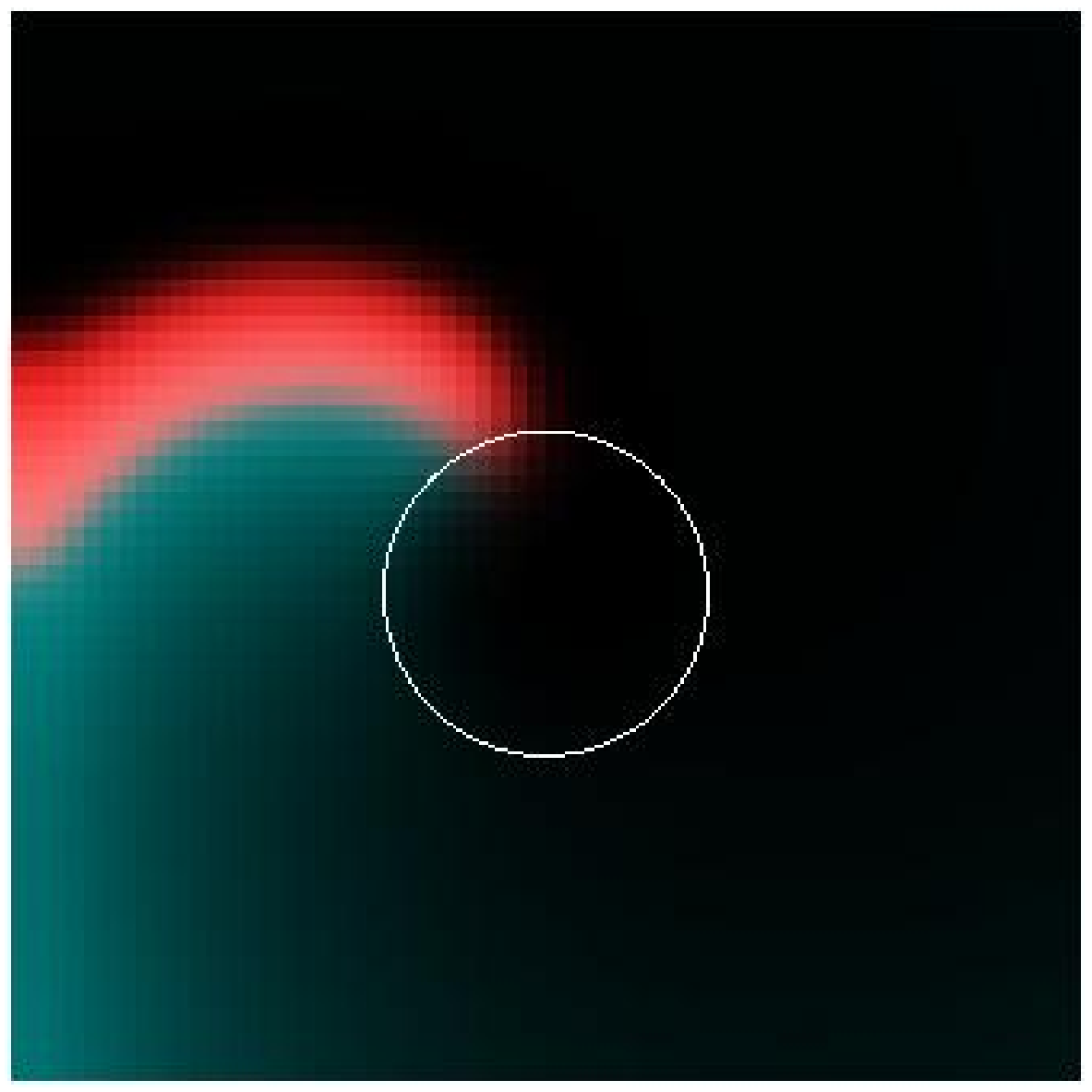}
\end{minipage}
\caption[Rigidly rotating spiral wave]{Snap snots from the simulation of a rigidly rotating spiral wave in a laboratory frame of reference, with the simulation starting from the left.}
\label{fig:ezf_rw}
\end{center}
\end{figure}

Once the simulation, during which advection was switched on, is brought to an end by whatever means (pressing keys \verb|q|, \verb|esc|, exiting the X-Window by physically closing it, or the number of time steps has been exhausted), a file called \verb|quot.dat| should be contained in the directory containing the program files.

\begin{center}
    \begin{tabular}{ | c | p{4.8cm} |}
    \hline
    Command & Commentary \\ \hline\hline
    \verb|make int| & This makes the program \verb|int| which is an integrator tool designed to reconstruct the tip trajectory from the advection coefficients.\\ \hline
    \verb|./int quot.dat 10| & Executes the program \verb|int| on the file \verb|quot.dat| starting from the file's $10^{{\mbox{\small {th}}}}$ line (the last argument is needed to eradicate any peculiarities in the initial transient).\\ \hline
    \verb|gnuplot| & Opens up Gnuplot (this is our plotter of choice. Please use one that you are more comfortable with if you want).\\ \hline
    \verb|p 'integrated_quot.dat' ev ::10 u 5:6 w lp,\| & plots the tip trajectories. Note that the\\
    \verb|'tip.dat' ev ::3000 u 2:3 w lp| & tip coords are located in columns 5 and 6 ($X$ and $Y$ respectively) for the reconstructed trajectory and columns 2 and 3 for the trajectory in the laboratory frame.\\ \hline
    \end{tabular}
\end{center}

You should observe a circular trajectory, corresponding to a rigidly rotating spiral wave. We show in Fig.(\ref{fig:ezf_rw_tips}) how the tip trajectories should look. This concludes the first example.

\begin{figure}[btp]
\begin{center}
\begin{minipage}[htbp]{0.49\linewidth}
\centering
\includegraphics[width=0.7\textwidth, angle=-90]{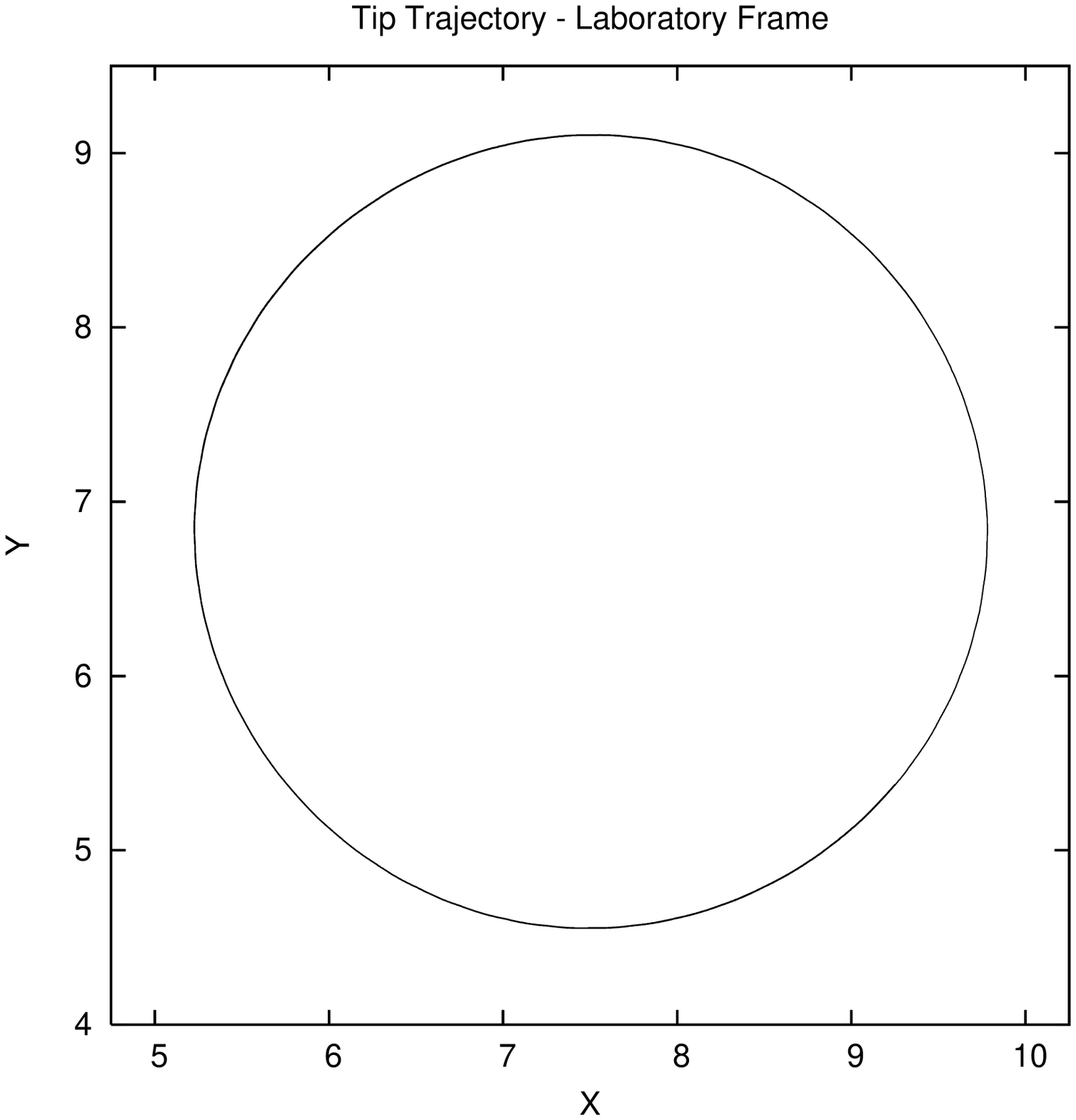}
\end{minipage}
\begin{minipage}[htbp]{0.49\linewidth}
\centering
\includegraphics[width=0.7\textwidth, angle=-90]{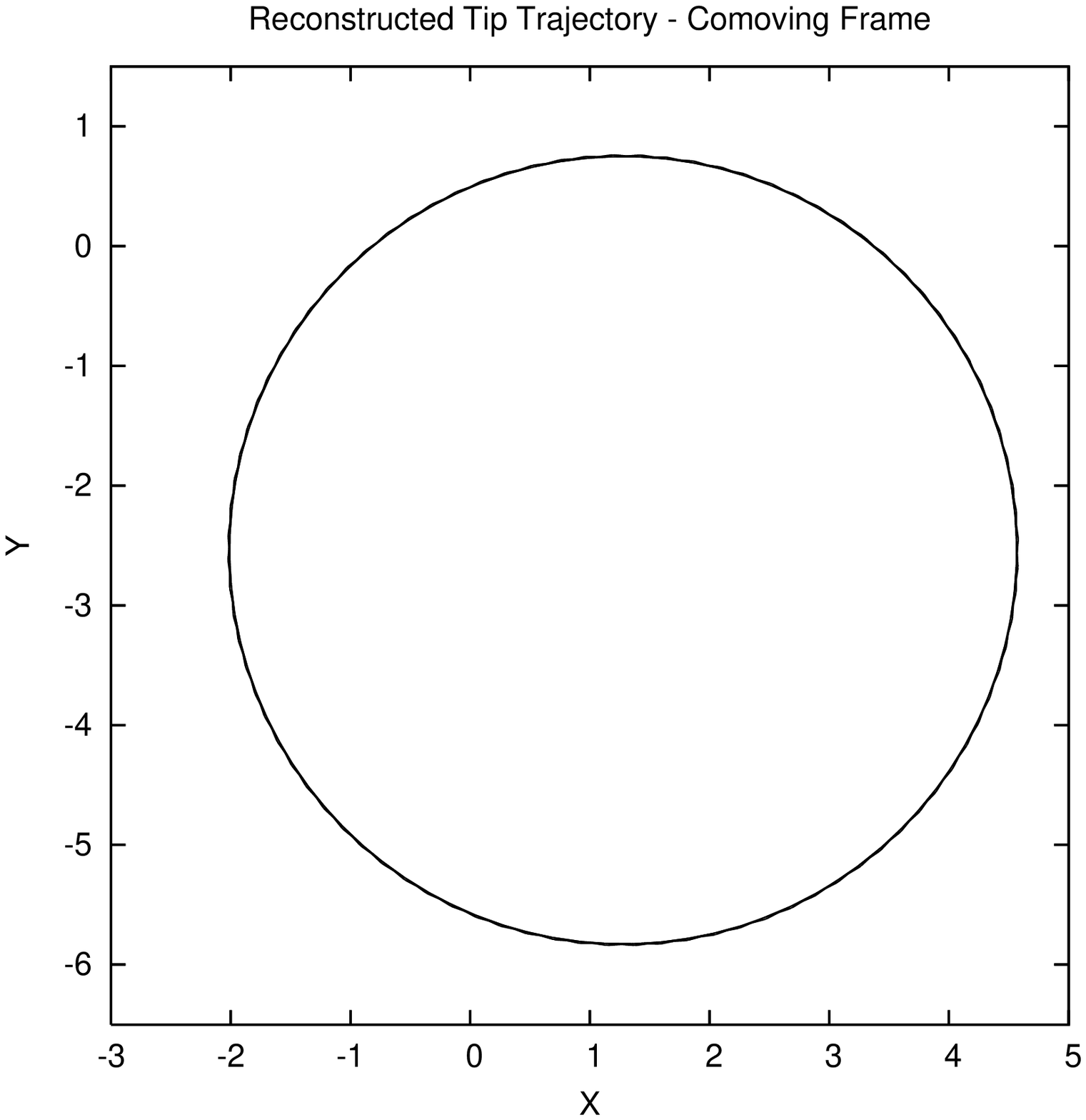}
\end{minipage}
\caption[Rigidly rotating spiral wave tip trajectory reconstruction]{(left) Tip trajectory in the laboratory frame of reference; (right) Reconstructed tip trajectory from quotient data in the comoving frame of reference.}
\label{fig:ezf_rw_tips}
\end{center}
\end{figure}

We will now look at a meandering wave using FHN. We note that there are two different methods of recompiling the program for different kinetics and these methods are described in Sec.(\ref{sec:ezf_Makefile}).

\begin{center}
    \begin{tabular}{ | c | p{7.8cm} |}
    \hline
    Command & Commentary \\ \hline\hline
    \verb|make ezfreeze KINETICS='FHN'| & This makes the program \verb|ezfreeze| using FHN kinetics and interactive graphics.\\ \hline
    \verb|./ezfreeze| & Executes the program \verb|ezfreeze|.\\ \hline
    Bring X-Window to top & Must be done in order to start the simulation and do a variety of other tasks.\\ \hline
    Press \verb|SPACE| bar & Initiates the simulation.\\ \hline
    Press \verb|t| key & Displays the tip path\\
    \hline
    \end{tabular}
\end{center}

This time, let the simulation self terminate, without having advection switched on. Again, we show some snaps of the simulation in Fig.(\ref{fig:ezf_mrw}).

\begin{figure}[btp]
\begin{center}
\begin{minipage}[htbp]{0.24\linewidth}
\centering
\includegraphics[width=0.7\textwidth]{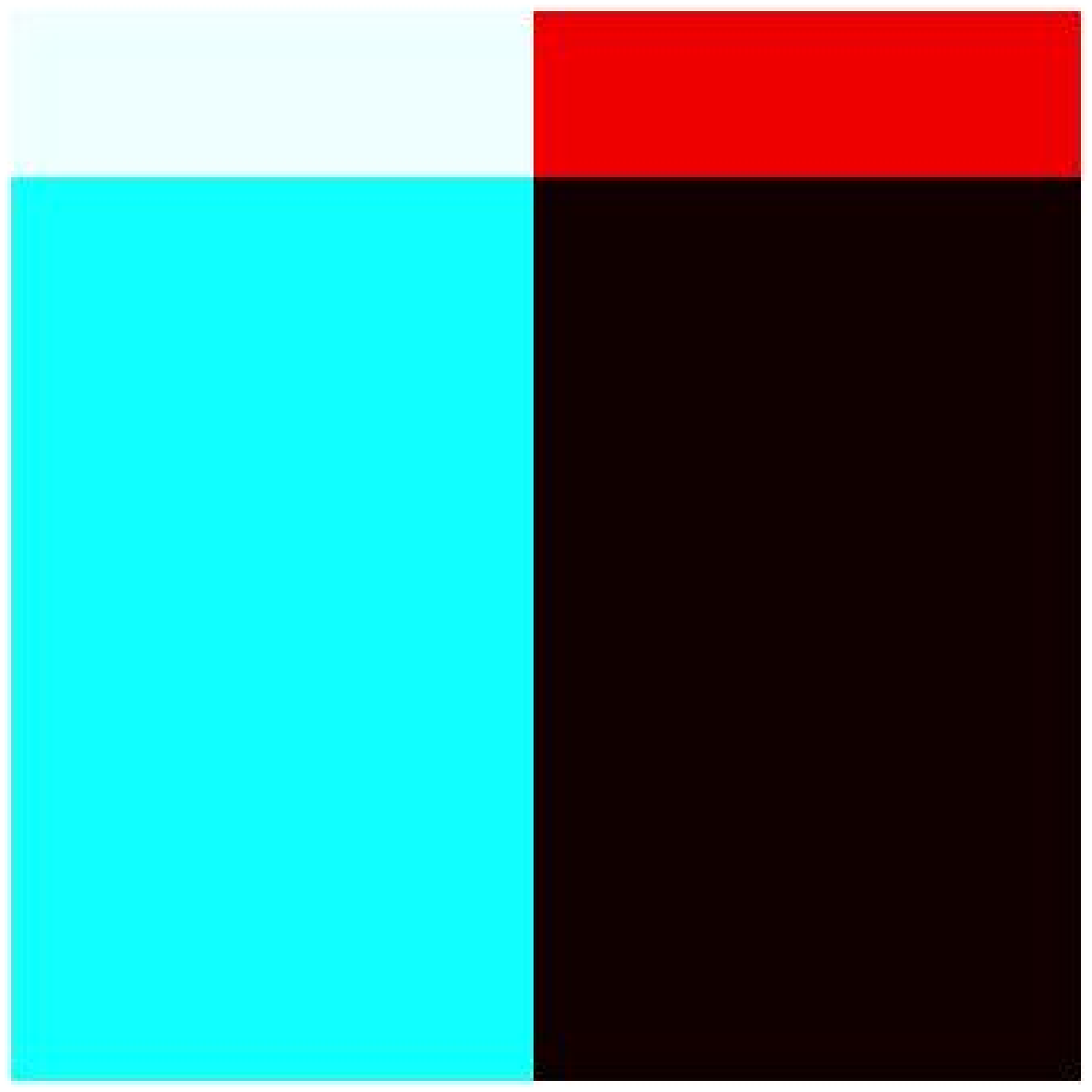}
\end{minipage}
\begin{minipage}[htbp]{0.24\linewidth}
\centering
\includegraphics[width=0.7\textwidth]{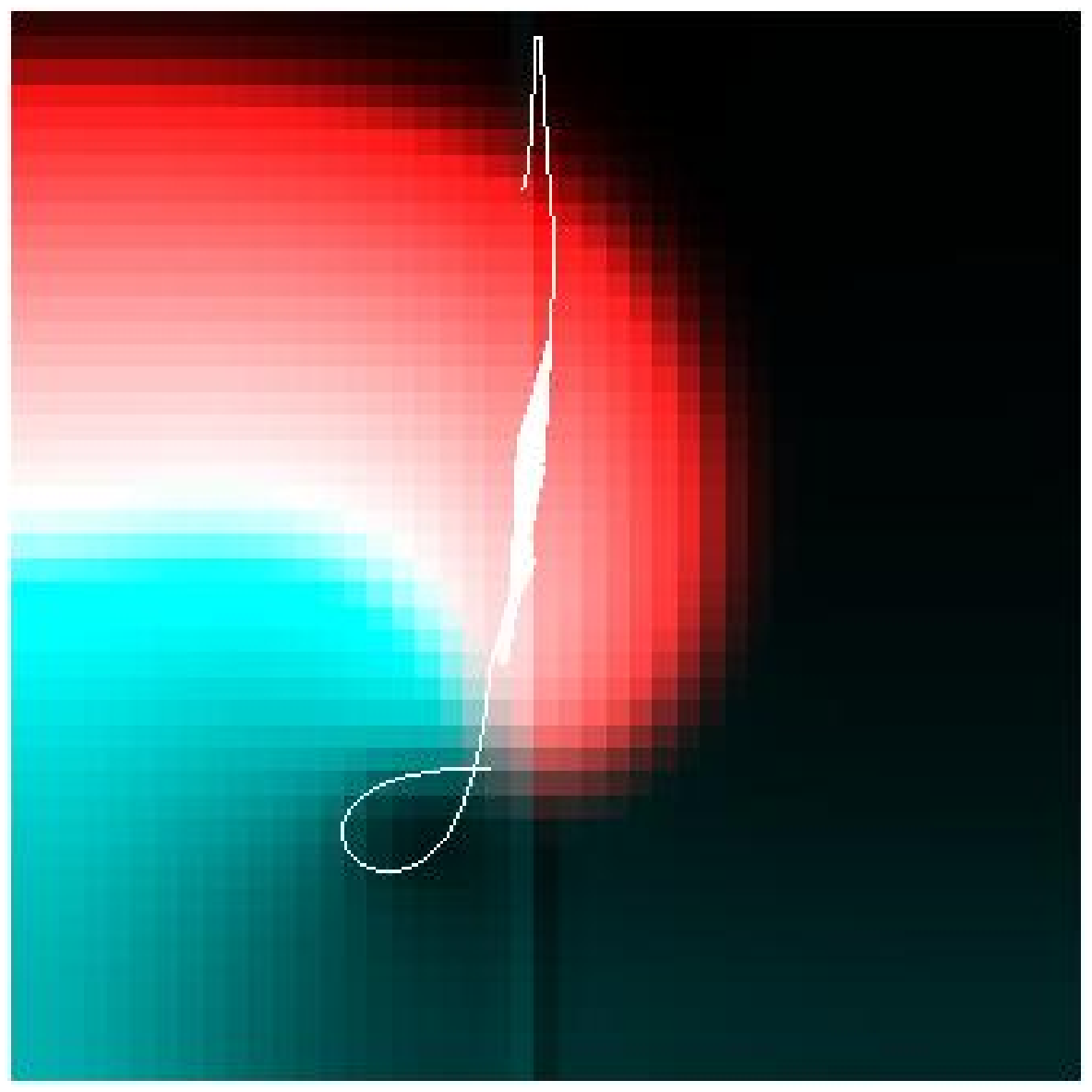}
\end{minipage}
\begin{minipage}[htbp]{0.24\linewidth}
\centering
\includegraphics[width=0.7\textwidth]{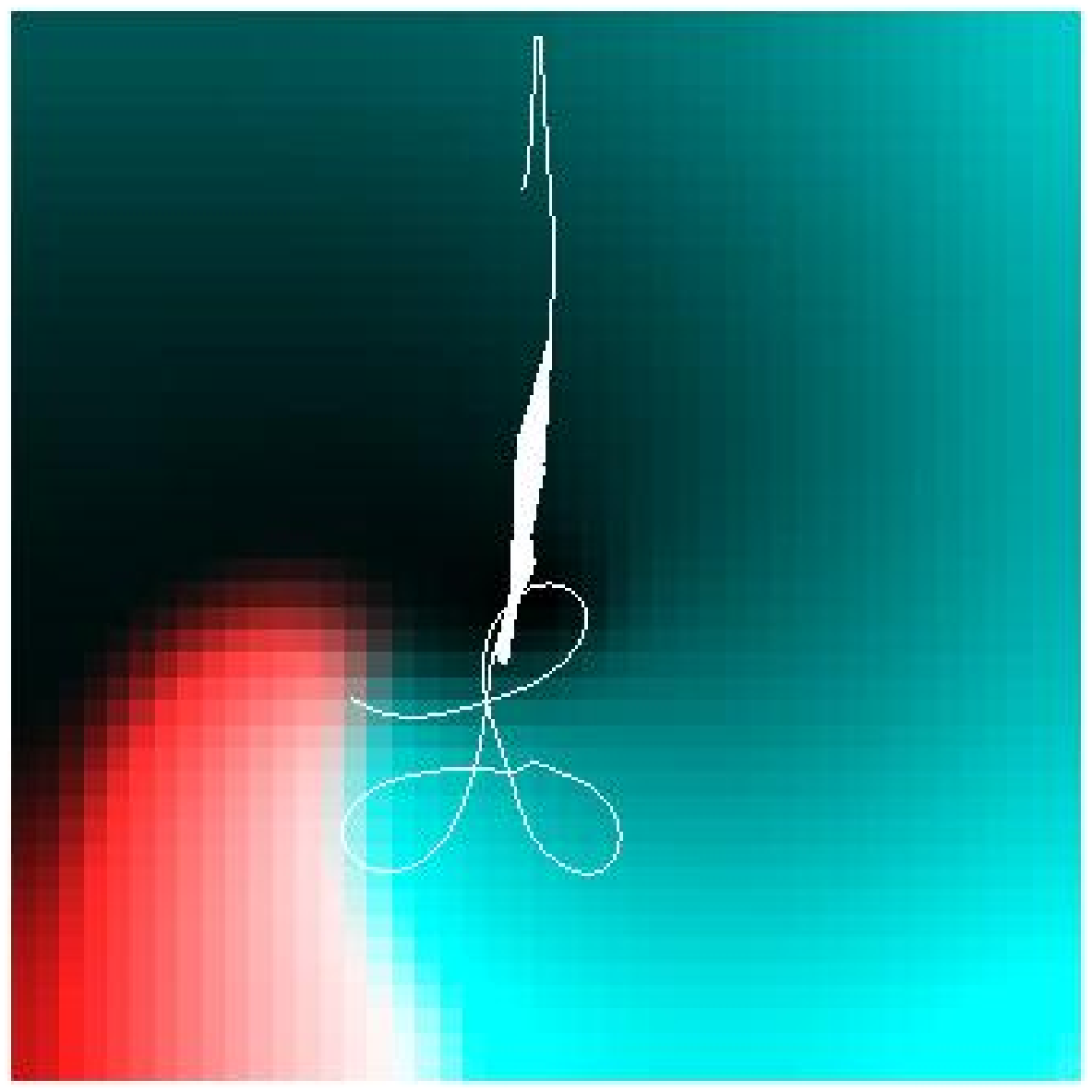}
\end{minipage}
\begin{minipage}[htbp]{0.24\linewidth}
\centering
\includegraphics[width=0.7\textwidth]{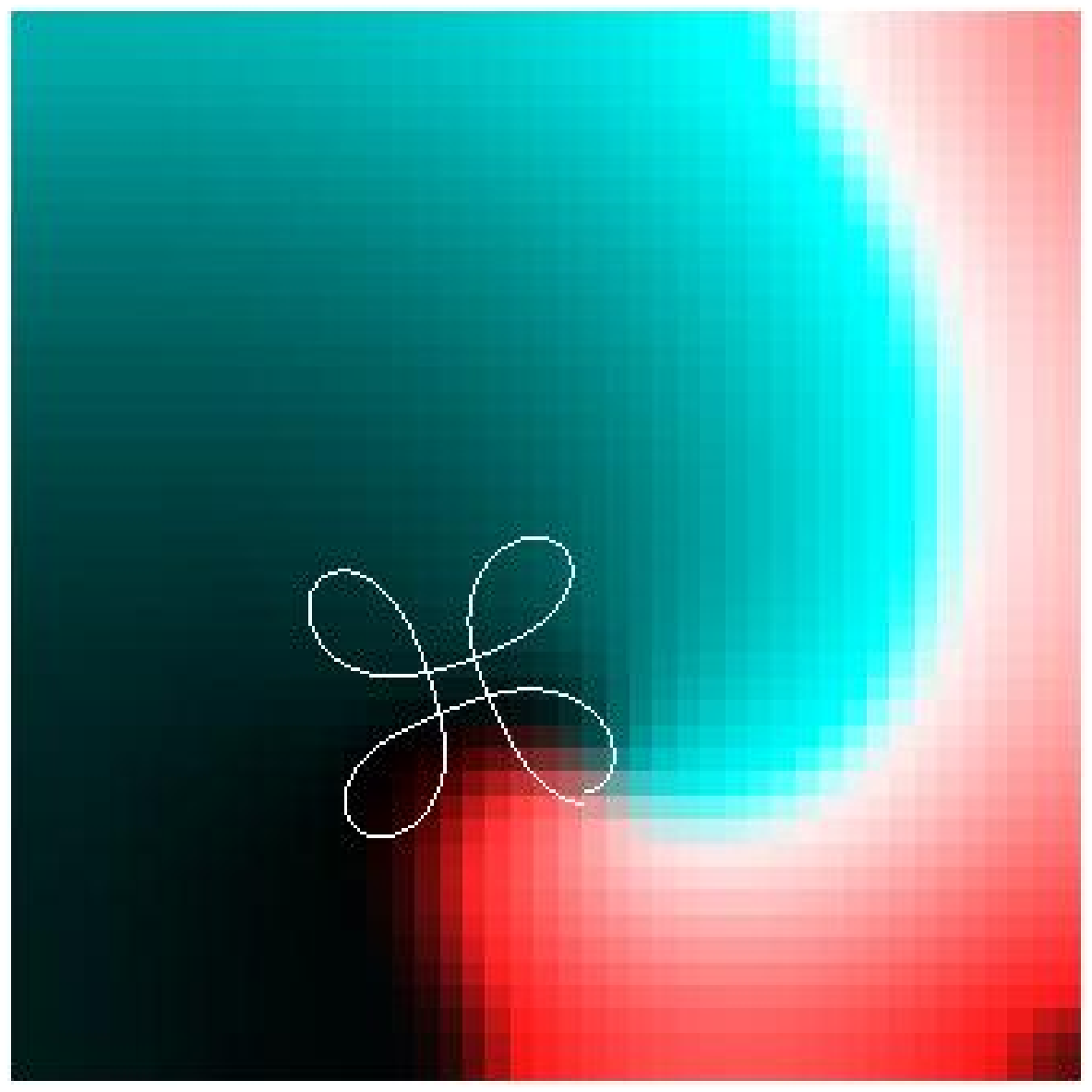}
\end{minipage}
\caption[A meandering spiral wave]{Snap snots from the simulation of a meandering spiral wave in a laboratory frame of reference, with the simulation starting from the left.}
\label{fig:ezf_mrw}
\end{center}
\end{figure}

We will now generate initial conditions by using the final conditions from the last run (saved in \verb|fc.dat|). Also, we will copy the tip file to another file so that the data generated (which we will use to plot the original tip trajectory), is not overwritten.

\begin{center}
    \begin{tabular}{ | c | p{8.9cm} |}
    \hline
    Command & Commentary \\ \hline\hline
    \verb|cp fc.dat ic.dat| & Copies the final conditions of the previous simulation to the initial conditions for future simulations.\\ \hline
    \verb|cp tip.dat tip_fhn.dat| & Copies the tip data file to another data file.\\ \hline
    \verb|./ezfreeze| & Executes the program \verb|ezfreeze|.\\ \hline
    Bring X-Window to top & Must be done in order to start the simulation and do a variety of other tasks.\\ \hline
    Press \verb|SPACE| bar & Initiates the simulation.\\ \hline
    Press \verb|z| key & Switches on advection (moves to a comoving frame of reference).\\ 
    \hline
    \end{tabular}
\end{center}

Allow the program to self terminate. Now integrate the quotient data and compare the reconstructed tip trajectory to the original tip in the laboratory frame:

\begin{center}
    \begin{tabular}{ | c | p{4.4cm} |}
    \hline
    Command & Commentary \\ \hline\hline
    \verb|./int quot.dat 10| & Executes the program \verb|int| on the file \verb|quot.dat| starting from the file's $10^{{\mbox{\tiny {th}}}}$ line (the last argument is needed to eradicate any peculiarities in the initial transient).\\ \hline
    \verb|gnuplot| & Open Gnuplot.\\ \hline
    \verb|p 'integrated_quot.dat' ev ::10 u 5:6 w lp,\| & plots the reconstructed tip trajectory\\
    \verb|'tip_fhn.dat' ev ::10000 u 2:3 w lp| & and also the original tip trajectory. These can then be compared.\\
    \hline
    \end{tabular}
\end{center}

You should get two flower patterns produced which are very similar. We show these pictures in Fig.(\ref{fig:ezf_mrw_tips}).

\begin{figure}[btp]
\begin{center}
\begin{minipage}[htbp]{0.49\linewidth}
\centering
\includegraphics[width=0.7\textwidth, angle=-90]{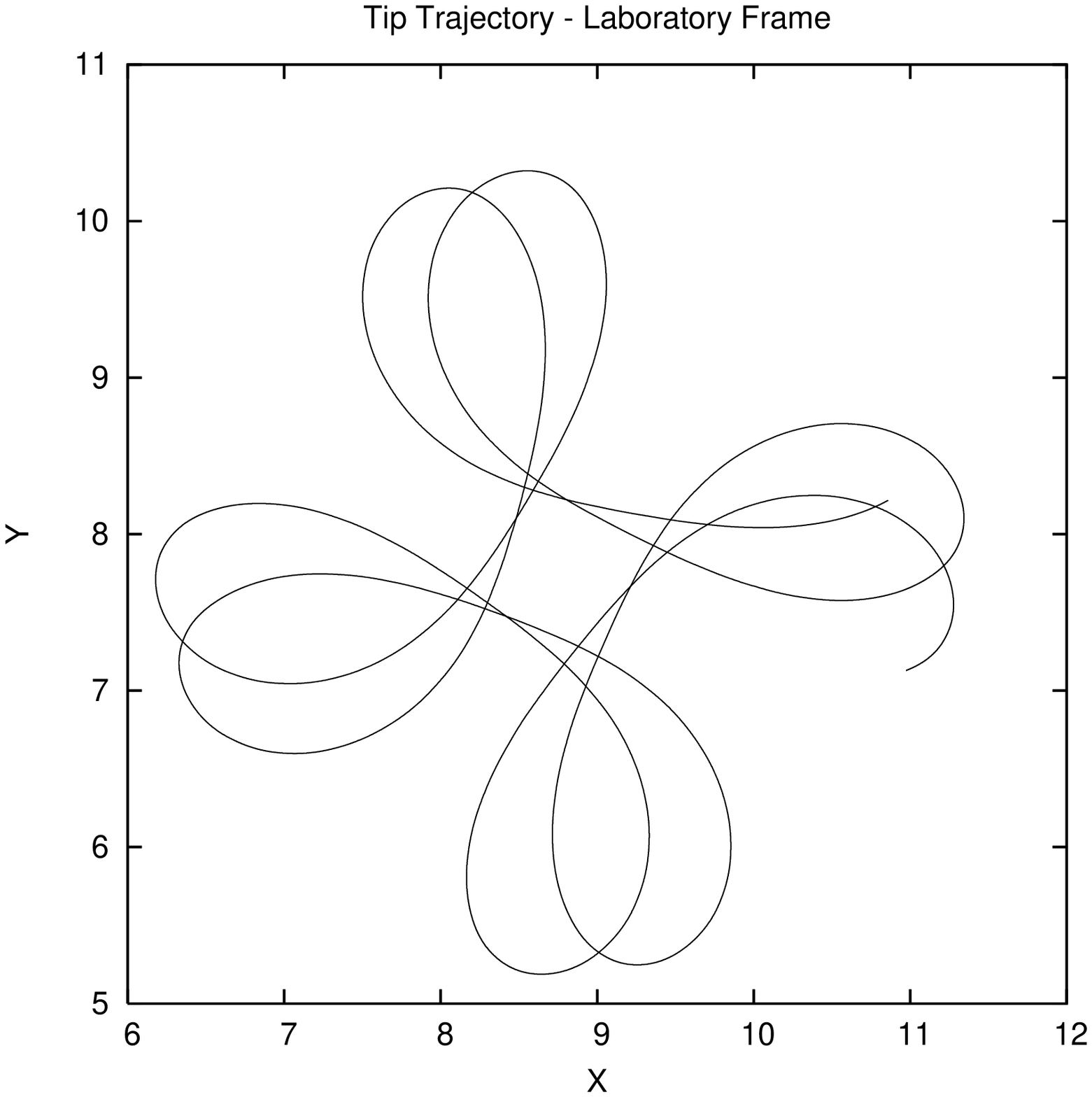}
\end{minipage}
\begin{minipage}[htbp]{0.49\linewidth}
\centering
\includegraphics[width=0.7\textwidth, angle=-90]{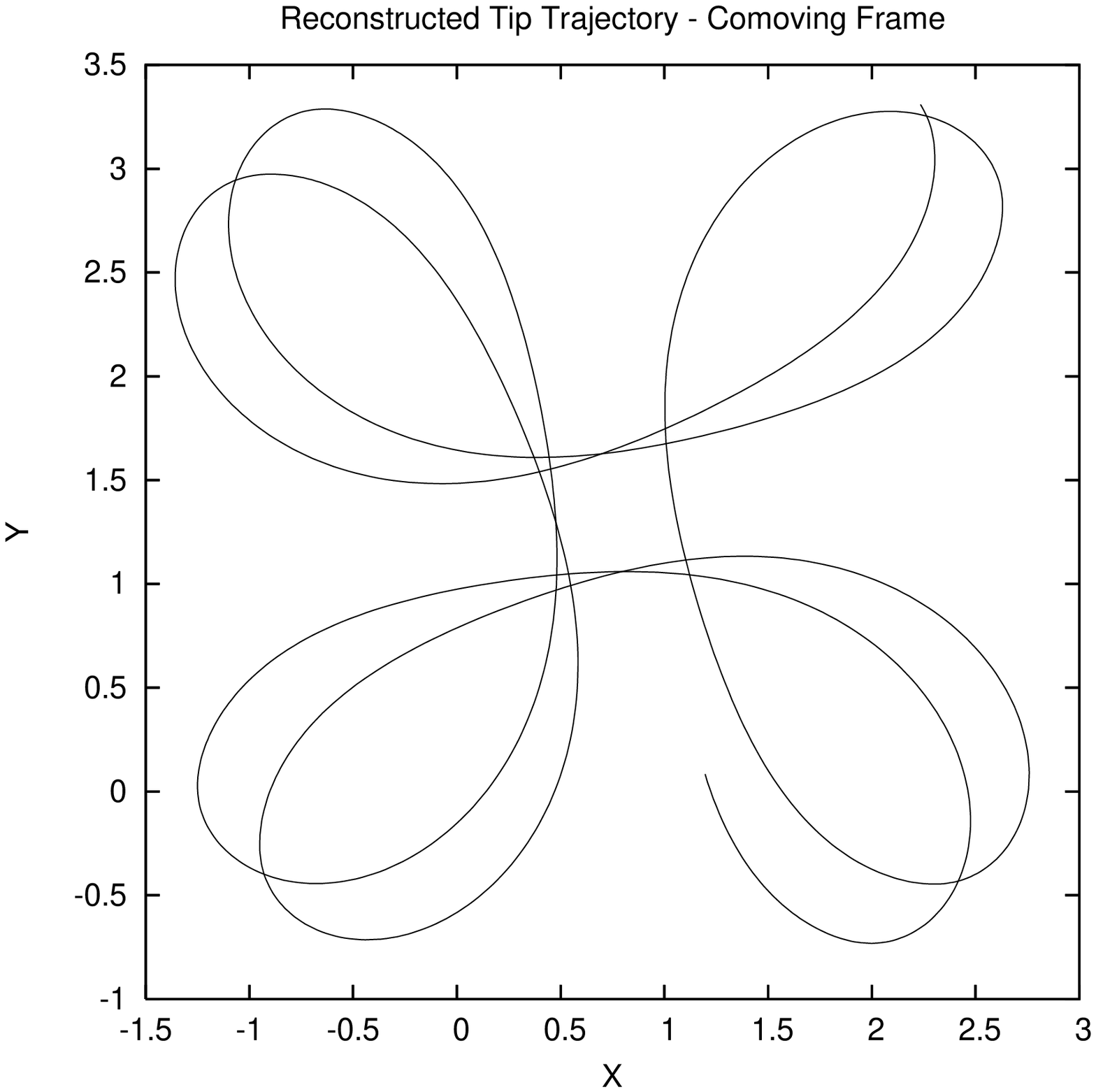}
\end{minipage}
\caption[Meandering spiral tip trajectory reconstruction]{(left) Tip trajectory in the laboratory frame of reference; (right) Reconstructed tip trajectory from quotient data in the comoving frame of reference.}
\label{fig:ezf_mrw_tips}
\end{center}
\end{figure}

We noted at the beginning of this section that we have chosen parameters which will illustrate just how fast this program can be. However, as we have seen in Figs.(\ref{fig:ezf_rw_tips})\&(\ref{fig:ezf_mrw_tips}) the tip trajectories can be different. In order to reduce this difference, we refine the numerical parameters. Open the file \verb|task_fhn.dat| for amending using your favourite editor and change the following:

\begin{center}
    \begin{tabular}{ | c | c | c |}
    \hline
    Parameter 			& Change From 	& Change To \\ \hline\hline
    Box Size (\verb|Lx|) 	& 15 		& 20\\ \hline
    Grid points (\verb|Nx|) 	& 46 		& 101\\ \hline
    Timestep ratio (\verb|ts|) 	& 0.2 		& 0.1\\ \hline
    Time steps per plot 	& 10 		& 200\\ \hline
    \end{tabular}
\end{center}

Now, repeat the simulation for the meandering spiral, starting from but not including the step when we made the program with FHN kinetics (no need to make the program again).

So, the above changes have reduced the space and time step, and also, by increasing the time steps per plot, we have reduced the amount of routines called compared to the original simulations. This results in a slower simulation but the results are more accurate. You should observe this when you plot the two trajectories.


\section{How it works - The Users Manual}

This section will detail how to perform certain tasks using \ez. We will not go into the detail on how the code actually works (this is covered in the next section) but just simply how to preform a task.

\subsection{Task Files}

There are two task files which hold the important parameters, viz. \verb|task_fhn.dat| for FHN kinetics and \verb|task_bark.dat| for Barkley Kinetics. The exact details will not be discussed fully in this section since the files themselves are very well commented.

{\bf NOTE}: For the remainder of this report, we will shorten the names of the task files to
 \verb|task_kin.dat|.

\subsection{Key Presses}

Once the program has been executed, there are several key presses that can be issued. It should be noted that in some systems, it is necessary to have the mouse arrow within the X-Window for the key press to take effect. The key presses are detailed in the table below (note - only lower case letters are shown as keys presses; capital key presses also apply here):

\begin{center}
\begin{tabular}{rcl}
key &\vline& action\\
\hline
\verb|space| 	&\vline& starts the simulation\\
\verb|p| 	&\vline& pause the simulation (restarts when \verb|space| is pressed)\\
\verb|q|     	&\vline& quits the simulation and saves the final conditions data\\
\verb|esc|   	&\vline& quits the simulation but doesn't save the final conditions data\\
\verb|u| 	&\vline& show the $u$ field\\
\verb|v| 	&\vline& show the $v$ field\\
\verb|f| 	&\vline& show the phase field (a combination of the $u$ and $v$ fields)\\
\verb|n| 	&\vline& show no field (useful for when you know a spiral wave is present and\\
         	&\vline& just want to see the tip trajectory (saves computational time))\\
\verb|t| 	&\vline& toggle tip on/off: show the tip positions from time of key press\\
\verb|z| 	&\vline& toggles advection terms on/off\\
\verb|b| 	&\vline& toggles between Neumann and Dirichlet Boundary Conditions\\
$\uparrow$, $\downarrow$, $\rightarrow$, $\leftarrow$ &\vline& manually shifts the spiral in a particular direction\\
\verb|s| 	&\vline& saves the current image on screen to a \verb|png| file\\
\verb|i| 	&\vline& saves a series of images, with each image saved to a \verb|png| file. \\
		&\vline& Starts when i is pressed and ends when either is pressed again or the\\
		&\vline& simulation is brought to an end.
\end{tabular}
\end{center}

\subsection{Output Files}

\subsubsection{tip.dat}

This file contains all the data relating to the tip position and orientation. There are four columns within the file relating to (from left to right) Time, $X$ (tip position in the $x$ direction), $Y$ (tip position in the $y$ direction), $\Theta$ (tip orientation).

\subsubsection{quot.dat}

This file contains the values of the advection coefficients (the quotient system) $c_x$, $c_y$ and $\omega$. It is organised as four columns as (from left to right) time, $c_x$, $c_y$ and $\omega$.

\subsubsection{integrated\_quot.dat}

This file is generated when the program \verb|int| is executed and contains all the data from the reconstruction of the tip trajectory from the quotient data. There are seven columns relating to (from left to right) Time, $c_x$, $c_y$ and $\omega$, $X$ (reconstructed tip position in the $x$ direction), $Y$ (reconstructed tip position in the $y$ direction), $\Theta$ (reconstructed tip orientation).

Note that the name of the file is determined by the file being integrated. For instance, if the \verb|quot.dat| file was renamed to \verb|quotient.dat| for instance, then the command \verb|int quotient.dat 10| would generate a file called \verb|integrated_quotient.dat|.

\subsubsection{history.dat}

This is used to track the values of $u$ and $v$ at a particular grid point and contains three columns time, $u$ and $v$.

\subsubsection{fc.dat}

This file contains a lot of important information. The first two lines contains the model parameters used in that simulation, and also the numerical parameters. The third line gives the date and time the file was formed. Line 4 contains the version of program being used. Lines 5-10 are intentionally blank. Lines 11 onward contain the $u$ and $v$ values of each grid point starting from bottom left and working right and upwards.

\subsection{Change Kinetics \& switch graphics on or off}
\label{sec:ezf_Makefile}

Changing both the kinetics and whether the program is run in the graphics mode is done via the \verb|Makefile|, so no need to amend the code. There are two different ways to do this according to your preference:

\subsubsection{Method 1: Command Line Compilation}

The Makefile contains two important macros which indicate which kinetics to use (\verb|KINETICS|) or whether graphics are to be used (\verb|GRAPHICS|). \verb|KINETICS| takes the values \verb|Barkley| or \verb|FHN|, which obviously compile the program using Barkley's or FHN's kinetics respectively. Whilst \verb|GRAPHICS| takes the values 1 for interactive graphics, or 0 for no graphics.

So, we can take advantage of overriding the values of the macros within the \verb|Makefile| via the command line as shown in the following table

\begin{center}
\begin{tabular}{rcl}
command &\vline& program compiled with:\\
\hline
\verb|make ezfreeze KINETICS='Barkley' GRAPHICS='1'| &\vline& Barkley and Interactive\\
&\vline& Graphics\\
\verb|make ezfreeze KINETICS='Barkley' GRAPHICS='0'| &\vline& Barkley and No Graphics\\
\verb|make ezfreeze KINETICS='FHN' GRAPHICS='1'|     &\vline& FHN and Interactive\\
&\vline& Graphics\\
\verb|make ezfreeze KINETICS='FHN' GRAPHICS='0'|     &\vline& FHN and No Graphics\\
\end{tabular}
\end{center}

If the command \verb|make ezfreeze| is issued, then the program is compiled with the value of the macros set within the \verb|Makefile|. So, by default, the program comes with the macros set at \verb|KINETICS=Barkley| and \verb|GRAPHICS=1|, i.e. Barkley kinetics and interactive graphics.

Also, if the user is not confident to amend the \verb|Makefile| to change the C compiler macros (\verb|CC|) from the default \verb|CC=cc| to their systems own compiler, for example to \verb|gcc|, then this can be done as:

\begin{center}
\verb|make ezfreeze KINETICS='Barkley' GRAPHICS='1' CC=gcc|
\end{center}

\textbf{NOTE}: To run \ez with no interactive graphics, you must already have a set of initial conditions. Therefore, it is recommended that you run the program with graphics, switch on advection, let it settle down (after an initial transient period) and copy the final conditions file to initial conditions (\verb|cp fc.dat ic.dat|).

\subsubsection{Method 2: Amending the Values of the Macros within the Makefile}

So, the second way is to physically change the values of the macros mentioned in method 1 to the values the user requires.

Open the \verb|Makefile| using your favourite text editor. To compile the program with Barkley kinetics, line 26 should be uncommented (i.e. no '\#' should be at the beginning of the line), and make sure that line 27 is commented out (i.e. there is a '\#' at the start of the line). Save and close the file and remake the program by issuing the command \verb|make ezfreeze|.

To change the graphics, you will need to change the value of \verb|GRAPHICS| in line 19 to 1 for interactive graphics or 0 for no graphics. Save and close the file and then issue the command \verb|make ezfreeze|.

\subsection{Initial Conditions}

To generate initial conditions from scratch, firstly remove any current initial conditions file (either rename the initial conditions file (e.g. \verb|mv ic.dat ic_new.dat|) or remove it completely (\verb|rm ic.dat|) - the first option is recommended if there is any chance you may be using that file again). Start \ez by issuing the command \verb|./ezfreeze| and pressing \verb|space|. You should see your spiral forming, obviously depending on your choice of parameters.

Ideally, you should try and get your spiral tip rotating around the center of the box (keeping it away from the boundaries). Therefore, you can adjust the initial position of the cross stimulation fields by one of two methods. You can firstly issue \verb|./ezfreeze| and then use the arrow keys to move the fields, such that the wave ends up, once started, rotating about a point close to the center, or you can use the parameters in the files \verb|task_kin.dat| in line 22 to move the field prior to the program being executed.

The next way to use initial conditions is to use the final conditions from the previous run as initial conditions for the next. This is done by issuing the command \verb|cp fc.dat ic.dat|.

\subsection{Change Box Size}

The box size is amended in \verb|task_kin.dat|. The parameters are called \verb|Lx| and \verb|Ly| and are located on line 4. Also, to perserve a particular spacestep, you will need to change \verb|Nx| and \verb|Ny| accordingly.

\subsection{Change Timestep}

The timestep can be changed by amending \verb|ts| in the \verb|task_kin.dat| file on line 8.

\subsection{History File}

If you feel that you need to observe a particular feature of the spiral wave by observing how $u$ and $v$ change over time, then you can activate the history file. This is done within \verb|task.dat|. 

Open up \verb|task_kin.dat|. On line 15, you can specify how many time steps per write to the history file (called \verb|history.dat|). If this is zero then no history file is written. Also, you can change the point which is being used as the history point on line 16.

\subsection{Changing Boundary Conditions}

This is done via pressing \verb|b|. The default BC's are set via the \verb|task_kin.dat| file  on line 23.

\subsection{Change Position of Second Pinning Point}

Within \verb|task_kin.dat| (on line 17) there are of parameters called \verb|x_inc| and \verb|y_inc|. Adjusting the values of these will change the position of the second pinning point.

{\bf WARNING}: Setting \verb|x_inc| and \verb|y_inc| to zero simultaneously will give a division by zero if advection is switched on (i.e. by pressing \verb|z|). When the program is executed, a warning is displaying on the terminal screen if they are zero. The program will still run in the stationary frame (Laboratory Frame) of reference, but will terminate once advection is switched on.

\subsection{Change Orientation of Tip in the Comoving Frame of Reference}

This follows on from the preceding subsection (Change Position of Second Pinning Point). By changing the second pinning point, you will actually change the fixed orientation of the tip in the comoving frame of reference. Note the warning in the above section.


\section{How it works - The Programmers Manual}

This section will describe how particular parts of the code work. We firstly detail the main differences between \ez and EZ-Spiral

\subsection{Differences between \ez and EZ-Spiral}

\begin{enumerate}
 \item Kinetics - FHN kinetics have been incorporated into \ez via \verb|ezstep.h|. Also, different parameters are needed for FHN compared to Barkley kinetics and therefore this is incorporated throughout the code.
 \item Boundary Conditions - Dirichlet BC's have been incorporated. Switching between Neumann and Dirichlet BC's is done by pressing \verb|b|.
 \item Advection Calculation - Advection terms are calculated via the function \verb|step1()| found in \verb|ezstep.c|.
 \item Phase Field - this is a combination of the u and v-fields, switched on by pressing \verb|f|.
 \item Output - \verb|tip.dat| now holds the orientation of the tip as well as its spatial coordinates; \verb|quot.dat|, created when advection is calculated, holds the values of the coefficients of the advection terms, i.e. $c_x$, $c_y$ and $\omega$.
 \item Saving Images - the user can now save just a single image to a \verb|png| file (pressing \verb|s|) or a series of images to a series of \verb|png| files (pressing \verb|i|).
 \item Task Files - there are now two task files (one for Barkley kinetics, one for FHN). The program will choose which is necessary upon compilation.
 \item Makefile - there are now options via the Makefile to make the program with a particular set of kinetics and whether interactive graphics are used.
\end{enumerate}

\subsection{Files}

\begin{center}
    \begin{tabular}{ | c | p{5.5cm} |}
    \hline
    File & Description \\ \hline\hline
    \verb|ezfreeze.c| & Contains the main function files, including reading in parameters from task files, allocation of memory, initialising the simulation.\\ \hline
    \verb|ezfreeze.h| & Contains all the global variables needed in the code and some of the major macros.\\ \hline
    \verb|ezstep.c| & Contains procedures to calculate the values of the variables $u(x,y,t)$ and $v(x,y,t)$ for each time step, as well as the boundary conditions\\ \hline
    \verb|ezstep.h| & Contains mainly the algorithms for the kinetics.\\ \hline
    \verb|ezgraphGL.c| & Contains the code for graphical simulations.\\ \hline
    \verb|ezgraphGL.h| & Contains the macros for the graphical simulations.\\ \hline
    \verb|eztip.c| & Contains the procedures to find the tip position and orientation.\\ \hline
    \verb|Makefile| & Makes the program, as well as clean up the files and compress them if the user wishes.\\ \hline
    \verb|task_fhn.dat|,\quad\verb|task_bark.dat| & Task files containing the physical and numerical parameters, as well as some important flags.\\ \hline
    \end{tabular}
\end{center}

It is intended that the user will not only use the program as it is, but will amend it to suit their needs. We therefore detail in the following subsections, the important functions that are used in \ez, including where they are found and what they do.

\subsubsection{Makefile}

The \verb|Makefile| contains several commands to compile \verb|ezfreeze| (the main simulating program) and \verb|int| (which can be used to integrate the quotient data). We refer the reader to Sec.(\ref{sec:ezf_Makefile}) for information on how to compile the program for particular kinetics and choice of graphics.

It is assumed that the user is fluent in UNIX commands and can therefore am mend any of the files in this package to suit their needs - in particular the command in \verb|Makefile| which refers to their systems particular C compiler. However, if the user is not fluent in UNIX commands the user can still compile the program via the command line without having to go into the \verb|Makefile| (see Sec.(\ref{sec:ezf_Makefile})).

The main method which is implemented in the \verb|Makefile| is the way in which the program \verb|ezfreeze| is compiled. The two major macros integrated within \verb|Makefile| are \verb|KINETICS| and \verb|GRAPHICS|, which determine which Kinetics are used and whether to use graphics respectively. The values of these macros are passed to the compilation command line via the macro \verb|CFLAGS| using the preprocessor command \verb|-DFHN=$(KIN)| and \verb|-DGRAPHICS=$(GRAPH)|, where \verb|KIN| is either 1 (for FHN kinetics) or 0 (for Barkley), and \verb|GRAPH| is either 1 (for graphics) or 0 (for no graphics).

It is assumed that the user is fluent in UNIX commands and can therefore amend any of the files in this package to suit their needs - in particular the command in \verb|Makefile| which refers to their systems particular C compiler.

Also in \verb|Makefile| are several cleaning up processes such as \verb|make tidy| (deletes the object files and executable \verb|ezfreeze|) and \verb|make clean| (does what \verb|make tidy| does, plus deletes several other files including any files created when the manual is formed and *.*\ensuremath{\sim} files).

\subsubsection{ezfreeze.c}
\label{sec:ezfreeze_c}

This file contains a number of important functions. The most important function is \verb|main|. Within the main function the whole program is initialised (via \verb|Initialise()|), and then the main loop is executed. Within the loop there are calls of the step functions which calculate the values of $u(x,y,t)$ and $v(x,y,t)$ (see Sec.(\ref{sec:ezstep_c})), and then the drawing routines are called.

Also in the \verb|main| function there is a timer which calculates the time the whole simulation has taken (in real time). This will prove useful when comparing the time taken for particular simulations.

Also, in \verb|ezfreeze.c|, there are various other functions. Two of these functions, \verb|cube()| and \verb|root()|, calculate the steady state in FHN, which are then used in the drawing procedure.

Another function is \verb|Write_quot(float cx, float cy, float om)|, which writes the quotient data ($c_x$, $c_y$ and $\omega$) to a file called \verb|quot.dat|. However, this function is only called when all three components are being calculated (see Sec.(\ref{sec:ezstep_c}) for more details).

Finally, the way that initial conditions are generated via \verb|Generate_ic()| has been amended slightly so that the position of the cross field stimulations can be controlled via the variables found in \verb|task_kin.dat| on line 22.

\subsubsection{ezfreeze.h}
\label{sec:ezfreeze_h}

This file is the main header file, used in all the other files, and contains mainly macros used throughout the program and also declares the global variables.

\subsubsection{ezstep.c}
\label{sec:ezstep_c}

There are three main functions contained in this file.

The first is \verb|step()|. This was originally contained in \verb|ezstep.c| used in EZ-Spiral, and it's purpose is to calculate the values of $u(x,y,t)$ and $v(x,y,t)$ for each time step for the Reaction-Diffusion part of the equation. Operator splitting can be used here (switched on via the macro \verb|SPLIT| found in \verb|ezfreeze.h|).

The next function is \verb|step1()|, which calculates the final values of $u(x,y,t)$ and $v(x,y,t)$ when advection is switched on. For a more detailed analysis, please see Sec.(\ref{sec:maths_num}).

Finally, we have amended the function \verb|Impose_boundary_conditions()| to include Dirichlet boundary conditions as well as Neumann boundary conditions.

\subsubsection{ezstep.h}
\label{sec:ezstep_h}

This relatively short file contains all the macros for the kinetics used in \verb|step()| and \verb|step1()|.

\subsubsection{eztip.c}
\label{sec:eztip_c}

This file is entirely devoted to finding the tip position. It has only been amended slightly from the \verb|eztip.c| used for EZ-Spiral. The main changes are firstly the calculation of the orientation of the tip, and also the variable \verb|MAX_TIPS| has been increased to 10,000,000 to allow for long simulations (this can be increased if needed).

\subsubsection{ezgraphGL.c}
\label{sec:ezgraphGL_c}

There have been several changes to this particular file compared to the file used in EZ-Spiral. The main changes are to the extra key presses. These include key presses to change the boundary conditions (key \verb|b|), toggling the switching on/off of the advection terms (key \verb|z|), toggling between the way the advection coefficients (quotient system) is calculated (key \verb|e|), save images (keys \verb|s| and \verb|i|), and change to a phase field (combination of $u$ and $v$-fields; key press \verb|f|).

Associates with these keys presses are functions which execute the action that the key press is for.

\subsubsection{ezgraphGL.h}
\label{sec:ezgraphGL_h}

Nothing has been amended in this file from that used in EZ-Spiral. This file contains a variety of macros used in the drawing routines of the spiral wave.


\section{Mathematical Background}

The original idea behind \ez was twofold. Firstly, we intended \ez to be used as a tool for which computational time can be significantly reduced. Secondly, for the first time, it can be used as a tool in order to study the limit cycles behind the dynamics of \emph{meandering} spiral waves. There are other uses of \ez such as calculation of the critical eigenvalues of the spiral, which are discussed in \cite{Foul08}.

With regards to the first use, computational time is reduced significantly because the tip of the wave is always located at the center of the computational box. This means that the tip of the wave \emph{never} reaches the boundary of the box. Therefore a smaller box size can be afforded. One of the applications of this technique, as detailed in \cite{Foul08} is to study 1:1 resonance in meandering spiral waves. In \cite{Foul08}, we show a study in which the box size is $L_x=L_y=15$s.u. in the comoving frame with $N_x=N_y=76$ grid points in each direction. But in the laboratory frame of reference, the trajectory should have been $L_x=L_y=700$s.u. with $N_x=N_y=3500$ grid points in each direction. Therefore, we see that the simulation in the comoving frame reduces the computational time significantly.

\subsection{Mathematics behind \ez}

The system of equations that we are solving is a 2 species Reaction-Diffusion-Advection system as shown below:

\begin{eqnarray}
\label{eqn:ezf_rda}
\pderiv{u}{t} &=& \nabla^2u+f(u,v)+(\textbf{c},\nabla)u+\omega\pderiv{u}{\theta}\\
\pderiv{v}{t} &=& \delta\nabla^2v+g(u,v)+(\textbf{c},\nabla)v+\omega\pderiv{v}{\theta}
\end{eqnarray}
\\
where $f(u,v)$ and $g(u,v)$ are the reaction terms (kinetics) of the systems. \ez implements two different types of kinetics - Barkley's and FHN's.

\[\begin{array}{lrclcrcl}
\mbox{Barkley:} & f(u,v) &=& \frac{1}{\epsilon}u(1-u)(u-\frac{v+b}{a}) &,& g(u,v) &=& (u-v)\\
\mbox{FHN:}     & f(u,v) &=& \frac{1}{\epsilon}(u-\frac{u^3}{3}-v)     &,& g(u,v) &=& \epsilon(u+\beta-\gamma v)
\end{array}\]

In both systems, $\delta$ is the ratio of diffusion coefficients (typically taken as $\delta$=0 or $\delta$=1), and the model parameters are $\beta$, $\gamma$ and $\epsilon$ for FHN, and $a$, $b$ and $\epsilon$ for Barkley. In both cases, $\epsilon$ determines the slow and fast times of the systems. We further refer to \cite{Bark94b} and \cite{Winf91} which show parametric portraits for each model fixing one parameter and varying the others. The user of \ez is urged to use these parametric portraits when trying out \ez to determine the parameters for a particular type of spiral wave.

In order to ``pin'' the tip to a particular position and with a particular orientation, we need three conditions (the three are for fixing the position in $x$ and $y$ directions and also the phase angle). We decided that due to the sensitivity of calculating the $\omega$, we would use two different sets of conditions:

\[\begin{array}{rclcrcl}
\mbox{Tip Pinning} &&&\vline& \mbox{Tip Pinning} &&\\
\mbox{\underline{Conditions 1}} &&&\vline& \mbox{\underline{Conditions 2}} &&\\
u(x_{tip},y_{tip}) &=& u_* &\vline& u(x_{tip},y_{tip}) &=& u_*\\
v(x_{tip},y_{tip}) &=& v_* &\vline& v(x_{tip},y_{tip}) &=& v_*\\
\pderiv{u}{x}(x_{tip},y_{tip})  &=& 0 &\vline& u(x_{tip}+x_{inc},y_{tip}+y_{inc}) &=& u_*
\end{array}\]
\\
where $(x_{tip},y_{tip})$ are the coordinates at the desired tip position and $x_{inc}$ and $y_{inc}$ are increments away from desired tip position. We note that in Conditions 2, when $x_{inc}=1$ and $y_{inc}=0$, then we get Conditions 1. We also impose that $x_{inc}\neq0$ and $y_{inc}\neq0$ in Conditions 2 otherwise we will get division by zero when calculating $\omega$.

Also, in \ez we implement these conditions such that the desired tip position is at the center of the medium.

\subsection{Numerical Methods}
\label{sec:maths_num}

Firstly, the calculation of the time derivatives are done using an explicit forward Euler method. Then, the spatial derivatives are done in two parts - the diffusion terms (solved using either a five or nine point Laplacian method - which to use is chosen by amending the macro \verb|NINEPOINT| (0 for five point and 1 for nine point) in \verb|ezfreeze.h|), and the advection terms which are calculated using a second order accurate upwind scheme as shown below:

\begin{eqnarray}
\label{eqn:adv_pos}
\left.\pderiv{u}{x}\right|_{(x,y)} &\approx& \frac{1}{2\Delta x}(-3u(x,y)+4u(x+\Delta x,y)-u(x+2\Delta x,y))\\
\label{eqn:adv_neg}
\mbox{or}\quad \left.\pderiv{u}{x}\right|_{(x,y)} &\approx& \frac{1}{2\Delta x}(3u(x,y)-4u(x-\Delta x,y)+u(x-2\Delta x,y))
\end{eqnarray}
\\
where we determine which scheme is used by looking at the sign of the coefficient in front of the derivative, i.e. if the coefficient is positive, we use scheme (\ref{eqn:adv_pos}) and vice versa. A first order accurate scheme is not used as we found it did not give accurate results. A discussion of this can be found in \cite{Foul08}.

Throughout the code, we use operator splitting, and in particular we will use the following scheme:

\begin{eqnarray}
\label{eqn:ezf_op_split_1}
 u^{n+\frac{1}{2}}_{i,j} &=& u^{n}_{i,j}+\Delta t\mathcal{RD}(u^{n}_{i,j},v^{n}_{i,j})\\
\label{eqn:ezf_op_split_2}
 u^{n+1}_{i,j} &=& u^{n+\frac{1}{2}}_{i,j}+\Delta t\mathcal{AD}(u^{n+\frac{1}{2}}_{i,j},v^{n+\frac{1}{2}}_{i,j})
\end{eqnarray}
\\
where $u^{n}_{i,j}$ is $u$ at the $n$-th time step and at the grid coordinate $(i,j)$, and $\mathcal{RD}$ and $\mathcal{AD}$ are the Reaction-Diffusion and Advection terms respectively, and $\Delta t$ is the time step. Therefore, we note that Eqn.(\ref{eqn:ezf_op_split_1}) is implemented into the program in function \verb|step()|, and Eqn.(\ref{eqn:ezf_op_split_2}) via the function \verb|step1()|. Similar equations also exist for $v$. 

By using operator splitting in this way, we can formulate a system of three linear equations in $c_x$, $c_y$, and $\omega$ which can easily be solved. It turns out that, after using the tip pinning conditions, we get the following linear system:

\begin{eqnarray}
c_x\pderiv{u}{x}+c_y\pderiv{u}{y}                                                                                      &=& \pderiv{u}{t}\\
c_x\pderiv{v}{x}+c_y\pderiv{v}{y}                                                                                      &=& \pderiv{v}{t}\\
c_x\pderiv{\tilde{u}}{x}+c_y\pderiv{\tilde{u}}{y}+\omega\left(\tilde{x}\pderiv{\tilde{u}}{y}-\tilde{y}\pderiv{\tilde{u}}{x}\right) &=& \pderiv{\tilde{u}}{t}
\end{eqnarray}
\\
where
\begin{eqnarray}
u       &=& u^{n}_{\frac{NX}{2},\frac{NY}{2}}\\
v       &=& v^{n}_{\frac{NX}{2},\frac{NY}{2}}\\
\tilde{u} &=& u^{n}_{\frac{NX}{2}+x\_inc,\frac{NY}{2}+y\_inc}\quad x\_inc,y\_inc>1
\end{eqnarray}

These are the main numerical procedures implemented into EZ-Freeze.

Finally, the integrator program \verb|int| solves the following equations:

\begin{eqnarray}
\label{eqn:ezf_tip_coord}
\deriv{R}{t} &=& ce^{i\Theta t}\\
\label{eqn:ezf_tip_orien}
\deriv{\Theta}{t} &=& \omega
\end{eqnarray}
\\
where $R=X+iY$ is the tip position, and $\Theta$ is it's orientation. $c=c_x+ic_y$ is the translational velocity of the tip and $\omega$ is its angular velocity.

Eqns.(\ref{eqn:ezf_tip_coord}) and (\ref{eqn:ezf_tip_orien}) are numerically integrated using the following scheme:

\begin{eqnarray}
\Theta^{n+1} &=& \Theta^n+\Delta_t\omega^n\\
X^{n+1} &=& X^n+\Delta_t(c_x^n\cos(\Theta)-c_y^n\sin(\Theta))\\
Y^{n+1} &=& Y^n+\Delta_t(c_x^n\sin(\Theta)+c_y^n\cos(\Theta))
\end{eqnarray}
\\
where the superscript represents what time step we are at and $\Delta_t$ is the value of the time step increment.

\bibliographystyle{plain}
\bibliography{thesis}
\addcontentsline{toc}{chapter}{Bibliography}

\end{document}